\documentclass[
twoside,openright,titlepage,numbers=noenddot,headinclude,
footinclude=true,cleardoublepage=empty,
BCOR=5mm,paper=a4,fontsize=11pt, 
american, aps,fleqn]{scrreprt}

\input{classicthesis-config}
\usepackage{tcolorbox}
\usepackage{cancel}   

\numberwithin{equation}{chapter}

\newcommand*{\colorboxed}{}
\def\colorboxed#1#{%
	\colorboxedAux{#1}%
}
\newcommand*{\colorboxedAux}[3]{%
	\begingroup
	\colorlet{cb@saved}{.}%
	\color#1{#2}%
	\boxed{%
		\color{cb@saved}%
		#3%
	}%
	\endgroup
}

\usepackage{tcolorbox}
\usepackage{cancel}   
\usetikzlibrary{calc} 

\def\trace#1#2{\mathinner{\text{Tr}_{#2}\left[{#1}\right]}}

\begin{document}
	
	\selectlanguage{american} 
	\pagenumbering{roman} 


\include{Titlepage_v2_CenterAligned_New} 
\cleardoublepage
\include{Titlepage_v2_CenterAligned_UAM} 
\cleardoublepage

\thispagestyle{empty}
\newgeometry{right=3cm,bottom=3cm}

\noindent
\begin{tikzpicture}[remember picture, overlay]
	
	\node[anchor=north west] at ($(current page.north west) + (2.5cm,-7.5cm)$) {%
		\begin{minipage}{10cm} 
			\textbf{Director}: \\[0.05em]
			\vspace{0.5em} 
			
			\noindent
			\begin{tabular}{p{7cm} p{8cm}}
				\hspace{-1.9mm}Dr. Carlos Sánchez Muñoz \dotfill & \hspace{-2.5mm}Instituto de Física Fundamental CSIC, Spain
			\end{tabular}\\[1em]
			
			\noindent
			\textbf{Tribunal}: \\[0.05em]
			\vspace{0.5em} 
			
			\noindent
			\begin{tabular}{p{7cm} p{8cm}}
				\hspace{-1.9mm}Prof. Johannes Feist \dotfill & \hspace{-2.5mm}Universidad Autónoma de Madrid, Spain \\
				\hspace{-1.9mm}Dr. Tomás Ramos \dotfill & \hspace{-2.5mm}Instituto de Física Fundamental CSIC, Spain\\
				\hspace{-1.9mm}Dr. Carlos Antón Solanas \dotfill & \hspace{-2.5mm}Universidad Autónoma de Madrid, Spain \\
				\hspace{-1.9mm}Prof. Erik Gauger \dotfill & \hspace{-2.5mm}Heriot-Watt University, United Kingdom \\
				\hspace{-1.9mm}Prof. Kai Müller \dotfill & \hspace{-2.5mm}Technische Universität München, Germany \\
			\end{tabular}
		\end{minipage}
	};
	
\end{tikzpicture}
\vfill
\vspace{5cm}

\noindent
Alejandro Vivas Viaña.\\
\textit{Nonclassical Driven-Dissipative Dynamics in Collective Quantum Optics,} \\
Cover picture by J. M. William Turner:\\
 \textit{Light and Colour (Goethe's Theory) – The Morning after the Deluge – Moses Writing the Book of Genesis.} 

\restoregeometry             
\cleardoublepage
\thispagestyle{empty}
\phantomsection
\pdfbookmark[1]{Dedication}{Dedication}
\newgeometry{margin=0cm} 

\vspace*{3cm}

\begin{tikzpicture}[remember picture, overlay]
	
	\node[anchor=north west] at ($(current page.north west) + (8cm,-5cm)$) {%
		\begin{minipage}{10cm}
				Have you gentlemen ever considered the question of the true nature of the University?  
				\\[0.5em]
				[...] It is an asylum or---what do they call them now?---a rest home, for the infirm, the aged, the discontent, and the otherwise incompetent. Look at us... we are the University. The stranger would not know that we have so much in common, but we know, don't we? We know well. 
				\\[0.5em]
				[...] And so providence, or society, or fate, or whatever name you want to give it, has created this hovel for us, so that we can go in out of the storm. It's for us that the University exists, for the dispossessed of the world; not for the students, not for the selfless pursuit of knowledge, not for any of the reasons that you hear. 
				\\[0.5em]
				[...] We do no harm, we say what we want, and we get paid for it; and that's a triumph of natural virtue, or pretty damn close to it. 
				\begin{flushright}
				--- \textit{Stoner,} John Williams.
				\end{flushright}
		\end{minipage}
	};
	
	\node[anchor=north west] at ($(current page.north west) + (8cm,-15.5cm)$) {%
		\begin{minipage}{10cm}
				Well now, this is what perplexes me and I cannot grasp adequately for myself, what knowledge---science---actually is? So now can we speak of this? What do you say? Which of us will speak first? 
				\begin{flushright}
				--- \textit{Theaetetus,} Plato.
		     	\end{flushright}
		\end{minipage}
	};
	
\end{tikzpicture}

\restoregeometry                 
\cleardoublepage
\setcounter{page}{1} 
\begingroup
\let\clearpage\relax
\let\cleardoublepage\relax
\pdfbookmark[1]{Agradecimientos}{Agradecimientos}
\chapter*{Agradecimientos}

Cinco años han pasado desde que comencé este periplo académico allá por octubre de 2020.
Usar aquí este término---periplo---puede parecer otra pedantería más de esta tesis; sin embargo, en este caso está más que justificado. 
Durante este tiempo he podido disfrutar de un viaje extraordinario, repleto de personas maravillosas, momentos de alegría, de tristeza y, en ocasiones, incluso de desesperación.
Ha sido mucho más que una travesía académica, ha sido toda una transformación vital. Cada circunstancia, cada experiencia y cada persona a lo largo de estos años ha dejado una marca indeleble, moldeando quién soy hoy y quién seré mañana. Por ello, esta tesis no es sólo mía, sino también de todas aquellas personas que, de una forma u otra, me han acompañado en este tortuoso camino, el cual aún dista mucho de terminar ya que apenas acaba de comenzar.

Es en este momento cuando uno debería dar paso a la \textit{lista infinita} de nombres, mencionando aquella anécdota, aquel momento que te hace recordar la importancia de estas personas. Obviamente, esto es exactamente lo que viene a continuación---tal como manda la tradición---, ya que una tesis no sólo representa un papel de transmisión de conocimiento a futuras generaciones, sino que también pone de manifiesto la relevancia de lo pasado y vivido en lo presente.
Los primeros en esta \textit{lista} son las personas que, con infinita paciencia, han soportado mis reflexiones existenciales y comentarios derrotistas (también momentos de diversión, claro) durante cada día en el despacho y en la universidad. 
Con su sentido del humor y su \textit{guasa} han sabido equilibrar la balanza. Quiero mencionar con especial cariño a Diego, Francisco, Alejandro, Carlos, Ruth, Anna, Miguel, Manuel y Juanjo.
Un agradecimiento especial también a Jacopo, Maksim y Petros (y más recientemente, a Frieder), con quienes no solo he compartido horas de trabajo, sino también momentos esenciales de la vida. 

Agradecer enormemente a toda la gente de QUINFOG por acogerme tan cálidamente durante este último año. Ha sido un tiempo breve pero realmente intenso. 
Quiero mencionar en especial a Carlos, Gabriel, Adrián, Tomás, Miguel, Alberto, Cristian y Paula, quienes soportaron mis \textit{controversias} y ocurrencias diarias con humor (veremos cuánto tiempo tardan en enviarme de vuelta a España).
Durante estos años también he tenido la posibilidad de disfrutar tres meses en la ciudad de Viena, donde tuve la genial oportunidad de trabajar en el grupo del Prof. Peter Rabl y conocer a Íñigo y Rocío. Sin lugar a dudas, sin ellos mi tiempo allí no podría haber sido el mismo. Además, quiero agradecer a la gente de Chalmers, a Ariadna y,  especialmente, a Jiaying Y. y a Simone G., con quienes he tenido la suerte de colaborar y aprender la \textit{física de verdad}, la del laboratorio.

La tesis ha transcurrido a lo largo de cinco largos años, pero el haber llegado hasta aquí, no sólo académicamente sino vitalmente, viene de mucho antes. Todo este viaje no habría sido posible sin mis grandes amigos de la \textit{patria cervantina} (Javier, Daniel, Sofía, Alejandro,  Álvaro y Santiago), los \textit{filósofos} (Roberto, Iago, Pablo y Gonzalo) y \textit{complutenses} (Irián, Eduardo, Gerardo, Germán y Pepe), quienes, entre todos, me ayudaron a expandir mis intereses, mis pasiones, y a construir quien soy ahora. 
Mención especial merecen Luis J. Garay, quien me permitió otear por primera vez qué es la academia, y  Alfredo L., con quien tantas veces he debatido sobre física y la vida. Ambos son responsables de mi querer a la Universidad.

Quiero reservar estas líneas para expresar mi gratitud a Antonio I. F. D., Diego M. C. y Alejandro G. T., cuya brillantez profesional sólo es superada por su calidad personal. Ha sido un privilegio conocerlos, discutir y trabajar con ellos.
Finalmente, no puedo cerrar estos agradecimientos sin mencionar a mi director, Carlos. Gracias por brindarme la oportunidad de formar parte de este \textit{refugio aislado de la tormenta de la realidad}, por compartir tu conocimiento y por nuestras extensas conversaciones. Sin tu guía esta tesis no habría sido posible.

Por último, y más importante aún, quiero dedicar estos agradecimientos a mi familia. 
\\[20pt]
\flushright{Alejandro Vivas Viaña \\
	Madrid, Febrero 2025}

\endgroup
\cleardoublepage



\begingroup
\let\clearpage\relax
\let\cleardoublepage\relax
\pdfbookmark[1]{Abstract}{Abstract}
\chapter*{Abstract}
\chaptermark{Abstract}

This Thesis is devoted to the study of emergent cooperative phenomena in collective light-matter systems.
We consider ensembles of interacting quantum emitters that are coherently driven by a laser field, introducing a hybrid description of light and matter as the emitters become \textit{dressed} by the laser. 
These emitters can also couple to a photonic structure, whose interactions are tailored to harness the collective effects arising from the light-matter dressing. 
The interplay among these three elements---quantum emitters, coherent driving, and photonic structures---constitutes the central theme of this Thesis, revealing a rich landscape of cooperative phenomena.
While the light-matter dressing induced by the laser field has been extensively studied in the context of a single quantum emitter, e.g., for the generation of non-classical states of light, or in the case of emitters with identical natural frequencies, here, we explore collective effects emerging in a system of two non-identical, interacting quantum emitters, which are modelled as two-level systems with unequal transition energies.
In the limit of strong interacting emitters, these form a dimer structure, exhibiting the characteristic superradiant and subradiant states---symmetric and antisymmetric combinations of the single-excitation states---, which are well-known hallmarks of quantum collective behaviour. 
Of particular relevance is the emergence of two-photon effects when the two interacting emitters are coherently driven.  The emitter-emitter interaction enables a nonlinear process that bypasses the single-excitation states and directly connects the ground and doubly-excited states via a two-photon resonance. 
This phenomenon constitutes the backbone of the Thesis, which is dedicated to explore both fundamental aspects and technological applications arising from this collective light-matter dressing and its interaction with structured photonic environments.

Through an analytical study of the system, we reveal how cooperative interactions reshape optical observables---such as the emission intensity, the photon statistics, or the fluorescence spectra---of the light emitted by the emitters, illustrating the intricate interplay of emitter-emitter interaction and coherent driving in the generation of quantum collective effects.
Beyond offering theoretical insights, these findings have direct implications for quantum metrology. 
Due to the high sensitivity of this two-photon effect  to parameters like the inter-emitter distance and the driving strength, it offers a promising route for high-precision parameter estimation and sub-wavelength imaging. 
Motivated by the aim of exploiting two-photon processes to generate stable entanglement via a phenomenologically proposed tailored system, we unveil an unexpected result:
\textit{off-resonant virtual states---conventionally considered unpopulated and just acting as mediators of interactions between quasi-resonant real states---may acquire sizeable population through dissipation in the long-time limit}. 
This unconventional behaviour implies that the regime in which virtual state remains unpopulated is merely metastable,  re-conceptualizing the notion of virtual states in dissipative scenarios. 
To characterize this phenomenon, we develop a hierarchical adiabatic elimination technique that provides a general framework for deriving analytical expressions for both the time evolution of the density matrix elements and the relaxation timescales in metastable open quantum systems.
Next, we proceed to exploit the idea of generating entanglement via two-photon processes by examining a configuration in which emitters are coupled to a lossy cavity.
Within this configuration, we identify five distinct mechanisms leading to both stable and metastable entanglement enabled by the rich interplay among the three main elements: the ensemble of interacting quantum emitters, the light-matter dressing provided by the laser field, and the lossy cavity, providing dissipative-induced processes within the emitters.
Notably, one of the mechanisms---the \textit{frequency-resolved Purcell effect}---is introduced here for the first time, while the others have been already reported in different models. In this particular mechanism, the cavity selectively enhances specific dressed-state energy transitions, thereby stabilizing emitters into superradiant or subradiant states and enabling scalable entanglement generation in multi-emitter systems. 
Our analysis offers a unifying and comprehensive perspective on  all these mechanisms of entanglement generation.
We also address key challenges related to detecting entanglement via signatures in optical observables, such as subradiance in the photon emission or antibunching in the photon statistics, and preserving entanglement in the presence of more realistic dissipative scenarios.
We emphasize that our models are general and apply to a wide diversity of possible implementations. However, by means of our parametrization and choices of parameter values, we place our focus specially on solid-state platforms---such as quantum dots and molecular aggregates---, where cooperative effects hold significant promise for both fundamental research and technological applications.
In these platforms, quantum emitters are artificially designed atoms, typically modelled as two-level systems. 
However, practical implementations face several challenges, including the fabrication of strongly interacting quantum emitters, inhomogeneous broadening---emitters with different resonant frequencies---, and decoherence.
The results presented in this Thesis address these challenges, demonstrating their validity in state-of-the-art solid-state platforms, and thus paving the way for novel quantum technological applications based on harnessing cooperative light-matter effects.

\endgroup

\cleardoublepage

\begingroup
\let\clearpage\relax
\let\cleardoublepage\relax
\pdfbookmark[1]{Resumen}{Resumen}
\chapter*{Resumen}
\chaptermark{Resumen}

Esta Tesis está dedicada al estudio de fenómenos cooperativos emergentes en sistemas colectivos de luz y materia.
Consideramos conjuntos de emisores cuánticos interactuantes que son excitados de forma coherente por un láser, introduciendo una descripción híbrida de luz y materia ya que los emisores se “visten”  con el láser.
Estos emisores también pueden acoplarse a una estructura fotónica, cuyas interacciones se diseñan para aprovechar los efectos colectivos que surgen del “vestimiento” luz-materia.
La interacción entre estos tres elementos—emisores cuánticos, excitación coherente y estructuras fotónicas—constituye el tema central de esta Tesis, revelando un amplio abanico de fenómenos cooperativos.

Mientras que el “vestimiento” luz-materia inducido por el láser ha sido ampliamente estudiado en el contexto de un solo emisor cuántico, por ejemplo, para la generación de estados no clásicos de la luz, o en el caso de emisores con frecuencias naturales idénticas, aquí exploramos los efectos colectivos emergentes en un sistema de dos emisores cuánticos interactuantes y no idénticos, los cuales se describen como sistemas de dos niveles con energías de transición desiguales.
En el límite de emisores fuertemente interactuantes, éstos forman una estructura de dímero, exhibiendo los característicos estados superradiantes y subradiantes---combinaciones simétricas y antisimétricas de los estados de una excitación---, los cuales son signos distintivos bien conocidos del comportamiento colectivo cuántico.
De particular relevancia es la aparición de efectos de dos fotones cuando los dos emisores interactuantes son excitados de forma coherente. La interacción entre emisores permite un proceso no-lineal que evita los estados de una excitation y conecta directamente los estados fundamental y doblemente excitado a través de una resonancia de dos fotones.
Este fenómeno constituye la columna vertebral de la Tesis, la cual se dedica a explorar tanto aspectos fundamentales como aplicaciones tecnológicas derivadas de este “vestimiento” colectivo de luz-materia y su interacción con entornos fotónicos estructurados.

A través de un estudio analítico del sistema, revelamos cómo las interacciones cooperativas remodelan los observables ópticos---tales como la intensidad de emisión, la estadística de fotones o los espectros de fluorescencia---de la luz emitida por los emisores, ilustrando la compleja interrelación entre la interacción emisor-emisor y la excitación coherente en la generación de efectos colectivos cuánticos.
Más allá de ofrecer conocimientos teóricos, estos resultados tienen implicaciones directas en la metrología cuántica.
Debido a la alta sensibilidad de este efecto de dos fotones a parámetros como la distancia entre emisores y la fuerza de la excitación, éste ofrece una vía prometedora para la estimación de parámetros de alta precisión y para la obtención de imágenes a sublongitud de onda.
Motivados por el objetivo de aprovechar procesos de dos fotones para generar entrelazamiento estable a través de un sistema ajustado propuesto fenomenológicamente, revelamos un resultado inesperado: \textit{los estados virtuales fuera de resonancia---convencionalmente considerados como no poblados y que actúan únicamente como mediadores de interacciones entre estados reales cuasi-resonantes---pueden adquirir una población considerable a través de la disipación en el estado estacionario}.
Este comportamiento poco convencional implica que el régimen en el que el estado virtual permanece sin poblar es meramente metastable, re-conceptualizando la noción de estados virtuales en escenarios disipativos.
Para caracterizar este fenómeno, desarrollamos una técnica de eliminación adiabática jerárquica que proporciona un marco general para derivar expresiones analíticas tanto para la evolución temporal de los elementos de la matriz densidad como para las escalas de tiempo de relajación en sistemas cuánticos abiertos metastables.
A continuación, procedemos a explotar la idea de generar entrelazamiento mediante procesos de dos fotones examinando una configuración en la que los emisores están acoplados a una cavidad disipativa.
Dentro de esta configuración, identificamos cinco mecanismos distintos que conducen tanto a entrelazamiento estable como metastable, posibilitado por la intercorrelación entre los tres elementos principales: el conjunto de emisores cuánticos interactuantes, el “vestimiento” luz-materia proporcionado por el láser, y la cavidad disipativa, que induce procesos disipativos en los emisores.
Cabe destacar que uno de los mecanismos---el \textit{efecto Purcell con resolución en frecuencia}---se introduce aquí por primera vez, mientras que los otros ya han sido reportados en diferentes modelos. En este mecanismo en particular, la cavidad aumenta selectivamente transiciones energéticas específicas entre estados vestidos, estabilizando así a los emisores en estados superradiantes o subradiantes y permitiendo la generación de entrelazamiento escalable en sistemas con múltiples emisores.
Nuestro análisis ofrece una perspectiva unificadora y detallada sobre todos estos mecanismos de generación de entrelazamiento.
También tratamos desafíos clave relacionados con la detección del entrelazamiento a través de patrones en observables ópticos, tales como la subradiancia en la emisión de fotones o la \textit{anti-coincidencia} en la estadística de fotones, así como la preservación del entrelazamiento en presencia de escenarios disipativos más realistas.

Enfatizamos que nuestros modelos son generales y pueden aplicarse a una gran variedad de implementaciones posibles. Sin embargo, mediante nuestra parametrización y elección de valores de parámetros, centramos especialmente nuestro enfoque en plataformas de estado sólido---tales como puntos cuánticos y agregados moleculares---, donde los efectos cooperativos tienen un potencial significativo tanto para la investigación fundamental como para aplicaciones tecnológicas.
En estas plataformas, los emisores cuánticos son átomos diseñados artificialmente, típicamente modelados como sistemas de dos niveles.
No obstante, las implementaciones prácticas enfrentan varios desafíos, incluyendo la fabricación de emisores cuánticos fuertemente interactuantes, el ensanchamiento inhomogéneo---emisores con diferentes frecuencias resonantes---, y la decoherencia.
Los resultados presentados en esta Tesis abordan estos desafíos, demostrando su validez en plataformas actuales de estado sólido, y abriendo así el camino para nuevas aplicaciones tecnológicas cuánticas basadas en el aprovechamiento de efectos cooperativos luz-materia.

\endgroup

\cleardoublepage
\pagestyle{scrheadings}    

\cleardoublepage
\pagestyle{scrheadings}
\pdfbookmark[1]{\contentsname}{tableofcontents}
\setcounter{tocdepth}{2} 
\setcounter{secnumdepth}{3} 
\markboth{\spacedlowsmallcaps{\contentsname}}{\spacedlowsmallcaps{\contentsname}}
\tableofcontents
\automark[section]{chapter}
\renewcommand{\chaptermark}[1]{\markboth{\spacedlowsmallcaps{#1}}{\spacedlowsmallcaps{#1}}}
\renewcommand{\sectionmark}[1]{\markright{\textsc{\thesection}\enspace\spacedlowsmallcaps{#1}}}
\clearpage
\begingroup
    \let\clearpage\relax
    \let\cleardoublepage\relax
    \pdfbookmark[1]{\listfigurename}{lof}
    \listoffigures

    \vspace{10ex}
	
	\newpage
    \pdfbookmark[1]{\listtablename}{lot}
    \listoftables


\endgroup

\cleardoublepage

\pagenumbering{arabic}

\frenchspacing 
\raggedbottom 

\cleardoublepage
\setcounter{page}{1}
\chapter{Introduction}
\label{chapter:introduction}

	\section{The role of light in science and philosophy}

	The study of light has been a source of fascination from the earliest days of human existence until the most recent times of our era. Throughout history, light has enchanted humanity, becoming a symbol of wonder, knowledge, and creation. 
	
\begin{SCfigure}[1][b!]
		\includegraphics[width=0.475\columnwidth]{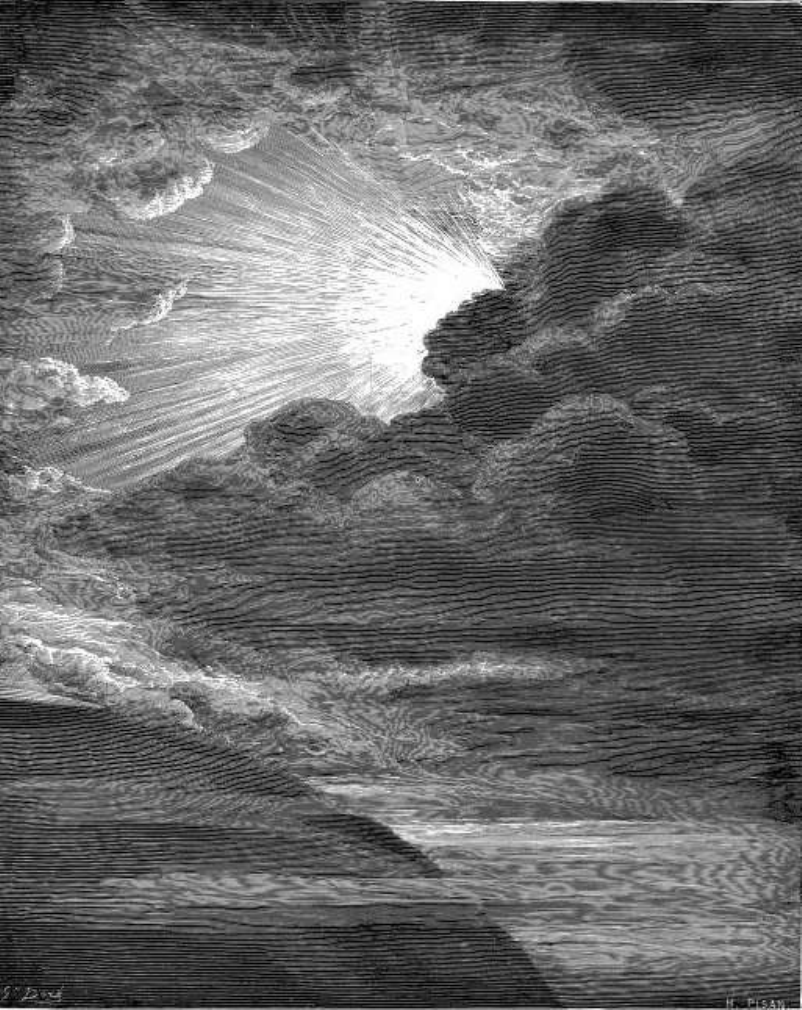}
		\captionsetup{justification=justified}
		\caption[The Creation of Light.]{
		\label{fig:FiatLux}		\textbf{\textit{The Creation of Light}, by Gustave Doré, 1866.} One of the illustrations (241 in total) performed by Gustave Doré in 1866 for \textit{La Grande Bible de Tours}.  ``The Creation of Light'' captures the rising sun as a victorious symbol, dispelling darkness and chaos.
	}
	\end{SCfigure}

	One of the earliest and most profound examples can be found in the first verses of Genesis, where God’s first divine words read, 
	\begin{quote}
		\emph{``Dixitque Deus: Fiat lux. Et facta est lux''}, \qquad (Genesis 1.3),
	\end{quote}
	translated as, \textit{``And God said, Let there be light: and there was light''}. This passage  marks the emergence of light from darkness, signifying the dawn of existence itself [see \reffig{fig:FiatLux}]. 
	From a scientific perspective, cosmology offers the theory of \textit{Big Bang}. This explains the origin and evolution of the universe, as described by the standard $\Lambda$CDM cosmological model~\cite{MukhanovPhysicalFoundations2005}.
	Approximately $13.8$ billion years ago, an \textit{ex-nihilo} big bang---interpreted as an energy density divergence in the Einstein equations---gave rise to the universe as a \textit{whole}, initiating the emergence of spacetime and matter.
	After a brief inflation process, during which the universe grew exponentially,  the expansion of the universe slowed, entering into a cooling phase. During this phase,  energy transformed into particles and photons. And later, atoms, leading to the formation of  stars, galaxies, and eventually, the large-scale structures observed today. 
 	The Big Bang theory is strongly supported by experimental data from satellites such as COBE, PLANCK, and WMAP, which have provided substantial insights into the \textit{cosmic microwave background radiation} (CMB)---the primordial radiation emitted during the recombination era, approximately $380.000$ years after the big bang, when matter and radiation decoupled as a consequence of the cooling process---.
	These \textit{relic photons} might be considered the earliest source of information about the universe, as they contain valuable insights into its history. However, due to the opacity of the universe prior to recombination, these photons carry no direct information from earlier epochs. While this imposes observational limitations, the Big Bang theory remains a robust and extensively validated framework for understanding the origin of the universe.
	%
	
%
%

\begin{SCfigure}[1][h!]
		\includegraphics[width=0.65\columnwidth]{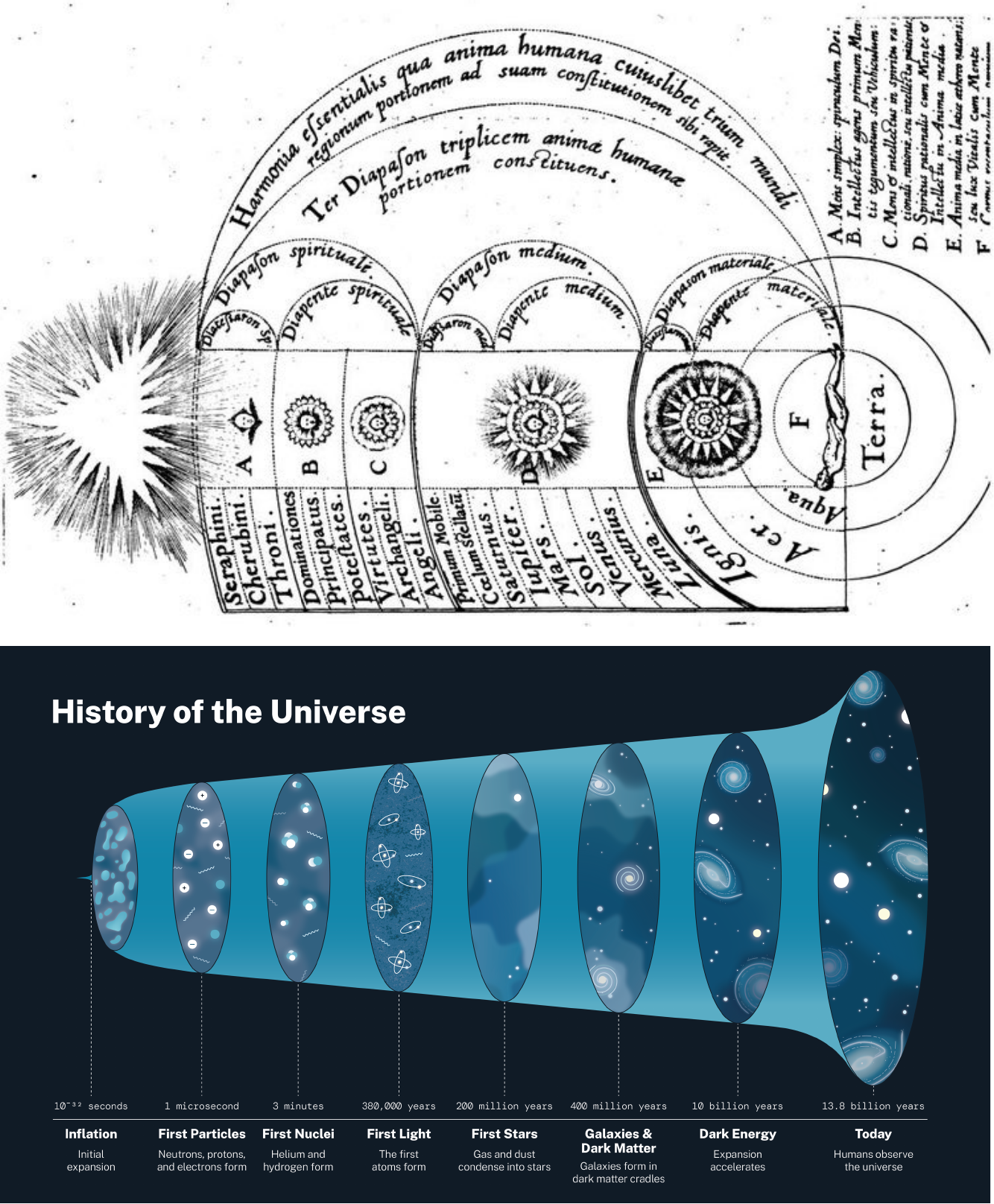}
		\captionsetup{justification=justified}
		\caption[Schematic comparison between the theological and scientific explanation of the universe.]{
			\label{fig:ComparisonBigBang}		\textbf{Schematic comparison between the theological and scientific explanation of the universe.}  The upper illustration depicts the \textit{Hierarchy of planets and angelic orders} by the English doctor and philosopher Robert Fludd in 1619. The lower panel shows a schematic representation of the history of the universe (extracted from the NASA webpage). Both representations essentially share an emergent structure of complexity in their evolution.
		}
	\end{SCfigure}

	This cosmological question remains one of the most fundamental and unsolved aspects in human understanding, shaped by both its metaphysical nature and technological challenges. Interestingly, a ``conceptual parallel'' emerges between both theological and scientific perspectives: the \textit{Big Bang}, as the generative principle of the existence of the universe, and the divine \textit{fiat lux}, where God creates the world. In \reffig{fig:ComparisonBigBang}, both illustrations reveal, \textit{in essence}, a complex emergent structure throughout its evolution.
	Though this comparison might be considered as a \textit{scientific transgression}, it highlights the intrinsic and inseparable connections between science and philosophy.	Physics, and the sciences in general, cannot be fully understood as isolated disciplines as these are constantly influenced by the cultural, religious, or philosophical contexts in which they are developed. 
	
	The history of sciences provides numerous examples of this dialectical relationship between science and philosophy.
	This interplay is evident in debates ranging from the Democritus-Aristotelian discussion on the nature of the ultimate constituents of matter---which continues permeating the conceptual interpretation of the elementary particles---to the ontological challenges originated at the beginning of the XX century. During this period, the \textit{aprioristic} conceptions of reality, epitomized by Leibniz's axiom \textit{natura non facit saltus}, were challenged by the groundbreaking experiments and theories in quantum mechanics and relativity$^{\textcolor{Maroon}{*}}$.
	\graffito{
	$^{*}$Similarly, the ongoing development of quantum theories of gravity remain deeply metaphysical, both due to the abstract nature of the subject and the lack of experiments to verify these theories to date.
	}
	 Here, light played a central  and transformative role, revealing the limitation of classical intuitions and reshaping our understanding of the fundamental nature of reality.
	Several key examples illustrate the significant impact of light:
	Planck's explanation of black body radiation~\cite{PlanckUeberGesetz1901}, Einstein's work on the photoelectric effect~\cite{EinsteinUberErzeugung1905}---despite the fact that a semiclassical theory later accounted for this effect~\cite{LambPhotoelectricEffect1968,MandelOpticalCoherence1995}---, 	and Einstein's special theory of relativity~\cite{EinsteinZurElektrodynamik1905}.
	All these examples illustrate how light served as a cornerstone in challenging classical physics, not only refuting the existence of a \textit{luminiferous aether}, but also introducing the concept of \textit{quanta} and revolutionizing the concept of space and time, leading to the unified framework of \textit{spacetime}.
	%
	%
	%
	
	It is particularly interesting to mention the \textit{philosophical atomism}~\cite{MariasHistoryPhilosophy1967,MosterinConceptosTeorias2000}, a framework rooted in the idea that all observable objects---i.e., the \textit{mundus adspectabilis}---are specific configurations  of real, unalterable, indivisible, and eternal elements. That is, atoms. This philosophical perspective---not scientifically supported until the early 19th century with the novel experiments in chemistry by Lavoisier~\cite{LavoisierTraiteElementaire1789,LavoisierElementsChemistry1984}---dominated the scientific view of the 17th century. It stood in clear opposition to the scholastic doctrines prevalent at the time. An illustrative example is found in Newton's \textit{Opticks}~\cite{NewtonOpticksTreatise1704} regarding the corpuscular interpretation of light---juxtaposed with  Goethe's Aristotelian perspective in his \textit{Theory of Colours}~\cite{GoetheTheoryColours1840}---:
	\begin{quote}
		\emph{
		``It seems probable to me, that God in the Beginning forme'd Matter in solid, massy, hard, impenetrable, moveable Particles, of such Sizes and Figures, and with such other Properties, and in such Proportion to Space, as most conduced to the End for which he formed them. And that these primitive Particles being Solids, are incomparably harder than any porous Bodies compounded of them; even so very hard, as never to wear or break in pieces; no ordinary Power being able to divide what God himself made one in the first Creation. 
		While the Particles continue entire,
		they may compose Bodies of one and the fame
		Nature and Texture in all Ages : But should
		they wear away, or break in pieces, the Nature
		of Things depending on them, would be chang'd.
		[...] 
		And therefore, that Nature
		may be lasting, the Changes of corporeal Things
		are to be placed only in the various Separations
		and new Associations and Motions of these permanent Particles.''
		}
	\end{quote}

	Our understanding and control of light have had singular, tangible impacts throughout history. From enabling early navigational techniques, such as the use of the sun and stars to chart unknown seas, to driving agricultural innovation by improving crop cultivation methods through the study of sunlight and seasonal patterns, the ``domination''---\textit{vexationes artium}, in words of Chancellor Bacon~\cite{MariasHistoryPhilosophy1967,BaconWorksFrancis1857}---of light has shaped human progress in numerous ways.
	Significant milestones include the early use of mirrors in Ancient Egypt (\textit{circa} 1900 B.C.E.), Aristophanes' reference to the burning glass in \textit{The Clouds} (424 B.C.E.), Alhazen's groundbreaking studies of the eye (\textit{circa} 1000 A.D.), the invention of refracting telescopes by Lippershey (1587-1619) and Galileo Galilei (1564-1642), and the advent of lasers in 1960, which revolutionized communication technologies.
	This brief overview merely scratches the surface, and readers are encouraged to explore the following \colorrefs{HechtOptics2017,DarrigolHistoryOptics2012,JuanCamiloLopezCarrenoExcitingQuantum2019, CalvoPadillaPioneroLuz2019} for a comprehensive historical account of optics.

{
		\sidecaptionvpos{figure}{t}
	\begin{SCfigure}[1][t!]
	\includegraphics[width=0.54\columnwidth]{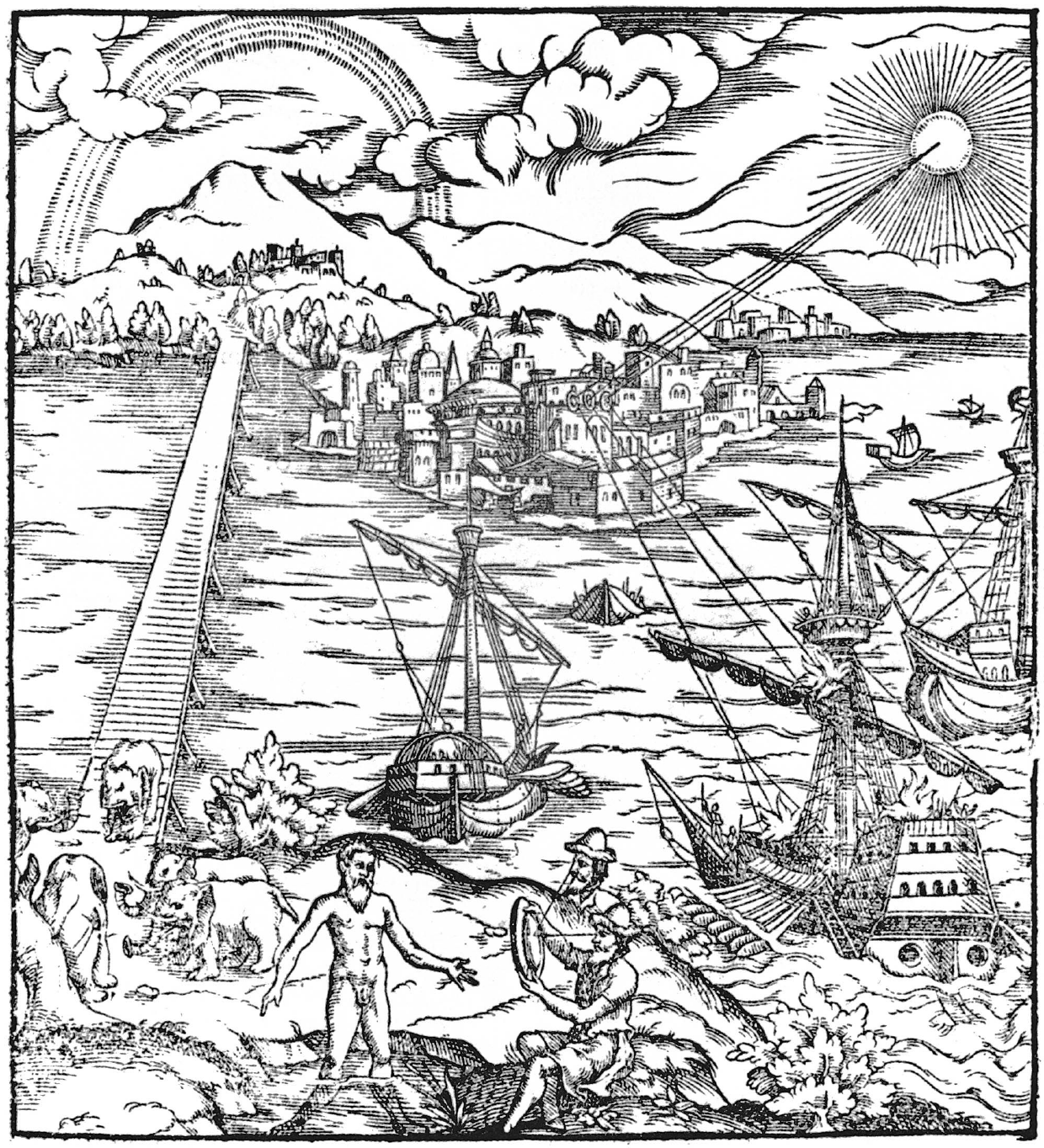}
	\captionsetup{justification=justified}
	\caption[Front page of the Latin edition of \textit{Opticae Thesaurus}.]{
		\label{fig:Thesaurus_opticus}	\textbf{Front page of the Latin edition of \textit{Opticae Thesaurus}.}  Edited and published by Friedrich Risner in 1572, this book features the works of Ibn al-Haitham (Alhazen) and Erazmus Ciołek Witelo (Vitello), pioneers in the field of optics, translated into Latin. The illustration depicts various optical phenomena already understood at the time, such as the rainbow, the vanishing perspective of distant objects (illustrated by the wooden bridge), the distortion of images caused by refraction in water (shown by the man standing in water at the bottom center), and a representation of Archimedes using parabolic mirrors to set Roman ships on fire during the siege of Syracuse.
	}
\end{SCfigure}}

	Beyond all these advancements lies a profound inquiry to consider: the very act of perceiving the world---our \textit{reality}---through our eyes or any optical apparatus. This, in turn, raises an even deeper and more intricate issue: is what we see with our own eyes truly \textit{real}? Or, more fundamentally, what do we actually mean by \textit{reality}?
	
	\subsection{Do we observe  the reality?}
	
	The question of distinguishing \textit{what is real and what is not}---or, in other words, \textit{what constitutes truth}---has been a central topic of discussion since the emergence of science in Ancient Greece, around the VI century B.C.E.
	The development of technology---whether conceived through language, mathematics, or engineering---provided the solid ground to advance the rational process, transitioning from \textit{mythos} to \textit{logos}. This radical change led humanity to question the ultimate truth, if any, about nature, as well as fundamental aspects of human existence, including ethics, politics, and aesthetics. In essence, it marked the birth of philosophy.
	It could be said that science and philosophy emerged simultaneously from the effort to understand and control nature itself~\cite{BuenoQueEs1995,MadridCasadoFilosofiaCosmologia2018}. 
	 \begin{SCfigure}[1][t!]
	 	\includegraphics[width=0.65\columnwidth]{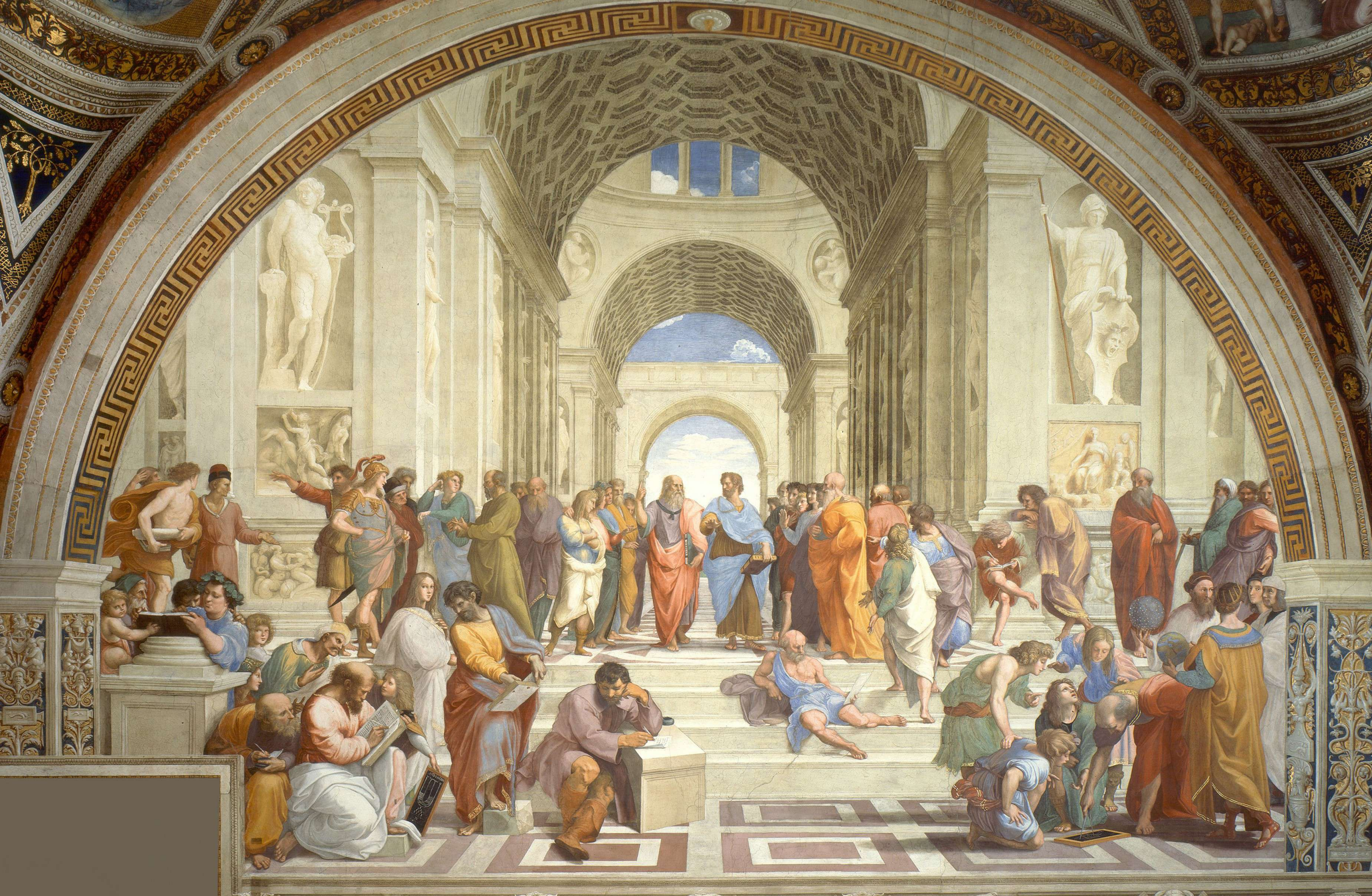}
	 	\captionsetup{justification=justified}
	 	\caption[School of Athens.]{
	 		\label{fig:School_of_Athens} \textbf{\textit{School of Athens}, by Raphael, 1509-1511.} The fresco serves as an allegory of knowledge, portraying a grand hall filled with the most prominent scientists, mathematicians, and philosophers of Ancient Greece, including Socrates, Pythagoras, Archimedes, and Euclid. The Epicureans, Stoics and Cynics are also present. At the center, Plato, holding \textit{Timaeus} and pointing to the sky, and Aristotle, holding \textit{Nicomachean Ethics} with his palm facing downward, embody the contrasting philosophical currents of idealism and empiricism that define Western thought.
	 	}
	 \end{SCfigure}
	 
	The formal study of geometry, exemplified by Euclid's Elements, Anaximander's first cartography of the known world---introducing the concept of the cosmos as an organized separation between the terrestrial and celestial spheres---, Archimedes' method of mechanical theorems, and his groundbreaking contributions to mathematics and physics, along with the exploration of the nature of matter and time by figures such as Democritus, Aristotle, Parmenides, and Zeno, are foundational examples to the development of Western thought~\cite{MariasHistoryPhilosophy1967,RussellHistoryWestern2004}.
	 These intellectual advances, emerging from direct observation of nature, not only yielded practical knowledge in fields like architecture, sailing, or medicine, among others, but also introduced novel ways of perceiving and understanding the world.
	From such examples, additional question emerge: \textit{is the world merely a representation of our ideas? Do we, in some sense, create the world ourselves? Or are we even capable of uncovering a final and definitive truth about nature?}

	\subsection{Looking  at science through materialism}
	
	These kind of questions prompted the development of numerous philosophical frameworks proposed throughout the history of knowledge, spanning Ancient Greece, the Medieval period, or the Modern Era. In fact, the possible answers to such questions constitutes the \textit{backbone} of the frameworks themselves. Yet, this \textit{backbone} is not unique. It can be interpreted through different and often radical perspectives, imbuing the concept of truth with an intrinsic \textit{obscureness}.
	The study of such questions falls within the domain of the so-called philosophy of science.
		
	At this point, we might ask ourselves 
	\graffito{
		$^{*}$It should be noted that the following philosophical digression reflects the author's personal perspective on the nature of science. Since it does not directly impact the understanding of this Thesis, readers may choose to skip it and proceed to the next section.
	}
	whether this philosophical discussion about the	 \textit{reality} of nature is actually necessary for this Thesis, or for sciences in general$^{\textcolor{Maroon}{*}}$. 
	On the surface, one might argue ``no, of course not, such disquisitions are unnecessary to compute quantum probability distributions or design environments that enhance quantum effects for potential technological applications''. 
	While this approach, which is aligned with the famous quote attributed to R. Feynman: ``\textit{shut up and calculate}'',  is generally valid and effective for the development of sciences, an important caveat must be noted. Adopting this mindset uncritically and without nuance would be a significant oversight if our goal is to cultivate a more holistic understanding of what nature might truly be. That is, develop a \textit{Weltanschauung}, a world-view.

	Science, understood as an \textit{intellectual tool}, might be defined as a set of rational, systematic, verifiable, and thus, fallible, ideas aimed at answering the question of \textit{what is real} in nature. However, providing such an answer is not only genuinely challenging from a technological perspective, but philosophical as well.  The neutrality of science is a chimera; it is inherently impossible. Every component of a scientific system---whether its object of study (e.g., the elementary particles, or atoms photons) or the formal elements of its theory (e.g., the Einstein field equations or the Schrödinger equation)---carries an ontological burden that vanish any attempt of neutrality. In other words, the very concepts themselves already encapsulate some \textit{a priori} understanding, imbued by the scientists who proposed them.
	A capital example of this lack of neutrality is the simultaneous development of the quantum theory by E. Schrödinger~\cite{SchrodingerQuantisierungAls1926} and W. Heisenberg~\cite{HeisenbergUeberQuantentheoretische1925}. While both formulations of quantum mechanics are experimentally and mathematically equivalent, they reflect an intrinsic ontological incompatibility. Schrödinger's approach centres on the wavefunction as the primary ontological entity, representing the complete state of a quantum system, whereas Heisenberg's formulation emphasizes the measurability of the system, shifting the ontological focus to the physical properties revealed through measurement.
	In the years that followed, multiple interpretations of quantum mechanics emerged~\cite{FayeCopenhagenInterpretation2024}, each offering radically different perspectives on the nature of quantum entities while producing identical empirical predictions.
	%
	%
	
	%
	Consequently, a philosophical \textit{reference frame} is needed to properly address the latest results offered from physics, or any other science, in order to avoid falling into conceptual contradictions or banalities, what might be termed as \textit{spontaneous philosophy} \cite{OrtegayGassetRebelionMasas1930}. Interestingly, by referring to philosophical systems as frameworks, we recognize that they function as reference frames, analogous to those used in the analysis of dynamics in classical mechanics. The philosophical system typically employed to address science is \textit{materialism}, which essentially seeks to define scientific truths while avoiding the invocation of metaphysical entities~\cite{BungeScientificMaterialism1981,BuenoQueEs1995,MosterinConceptosTeorias2000,DiezFundamentosFilosofia2018,MadridCasadoFilosofiaCosmologia2018}.

	From this framework, sciences do not unveil \textit{how the world is}, as if nature were simply ``hidden'' and scientists were playing a game of \textit{hide-and-seek} with a Spinozian natural God, as stated in Spinoza's \textit{Ethica, ordine geometrico demonstrata}~\cite{SpinozaSpinozaEthics2018}, 
		\begin{quote}
		\emph{
			``For the eternal and infinite Being, which we call God or Nature, acts by the same necessity as that whereby it exists. [...] The reason or cause why God or Nature exists, and the reason why he acts, are one and the same.''
		}
	\end{quote}
	Rather, \textit{sciences generate the world}. Through a dialectical and constructive process between theoretical speculation and experimental verification, scientific truths are established in an operative way via apparatus (blackboards, computers, measurement devices, ...), and not solely through theoretical thoughts. This view consider sciences as a \textit{savoir-faire}~\cite{BuenoQueEs1995}, and we can summarize by slightly modifying the scholastic axiom from Thomas Aquinas~\cite{AquinasQuaestionesDisputatae1256}: 
	%
	%
\emph{Nihil est in intellectu quod prius non fuerit in apparatu},
	translated as \textit{nothing is in the intellect that was not first in the instruments}.
	This \textit{savoir faire} perspective, thus, conceive the instruments---e.g., the measurement devices---as the responsible elements for advancing in the acquisition of scientific truths. The development of novel technological instruments through history has not only enabled technological advancements in society and more precise measurements but has also provided ways to explore novel physical phenomena, ultimately revolutionizing our understanding of the world.

	There exists alternative philosophical frameworks for discerning what constitutes scientific truth~\cite{MadridCasadoFilosofiaCosmologia2018}, such as an \textit{idealism} perspective where theoretical results hold a degree of validity even without experimental verification, or a purely \textit{pragmatism} perspective, where the functionality of theories is what truly matters. 
	We readily acknowledge that this fundamental issue exceeds the scope of both this introduction and the Thesis itself. However, having a general overview—a big-picture perspective—provides a valuable path to guide our exploration. In our case, we adopt a \textit{materialist perspective}~\cite{BuenoQueEs1995,MadridCasadoFilosofiaCosmologia2018}, which we consider the most successful approach to expanding our knowledge of nature without falling into non-scientific \textit{entelechies}.

\begin{SCfigure}[1][t!]
	\includegraphics[width=0.65\columnwidth]{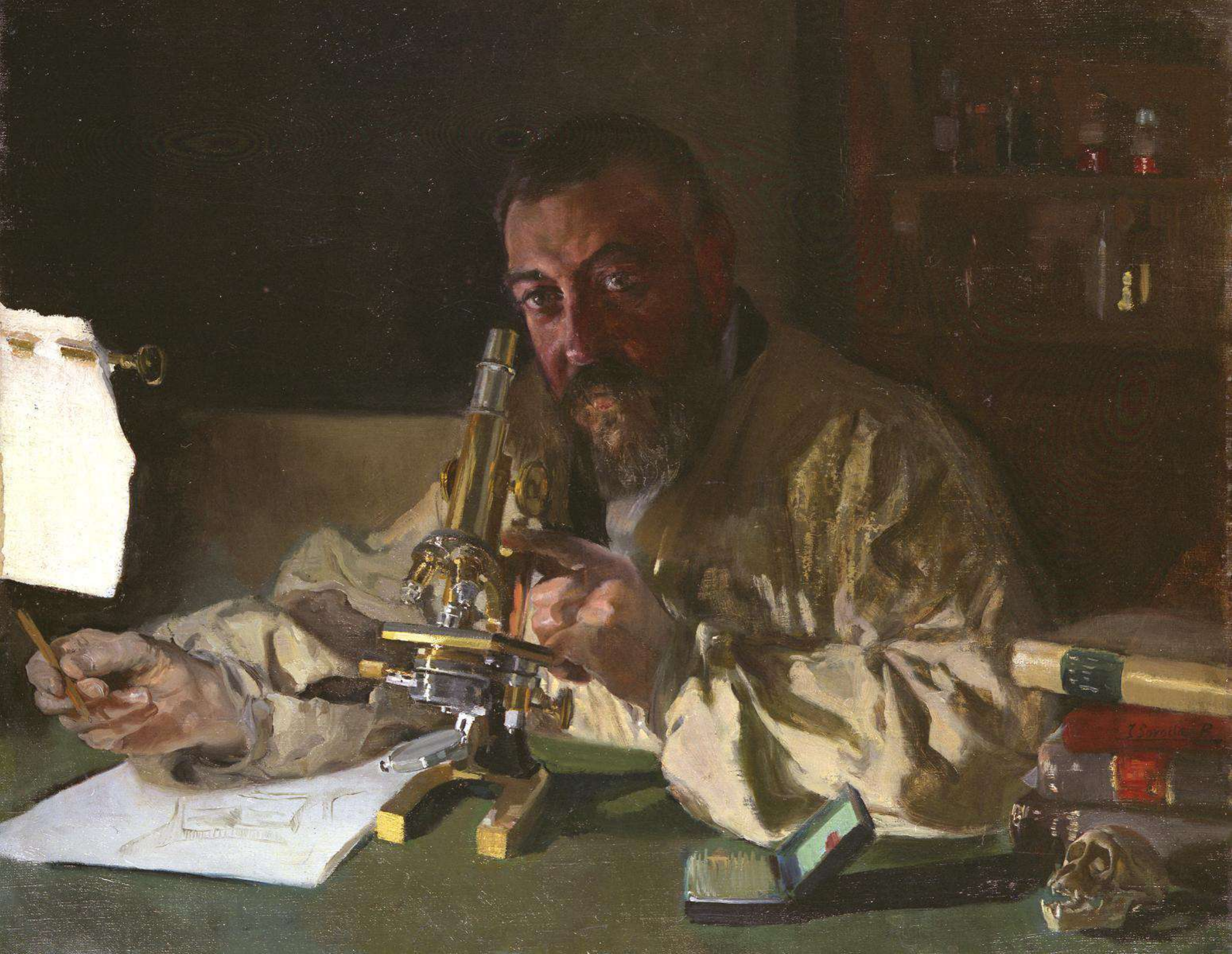}
	\captionsetup{justification=justified}
	\caption[Portrait of Dr. Simarro at the microscope.]{
		\label{fig:Sorolla}		\textbf{\textit{Portrait of Dr. Simarro at the microscope}, by Joaquín Sorolla, 1897.} 
		This oil painting portrays Dr. Luis Simarro, a Spanish neurologist active in the late 19th and early 20th centuries. The painting vividly captures the scientific \textit{savoir-faire}: the scientist develops new insights into nature through the use of contemporary apparatus (in this case, a microscope), enabling him to explore and create a new world of phenomena. 
	}
\end{SCfigure}

	\section{Controlling light-matter interactions}

	\subsection{Lightening the truth}
	%
	%
	Following the example that opened the previous philosophical discussion, the development of alternative ways to perceive the world---ways that go beyond our natural eyesight---has profoundly transformed our understanding of nature. And this change is fundamentally tied to the understanding and control of \textit{light}, not only within the visible spectrum but across the entire electromagnetic spectrum [see \reffig{fig:EMspectrum}].
\begin{SCfigure}[1][h!]
		\includegraphics[width=1.\columnwidth]{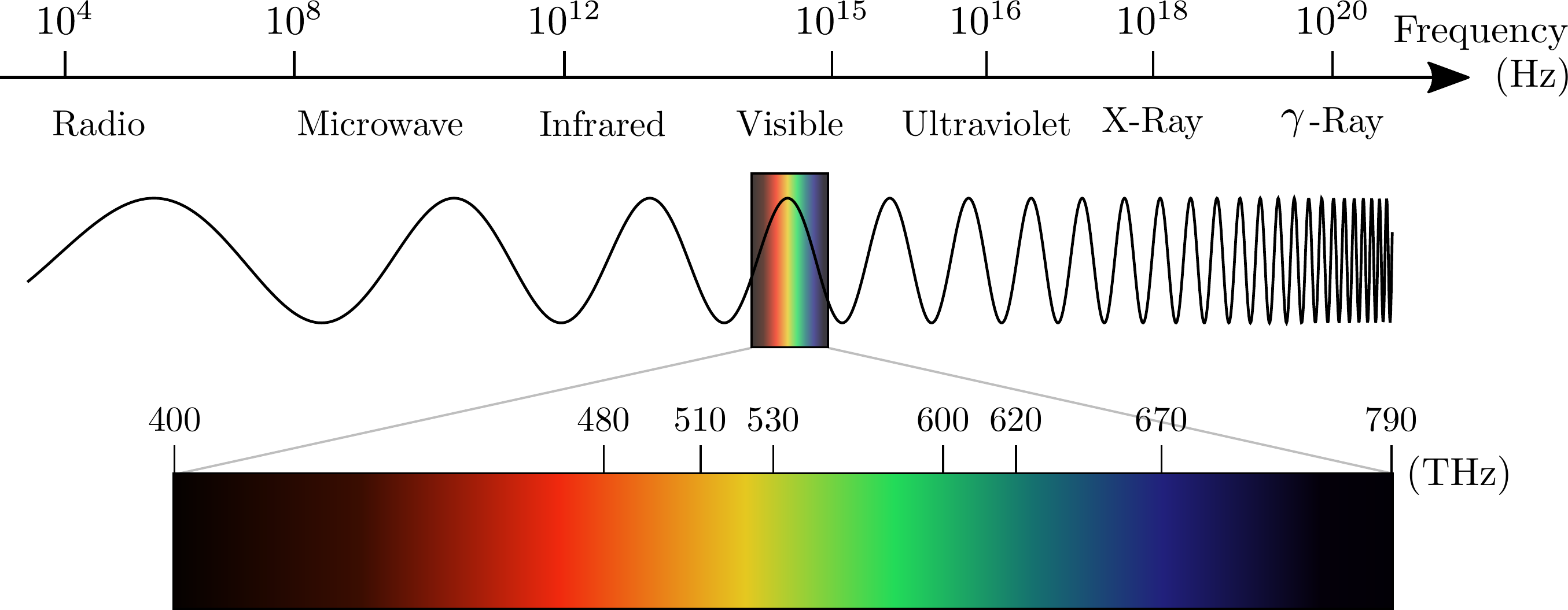}
		\captionsetup{justification=justified}
\end{SCfigure}
\addtocounter{figure}{-1}
	\graffito{\vspace{-5.1cm}
		\captionof{figure}[Diagram of the electromagnetic spectrum.]{\label{fig:EMspectrum}		
			\textbf{Diagram of the electromagnetic spectrum.} 
			Diagram of the electromagnetic spectrum across frequencies (in Hz), ranging from radio waves to $\gamma$-rays. The bottom section illustrates the visible light range detected by the human eye, ranging between $400-790$ THz.
		}
	}

	From this \textit{materialist perspective}, in which scientific truth vitally depends on the measurement apparatus, the human eye serves as an instrument to \textit{measure reality}, but only  while the object to be tested is macroscopic. 
	Nevertheless, when we consider objects with smaller sizes than what can be resolved by human vision~\cite{HechtOptics2017},
	%
	%
	 we turn to alternatives ways to discriminate reality through artificial extensions of our sight. At these scales, we use powerful electronic microscopes that send focused light to interact with microscopic objects, and then the scattered light is analysed. This has expanded our definition of what can be observed, and, by extension, what can be considered real.

	In this way, \textit{mutatis mutandis}, light serves as a cornerstone for exploring the quantum realm, acting as both a probe and an instrument. By sending light, whether in the form of quantum pulses or coherent light fields like lasers, we can interact with quantum matter---such as natural atoms or artificial quantum emitters---and analyse the scattered light, to extract otherwise inaccessible information. Examples of this include determining the eigenenergy structure of quantum systems via fluorescence measurements~\cite{MollowPowerSpectrum1969,Cohen-TannoudjiAtomPhotonInteractions1998},  estimating internal system parameters (e.g., the natural energy of a two-level system) via photon-counting measurements~\cite{DelaubertQuantumLimits2008,DowlingQuantumOptical2015,PolinoPhotonicQuantum2020,BarbieriOpticalQuantum2022}, and resolving photon number state via qubit spectroscopy techniques~\cite{SchusterResolvingPhoton2007,KonoNonclassicalPhoton2017,RomlingResolvingNonclassical2023}. However, the significance of light plays an even more fundamental role in the quantum regime. Its control enables the exploration of realms beyond the reach of the naked eye and drives the development of groundbreaking technological devices based on quantum light, commonly referred to as \textit{photonic quantum technologies}~\cite{OBrienPhotonicQuantum2009}.

	Photons, distinguished by their high mobility and robustness against decoherence due to low interaction with the environment, are ideally suited for tasks such as quantum communication~\cite{GisinQuantumCryptography2002}, optical quantum computing~\cite{OBrienOpticalQuantum2007} and metrology~\cite{KimbleQuantumInternet2008,PezzeQuantumMetrology2018}. 
	Photonic technologies enable the development of devices for the generation, manipulation, and detection of quantum states encoded in the various degrees of freedom of light~\cite{WangIntegratedPhotonic2020,ElshaariHybridIntegrated2020}, driving advancements in novel quantum technologies~\cite{AcinQuantumTechnologies2018}.
	These include the design of metrological protocols that allow information acquisition in ways impossible classically~\cite{GiovannettiAdvancesQuantum2011,BraskImprovedQuantum2015,Demkowicz-DobrzanskiQuantumLimits2015,DowlingQuantumOptical2015,PezzeQuantumMetrology2018,PirandolaAdvancesPhotonic2018,PolinoPhotonicQuantum2020,BarbieriOpticalQuantum2022}, photonic-based quantum computing and communication~\cite{KnillSchemeEfficient2001,DuanLongdistanceQuantum2001,BriegelMeasurementbasedQuantum2009,PaulischAdiabaticElimination2014,PaulischUniversalQuantum2016, LodahlQuantumdotBased2018,LabonteIntegratedPhotonics2024}, as well as the development of fluorescence-based medical non-invasive imaging techniques~\cite{DenkTwoPhotonLaser1990,HellFluorescenceNanoscopy2003,ZipfelNonlinearMagic2003,MaestroCdSeQuantum2010,LiHighspeedTwophoton2024}.

	Building on the concept of light as an \textit{instrument} for creating novel scientific truths in the quantum realm, understanding and controlling its interaction with quantum matter becomes crucial.  The study of light-matter interactions, grounded in fundamental principles, not only provides new theoretical insights but also propels advancements in technological applications. This interplay creates a dialectical---or synergistic---connection between theoretical research and experimental, or technological, innovations. Such a relationship drives the exploration of novel quantum phenomena, and may even pave the way for potential groundbreaking scientific revolutions.
	Therefore, researching light-matter interactions within state-of-the-art quantum technological platforms fosters mutual progress in both domains, bridging the gap between theory and practice.

	This Thesis is devoted to the theoretical study of quantum systems, specifically quantum emitters described as two-level systems, interacting with both classical and quantum light. We explore the potential of these light-matter systems to generate quantum entanglement and their applications in quantum metrology, particularly in the context of quantum parameter estimation. The results presented here aim to modestly contribute the understanding of light-matter interactions in quantum optics, and open up possibilities for using quantum matter and light in both fundamental research and technological applications. %

	\subsection{Tailoring light-matter interactions}
	
	When studying quantum light-matter interactions in realistic scenarios, it is essential to consider their two principal ingredients: matter and light. 
	In the context of quantum optics~\cite{GardinerQuantumNoise2004}, the \textit{matter} component typically consists of a finite number of quantum emitters, each characterized by at least one well-defined energy transition.
	We will adopt this description of matter; however, it is important to note that in other contexts, such as  in condensed matter physics~\cite{ChaikinPrinciplesCondensed1995} or quantum field theory~\cite{RyderQuantumField1996}, the description of matter can be far more complex or even ontologically different.
	 The \textit{light} component, on the other hand, is described by photonic modes, which may exhibit either discrete or continuous energy spectrum depending on the boundary conditions~\cite{Cohen-TannoudjiPhotonsAtoms1997, Cohen-TannoudjiAtomPhotonInteractions1998}.
	Physical systems integrating these two elements are commonly referred as \textit{light-matter interfaces}~\cite{Gonzalez-TudelaLightMatter2024}.
	
\begin{SCfigure}[1][h!]
	 	\includegraphics[width=1.\columnwidth]{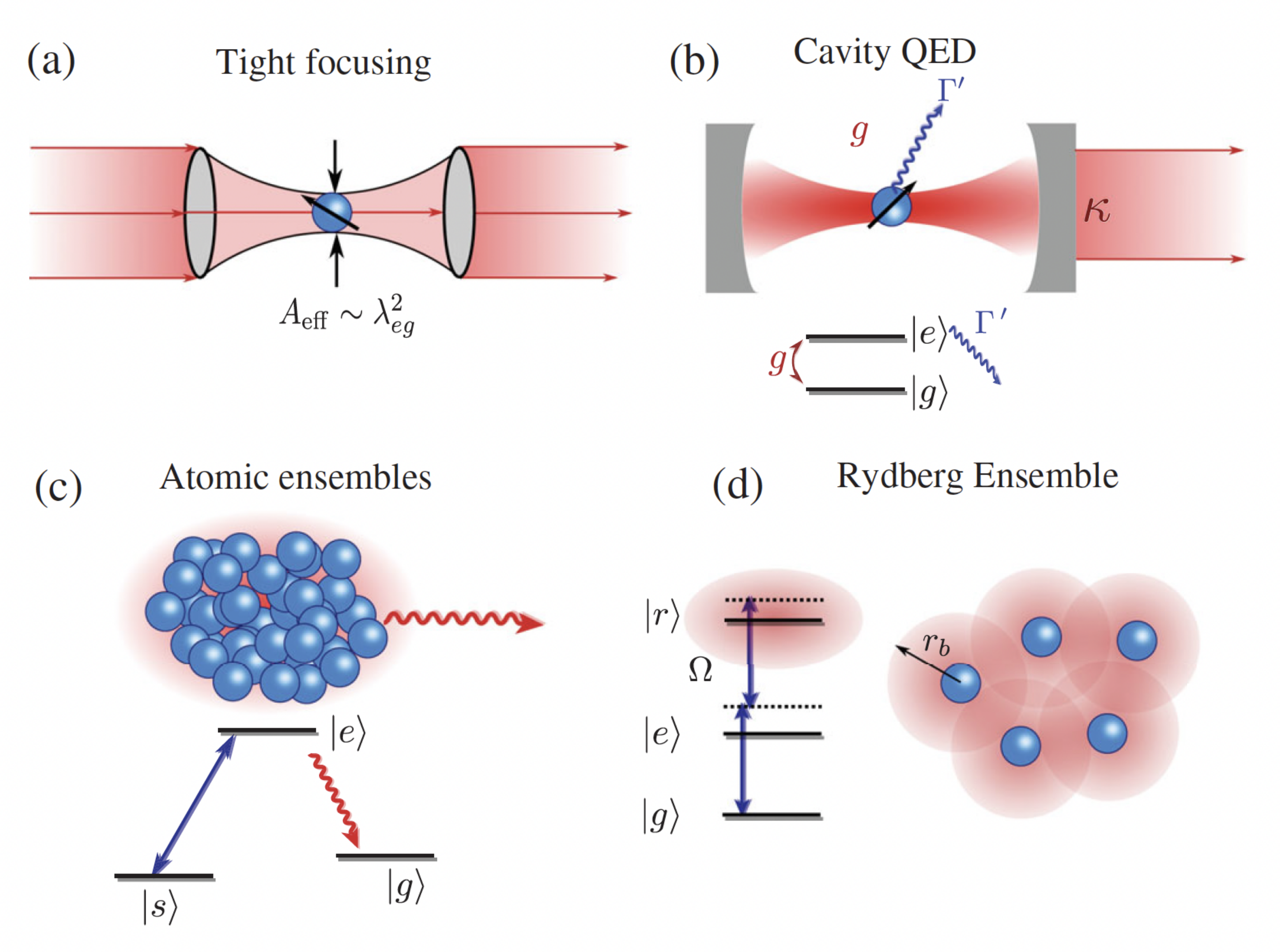}
	 	\captionsetup{justification=justified}
	 	\caption[Conventional approaches in quantum optics to achieve strong atom-photon interactions.]{
	 		\label{fig:ConventionalApproaches}		\textbf{Conventional approaches in quantum optics to achieve strong atom-photon interactions. } 
	 		(a) Diffraction-limited focusing of an optical beam onto a tightly trapped atom, (b) cavity QED, where the interaction is enhanced by a large number of photon round-trips, (c) atomic ensemble, where a large atom numberresults in high probability of interaction with a single photon, and(d) atomic ensemble of Rydberg atoms. Reproduced with permission from \colorref{ChangColloquiumQuantum2018}.
	 	}
	 \end{SCfigure}

	Photon-based quantum technologies rely on efficient light-matter interfaces,  where  efficiency is defined as a high probability of  interaction between a single photon and a single emitter. In other words, achieving  deterministic light-matter interactions. Nevertheless, this remains a significant challenge since such interactions do not naturally occurs with high probability.
	In free-space setups, a fundamental problem emerges: optical diffraction. This phenomenon constrains the probability of single-photon scattering with a single emitter by imposing a minimum light confinement determined by the emitter's wavelength~\cite{DeVriesPointScatterers1998}. Additionally, diffraction also reduces the probability of a photon emitted by an emitter and absorbed by another at a different position. 
	To overcome this limitation, various strategies have been developed over the past two decades [see \reffig{fig:ConventionalApproaches}]:
	\begin{enumerate}[label=\textcolor{Maroon}{(\roman*)}]
		\item \textcolor{Maroon}{Tight focusing}.  Single atoms are trapped and free-space beams are focused onto them using high-numerical-aperture optics~\cite{VanEnkStronglyFocused2001,DarquieControlledSinglePhoton2005,TeyStrongInteraction2008,HetetSingleAtom2011}. 
		\item  \textcolor{Maroon}{Cavity QED. }
		This approach involves confining emitters between highly reflective mirrors---what it is usually termed as cavity QED---, which increases photon-emitter interaction rates by repeatedly reflecting photons within the mirrors~\cite{RaimondManipulatingQuantum2001,HarocheExploringQuantum2006,HarocheNobelLecture2013}.
		\item  \textcolor{Maroon}{Atomic ensembles. } Atomic ensembles are employed to enhance the interaction probabilities by increasing the chance that a photon interacts with at least one emitter in the ensemble~\cite{HammererQuantumInterface2010}.
		\item \textcolor{Maroon}{Rydberg gases. } In Rydberg gases~\cite{LukinDipoleBlockade2001,SaffmanQuantumInformation2010}, strong nonlinear interactions are achieved by using highly excited atomic states (Rydberg states) to significantly enhance the dipole moments of the emitters, thereby strengthening light-matter interactions.
	\end{enumerate}
	Despite these advances, optical diffraction continues to dictate both the strength and spatial characteristics of light-matter interactions.
	 
	An alternative to free-space approaches involves interfacing quantum emitters with photons---or photon-like excitations---confined in subwavelength-engineered devices, a field known as \textit{quantum nanophotonics}~\cite{LodahlInterfacingSingle2015,ChangColloquiumQuantum2018,Gonzalez-TudelaLightMatter2024}. These \textit{photonic tailored environments}, including photonic or plasmonic cavities and waveguides~\cite{ChangColloquiumQuantum2018,PoyatosQuantumReservoir1996,LiuComparingCombining2016}, enable the confinement of photons on subwavelength scales. This leads to configurations with smaller spatial dimensions, stronger light-matter interactions, and faster dynamics. These attributes open pathways for integrating quantum optics and atomic physics on scalable platforms, paving the way for robust, chip-based quantum technologies~\cite{AcinQuantumTechnologies2018}.

	In addition to these advantages, tailored photonic environments provide a suitable strategy to counter the loss of \textit{quantumness} and fight decoherence in driven-dissipative scenarios by deliberately engineering the interactions between the system and the environment~\cite{ChangColloquiumQuantum2018}. 
	This \textit{reservoir engineering} technique is an essential ingredient for the development of quantum technologies in driven-dissipative setups~\cite{VerstraeteQuantumComputation2009,MirrahimiDynamicallyProtected2014}.
	Notably, reservoir engineering allows the generation and stabilization of entangled states within ensembles of emitters~\cite{PlenioCavitylossinducedGeneration1999,Gonzalez-TudelaEntanglementTwo2011,Martin-CanoDissipationdrivenGeneration2011,Gonzalez-TudelaMesoscopicEntanglement2013,RamosQuantumSpin2014,PichlerQuantumOptics2015,HaakhSqueezedLight2015,ReitzCooperativeQuantum2022,KastoryanoDissipativePreparation2011} that survive in timescales much longer than the usual timescales of decoherence, with key applications in quantum computing~\cite{PoyatosQuantumReservoir1996,VerstraeteQuantumComputation2009,MirrahimiDynamicallyProtected2014}, quantum information~\cite{KimbleQuantumInternet2008,NielsenQuantumComputation2012,NarlaRobustConcurrent2016}, quantum communications~\cite{Hai-JingQuantumSecure2006,JooQuantumTeleportation2003,JungGreenbergerHorneZeilingerStates2008} and quantum metrology and sensing~\cite{BraskImprovedQuantum2015,PezzeQuantumMetrology2018}.
	Examples of this strategy applied to the stabilization of entangled states have been demonstrated in platforms such as superconducting circuits~\cite{LeghtasStabilizingBell2013,ShankarAutonomouslyStabilized2013,LiuComparingCombining2016}, atomic ensembles~\cite{KrauterEntanglementGenerated2011} or trapped ions~\cite{LinDissipativeProduction2013}. 
	These implementations highlight the versatility and promise of quantum nanophotonics in advancing next-generation quantum technologies. We note that many of the results presented in this Thesis fall within the framework of \textit{reservoir engineering}.

	Beyond  their ability to enhance free-space characteristics---such as modifying the photonic density of states around the emitter, and changing thus the emitter's decay rate via the Fermi's golden rule~\cite{FermiQuantumTheory1932}---, tailored photonic structures unlock opportunities to explore fundamentally novel regimes of quantum light-matter interactions that have no direct analogue in free-space systems.
	The unique polarization properties, dispersion relations, and spatial field characteristics in these engineered environments enable the observation and control of exotic quantum phenomena. These include examples such as chiral interactions between atoms and light~\cite{LodahlChiralQuantum2017,RamosQuantumSpin2014,PichlerQuantumOptics2015,SollnerDeterministicPhoton2015,OstfeldtOnDemandSource2022,Suarez-ForeroChiralQuantum2024}, atom-photon bound states~\cite{BykovSpontaneousEmission1975,KurizkiTwoatomResonant1990,JohnQuantumOptics1991,ScigliuzzoControllingAtomPhoton2022}, and the emerging fields of giant atoms~\cite{GustafssonPropagatingPhonons2014,FriskKockumQuantumOptics2021,SoroChiralQuantum2022,SoroInteractionGiant2023} and topological photonics~\cite{LuTopologicalPhotonics2014,PerczelTopologicalQuantum2017,OzawaTopologicalPhotonics2019}. 
	

	\paragraph{Cavity QED. }In this Thesis, we are particularly
	\graffito{	${}^{*}$Similarly, \textit{waveguide QED}~\cite{SheremetWaveguideQuantum2023} refers to a setup in which quantum emitters interact with a continuum of confined light modes propagating along one or two dimensions. The light-matter description in this case is slightly different with respect the cavity QED setup, since the band structure of the waveguide is crucial to understand the system dynamics.

		Nevertheless, the ``bad-cavity limit'' and waveguide QED have many similarities in the type of effective dynamics that channelled losses induce into the quantum emitters.
	}
	 interested in the regime where a collection of quantum emitters interact with a single localized photonic mode. This regime, known as \textit{cavity quantum electrodynamics} (cavity QED)$^{\textcolor{Maroon}{*}}$~\cite{RaimondManipulatingQuantum2001,HarocheExploringQuantum2006,HarocheNobelLecture2013,ReisererCavitybasedQuantum2015}, 	naturally emerges in different nanophotonics setups, e.g., due to photonic bandgap engineering or to geometric resonance conditions.
	Examples includes localized defects in photonic crystal waveguides~\cite{HennessyQuantumNature2007,LodahlInterfacingSingle2015}, or in plasmonic~\cite{ChikkaraddySinglemoleculeStrong2016,BenzSinglemoleculeOptomechanics2016} or hybrid plasmon-dielectric nanoantennas~\cite{PschererSingleMoleculeVacuum2021}.  
	In this limit, the coherent dynamics is easily described by the Tavis-Cummings Hamiltonian~\cite{TavisExactSolution1968}, 
\begin{equation}
	\hat H_{\text{TC}}= \sum_i g (\hat \sigma_i \hat a^\dagger +\text{H.c.}),
			\label{eq:TCmodel}
\end{equation}
	where $g$ is the coupling strength between the emitters and the photonic mode. Note that the Tavis-Cummings Hamiltonian reduces to the Jaynes-Cummings Hamiltonian~\cite{JaynesComparisonQuantum1963} when only a single emitter is considered,
	\begin{equation}
		\hat H_{\text{JC}}=  g (\hat \sigma \hat a^\dagger +\text{H.c.}),
		\label{eq:JCmodel}
	\end{equation}
	%
	%
	%
	%
	These two models, \eqref{eq:TCmodel} and \eqref{eq:JCmodel}, which are central in quantum optics, will be extensively used in this Thesis. A detailed discussion at the Hamiltonian level can be found in \refsec{Section:RabiJaynesModels}.

 	In realistic scenarios, a quantum system inevitably interacts with its environment. As a consequence, this interaction sets up a competition between the coherent exchange of excitations and the various decoherence processes. For the emitter, these processes, which may include spontaneous emission or pure dephasing, are collectively characterized by a decoherence rate  $\Gamma$, while for the cavity, decoherence primarily arises from photon leakage through the cavity mirrors, quantified by the rate $\kappa$.
 	The interplay between these coherent and dissipative processes defines the regimes of cavity QED~\cite{GardinerQuantumNoise2004}: 
 	\begin{enumerate}[label=\textcolor{Maroon}{(\roman*)}]
 		\item \textcolor{Maroon}{Weak-coupling regime. }
 		In the weak-coupling regime, the coherent light-matter interaction is dominated by dissipation, i.e., $g<\kappa$. Here, the excitation shared between the emitter and the photonic mode is rapidly lost since the excitation is rapidly channelled through the photonic structure.
 		The emitter experiences a mostly irreversible dynamics in which the structure alters the 
 		lifetime of the emitter in accordance with Fermi's golden rule~\cite{FermiQuantumTheory1932}, becoming an additional decay channel for the emitter. This phenomenon is known as \textit{Purcell effect}~\cite{PurcellResonanceAbsorption1946}, provoking an enhancement of the spontaneous decay rate and a complementary linewidth broadening.

 		Despite its dissipative nature, the weak-coupling regime holds significant promise for quantum technologies, particularly when cavity-induced decay dominates over other loss mechanisms. One notable application is the generation of non-classical light, such as single photons, by leveraging the saturable behavior of the emitter in the quantum nonlinear regime~\cite{ChangQuantumNonlinear2014}.  In this Thesis, we will be particularly interested in this regime.
 		\item \textcolor{Maroon}{Strong-coupling regime. }
 		The strong-coupling regime emerges when the light-matter interaction strength surpasses the emitter and photonic dissipation rates, $g\gtrsim \kappa, \Gamma$.
 		In this regime, light and matter coherently exchange energy multiple times before the excitation is lost to the environment,   forming a new \textit{quasi-particle} composed by a mixture of light and matter, usually called \textit{polariton}~\cite{KavokinMicrocavities2017}.	
		This results in the hybridization of the emitter and photonic states, in which light and matter cease to be independent entities. 

 		This light-matter regime is technologically promising, enabling applications such as the generation of multi-photon states of light~\cite{MunozEmittersNphoton2014}.
 	\end{enumerate} 	
\begin{SCfigure}[1][h!]
		\includegraphics[width=1.\columnwidth]{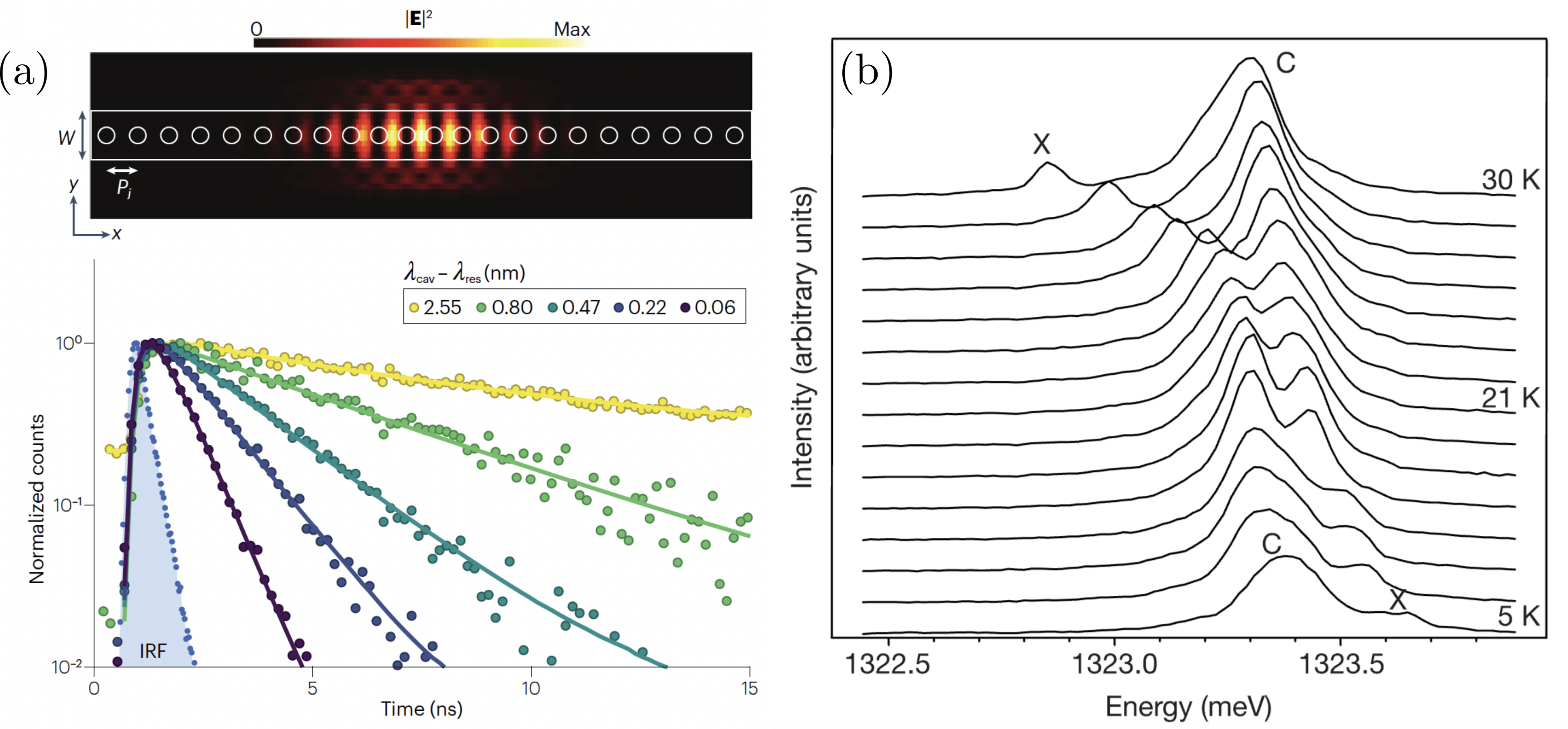}
		\captionsetup{justification=justified}
\end{SCfigure}
	\graffito{\vspace{-6.25cm}
		\captionof{figure}[Signatures of light-matter regimes in Cavity QED.]{	
			\label{fig:WeakStrongCoupling}		\textbf{ Signatures of light-matter regimes in Cavity QED.} 
			(a)  Weak coupling.  Excited-state lifetime of the $\text{SnV}^-$ centre for different cavity resonance wavelengths. Figure extracted from  \colorref{RugarQuantumPhotonic2021}. (b) Strong coupling. Seminal observation of strong coupling with a quantum dot in a microcavity. Reproduced with permission from Springer Nature from \colorref{ReithmaierStrongCoupling2004}.  
		}
	}

 	Interestingly, cavity QED exhibits an additional light-matter regime: the ultrastrong coupling regime~\cite{DeBernardisBreakdownGauge2018,FriskKockumUltrastrongCoupling2019,Forn-DiazUltrastrongCoupling2019}. 
 	In contrast to the weak- (WC) and the strong-coupling (SC) regimes, the ultrastrong coupling (USC) regime is achieved when light-matter coupling is comparable to, or exceeds, the resonant frequency of the emitters. In this regime, the commonly used rotating-wave approximation for light-matter interaction breaks down, leading to novel exotic phenomena, such as the emergence of ground states with virtual excitations and enhanced nonlinear optics~\cite{NiemczykCircuitQuantum2010,FriskKockumUltrastrongCoupling2019,Forn-DiazUltrastrongCoupling2019}.

	\subsection{Solid-state quantum emitters and cooperative effects} 
	
	In this Thesis, we focus on the study of cooperative phenomena among quantum emitters, with a particular emphasis on solid-state platforms. Specifically, we consider \textit{artificially designed atoms} in solid-state platforms as our quantum emitters, which might be coherently coupled among them or deterministically interacting with a photonic mode. These solid-state quantum emitters (QEs) are a promising alternative to natural atoms---which are already the quantum emitters by excellence---, since they exhibit some technological advantages in contrast to their atomic counterpart~\cite{EsmannSolidStateSinglePhoton2024}, such as control in the fabrication process, possibility of placing them permanently at fixed positions, control generation of quantum light (e.g., single-photon sources), or their prospect for scalability. 
	Nevertheless, significant technological challenges arise in these platforms, such as the fabrication of interacting QEs with identical natural frequencies. A more detailed discussion of these challenges is provided below.

\begin{SCfigure}[1][h!]
	\includegraphics[width=1.\columnwidth]{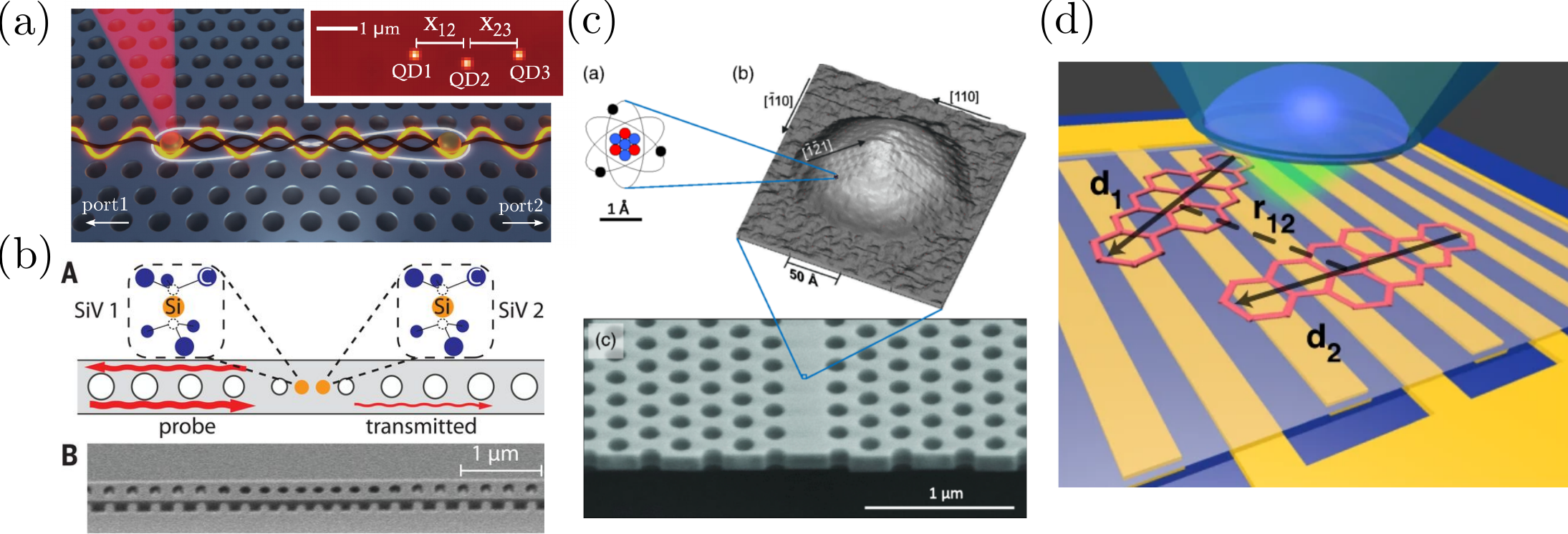}
\end{SCfigure}
\addtocounter{figure}{-2}
\graffito{\vspace{-4.75cm}
	\captionof{figure}[Scheme of different solid-state platforms.]{\label{fig:ExperimentalSetups}		\textbf{Scheme of different solid-state platforms. } 
		(a) 
		Two QDs photon-coupled embedded in a photonic-crystal waveguide (from \colorref{TiranovCollectiveSuper2023} reprinted with permission from AAAS ).
		(b)
		Diamond nanocavity containing two SiV centers  (from \colorref{EvansPhotonmediatedInteractions2018} reprinted with permission from AAAS).
		(c)
		Quantum dot embedded in a photonic crystal waveguide (reproduced with permission from Refs.\cite{MarquezAtomicallyResolved2001,LodahlInterfacingSingle2015}).
		(d)
		Sample of a highly doped napthalene crystal doped with DBATT molecules (extracted from Ref.\cite{TrebbiaTailoringSuperradiant2022}).
	}
}

	There are numerous types of solid-state quantum emitters that can be coupled to nanophotonic structures, including semiconductor quantum dots (QDs)~\cite{LodahlInterfacingSingle2015}, molecules in a crystalline host~\cite{WangTurningMolecule2019,HettichNanometerResolution2002,ToninelliSingleOrganic2021,TrebbiaTailoringSuperradiant2022,LangeSuperradiantSubradiant2024}, or  defect centres in semiconductors~\cite{CastellettoSiliconCarbide2020,DurandBroadDiversity2021,GritschNarrowOptical2022}. This Thesis explores how such emitters can exhibit cooperative effects, potentially leading to new physical insights and applications in photonic quantum technologies.
	%
	%
	%
	%

	 \paragraph{Cooperative effects. }
	
	Cooperative effects give rise to some of the most intriguing phenomena in nature, whether in the classical or quantum realm, as their uniqueness stems from the interplay among individual constituents rather than from the isolated dynamics of the components themselves~\cite{ReitzCooperativeQuantum2022}.

	Besides being a topic of great relevance from a fundamental perspective,  interest in controlling and harnessing quantum cooperative effects has grown significantly in recent years, particularly for their potential in emerging quantum technologies~\cite{DeutschHarnessingPower2020}, e.g., in quantum computation~\cite{OBrienOpticalQuantum2007,PaulischUniversalQuantum2016}, quantum communication~\cite{GisinQuantumCryptography2002,BriegelMeasurementbasedQuantum2009,KimbleQuantumInternet2008}, or quantum metrology~\cite{FacchinettiStoringLight2016,OstermannProtectedState2013,ManzoniOptimizationPhoton2018}.
	Notably, light-matter platforms, such as cavity/waveguide QED systems~\cite{HarocheExploringQuantum2006} or quantum nanophotonic devices~\cite{Gonzalez-TudelaLightMatter2024}, provide an ideal playground for observing and exploiting quantum cooperative effects. In these systems, quantum light---whether described  by the quantum electromagnetic vacuum or a single (multi-) mode confined in a photonic structure---interacts with multiple quantum emitters, leading to non-trivial cooperative light-matter effects. 
	The cooperative phenomena emerging from such systems is one of the central topics in quantum optics~\cite{DickeCoherenceSpontaneous1954,FicekQuantumInterference2005,GarrawayDickeModel2011,KirtonIntroductionDicke2019,ReitzCooperativeQuantum2022}. 
	
	The minimal implementation of this paradigm, i.e., \textit{two quantum emitters interacting with a common electromagnetic environment}, already captures the essential features of collective phenomena such as superradiant emission and the emergence of dark states~\cite{FicekQuantumInterference2005,LehmbergRadiationAtom1970}.
	Minimal models of two and three quantum emitters have been studied extensively in the literature~\cite{LehmbergRadiationAtom1970,RiosLeiteLineshapeCooperative1980,RichterPowerBroadening1982,FicekEffectInteratomic1983,PalmaPhasesensitivePopulation1989,VaradaTwophotonResonance1992,ItanoPhotonAntibunching1988,TanasStationaryTwoatom2004,BeigeTransitionAntibunching1998,LembessisTwoatomSystem2013,LakshmiEffectPairwise2014,AhmedCompetitionEffects2014,PengFilteredStrong2019,DarsheshdarPhotonphotonCorrelations2021,PengDirectionalNonclassicality2020,WangDipolecoupledEmitters2020}, revealing rich emergent phenomena including superradiance~\cite{DeVoeObservationSuperradiant1996,MlynekObservationDicke2014}, qubit entanglement generation~\cite{Gonzalez-TudelaEntanglementTwo2011,TanasEntanglingTwo2004,AlharbiDeterministicCreation2010}, spin and light squeezing~\cite{FicekPhotonAntibunching1984,FicekSqueezingTwoatom1994,HaakhSqueezedLight2015}, non-classical photon correlations~\cite{ItanoPhotonAntibunching1988,PengFilteredStrong2019,DarsheshdarPhotonphotonCorrelations2021,PengDirectionalNonclassicality2020}, entangled photon emission~\cite{WangDipolecoupledEmitters2020}, and molecule localization with nanometer resolution~\cite{HettichNanometerResolution2002,ZhangVisualizingCoherent2016}.
	The insights provided by these minimal theoretical models apply to a large variety of physical systems, including coupled quantum dots~\cite{GerardotPhotonStatistics2005,ReitzensteinCoherentPhotonic2006,LauchtMutualCoupling2010,PatelTwophotonInterference2010}, trapped ions~\cite{DeVoeObservationSuperradiant1996,EschnerLightInterference2001}, Rydberg atoms~\cite{AtesAntiblockadeRydberg2007,AmthorEvidenceAntiblockade2010,PritchardCorrelatedPhoton2012}, molecular systems~\cite{HettichNanometerResolution2002,ZhangVisualizingCoherent2016,TrebbiaTailoringSuperradiant2022,LangeSuperradiantSubradiant2024}, and superconducting qubits~\cite{LambertSuperradianceEnsemble2016,MlynekObservationDicke2014,VanLooPhotonMediatedInteractions2013}.
	Interest on the quantum optical properties of systems of few interacting emitters has been further propelled by the development of photonic nanostructures that mediate and enhance emitter-emitter interactions~\cite{LodahlInterfacingSingle2015,ChangColloquiumQuantum2018,HaakhSqueezedLight2015,ReitzensteinCoherentPhotonic2006,LauchtMutualCoupling2010,Gonzalez-TudelaEntanglementTwo2011,MlynekObservationDicke2014,VanLooPhotonMediatedInteractions2013}.

%


%
\begin{SCfigure}[1][h!]
	\includegraphics[width=0.92\columnwidth]{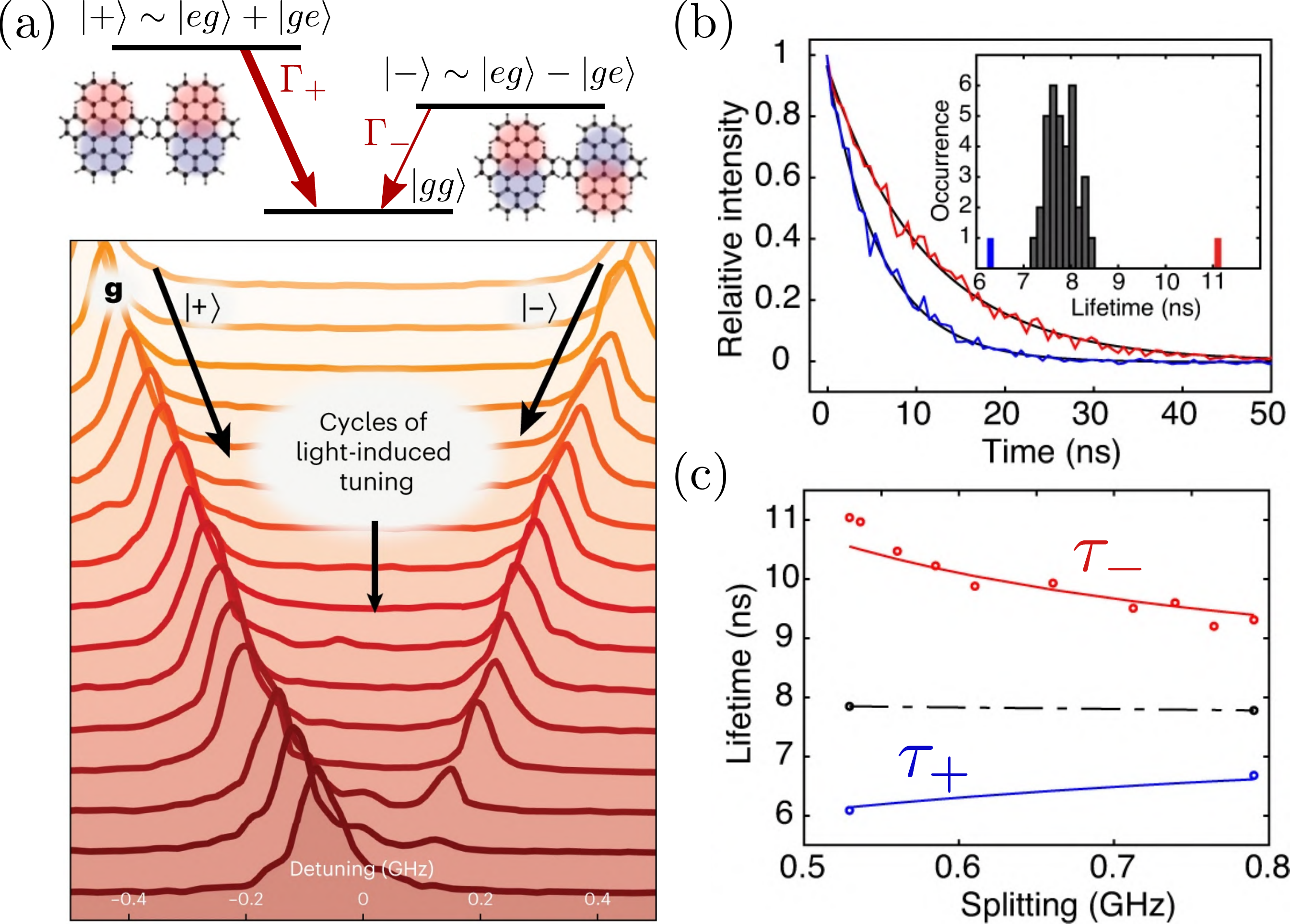}
	\captionsetup{justification=justified}
	\caption[Experimental observation of collective effects in solid-state emitters.]{\label{fig:SuperradianceExperimental}		\textbf{Experimental observation of collective effects in solid-state emitters. } 
		(a) Fluorescence spectra of two DBT molecules as they are tuned into resonance (reproduced with permission from Springer Nature from \colorref{LangeSuperradiantSubradiant2024}). The upper panel illustrate the energy diagram, where $|\pm\rangle $ are the super- and subradiant states. 
		(b) Decay curves of the subradiant (in red) and superradiant states (in blue), and (c) evolution of the lifetime of the subradiant (in red) and superradiant (in blue) states when the molecular detuning is shifted.
		 Figures (b, c) extracted from Ref.\cite{TrebbiaTailoringSuperradiant2022} 
		}
\end{SCfigure}

\paragraph{Challenges in solid-state platforms. }
Platforms in which optical photons are interfaced with solid-state quantum emitters (QEs)---see \reffig{fig:ExperimentalSetups}---, such as quantum dots (QDs)~\cite{LodahlInterfacingSingle2015}, molecules~\cite{ToninelliSingleOrganic2021} or colour centres~\cite{SipahigilIntegratedDiamond2016,AwschalomQuantumTechnologies2018}, are considered particularly promising for quantum technologies, given their potential for fast operation, on-chip integration, low-noise, and capability of transmission over long distances~\cite{OBrienPhotonicQuantum2009,WangIntegratedPhotonic2020}. In these solid-state platforms, proposals that can be considered variants of reservoir engineering have been made for the formation of steady-state entanglement between qubits through dissipative coupling  mediated by photonic nanostructures~\cite{Gonzalez-TudelaEntanglementTwo2011,Martin-CanoDissipationdrivenGeneration2011,Gonzalez-TudelaMesoscopicEntanglement2013,RamosQuantumSpin2014,PichlerQuantumOptics2015,HaakhSqueezedLight2015,ChangColloquiumQuantum2018,ReitzCooperativeQuantum2022}. 

Nevertheless, decoherence, while fundamental and ubiquitous, is not the only challenge quantum systems face in generating quantum effects, such as stable entanglement. When considering practical implementations of solid-state quantum emitters embedded in photonic environments, two additional challenges must be addressed due to technological limitations during the fabrication process:
\begin{enumerate}[label=\textcolor{Maroon}{(\roman*)}]
	\item \textcolor{Maroon}{Fabrication of strongly interacting quantum emitters. }
	There has been experimental progress in this direction, with reports of the formation of  superradiant and subradiant states---corresponding to the symmetric triplet and antisymmetric singlet Bell states---in  systems consisting of two solid state QEs, such as closely spaced molecules interacting via dipole coupling~\cite{HettichNanometerResolution2002,TrebbiaTailoringSuperradiant2022,LangeSuperradiantSubradiant2024} or coupled via a ring resonator~\cite{RattenbacherOnchipInterference2023}, SiV centres in a waveguide~\cite{SipahigilIntegratedDiamond2016,MachielseQuantumInterference2019} or semiconductor QDs coupled to cavities~\cite{ReitzensteinCoherentPhotonic2006,LauchtMutualCoupling2010} or waveguides~\cite{KimSuperRadiantEmission2018,GrimScalableOperando2019,TiranovCollectiveSuper2023,ChuIndependentElectrical2023}, as well as photon-mediated coherent interactions between color centers in a nanocavity~\cite{EvansPhotonmediatedInteractions2018}.
	This progress in synthesizing strongly interacting quantum emitters concurs with advancements in the design of light-matter interfaces with unprecedented radiative rate enhancements, which includes technologies such as plasmonic nanoantennas~\cite{ChikkaraddySinglemoleculeStrong2016,HoangUltrafastSpontaneous2015,HoangUltrafastRoomTemperature2016} or hybrid cavity-antenna systems~\cite{GurlekManipulationQuenching2018,ShlesingerHybridCavityantenna2023,GuskenEmissionEnhancement2023}.
	Arguably, light-matter platforms enabling the coupling between quantum emitters while suppressing undesired emission, such as emitter interactions
	within a photonic band gap~\cite{ChangColloquiumQuantum2018,ArcariNearUnityCoupling2014}, giant atoms~\cite{KockumDecoherenceFreeInteraction2018,KannanWaveguideQuantum2020} or subwavelength arrays of quantum emitters~\cite{ReitzCooperativeQuantum2022,BettlesCooperativeOrdering2015,GlicensteinCollectiveShift2020,ParmeeSignaturesOptical2020,SrakaewSubwavelengthAtomic2023,MassonDickeSuperradiance2024,HolzingerCooperativeSubwavelength2022}  stand out as privileged platforms to achieve strong interaction regimes. Indeed, in recent years, subwavelength atomic arrays trapped by optical tweezers
	have garnered attention as notable platforms for researching and implementing cooperative phenomena.
	
	Particularly, molecular aggregates stand out as promising candidates to achieve this goal~\cite{SaikinPhotonicsMeets2013}.
	Crucially, these aggregates can be synthetically engineered, with dimers of chromophores coupled by insulating
	bridges enabling precise control of the interchromophoric distance within just a few nanometers~\cite{DiehlEmergenceCoherence2014}, making
	them promising platforms for the study and implementation of cooperative quantum phenomena in the solid
	state~\cite{ReitzCooperativeQuantum2022,HolzingerCooperativeSubwavelength2022}. Relevant progress also include the synthesis of nanocrystals, allowing to embed two molecules
	within nanometer distance~\cite{PazzagliSelfAssembledNanocrystals2018,LangeSuperradiantSubradiant2024}. 

	This aligned development has called for further exploration into the phenomenology of strongly interacting quantum emitters when they are coupled to photonic structures~\cite{PlankensteinerCavityAntiresonance2017, PlankensteinerEnhancedCollective2019}, with prospects such as enabling novel quantum-optical phases~\cite{ParmeeSignaturesOptical2020,PerczelTopologicalQuantum2017,ReitzCooperativeQuantum2022} or the fast and robust preparation of the emitter ensemble into its entangled eigenstates~\cite{NicolosiDissipationinducedStationary2004, Gonzalez-TudelaEntanglementTwo2011}.

	\item \textcolor{Maroon}{Inhomogeneous broadening in solid-state emitters. }
	In the case of molecules and QDs, it is recognized that the main obstacle faced is inhomogeneous broadening~\cite{GrimScalableOperando2019}, i.e., the differences in the natural frequencies of the QEs due to different size, strain and composition of QDs~\cite{OBrienPhotonicQuantum2009,PatelTwophotonInterference2010,GieszCoherentManipulation2016, EvansPhotonmediatedInteractions2018,LukinIntegratedQuantum2020,LukinTwoEmitterMultimode2023}---in other words, the fabrication of identical emitters---, or the different local matrix environments of the molecules, which are also difficult to position selectively~\cite{ToninelliSingleOrganic2021,TrebbiaTailoringSuperradiant2022}. Nevertheless, recent progress in the identification of closely spaced molecules
	embedded in organic crystals also make these viable setups~\cite{LangeSuperradiantSubradiant2024,HettichNanometerResolution2002,TrebbiaTailoringSuperradiant2022}, with recent developments allowing to address 
	inhomogeneous broadening~\cite{GrimScalableOperando2019} by permanently tuning different emitters into resonance via optical drives~\cite{ColauttiLaserInducedFrequency2020}. 
\end{enumerate}

In this Thesis, we focus on quantum emitters with non-identical frequencies, addressing the challenge posed by the inhomogeneous broadening of the quantum emitters (QEs).
We begin with the simpler case of free-space interactions and then extend the analysis to scenarios where the emitters are coupled to a lossy photonic structure, specifically a single-mode cavity. 
In both cases, we gain valuable insights into cooperative phenomena. The second scenario, in particular, underscores technological applications, such as the dissipative generation of steady-state entanglement among the emitters. These studies pave the way for new approaches to understanding and controlling quantum emitter interactions in both ideal and state-of-the-art technological configurations.

\section{Summary of contents}

This Thesis is structured as follows.

In \refch{ch:TheoreticalBackground}, we provide a comprehensive background on the theoretical concepts central to this Thesis, focusing on light-matter interactions and open quantum systems. 
The Chapter begins with a review of the quantization of the electromagnetic field and the description of quantum light-matter interactions, followed by an introduction to the theoretical frameworks for describing open quantum systems in both the Schrödinger and Heisenberg pictures. Particular attention is given to the master equation and quantum Langevin equation approaches.
We also discuss cascaded quantum systems and the quantum trajectories framework for modelling dissipative systems. Additionally, the Chapter examines the spectral properties of photon emission, methods for calculating photon correlation functions, and techniques for quantum parameter estimation, with a focus on  the classical Fisher information. The Chapter concludes with an overview of relevant theoretical and computational methods.

In \refch{ch:TwoPhotonResonance}, we introduce one of the central models of this Thesis for studying collective effects: two nonidentical interacting quantum emitters driven by a coherent field. We analyse the fundamental properties of this model by considering a specific realization of the interaction mediated by the surrounding vacuum electromagnetic field while the emitters are driven at the two-photon resonance.
Analytical expressions for the stationary density matrix are derived using effective models. Additionally, we explore the fluorescence spectrum across the parameter space and demonstrate the potential of resonance fluorescence measurements for estimating the distance between emitters using quantum parameter estimation techniques.
The results presented in this chapter have been published in Physical Review Research~\cite{Vivas-VianaTwophotonResonance2021}.

In \refch{chapter:Unconventional}, we take a brief detour from the central topic of this Thesis---harnessing collective effects---to present a mechanism that challenges the intuitive notion that a virtual state remains unpopulated in the presence of dissipation. This leads to a novel and unconventional description of virtual states under dissipative dynamics.
We revisit the concept of virtual states in quantum mechanics and introduce a toy model consisting of three levels to illustrate what we term as \textit{unconventional mechanism of virtual state population through dissipation}.
To analyse this phenomenon, we develop a technique called hierarchical adiabatic elimination (HAE), which provides analytical approximations for the time-dependent elements of the  density matrix of the system and expressions for the characteristic timescales. By examining quantum trajectories, we offer essential intuition about the underlying mechanism of population through dissipation.
%
%
The results presented in this chapter have been published in Physical Review A~\cite{Vivas-VianaUnconventionalMechanism2022}.

In \refch{chapter:Entanglement}, we revisit the model introduced in \refch{ch:TwoPhotonResonance}---now allowing the emitters to be either coherently or incoherently driven---and extend it by embedding the emitters within a photonic structure modelled as a lossy single-mode cavity. We use this extended model to explore the generation of steady-state entanglement between the emitters.
We introduce the extended system and derive an effective master equation by tracing out the cavity mode, employing the Nakajima-Zwanzig approach to describe the system dynamics.
In this framework, we identify and analyse five different mechanisms for generating both stable and metastable entanglement between the emitters, spanning the parameter space of both the emitters and the cavity. A particular focus is placed on a mechanism we term as \textit{frequency-resolved Purcell effect}, where the cavity selectively enhances specific eigenenergy transitions, stabilizing the emitters into either superradiant or subradiant eigenstates. We prove that this driven-dissipative strategy can be implemented using both coherent and incoherent drives for the case of $N=2$ emitters. Furthermore, we show that, under incoherent excitation, this strategy can be scaled up to $N$ interacting emitters.
Additionally, we address two critical challenges associated with quantum entanglement: its detection, based on the quantum and classical properties of the emitted light, and its robustness against additional decoherence channels, such as pure dephasing. Finally, we provide detailed characterization of the emergent entanglement-generation mechanisms within the parameter space, offering novel physical insights and analytical descriptions.
The results presented in this chapter have been published in Physical Review Letters~\cite{Vivas-VianaFrequencyResolvedPurcell2024} and Physical Review Research~\cite{Vivas-VianaDissipativeStabilization2024}.

In \refch{chapter:Outlook}, we present our latest results on the generation of entanglement between photonic modes, along with a brief perspective on ongoing work in quantum metrology using spectral measurements. 
The chapter focuses on the fundamental properties of the quantum state of the output field emitted by a quantum emitter. Using the cascaded formalism, we explore how capturing these fields reveals entangled states between two temporally overlapping but spectrally orthogonal photonic modes. Furthermore, we demonstrate that frequency correlations between photonic modes can provide metrological enhancements compared to single-photon measurements.
We begin with an introduction to another central model: a coherently driven two-level system. This includes discussions of the Heitler and Mollow regimes, as well as an overview of the landscape of multi-photon processes in the frequency domain.
Next, we present theoretical and experimental results on the generation of entanglement between photonic modes, with a particular emphasis on the theoretical framework and its implications. These  results were performed in a joint experimental-theoretical collaboration with the experimental group led by Prof. Simone Gasparinetti at Chalmers University.
Finally, we provide an outlook on recent theoretical advancements in quantum metrology using spectral measurements. Specifically, we discuss how system parameters can be estimated via the classical Fisher information, with an upgrade to a frequency-resolved formulation of this quantity, highlighting its potential for enhanced precision.
The results on the generation of entanglement between photonic modes presented in this Chapter have been published in npj Quantum Information~\cite{YangEntanglementPhotonic2025}, while those on quantum metrology represent ongoing work$^{\textcolor{Maroon}{*}}$. 
\graffito{
	$^{*}$  This work is now published as an arXiv preprint~\cite{Vivas-VianaQuantumMetrology2025}.
}

\cleardoublepage

\chapter{Theoretical Background}\label{ch:TheoreticalBackground}


\section{Introduction}

This chapter is dedicated to the essential theoretical techniques used in this Thesis, covering fundamental concepts and computational methods to provide a self-contained material.

We begin with the quantization of the electromagnetic field and the description of light-matter interactions.
Next, we discuss the theoretical description of open quantum systems in both the Schrödinger and Heisenberg pictures, focusing on the master equation formalism and quantum Langevin equations.
Additionally, we introduce the concept of cascaded quantum systems and provide an alternative description of dissipative system using the quantum trajectories approach. The discussion is further extended to discuss the spectral properties of photon emission and introduce methods for computing photon correlation functions.
A concise introduction to quantum entanglement is also provided, along with a detailed presentation of two particular entanglement measures: concurrence and negativity.
Moreover, we offer an overview of quantum parameter estimation, with a focus on the classical Fisher information.
Finally, we present theoretical and computational techniques useful for the computations in later chapters.



\section[Canonical quantization of the  free electromagnetic field]{Canonical quantization of the  free \protect\newline electromagnetic field}

A fundamental aspect of quantum optics is the quantization of the electromagnetic field and its interaction with matter.
In this section, we outline the derivation of the canonical quantization of the free electromagnetic field, building upon the detailed analyses commonly found in textbooks on quantum optics and quantum field theory~\cite{Cohen-TannoudjiPhotonsAtoms1997,LoudonQuantumTheory2000,GerryIntroductoryQuantum2004,WallsQuantumOptics2008,SteckQuantumAtom,RyderQuantumField1996,ItzyksonQuantumField2005,MaggioreModernIntroduction2005,DiracLecturesQuantum2001,HenneauxQuantizationGauge1992}. 

\vspace{-3.5mm}
\subsection{Maxwell equations }
\vspace{-3.5mm}
The description of light
\graffito{
	$^{*}$This set of equations captures the fundamental nature of light.\newline
	\eqref{eq:Maxwell_1}: a varying magnetic field produces an electric field.\newline
	\eqref{eq:Maxwell_2}:  there are no magnetic charges, i.e., magnetic monopoles.
	\newline
	\eqref{eq:Maxwell_3}: the total charge inside a closed surface may be obtained by integrated the normal component of the electric field over the surface. \newline
	\eqref{eq:Maxwell_4}: changing electric fields produce magnetic fields.}
in terms of electric and magnetic vector fields, $\mathbf{E}\equiv\mathbf{E}(\mathbf{r},t)$ and $\mathbf{B}\equiv\mathbf{B}(\mathbf{r},t)$, respectively, is formulated by the following set of partial differential equations, the Maxwell equations of motion$^{\textcolor{Maroon}{*}}$,
\begin{subequations}
	\begin{align}
		&\nabla \times \mathbf{E}=-\frac{\partial  \mathbf{B}}{\partial t} ,  &&\text{(Faraday's law of induction)} 
		\label{eq:Maxwell_1}\\
		&\nabla \cdot \mathbf{B}=0,  &&\text{(Gauss's law for the magnetic field)} \label{eq:Maxwell_2}\\
		&\nabla \cdot \mathbf{E}=\frac{\rho}{\varepsilon_0},  &&\text{(Gauss's law)}  \label{eq:Maxwell_3}\\
		&\nabla \times \mathbf{B} = \frac{1}{c^2}\frac{\partial \mathbf{E}}{\partial t} + \mu_0 \mathbf{j},  &&\text{(Ampère's circuital law)} \label{eq:Maxwell_4}
	\end{align}
\end{subequations}
where $\rho$ is the electric charge density, $\mathbf{j}$ is the current density, $\varepsilon_0$ is the vacuum permittivity, $\mu_0$ is the magnetic constant, and $c=(\varepsilon_0 \mu_0)^{-1/2}$ is the speed of light.

The description of the electromagnetic field
\graffito{
	$^{*}$Using the explicit covariant description of the Maxwell equations via the  electromagnetic tensor,
	$$F_{\mu \nu}= \partial_\mu A_\nu - \partial_\nu A_\mu,$$
	and
	the four vector potential,  $A_\mu=(\phi,\vec A)$, the resulting Maxwell equations are obtained by two sets of equations. The first two equations,~\cref{eq:Maxwell_1,eq:Maxwell_2}, come from a geometrical restriction, the so-called \textit{Bianchi identity}~\cite{RyderQuantumField1996,NakaharaGeometryTopology2018}, 
	$$\partial_\mu F_{\nu \rho}+\partial_\nu F_{ \rho \mu}+\partial_\rho F_{\mu \nu}=0,$$
	whilst the other two equations,~\cref{eq:Maxwell_3,eq:Maxwell_4}, are physical restriction by imposing the variational principle into the electromagnetic action via the Noether's theorem: 
	$$
	\partial_\mu F^{\mu \nu}=j^\nu,
	$$
	where $j^\nu=(\rho, \mathbf{j})$.
}
 is then complete by considering the local conservation of the charge---$\text{U}(1)$ symmetry---as a consequence of~\cref{eq:Maxwell_1,eq:Maxwell_4}:   
\begin{equation}
	\frac{\partial \rho}{\partial t} + \nabla \cdot \mathbf{j}=0.
\end{equation}
The structure of Maxwell equations in~\cref{eq:Maxwell_1,eq:Maxwell_2,eq:Maxwell_3,eq:Maxwell_4} suggest that the electric and magnetic fields can always be expressed in terms of the electromagnetic potentials$^{\textcolor{Maroon}{*}}$:  %
\vspace{-2.5mm}
\begin{subequations}
		\label{eq:EMpotentials}
	\begin{align}
			\mathbf{E}&=-\frac{\partial \mathbf{A}}{\partial t}  - \nabla \phi, \\
			 \mathbf{B}&=\nabla \times \mathbf{A},
	\end{align}
\end{subequations}
where $\mathbf{A}$ is the vector potential, and $\phi$ the scalar potential, both defined in $(\mathbf{r},t)$. Given this definition, the new set of variables, $(\phi, \mathbf{A})$, automatically satisfies~\cref{eq:Maxwell_2,eq:Maxwell_3}, reducing the complexity of the problem to~\cref{eq:Maxwell_1,eq:Maxwell_4} in terms of the electromagnetic potentials:
\begin{subequations}
	\begin{align}
		\label{eq:MaxwellScalarPotential}
		&\Delta \phi =-  \left(\frac{\rho}{\varepsilon_0} +\nabla \cdot \frac{\partial  \mathbf{A}}{\partial t} \right)
		,  \\
		\label{eq:MaxwellVectorPotential}
		&\square \mathbf{A} 
		= \mu_0 \mathbf{j} - \nabla\left[ \nabla\cdot \mathbf{A} +\frac{1}{c^2}  \frac{\partial \phi}{\partial t}  \right]. 
	\end{align}  
\end{subequations}
\vspace{-9.5mm}
\subsection{Gauge invariance and the Coulomb gauge}
\vspace{-1.5mm}
The state of the EM field is determined by specifying the initial condition for $\mathbf{A}$ and $\partial \mathbf{A}/\partial t$ at all points $\mathbf{r}$ in \cref{eq:MaxwellScalarPotential,eq:MaxwellVectorPotential}. However, the solution is not uniquely determined since there is a local symmetry that leaves the electric and magnetic fields invariant under gauge transformations:
\vspace{-1.5mm}
\begin{subequations}
	\label{eq:GaugeTransformation}
	\begin{align}
	&\mathbf{A} \rightarrow 	\mathbf{A'}= \mathbf{A}+\nabla  \chi , \\
	&\phi \rightarrow \phi' = \phi - \frac{\partial \chi }{\partial t},
	\end{align}
\end{subequations}
where $\chi =\chi (\mathbf{r},t)$ is an arbitrary function. The existence 
\graffito{
	$^{**}$	
	Gauge transformations establish an equivalence relation within the set of electromagnetic potentials, highlighting the inherent gauge redundancy, or gauge ambiguity. This redundancy reflects the freedom to select any member of an equivalence class to represent the electromagnetic field. By imposing additional constraints, known as gauge fixing conditions, we eliminate this mathematical redundancy, retaining only the physical degrees of freedom~\cite{HenneauxQuantizationGauge1992}.
}
of this local symmetry results in a redundant description---a gauge redundancy$^{\textcolor{Maroon}{**}}$---
of the electromagnetic field, introducing extra degrees of freedom that are not physical.  Therefore, we must \textit{fix the gauge} to constraint the definition of the electromagnetic potentials. 
A commonly used gauge fixing condition is the Coulomb gauge (or radiation gauge, or transversal), in which we set:
\begin{equation}
\nabla \cdot \mathbf{A}=0.
\end{equation} 
This choice considerably simplifies the Maxwell equations from \cref{eq:MaxwellScalarPotential,eq:MaxwellVectorPotential} to:
\begin{subequations}
		\label{eq:MaxwellEqtsCoulombGauge}
	\begin{align}
			&\Delta \phi=-\frac{\rho}{\varepsilon_0},  		\label{eq:MaxwellEqtsCoulombGaugePotential} \\
			&\square \mathbf{A}= \mu_0\mathbf{j}.
					\label{eq:MaxwellEqtsCoulombGaugeVector}
	\end{align}
\end{subequations}
Consequently, the scalar potential is determined by the Poisson's equation of electrostatics, resulting in the instantaneous Coulomb potential, while the vector potential is ruled by an inhomogeneous wave equation whose solution is given in terms of plane waves and retarded and advanced propagators. We note that in this gauge, the vector potential $\mathbf{A}$ is transverse by definition, i.e., $\nabla \cdot \mathbf{A}=0$. Then, according to Helmoltz's theorem, we can split any arbitrary field $\mathbf{C}$ as a unique decomposition of a transverse and longitudinal components, $\mathbf{C}=\mathbf{C}_T+\mathbf{C}_L$, such that:
\begin{equation}
	\nabla \cdot \mathbf{C}_T=0 \quad \text{and} \quad \nabla \times \mathbf{C}_L=0.
\end{equation} 
As a result, the electric and magnetic fields are exclusively described by the physical degrees of freedom remaining after the gauge fixing process.
\begin{subequations}
	\begin{empheq}[box=\colorboxed{Maroon}]{align}
		&\mathbf{E}_T = -\frac{\partial \mathbf{A}}{\partial t}, \quad \hspace{0.235cm} \text{and}\quad \mathbf{E}_L = -\nabla \phi, 
		\label{eq:ElectricFieldRelations}
		\\
		&\mathbf{B}_T = \nabla \times \mathbf{A}, \quad \text{and}\quad \mathbf{B}_L = 0.
		\label{eq:MagneticFieldRelations}
	\end{empheq}
\end{subequations}
We must note that this gauge, in contrast to the 
\graffito{
	$^*$The choice of gauge often introduces conceptual challenges, and, at times, might lead to physical misunderstandings. Nevertheless, the uniqueness of physics is guaranteed as long as we are aware of the chosen gauge  and the transformation being applied. Furthermore, according to the gauge principle, all physical observables must be gauge-independent.}
Lorentz gauge, ($\partial \phi /\partial t + \nabla \cdot \mathbf{A}=0$), is not explicitly Lorentz invariant$^{\textcolor{Maroon}{*}}$. Covariance is explicitly lost in the Coulomb gauge, however, this gauge brings the theoretical advantage that the dynamical information of the electromagnetic field is absolutely described by the transversal component of the vector potential, $\mathbf{A}_T$, reducing the complexity of the problem from the initial six degrees of freedom, corresponding to the electric and magnetic field components, up to only two degrees of freedom, corresponding to the light polarization. Additionally, longitudinal components of the electromagnetic field emerge from spatial changes on the scalar potential $\phi$.
\vspace{-2mm}
\paragraph{Canonical Quantization.}
In order to quantise the electromagnetic field, we follow the canonical quantization program~\cite{DiracLecturesQuantum2001,HenneauxQuantizationGauge1992}. We must obtain a description of the EM field in terms of a classical Hamiltonian and identify the canonical coordinates of the theory, $(q_k,p_k)$, satisfying the fundamental Poisson brackets $\{q_k,p_{k'}\}=\delta_{k,k'}$. The quantization is achieved by upgrading the classical coordinates to quantum operators and replacing the Poisson brackets to commutators multiplied by $(i\hbar)^{-1}$: 
\begin{equation}
	  \{q_k,p_{k'}\}=\delta_{k,k'}\quad \longrightarrow \quad   [\hat q_k, \hat p_{k'}] =i\hbar \delta_{k,k'}.
	 \label{eq:CanonicalQuantization}
\end{equation}

In the absence of charges and current ($\rho=0,\mathbf{j}=\mathbf{0}$), the Maxwell equations from \cref{eq:MaxwellEqtsCoulombGaugePotential,eq:MaxwellEqtsCoulombGaugeVector} reduce to a simplified equation for the vector potential:
\begin{equation}
	\label{eq:MaxwellEqCoulombFree}
\square \mathbf{A}=0.
\end{equation}
To solve this wave equation, we assume that the electromagnetic field is confined within a finite cube of size $L=V^{1/3}$ and volume $V=L^3$, although this assumption becomes irrelevant once we take the limit $L\rightarrow \infty$. The purpose of introducing this artificial cube is to enable the imposition of periodic boundary conditions on its faces. Hence, we can generally write a formal solution of \eqref{eq:MaxwellEqCoulombFree} in terms of spatiotemporal modes,
\begin{equation}
	\mathbf{A}(\mathbf{r},t)=\sum_{\mathbf{k}} \sum_{s=1,2} \mathbf{e}_{\mathbf{k},s} \left[ A_{\mathbf{k},s} e^{-i \omega_k t + i \mathbf{k}\cdot \mathbf{r}} + A_{\mathbf{k},s}^* e^{i \omega_k t - i \mathbf{k}\cdot \mathbf{r}}\right] ,
\end{equation}  
where $\mathbf{k}$ is the wavevector, $s$ denotes the two independent polarizations,  $A_{\mathbf{k},s}$ is the complex amplitude of the mode, $\mathbf{e}_{\mathbf{k},s}$ is the polarization vector, and $\omega_k=c |\mathbf{k}|$ is the mode frequency. 
Given the periodic boundary conditions, and in accordance with the wave equation and the Coulomb gauge, the wavevector $\mathbf{k}$ takes the discrete values:
\begin{equation}
	\mathbf{k}=(k_x,k_y,k_z)=\frac{2\pi}{L} (n_1,n_2,n_3), \quad \text{with} \quad n_i \in \mathbb{Z},
\end{equation}
while the polarization vector satisfies:
\begin{subequations}
	\begin{align}
		&\mathbf{e}_{\mathbf{k},s} \cdot \mathbf{e}_{\mathbf{k},s'}=\delta_{s,s'}, \quad \text{(Orthogonality condition)} 
		\label{eq:PolarizationProp1}\\
		&\mathbf{k} \cdot \mathbf{e}_{\mathbf{k},s}=0, \quad \quad \quad \  \text{(Transversality condition)}, 
		\label{eq:PolarizationProp2}\\
		&\sum_{s} e^i_{\mathbf{k},s} e^j_{\mathbf{k},s}=\delta_{ij}-\frac{k_i k_j}{|\mathbf{k}|^2}, \quad \text{with} \quad i,j=1,2,3,
		\label{eq:PolarizationProp3}
	\end{align}
\end{subequations}
where $e^i_{\mathbf{k},s}$ is the $i$th component of the vector $\mathbf{e}_{\mathbf{k},s}$. Additionally, we note that the polarization vectors form a right-handed system such that they define a propagation vector, $	\bm{\kappa}=\mathbf{e}_{\mathbf{k},1}\times \mathbf{e}_{\mathbf{k},2}=\mathbf{k}/|\mathbf{k}|.$
Recalling the relations in ~\cref{eq:ElectricFieldRelations,eq:MagneticFieldRelations}, we then derive the well-known expression for the classical Hamiltonian of a confined electromagnetic field, expressed as a sum of independent harmonic oscillators of unit mass:
\begin{equation}
	H=\frac{1}{2} \int_V dV (\varepsilon_0 \mathbf{E}^2+\frac{1}{\mu_0}\mathbf{B}^2)=\frac{1}{2}\sum_{\mathbf{k},s} (p_{\mathbf{k},s}^2+\omega_k^2 q_{\mathbf{k},s}),
	\label{eq:ClassicalHamiltonian}
\end{equation}
where $(q_{\mathbf{k},s},p_{\mathbf{k},s})$ are the canonical coordinates of the theory, defined via the relations
\begin{subequations}
	\begin{align}
		A_{\mathbf{k},s}&= \frac{1}{2\omega_k \sqrt{\varepsilon_0 V}}(\omega_k q_{\mathbf{k},s}+ip_{\mathbf{k},s}), \\
		A_{\mathbf{k},s}^*&= \frac{1}{2\omega_k \sqrt{\varepsilon_0 V}}(\omega_k q_{\mathbf{k},s}-ip_{\mathbf{k},s}),
	\end{align}
\end{subequations}
and satisfying the Poisson bracket relations: 
\begin{equation}
	\{ q_{\mathbf{k},s},p_{\mathbf{k'},s'} \}=\delta_{\mathbf{k},\mathbf{k'}}\delta_{s,s'}
\end{equation}

At this point, we can proceed with the canonical quantization program, as outlined in \eqref{eq:CanonicalQuantization}, by promoting the canonical coordinates to quantum operators that obey the canonical commutation relations: 
\begin{equation}
	[ \hat q_{\mathbf{k},s},\hat p_{\mathbf{k'},s'} ]=i\hbar \delta_{\mathbf{k},\mathbf{k'}}\delta_{s,s'}.
\end{equation}
Since the resulting Hamiltonian in \eqref{eq:ClassicalHamiltonian} corresponds to an infinite sum of quantum harmonic oscillators, we can define a new set operators that diagonalize it. These operators receive the name of annihilation ($\hat a_{\mathbf{k},s}$) and creation ($\hat a^\dagger_{\mathbf{k},s}$) operators:
\vspace{-1mm}
\begin{subequations}
		\label{eq:CreationOp}
\begin{align}
	\hat a_{\mathbf{k},s}=\frac{1}{\sqrt{2\hbar \omega_k}}(\omega_k \hat q_{\mathbf{k},s} + i \hat p_{\mathbf{k},s}), \\
	\hat a^{\dagger}_{\mathbf{k},s}=\frac{1}{\sqrt{2\hbar \omega_k}}(\omega_k \hat q_{\mathbf{k},s} - i \hat p_{\mathbf{k},s}),
\end{align}
\end{subequations}
satisfying the commutation relation:
\begin{equation}
	 [ \hat a_{\mathbf{k},s},\hat a^\dagger_{\mathbf{k'},s'} ]=\delta_{\mathbf{k},\mathbf{k'}}\delta_{s,s'}.
\end{equation}
As a consequence, the Hamiltonian of the electromagnetic field from \eqref{eq:ClassicalHamiltonian} in terms of the new operators reads
\begin{equation}
\hat H =\sum_{\mathbf{k},s} \hbar \omega_k \left(\hat a^\dagger_{\mathbf{k},s} \hat a_{\mathbf{k},s}+ \frac{1}{2}\right).
\end{equation}
The electromagnetic field is then described by a collection of quantum harmonic oscillators with frequency $\omega_k$, spanning a Hilbert space, $\mathcal{H}_{\text{EM}}$, which is formed by the tensor product of infinite-dimensional individual Hilbert spaces for each mode, $\mathcal{H}_{\text{EM}}^i$,  
\begin{equation}
	\mathcal{H}_{\text{EM}}=\bigotimes_{i=1}^\infty  \mathcal{H}_{\text{EM}}^i.
\end{equation}
Finally, we note that the mode amplitude has been promoted to a quantum operator, $\hat A_{\mathbf{k},s}= \sqrt{\hbar/2\omega_k \varepsilon_0 V} \hat a_{\mathbf{k},s}$
giving rise to the quantized version of the vector potential
\begin{equation}
	\mathbf{\hat  A}(\mathbf{r},t)=\sum_{\mathbf{k},s}   \sqrt{\frac{\hbar}{2\omega_k \varepsilon_0 V}}  \mathbf{e}_{\mathbf{k},s} \left[ \hat a_{\mathbf{k},s} e^{-i \omega_k t + i \mathbf{k}\cdot \mathbf{r}} + \hat a^\dagger_{\mathbf{k},s} e^{i \omega_k t - i \mathbf{k}\cdot \mathbf{r}}\right],
\end{equation}
and the electric and magnetic field operators:
\begin{subequations}
	\begin{empheq}{align}
		&\text{ \fontsize{10}{10}\selectfont $
	\mathbf{\hat E}(\mathbf{r},t)=i\sum_{\mathbf{k},s}  \sqrt{\frac{\hbar\omega_k}{2 \varepsilon_0 V}}  \mathbf{e}_{\mathbf{k},s} \left[ \hat a_{\mathbf{k},s} e^{-i \omega_k t + i \mathbf{k}\cdot \mathbf{r}} - \hat a^\dagger_{\mathbf{k},s} e^{i \omega_k t - i \mathbf{k}\cdot \mathbf{r}}\right], $}
	\label{eq:ElectricFieldOp}\\
		&\text{ \fontsize{10}{10}\selectfont $\mathbf{\hat B}(\mathbf{r},t)=\frac{i}{c}\sum_{\mathbf{k},s} \sqrt{\frac{\hbar\omega_k}{2 \varepsilon_0 V}}  (\bm{\kappa} \times \mathbf{e}_{\mathbf{k},s}) \left[ \hat a_{\mathbf{k},s} e^{-i \omega_k t + i \mathbf{k}\cdot \mathbf{r}} - \hat a^\dagger_{\mathbf{k},s} e^{i \omega_k t - i \mathbf{k}\cdot \mathbf{r}}\right].$}
	\label{eq:MagneticFieldOp}
	\end{empheq}	
\end{subequations}

\section{Quantum light-matter interaction}

The discussion so far has concerned with the quantization of the free electromagnetic field, that is, the field where there is no matter to interact with the radiation. In this section, we introduce the notion of quantum emitters and their interaction with the quantum electromagnetic field. 

\subsection{Minimal coupling hamiltonian}
 
We describe matter as a collection of $N$ charged particles of mass $m_i$ and charge $q_i$, such that, within the atomic Bohr approximation, the matter Hamiltonian reads
\begin{equation}
	\label{eq:MatterHamiltonian}
	\hat H_{M}=\sum_i^N \frac{\mathbf{\hat p}_i^2}{2 m_i}+V, \quad \text{with}\quad 	V= \frac{1}{2} \sum_{i\neq j}^N  \frac{1}{4\pi \varepsilon_0}  \frac{q_i q_j}{|\mathbf{\hat r}_i-\mathbf{\hat  r}_j|}.
\end{equation} 
The first term corresponds to the kinetic energy, and the second denotes a Coulomb interaction term between particles. Here, $ \mathbf{\hat r }_i$ and $\mathbf{\hat p}_i=-i\nabla_i$ are the position and momentum operators for the $i$th particle, satisfying the canonical commutation relations $[ \mathbf{\hat r}_i, \mathbf{\hat p}_j]=i \hbar \delta_{ij}$. These operators act in the matter Hilbert space, $\mathcal{H}_\text{M}$, composed by the tensor product of the $N$ individual particles, $ \mathcal{H}_\text{M}^i$: 
\begin{equation}
	\mathcal{H}_\text{M}=\bigotimes_{i=1}^N \mathcal{H}_\text{M}^i.
\end{equation}
%
%
%

The interaction between the quantized light and matter is essentially given by the minimal-coupling Hamiltonian:
\begin{equation}
\colorboxed{Maroon}{
	\hat H= \sum_i \frac{\left[\mathbf{\hat p}_i-q_i \mathbf{\hat A}(\mathbf{r}_i)\right]^2}{2 m_i} +V + \hat H_{\text{EM}},
}
	\label{eq:LightMatterHamiltonian}
\end{equation}
where the momentum operator has been replaced by
\begin{equation}
	\mathbf{\hat p}_i \rightarrow \mathbf{\hat p}_i-q_i \mathbf{\hat A}(\mathbf{r}_i),
\end{equation}
such that the resulting light-matter Hamiltonian is gauge invariant. In other words, the interaction is determined by the transformation properties of the symmetry group of the electromagnetic field, the group $\text{U}(1)$. 
We note that this Hamiltonian can be seen as the non-relativistic limit of the Dirac equation in relativistic QED~\cite{Cohen-TannoudjiPhotonsAtoms1997,RyderQuantumField1996,ItzyksonQuantumField2005,MaggioreModernIntroduction2005}, where, additionally, we have also assumed that the spin-magnetic interaction term is negligible. 

The total Hilbert space $\mathcal{H}_T$ is the tensor product of the matter Hilbert space, $\mathcal{H}_\text{M}$, with the electromagnetic Hilbert space, $\mathcal{H}_{\text{EM}}$, such that
\begin{equation}
	\mathcal{H}_\text{T}=\mathcal{H}_\text{M} \otimes \mathcal{H}_{\text{EM}} =\left( \bigotimes_{i=1}^N  \mathcal{H}^i_{\text{M}}\right) \otimes\left(\bigotimes_{j=1}^\infty  \mathcal{H}^j_{\text{EM}}\right).
\end{equation}

\paragraph{Electric-dipole approximation.}

In quantum optics, the optical radiation wavelengths, $\lambda$, are in the range of $400-700$ nm [see \reffig{fig:EMspectrum}], while the atomic dimensions, $|\mathbf{r}_i|$, are of the order of the Bohr radius, $a_0\sim 1 \text{\AA}$. Consequently, the spatial variation of the field over the atomic system is considered to be negligible, such that $ \mathbf{|r|}\ll \lambda/2\pi$, resulting in 
\begin{equation}
	 \mathbf{\hat A}(\mathbf{r})\approx \mathbf{\hat A}(\mathbf{R}),
\end{equation} 
where $\mathbf{R}$ is a fixed point in the interior of the system of charges, which we choose as the origin of coordinates $\mathbf{R}=\bm 0$. This is called the electric-dipole approximation, or long-wavelength approximation.   

Considering this approximation, we can transform the Hamiltonian in \eqref{eq:LightMatterHamiltonian}, which is still expressed in terms of the vector potential, by applying the Power-Zienau-Woolley$^{\textcolor{Maroon}{*}}$ (PZW) 
\graffitoshifted[-0.875cm]{
	$^*$We note that by considering the electric-dipole approximation, we have neglected magnetic field interactions, and higher order multipole couplings. A systematic analysis of these terms can be done by considering the general Power-Zienau-Woolley transformation~\cite{PowerCoulombGauge1959,WoolleyMolecularQuantum1971,WoolleyPowerZienauWoolleyRepresentations2020}, defined as
	$$
	\hat U=\text{exp}\left[-i\int d^3 r \mathbf{\hat P}\cdot \mathbf{\hat A}\right],
	$$
	where $\mathbf{\hat P}$ is the polarization operator. However, the electric-dipole interaction term is enough to describe the light-matter interaction at the typical energy scales of quantum optical systems.
}
unitary transformation~\cite{PowerCoulombGauge1959,WoolleyMolecularQuantum1971,WoolleyPowerZienauWoolleyRepresentations2020,VukicsGaugeinvariantLagrangian2021} under the electric-dipole approximation:
\begin{equation}
	\hat U_{\text{PZW}}=e^{-i \mathbf{\hat d}\cdot \mathbf{\hat A}(0) }, 
\end{equation}
where $\mathbf{\hat d} \equiv \sum_i  q_i \mathbf{\hat r}_i$ is the dipole moment of the charge distribution. Then, by noting that the transformation is time-independent, the transformed Hamiltonian in the PZW picture, $\hat H'=\hat U_{\text{PZW}}^\dagger \hat H \hat U_{\text{PZW}}$, reads
\begin{equation}
	\hat H'= \sum_i \frac{\mathbf{\hat p}_i^2}{2 m_i} +\sum_{\mathbf{k},s}\frac{1}{2  \varepsilon_0 V} \left(\mathbf{\hat d}\cdot \mathbf{e}_{\mathbf{k},s}\right)^2+ \hat V + \hat H_{\text{EM}}-\mathbf{\hat d}\cdot \mathbf{\hat E}(0) .
	\label{eq:LightMatterHamiltonianPZW}
\end{equation}
In this picture, the second term represents a (divergent) dipole self-energy term, which can be absorbed as a renormalization shift to the atomic energy levels, and the last term is the familiar electric-dipole interaction. Note that the PZW transformation is the quantized version of the Göpert-Mayer transformation~\cite{Goppert-MayerUberElementarakte1931,Cohen-TannoudjiPhotonsAtoms1997}, where the EM field is treated as a classical dynamical variable, and thus the self-energy term does not appear. By absorbing the self-energy term in the definition of the matter Hamiltonian, and renaming the Hamiltonian simply as $\hat H$, we obtain the electric-dipole Hamiltonian:
\begin{equation}
	\colorboxed{Maroon}{\hat H=\hat H_M + \hat H_{EM}-\mathbf{\hat d}\cdot \mathbf{\hat E}.}
	\label{eq:LightMatterHamiltonianDipoleApprox}
\end{equation}

We note that the
\graffito{
	$^{**}$The parity operator $\hat \Pi$ flips the sign of $ \mathbf{\hat r}$, defined as
	$$
	\hat \Pi^\dagger \mathbf{\hat r} \hat \Pi=- \mathbf{\hat r},
	$$
	making it unitary with eigenvalues $\pm 1$ since $\hat \Pi^2=\mathbb{I}$. 
	Rearranging this definition as $\{ \hat \Pi , \mathbf{\hat r} \}=0$, and applying it into two states with well-defined parity, we arrive at
	$$
	(\pi_a+\pi_b)\langle a |  \mathbf{\hat r}  |b\rangle =0,
	$$
	where $\pi_{a/b}$ are the eigenvalues of $\hat \Pi$. For this expression to hold, either $\pi_a+\pi_b=0$ or $\langle a |  \mathbf{\hat r}  |b\rangle =0$. 
	Consequently, diagonal elements of $\mathbf{\hat d}$ vanish, while off-diagonal elements are non-zero only between states of opposite parity~\cite{SteckQuantumAtom}. 
}
 dipole operator, $\mathbf{\hat d}$, only acts on the particle subspace, allowing us to recast it in terms of its eigenbasis. In the case of one particle, we have that $\hat H_{M} |m\rangle =E_m |m \rangle$, enabling the dipole operator to be written as
\begin{equation}
	\mathbf{\hat d}=\sum_{m,n} \langle m |  \mathbf{\hat d} | n \rangle |m \rangle \langle n|= \sum_{m, n} \mathbf{ d}_{mn}  |m \rangle \langle n|,
\end{equation}
where $d_{mn}\equiv  \langle m |  \mathbf{\hat d} | n \rangle$ represents the matrix element of the dipole moment operator. Particularly, when the eigenstates exhibits a centrosymmetric charge distribution~\cite{JacksonClassicalElectrodynamics1999}, the dipole moment operator, due to parity symmetry$^{\textcolor{Maroon}{**}}$,  only connects states of opposite parity, implying that $\mathbf{d}_{mn} = 0$ when $m = n$. As a consequence, the summation is restricted to terms where $m \neq n$, reflecting that the dipole moment operator cannot induce any first-order energy shifts due to symmetry constraints. 

\paragraph{Quantum emitters: the two-level system approximation.}
Quantum emitters (QEs) are quantum systems with discrete energy spectra, such as atoms~\cite{HammererQuantumInterface2010,ChangQuantumNonlinear2014}, molecules~\cite{ToninelliSingleOrganic2021}, or quantum dots~\cite{LodahlInterfacingSingle2015}, characterized by an anharmonic energy structure where transitions  are not all identical, though they need not be entirely distinct.
In practice, the energy spectrum of a QE is often reduced to just a few energy levels that are relevant to the dynamics. 
This reduction occurs when a small, discrete set of energy levels can be effectively isolated from the rest.

A particularly important case of energy-level reduction is the two-level approximation, where the dynamics of the system is effectively restricted to just two energy states: a ground state $|g\rangle$ and an excited state $|e\rangle$. 
For instance, this scenario arises when the system is driven by a laser field resonant with the energy transition between these two states. Under such conditions, other energy levels can be neglected as they do not significantly influence the system dynamics. This leads to an effective Hamiltonian expressed in terms of the two relevant states
\begin{equation}
	\hat H_M= \sum_m E_m |m\rangle \langle m| \approx E_g |g\rangle \langle g| + E_e |e\rangle \langle e| .
\end{equation}
In this case, 
\graffito{
	$^*$We observe that when the  assumption of a centrosymmetric charge distribution is violated due to asymmetries in the particle ground state, a permanent dipole moment arises. In this case,  the dipole moment operator develops an additional term
	$$
	\begin{aligned}[b]
		\mathbf{\hat d}= \mathbf{ d}_{ee}&(\mathbb{I}_2+\hat \sigma_z)/2 \\
		&+ \mathbf{ d}( \hat \sigma + \hat \sigma^\dagger),
	\end{aligned}
	$$
	where $ \mathbf{ d}_{ee}$ describes the permanent dipole component. For instance, this modification enables the generation of tunable quantum THz radiation~\cite{ChestnovTerahertzLasing2017,GroiseauSinglePhotonSource2024}.
}
the system is typically referred to as a two-level system (TLS), or a qubit.  Then, assuming again the particle exhibits a centrosymmetric charge distribution without a permanent dipolar moment$^{\textcolor{Maroon}{*}}$---that is, the diagonal elements of $\mathbf{\hat d}$ vanish---, the dipole moment operator  can be easily expressed as a $2\times 2$ matrix given by
\begin{equation}
	\mathbf{\hat d}=\langle g | \mathbf{\hat d} |e \rangle | g\rangle \langle e | + \langle e | \mathbf{\hat d} |g \rangle | e\rangle \langle g |=  \mathbf{ d}( \hat \sigma + \hat \sigma^\dagger),
\end{equation}
where we choose the dipole matrix element $\mathbf{d}\equiv \langle g | \mathbf{\hat d}|e\rangle$ to be real, and $\hat \sigma $ and $\hat \sigma^\dagger$ are the annihilation and creation operators for the TLS:
\begin{equation}
	\hat \sigma = |g\rangle \langle e |, \quad \hat \sigma^\dagger=|e\rangle \langle g|.
\end{equation}
By identifying the Hilbert space of the TLS, $\{ |g\rangle, |e\rangle \}$, with states of a fictitious spin-1/2 system, a \textit{pseudo-spin}, we can use all the theoretical machinery from group theory~\cite{TungGroupTheory1985}. Therefore, it is natural to define the Pauli matrices
\begin{equation}
	\hat \sigma_x= \hat \sigma + \hat \sigma^\dagger, \quad \hat \sigma_y=i (\hat \sigma - \hat \sigma^\dagger), \quad \hat \sigma_z=2 \hat \sigma^\dagger \hat \sigma - \mathbb{\hat I}_2,
\end{equation}
where $\mathbb{\hat I}_2$ is the identity of the two-dimensional Hilbert space. The set $\{ \hat \sigma_x,\hat \sigma_y,\hat \sigma_z \}$ forms the fundamental representation of the $\mathfrak{su}(2)$ algebra, satisfying the following relations:
\begin{subequations}
	\begin{align}
		&\left[ \hat \sigma_j, \hat \sigma_k \right]= 2i \epsilon_{jkl} \hat \sigma_l, 
		\label{eq:PauliRelations1} \\
		&\left\{ \hat \sigma_j, \hat \sigma_k \right\}= 2\delta_{jk}\mathbb{\hat I}_2 ,
		\label{eq:PauliRelations2} \\
		&\hat \sigma^\nu \hat \sigma^{\dagger \mu}= \mu \nu \mathbb{\hat I}_2 + (1-2\mu \nu) \hat \sigma^{\dagger \mu}  \hat \sigma^\nu, 
		\label{eq:PauliRelations3}
		%
		%
	\end{align}
\end{subequations}
where $\epsilon_{jkl}$ is the
\graffito{
$^{**}$The quantum statistics obeyed by indistinguishable fundamental particles depend on their spin, as stated by the spin-statistics theorem. According to this theorem, particles with half-integral spin follow Fermi-Dirac statistics, while those with integral spin follow Bose-Einstein statistics.
}
 completely antisymmetric Levi-Civita tensor, $j,k,l\in \{x,y,z\}$ and $\mu,\nu\in \{0,1\}$. It is important to note that qubits do not strictly follow either  fermion or boson statistics$^{\textcolor{Maroon}{**}}$; instead, they obey a mixture of both, referred to as parafermion statistics~\cite{WuQubitsParafermions2002}, which represent a hybrid boson-fermion quantum statistics. This distinction arises because qubits are not fundamental particles themselves, but rather effective descriptions of them. In any case, for simplicity, we will treat qubits as fermionic particles, allowing us to define the simplest theoretical nonlinear medium.

In analogy with the relations for the Pauli matrices in \cref{eq:PauliRelations1,eq:PauliRelations2,eq:PauliRelations3}, we can derive similar expressions for bosonic operators by means of the Wick theorem~\cite{BlasiakGeneralBoson2003,BlasiakCombinatoricsBoson2007}, such that
\begin{subequations}
	\begin{align}
		&[\hat a, \hat a^\dagger]=1, \\
		&\hat a^n \hat a^{\dagger m}=\sum_{k=0}^{\text{min}(n,m)} k!
		\binom{m}{k} \binom{n}{k}
		a^{\dagger m-k}a^{n-k}, \\
		&[ \hat a^{\dagger m} \hat a^n, \hat a^\dagger \hat a]=(n-m)\hat a^{\dagger m} \hat a^n.
	\end{align}
\end{subequations}

\subsection{The quantum Rabi model and the Jaynes-Cummings model}
\label{Section:RabiJaynesModels}

The Hamiltonian derived in \eqref{eq:LightMatterHamiltonianDipoleApprox} is a general expression valid for the case of $N$ quantum emitters interacting with $M$ quantum light modes under the electric-dipole approximation. Nevertheless, in quantum optics, we are usually interested in the dynamics that emerges from the interaction of a single quantum emitter, or a collection of them, with a radiation field that is close to resonance with the QEs, such that it can be described by a single quantum light mode.
\paragraph{The Quantum Rabi model. } Let us consider the  particular case of a single quantum emitter (QE), modelled as a two-level system (TLS) with natural frequency $\omega_\sigma$, interacting with a single light mode.  This light mode is characterized by a wavevector $\mathbf{k}$ and polarization $s$, and is described by a quantum harmonic oscillator with frequency $\omega_a$ and annihilation operator $\hat a \equiv \hat a_{\mathbf{k},s}$. Then, following the general Hamiltonian from \eqref{eq:LightMatterHamiltonianDipoleApprox}, the interaction between the QE and the light mode results in the following light-matter Hamiltonian (using natural units, where $\hbar = 1$),  the quantum Rabi model~\cite{RabiProcessSpace1936,RabiSpaceQuantization1937}:
\begin{equation}
	\colorboxed{Maroon}{
		\hat H_\text{Rabi}= \omega_a \hat a^\dagger \hat a + \omega_\sigma \hat \sigma^\dagger \hat \sigma + g (\hat \sigma+\hat \sigma^\dagger)(\hat a^\dagger +\hat a),}
		\label{eq:RabiHamiltonian}
\end{equation}
where the light-matter coupling rate is defined by
\begin{equation}
	 g\equiv g_{\mathbf{k},s}=  -i\sqrt{\frac{ \omega_k}{2\varepsilon_0 V}}  e^{i \mathbf{k}\cdot \mathbf{r}}\mathbf{e}_{\mathbf{k},s}\cdot \mathbf{d} ,
	 \label{eq:lightmattercoupling}
\end{equation}
with $\mathbf{r}$ denoting the position of the dipole at which the potential vector is considered to be spatially constant. Additionally, we note that, in what follows, for the sake of simplicity, we will assume that the dipole configuration is such that the light-matter coupling is purely real and positive, $g\in \mathbb{R}^+$.
We also note that thorough the Thesis we will use natural units, $\hbar=1$, such that there is no physical distinction between energy and frequency.

The resulting Hamiltonian in \eqref{eq:RabiHamiltonian} is considered as the canonical quantum model of light-matter interaction.
It has a global $\mathbb{Z}_2$ symmetry related to the parity conservation of the total number of excitations, but not to the \textit{total number of excitations}, resulting in a complex analytical integrability~\cite{BraakIntegrabilityRabi2011,XieQuantumRabi2017}. 

\paragraph{The Jaynes-Cummings model. } 
For nearly resonant frequencies, $\omega_\sigma \approx \omega_a$, and weak light-matter coupling, $g\sqrt{\hat a^\dagger \hat a}\ll \omega_\sigma,\omega_a$, we can perform a rotating wave approximation (RWA) that neglects the counter-rotating terms~\cite{BurgarthTamingRotating2024}, that is, eliminates those terms that oscillate rapidly compared with the coupling rate:  $ \hat \sigma^\dagger \hat a^\dagger e^{i(\omega_a+\omega_\sigma)}$ and $ \hat \sigma \hat a  e^{-i(\omega_a+\omega_\sigma)}$.
This leads to the so-called Jaynes-Cummings Hamiltonian~\cite{JaynesComparisonQuantum1963},
\begin{equation}
	\colorboxed{Maroon}{
		\hat H_\text{JC}= \omega_a \hat a^\dagger \hat a + \omega_\sigma \hat \sigma^\dagger \hat \sigma + g (\hat \sigma \hat a^\dagger+\hat \sigma^\dagger \hat a).}
		\label{eq:JaynesCummingsHamiltonian}
\end{equation}
This model, in contrast to the quantum Rabi model, has a continuous $\text{U}(1)$ symmetry associated to the conservation of the number of excitations, making it diagonalizable in terms of \textit{dressed states}~\cite{Cohen-TannoudjiAtomPhotonInteractions1998}, i.e., hybrid light-matter states. This Hamiltonian will be extensively used throughout the rest of this Thesis.


\section[Description of open quantum systems: Schrödinger picture]{Description of open quantum systems: \protect\newline Schrödinger picture}
\label{Section:OPS}
The time evolution of physical systems is of crucial relevance in order to understand their nature and properties. In classical mechanics, the time evolution of a given system is described in terms of differential equations, e.g., Euler-Lagrange's equations or Hamilton's equations. In analogy to this, the evolution of a closed quantum system that is described by a density matrix, $\hat \rho$, is governed by the \textit{Liouville-von Neumann equation}, $d \hat \rho (t)/dt=-i [\hat H,\hat \rho]$, such that
\graffitoshifted[-1.3cm]{When the Liouvillian superoperator is time-dependent,
	$$
	\mathcal{L}(t)=-i [\hat H(t), (\cdot) ],
	$$
	the evolution can be formally described as a Dyson expansion
	$$
	\hat \rho (t)=\mathcal{T}e^{\int_{t_0}^t d\tau \mathcal{\hat L}(\tau)}[\hat \rho (t_0)],
	$$
	where $\mathcal{T}$ is the \textit{chronological} time-ordering operator. 
}
\begin{equation}
	  \hat \rho (t)=e^{ \mathcal{\hat L}(t-t_0)}[\hat \rho (t_0)],
	  \label{eq:VonNeumannEq}
\end{equation}
where $ \mathcal{\hat L}\equiv -i [\hat H, (\cdot) ]$ is the \textit{Liouville superoperator}---acting on operators to yield another operator---, $\hat H$ is a time-independent Hamiltonian, and $t_0$ is the initial time$^{\textcolor{Maroon}{*}}$.

The time evolution described above is always valid as long as we have access to all the information in the system. When this condition does not hold, we say that the system is an open quantum system, in the sense that our quantum system---e.g., an atom---is interacting with an environment---e.g., an electromagnetic field---which has too many degrees of freedom to be monitored. As a consequence, the time evolution of the quantum system is not given any more by a unitary operator, but by a more complex object that takes into account possible system-environment correlations. 
The theory of open quantum systems addresses this problem by providing an effective description of the quantum system alone, where the environment degrees of freedom have been traced out.
Essentially,  the description of the system is given in terms of mixed states via the master equation in the Schrödinger picture, while the dynamics is governed by stochastic differential equations via the quantum Langevin equations in the Heisenberg picture.

In what follows, we discuss both formalisms by adapting the derivations from \textcolor{Maroon}{Refs.}~\cite{CarmichaelOpenSystems1993,GardinerQuantumNoise2004,GardinerQuantumWorld2014,GardinerQuantumWorld2015,BreuerTheoryOpen2007,RivasOpenQuantum2012}.

\subsection{Master equation formalism}

In this section, we introduce the formalism of open quantum systems in the Schrödinger picture, resulting in a series of differential equations for the density matrix of the system whose validity depends on the physical assumptions of the environment.

The following derivations are adapted from \textcolor{Maroon}{Refs.}~\cite{CarmichaelOpenSystems1993,BreuerTheoryOpen2007,RivasOpenQuantum2012}.

\paragraph{Born approximations.}
Let us assume an open quantum system composed by a quantum system $S$ that is weakly coupled to an environment $E$, such that the total system belongs to a Hilbert space $\mathcal{H}_T=\mathcal{H}_S\otimes \mathcal{H}_E$, where $\mathcal{H}_{S/E}$ are the Hilbert spaces for the system and the environment, respectively. The total Hamiltonian is given by
\begin{equation}
	\hat H=\hat H_S +\hat H_E +\alpha \hat H_{\text{int}}, 
\end{equation}
where $\alpha$ is a dimensionless parameter quantifying the strength of the coupling between the system and environment. The interaction Hamiltonian $\hat H_{\text{int}}$ is defined in the most general way as
\begin{equation}
	\hat H_{\text{int}}=\sum_i \hat S_i \otimes \hat E_i,
	\label{eq:InteractHamiltonian}
\end{equation}
where $\hat S_i \in \mathcal{H}_S$ and $\hat E_i \in \mathcal{H}_E$. The evolution of the total density matrix in the interaction picture, $\rho^{\text{I}}_T(t)=e^{i(\hat H_S+\hat H_E)t}\hat \rho_T(t) e^{-i(\hat H_S+\hat H_E)t}$, is governed by the Liouville-von Neumann equation,
\begin{equation}
		\frac{d \hat \rho^{\text{I}}_T (t)}{dt}=-i \alpha  [\hat H_{\text{int}}(t),\hat \rho^{\text{I}}_T(t)],
		\label{eq:vonNeumannEq}
\end{equation}
such that its formal solution in integral form is
\begin{equation}
	\hat \rho^{\text{I}}_T(t)=\hat \rho^{\text{I}}_T(t_0)-i \alpha \int_{t_0}^t ds [\hat H_{\text{int}}(s),\hat \rho^{\text{I}}_T(s)]. 
\end{equation}
Without losing 
\graffitoshifted[-1cm]{ 
	$^*$The first term in \eqref{eq:BornMarkov1} can be set to zero since the initial state of the environment is thermal, satisfying
	$$
	\langle \hat E_i\rangle=\text{Tr}_E[\hat E_i \hat \rho_E(0)]=0,
	$$
	for all values of $i$. Alternatively, we can redefine the system Hamiltonian as 
	$
	\hat H'_S \equiv \hat H_S+ \sum_i \alpha \hat S_i \langle \hat E_i\rangle 
	$
	and the interaction Hamiltonian as
	$
	\hat H'_{\text{int}}\equiv \sum_i \hat S_i \otimes (\hat E_i-\langle \hat E_i\rangle),
	$
	such that 
	$$\hat H=\hat H'_S +\hat H_E +\alpha \hat H'_{\text{int}}.
	$$
	Under this transformation, we find the trivial result
	$$
	\text{Tr}_E[(\hat E_i-\langle \hat E_i \rangle ) \hat \rho_E(0)]=0.
	$$ 
}
generality, we set $t_0=0$. Now, we insert again the formal solution in \eqref{eq:vonNeumannEq}, leading to
\begin{equation}
	\frac{d \hat \rho^{\text{I}}_T (t)}{dt}=-i\alpha[\hat H_{\text{int}}(t),\hat \rho^{\text{I}}_T(0)]\\
	- \alpha^2 \int_{0}^t ds  [\hat H_{\text{int}}(t), [\hat H_{\text{int}}(s),\hat \rho^{\text{I}}_T(s)]].
	\label{eq:BornMarkov1}
\end{equation}
Since we are considering that the system is weakly interacting with the environment, terms of the order of $ \mathcal{O}(\alpha^3)$ or higher can be neglected. This is called the \textit{first Born approximation}. 
Additionally, we can always assume that at initial times, $t=0$, the system and the environment are uncorrelated, i.e., $\hat \rho_T (0)=\hat \rho_S (0) \otimes \hat \rho_E (0)$, where $\hat \rho_{S/E}\equiv \text{Tr}_{S/E}[\hat \rho_T]$, respectively. Under this assumption, we can trace out the environment and set the first term to zero$^{\textcolor{Maroon}{*}}$. Hence, we get an integro-differential equation for the system:
\begin{equation}
	\frac{d \hat \rho^{\text{I}}_S (t)}{dt}=- \alpha^2 \int_{0}^t ds \text{Tr}_E [\hat H_{\text{int}}(t), [\hat H_{\text{int}}(s),\hat \rho^{\text{I}}_T(s)]].
\end{equation}
However, this expression still depends on the total density matrix $\hat \rho^{\text{I}}_T(t)$.  To unravel the system from the environment, we need to perform a second approximation---\textit{the second Born approximation}---in which we assume that the system and the environment remain uncorrelated at all times. Since we are considering a weak coupling regime, $\alpha\ll 1$, we can assume that the influence of the system on the environment is negligible. Therefore, the total density matrix at time $t$ may be approximately described by a tensor product at zeroth order in the interaction Hamiltonian:
\begin{equation}
	\hat \rho^{\text{I}}_T(t) \approx \hat \rho^{\text{I}}_S(t)\otimes \hat \rho_E,
\end{equation} 
where $\hat \rho_E$ is assumed to be a stationary state of the environment, such that $[\hat H_E, \hat \rho_E]=0.$ Hence, \eqref{eq:BornMarkov1} transforms into the  \textit{Born master equation}
\begin{equation}
	\frac{d \hat \rho^{\text{I}}_S (t)}{dt}=- \alpha^2 \int_{0}^t ds \text{Tr}_E [\hat H_{\text{int}}(t), [\hat H_{\text{int}}(s),\hat \rho^{\text{I}}_S(s)\otimes \hat \rho_E]].
	\label{eq:BornMarkov2}
\end{equation}
\paragraph{Markovian approximation.} At this stage, the equation is completely defined in terms of the system state. However, it still depends on the past values of the system through the term $\hat \rho^{\text{I}}_S(s)$ in the integral, meaning that there are memory effects in the dynamics since the environment is affecting back the system, resulting in non-Markovian dynamics.

We can simplify \eqref{eq:BornMarkov2} by assuming that the correlation functions between the system and the environment decay fast enough such that any change imprinted by the system into the environment is lost  over a time $\tau_E$ which is small compared to the relaxation time of the system $\tau_R$, that is, $\tau_E\ll \tau_R$. 
This is called \textit{Markovian  approximation}, which involves the following steps: replacing 
$\hat \rho^{\text{I}}_S(s) \rightarrow \hat \rho^{\text{I}}_S(t)$, substituting $s\rightarrow t-s$, and extending  the upper limit of the integral  to infinity. This extension is justified provided the integrand decays sufficiently fast for $s\gg \tau_E$, where $\tau_E$ is the characteristic timescale of the environment. Then, we arrive to the \textit{Born-Markov master equation}, or also called \textit{Bloch-Redfield master equation} 
\vspace{-1mm}
\begin{equation}
	\colorboxed{Maroon}{\frac{d \hat \rho^{\text{I}}_S (t)}{dt}=- \alpha^2 \int_{0}^\infty ds \text{Tr}_E [\hat H_{\text{int}}(t), [\hat H_{\text{int}}(t-s),\hat \rho^{\text{I}}_S(t)\otimes \hat \rho_E(0)]].}
	\label{eq:BornMarkovMaster}
\end{equation}
\vspace{-3mm}

This equation can be cast in a suitable form by considering the following spectral expansion of the interaction Hamiltonian $\hat H_{\text{int}}$ in \eqref{eq:InteractHamiltonian} as
\begin{equation}
		\label{eq:spectralIntHamiltonian}
	\hat H_{\text{int}}(t)=\sum_{m, \omega} e^{-i \omega t} \hat S_m (\omega)\otimes \hat E_m(t)=\sum_{m, \omega} e^{i \omega t} \hat S^\dagger_m (\omega)\otimes \hat E^\dagger_m(t),
\end{equation}
\\[-1em]
where we are describing the $m$th system operator, $\hat S_m(t)$, in the eigenbasis of the system Hamiltonian, that is,
\begin{equation}
	\hat S_m (t) = \sum_\omega e^{- i\omega t} \hat S_m(\omega) \quad \text{with} \quad [\hat H_S, S_m(\omega)]=-\omega \hat S_m (\omega).
\end{equation}
\\[-1em]
Additionally, we collect the environment effects via the one-sided Fourier transform of the environment correlation function:
\begin{equation}
	\Gamma_{mn}(\omega) \equiv  \int_0^\infty ds e^{i \omega s} \langle \hat E^\dagger_m (t) \hat E_n(t-s) \rangle.
	\label{eq:onesidedFourier}
\end{equation}
Note that the environment correlator simplifies to 
\begin{equation}
	\langle \hat E^\dagger_m (t) \hat E_n(t-s) \rangle= \langle \hat E^\dagger_m (s) \hat E_n(0) \rangle,
\end{equation}
when $\hat \rho_E$ is a stationary state (second Born approximation). Then, by inserting the spectral decomposition of the interaction Hamiltonian [\eqref{eq:spectralIntHamiltonian}] and the environment correlations [\eqref{eq:onesidedFourier}], as well as expanding the commutators and re-arranging the terms in \eqref{eq:BornMarkovMaster}---we also drop the perturbative parameter $\alpha$---, we arrive at the standard form of the Bloch-Redfield master equation~\cite{RedfieldTheoryRelaxation1965}:
\begin{equation}
\colorboxed{Maroon}{	\begin{aligned}[b]
		\frac{d \hat \rho^{\text{I}}_S (t)}{dt}= \sum_{\omega,\omega'}\sum_{m,n} 
		&\left\{
		e^{i(\omega'-\omega)t} \Gamma_{mn}(\omega) [\hat S_n(\omega) \hat \rho^{\text{I}}_S (t), \hat S^\dagger_m(\omega')  ]
		\right. \\
		&\left. 
		\ +	e^{i(\omega-\omega')t} \Gamma_{nm}^*(\omega') [ \hat S_n(\omega),\hat \rho^{\text{I}}_S (t) \hat S^\dagger_m(\omega') ]
		\right\}.
\end{aligned}}
\label{eq:BlochRedfieldEq}
\end{equation}

\paragraph{Secular approximation.}
Generally, the Bloch-Redfield master equation may not guarantee the positivity of the density matrix throughout the time evolution and can thus lead to unphysical density matrices~\cite{DaviesMarkovianMaster1974,DumckeProperForm1979}. Therefore, to ensure a complete positive dynamical map, we need to perform an additional approximation, the \textit{secular approximation}. This consists in averaging over the rapidly oscillating terms in the  Bloch-Redfield master equation, in a similar way we performed a rotating-wave approximation to obtain the Jaynes-Cummings Hamiltonian in \eqref{eq:JaynesCummingsHamiltonian}. 

We denote $\tau_S$ as the \textit{intrinsic time scale} of the system, of the order of $\tau_S \sim |\omega - \omega'|^{-1}$ for $\omega \neq \omega'$, and $\tau_R$ as the relaxation rate of the system. Then, when $\tau_S \ll \tau_R$ and $\tau_S \gg \alpha^2$, oscillating terms proportional to $e^{i (\omega'-\omega)t}$ for $\omega \neq \omega'$ can be neglected since they oscillate very rapidly during the typical time $\tau_R$ over which the system varies appreciably, and hence not contribute effectively to the evolution. Consequently, we just keep the resonant terms $\omega=\omega'$, transforming \eqref{eq:BlochRedfieldEq} into
\begin{multline}
	\frac{d \hat \rho^{\text{I}}_S (t)}{dt}= \sum_{\omega, m,n} 
	\left\{
	\Gamma_{mn}(\omega) [\hat S_n(\omega) \hat \rho^{\text{I}}_S (t), \hat S^\dagger_m(\omega)  ]
	\right. \\
	\left. 
	+ \Gamma_{nm}^*(\omega) [ \hat S_n(\omega),\hat \rho^{\text{I}}_S (t) \hat S^\dagger_m(\omega) ]
	\right\}.
\end{multline}
Now, it is convenient to decompose $\Gamma_{mn}$ as
\begin{equation}
	\Gamma_{mn} (\omega)=\frac{1}{2} \gamma_{mn}(\omega) + i \Lambda_{mn} (\omega),
	\label{eq:GammaDecomposition}
\end{equation}
where the coefficients are given by the following Hermitian matrices
\begin{subequations}
	\begin{align}
		\gamma_{mn}(\omega)&\equiv\Gamma_{mn}(\omega)+\Gamma_{nm}^*(\omega)=\int_{-\infty}^{\infty}ds e^{i\omega s} \langle \hat E^\dagger_m(s)\hat E_n(0) \rangle, 
		\label{eq:gamma_mn}\\
		\Lambda_{mn}(\omega)&\equiv \frac{1}{2i} [\Gamma_{mn}(\omega)-\Gamma_{nm}^* (\omega)].
		\label{eq:Lambshift}
	\end{align}
\end{subequations}
The matrix $\gamma=(\gamma_{mn})$ in \eqref{eq:gamma_mn}, referred to as  \textit{Kossakowski matrix}~\cite{BenattiOpenQuantum2005,BenattiAsymptoticEntanglement2011}, encodes the dissipation of the system through the environment via the dissipators of the master equation, as we will see below.
To ensure a well-defined time-evolution generator, the Kossawkoski matrix must be positive semidefinite~\cite{GoriniCompletelyPositive1976}. This property allows its diagonalization using an appropriate unitary transformation, such that the resulting diagonal terms are non-negative, $\gamma_i(\omega)\geq 0$.  On the other hand, the matrix $\Lambda=(\Lambda_{mn}$) in \eqref{eq:Lambshift} generates an effective Hamiltonian of the form
\\[-1em]
\begin{equation}
	\hat H_{LS}=\sum_{\omega, m,n} \Lambda_{mn}(\omega) \hat S_m (\omega) \hat S_n (\omega),
\end{equation}
\\[-1em]
that renormalizes the unperturbed energy levels induced by the system-environment coupling, receiving the name of \textit{Lamb shift Hamiltonian}. 

Finally, by introducing the new set of operators that diagonalize the Kossawkoski matrix, defined as $\hat L_i (\omega)$, and coming back to the Schrödinger picture, we arrive into the diagonal form of the celebrated \textit{Linblad-Gorini-Kossakowski-Sudarshan master equation}~\cite{LindbladGeneratorsQuantum1976}:
\\[-1.5em]
\begin{multline}
			\frac{d \hat \rho(t)}{dt }= -i[\hat H+ \hat H_{LS}, \hat \rho(t)]\\
			+ \sum_{i,\omega} \frac{\gamma_i(\omega)}{2} \left[
		2 \hat L_i (\omega) \hat \rho(t) \hat L_i^\dagger(\omega) - \left\{\hat L_i^\dagger (\omega) \hat L_i(\omega),  \hat \rho(t)\right\} 
		\right],
\end{multline}
\\[-1.5em]
which, in the simplest case of just one relevant frequency $\omega$, simplifies to:
\begin{equation}
 	\colorboxed{Maroon}{
 		\frac{d \hat \rho_S (t)}{dt}=-i[\hat H_S+\hat H_{LS}, \hat \rho_S(t)]
 		+\sum_i \frac{\gamma_i}{2} \mathcal{D}[\hat L_i] \hat \rho_S(t)= \mathcal{\hat L}\hat \rho_S(t),
 	}
 	\label{eq:LindbladMasterEq}
\end{equation}
where $\mathcal{\hat L}$ is the \textit{Liouvillian superoperator}, $\hat L_i$ receives the name of \textit{jump operator}, and $\mathcal{D}[\hat L_i] \hat \rho_S(t)$ is the \textit{Lindblad term}, or \textit{dissipator},
\begin{equation}
	\mathcal{D}[\hat L_i] \hat \rho_S(t)\equiv  2 \hat L_i \hat \rho_S(t)\hat L_i^\dagger - \left\{\hat L_i^\dagger \hat L_i, \hat \rho_S(t)\right\}  .
	\label{eq:LindbladTerms}
\end{equation}
Lindblad terms account for some kind of incoherent process induced in the system via the system-environment interaction, such as spontaneous emission, incoherent excitation or local pure dephasing. 
\\[-1.5em]

\paragraph{Adjoint master equation.}
In full analogy
\graffitoshifted[-5.5cm]{
	$^*$We can compute the equation of motion for the operator $d \langle \hat X(t)\rangle/dt$ by noting that
	$$
	\frac{d}{dt} \text{Tr}[\hat X \hat \rho(t)] =\text{Tr}[\hat X \frac{d \hat \rho(t)}{dt}].
	$$
	Using the definition of the master equation in \eqref{eq:LindbladMasterEq}, we can rearrange the terms as follows:
	$$
	\text{Tr}[-i\hat X[\hat H, \hat \rho] ]=i\langle [\hat H, \hat X]\rangle,
	$$
	and
	$$
	\begin{aligned}[b]
		\text{Tr}&[\hat X \mathcal{D}[\hat L_i]\hat \rho]=\\
		&\langle 2\hat L_i^\dagger \hat X \hat L_i -\{ \hat L_i^\dagger \hat L_i, \hat X \}  \rangle.
	\end{aligned}
	$$
	From these results, we directly arrive to \eqref{eq:AdjointMasterEq}, as this derivation is valid for any quantum state $\hat \rho$.
}
 with close quantum systems, it is possible to define the time evolution equation for a system operator in the Heisenberg picture, what it is called, \textit{Heisenberg equation} or \textit{adjoint master equation}~\cite{BreuerTheoryOpen2007}. 
Given a general Lindblad master equation [\eqref{eq:LindbladMasterEq}] and using the property of the trace, the equation of motion for an arbitrary system operator $\hat X$ is  given by$^{\textcolor{Maroon}{*}}$
\\[-1.5em]
\begin{equation}
\frac{d \hat X(t)}{dt }= i[\hat H, \hat X(t)]
+ \sum_i \frac{\gamma_i}{2} \left( 
2 \hat L_i^\dagger   \hat X(t) \hat L_i- \left\{\hat L_i^\dagger \hat L_i,  \hat X(t)\right\} 
\right).
\label{eq:AdjointMasterEq}
\end{equation}
\\[-1.5em]
This expression is formally valid only when applied to the expectation value of the operator$^{\textcolor{Maroon}{**}}$, $\langle \hat X(t) \rangle$, in which case it simplifies to
\graffitoshifted[-1.0cm]{
	$^{**}$The general operator equation in the Heisenberg picture will be presented later in \refsec{Section:OPS_Heisenberg}, where the environment degrees of freedom appear explicitly through quantum noise operators, also known as \textit{quantum Wiener increments}.
}
\\[-1.5em]
\begin{equation}
	\colorboxed{Maroon}{	\frac{d \langle \hat X(t) \rangle}{dt }= i\langle [\hat H, \hat X(t)] \rangle+ \sum_i \frac{\gamma_i}{2} \left( 
\langle [\hat L_i^\dagger, \hat X(t)] \hat L_i \rangle 	+\langle \hat L_i^\dagger [ \hat X(t),\hat L_i]  \rangle 
		\right).}
\end{equation}

\subsection{Example: a two-level system in a thermal environment}
\label{Section:SpontaneousEmission}
\vspace{-1mm}

Once we have developed the master equation formalism, let us consider a specific example that will be of particular interest in this Thesis: the interaction of a single quantum emitter with a thermal environment  composed of harmonic oscillators.

The total Hamiltonian of the system is given by 
\begin{equation}
	\hat H=\hat H_S+ \hat H_E +\hat H_{\text{int}},
\end{equation}
where $\hat H_S$ is the free Hamiltonian of the quantum emitter (system), $\hat H_E$ is the free Hamiltonian of the electromagnetic field (environment), and $\hat H_{\text{int}}$ is the light-matter interaction Hamiltonian derived in \eqref{eq:JaynesCummingsHamiltonian}:
\begin{subequations}
	\begin{align}
		\hat H_S&=\omega_\sigma \hat \sigma^\dagger \hat \sigma, \\
		\hat H_E&=\sum_{\mathbf{k},s}\omega_k \hat a_{\mathbf{k},s}^\dagger \hat a_{\mathbf{k},s} , 
		\label{eq:EMHamiltona}\\
		\hat H_{\text{int}}&= - \mathbf{\hat d} \cdot  \mathbf{\hat E}\approx  \sum_{\mathbf{k},s} g_{\mathbf{k},s} (\hat \sigma^\dagger \hat a_{\mathbf{k},s}+ \hat \sigma \hat a^\dagger_{\mathbf{k},s} ).
	\end{align}
\end{subequations}
\\[-1em]
The light-matter 
\graffito{
	$^*$We note that the identification of system-environment operators in~\cref{eq:identification1,eq:identification2} is not unique. We could have defined them as
	$$
	\hat S_1= \mathbf{d} \hat \sigma, \quad \hat S_2= \mathbf{d}^* \hat \sigma^\dagger,
	$$
	and
	$$ 
	\begin{aligned}
		&\hat E_1=\sum_{\mathbf{k},s} \sqrt{\frac{ \omega_k}{2\varepsilon_0 V}}  e^{i \mathbf{k}\mathbf{r}} \mathbf{e}_{\mathbf{k},s}  \hat a_{\mathbf{k},s}^\dagger, \\
		&\hat E_2= \sum_{\mathbf{k},s}\sqrt{\frac{ \omega_k}{2\varepsilon_0 V}}  e^{-i \mathbf{k}\mathbf{r}} \mathbf{e}_{\mathbf{k},s}  \hat a_{\mathbf{k},s}.
	\end{aligned}
	$$
	This alternative identification just leads to a slightly different formulation of the Bloch-Redfield master equation in comparison with \eqref{eq:MasterEqExample1}. Nevertheless, the resulting Lindblad master equation is ultimately equivalent.
}
coupling, $g_{\mathbf{k},s}$, is defined as in \eqref{eq:lightmattercoupling}, and we have also assumed that it is a purely real and positive quantity.
Then, we can easily identify the system and environment operators$^{\textcolor{Maroon}{*}}$  following the general description of the interaction Hamiltonian (before the RWA is applied) in \eqref{eq:spectralIntHamiltonian}:
\begin{equation}
	\hat H_{\text{int}}=\sum_{m,\omega} e^{-i\omega t} \hat S_m(\omega) \otimes \hat E_m=\sum_\omega  e^{-i\omega t} \mathbf{\hat S}(\omega) \otimes \mathbf{\hat E}(t), 
	\label{eq:identification1}
\end{equation}
where  $ \mathbf{\hat E}(t)$ is the electric field operator defined in \eqref{eq:ElectricFieldOp}, and $\omega=\pm \omega_\sigma$, such that the system operators obey
\begin{equation}
	\mathbf{\hat S }(\omega_\sigma)= \mathbf{d} \hat \sigma, \quad 	\text{and}\quad \mathbf{\hat S }(-\omega_\sigma)= \mathbf{d}^* \hat \sigma^\dagger.
	\label{eq:identification2}
\end{equation}

Given this information, we can use the techniques presented in the previous section in order to obtain an effective description of the quantum emitter in terms of the following Bloch-Redfield equation: 
\begin{equation}
	\frac{d \hat \rho^{\text{I}}_S (t)}{dt}
	\approx 	\sum_{\omega} \sum_{i,j} 	\Gamma_{ij}(\omega) [\hat S_j(\omega) \hat \rho^{\text{I}}_S (t),\hat S^\dagger_i(\omega) ]
	+
	\text{H.c.}	 ,
\label{eq:MasterEqExample1}
\end{equation}
\\[-1em]
where we have performed a rotating wave approximation, keeping only the diagonal terms $\omega=\omega'$ in the frequency sum [see \eqref{eq:BlochRedfieldEq}].
The one-sided Fourier transform of the environment correlation function defined in \eqref{eq:onesidedFourier}, $\Gamma_{ij}(\omega)$, which in this case referred to as the \textit{spectral correlation tensor}~\cite{BreuerTheoryOpen2007}, is expressed in terms of the electric field operator via
\begin{multline}
	\Gamma_{ij}(\omega)
=\int_0^\infty ds e^{i\omega s}\langle \hat E_i(s) \hat E_j(0) \rangle=\sum_{\mathbf{k},s}\sum_{\mathbf{k'},s'} \sqrt{\frac{\omega_k}{2\varepsilon_0 V}}  \sqrt{\frac{\omega_{k'}}{2\varepsilon_0 V}} e_{\mathbf{k},s}^i e_{\mathbf{k'},s'}^j \\
	 \times \int_0^\infty  ds 
	\left[	
		\langle \hat a_{\mathbf{k},s} \hat a^\dagger_{\mathbf{k'},s'} \rangle e^{-i(\omega_k-\omega s)t}
		+
			\langle \hat a^\dagger_{\mathbf{k},s} \hat a_{\mathbf{k'},s'} \rangle e^{i(\omega_k+\omega s)}
			\right.\\
			\left. -
				\langle \hat a_{\mathbf{k},s} \hat a_{\mathbf{k'},s'} \rangle e^{-i(\omega_k-\omega s)t}
				-
					\langle \hat a^\dagger_{\mathbf{k},s} \hat a^\dagger_{\mathbf{k'},s'} \rangle e^{+i(\omega_k+\omega s)t}
		\right].
		\label{eq:spectralcorrtensor}
\end{multline}
In order to compute these environment correlators, we assume the electromagnetic field to be in thermal equilibrium at temperature $T$, corresponding to the state
\begin{equation}
	\hat \rho_{E}=\frac{e^{-\hat H_E/k_B T}}{\text{Tr}[e^{-\hat H_E/k_B T}]} = \bigotimes_{\omega_k} \hat \rho_{\text{th}}[\bar n(\omega_k)],
\end{equation}
where $\hat \rho_{\text{th}}[\bar n(\omega_k)]$ is a thermal Gaussian state for each mode, with an average number of excitations given by the Bose-Einstein distribution,
\begin{equation}
	\bar n(\omega_k)= \frac{1}{e^{\omega_k/k_B T}-1}.
\end{equation}
As a consequence, the correlators evaluated in this state read:
\begin{subequations}
	\begin{align}
		\langle \hat a_{\mathbf{k},s} \hat a^\dagger_{\mathbf{k'},s'} \rangle &= \delta_{\mathbf{k},\mathbf{k'}} \delta_{s,s'} [1+ \bar n(\omega_k)], \\
		\langle \hat a^\dagger_{\mathbf{k},s} \hat a_{\mathbf{k'},s'} \rangle &= \delta_{\mathbf{k},\mathbf{k'}} \delta_{s,s'} \bar n(\omega_k), \\
		\langle \hat a_{\mathbf{k},s} \hat a_{\mathbf{k'},s'} \rangle&=\langle \hat a^\dagger_{\mathbf{k},s} \hat a^\dagger_{\mathbf{k'},s'} \rangle=0.
	\end{align}
\end{subequations}

At this point, we perform the limit to the continuum by considering the bath to be infinitely large, such that the set of modes becomes infinitely dense in frequency space. This approximation translates into the substitution of the sum in $\mathbf{k}$ by the following integral 
%
%
%
	%
	%
%
\begin{equation}
	\sum_{\mathbf{k}} \rightarrow V\int \frac{d^3 k}{(2\pi )^3}= \frac{V}{(2\pi)^3 c^3 } \int_0^\infty d\omega_k \omega_k^2 \int d\Omega.
	\label{eq:ContinuumRelation}
\end{equation}
%
%
The integration over the solid angle $d\Omega$ of the wavevector $\mathbf{k}$  can be easily computed with the help of the polarization vector properties presented in ~\cref{eq:PolarizationProp1,eq:PolarizationProp2,eq:PolarizationProp3}:
\begin{equation}
	\int d\Omega \left(\delta_{ij}- \frac{k_i k_j}{k^2} \right)= \frac{8\pi}{3}\delta_{ij}.
\end{equation}
%
%
%
%
%
%
%
%
%
Then, the spectral correlation tensor in the continuum reads,
\\[-1em]
\begin{multline}
	\Gamma_{ij}(\omega)= \frac{1}{3\pi^2\varepsilon_0 c^3}\delta_{ij} \int_0^\infty d\omega_k \omega_k^3 \left[
	[1+\bar n(\omega_k)]\int_0^\infty ds e^{-i(\omega_k-\omega)s}\right. \\
	\left. +
	\bar n(\omega_k) \int_0^\infty ds e^{i(\omega_k+\omega)s}
	\right],
\end{multline}
\\[-1em]
where we have to deal with the diverging integrals of the complex exponentials. To address them, we make use of the Sokhotski-Plemelj formula, 
\\[-1em]
\begin{equation}
	\lim_{\varepsilon \rightarrow 0^+}\int_0^\infty dx e^{i \alpha x \pm \varepsilon x}=-\lim_{\varepsilon \rightarrow 0^+} \frac{i}{\alpha \pm i \varepsilon}=\mp \pi \delta(\alpha) - i \text{P.V.}\left(\frac{1}{\alpha}\right),
\end{equation} 
\\[-1em]
where $\text{P.V.}$ denotes the Cauchy Principal value. Then, by using the decomposition of the spectral correlation tensor in \eqref{eq:GammaDecomposition}, we arrive to
\begin{equation}
	\Gamma_{ij}(\omega)=\delta_{ij}  \left[\frac{1}{2}\gamma(\omega)+i \Lambda(\omega) \right],
\end{equation}
where the corresponding quantities are defined as
\begin{subequations}
	\begin{align}
		\gamma(\omega)&\approx\frac{\omega^3}{3 \pi \varepsilon_0 c^3}[1+\bar n(\omega)], 
		\label{eq:gammasigma}\\
		\Lambda(\omega)&\approx\frac{1}{6\pi^2 \varepsilon_0 c^3}\text{P.V.} \int_{0}^\infty d\omega_k \omega_k^3 \left[
		\frac{1+\bar n(\omega)}{\omega-\omega_k}+ \frac{\bar n(\omega)}{\omega+\omega_k}
		\right].
		\label{eq:Lambdasigma}
	\end{align}
\end{subequations}
The approximation symbols in~\cref{eq:gammasigma,eq:Lambdasigma} arise because we have assumed that $\bar n(\omega_k)\approx \bar n(\omega)\equiv \bar n$ since only a narrow band of frequencies around the system resonance $\omega_\sigma$ contribute to the dynamics. Finally, by inserting these terms into \eqref{eq:MasterEqExample1} and summing over the frequencies $\omega$, we obtain the quantum optical master equation:
\begin{equation}
\colorboxed{Maroon}{	
\begin{aligned}[b]
	\frac{d \hat \rho_S (t)}{dt}= -i[\hat H_S&+\hat H_{LS}, \hat \rho_S(t)]\\
&+ \frac{\Gamma}{2} (1+\bar n) \mathcal{D}[\hat \sigma]\hat \rho_S (t) +  \frac{\Gamma}{2}\bar n \mathcal{D}[\hat \sigma^\dagger]\hat \rho_S (t),
\end{aligned}
}
\label{eq:OpticalMasterEquation}
\end{equation}
where $\hat H_{LS}$ is the Lamb shift Hamiltonian,
\begin{equation}
	\hat H_{LS}=\Delta_\sigma  \hat \sigma^\dagger \hat \sigma, \quad \text{with }\quad \Delta_\sigma\equiv \Lambda(\omega_\sigma)|\mathbf{d}|^2.
\end{equation}
This additional term in the Hamiltonian leads to a renormalization of the system induced by the vacuum fluctuations of the radiation field (Lamb shift) and by thermally induced process (Stark shift)~\cite{BreuerTheoryOpen2007}. In what follows, we will omit this term by absorbing it in the definition of $\hat H_S$.  Additionally, the parameter $\Gamma$ corresponds to the spontaneous emission rate of the qubit in free space, defined as
\begin{equation}
	\Gamma\equiv \frac{\omega_\sigma^3  |\mathbf{d}|^2}{3\pi  \varepsilon_0 c^3}.
	\label{eq:GammaSpont}
\end{equation}
We note that the dissipation rate obtained in \eqref{eq:GammaSpont} is equivalent to the prediction of the Einstein A coefficient derived from the Wigner-Weisskpf theory of natural linewidth~\cite{WeisskopfBerechnungNatuerlichen1930,CarmichaelOpenSystems1993}.
The dissipators of the master equation in \eqref{eq:OpticalMasterEquation} describe spontaneous emission at a rate $\Gamma$, as well as thermally induced emission and absorption processes at rate $\Gamma \bar n$.
We also note that $\bar{n} \approx 0$ when optical transitions are considered, which is analogous to assuming the system is effectively in the zero temperature limit, $T \approx 0$.
\\[-2em]

\section[Description of open quantum systems: Heisenberg picture]{Description of open quantum systems: \protect\newline Heisenberg picture}
\label{Section:OPS_Heisenberg}
\vspace{-1mm}
\subsection{Quantum Langevin equations}
\vspace{-1mm}
In the previous section we presented the formalism of open quantum systems in the Schrödinger picture for the dynamics of the system density matrix. 
Here, in this section, we present an alternative description in terms of equations of motion for the system operators, receiving the name of quantum Langevin equations in analogy with stochastic mechanics~\cite{GardinerQuantumNoise2004,GardinerStochasticMethods2009}. 

The following derivations are adapted from \textcolor{Maroon}{Refs.}~\cite{GardinerQuantumNoise2004,GardinerQuantumWorld2014,GardinerQuantumWorld2015,Navarrete-BenllochIntroductionQuantum2022}.

\paragraph{Forwards Quantum Langevin Equation.}
Following the same procedure of the previous sections, let us consider a total Hamiltonian describing a system $S$ interacting with an environment $E$, such that
\\[-1em]
\begin{equation}
	\hat H= \hat H_S +\hat H_E +\hat H_{\text{int}},
\end{equation}
\\[-1em]
where $\hat H_S$ is the system Hamiltonian, and $\hat H_E$ is the environment Hamiltonian describing the free electromagnetic field,
\\[-1em]
\begin{equation}
	\hat H_E= \sum_{\mathbf{k}} \omega_{\mathbf{k}} \hat b^\dagger_{\mathbf{k}} \hat b_{\mathbf{k}}.
\end{equation}
\\[-1em]
To simplify the derivation,
\graffito{
	$^*$The quasi-1D approximation is usually applied in one-dimensional photonic crystal waveguides, Fabry-Perot cavities, or in plasmonic nanoantennas, where the EM field is essentially confined in one dimension~\cite{ChangColloquiumQuantum2018,SheremetWaveguideQuantum2023,Gonzalez-TudelaLightMatter2024}.
}
we have assumed the quasi-1D approximation$^{\textcolor{Maroon}{*}}$, such that the EM field propagates along one direction $\mathbf{k}=k$ (e.g., the $z-$axis), with linear dispersion relation $\omega_\mathbf{k}=c|\mathbf{k}|$, and in a concrete polarization $s$ that does not change upon propagation. 
Additionally, we consider that the system-environment interaction term is described by
\begin{equation}
	\hat H_{\text{int}}= i \sum_{\mathbf{k}} g_{\mathbf{k}} \left( \hat b_{\mathbf{k}}^\dagger \hat c - \hat b_{\mathbf{k}} \hat c^\dagger\right), 
\end{equation}
where $\hat c$ stands for the system annihilation operator, such that its evolution is given by the action of $\hat H_S$, $\hat c(t)=e^{-i\omega_0}\hat c(0)$, being $\omega_0$ its resonant frequency. In what follows, we are not going make any assumption about the nature of this operator, hence leaving the derivation general. If the system Hamiltonian describes a bosonic mode, then $\hat c = \hat a$, while if it is a fermionic mode, namely, a TLS, then $\hat c= \hat \sigma$. 
As we have done in previous derivations, we assume that the coupling rate is purely real and positive, $g_{\mathbf{k}}\in \mathbb{R}^+$, and that the system-environment interaction follows a Jaynes-Cummings Hamiltonian as in \eqref{eq:JaynesCummingsHamiltonian}. 

From this information, we derive the Heisenberg equations for the environment operator, $\hat b_{\mathbf{k}}$, and for an arbitrary system operator, $\hat X$, given by
\\[-1em]
\begin{subequations}
	\begin{align}
		\frac{d \hat b_{\mathbf{k}} }{d t}&=-i\omega_{\mathbf{k}} \hat b_{\mathbf{k}}+ g_{\mathbf{k}} \hat c, 
		\label{eq:HeisenbergEqEnvironment}\\
		\frac{d \hat X}{d t}&=-i[\hat X, \hat H_S ]+  \sum_{\mathbf{k}} g_{\mathbf{k}} \left( \hat b_{\mathbf{k}}^\dagger [\hat X,\hat c] - \hat b_{\mathbf{k}} [\hat X,\hat c^\dagger]  \right).
		\label{eq:HeisenbergEqSystem}
	\end{align}
\end{subequations}
\\[-1em]
We aim to derive a dynamical equation solely for the system operator. To achieve this, we formally solve \eqref{eq:HeisenbergEqEnvironment}:
\\[-1em]
\begin{equation}
	 \hat b_{\mathbf{k}}(t)= e^{-i\omega_{\mathbf{k}}(t-t_0)}\hat b_{\mathbf{k}} (t_0) + g_{\mathbf{k}}  \int_{t_0}^t dt' e^{-i \omega_\mathbf{k} (t-t')} \hat c(t').
	 \label{eq:EnvOpForwards}
\end{equation}
\\[-1em]
Then, by introducing the solution for $\hat b_{\mathbf{k}}(t)$ in~\eqref{eq:HeisenbergEqSystem}, the Heisenberg equation for $\hat X$ results in
\\[-1em]
\begin{multline}
	\frac{d \hat X}{d t}=
	-i[\hat X, \hat H_S ]\\
	+\sum_{\mathbf{k}} g_{\mathbf{k}} 
	\left\{  e^{i\omega_{\mathbf{k}} (t-t_0)}\hat b^\dagger_{\mathbf{k}} (t_0) [\hat X,\hat c]  -   [\hat X,\hat c^\dagger] e^{-i\omega_{\mathbf{k}} (t-t_0)}\hat b_{\mathbf{k},s} (t_0) \right\}\\
	+\sum_{\mathbf{k}} g_{\mathbf{k}}^2 \int_{t_0}^t dt'
	\left\{  e^{i\omega_{\mathbf{k}} (t-t')}\hat c^\dagger (t') [\hat X,\hat c] -   [\hat X,\hat c^\dagger] e^{-i\omega_{\mathbf{k}} (t-t')}\hat c(t') \right\}.
	\label{eq:HeisenbergEqSystem2}
\end{multline}
\\[-1em]
Note that for lightening the notation, we do not explicitly include the time dependency on $t$ for the operators, e.g., $\hat X\equiv X(t)$, and equivalently for the rest.  

Following the same methodology used in \refsec{Section:SpontaneousEmission}, we perform the limit to the continuum by considering that the environment is composed by modes that are infinitely dense in frequency space. As a consequence, we must replace the sum in $\mathbf{k}$ by the integral:
\begin{equation}
	\lim_{L\rightarrow \infty}\sum_{\mathbf{k}} \rightarrow   \int_0^\infty d\omega  \rho(\omega),
\end{equation}
where $\rho (\omega)=L/2\pi c$ is the mode density 
\graffito{
	$^*$The spectral density totally determines the light-matter dynamics, being of great relevance in quantum optics and nanophotonics, where tailored environment can produce non-trivial interactions. 
}
in a given $d\omega$, and $L$ is the length of the one-dimensional environment. This magnitude allows us to define the spectral density$^{\textcolor{Maroon}{*}}$, $J(\omega)$, as
\begin{equation}
	J(\omega)\equiv\rho(\omega) g(\omega)^2,
\end{equation} 
where $g(\omega)$ is the extension to the continuum of the coupling rate $g_{\mathbf{k}}$. Accordingly, the Kronecker delta converges to a Dirac delta,
\begin{equation}
	\lim_{L\rightarrow \infty}\delta_{k,k'} =\frac{2\pi c}{L} \delta(\omega-\omega').
\end{equation}
Then, it is convenient to defined a continuous set of bosonic operators$^{\textcolor{Maroon}{**}}$:
\graffito{
	$^{**}$The definition of normal modes in the continuum limit requires a subtler description when more general setups are considered, such as an EM field confined in higher dimensions, or in nanophotonic environments~\cite{Navarrete-BenllochContributionsQuantum2015,FeistMacroscopicQED2020}. In these cases, the formal structure of the light-matter interaction is crucial to define a consistent continuum bosonic operator.  
}
\begin{equation}
	\hat b(\omega)=\sqrt{\frac{L}{2\pi c}} \lim_{L\rightarrow \infty} \hat b_{k}
\end{equation}
which satisfies the \textit{canonical commutation relations}:
\begin{equation}
	[\hat b(\omega),\hat b(\omega')]=0, \quad 	[\hat b(\omega),\hat b^\dagger(\omega')]=\delta(\omega-\omega').
\end{equation}
Consequently, the bosonic operator in \eqref{eq:EnvOpForwards} now reads:
\begin{equation}
	\hat b(\omega)= e^{-i \omega (t-t_0)} \hat b_0(\omega)-g(\omega) \int_{t_0}^{t}dt' e^{-i \omega (t-t')}\hat c(t'),
	\label{eq:EnvOpForwardsFreq}
\end{equation}
where we have defined $\hat b_0(\omega)\equiv \hat b(\omega,t_0)$.
These ingredients allow us to redefine the environment and interaction Hamiltonians as
\begin{subequations}
	\begin{align}
		\hat H_E&= \int_{0}^\infty d\omega \omega \hat b^\dagger(\omega) \hat b(\omega), \\
		\hat H_{\text{int}}&= i\int_{0}^\infty d\omega \sqrt{J(\omega)} [\hat b^\dagger(\omega) \hat c - \hat b(\omega) \hat c].
	\end{align}
\end{subequations}

Returning to our problem, \eqref{eq:HeisenbergEqSystem2} is now expressed in the continuum limit by 
\begin{multline}
	\frac{d \hat X}{d t}=
	-i[\hat X, \hat H_S ]\\
	+\int_{-\infty}^\infty d\omega \sqrt{J(\omega)} 
	\left\{  e^{i\omega (t-t_0)}\hat b^\dagger_0(\omega) [\hat X,\hat c]  -   [\hat X,\hat c^\dagger] e^{-i\omega (t-t_0)}\hat b_0 (\omega) \right\}\\
	+\int_{-\infty}^\infty d\omega J(\omega) \int_{t_0}^t dt'
	\left\{  e^{i\omega (t-t')}\hat c^\dagger (t') [\hat X,\hat c] -   [\hat X,\hat c^\dagger] e^{-i\omega (t-t')}\hat c(t') \right\},
	\label{eq:HeisenbergEqSystem3}
\end{multline}
where we have extended the lower limit in the integrals from $0\rightarrow -\infty$ to simplify upcoming integrals. This is justified since only environment frequencies near the system resonance $\omega\approx \omega_0$ contribute significantly to the dynamics.	Any fictitious unphysical negative frequency mode $\hat b(-|\omega|)$ and physical modes $\hat b(|\omega|)$ with $|\omega|\gg \omega_0$ will not affect the physics as they are very off-resonant.

Now, we can perform the Markovian approximation by assuming that the spectral density is approximately constant around the resonant frequency, e.g., the resonance linewidth in the case of an optical cavity: 
\begin{equation}
J(\omega)\approx J(\omega_0)\equiv \frac{\gamma}{2\pi}.
\label{eq:SpectralDensityFlat}
\end{equation}
Then, we can define the \textit{input field} as
\begin{equation}
	\hat b_{\text{in}}(t) \equiv \frac{1}{\sqrt{2\pi}} \int_{-\infty}^\infty d\omega e^{-i\omega(t-t_0)}\hat b_0(\omega),
	\label{eq:InputOp}
\end{equation}
satisfying the continuous canonical commutation relation:
\begin{equation}
[\hat b_{\text{in}}(t),\hat b_{\text{in}}(t')]=0,\quad 	[\hat b_{\text{in}}(t),\hat b^\dagger_{\text{in}}(t')]=\delta(t-t').
\end{equation}

Finally, by inserting the
\graffito{
	$^*$To get the final expression for the quantum Langevin equation in \eqref{eq:LangevinEq}, we must use the following property of the delta function when the peak is at the end of the integration interval:
	$$
	\resizebox{0.32\textwidth}{!}{$\int_{\tau_-}^{\tau_+} d\tau \, f(\tau)\delta(\tau-\tau_\pm) = \frac{1}{2}f(\tau_\pm)$}
	$$
	valid for any function $f(\tau)$ that is continuous in all the integration domain. It can be understood as \textit{taking just half of the peak}.
}
 definitions from~\cref{eq:SpectralDensityFlat,eq:InputOp} into \eqref{eq:HeisenbergEqSystem3}, we arrive to the \textit{quantum Langevin equation}$^{\textcolor{Maroon}{*}}$:
\begin{equation}
\colorboxed{Maroon}{\begin{aligned}[b]
		\frac{d \hat X}{d t}=
		-i[\hat X, \hat H_S ]
		- [\hat X, \hat c^\dagger] \left(
		\frac{\gamma}{2} \hat c \right. + & \left. \sqrt{\gamma} \hat b_{\text{in}} 
		\right)
		\\
		&+\left(
		\frac{\gamma}{2} \hat c^\dagger + \sqrt{\gamma} \hat b^\dagger_{\text{in}} 
		\right) [\hat X, \hat c].
\end{aligned}}
\label{eq:LangevinEq}
\end{equation}
The quantum Langevin equation presented in \eqref{eq:LangevinEq} is valid without requiring any further assumptions about the initial state. The input operator, $\hat b_{\text{in}}(t)$, in complete analogy with the classical Langevin equations, is interpreted as a noise operator that drives the system at each instant, and whose nature its determined at the origin of time given a concrete initial quantum state of the environment. We note that this freedom for determining the nature of the input field is a consequence of assuming the system to be initially factorized, so that there are no correlations between the system and the environment.

\paragraph{Time-Reversed Quantum Langevin Equation.}

The resulting quantum Langevin equation in \eqref{eq:LangevinEq} contains the input field operator $\hat b_{\text{in}}(t)$ defined in terms of the values of the field operator $b_0$ at the origin of times $t_0$, which is implicitly understood to be in the \textit{remote past}. Similarly, we can formulate this equation in terms of the values of the field operators at a time $t_1>t$, implicitly understood to be in the \textit{remote future}. Hence, we can perform a backwards integration of the environment operator [\eqref{eq:HeisenbergEqEnvironment}],
\begin{equation}
	\hat b(\omega)= e^{-i \omega (t-t_1)} \hat b_1(\omega)-g(\omega) \int_t^{t_1}dt' e^{-i \omega (t-t')}\hat c(t'),
	\label{eq:EnvOpBackwardsFreq}
\end{equation}
such that, performing an analogous analysis by inserting this solution into \eqref{eq:HeisenbergEqSystem}, we can define the \textit{output field} operator as
\begin{equation}
	\hat b_{\text{out}}(t)= \frac{1}{\sqrt{2\pi}} \int_{-\infty}^\infty d\omega e^{-i\omega(t-t_1)}\hat b_1(\omega),
\end{equation}
where $\hat b_1(\omega)\equiv \hat b(\omega,t_1)$, following the notation introduced before. Analogously as with the input case, the output field operator satisfies continuous canonical commutation relations:
\begin{equation}
	[\hat b_{\text{out}}(t),\hat b_{\text{out}}(t')]=0,\quad 	[\hat b_{\text{out}}(t),\hat b^\dagger_{\text{out}}(t')]=\delta(t-t').
\end{equation}
Therefore, we arrive to the \textit{time-reversed quantum Langevin equation}:
\begin{equation}
	\colorboxed{Maroon}{\begin{aligned}[b]
			\frac{d \hat X}{d t}=
			-i[\hat X, \hat H_S ]
			- [\hat X, \hat c^\dagger] \left(
			\frac{\gamma}{2} \hat c \right. - & \left. \sqrt{\gamma} \hat b_{\text{out}} 
			\right)
			\\
			&+\left(
			\frac{\gamma}{2} \hat c^\dagger - \sqrt{\gamma} \hat b^\dagger_{\text{out}} 
			\right) [\hat X, \hat c].
	\end{aligned}
}
	\label{eq:LangevinEqTimeReversed}
\end{equation}
We note that the signs of the last term within both parenthesis are interchanged in comparison to the result in terms of input operators in \eqref{eq:LangevinEq}. This is natural once one realizes that, for causal consistency, this equation must be solved backwards in time, starting from a final condition at $t_1$. Therefore, what it is a damping term, e.g.,  $-\gamma/2 [\hat a,\hat c^\dagger] \hat c$, in the quantum Langevin equation becomes an amplification term, $\gamma/2 [\hat a,\hat c^\dagger] \hat c$, in the time-reversed quantum Langevin equation.

\subsection{Input-output theory}
\label{sec:input_output_theory}

Either integrating forwards or backwards, we have an expression for the environment operator $\hat b(\omega)$, meaning that both \eqref{eq:EnvOpForwardsFreq} and \eqref{eq:EnvOpBackwardsFreq} must be identical. In particular, by integrating these equation over all frequencies and comparing them, we arrive to the so-called \textit{input-output relations}:
\begin{equation}
\colorboxed{Maroon}{	\hat b_{\text{out}}(t)=\sqrt{\gamma} \hat c(t)+\hat b_{\text{int}}(t).}
\label{eq:InputOutputRelation}
\end{equation} 
Despite its formal simplicity, this expression encodes a powerful theoretical result: the field coming out from a given system is a superposition of the input field and the scattered field emerging as a consequence of the interaction with the system.  	The input-output relation can be viewed as a boundary condition, relating input, output, and system modes, that respects quantum causality: only the future motion of the system is affected by the present input, and only the future value of the output field is affected by the present values of the system operators.

The general strategy using quantum Langevin equations is, then, summarized in: \textcolor{Maroon}{(i)} specify the nature of the input field $\hat b_{\text{in}}(t)$, \textcolor{Maroon}{(ii)} determine the influence of the input on a system operator $\hat X(t)$, and \textcolor{Maroon}{(iii)} use the input-output relation to determine the output field $\hat b_{\text{out}}(t)$, which is going to be measured in practice.
Therefore, if we consider that the  input field is in vacuum, any normal-ordered correlator of the output field depends only on the system operator $\hat c$. This approach allows us to link the output electric field---what is actually observed in the laboratory---to the system operator, up to some constants: $\hat{E} \rightarrow \hat{c}$.
Even if the situation is physically more complex, we can always consider that the input field is in vacuum by trivially extending the derivations of the quantum Langevin equations to the case of coupling to several environments, and consider one of them to be in vacuum, in which we perform the measurement~\cite{CarmichaelOpenSystems1993}.

\paragraph{Consistency with quantum mechanics.}
We note that the quantum Langevin equation in its standard form [\eqref{eq:LangevinEq}] can be rewritten as an adjoint master equation:
\begin{equation}
	\colorboxed{Maroon}{\begin{aligned}[b]
	\frac{d \hat X}{d t}=-i[\hat X, \hat H_S ]+ \frac{\gamma}{2}\left( 2 \hat c^\dagger \hat X \hat c\right.& \left.-\{ \hat c^\dagger \hat c,\hat X \} \right)\\
	&+\sqrt{\gamma} \left(
	\hat b_{\text{in}}^\dagger [\hat X, \hat c] - [\hat X, \hat c^\dagger]\hat b_{\text{in}}
	\right),
\end{aligned}
}
\end{equation}
but with an additional term that does not appear in \eqref{eq:AdjointMasterEq} since we have already traced out the environment degrees of freedom during the derivation of the master equation.  This difference has an impact when computing the canonical commutation relation, for instance, of a bosonic mode $\hat{a}$ interacting with an environment at thermal equilibrium:
\begin{subequations}
	\begin{align}
		[\hat a(t),\hat a^\dagger (t)]_{\text{Adjoint Master Eq.}}= e^{-\gamma t},
		\label{eq:CommutatorAdjoint} \\
		[\hat a(t),\hat a^\dagger (t)]_{\text{Quantum Langevin Eq.}}=1,
		\label{eq:CommutatorLangevin}
	\end{align}
\end{subequations}
At long times, \eqref{eq:CommutatorAdjoint} suggests a breakdown of the commutation relation, leading to an apparent inconsistency with quantum mechanics. This inconsistency, however, can be resolved by noting that the master equation formalism has already traced out the environment degrees of freedom, while the quantum Langevin equation remains explicitly \textit{environment-dependent}, preserving the quantum properties of the full system as long as no expectation values are taken. In fact, if we compute the expectation value of $\hat a(t)$ obtained via the quantum Langevin equations, and then we compute the canonical commutator, we recover the same result as in \eqref{eq:CommutatorAdjoint}.  

\paragraph{Difference between Schrödinger and Heisenberg pictures in open quantum systems. } This distinction between both formalisms---the master equation and the quantum Langevin equation---evidenced in \cref{eq:CommutatorAdjoint,eq:CommutatorLangevin} highlights the relevance of retaining the information of the environment for specific computations, such as the canonical commutators of a bosonic operator.
However, both formalisms provide the same physical results once the nature of the environment is specified and expectation values are taken. Therefore, we can use them indistinctly for computing physical observables.

\subsection{Example: Injection of a laser}
\label{Section:CoherentExcitation}

Let us consider the injection of a classical beam, i.e., a laser, that drives the system as an input field. In this case, the environment is supposed to be in a coherent state
\begin{equation}
	|\psi_E\rangle = \text{exp}\left[
	\int_{-\infty}^\infty d\omega (\alpha(\omega)\hat b^\dagger (\omega)-\alpha^*(\omega)\hat b(\omega))
	 \right] \bigotimes_\omega |0\rangle \equiv 	|\alpha(\omega)\rangle,
\end{equation}
where each external mode of frequency $\omega$ is described as a coherent state $|\alpha(\omega)\rangle$, fulfilling the properties of a coherent state:
\begin{subequations}
	\begin{align}
		&\langle \hat b_0 (\omega)\rangle=\alpha (\omega), \\
		&\langle \hat b_0 (\omega)\hat b_0 (\omega')\rangle= \alpha (\omega)\alpha (\omega'), \\
		&\langle \hat b^\dagger_0 (\omega)\hat b_0 (\omega')\rangle= \alpha^* (\omega)\alpha (\omega').
	\end{align}
\end{subequations}
As a consequence, the statistical properties of the input operators are easily given by:
\begin{subequations}
	\begin{align}
		&\langle  \hat b_{\text{in}}(t)\rangle= \frac{\tilde{\alpha}(t)}{\sqrt{2\pi}},\\
		&\langle  \hat b_{\text{in}}(t)\hat b_{\text{in}}(t')\rangle=\frac{\tilde{\alpha}(t)\tilde{\alpha}(t')}{2\pi}, \\
		&\langle  \hat b^\dagger_{\text{in}}(t)\hat b_{\text{in}}(t')\rangle=\frac{\tilde{\alpha}^*(t)\tilde{\alpha}(t')}{2\pi}.
	\end{align}
\end{subequations}
having defined $\tilde{\alpha}(t)\equiv \int_{-\infty}^\infty d\omega e^{-i\omega t}\alpha(\omega)$. That is, each of the input modes is described by a coherent state $|\alpha_{\text{in}}(t)\rangle$ with an amplitude $\alpha_{\text{in}}=\tilde{\alpha}(t)/\sqrt{2\pi}$. In cases where the first moment of the input field is not zero, it is helpful to introduce a new input operator:
\begin{equation}
	\hat a_{\text{in}}(t)\equiv \hat b_{\text{in}}(t)- \langle \hat b_{\text{in}}(t)\rangle ,
\end{equation} 
such that reproduces the behaviour of a vacuum external mode with $\langle \hat a_{\text{in}}\rangle=0$. 
Additionally, we assume that the laser field contains $N$ frequency components $\{ \omega_i\}_{i=1,2,\ldots,N}$, along with their respective field amplitude components $\alpha_i$. Then, the spectral decomposition of $\alpha(\omega)$ can be written as
\begin{equation}
	\alpha(\omega)=\sum_{l} \alpha_l \delta(\omega-\omega_l).
	\label{eq:alpha_laser}
\end{equation}
If only one frequency significantly influences the dynamics---e.g., when the system is able to resolve each spectral component since their spectral separation is larger than their linewidths---we can approximate \eqref{eq:alpha_laser} to just 
\begin{equation}
	\alpha(\omega)\approx \alpha_L \delta(\omega-\omega_L).
\end{equation}
In other words, the external field is to be considered as a monochromatic laser.

Hence, considering again the case of a single bosonic mode with Hamiltonian $\hat H_S=\omega_a \hat a^\dagger \hat a$, the quantum Langevin equation [see \eqref{eq:LangevinEq}] for the bosonic mode reads:
\begin{equation}
	\frac{d \hat a}{d t}= i\Omega_L e^{-i\omega_L t} -(\gamma/2+i \omega_a)\hat a(t)+\sqrt{\gamma}\hat a_{\text{in}}(t),
	\label{eq:LaserLangevin}
\end{equation}
where we have defined 
\begin{equation}
	\Omega_L \equiv  \sqrt{\frac{\gamma}{2\pi}}\alpha_L.
\end{equation}
Then, given the structure of \eqref{eq:LaserLangevin}, we realize that the injection of a laser field is equivalent to the addition of an extra time-dependent term in the Hamiltonian, which will be simply called as \textit{driving Hamiltonian}:
\begin{equation}
	\colorboxed{Maroon}{\hat H_d(t)= \Omega_L (\hat a e^{-i \omega_L t} +\hat a^\dagger e^{i \omega_L t} ),}
	\label{eq:LaserHamiltonian}	
\end{equation}
where we have assumed that $\alpha_L$ is such that the driving strength, or \textit{Rabi frequency}, $\Omega_L$ is purely real.  The resulting Hamiltonian, $\hat H_d$, corresponds to a time-dependent driving term that excites the system coherently, offering an approximated description of the injection of a laser field with an intensity $I \propto \Omega_L^2$. We note that the Rabi frequency depends on the decay rate of the system that is being excited, $\Omega_L\propto \sqrt{\gamma}$, in agreement with the fact that a system that cannot emit cannot be excited either~\cite{CarrenoExcitationQuantum2016a,CarrenoExcitationQuantum2016,KavokinMicrocavities2017}. We will make use of this model extensively throughout this Thesis.

\section{Cascaded formalism}
\label{Section:Cascaded}

In the previous section, we presented the input-output formalism in the simplest situation: an input field drives a system, and an output field emerges from this interaction, carrying the information of the system and the input field via the \textit{input-output relation}.  
This idea can be generalized to the case in which the output field of a system (the \textit{source}) is fed into the input of another system (the \textit{target}), without any back-action from the target to the source, allowing thus the description of more general setups which are driven by a general source of light. Systems linked in this way are known as \textit{cascaded quantum systems}, and are formally described by the cascaded formalism~\cite{KolobovInterferencePhysical1987,GardinerDrivingQuantum1993,GardinerDrivingAtoms1994,CarmichaelQuantumTrajectory1993}.

In this section, we present the cascaded quantum master equation by adapting the derivations from \textcolor{Maroon}{Refs.}~\cite{GardinerQuantumNoise2004,GardinerStochasticMethods2009,GardinerQuantumWorld2014,GardinerQuantumWorld2015}.

\subsection{Cascaded Quantum Langevin equation}
The description of a cascaded quantum system is easily understood in terms of quantum Langevin equations, where a driving field $\hat b_{\text{in}}(1,t)$ drives the \textit{source}, giving rise to an output field $\hat b_{\text{out}}(1,t)$, which, after a propagation time $\tau$, becomes the input field $\hat b_{\text{in}}(2,t)$ of the second system.  

Let $\hat a_1, \hat a_2$ be operators for the two systems, and let $\hat H=\hat H_1 + \hat H_2$ be the sum of their independent free Hamiltonians. In this configuration, the quantum Langevin equation [see \eqref{eq:LangevinEq}] for  $\hat a_i$ (with $i=1,2$) reads:
\begin{multline}
	\frac{d \hat a_i}{d t} = -i[\hat a_i, \hat H_i]-[\hat a_i, \hat c^\dagger_i] \left( \frac{\Gamma_i}{2} \hat c_i + \sqrt{\Gamma_i}\hat b_{\text{in},i} \right)  \\
	+\left( \frac{\Gamma_i}{2} \hat c^\dagger_i + \sqrt{\Gamma_i}\hat b^\dagger_{\text{in},i} \right)[\hat a_i, \hat c_i].
\end{multline}
The cascaded coupling implies that the output field from the first system, $\hat b_{\text{out},1}=\hat b_{\text{in},1}+\sqrt{\Gamma_1}\hat c_1$, feds the second system after a time $\tau$.  We can express this cascaded effect by writing the input field of system $2$ as
\begin{equation}
	\hat b_{\text{in},2}=\hat b_{\text{out},1}(t-\tau)=\hat b_{\text{in},1}(t-\tau)+\sqrt{\Gamma_1}\hat c_1(t-\tau),
	\label{eq:CascadedInputRelation}
\end{equation}
so that the output of the combined system reads:
\begin{multline}
	\hat b_{\text{out},2}=\hat b_{\text{in},2}(t-\tau)+\sqrt{\Gamma_2}\hat c_2(t-\tau)\\
	=b_{\text{in},1}(t-\tau) +\sqrt{\Gamma_2}\hat c_2(t-\tau)+\sqrt{\Gamma_1}\hat c_1(t-\tau).
	\label{eq:CascadedOutputRelation}
\end{multline}
We now have operators evaluated at two different times, $t$ and $t-\tau$.  However, since we are considering a one-way driving scenario and $\tau > 0$, the effect of this \textit{retarded time} is merely a time shift of the origin of the time axis for the second system. Thus, we can set $\tau \rightarrow 0^+$ without losing generality while maintaining causality. We also note that, given the \textit{cascaded input-output relations} in~\cref{eq:CascadedInputRelation,eq:CascadedOutputRelation}, we can identify the input and output fields of the combined systems just as $\hat b_{\text{in}}\equiv \hat b_{\text{in},1}$ and $\hat b_{\text{out}}\equiv \hat b_{\text{out},2}$.

Finally, by considering a general operator $\hat a$, which may act either in the Hilbert space of the source or the target, we obtain the cascaded quantum Langevin equation:
\begin{equation}
\colorboxed{Maroon}{\begin{aligned}[b]
		\frac{d \hat a}{d t} &= -i[\hat a, \hat H] - \sqrt{\Gamma_1 \Gamma_2} \left( [\hat a, \hat c^\dagger_2]\hat c_1- \hat c_1^\dagger[\hat a, c_2] \right)
		\\
		&+\sum_{i=1}^2 \left[ 
		-[\hat a, \hat c_i^\dagger]\left( \frac{\Gamma_i}{2} \hat c_i + \sqrt{\Gamma_i}\hat b_{\text{in}} \right)
		+
		\left( \frac{\Gamma_i}{2} \hat c^\dagger_i + \sqrt{\Gamma_i}\hat b^\dagger_{\text{in}} \right)[\hat a, \hat c_i]
		\right],
\end{aligned}}
\label{eq:CascadedLangevin}
\end{equation}
where the second term proportional to $\sqrt{\Gamma_1 \Gamma_2}$ stands for the asymmetric cascaded coupling by which system $1$ drives system $2$ but not reversely.

Alternatively, we can express \eqref{eq:CascadedLangevin} in the form of a standard quantum Langevin equation [see \eqref{eq:LangevinEq}]. This follows from the structure of the output relation in \eqref{eq:CascadedOutputRelation}, which indicates that the effective coupling of the cascaded system to the input field is given by the collective operator 
$	\hat C= \sqrt{\Gamma_1}\hat c_1+\sqrt{\Gamma_2}\hat c_2$, such that
\begin{multline}
	\frac{d \hat a}{d t} = -i[\hat a, \hat H+\frac{i}{2}\sqrt{\Gamma_1\Gamma_2}(\hat c_1^\dagger \hat c_2 -\hat c_2^\dagger \hat c_1)] 
	\\
-[\hat a, \hat C^\dagger]\left( \frac{1}{2}\hat C + \hat b_{\text{in}} \right)
+
\left( \frac{1}{2} \hat C^\dagger + \hat b^\dagger_{\text{in}} \right)[\hat a, \hat C].
\label{eq:CascadedLangevin1}
\end{multline}
This form explicitly shows the cascaded coupling as an additional Hamiltonian term
\begin{equation}
	\hat H_c= \frac{i}{2}\sqrt{\Gamma_1\Gamma_2}(\hat c_1^\dagger \hat c_2 -\hat c_2^\dagger \hat c_1),
\end{equation}
which exchanges the excitation between both systems but that is antisymmetric in indices. Additionally, in this form, the combined cascaded system behaves as \textit{single system}, with a single coupling term to the radiation field given by the combination $\sqrt{\Gamma_1}\hat c_1+\sqrt{\Gamma_2}\hat c_2$. 

\subsection{Cascaded Quantum Master equation}
Although the cascaded quantum Langevin equation gives an elegant description of the physics involved, it is necessary to convert this into the appropriate \textit{Quantum Ito equation}, and then obtain the corresponding master equation. 

To achieve this, we must treat the input operator as a noise operator by replacing
\begin{equation}
	\hat b_{\text{in}}(t)dt \rightarrow d\hat B(t),
\end{equation} 
where $ d\hat B(t)$ is the \textit{quantum Wiener increment}, fulfilling the \textit{quantum Ito algebra}~\cite{GardinerQuantumNoise2004,GardinerStochasticMethods2009,WisemanQuantumMeasurement2014}:
\begin{subequations}
	\begin{align}
		&d\hat B(t)  d\hat B^\dagger(t)=dt, \\
		& d\hat B^\dagger(t) d\hat B(t)=0, \\
		& d\hat B(t)d\hat B(t)=d\hat B^\dagger(t)d\hat B^\dagger(t)=0.
	\end{align}
\end{subequations}
All other products, including $dt d\hat B(t)$ or $dt d\hat B^\dagger(t)$, and higher orders are set to zero. We note that these quantum Ito rules are valid when the expected value is applied. Hence, using the quantum Ito algebra and taking the expected value of the quantum Langevin equation in \eqref{eq:CascadedLangevin}, we obtain the cascaded master equation:
\begin{equation}
	\colorboxed{Maroon}{
	\begin{aligned}[b]
		\frac{d \rho(t)}{d t}=-i [\hat H,\hat \rho] + \sum_i^2 \frac{\Gamma_i}{2}&\mathcal{D}[\hat c_i]\hat \rho\\
		&- \sqrt{\Gamma_1 \Gamma_2}\left( [\hat c_2^\dagger , \hat c_1 \hat \rho]+ [\hat \rho \hat c_1^\dagger, \hat c_2] \right).
	\end{aligned}
	}
	\label{eq:CascadedMaster}
\end{equation}
This equation contains the standard terms of a Lindblad master equation [\eqref{eq:LindbladMasterEq}]:  a term for the free evolution governed by the Hamiltonian $\hat H$, and Lindblad terms for both the source and the target, with corresponding dissipative rates $\Gamma_1$ and $\Gamma_2$, respectively. Additionally, it features an extra term that accounts for the asymmetric cascaded coupling between the source and the target.
Analogously to \eqref{eq:CascadedLangevin1}, the cascaded master equation can also be expressed in terms of the collective operator $\hat C$ as
\begin{equation}
	\frac{d \rho(t)}{d t}
	= -i \left[\hat H+i\frac{\sqrt{\Gamma_1 \Gamma_2}}{2} ( \hat c^\dagger_1  \hat c_2 - \hat c^\dagger_2 \hat c_1),\hat \rho\right]+\frac{1}{2}\mathcal{D}[\hat C]\hat \rho.
		\label{eq:CascadedMaster2}
\end{equation}
Both descriptions, \eqref{eq:CascadedMaster} and \eqref{eq:CascadedMaster2}, are entirely equivalent. Therefore, the choice of which to use depends on the specific nature of the problem at hand.

We must note that the cascaded coupling rate, $\propto \sqrt{\Gamma_1 \Gamma_2}$, is totally determined by the dissipative decay rates of the source and target. 
Consequently, a strong cascaded coupling induces a very lossy regime either for one or both systems, producing that the quantum state in the target will not live long as a consequence of a fast decay.
Therefore, there is a trade-off between achieving efficient cascaded coupling to excite the target and maintaining a sufficiently long relaxation time to observe quantum effects. 
We also note that arbitrary coupling strengths are not allowed, otherwise, the target would feature unphysical states~\cite{CarrenoExcitationQuantum2016a,CarrenoExcitationQuantum2016}.

\subsection{Imperfect coupling}
In the derivation of the cascaded master equation we have assumed that the cascaded coupling is perfect: the whole output of the first system is used as the input of the second. However, this is usually not realistic since, for instance, there may be losses in the transmission, or we just want to retain the scattered light from a coherently driven system.

This problem can be addressed by introducing two additional input-output channels, with amplitudes $\varepsilon_1$ and $\varepsilon_2$ for the source and target, respectively, to account for imperfect physical transmission. These channels must satisfy the conditions $\sum_i \varepsilon_i = 1$ and $\varepsilon_i \leq 1$ for physical consistency. The inclusion of these terms leads to the imperfect cascaded master equation~\cite{GardinerQuantumNoise2004}:
\begin{equation}
	\colorboxed{Maroon}{
		\begin{aligned}[b]
			\frac{d \rho(t)}{d t}=-i [\hat H,\hat \rho] + \sum_i^2 \frac{\Gamma_i}{2}&\mathcal{D}[\hat c_i]\hat \rho\\
			&- \sqrt{\eta \Gamma_1 \Gamma_2}\left( [\hat c_2^\dagger , \hat c_1 \hat \rho]+ [\hat \rho \hat c_1^\dagger, \hat c_2] \right).
		\end{aligned}
	}
	\label{eq:CascadedMasterImperfect}
\end{equation}
where we have replaced $\sqrt{\Gamma_1 \Gamma_2} \rightarrow \sqrt{\eta \Gamma_1 \Gamma_2}$, with $\eta \equiv (1-\varepsilon_1) (1-\varepsilon_2)$.  The factor $\eta$ quantifies the amount of signal that each part, source and target, generates on its own. Clearly, due to the imperfect coupling, the cascaded coupling is less efficient than in the ideal case. We observe that by setting $\varepsilon_1 = \varepsilon_2 = 0$, we recover the perfect coupling scenario in \eqref{eq:CascadedMaster}.

If we consider the case in which the source is driven by a coherent field [see \refsec{Section:CoherentExcitation}], the driving Hamiltonian from \eqref{eq:LaserHamiltonian} gets modified to
\begin{equation}
	\hat H_d= \sqrt{\varepsilon_1}\Omega_L (\hat c_1^\dagger + \hat c_1),
\end{equation}
 ensuring that the target is not directly excited by the coherent field but only through the scattered emission from the source. Nevertheless, in practice, we can simply redefine the Rabi frequency as $\Omega_L \equiv \sqrt{\varepsilon_1}\Omega_L $, and concentrate  on the imperfect coupling factor affecting the target.

\section{Quantum trajectories formalism}
\label{Section:QuantumTraject}

In the context of quantum optics, there is an alternative approach to the formalisms presented in \refsec{Section:OPS} and \refsec{Section:OPS_Heisenberg}, named Monte Carlo method of quantum trajectories, or simply quantum trajectories formalism~\cite{CarmichaelPhotoelectronWaiting1989,MolmerMonteCarlo1993,MolmerMonteCarlo1996,PlenioQuantumjumpApproach1998}. It is based on simulating open quantum systems by expressing the evolution of the system's density matrix in terms of a pure wave function that evolves under stochastic, non-unitary dynamics. 
This approach features two main benefits: \textcolor{Maroon}{(i)}  it provides insights about the physical mechanisms underlying the evolution of the density matrix by studying individual quantum trajectories (which are averaged out by the master equation), and \textcolor{Maroon}{(ii)} it can be computationally faster when the Hilbert space of the quantum system is very large ($N\gg 1$) since this method involves the manipulation of a wavevector of dimension $N$ instead of a density matrix of dimension $N^2$.

The following derivations are adapted from \textcolor{Maroon}{Refs.}~\cite{MolmerMonteCarlo1993,MolmerMonteCarlo1996,GerryIntroductoryQuantum2004}.

\subsection{Monte Carlo method of quantum trajectories}
Let us consider a quantum system at time $t$, described by $|\psi (t)\rangle$, that is governed by the \textit{non-Hermitian Hamiltonian}, $\tilde H$,
\begin{equation}
	\tilde H= \hat H_S -i \sum_\mu \frac{ \Gamma_\mu}{2} \hat L_\mu^\dagger \hat L_\mu,
	\label{eq:NonHermHamiltonian}
\end{equation}
which encodes the interplay between the coherent and dissipative processes by means of the system Hamiltonian, $\hat H_S$, and the jump operators, $\hat L_\mu$. Then, the wavefunction at a time $t+\delta t$ is given by
\begin{equation}
	|\psi (t+\delta t) \rangle=\frac{e^{-i\delta t \tilde H} |\psi(t)\rangle}{(\langle \psi(t)|e^{i\delta t \tilde H^ \dagger}e^{-i\delta t \tilde H} |\psi(t)\rangle)^{1/2} } 
	\approx 
	\frac{(\mathbb{I}-i\delta t \tilde H) |\psi(t)\rangle}{(1-\delta p)^{1/2} },
	\label{eq:NonHermEvolution}
\end{equation}
where the approximation holds to order $\delta t$, and $\delta p$ is defined as
\begin{equation}
\delta p\equiv i  \delta t \langle \psi(t)| \tilde H-\tilde H^ \dagger |\psi(t)\rangle = \sum_\mu \delta p_\mu.  \label{eq:ProbNoQuantumJump}
\end{equation}
Here, $\delta p_\mu$ is the probability that during the evolution from $t$ to $t+\delta t$ the system undergoes a quantum jump induced by the jump operator $\hat L_\mu$,
\begin{equation}
	\delta p_\mu\equiv \delta t \Gamma_\mu \langle \psi(t)|\hat L_\mu^\dagger \hat L_\mu |\psi(t)\rangle \geq 0.
	\label{eq:ProbQuantumJump}
\end{equation}
The value of $\delta t$ is adjusted so that $\delta p\ll 1$ and such that the approximated evolution remains valid in \eqref{eq:NonHermEvolution}. 

In order to decide whether the system undergoes a quantum jump or not, we follow a \textit{gedanken} measurement process by generating random numbers (as many as jump operators), $r_\mu$, uniformly distributed between $0$ and $1$, and compare them with $\delta p_\mu$. Hence, we distinguish between two cases:
\begin{subequations}
	\begin{align}
		\label{eq:TrajEmit}
		&|\psi (t+\delta t) \rangle_\text{emit} = \frac{	\hat L_\mu |\psi (t) \rangle}{(\delta p_\mu/\delta t)^{1/2}}, \quad \text{if}\quad r_\mu < \delta p_\mu, \\
		\label{eq:TrajNoEmit}
		&|\psi (t+\delta t) \rangle_\text{no-emit} \approx 
		\frac{(\mathbb{I}-i\delta t \tilde H) |\psi(t)\rangle}{(1-\delta p)^{1/2} }, \quad \text{if}\quad r_\mu > \delta p.
	\end{align}
\end{subequations}
We note that in the exceptional case where there are more than one $\mu$ satisfying $ r_\mu < \delta p_\mu$, we need to design a strategy to select the operator that induces the jump. For instance, one can select the operator for which the difference $|p_\mu-r_\mu|$ is the largest. On the other hand, we note that when $r_\mu > \delta p=\sum_\mu \delta p_\mu$, which occurs in most cases since $\delta p\ll 1$, no quantum jumps occurs and the state evolves with the non-unitary Hamiltonian from \eqref{eq:NonHermHamiltonian}. 
We repeat this process to obtain the history of the trajectory that describe its conditional evolution while recording the jump events at times $t_1,t_2,\ldots$ 

\subsection{Equivalency to the master equation formalism}

We note that the present method is completely equivalent to the master equation formalism when taking averages in the limit of many realizations and $\delta t\rightarrow 0$.

Let us consider a vector state at a time $t+\delta t$, described by $|\psi(t+\delta t)\rangle$. The average of this state over different realization causes two alternatives: \textcolor{Maroon}{(i)} jump with a probability $\delta p_\mu$, or \textcolor{Maroon}{(ii)} evolve with a probability $1-\delta p$. This is easily translated in terms of density matrices as:
\begin{multline}
	|\psi(t+\delta t)\rangle \langle \psi(t+\delta t)| 
	=  \delta p |\psi (t+\delta t) \rangle_\text{emit} \langle \psi (t+\delta t)| \\
	+ (1-\delta p)|\psi (t+\delta t) \rangle_\text{no-emit} \langle \psi (t+\delta t)|.
\end{multline} 
By inserting the definitions of $|\psi (t+\delta t) \rangle_\text{emit}$ and $|\psi (t+\delta t) \rangle_\text{no-emit}$ from \cref{eq:TrajEmit,eq:TrajNoEmit}, and considering that $\delta p/\delta p_\mu \approx 1$ for the $\mu$ responsible of the jump, we arrive into
\begin{equation}
	\hat \rho(t+\delta t)=\hat \rho(t)-i\delta t [\hat H_S,\hat \rho(t)]+ \sum_\mu \frac{\Gamma_\mu}{2}\delta t\mathcal{D}[\hat L_\mu] \hat \rho(t),
\end{equation}
where $\hat \rho(t)$ is the density matrix of the ensemble. Then, by taking the limit  $\delta t \rightarrow 0$ and rearranging the expression, we prove that the quantum trajectories formalism is equivalent to a master equation description [see \eqref{eq:LindbladMasterEq}]
\begin{equation}
	\lim_{\delta t \rightarrow 0} \frac{	\hat \rho(t+\delta t)-	\hat \rho(t)}{\delta t}=-i [\hat H_S,\hat \rho(t)]+ \sum_\mu \frac{\Gamma_\mu}{2}\mathcal{D}[\hat L_\mu] \hat \rho(t).
\end{equation}

\section{Spectral properties in the photon emission}
 \label{Section:SpectralProperties}
 
In this Thesis, we are interested on the study of quantum optical system that are surrounded by an environment, giving rise to decoherence processes.
Besides introducing leakage of radiation, the environment also introduces a way to extract information from the system through the output field [see \eqref{eq:InputOutputRelation}]. Therefore, by collecting all the photons coming out from the system with a frequency-sensitive detector we can unveil some inner properties of the quantum system. This sort of measurement receives the name of spectral measurements.

In this section we introduce a brief discussion about the role of the detection process, and define the $N$-photon correlation function, showing special emphasis on the one- and two-photon correlation functions. We also present a computational method to easily compute these magnitudes. 

To address this section, we mainly adapt the discussions from \textcolor{Maroon}{Refs.}~\cite{DelValleTheoryFrequencyFiltered2012,KavokinMicrocavities2017}.

\subsection{Role of the detection}
To provide a realistic description 
\graffito{
	$^*$Since time and frequency are conjugate variables by means of the Mandelstam time-energy uncertainty~\cite{MandelstamUncertaintyRelation1991}, if we are able to determine with absolute precision the time at which the photon is detected, it will be impossible to determine its frequency, and \textit{viceversa}.
}
of the emission spectrum, we must incorporate the detection process in the theoretical model since previous methods based on the Wiener-Khinchin theorem relied on abstract and unrealistic assumptions, such as the stationarity of the electric field~\cite{MandelOpticalCoherence1995}. This leads us to the concept of the \textit{physical spectrum of light} proposed by Eberly and Wodkiewicz~\cite{EberlyTimedependentPhysical1977}. Additionally, to have a faithful physical description, we must also account for the uncertainties of detection in both time and frequency$^{\textcolor{Maroon}{*}}$.

We tackle this problem by including a filter function that reflects the spectral limitations of the detectors. This is easily implemented through a redefinition of the electric field operator as $\hat E(t)\rightarrow f(t) \hat E(t) $, with $f(t)$ an arbitrary filter function.
A commonly used filter function is the \textit{Lorentzian band-pass filter}, defined in the time domain as 
\begin{equation}
	f_{\omega_0}(t)=\frac{\Gamma}{2}e^{-(\Gamma/2+i\omega_0)t}\theta (t),
\end{equation}
where $\Gamma$ is the Full Width at Half Maximum (FWHM), representing the detector's linewidth, $\omega_0$ is the central frequency, and $\theta(t)$ is the Heaviside step function. By Fourier transform $	f_{\omega_0}(t)$, we arrive to the response function in the frequency domain:
\begin{equation}
	f_{\omega_0}(\omega)= \frac{\Gamma/2}{\Gamma/2 + i (\omega-\omega_0)}.
\end{equation}

\subsection{One and two-photon correlation functions}
From this definition of the filtered electric field, denoted as $\hat E_{\Gamma,\omega_0}(t)\equiv f_{\omega_0}(t)\hat E(t) $, we define the time-dependent spectrum as:
\begin{multline}
	S_\Gamma(\omega,t) \equiv  \langle \hat E^\dagger_{\Gamma,\omega}(t) \hat E_{\Gamma,\omega}(t) \rangle \\
	 =\left(\frac{\Gamma}{2}\right)^2 \int_{-\infty}^t \int_{-\infty}^t dt_1 dt_2 e^{-\Gamma/2(t-t_1)}e^{-\Gamma/2(t-t_2)} e^{i\omega (t_2-t_1)}    \langle \hat E^\dagger(t_1) \hat E(t_2) \rangle,
\end{multline}
such that, in the limit of an ideal detector ($\Gamma \rightarrow 0$) and in the steady-state regime, $t\rightarrow \infty$, having defined $\tau\equiv t_2-t_1$, we recover the Wiener-Khinchin theorem~\cite{MandelOpticalCoherence1995},
\begin{equation}
	\colorboxed{Maroon}{	S(\omega)\equiv \lim_{\substack{\Gamma \rightarrow 0 \\ t \rightarrow \infty}}
		S_\Gamma(\omega,t)  = \frac{1}{\pi } \frac{1}{n_c }\text{Re}\int_0^\infty d\tau e^{i\omega \tau} \langle \hat c^\dagger (0)\hat c(\tau) \rangle .
		}
		\label{eq:Spectrum}
\end{equation}
We note that the (normalized) spectrum is written in terms of the system operators as $\hat E^\dagger (t) \propto \hat c(t) $ by means of the input-output relation [see \eqref{eq:InputOutputRelation}]. Therefore, \eqref{eq:Spectrum} denotes the fluorescence spectrum of the light emitted by the mode $\hat c$, with $n_c$ being the steady-state population of the mode,
\begin{equation}
	n_c=\int_{-\infty}^\infty d\omega S(\omega)=\langle \hat c^\dagger \hat c \rangle =\text{Tr}[\hat c^\dagger \hat c \hat \rho_{\text{ss}}].
\end{equation}

Analogously to \eqref{eq:Spectrum}, we can obtain a normalized expression for the two-photon correlation function from different frequencies~\cite{ArnoldusPhotonCorrelations1984,KnollTheoryTimeresolved1984,KnollTheoryNfold1986,KnollQuantumNoise1986,CresserIntensityCorrelations1987}, %
\begin{equation}
	\colorboxed{Maroon}{	g^{(2)}_{\Gamma_1,\Gamma_2}(\omega_1,t_1; \omega_2,t_2) \equiv  \frac{\langle :\mathcal{T} [
			\prod_{i=1}^2  	\hat E^\dagger_{\Gamma_i,\omega_i}(t_i) \hat E_{\Gamma_i,\omega_i}(t_i)
			] : \rangle}{\prod_{i=1}^2 \langle 	\hat E^\dagger_{\Gamma_i,\omega_i}(t_i) \hat E_{\Gamma_i,\omega_i}(t_i) \rangle},
		}
		\label{eq:FreqResolvedFun}
\end{equation}
where $: \mathcal{T}[\cdot]:$ denotes the normal ordering of time-ordered products of operators.  
This quantity, also named frequency-resolved correlation function~\cite{DelValleTheoryFrequencyFiltered2012}, provides the density of joint photo-detection at frequencies $\omega_1$ and $\omega_2$, at their respective times $t_1$ and $t_2$, with detectors of widths $\Gamma_1$ and $\Gamma_2$. In the steady-state regime, the time-dependency reduces to the delay time, $\tau=t_1-t_2$, such that 
$g^{(2)}_{\Gamma_1,\Gamma_2}(\omega_1,t_1; \omega_2,t_2) \rightarrow g^{(2)}_{\Gamma_1,\Gamma_2}(\omega_1, \omega_2; \tau)$. In the case of zero delay, we will adopt the following notation
\begin{equation}
	g^{(2)}_{\Gamma_1,\Gamma_2}(\omega_1, \omega_2) \equiv g^{(2)}_{\Gamma_1,\Gamma_2}(\omega_1, \omega_2; \tau=0).
\end{equation}
Interestingly, from the definition of the normalized two-photon spectrum in \eqref{eq:FreqResolvedFun}, we obtain two limiting cases: 
\begin{subequations}
	\begin{align}
		&\lim_{|t_1-t_2|\rightarrow \infty} g^{(2)}_{\Gamma_1,\Gamma_2}(\omega_1,t_1; \omega_2,t_2)=1, \\
		&\lim_{\Gamma_1,\Gamma_2\rightarrow \infty} g^{(2)}_{\Gamma_1,\Gamma_2}(\omega_1,t_1; \omega_2,t_2)= g^{(2)}(t_1; t_2),
	\end{align}
\end{subequations}
corresponding to the case of uncorrelated photons at infinite delays, and color-blind detectors, respectively.  

Further generalization to the $N$-photon spectrum is theoretically straightforward by adding pairs of operators (with their corresponding integrals and detectors), as noted in \textcolor{Maroon}{Refs.}~\cite{KnollTheoryNfold1986,KnollSpectralProperties1990}. However, it becomes computationally intractable due to the all possible time orderings of the $2N$-time correlator which result in $(2N-1)!!2^{N-1}$ independent regions of integration~\cite{KavokinMicrocavities2017}. Hence, the complexity of the task makes a direct computation impractical, although it can be experimentally straightforward by detecting photon clicks as a function of time and energy, for instance, using of a streak camera photodetector~\cite{WiersigDirectObservation2009,SilvaColoredHanbury2016}.

\subsection{The sensor method}
\label{sec:sensor_method}

To overcome this computational problem, \textit{del Valle et al.}~\cite{DelValleTheoryFrequencyFiltered2012} developed an efficient method to compute the $N$-photon correlations for arbitrary time delays and frequencies, applicable to any quantum system.

Essentially, the method consists in the addition of $N$ two-level systems to the dynamics, acting as artificial sensors, each with annihilation operator $\hat \xi_k$, frequency $\omega_{\xi_k}$, and linewidth $\Gamma_{\xi_k}$. The coupling rate of each sensor to the system, $g_{\xi_k}$, is small enough such that we can avoid back-action interaction since the system dynamics must remain unaltered by their presence. As a consequence of this, the population of each sensor is vanishingly small. Hence, the general description of an arbitrary system with $M$ Lindblad terms along with $N$ detectors is given by
\begin{equation}
	\frac{\partial \hat \rho}{\partial t}=-i[\hat H+\hat H_\xi + \hat H_I , \hat \rho]+ \sum_{i=1}^M \frac{\gamma_i}{2} \mathcal{D}[\hat L_i]\hat \rho + \sum_{k=1}^N \frac{\Gamma_{\xi_k}}{2}\mathcal{D}[\hat \xi_k]\hat \rho,
\end{equation}
where $\hat H$ is the Hamiltonian of the system, and $\hat H_\xi$ and $ \hat H_I$ are the free energy and the interaction Hamiltonians of the detectors, respectively, which read:
\begin{subequations}
	\begin{align}
		\hat H_\xi&= \sum_k \omega_{\xi_k} \hat \xi_k^\dagger \hat \xi , \\
		\hat H_I&= \sum_k g_{\xi_k} (\hat \xi_k \hat c^\dagger +\hat \xi_k^\dagger \hat c ).
	\end{align}
\end{subequations}
Following the notation used throughout this chapter, $\hat c$ denotes the annihilation operator of the quantum system. Therefore, the $N$-photon frequency resolved correlation function simply reads:
\begin{equation}
	\colorboxed{Maroon}{
		\begin{aligned}[b]
			g^{(N)}_{\Gamma_1\ldots \Gamma_N}(\omega_1,&t_1; \ldots; \omega_N,t_n)\\
		&= \lim_{g_1,\ldots,g_N \rightarrow 0}\frac{\langle :  \hat \xi_1^\dagger (t_1)\hat \xi_1(t_1) \ldots \hat \xi_N^\dagger (t_N) \hat \xi_N (t_N) : \rangle}{\prod_{i=k}^N \langle \xi_k^\dagger (t_k) \hat \xi_k (t_k) \rangle}.
		\end{aligned}
	}
\end{equation}
As a consequence, the computation of $N$-photon correlations gets easily simplified since now it is just described by intensity correlations of the sensors. The proof of the formal equivalency in the computation of normalized correlators between the spectrally-filtered photo-detection theory and the sensors method can be found in the supplementary material of the original paper~\cite{DelValleTheoryFrequencyFiltered2012}.

Particularly, the zero-delay $N$-photon frequency-resolved correlation function reduces to
\begin{equation}
	g^{(N)}_{\Gamma_1\ldots \Gamma_N}(\omega_1, \ldots ,\omega_N)
	= \lim_{g_{\xi_1},\ldots,g_{\xi_N} \rightarrow 0}
	\frac{ \langle 
	: \prod_{i=1}^N  \hat \xi_i^\dagger \hat \xi_i:
	\rangle
	}{\prod_{i=1}^N  \langle \hat \xi_i^\dagger \hat \xi_i \rangle},
\end{equation}
and the $N$-photon frequency-resolved autocorrelation function:
\begin{equation}
	g^{(N)}_{\Gamma_\xi}(\omega_\xi)
	= \lim_{g_\xi \rightarrow 0}
	\frac{ \langle 
		   (\hat \xi^\dagger)^N \hat \xi^N
		\rangle
	}{  \langle \hat \xi^\dagger \hat \xi \rangle^N},
\end{equation}
for which we just use a single detector of frequency $\omega_\xi$ and linewidth $\Gamma_\xi$. We note that in this case, we must promote the TLS sensor to a bosonic sensor truncated up to $N$ excitations. For example, the spectrum introduced in \eqref{eq:Spectrum} can now be efficiently computed using the sensor method through the relation:
\begin{equation}
	S(\omega)=\frac{1}{2\pi} \frac{\Gamma_\xi}{g_\xi} \langle \hat \xi^\dagger \hat \xi \rangle.
	\label{eq:spectrum_sensor}
\end{equation}

\subsection{The cascaded-sensor method}
\label{sec:cascaded_sensor}
Recently, it was proven that \textit{del Valle's} sensors method for computing $N$-photon correlation functions is equivalent to use a cascaded system setup, and, consequently, by logical transitivity, the cascaded-sensor method is formally equivalent to the spectrally-filtered theory of photo-detection~\cite{CarrenoExcitationQuantum2016a,CarrenoExcitationQuantum2016,JuanCamiloLopezCarrenoExcitingQuantum2019}.
This alternative description based on the cascaded formalism [see \refsec{Section:Cascaded}] allows the extraction of photons from any frequency window of a quantum source using all the signal theoretically available, while physically accounting the effect of detection. Additionally, since, by definition, the coupling is cascaded, i.e., there is not back-action, the theory is not restricted to vanishing coupling and only restricted to physical cascaded couplings. 

The cascaded-sensor master equation for one sensor, enabling the extraction of information within a single frequency window, is easily adapted from \eqref{eq:CascadedMaster}:
\begin{equation}
	\colorboxed{Maroon}{
		\begin{aligned}[b]
	\frac{d \rho(t)}{d t}=-i [\hat H+&\omega_\xi \hat \xi^\dagger \hat \xi ,\hat \rho] +  \frac{\gamma}{2}\mathcal{D}[\hat c]\hat \rho \\
	&+ \frac{\Gamma_\xi}{2}\mathcal{D}[\hat \xi] \hat \rho- \sqrt{ \gamma \Gamma_\xi}\left( [\hat \xi^\dagger , \hat c \hat \rho]+ [\hat \rho \hat c^\dagger, \hat \xi] \right).
\end{aligned}}
\end{equation}
In \colorref{JuanCamiloLopezCarrenoExcitingQuantum2019}, it is formally proven how both methods, sensor method and cascaded-sensor method, provide exactly the same normalized correlators of the spectrally-filtered theory of photo-detection~\cite{ArnoldusPhotonCorrelations1984,KnollTheoryTimeresolved1984,KnollTheoryNfold1986,KnollQuantumNoise1986,CresserIntensityCorrelations1987} for vanishingly small couplings in the case of sensors and exact in all cases with the cascaded formalism.

\section{Quantum entanglement}
\label{Sec:QuantumEntanglement}
%
%
In this Thesis, we are also particularly interested on the generation of entanglement both between emitters or photonic modes, covering a considerable part of the present text.
Although quantum entanglement is often invoked in an operational sense---using an entanglement measure that conveniently suits a particular problem---, defining and quantifying entanglement is a subtle task, even in the simplest scenario of a bipartite system~\cite{PlenioIntroductionEntanglement2007,HorodeckiQuantumEntanglement2009}. Therefore, in this section, we delve deeper into the concept of entanglement and discuss some of its main challenges. We also introduce two entanglement measures that are widely use in this Thesis: the concurrence~\cite{WoottersEntanglementFormation1998,WoottersEntanglementFormation2001} and the (logarithmic) negativity~\cite{VidalComputableMeasure2002,PlenioLogarithmicNegativity2005}.

%
%
\subsection{Entanglement - Conceptual definition}

The concept of \textit{quantum entanglement} is one of the most non-classical manifestations of quantum mechanics. It reveals the non-local character of the quantum theory---without violating causality---by conceiving a holistic description of composite systems in which there exists global states that cannot be factorized into products of its components~\cite{PlenioIntroductionEntanglement2007,HorodeckiQuantumEntanglement2009}. As expressed by  Schrödinger in his seminal paper from 1935~\cite{SchroedingerGegenwaertigeSituation1935,WheelerQuantumTheory1983}:
\newline

\emph{``Thus one disposes provisionally (until the entanglement is resolved by actual observation) of only a common description of the two in that space of higher dimension. This is the reason that knowledge of the individual system can decline to the scantiest, even to zero, while that of the combined system remains continually maximal. The best possible knowledge of a whole does not include the best possible knowledge of its parts---and this is what keeps coming back to haunt us.''}
\newline

\noindent
This counter-intuitive idea is crucial to understand quantum phenomena such as quantum teleportation in non-relativistic quantum mechanics~\cite{BennettTeleportingUnknown1993,NielsenQuantumComputation2012},  quantum vacuum in quantum field theories~\cite{RyderQuantumField1996,WittenAPSMedal2018}, or even recent theories of quantum gravity where, e.g., quantum entanglement is \textit{dual} to Einstein-Rosen bridges---that is, wormholes---via the ER=EPR conjecture~\cite{MaldacenaCoolHorizons2013}. This way we note that quantum entanglement is an ubiquitous feature in the quantum theory.
Since this notion of quantum non-locality was originally reported by Schrödinger~\cite{SchroedingerGegenwaertigeSituation1935} and Einstein, Podolsky and Rosen (EPR)~\cite{EinsteinCanQuantumMechanical1935} in 1935, both physical and philosophical questions have arisen regarding the intrinsic understanding of nature. These questions include paradigmatic debates about the existence of free will, the (in-)completeness of the quantum theory, or the actual properties of a physical system prior measurements.
Providing definitive answers to such questions is no trivial task, given their deeply philosophical nature, so that they can be debated \textit{ad infinitum}. 

Nevertheless, from a scientific perspective, the pioneered experiments of J. Bell in 1964~\cite{BellEinsteinPodolsky1964}, followed by the groundbreaking experiments of A. Aspect \textit{et al.} in 1980-82~\cite{AspectTimeCorrelations1980,AspectExperimentalTests1981,AspectExperimentalTest1982}, Greenberger, Horne, and Zeilinger (GHZ) in 1989~\cite{GreenbergerGoingBells1989}, and subsequent experiments performed from late the 1980s to the present days ~\cite{OuViolationBells1988,KwiatNewHighIntensity1995,TittelViolationBell1998,TittelLongdistanceBelltype1999,WeihsViolationBells1998,RoweExperimentalViolation2001,HasegawaViolationBelllike2004,BovinoExperimentalNoiseresistant2006,SakaiSpinCorrelations2006,UrsinEntanglementbasedQuantum2007,HensenLoopholefreeBell2015,HandsteinerCosmicBell2017,LiuViolationBell2024}, have offered significant insights regarding the completeness of the quantum theory, the so-called 
EPR paradox$^{\textcolor{Maroon}{*}}$~\cite{EinsteinCanQuantumMechanical1935}. 
\graffito{
	$^*$The EPR paradox is a \textit{gedanken} experiment that states the quantum theory does not offer a complete description of reality, and that some hidden variables must exist to recover the deterministic description of nature.}
These experiments confirmed the predictions of the quantum theory by showing that the quantum statistical correlations exhibited by an EPR pair, i.e., a bipartite entangled state, violate the constraints imposed by a deterministic local hidden variables theory (LHVT), the so-called Bell inequalities~\cite{BellEinsteinPodolsky1964}, supporting the conclusion that nature appears to be intrinsically quantum.  

The question regarding the completeness of the quantum theory exceeds the topics of this Thesis, nevertheless, we note that it is still a matter of debate in both the scientific and philosophical communities~\cite{KarimiClassicalEntanglement2015,DelSantoIndeterminismCausality2021,AresSemiclassicalViolation2023}. In fact, given the metaphysical character of the question, only depending on our philosophical framework of beliefs~\cite{BungeScientificMaterialism1981,BuenoQueEs1995,MosterinConceptosTeorias2000,MadridCasadoFilosofiaCosmologia2018,DiezFundamentosFilosofia2018}, we can be satisfied about the possible answers. 

\paragraph{Entanglement as a resource.}
The experimental progress and the advent of quantum technologies has promoted the initial debates regarding the ``metaphysical'' question of \textit{what is entanglement}, focused on the qualitative features that distinguish it from the classical intuition, into a more ``material'', or practical, perspective. In this view,  entanglement is now conceived as a \textit{technological resource} for tasks that are either impossible or very inefficient to be performed by means of classical resources. Examples include, quantum computation~\cite{GalindoInformationComputation2002}, quantum communication~\cite{KimbleQuantumInternet2008,NarlaRobustConcurrent2016} or quantum metrology~\cite{PezzeQuantumMetrology2018,BraskImprovedQuantum2015}. 
This paradigmatic change of conception of quantum entanglement as a resource was the basis for the development of a novel theoretical framework, called entanglement theory~\cite{HorodeckiQuantumEntanglement2009,PlenioIntroductionEntanglement2007,StreltsovColloquiumQuantum2017}, that tries to address fundamental questions as:
\begin{enumerate}[label=\textcolor{Maroon}{(\roman*)}]
	\item How to define, control and quantify entanglement.
	\item How to detect entanglement.
	\item How to protect entanglement against decoherence.
\end{enumerate}

\subsection[Definition and quantification of entanglement:\protect \newline concurrence and negativity]{Definition and quantification of entanglement: \protect \newline concurrence and negativity}
\label{sec:Conc_and_Negat}

In  a composite Hilbert space, $\mathcal{H}=\otimes_{i=1}^N \mathcal{H}_i$, entangled states are typically identified by the following \textit{negative definitions}~\cite{HorodeckiQuantumEntanglement2009,PlenioIntroductionEntanglement2007}: a pure quantum state is considered entangled if it cannot be described as a tensor product of its components
\begin{equation}
	|\psi \rangle= \sum_{i_1,i_2,\ldots,i_N} c_{i_1,i_2,\ldots,i_N} |i_1 \rangle \otimes |i_2\rangle \otimes \ldots \otimes |i_N\rangle \neq |\psi_1 \rangle \otimes |\psi_2\rangle \otimes \ldots \otimes |\psi_N\rangle,
	\label{eq:def1}
\end{equation}
while a general mixed state is said to be entangled if it cannot be expressed as a convex combination of product states
\begin{equation}
	\hat \rho \neq \sum_i p_i \hat \rho_1^i \otimes \ldots \otimes \hat \rho_N^i.
	\label{eq:def2}
\end{equation}
From these negative definitions, as expressed in \cref{eq:def1,eq:def2}, states that are not entangled  are commonly referred as separable states. 

Besides the pedagogical examples for distinguishing entangled and separable states in bipartite systems $\mathcal{H}=\mathcal{H}_A\otimes \mathcal{H}_B$, such as
\begin{subequations}
	\begin{align}
		\label{eq:entangled}
		|\psi\rangle&=\frac{1}{\sqrt{2}}(|0\rangle_A |0\rangle_B +|1\rangle_A |1\rangle_B  )  \quad \text{(entangled)}, \\
		\label{eq:separable}
		|\phi\rangle&= |1\rangle_A |1\rangle_B \quad\quad\quad\quad\quad\quad\quad \quad \ \ \text{(separable)},
	\end{align}
\end{subequations}
in the simplest case of $\text{dim}\ \mathcal{H}_{A/B}=2$, a  formal quantifier of the degree of entanglement between arbitrary subsystems remains a fundamental challenge in the theory of entanglement~\cite{PlenioIntroductionEntanglement2007}, as there is no a unique method to tackle it beyond distinguishing the extreme cases of separable or maximally entangled states---if such a notion can even be defined---. 
A fully general solution to this problem is still a matter of active debate, specially in the context of multi-partite systems~\cite{PlenioIntroductionEntanglement2007,AmicoEntanglementManybody2008,HorodeckiQuantumEntanglement2009,EisertColloquiumArea2010}. 
For instance, to address this issue, in \colorref{PlenioIntroductionEntanglement2007} an axiomatic approach based on the LOCC paradigm---Local Operations and Classical Communications (LOCC)---is presented to define well-behaved entanglement monotones for bipartite systems, such as the entanglement of formation~\cite{WoottersEntanglementFormation1998,WoottersEntanglementFormation2001,PlenioIntroductionEntanglement2007,HorodeckiQuantumEntanglement2009}, or the (logarithmic) negativity~\cite{VidalComputableMeasure2002,PlenioIntroductionEntanglement2007,HorodeckiQuantumEntanglement2009}. 

Over the last decades, a zoo of entanglement measures has been proposed in the literature, either physically or axiomatically motivated; hence, the choice of which entanglement measure to use often depends on the physical situation or just up to our decision. In general, any well-defined measure is suitable for quantifying the degree of entanglement, from separable to maximally state (if possible).

\paragraph{Concurrence.} 
In this Thesis, we will extensively use the concurrence as the  measure for quantifying the degree of entanglement between two quantum emitters. Therefore, given its relevance, we will now provide a more in-depth understanding of this measure and offer additional physical insights into what it represents. This will help clarify its role in characterizing entanglement and its implications for the system under study.

The two-qubit concurrence, $\mathcal{C}$, is, in fact, a closed-form solution of a variational problem that defines an actual entanglement measure, the entanglement of formation~\cite{WoottersEntanglementFormation1998,WoottersEntanglementFormation2001,PlenioIntroductionEntanglement2007,HorodeckiQuantumEntanglement2009}, $E_F$, defined as
\begin{equation}
	E_F\equiv \text{inf}\{ \sum_i p_i E(|\psi_i\rangle\langle\psi_i|) \quad \text{with}\quad  \hat \rho=\sum_i p_i |\psi_i\rangle\langle\psi_i| \}.
	\label{eq:Entanglement_of_Form}
\end{equation}
It represents the minimal possible average entanglement over all pure state decomposition of a mixed state $\hat \rho$, where $E(\cdot)$ denotes the \textit{entropy of entanglement}~\cite{BennettConcentratingPartial1996}---a measure of entanglement for pure states---, defined as
\\[-1em]
\begin{equation}
	E(|\psi\rangle\langle\psi|)\equiv S(\text{Tr}_A |\psi\rangle\langle\psi|)=S(\text{Tr}_B |\psi\rangle\langle\psi|),
\end{equation}
\\[-1em]
with $S$ denoting the von-Neumann entropy $S(\hat \rho)=-\text{Tr}[\hat \rho \log_2\hat \rho]$, and $\text{Tr}_{A/B}$ denoting the partial trace over subsystem $A/B$, respectively. Generally, the variational problem in \eqref{eq:Entanglement_of_Form} is difficult to solve and numerical methods are needed~\cite{AudenaertVariationalCharacterizations2001}, nevertheless, for general bipartite two-level systems, this measure adopts a tractable analytical expression given by
\begin{equation}
	E_F\equiv s\left[(1+\sqrt{1-\mathcal{C}^2})/2\right],
\end{equation}
with $s(x)\equiv -x \log_2 x - (1-x)\log_2 (1-x)$, and $\mathcal{C}$ denoting the concurrence,
\begin{equation}
\colorboxed{Maroon}{
	\mathcal{C}=\text{Max}[0,\sqrt{\lambda_1}-\sqrt{\lambda_2}-\sqrt{\lambda_3}-\sqrt{\lambda_4}],
	}
	\label{eq:Concurrence}
\end{equation}
where  $\lambda$'s are the eigenvalues---in decreasing order---of the matrix 
\begin{equation}
	\lambda=\text{eigenvalues}[\hat \rho  (\hat \sigma_y \otimes \hat \sigma_y)\hat \rho^* (\hat \sigma_y\otimes \hat \sigma_y)],
	\label{eq:eigenvalues-concurrence}
\end{equation}
with $\hat \sigma_y$ the y-component of the Pauli matrices. 
Recalling that $\hat \sigma_y$ is associated to spin-flip transformation~\cite{NielsenQuantumComputation2012,SakuraiModernQuantum2017}, we can geometrically interpret the present entanglement measure as a way to quantify how invariant the quantum state is under spin-flipping both qubits in their respective Bloch sphere representations. Let us consider two extreme situations to illustrate this insight:
\begin{enumerate}[label=\textcolor{Maroon}{(\roman*)}]
	\item \textcolor{Maroon}{Maximally entangled states.} For maximally entangled states,  e.g., the state in \eqref{eq:entangled}, $|\psi\rangle=\frac{1}{\sqrt{2}}(|0\rangle_A |0\rangle_B +|1\rangle_A |1\rangle_B  )$ , the composite spin-flip transformation leaves the state unchanged up to a phase:
	\begin{equation}
		(\hat \sigma_{y,A}\otimes \hat \sigma_{y,B})|\psi\rangle=-\frac{1}{\sqrt{2}}(|1\rangle_A |1\rangle_B +|0\rangle_A |0\rangle_B  ).
	\end{equation}
	Then, the eigenvalues of the matrix $\hat \rho (\hat \sigma_y \otimes \hat \sigma_y)\hat \rho^* (\hat \sigma_y\otimes \hat \sigma_y)$, which for pure states is just an overlap between the original and the spin-flipped states, reduces to $\lambda_1=1$ and zero otherwise, $\lambda_{2,3,4}=0$, so that concurrence is maximum, $\mathcal{C}=1$.
	\item \textcolor{Maroon}{Separable states.} On the other hand, for separable states, e.g., the state in \eqref{eq:separable}, $|\phi \rangle=|1\rangle_A |1\rangle_B$, the spin-flip transformation generates a state that is orthogonal to the original one,
	\begin{equation}
		(\hat \sigma_{y,A}\otimes \hat \sigma_{y,B})|\phi\rangle=-|0\rangle_A |0\rangle_B.
	\end{equation}
	Consequently, the matrix $\hat \rho (\hat \sigma_y \otimes \hat \sigma_y)\hat \rho^* (\hat \sigma_y\otimes \hat \sigma_y)$ is filled with zeros since the original and the transformed states do not overlap, 	yielding $\lambda_{1,2,3,4}=0$, and, consequently, the expected result of zero concurrence, $\mathcal{C}=0$.
\end{enumerate}
Then, extending this discussion to general states, we may understand concurrence, and thus entanglement of formation, to the  geometrical comparison of a state and its spin-flip reflection in the state space, introducing a kind of geometrical distance from separable to maximally entangled states. 

We finally note that, since both the entanglement of formation, $E_F$, and the two-qubit concurrence, $\mathcal{C}$, are monotonically related, in practice both magnitudes can be used for quantifying the degree of entanglement present in bipartite two-level systems. Nevertheless, it is important to highlight that the significance of concurrence as an entanglement measure emerges from its connection to the entanglement of formation, and not \textit{vice versa}~\cite{PlenioIntroductionEntanglement2007}. 

\paragraph{Negativity and logarithmic negativity.}

While concurrence is a widely used entanglement measure for quantifying the degree of entanglement between two quantum emitters, it becomes less relevant when the Hilbert space of the subsystems exceeds dimension $2$. In such cases, other entanglement measures are more appropriate. One particularly useful measure is the (logarithmic) negativity~\cite{VidalComputableMeasure2002,PlenioLogarithmicNegativity2005}.

Consider a bipartite composite Hilbert space, $\mathcal{H}=\mathcal{H}_A\otimes \mathcal{H}_B$, where the dimension of the subsystems $\mathcal{H}_{A}$ and  $\mathcal{H}_B$  are arbitrary. A general quantum state in this space can be described by 
\begin{equation}
	\hat \rho=\sum_{ij,mn} \rho_{ij,mn} |i\rangle \langle j| \otimes |m\rangle \langle n|,
\end{equation}
where $\rho_{ij,mn}\equiv \langle i, m | \hat \rho | j, n \rangle$, having adopted the notation $|i,m \rangle\equiv |i\rangle \otimes |m\rangle$.
The partial transposition of $\hat \rho$ with respect one of the subsystems, e.g., $A$, is defined as
\begin{equation}
	\hat \rho^{\text{T}_A}=\sum_{ij,mn} \rho_{ij,nm} |i\rangle \langle j| \otimes |n\rangle \langle m|,
\end{equation}
where $\text{T}_A$ denotes the partial transposition operation over system $A$, leaving subsystem $B$ unaltered.  

The Peres-Horodecki criterion, also known as the positive partial transpose (PPT) criterion~\cite{PeresSeparabilityCriterion1996}, states that for a separable quantum state $\hat \rho=\hat \rho_A\otimes \hat \rho_B$, the partial transposed operator $\hat \rho^{\text{T}_A}$ must describe a valid quantum state. Specifically, this means that $\hat \rho^{\text{T}_A}$ has a non-negative spectrum, guaranteeing the semi-definite positivity of the state. Conversely, if $\hat \rho^{\text{T}_A}$ has at least one negative eigenvalue, the composite system is said to be entangled.
This provides a necessary and sufficient conditions for separability in low dimensional cases, $2\otimes 2$ and $3\otimes 3$, and sufficient condition in higher dimensions.

In order to quantify the degree of entanglement,  we introduce the negativity~\cite{VidalComputableMeasure2002,PlenioLogarithmicNegativity2005}, a measure that captures the extent to which the spectrum of $\hat \rho^{\text{T}_A}$ violates positivity. The negativity is defined as
\begin{equation}
\colorboxed{Maroon}{
	\mathcal{N}\equiv \frac{||\hat \rho^{\text{T}_A}||_1-1}{2}= \sum_{\mu \in \sigma^-} |\lambda_\mu| \geq 0,
}
	\label{eq:Negativity1}
\end{equation}
where $||\hat X||_1\equiv \text{Tr}\sqrt{\hat X^\dagger \hat X}$ is the trace norm operation. In the second equality we used the fact that for $\hat \rho^{\text{T}_A}$, the trace norm can be expressed as
\begin{equation}
||\hat \rho^{\text{T}_A}||_1= \text{Tr}\sqrt{(\hat \rho^{\text{T}_A})^\dagger \hat \rho^{\text{T}_A}}=\sum_\mu |\lambda_\mu| =  \sum_{\mu \in \sigma^+} \lambda_\mu + \sum_{\mu \in \sigma^-} |\lambda_\mu|,
\end{equation}
where $\lambda_\mu$ are the eigenvalues of $\hat \rho^{\text{T}_A}$  and $\sigma^\pm$ denotes the set of positive and negative eigenvalues, respectively. Then, by noting that the partially transposed density matrix satisfies the trace-preserving property $\text{Tr}[\hat \rho^{\text{T}_A}]=\sum_{\mu \in \sigma^+} \lambda_\mu+\sum_{\mu \in \sigma^-} \lambda_\mu=1 $, the negativity can be simplified to $	\mathcal{N}=  \sum_{\mu \in \sigma^-} |\lambda_\mu|$.

Another commonly used measured of entanglement is the logarithmic negativity, which is directly related to the trace norm of $\hat \rho^{\text{T}_A}$
\begin{equation}
\colorboxed{Maroon}{
	E_{\mathcal{N}}= \log_2 ||\hat \rho^{\text{T}_A}||_1= \log_2 (1+2   \sum_{\mu \in \sigma^-} |\lambda_\mu| )\geq 0.
}
	\label{eq:LogNegativity1}
\end{equation}
The equality in \cref{eq:Negativity1,eq:LogNegativity1} holds for separable states, $\hat \rho=\hat \rho_A\otimes \hat \rho_B$. In that case, $\hat \rho^{\text{T}_A}=\hat \rho$, the partial transpose operation leaves the quantum state unaltered, remaining positive semi-definite. Therefore, both the negativity and the logarithmic negativity trivially vanish, $\mathcal{N}=0$ and $E_\mathcal{N}=0$, since there are no negative eigenvalues.

Both the negativity, $\mathcal{N}$, and the logarithmic negativity, $E_\mathcal{N}$, quantify how much $\hat \rho^{\text{T}_A}$ fails to be positive semi-definite, serving as well-defined measures of entanglement~\cite{PlenioIntroductionEntanglement2007,HorodeckiQuantumEntanglement2009}. These measures are positive, bounded from below by zero, and provide an operationally meaningful way to characterize entanglement.
In this Thesis, we adopt the logarithmic negativity  $E_\mathcal{N}$ as the entanglement measure to quantify the degree of quantum correlations between photonic modes. Its computational simplicity and connection to the PPT criterion make it a practical and insightful choice for analyzing entanglement in quantum systems.

\subsection{Witnessing entanglement}
The quantification of entanglement via entanglement measures~\cite{PlenioIntroductionEntanglement2007,HorodeckiQuantumEntanglement2009} relies solely on the mathematical properties of the density matrix that describes the composite system.
For instance, one may use the eigenvalues of the matrix $\hat \rho (\hat \sigma_y \otimes \hat \sigma_y)\hat \rho^* (\hat \sigma_y\otimes \hat \sigma_y)$ in the concurrence~\cite{WoottersEntanglementFormation1998,WoottersEntanglementFormation2001,PlenioIntroductionEntanglement2007,HorodeckiQuantumEntanglement2009} in \eqref{eq:Concurrence}, or the eigenvalues of the partially transpose density matrix in the (logarithmic-) negativity~\cite{VidalComputableMeasure2002,PlenioIntroductionEntanglement2007,HorodeckiQuantumEntanglement2009} in \cref{eq:Negativity1,eq:LogNegativity1}. 
However, in practice, quantifying entanglement is challenging because one usually does not have access to the full quantum state. Instead, one may reconstruct the state from measurement data—when possible—via $n$th-order moments and quantum correlators, a process known as quantum state tomography~\cite{LvovskyContinuousvariableOptical2009,StrandbergSimpleReliable2022,AhmedClassificationReconstruction2021,YangEntanglementPhotonic2025}. 
Despite this technique provides an effective approach for obtaining the actual description of a quantum state in an experiment, reconstructing the state with perfect fidelity remains challenging. This is because many experimental measurements may be required, and some of them might not be accessible, e.g.,  due to experimental limitations.

To address this challenge in the context of quantum optics, \textit{entanglement witnesses} have been proposed as a practical method for detecting whether a composite state is entangled or separable by analysing the optical properties of the emitted light.
Note that, unlike entanglement measures, entanglement witnesses merely \textit{signal} the presence of entanglement without quantifying its magnitude.
For instance, based on the Peres-Horodecki criterion~\cite{PeresSeparabilityCriterion1996,HorodeckiSeparabilityMixed1996}, we can \textit{witness} entanglement between photonic modes by checking for  inequality violations in matrix moments~\cite{ShchukinInseparabilityCriteria2005, VogelNonclassicalityEntanglement2007,HilleryEntanglementConditions2006,MiranowiczInseparabilityCriteria2009,GrunwaldEntanglementAtomic2010}. Additionally, we can also detect entanglement by performing a Hanbury-Brown and Twiss (HBT) experiment~\cite{HanburyBrownCorrelationPhotons1956,HanburyBrownQuestionCorrelation1956}, where second- and fourth-order moments from the photon emission reveal a rich landscape of multiphoton processes in the frequency domain that evidence the generation of nonclassical correlations~\cite{UlhaqCascadedSinglephoton2012,SanchezMunozViolationClassical2014,PeirisTwocolorPhoton2015,PeirisFransonInterference2017,LopezCarrenoPhotonCorrelations2017,ZubizarretaCasalenguaConventionalUnconventional2020,LopezCarrenoEntanglementResonance2024,YangEntanglementPhotonic2025}, for example, via violation of Cauchy-Schwarz (CSI) or Bell (BI) inequalities~\cite{ClauserProposedExperiment1969,ReidViolationsClassical1986,SanchezMunozViolationClassical2014}, reformulated as
\begin{subequations}
\begin{align}
		&\text{ \fontsize{9.3}{10}\selectfont $R_{\Gamma} (\omega_1, \omega_2) = \frac{g_\Gamma^{(2)}(\omega_1,\omega_2)}{[g_\Gamma^{(2)}(\omega_1,\omega_1)  g_\Gamma^{(2)}(\omega_2,\omega_2) ]} \leq  1,$} \\
		&\text{ \fontsize{9.3}{10}\selectfont $B_\Gamma(\omega_1,\omega_2)= \sqrt{2} \left|
		\frac{\langle \hat a_1^{\dagger 2} a_1^2   \rangle + \langle \hat a_2^{\dagger ^2} a_2^2   \rangle -4\langle \hat a_1^\dagger \hat a_2^\dagger \hat a_2 \hat a_1  \rangle - \langle \hat a_1^{\dagger 2}a_2^2   \rangle- \langle \hat a_2^{\dagger 2} a_1^2   \rangle}{\langle \hat a_1^{\dagger 2} a_1^2+\langle \hat a_2^{\dagger 2} a_2^2 + 4\langle \hat a_1^{\dagger } \hat a_2^{\dagger } \hat a_2 \hat a_1\rangle} \right| \leq 2 $},
\end{align}
\end{subequations}
where $g_\Gamma^{(2)}(\omega_1,\omega_2)$ is the two-photon correlation function from \eqref{eq:FreqResolvedFun}. Note that the operators $\hat a_1$ and $\hat a_2$ can be replaced by the sensor operators $\hat \xi_1$ and $\hat \xi_2$ introduced in \refsec{sec:sensor_method}.

Another direct approach to detect entanglement involves examining the eigenstructure of strongly interacting quantum emitters and its influence on the optical properties of their emission, as we will do in \refch{chapter:Entanglement}. In this context, the observation of superradiance and subradiance emission is the most sought-for signature revealing the formation entanglement, as these phenomena are direct manifestations of the collective behaviour of the emitters forming maximally symmetric or antisymmetric entangled states.
Examples include systems of two quantum dots (QDs) embedded in a photonic waveguide~\cite{TiranovCollectiveSuper2023}, where time-resolved emission measurements reveal superradiant and subradiant effects, and DBATT molecular aggregates~\cite{TrebbiaTailoringSuperradiant2022}, where the lifetimes of these states are extracted from photon counting measurements, as shown in \reffig{fig:SuperradianceExperimental}~\textcolor{Maroon}{(b,c)}.

\section{Quantum Parameter Estimation}

In quantum mechanics, the process of extracting meaningful information about a physical system relies on the fundamental ability to measure and distinguish between \textit{causes} and \textit{effects}. This is what an experiment fundamentally does in a controlled way, where we define which is the entity under study---the system---and which are the entities responsible of the interaction---the environment or external fields---.
The measured quantities, such as the emitted photons or the quadratures of the electromagnetic field, carry information of the parameters governing the interaction, allowing us to infer key properties of the system or its environment.
%
%
%
The efficiency of these measurements are constrained by both theoretical and practical requirements. However, once all instrumental causes of error and disturbances have been removed, we are left with the unavoidable fluctuations inherent to quantum mechanics. The theory of \textit{quantum metrology} tackles this question by the understanding, manipulation, and control of these fundamental restrictions~\cite{ParisQUANTUMESTIMATION2009,WisemanQuantumMeasurement2014,Demkowicz-DobrzanskiQuantumLimits2015,PezzeQuantumMetrology2018,PolinoPhotonicQuantum2020,BarbieriOpticalQuantum2022}.
%
%

The goal of a measurement is to associate a value with a physical quantity, providing an estimate of it. However,  in most of the cases, many quantities of interest are not directly accessible, either because of fundamental or experimental difficulties. In such situations, we must infer the value of the parameter of interest by inspecting the probabilistic outcomes coming from indirect measurements. This leads to a statistical inference problem, which can be systematically addressed within the framework of \textit{quantum estimation theory}. In this Thesis we will focus on the application of the Fisher information to address the estimation problem.
%
%

The following derivations are adapted from \textcolor{Maroon}{Refs.}~\cite{ParisQUANTUMESTIMATION2009,Demkowicz-DobrzanskiQuantumLimits2015,PolinoPhotonicQuantum2020,BarbieriOpticalQuantum2022,JaynesProbabilityTheory2003}.

\vspace{-0.15cm}
\subsection{Classical Fisher information and the Cramér-Rao bound}
\vspace{-0.15cm}

The estimation process of inferring a parameter, $\theta$, that is encoded in a physical system, reduces to finding an estimator, i.e., a function mapping $\hat \theta=\hat \theta(\{ x\})$ from the set of possible measurements outcomes $\{ x\}=\{x_1,x_2, \ldots\}$ into the parameter space. The performance of an estimator from the actual value $\theta$ can be quantified by the \textit{Mean Square Error} (MSE) deviation, denoted by $\Delta^2 \theta$,
\vspace{-0.1cm}
\begin{equation}
	\Delta^2 \theta= \mathbb{E}[(\hat \theta(\{ x\})-\theta)^2]=\sum_{ x} [\hat \theta(x)-\theta]^2 p( x|\theta),
	\label{eq:MSEdeviation}
\end{equation}
\vspace{-0.1cm}
where $\mathbb{E}[(*)]$ denotes the expectation value, and  $p(x|\theta)$  is the conditional probability of observing the outcome $x$ given the parameter $\theta$. Here we have assumed $x$ to be a discrete variable, but the continuous case is straightforward by replacing $\sum_x \rightarrow \int dx$.
%

An often desired property for an estimator is that it is \textit{unbiased}$^{\textcolor{Maroon}{*}}$, such
\graffito{
	$^*$However, in some cases, a biased estimator may actually result in a lower error compared to an unbiased one, offering better overall performance~\cite{JaynesProbabilityTheory2003,RubioQuantumMetrology2019,RinaldiParameterEstimation2024,RubioFirstprinciplesConstruction2024}.
}
 that its average coincides with the unknown parameter, $\mathbb{E}[\hat \theta ]=\theta$, minimizing thus the MSE deviation.
In this context, optimal unbiased estimators are those that saturate the \textit{Cramér-Rao bound} (CRB) inequality
\begin{equation}
	\colorboxed{Maroon}{\Delta^2 \theta \geq \frac{1}{M F[p(x|\theta)]},}
	\label{eq:CRB}
\end{equation}
where $M$ is the number of independent repetitions measurements, and $F[p(x|\theta)]$ is the \textit{(classical) Fisher information} (FI) defined as
\begin{equation}
	\colorboxed{Maroon}{F[p(x|\theta)]\equiv \mathbb{E}\left[ \left(\frac{\partial \ln p(x|\theta)}{\partial \theta}\right)^2 \right]= \sum_{x} \frac{1}{p(x|\theta)} \left(\frac{\partial p(x|\theta)}{\partial \theta}\right)^2.}
	\label{eq:FisherInform}
\end{equation}
The Fisher information quantifies, on average, the sensitivity of the conditional probability distribution to small changes on the parameter to infer. In other words, the bigger the FI is the higher estimation precision may be expected. 
Note that there are several approaches to obtaining an unbiased estimator that can, in principle, saturate the Cramér-Rao bound.  
A typical example is the \textit{Maximum Likelihood Estimator} (MLE), which becomes asymptotically unbiased and efficient as the sample size increases.  Specifically, as the number of samples  $M$ tends to infinity, the likelihood function
\begin{equation}
	\mathcal{L}(\{x\}|\theta)\equiv  \prod_{x}^M p(x|\theta),
\end{equation}
leads to an estimator of $\theta$ given by 
\begin{equation}
	\hat \theta (\{x\})= \text{arg} \max_{\theta} 	\mathcal{L}(\{x\}|\theta),
\end{equation}
with the property that $\mathbb{E}[\hat \theta ]\rightarrow \theta$ as $M\rightarrow \infty$~\cite{JaynesProbabilityTheory2003}.

In quantum estimation scenarios, the parameter $\theta$ is encoded in the quantum state, $\hat \rho_\theta$, which is subject to positive operator-valued measurements (POVMs), $\hat \Lambda$, consisting of a set of operators $\{ \hat \Lambda_\mu\}$, where the index $\mu\in \{1,2,\ldots,M\}$ denotes all possible outcomes. The elements of the POVM are constrained by $\hat \Lambda_\mu\geq 0$ and $\sum_\mu \hat \Lambda_\mu^\dagger \hat \Lambda_\mu =\mathbb{I}$, such that the probability for each measurement outcome is given by $p(x|\theta)=\text{Tr}[\hat \Lambda_x \hat \rho_\theta]\geq 0$. 
Hence, the classical FI associated to the combination of the state $\hat \rho_\theta$ and the POVMs can be expressed as
\begin{equation}
F[\hat \rho_\theta]= \sum_{x} \frac{ \left(\text{Tr}[ \partial_\theta \hat \rho_\theta \hat \Lambda_x] \right)^2}{\text{Tr}[\hat \rho_\theta \hat \Lambda_x]}.
\end{equation}

\subsection[Quantum Fisher information and the quantum Crámer-Rao bound]{Quantum Fisher information and  the quantum Crámer-Rao bound}
The Fisher information provides a lower bound on the asymptotic precision subject to a specific measurement strategy fixed by the choice of POVMs.  Consequently, different strategies lead to different CRBs, each extracting different amount of information. This distinction highlights the inherent complexity of determining the optimal measurement scheme for a specific estimation scenario.
Therefore, to provide an ultimate theoretical bound that is independent of the measurement scheme,  we can maximize the FI over all possible POVMs, leading to the so-called \textit{Quantum Fisher information} (QFI):
\begin{equation}
\colorboxed{Maroon}{	F_Q[\hat \rho_\theta]=\text{max}_{\{\Lambda_x\}} F[\hat \rho_\theta]=\text{Tr}[\hat \rho_\theta \hat L_\theta^2],}
\end{equation}
where $\hat L_\theta$ is a Hermitian operator called \textit{symmetric logarithmic derivative} (SLD) which relates to the variation of the state $\hat \rho_\theta$ under infinitesimal changes in the value of the unknown parameter $\theta$, 
\begin{equation}
	\partial_\theta \hat \rho_\theta =\frac{1}{2}(\hat \rho_\theta \hat L_\theta +\hat L_\theta \hat \rho_\theta).
\end{equation}

The QFI is determined exclusively by the dependence of $\hat \rho_\theta$ on the estimated parameter, allowing us to analyse the parameter sensitivity without considering any particular measurement scheme. By definition, $	F_Q[\hat \rho_\theta]\geq 	F[\hat \rho_\theta]$, and thus the CRB can be extended to the \textit{Quantum Cramér-Rao bound} (QCRB):
\begin{equation}
	\colorboxed{Maroon}{\Delta^2\theta \geq \frac{1}{N 	F[\hat \rho_\theta]} \geq \frac{1}{N 	F_Q[\hat \rho_\theta]}.}
\end{equation}
We note that there always exist a measurement strategy, i.e., a POVM set in the eigenbasis of the SLD, for which the FI equals the QFI, and, consequently, the QCRB and the CRB coincide~\cite{BraunsteinStatisticalDistance1994}. Finally, both the classical and quantum Fisher information satisfy the following properties:
\begin{subequations}
\begin{align}
	& F[\hat \rho_\theta] \geq 0, \\
	&F[\sum_i c_i \hat \rho_{\theta}^i] \leq \sum c_i F[\hat \rho_{\theta}^i], \\
	& F[\otimes_i \hat \rho^i_{\theta}]=\sum_i  F[ \hat \rho^i_{\theta}] ,
\end{align}
\end{subequations}
that is, they are non-negative, convex, and additive for uncorrelated events.

\subsection{Photon-counting measurements }
\label{sec:photon_counting}
Although the QFI provides the ultimate theoretical bound of the MSE deviation, it does not explicitly clarify which measurement scheme is the optimal to saturate the QCRB. Hence, in order to have a more realistic connection to the underlying physics in typical quantum optics setups, in this Thesis we focus on the classical Fisher information associated with photon-counting measurements, where  the POVMs are defined by
\begin{equation}
\hat \Lambda_n =\{|n \rangle \langle n |\},
\end{equation}
with $|n\rangle$ denoting the photon number eigenstates, e.g., of a cavity field. In this situation, the conditional probability is simply the diagonal of the density matrix,
\begin{equation}
	p(n|\theta)=\text{diag}[\hat \rho_\theta].
\end{equation}
For example, $\theta$ could represent a Hamiltonian or a Lindblad parameter from the master equation, such as the driving strength $\Omega$ of a laser interacting with a quantum emitter, or the decay rate of the emitter $\Gamma$. The density matrix $\hat \rho_\theta$ may correspond to the steady-state of the system governed by that specific master equation.
 


\section{Theoretical and computational techniques}

\subsection{Liouvillian properties}
\label{sec:LiouvProp}

In this section, we introduce several useful theoretical and computational techniques for calculating the density matrix and its derivatives, as well as for performing the spectral decomposition of the Liouvillian.

\paragraph {Vectorization.}
Let us assume an arbitrary open quantum system whose dynamics is described by a Lindblad master equation:
\begin{equation}
	\frac{d \hat \rho}{dt }= -i[\hat H, \hat \rho]+ \sum_i \mathcal{D}[\hat L_i] \hat \rho \equiv \mathcal{\hat L} \hat \rho,
\end{equation}
whose solution, in the simplest case in which $\mathcal{\hat L}$ is time-independent, is easily given by
\begin{equation}
	\hat \rho(t)=e^{\mathcal{\hat L}t}[\hat \rho(0)].
\end{equation}

Given this structure, in which the Liouvillian superoperator 
\graffito{
	$^*$We can vectorize a matrix following either a row- or column order. We choose the first option, row-ordered vectorization. While both approaches are mathematically equivalent, it is essential to remain consistent in the chosen order throughout calculations. This consideration is particularly important when working with different computational libraries.
}
acts linearly on the density matrix, we can transform the notation to a vector-like structure in which we replace the density matrix by a vector$^{\textcolor{Maroon}{*}}$:
\begin{equation}
	\hat \rho =\begin{pmatrix}
		\rho_{11} & \rho_{12} & \ldots & \rho_{1M} \\
		\rho_{21} & \rho_{22} & \ldots & \rho_{2M} \\
		\vdots & \vdots & \ddots & \vdots \\
		 \rho_{M1} & \rho_{M2} & \ldots & \rho_{MM}
	\end{pmatrix} \rightarrow | \hat \rho \rangle = 
	(	\rho_{11}, \ldots ,\rho_{1M} ,  \rho_{21},  \ldots ,\rho_{MM} )^T,
\end{equation}
such that the inner product of two vectorized matrices is defined as $\langle \hat  \phi | \hat \rho\rangle \equiv \text{Tr}[\hat \phi^\dagger \hat \rho]$. Since the Liouvillian operator consists of left- and right- multiplication over the density matrix, we make use of the matrix-to-vector relation
\begin{equation}
	\hat A \hat B \hat C \rightarrow (\hat A\otimes  \mathbb{\hat I}_d)( \mathbb{\hat I}_d \otimes \hat C^T) |\hat B\rangle = (\hat A \otimes \hat C^T)|\hat B\rangle,
\end{equation}
such that the vectorized master equation reads
\begin{equation}
\colorboxed{Maroon}{
\begin{aligned}[b]
		\partial_t  &|\hat \rho\rangle= \left[ -i \left( \hat H \otimes \mathbb{\hat I}_d -\mathbb{\hat I}_d \otimes \hat H^T \right) \right. \\
	&\left. +
	\sum_i \frac{\gamma_i}{2}\left( 2 \hat L_i \otimes (\hat L^\dagger_i)^T  -\hat L^\dagger_i \hat L_i \otimes \mathbb{\hat I}_d - \mathbb{\hat I}_d \otimes (\hat L^\dagger_i \hat L_i)^T\right) \right]|\hat \rho \rangle,
\end{aligned}
}
\end{equation} 
where $T$ denotes the transpose operation and $d$ is the dimension of the Hilbert space. This alternative definition of the master equation will be beneficial both from a conceptual and computational perspectives.

\paragraph{Spectral decomposition.}
The spectrum of the Liouvillian superoperator, analogous to the Hamiltonian spectrum in the context of closed quantum systems, encodes essential information about the system dynamics it describes. However, since the Liouvillian is a non-Hermitian matrix, the spectral theorem cannot be directly applied. Hence, we must consider both the left- and right-eigenvectors$^{\textcolor{Maroon}{*}}$, 
\graffitoshifted[-2.4cm]{
	$^*$In practice, we just need to compute the right-eigenvectors of the Liouvillian since the left-eigenvectors are easily obtained by the orthogonality relation $\langle \hat  \Lambda_\mu^L|\hat  \Lambda_\nu^R\rangle = \delta_{\mu \nu}$. In matrix language, this relation translates to
	$$
	(\hat  \Lambda^L)^T \hat  \Lambda^R= \mathbb{\hat I},
	$$
	that is, computing the inverse of the right-eigenmatrix:
	$$
	(\hat  \Lambda^L)^T=(\hat  \Lambda^R)^{-1}.
	$$
}
 $|\Lambda_\mu^L\rangle$ and $|\Lambda_\mu^R\rangle$, along with their corresponding complex eigenvalues  $\{\Lambda_\mu  \in \mathbb{C},\ \mu=1,2,\ldots, d^2\}$, such that, for a given eigenvalue $\Lambda_\mu$, the eigenvector relations read
\begin{align}
	\mathcal{\hat  L}|\hat  \Lambda_\mu^R\rangle &=\Lambda_\mu |\hat  \Lambda_\mu^R\rangle, \\
\langle \hat  \Lambda_\mu^L|	\mathcal{\hat L} &=\Lambda_\mu \langle \hat  \Lambda_\mu^L|,
\end{align}
satisfying the orthogonality relation $\langle \hat  \Lambda_\mu^L|\hat  \Lambda_\nu^R\rangle = \delta_{\mu \nu}$. Without losing any generality, the complex eigenvalue $\Lambda_\mu$ accepts a decomposition into its real and imaginary parts such that
\begin{subequations}
		\label{eq:EigenDecomposition}
\begin{align}
	&\Lambda_\mu= \lambda_\mu + i \omega_\mu, \\
	&\lambda_\mu=\text{Re}[\Lambda_\mu], \\
	&\omega_\mu= \text{Im}[\Lambda_\mu],
\end{align}
\end{subequations} 
where the real part $\lambda_\mu\leq 0$ as a consequence of the Kossakowski conditions~\cite{BreuerTheoryOpen2007,RivasOpenQuantum2012}: the dynamics produced by the Liouvillian must be completely positive and trace preserving. 
Additionally, by means of the Evan's theorem~\cite{EvansIrreducibleQuantum1977,EvansGeneratorsPositive1979}, any finite system presents at least one eigenvalue that is zero, $\Lambda_1=0$, corresponding to the steady-state of the system, 
\begin{equation}
	\mathcal{\hat L}	|\hat \rho_{ss}\rangle=0, \quad  \text{with}\quad |\hat \rho_{ss}\rangle \equiv	|\hat \Lambda_1^R\rangle.
\end{equation}
For convenience, we assume the existence 
\graffito{
	$^{**}$In exceptional cases, the steady-state may be degenerated due to the presence of symmetries in the system~\cite{BucaNoteSymmetry2012,AlbertSymmetriesConserved2014,ThingnaDegeneratedLiouvillians2021}, such that there exist multiple steady-states towards which the system can evolve depending on the initial conditions. 
}
of a unique stationary state$^{\textcolor{Maroon}{**}}$ and sort the eingenvalues in decreasing order of the real part, such that $\lambda_\mu\geq \lambda_{\mu+1}$.
From this definition, we identify another relevant magnitude in open quantum systems, the Liouvillian gap, $|\lambda_2|$, giving the relaxation time necessary to reach the steady-state, $\tau_{ss}=1/|\lambda_2|$.

Hence, we can generally express the solution of the master equation at a time $t$ as
\begin{equation}
	|\hat \rho(t)\rangle=	|\hat \rho_{ss}\rangle+ \sum_{\mu=2}^{d^2} e^{\Lambda_\mu  t} \langle \hat \Lambda_\mu^L|\hat \rho(0)\rangle | \hat  \Lambda_\mu^R\rangle .
\end{equation}
Returning to the usual matrix notation,  $|\hat{\rho}(t)\rangle \rightarrow \hat{\rho}(t)$, this expression becomes:
\begin{equation}
	\colorboxed{Maroon}{\hat \rho(t)= \hat \rho_{ss}+\sum_{i=2}^{d^2} e^{\Lambda_\mu t} \text{Tr}[\hat \Lambda_\mu^L \hat \rho(0)] \hat \Lambda_\mu^R .}
	\label{eq:MasterEqSpectralDecomposition}
\end{equation}
Finally, we note that from this spectral decomposition, and recalling that $\text{Re}[\Lambda_\mu]<0$, in the long-time limit the system reaches the steady-state
\begin{equation}
	\lim_{t\rightarrow \infty} \hat \rho(t)=\hat \rho_{ss}.
\end{equation}

\paragraph{Computation of the steady-state.}
As we have just seen, the steady-state of the system is determined by the nullspace of the Liouvillian superoperator, satisfying
\begin{equation}
	\mathcal{\hat L} \hat \rho_{ss}=0,
	\label{eq:SteadyStateRelation}
\end{equation}
where the steady-state matrix is defined as $\hat \rho_{ss}\equiv\hat \Lambda_1^R$. However, the diagonalization of the Liouvillian may be computationally demanding, especially for large Hilbert spaces. 

We can implement a more efficient method to compute steady-states by considering the physical constraint of trace preservation: $\text{Tr}[\hat \rho_{ss}]=1$.  This constraint allows us to reduce the computational complexity by replacing, for instance, the first row of the Liouvillian with another with $1$s in the elements 
\begin{equation}
	\mathcal{\hat L}_{1,i}= \delta_{1,i} \quad \text{when} \quad i= d (\alpha-1)+\alpha \quad \text{for} \quad \alpha=1,2, \ldots d,
\end{equation}
where $d$ is the dimension of the Hilbert space. That is, there will be a $1$ in those entries that multiply an element of the diagonal of the density matrix, and $0$ otherwise. This yields an effective Liouvillian matrix, $\mathcal{\hat L}_{\text{eff}}$, satisfying the following simple linear equation,
\begin{equation}
\mathcal{\hat L}_{\text{eff}} \hat \rho_{ss}=\mathbf{c} \quad \text{with}\quad \mathbf{c}=(1,0,\ldots,0)^T,
\end{equation}
 such that the steady-state is simply given by
\begin{equation}
 \colorboxed{Maroon}{\hat \rho_{ss}=	\mathcal{\hat L}_{\text{eff}}^{-1} \mathbf{c}.}
 \label{eq:FormalSteadyState}
\end{equation}

\paragraph{Computation of the differentiation of the steady-state.} The formal relation from \eqref{eq:FormalSteadyState} can be extended to compute the differentation of the steady-state density matrix with respect to a system parameter $\theta$, denoted as $\partial_\theta \hat \rho_{\text{ss}}$. 
This is achieved by taking the derivative with respect $\theta$ in \eqref{eq:SteadyStateRelation}, such that
\begin{equation}
	\partial_\theta [\mathcal{\hat L} \hat \rho_{ss}]= (\partial_\theta  \mathcal{\hat L} ) \hat \rho_{\text{ss}}+ \mathcal{\hat L} \partial_\theta \hat \rho_{\text{ss}}=0, 
\end{equation}
and get the relation
\begin{equation}
	\partial_\theta \hat \rho_{\text{ss}}=-  \mathcal{\hat L}^{-1}(\partial_\theta  \mathcal{\hat L} ) \hat \rho_{\text{ss}}.
\end{equation}
Analogously to the steady-state computation, we can reduce the computational complexity by applying the preservation of the trace, and obtain a similar relation but in terms of the effective Liouvillian superoperator:
\begin{equation}
\colorboxed{Maroon}{	\partial_\theta \hat \rho_{\text{ss}}=-  \mathcal{\hat L}_{\text{eff}}^{-1}(\partial_\theta  \mathcal{\hat L}_{\text{eff}} ) \hat \rho_{\text{ss}}.}
\label{eq:DiffSteadyStateRelation}
\end{equation}
Note that before computing the differentiation of the steady state with respect to a parameter $\theta$, the steady state itself must first be determined. Using the formal relation for computing the steady-state in \eqref{eq:SteadyStateRelation}, we can rewrite \eqref{eq:DiffSteadyStateRelation} solely in terms of the effective Liouvillian,
\begin{equation}
	\partial_\theta \hat \rho_{\text{ss}}=-  \mathcal{\hat L}_{\text{eff}}^{-1}(\partial_\theta  \mathcal{\hat L}_{\text{eff}} )	\mathcal{\hat L}_{\text{eff}}^{-1} \mathbf{c}\quad \text{with}\quad \mathbf{c}=(1,0,\ldots,0)^T.
	\label{eq:DiffSteadyStateRelation1}
\end{equation}

\subsection{Change of picture: the rotating frame}
\label{sec:rotating_frame}

In previous sections, we have we used the Schrödinger and Heisenberg pictures, which are specific cases of a broader set of \textit{intermediate pictures} where both states and operators evolve.   Intermediate pictures are particularly useful for removing time-dependent contributions that induce trivial fast dynamics, overshadowing more subtler slowly-varying phenomena. For instance, the case of a coherently driven system:
\begin{equation}
	\hat H_d= \omega \hat c^\dagger \hat c + \Omega (\hat c e^{i\omega_L t} + \hat c^\dagger e^{-i\omega_L t}),
		\label{eq:CoherentlyDrivenSystem}
\end{equation}
where the laser introduces a time-dependency through oscillations with the frequency of the driving field. In such cases, it is convenient to change to an intermediate picture, a rotating frame, in which these time-dependent terms disappear from the Hamiltonian.

\paragraph{The intermediate picture.}
Let us consider a completely general unitary operator $\hat U_c(t)$ that encodes the change to an intermediate picture. Starting from the Schrödinger picture, the expectation value of a generic operator $\hat A$ at any time is given by 
\begin{equation}
	\langle \hat A (t)\rangle = \text{Tr}[ \hat A \hat \rho(t) ]=\text{Tr}[\hat U^\dagger_c(t) \hat A  \hat U_c(t) \hat U^\dagger_c(t)\hat \rho(t) \hat U_c(t)]
\end{equation}
where we have used $\hat U_c(t) \hat U^\dagger_c(t)= \mathbb{\hat I}$ and the cyclic property of the trace. Therefore, we can define the state and the operator $\hat A$ in the new picture as:
\begin{subequations}
	\begin{align}
		\hat \rho_I(t)&=\hat U_c^\dagger \hat \rho(t)\hat U_c(t), 
		\label{eq:StateIntermediate}\\
		\hat A_I(t)&=\hat U_c^\dagger \hat A \hat U_c(t),
		\label{eq:OperatorIntermediate}
	\end{align}
\end{subequations}
where $I$ stands for intermediate (picture) 
\graffito{
	$^*$It is trivial to prove that by taking $\hat U_c(t)=\mathbb{\hat I}$ we recover the Schrödinger picture, while taking $\hat U_c(t)=\hat U(t)$ we move to the Heisenberg picture. 
}
and $\hat \rho(t)=\hat U(t)\rho(0)\hat U^\dagger(t)$, with $\hat U(t)$ the total evolution operator from a Hamiltonian $\hat H(t)$. Then, the unitary operator that transform the state into the intermediate picture is defined as$^{\textcolor{Maroon}{*}}$
\begin{equation}
	\hat U_I(t)= \hat U_c^\dagger(t)  \hat U(t),
\end{equation} 
where we are removing $\hat U_c(t)$ from the total evolution $\hat U (t)$, which is applied to the operator, according to~\cref{eq:StateIntermediate,eq:OperatorIntermediate}.
Now, in this new picture, the state $\hat \rho_I(t)$ satisfies the following Liouville-von Neumann equation:
\begin{equation}
	\frac{\partial \hat \rho_I }{\partial t}=-i[\hat H_I(t),\hat \rho_I], 
\end{equation}
with
\begin{equation}
	H_I(t)=\hat U_c^\dagger \hat H(t) \hat U_c(t)-i \hat U_c^\dagger  \frac{\hat U_c(t)}{\partial t}.
\end{equation}
Note that the states still evolve with a modified Hamiltonian $\hat H_I(t)$ that, in general, is time dependent. The particular case in which we choose $\hat U_c(t)=\text{exp}(-i \hat H_c t)$, with $\hat H_c$ an arbitrary Hermitian time-independent operator with energy units, the Hamiltonian in the intermediate picture reads:
\begin{equation}
\colorboxed{Maroon}{	\hat H_I(t)= \hat U_c^\dagger \hat H(t) \hat U_c(t)-\hat H_c.}
	\label{eq:RotatingFrameTransf}
\end{equation}

\paragraph{Rotating frame of a coherent drive.}
Let us consider the Hamiltonian presented in \eqref{eq:CoherentlyDrivenSystem},
\begin{equation}
	\hat H= \omega \hat c^\dagger \hat c + \Omega (\hat c e^{i\omega_L t} + \hat c^\dagger e^{-i\omega_L t}),
\end{equation}
In order to remove the time-dependency, we consider the unitary operator
\begin{equation}
	\hat U_c(t)=e^{-i \omega_L t \hat c^\dagger \hat c},
\end{equation}
such that the annihilation and creation operators transform as
\begin{equation}
		\hat U^\dagger_c(t) \hat c 	\hat U_c(t)= \hat c e^{-i \omega_L t} \quad \text{and} \quad   	\hat U^\dagger_c(t) \hat c^\dagger 	\hat U_c(t)= \hat c^\dagger e^{i \omega_L t},
		\label{eq:TransformationRules}
\end{equation}
where we have used the commutation relation $[\hat c^\dagger \hat c, \hat c]=-\hat c$ (valid for either bosonic or fermionic operators), along with the Baker-Haussdorff lemma~\cite{SakuraiModernQuantum2017}, that for two general operators $\hat A$ and $\hat B$ reads:
\begin{multline}
	e^{\hat B}  \hat A e^{- \hat B} = \hat A + [\hat B, \hat A]+ \frac{1}{2!} [\hat B,[\hat B,\hat A]]+ \ldots \\
	=  \sum_{n=0}^{\infty} \frac{1}{n!} \underbrace{[\hat B, [\hat B, \ldots[\hat B,}_{n}\hat A\underbrace{]\ldots]]}_{n}.
\end{multline}
Finally, using \eqref{eq:RotatingFrameTransf}, the Hamiltonian $\hat H$ in the intermediate picture reads
\begin{equation}
	\hat H_{I}=(\omega-\omega_L)\hat c^\dagger \hat c + \Omega(\hat c^\dagger + \hat c),
\end{equation}
where the time-dependency has been removed and the effective energy of the mode $\hat c$ is now given by the detuning $\Delta\equiv (\omega-\omega_L)$ since we are in the rotating frame of the laser.  We note that Lindblad terms [see \eqref{eq:LindbladTerms}] are unaffected by this transformation because the jump operators transform as \eqref{eq:TransformationRules}, introducing conjugate phases $e^{\pm i \omega_L t}$ that cancel out. 

We can generalize this result for the case of two interacting modes $\hat b$ and $\hat c$, in which only one of them is being coherently excited,
\begin{equation}
	\hat H=\omega_b\hat b^\dagger \hat b+\omega_c\hat c^\dagger \hat c +g(\hat b \hat c^\dagger + \hat b^\dagger \hat c)+ \Omega(  \hat c e^{i \omega_L t} + \hat c^\dagger  e^{-i \omega_L t}).
	\label{eq:TwoModesDriven}
\end{equation}
In this case, the unitary transformation that move the system to the rotating frame of the laser reads
\begin{equation}
	\hat U_c(t)=e^{-i \omega_L t (\hat b^\dagger \hat b+\hat c^\dagger \hat c)},
\end{equation}
producing an energy shift of $\omega_L$ in both modes,
\begin{multline}
	\hat H_{I}=(\omega_b-\omega_L)\hat b^\dagger \hat b+(\omega_c-\omega_L)\hat c^\dagger \hat c \\
	+g(\hat b \hat c^\dagger + \hat b^\dagger \hat c)+ \Omega(\hat c^\dagger + \hat c).
	\label{eq:TwoModesDriven_Rotating}
\end{multline}
 We must note the time-dependency in \eqref{eq:TwoModesDriven_Rotating} has been removed since the interaction between modes do not do not present counter-rotating terms. In general situations,  we need to use more sophisticated approaches to deal with time-dependent terms, e.g., by means of the method of continued fractions using Floquet theory~\cite{MajumdarProbingSingle2011, PapageorgeBichromaticDriving2012,MaragkouBichromaticDressing2013}.

We will assume that the rotating frame is the starting point in this Thesis whenever a coherent drive appears in the dynamics. Therefore, to ease the notation, we remove the subindex $I$ from the Hamiltonians and states.

\subsection{Multi-time correlators}

The master equation formalism provides a useful tool to obtain the description of the density matrix either in time or in the steady-state, allowing us to compute time-dependent expectation values for any system operator. 
However, the computation of multi-time correlators, e.g., $\langle \hat A(t)\hat B(t')\hat C(t)\rangle$, requires a more subtle description since averages involving two times (or more) cannot be calculated directly from the master equation. 
In order to compute this kind of correlators in the framework of the master equation, we invoke the so-called \textit{Quantum Regression Theorem} (QRT)~\cite{LaxFormalTheory1963,LaxQuantumNoise1967}, a theoretical link that relates the computation of time-dependent correlation function in the Heisenberg picture to its counterpart in the Schrödinger picture for systems interacting with Markovian environments.  
  
The following derivations are adapted from \textcolor{Maroon}{Refs.}~\cite{CarmichaelOpenSystems1993,Navarrete-BenllochIntroductionQuantum2022}.

\paragraph{The Quantum Regression Theorem. } Let us consider the general case of the multi-time correlation function given by
\begin{equation}
	\langle \hat A(t)\hat B(t')\hat C(t)\rangle, \quad \text{with}\quad t'>t,
	\label{eq:MultiTimeCorrelator}
\end{equation}
where $\{\hat A ,\hat B, \hat C \}$ are three different system operators. These operator evolve according to their Heisenberg equations of motions:
\begin{equation}
	\partial_t \hat X(t)=-i[\hat X(t),\hat H] \quad \rightarrow \quad \hat X(t)=e^{i\hat H t} \hat X(0) e^{-i \hat H t},
\end{equation}
where $\hat X \in \{\hat A ,\hat B, \hat C \}$, $\hat H$ is the total Hamiltonian of the system and the environment, such that the evolution operator is defined by $\hat U(t) = e^{-i \hat H t}$.  Introducing this formal solution into \eqref{eq:MultiTimeCorrelator}, denoting $ \hat \rho_{SE}$ as the density matrix of the total system-environment, and using the cyclic property of the trace, we get:
\begin{multline}
	\langle \hat A(t)\hat B(t')\hat C(t)\rangle\\
	= \text{Tr}_{S\otimes E}\left[
\hat U^\dagger(t) \hat A \hat U(t) \hat U^\dagger(t') \hat B\hat U(t') \hat U^\dagger(t)  \hat C\hat U(t) \hat \rho_{SE}
	\right] \\
	=  \text{Tr}_{S\otimes E}\left[
  \hat B\hat U(t'-t)  \hat C\hat U(t) \hat \rho_{SE} 	\hat U^\dagger(t)\hat A  \hat U^\dagger(t'-t)
	\right] \\
	= \text{Tr}_S\left\{
	\hat B \text{Tr}_E\left[\hat U(t'-t) \left(\hat C \hat \rho_{SE}(t) \hat A  \right) \hat U^\dagger(t'-t) \right]
	\right\}
	\\
	\approx \text{Tr}_S \left\{  
	\hat B e^{\mathcal{\hat L}(t'-t)} \text{Tr}_E \left[\hat C \hat \rho_{SE}(t) \hat A  \right]
	\right\} = \text{Tr}_S \left[\hat B e^{\mathcal{\hat L}(t'-t)} \left(\hat C \hat \rho_{S}(t) \hat A \right)   \right],
\end{multline} 
where we used the composition property of the time-evolution operator, $\hat U(t')\hat U^\dagger(t)=\hat U(t'-t)$, and in the last line we considered the Born-Markov approximations. Hence, by defining $\tau\equiv t'-t$, we arrive to the well-known expression of the quantum regression theorem:
\begin{equation}
	\colorboxed{Maroon}{	\langle \hat A(t)\hat B(t+\tau)\hat C(t)\rangle= \text{Tr}_S \left[\hat B e^{\mathcal{\hat L}\tau} \left(\hat C \hat \rho_{S}(t) \hat A \right)   \right].}
	\label{eq:QRT}
\end{equation}
Additionally, we can make use of the spectral decomposition in \eqref{eq:MasterEqSpectralDecomposition}, and express the quantum regression formula as a sum of different traces of operators:
\begin{equation}
	\colorboxed{Maroon}{
		\begin{aligned}[b]
			\langle \hat A(t)\hat B(t & +\tau)\hat C(t)\rangle = 
			\\
			&\sum_{\mu,\nu=1}^{d^2} e^{ \Lambda_\mu t+\Lambda_\nu \tau} \text{Tr}[\hat \Lambda_\mu^L \hat \rho(0)] \text{Tr}[\hat \Lambda^L_\nu (\hat C \hat \Lambda^R_\mu \hat A)]\text{Tr}[\hat B \hat \Lambda_\nu^R].
		\end{aligned}
	}
	\label{eq:QRTspectral}
\end{equation}
This result highlights the power of the spectral decomposition of the Liouvillian superoperator.  With the left- and right-eigenvectors, and their corresponding eigenvalues, for a given parameter set, any quantity can be computed as a sum of matrix multiplications involving the initial condition and system operators. Furthermore, we note that time dependency is totally encoded in the complex exponential, which simplifies further manipulations.

\paragraph{Calculation of the spectrum.}
Given the spectrum of the light emitted by a mode $\hat c (t)$,  \eqref{eq:Spectrum}, we can recast it using the quantum regression theorem (QRT) from \eqref{eq:QRT} as:
\begin{multline}
	S(\omega)=\frac{1}{\pi }\frac{1}{n_c} \text{Re}\int_0^\infty d\tau e^{i\omega \tau} \langle \hat c^\dagger(0) \hat c(\tau) \rangle\\
	=\frac{1}{\pi }\frac{1}{n_c} \text{Re} \left\{\text{Tr}\left[
	\int_0^\infty d\tau  \hat c e^{(\mathcal{\hat L}+i\omega )\tau}[\hat \rho_{\text{ss}} \hat c^\dagger]
	\right] \right\}.
	\label{eq:SpectrumQRT}
\end{multline}

The fluorescence spectrum of a given system is typically decomposed into its elastic and  inelastic components~\cite{ScullyQuantumOptics1997,WallsQuantumOptics2008}. 
The elastic component does not carry information about the system since it just emits light at the frequency of the input field--- e.g., the frequency of the laser in case we drive the system with a coherent field---, while all the non-trivial information is contained in the inelastic component.
Therefore, it is convenient to formally separate these two contributions by replacing $\hat \rho_{\text{ss}} \hat c^\dagger \rightarrow \hat \rho_{\text{ss}}  (\hat c^\dagger+ \langle c^\dagger\rangle-\langle c^\dagger\rangle)$ in \eqref{eq:SpectrumQRT}, such that
\begin{multline}
	S(\omega)=\frac{1}{\pi }\frac{1}{n_c} \text{Re} \left\{ \text{Tr}\left[
	\int_0^\infty d\tau  \hat c e^{(\mathcal{\hat L}+i\omega )\tau}[\hat \rho_{\text{ss}} (\hat c^\dagger-\langle c^\dagger\rangle) ]
	 \right.  \right.\\
	\left. \left. +
	\int_0^\infty d\tau  \hat c e^{(\mathcal{\hat L}+i\omega )\tau}[\hat \rho_{\text{ss}}\langle c^\dagger\rangle ]
	\right] \right\}.
	\label{eq:Spectrum0}
\end{multline}
We will now show that the first line of this equation describes 
\graffito{
	$^*$The Liouvillian superoperator is trace-preserving, implying that
	$$
	\text{Tr}[e^{\mathcal{\hat L}\tau}[\hat \rho]]=\text{Tr}[\hat \rho]=1,
	$$
	for any $\tau$ and any density matrix $\hat \rho(\tau)$. Then, considering an arbitrary operator $\hat A$, we would get
	$$
	\text{Tr}[e^{\mathcal{\hat L}\tau}[\hat \rho \hat A]]=\text{Tr}[\hat \rho \hat A]=\langle \hat A \rangle.
	$$
	As a consequence, 
	$$
	\lim_{\tau \rightarrow \infty} e^{\mathcal{\hat L}\tau}[\hat \rho \hat A]=\hat \rho \langle \hat A \rangle
	$$
}
the inelastic contribution of the spectrum, while the second is the elastic. 
We can formally integrate \eqref{eq:Spectrum0} by using the fact that the Liouvillian is trace-preserving$^{\textcolor{Maroon}{*}}$~\cite{BreuerTheoryOpen2007,RivasOpenQuantum2012}, so that we can prove that
\begin{equation}
	\lim_{\tau \rightarrow \infty} e^{\mathcal{\hat L}\tau}[\hat \rho_{\text{ss}}(\hat c^\dagger-\langle \hat c \rangle]=0,
\end{equation}
and using again the Sokhotski-Plemelj formula
\begin{multline}
	\int_0^\infty d\tau  \hat c e^{(\mathcal{\hat L}+i\omega )\tau}[\hat \rho_{\text{ss}}\langle c^\dagger\rangle ] \\
	=\hat c \hat \rho_{\text{ss}}\langle c^\dagger\rangle  \int_0^\infty d\tau   e^{+i\omega \tau}	=\hat c \hat \rho_{\text{ss}}\langle c^\dagger\rangle  \left(\pi \delta(\omega)-i \text{P.V.}\frac{1}{\omega}\right),
	\label{eq:secondline}
\end{multline}
where the Cauchy principal value is ignored since only the real part contributes. 
Then, considering that the second line in \eqref{eq:Spectrum0} is straightforwardly integrated yielding a Dirac delta as a consequence of  \eqref{eq:secondline}, the fluorescence spectrum is given by
\begin{equation}
\colorboxed{Maroon}{	S(\omega)=\frac{1}{\pi }\frac{1}{n_c} \text{Re}\text{Tr}\left[
	-  \hat c \frac{1}{\mathcal{\hat L}+i\omega} [\hat \rho_{\text{ss}} (\hat c^\dagger-\langle c^\dagger\rangle) ] \right]
	+ \frac{|\langle c \rangle|^2}{n_c} \delta(\omega).}
	\label{eq:SpectrumResult}
\end{equation}
In this form$^{\textcolor{Maroon}{**}}$, 
\graffitoshifted[-1cm]{
	$^{**}$When calculating the spectrum for multiple values of $\omega$ and the Hilbert space is not too large, it might be computationally advantageous to express the spectrum in terms of the diagonalized Liouvillian, 
	$$
	\mathcal{\hat D}=( \hat  \Lambda^L)^T \mathcal{\hat L} \hat  \Lambda^R,
	$$
	such that
	$$
	\frac{1}{\mathcal{\hat L}+i\omega}=(\hat \Lambda^L)^T \frac{1}{\mathcal{\hat  D}+i\omega} \hat  \Lambda^R.
	$$
	Hence, we reduce the problem to compute once the eigendecomposition of the Liouvillian, and simple algebraic computations. 
}
we clearly see that the spectrum consists of two terms. The first term---the inelastic contribution---accounts for the absorption and emission processes between the emitter and the electromagnetic field. The second term---the elastic contribution---, known as Rayleigh scattering, corresponds to the elastic scattering by the system when driven by a laser (if present).
Since the photons are not absorbed by the system, the lineshape is inherited from the source. For instance, an ideal laser, as the one described in \eqref{eq:LaserHamiltonian}, would feature an infinitely narrow linewidth.

On the other hand, we could have applied the QRT in its spectral form [see \eqref{eq:QRTspectral}] to formally compute the fluorescence spectrum, which, in fact, will give us much more physical insights about this quantity. 
Since the spectral decomposition from \eqref{eq:MasterEqSpectralDecomposition} is valid for any operator, not necessarily Hermitian or positive-definite, we can use \eqref{eq:QRTspectral} to express the integrand of  the spectrum from \eqref{eq:SpectrumQRT} as
\begin{multline}
	\langle \hat c^\dagger(0) \hat c(\tau) \rangle = \text{Tr}\left[
 \hat c e^{\mathcal{\hat L}\tau}[\hat \rho_{\text{ss}} \hat c^\dagger] \right] \\
 = \sum_\mu e^{\Lambda_\mu \tau} \text{Tr}[\hat \Lambda_\mu^L (\hat \rho_{\text{ss}} \hat c^\dagger)] \text{Tr}[ \hat c \hat \Lambda_\mu^R ],
\end{multline}
where  we have replaced $\hat \rho(0) \rightarrow \hat \rho_{\text{ss}} \hat c^\dagger$. This result allows us to simply write the spectrum as~\cite{SanchezMunozSymmetriesConservation2019}
\begin{equation}
	S(\omega)= \frac{1}{\pi }\frac{1}{n_c} \text{Re} \left\{  
	\sum_\mu \int_0^\infty d\tau e^{(\Lambda_\mu +i\omega)\tau} \text{Tr}[\hat \Lambda_\mu^L (\hat \rho_{\text{ss}} \hat c^\dagger)] \text{Tr}[ \hat c\hat  \Lambda_\mu^R ]
	\right\}.
\end{equation}
Now, we can formally integrate this expression by defining
\begin{subequations}
	\label{eq_spectral_coeff}
	\begin{align}
		\gamma_\mu &\equiv -2\text{Re}\Lambda_\mu , \\
		\omega_\mu & \equiv - \text{Im}\Lambda_\mu, \\
		  L_\mu&\equiv \text{Re}[  Z_\mu ], \\
		  K_\mu &\equiv \text{Im}[   Z_\mu ], \\
			\label{eq_spectral_coeff_Z_mu}
		 Z_\mu&\equiv  \text{Tr}[\hat \Lambda_\mu^L (\hat \rho_{\text{ss}} \hat c^\dagger)] \text{Tr}[ \hat c \hat  \Lambda_\mu^R ]  ,
	\end{align}
\end{subequations}
such that the spectrum reads
\begin{equation}
	\colorboxed{Maroon}{
		\begin{aligned}[b]
			S(\omega)=   &\frac{1}{\pi }\frac{1}{n_c} \sum_{\mu ,\text{Re} \Lambda_\mu \neq 0} \frac{ (\gamma_\mu/2) L_\mu -(\omega-\omega_\mu) K_\mu }{(\omega-\omega_\mu)^2+(\gamma_\mu/2)^2} \\
			&+\sum_{\mu ,\text{Re} \Lambda_\mu = 0}  \left[ \frac{L_\mu}{n_c} \delta(\omega-\omega_\mu) + \frac{1}{\pi}\frac{K_\mu}{n_c} \text{P.V.} \left(\frac{1}{\omega-\omega_\mu}\right) \right] 
		\end{aligned}
	}
	\label{eq:FormalSpectrum}
\end{equation}
%
%
This expression shows clearer than \eqref{eq:SpectrumResult} that the spectrum is composed of two contributions:
\begin{enumerate}[label=\textcolor{Maroon}{(\roman*)}]
	\item \textcolor{Maroon}{Inelastic scattering. }
	Incoherent (fluorescence) scattering that consists in a sum of Lorentzian profiles---proportional to $ L_\mu$---and dispersive contributions---proportional to $ K_\mu$---that break the symmetry of the Lorentzian around the frequency $\omega_\mu$  
	\item \textcolor{Maroon}{Elastic scattering. } Elastic (Rayleigh) scatterings with infinitely narrow shapes and proportional to $ L_\mu$ when $\text{Re}[\Lambda_\mu]=0$.
\end{enumerate}
Additionally, the last term in the second line of \eqref{eq:FormalSpectrum} denotes the Cauchy principal value at the resonant frequencies. Then, the existence of degenerated steady-states translates into the emergence of multiple $\delta$ peaks that are visible in the spectrum~\cite{SanchezMunozSymmetriesConservation2019}. Note that these infinitely narrow peaks acquire a finite width when we include the description of the filter along with its linewidth.  In the case of having a unique steady-state, this term vanishes since $\hat \Lambda_1^L=\mathbb{I}$ and $ K_1=\text{Im} |\langle \hat c \rangle|^2=0$, recovering thus the same expression for the elastic term in \eqref{eq:SpectrumResult}.

\section{Adiabatic elimination of physical degrees of freedom}

We dedicate this section to present a systematic method, based on projector operators, to eliminate spurious degrees of freedom and obtain a reduced description of the dynamics for the system of interest in a general way. 

The following derivations are adapted from \textcolor{Maroon}{Refs.}~\cite{BreuerTheoryOpen2007,RivasOpenQuantum2012,Navarrete-BenllochIntroductionQuantum2022,Gonzalez-BallesteroTutorialProjector2024}.

\subsection{The Nakajima-Zwanzig equation}
\label{sec_Nakajima}
The basic idea of the projector adiabatic elimination approach is to separate the total system into two subsystems, that we call the system and the environment, and obtain the effective dynamics of the system. This is done by defining a projector superoperator $\mathcal{P}$ that projects the total density matrix onto the subspace of the system. Correspondingly, the complementary projector superoperator is defined as $\mathcal{Q}=\mathcal{I}-\mathcal{P}$, with $\mathcal{I}$ representing the identity superoperator. Since they are orthogonal projection superoperators, they satisfy the following properties:
\begin{subequations}
	\begin{align}
		&\mathcal{P}^2=\mathcal{P},
		\label{eq:ProjectorProp1}\\
		&\mathcal{Q}^2=\mathcal{Q},
		\label{eq:ProjectorProp2} \\
		&\mathcal{P}\mathcal{Q}=\mathcal{Q}\mathcal{P}=0,
		\label{eq:ProjectorProp3} \\
		&\mathcal{P}+\mathcal{Q}=\mathcal{I}. \label{eq:ProjectorProp4}
	\end{align}
\end{subequations}

Let us consider a general quantum system described by a density matrix $\hat  \rho$, whose dynamics is governed by a Liouville-von Neumann equation
\begin{equation}
	\frac{d \hat  \rho(t)}{dt}=\mathcal{\hat L} \hat  \rho(t).
	\label{eq:LiouvNeumannEq}
\end{equation}
By applying the projection superoperators both to the left and to the right of the Liouvillian in \eqref{eq:LiouvNeumannEq}, we get the following set of coupled differential equations for the system and the environment parts of the density matrix
\begin{subequations}
	\begin{align}
		\partial_t[\mathcal{P}\rho(t)]&=\mathcal{P}\mathcal{L}\mathcal{P}\rho(t)+\mathcal{P}\mathcal{L}\mathcal{Q}\rho(t),\label{eq:Relevant}\\
		\partial_t[\mathcal{Q}\rho(t)]&=\mathcal{Q}\mathcal{L}\mathcal{P}\rho(t)+\mathcal{Q}\mathcal{L}\mathcal{Q}\rho(t)\label{eq:Irrelevant}
	\end{align}
\end{subequations}
Here we have assumed that the projectors are time-independent
\graffito{
	$^*$The generalization to a time-dependent Liouvillian requires the introduction of propagators and a subtler description in terms of chronological time-orderings. Besides this, the analysis described here is completely analogous. We refer the reader to \colorref{Gonzalez-BallesteroTutorialProjector2024} for the time-dependent derivation.}
so that $\mathcal{P}\dot\rho=d/dt(\mathcal{P}\rho)$. 
To get a closed equation for the system we formally solve \eqref{eq:Irrelevant} and insert the solution into \eqref{eq:Relevant}, such that we obtain the Nakajima-Zwanzig equation~\cite{NakajimaQuantumTheory1958,ZwanzigEnsembleMethod1960}
\begin{equation}
\colorboxed{Maroon}{	\partial_t[\mathcal{P}\hat  \rho(t)]=\mathcal{P}\mathcal{L}\mathcal{P} \hat \rho(t)
	+\mathcal{P}\mathcal{L}\int_{0}^{t} d\tau e^{\mathcal{Q}\mathcal{L}\tau}\mathcal{Q}\mathcal{L}\mathcal{P} \hat \rho(t-\tau),}
\end{equation}
where the Liouvillian $\mathcal{L}$ is assumed to be time-independent$^{\textcolor{Maroon}{*}}$. This integro-differential equation is exact and holds for all initial conditions and for almost arbitrary systems and interactions, describing non-Markovian memory effects. However, it is still generally unsolvable and perturbative expansions need to be applied.

In order to obtain an effective master equation for the system density matrix we consider the following assumptions: 
\begin{enumerate}[label=\textcolor{Maroon}{(\roman*)}]
	\item We define the projector superoperator 
	\begin{equation}
		\mathcal{P}(*)=\text{Tr}_E[(*)]\otimes \hat  \rho_E,
	\end{equation}
	and decompose the total Liouvillian as
	\begin{equation}
	\mathcal{L}=\mathcal{L}_S+\mathcal{L}_E+\varepsilon \mathcal{L}_{\text{int}}
\end{equation}
	where $\varepsilon \ll 1$ is a small parameter quantifying the weak interaction between the system and the environment. Additionally, the environment is assumed to be in its asymptotic state, such that  $\mathcal{\hat L}_E \hat  \rho_E=0$.
	\item  	We generally define $\mathcal{\hat L}_S,\mathcal{\hat L}_E$ and $\mathcal{\hat L}_{\text{int}}$ as
\begin{subequations}
		\begin{align}
		&\mathcal{L}_{S}[(*)]=-i[\hat H_S,(*)]+\mathcal{D}_S[(*)], \\
		&\mathcal{L}_{E}[(*)]=-i[\hat H_E,(*)]+\mathcal{D}_E[(*)],\\
		&\varepsilon \mathcal{L}_{\text{int}}[(*)]=-i[\hat H_{\text{int}},(*)],
		\label{eq:InteractHamil1}
	\end{align} 
\end{subequations}
	where $\mathcal{D}_{S/E}[(*)]$ denotes arbitrary dissipators with energy scales $\gamma_{S/E}$. We also assume a polynomial expansion in terms of products of system and environment operators in the interaction Liouvillian with $\hat S_m/\hat E_m \in \mathcal{H}_{S/E}$, 
	\begin{equation}
		\hat H_{\text{int}}=\sum_{m=1}^M g_m \hat S_m\otimes \hat E_m.
	\end{equation}
\end{enumerate}
%
%
	%
	%
		%
		%
		%
		%
		%
		%
	%
	%
%
%
In consequence, upon expanding over the small parameter $\varepsilon$, and considering the properties of the projector operators in~\cref{eq:ProjectorProp1,eq:ProjectorProp2,eq:ProjectorProp3,eq:ProjectorProp4}, we obtain the effective master equation~\cite{Navarrete-BenllochIntroductionQuantum2022}
\begin{equation}
	\colorboxed{Maroon}{
		\begin{aligned}[b]
			\partial_t &\hat  \rho_S(t)\approx \mathcal{L}_S[\hat  \rho_S]
			\\
			&+\Bigg(
			\sum_{m,n=1}^M  g_m g_n \int_0^{\infty} d\tau\left[  
			C_{nm}(\tau) \hat S_m \hat  \rho_S(t)\tilde S_n(\tau) \right. \\
			& \qquad \qquad \qquad \qquad  \qquad  \left. -K_{mn}(\tau)\hat S_m\tilde S_n (\tau) \hat  \rho_S(t)
			\right]
			+\text{H.c.}\Bigg),
		\end{aligned}
			\label{eq:EffectiveMasterEq}
	}
\end{equation}
where we have defined the asymptotic two-time correlators for the environment operators
\begin{subequations}
	\begin{align}
		C_{nm}(\tau)&\equiv \lim_{t\rightarrow \infty} \langle \hat E_n(t)\hat E_m(t+\tau) \rangle_E ,
		\label{eq:CorrCnm} \\
		K_{mn}(\tau)&\equiv \lim_{t\rightarrow \infty} \langle \hat E_m(t+\tau)\hat E_n(t) \rangle_E.
		\label{eq:CorrKmn}
	\end{align}
\end{subequations}
In addition, we assumed that the correlation functions $C_{nm}(\tau)$ and $K_{mn}(\tau)$ are either zero or vanish at a much faster rate than any other process within the system by means of the Markov approximation. Then, the system dynamics is governed by its coherent evolution
\begin{equation}
	\hat 	\rho_S(t-\tau)\approx e^{i\hat H_S \tau}\hat  \rho_S(t)e^{-i\hat H_S \tau},
\end{equation}
so that
\begin{equation}
	e^{\mathcal{L}_S \tau}[\hat S]\approx e^{-i\hat H_S\tau}\hat S e^{i\hat H_S\tau}\equiv \tilde  S(\tau) .
\end{equation}
Note that, in general, this effective master equation, which is just an alternative description of the Bloch-Redfield master equation presented in \eqref{eq:BlochRedfieldEq}, is time dependent, even if we started with a time-independent problem. 

\subsection{Effective Hamiltonians} 

The Nakajima-Zwanzig formalism is a powerful framework for deriving an effective description of the reduced dynamics of a system,  particularly in the study of open quantum systems. However, in simpler cases, such as when focusing on the dynamics within a purely Hamiltonian context, this formalism can become unnecessarily complex. Below, we outline how a similar procedure can be simplified and applied in these scenarios~\cite{SanzAdiabaticElimination2016}.

Let us consider a Hamiltonian, $\hat H$, acting on a Hilbert space $\mathcal{H}= \mathcal{H}_A \otimes \mathcal{H}_B$, and let $\mathcal{P}$ and $\mathcal{Q}=\mathcal{I}-\mathcal{P}$ projector operators onto subspaces $A$ and $B$, respectively.
In a Hamiltonian context, we can follow a similar procedure with respect to the previous section and recast the Schrödinger equation, $\hat H | \psi \rangle=E|\psi \rangle$, as a matrix problem:
\begin{equation}
	\begin{pmatrix}
		\mathcal{P}\hat H \mathcal{P} & 	\mathcal{P}\hat H \mathcal{Q} \\
		\mathcal{Q}\hat H \mathcal{P} & \mathcal{Q}\hat H \mathcal{Q}  
	\end{pmatrix} 
	\begin{pmatrix}
		| \psi_A \rangle \\ | \psi_B \rangle
	\end{pmatrix}
	=E		\begin{pmatrix}
		| \psi_A \rangle \\ | \psi_B \rangle
	\end{pmatrix},
\end{equation}
where $\mathcal{P}\hat H \mathcal{P}$ is $n\times n$ matrix denoting the projected Hamiltonian on the subspace $A$, $\mathcal{Q}\hat H \mathcal{Q}$ is a $N\times N$ matrix denoting the projected Hamiltonian on the subspace $B$, and $\mathcal{P}\hat H \mathcal{Q}$ is a $n\times N$ matrix denoting an interaction term connecting both subspaces. In addition, $\mathcal{P} |\psi\rangle\equiv| \psi_A \rangle$ and $\mathcal{Q}|\psi\rangle\equiv| \psi_B \rangle$ are $n\times 1$ and $N\times 1$ vectors, respectively.

By noting that the present matrix problem features a block structure, we can straightforwardly obtain an effective description for subspace $A$, $\hat H_{\text{eff}} |\psi_A\rangle =E |\psi_A\rangle$, in terms of an effective Hamiltonian by means of the matrix inversion lemma:
\begin{equation}
	\hat H_{\text{eff}} \equiv \mathcal{P}\left(\hat H+\hat H \mathcal{Q}\frac{1}{E-\mathcal{Q}\hat H \mathcal{Q}} \mathcal{Q} \hat H \right) \mathcal{P}.
\end{equation}
This expression takes a more familiar form if we rename the terms as: $\mathcal{P}\hat H \mathcal{P}\equiv \hat H_A$, $\mathcal{Q}\hat H \mathcal{Q}\equiv\hat H_B$ and $\mathcal{P}\hat H \mathcal{Q}\equiv V$,
\begin{equation}
	\hat H_{\text{eff}} = \hat H_A +\hat V \frac{1}{E-\hat H_B}\hat V^\dagger.
\end{equation}
Hence, the eigenvalue problem for the full Hamiltonian is equivalent to finding the eigenvalues of a smaller $n\times n$ matrix, $\hat H_{\text{eff}}$. Now, as usually stated in degenerated time-independent perturbation theory, we consider $\hat H_A$ to be a diagonal Hamiltonian in a subspace of (quasi-)degenerated levels, with energy $E_0$, and $\hat V$ is a perturbation. Then, expanding up to second order in the perturbative parameter, the effective Hamiltonian is easily given by
\begin{equation}
	\colorboxed{Maroon}{\hat H_{\text{eff}} \approx \hat H_A +\hat V \frac{1}{E_0-\hat H_B}\hat V^\dagger.}
	\label{eq:EffectiveHamiltonian}
\end{equation}

\subsection{Example: Purcell enhancement of the atomic decay}
\label{sec:AE_singleEmitter}
\paragraph{Nakajima-Zwanzig approach. }
To illustrate the Nakajima-Zwanzig method, let us consider a canonical dissipative scenario in quantum optics: a two-level system coupled to a lossy cavity. To address this problem, we assume that the system is described by a master equation:
\begin{equation}
\frac{d \hat \rho}{dt}= \mathcal{\hat L}_S[\hat \rho]+\mathcal{\hat L}_E[\hat \rho] +\mathcal{\hat L}_{\text{int}}[\hat \rho],
\end{equation}
where $\mathcal{\hat L}_S,\mathcal{\hat L}_E$ and $\mathcal{\hat L}_{\text{int}}$ are defined by
\begin{subequations}
	\begin{align}
		&\mathcal{\hat L}_S[\hat \rho]\equiv\frac{\gamma(\bar n+1)}{2}\mathcal{D}[\hat \sigma]\hat \rho+\frac{\gamma \bar n}{2}\mathcal{D}[\hat \sigma^\dagger]\hat \rho, \\
		&\mathcal{\hat L}_S[\hat \rho]\equiv -i[\Delta_a \hat a^\dagger \hat a,\hat \rho ]+\frac{ \kappa}{2}\mathcal{D}[\hat a]\hat \rho \\
		&\mathcal{\hat L}_{\text{int}}[\hat \rho]\equiv-i[g(\hat a^\dagger  \hat \sigma+ \hat a  \hat \sigma^\dagger ), \hat \rho ].
	\end{align}
\end{subequations}
We work in the rotating frame of the TLS, defined by the transformation operator $\hat U_c(t)=\text{exp}[\omega_q (\hat a^\dagger \hat a+\hat \sigma^\dagger \hat \sigma)]$, such that $\Delta_a\equiv \omega_a-\omega_q $ is the energy detuning of the cavity with respect the TLS.  The spontaneous decay rate of the TLS is given by $\gamma$, $\bar n$ is the thermal distribution, and the cavity decay rate by $\kappa$, while $g$ is the light-matter coupling rate.
Following the general expression of the interaction Hamiltonian in \eqref{eq:InteractHamil1}, $\hat H_{\text{int}}=\sum_m g_m \hat S_m\otimes \hat E_m$, we identify: $g_1=g_2=g$, $\hat E_1= \hat a^\dagger=\hat E_2^\dagger$, and $\hat S_1=\hat \sigma= \hat S_2^\dagger$. 

When $\kappa\gg g$, we can assume that no reversible dynamics occurs since any photon will leave the cavity before the TLS can reabsorb it, and then the cavity remains in vacuum: $\hat \rho_E \approx |0\rangle \langle 0|$. Moreover, we can also assume that $g\gg \gamma$ so that any atomic excitation is transferred to the cavity before the atom thermalizes. Hence, all excitations are leaked through the cavity provided the hierarchy $\kappa \gg g \gg \gamma$, usually termed the bad-cavity limit~\cite{Navarrete-BenllochIntroductionQuantum2022,SavageStationaryTwolevel1988,CiracInteractionTwolevel1992,ZhouDynamicsDriven1998}. In this limit, the cavity is considered to be decoupled from the TLS dynamics and, thus, we can adiabatically eliminate it following the Nakajima-Zwanzig procedure.

Firstly, we need to compute the environment two-time correlation functions $C_{nm}$ and $K_{mn}$ from~\cref{eq:CorrCnm,eq:CorrKmn} by invoking the quantum regression theorem (QRT)~\eqref{eq:QRT}: 
\begin{align}
	C_{nm}(\tau)= e^{-(\kappa/2-i\Delta_a)\tau} \delta_{m1} \delta_{n2}, \\
	K_{mn}(\tau)= e^{-(\kappa/2+i\Delta_a)\tau} \delta_{m2} \delta_{n1},
\end{align}
where we have used the fact that the bosonic operator forms a closed set,
\begin{equation}
	\frac{d \langle \hat a (\tau)\rangle_E }{d \tau}=\text{Tr}_E[\hat a \mathcal{\hat L}_E [\hat \rho(t)]]=-\left(\frac{\kappa}{2}+i\Delta_a\right)\langle \hat a (\tau)\rangle_E,
\end{equation}
that can be easily integrated in time to give $\langle \hat a (\tau)\rangle_E= \langle \hat a (0)\rangle_E e^{-(\kappa/2+i\Delta_a)\tau}$. We included a subscript $E$ (denoting environment) to remark that we are neglecting any effect of the system into the environment. 

Hence, assuming that the correlation functions are either zero or decay at a much faster rate that any other dissipative process affecting the system---i.e., $1/\kappa$ is the shortest dissipative timescale---and following the general structure of \eqref{eq:EffectiveMasterEq}, we obtain the effective master equation (in the Schrödinger picture) for the TLS dynamics:
\begin{equation}
\colorboxed{Maroon}{
\begin{aligned}[b]
		&\frac{d \hat \rho_S}{dt}\approx \mathcal{\hat L}_S[\hat \rho_S]+\left( \frac{g^2}{\kappa/2-i\Delta}[\hat \sigma \hat \rho_S, \hat \sigma^\dagger]+ \text{H.c.} \right)\\
	&= -i [(\omega_q-\Delta_q)\hat \sigma^\dagger \hat \sigma, \hat \rho_S]+ \frac{\Gamma_{\text{eff}}(\bar n_{\text{eff}}+1) }{2}\mathcal{D}[\hat \sigma]\hat \rho_S + \frac{\Gamma_{\text{eff}}\bar n_{\text{eff}} }{2}\mathcal{D}[\hat \sigma^\dagger]\hat \rho_S,
\end{aligned}
}
\label{eq:EffectiveMasterEqPurcell}
\end{equation}
where $\Delta_q,\Gamma_{\text{eff}},$ and $\bar n_{\text{eff}}$ are an atomic shift, and effective descriptions of the decay rate and the thermal distribution:
\begin{subequations}
	\begin{align}
		&\Delta_q \equiv \frac{g^2 \Delta_a}{\kappa^2/4+\Delta_a^2},
		\label{eq:LambShiftPurcell} \\
		&\Gamma_{\text{eff}}\equiv \gamma+ \frac{g^2 (\kappa/2)}{\kappa^2/4+\Delta_a^2}, 
		\label{eq:PurcellRate} \\
		&\bar n_{\text{eff}}\equiv \frac{ \gamma \bar n}{\gamma+\frac{g^2 (\kappa/2)}{\kappa^2/4+\Delta_a^2}}.
		\label{eq:ThermalEffect} 
	\end{align}
\end{subequations}
In this form, we observe that the TLS now experiences an effective thermal environment, inducing modified dissipative decay rates and thermal excitations. We note that the effective dissipative decay, $\Gamma_{\text{eff}}$, is composed of two terms, the first one is related to the intrinsic spontaneous decay rate of the atom, while the second comes from losses through the cavity into the environment of the cavity. Thus, the emission is mostly redirected in the direction of emission of the cavity. This is known as Purcell effect, and 
\begin{equation}
\colorboxed{Maroon}{	\Gamma_P \equiv \frac{g^2 (\kappa/2)}{\kappa^2/4+\Delta_a^2}}
\label{eq:Purcell_intro}
\end{equation}
is usually called Purcell rate~\cite{PurcellResonanceAbsorption1946}, which takes its familiar form at resonance, $\Gamma_P= 4g^2/\kappa$. From this definition, it is common to define the cavity cooperativity,
\begin{equation}
\colorboxed{Maroon}{	C \equiv \frac{4g^2}{\kappa \gamma},}
\label{eq:coop_intro}
\end{equation}
addressing to what extent the losses in the system are channelled through the cavity. 
Then, a value $C>1$ indicates that the losses in the system are predominantly channelled through the cavity mode rather than through spontaneous emission, effectively enhancing the cooperative effects between the emitters and the cavity even though the emitter is weakly coupled to the cavity~\cite{ReitzCooperativeQuantum2022,PlankensteinerEnhancedCollective2019}.
Both quantities, the Purcell rate, $\Gamma_P$, and the cooperativity, $C$, will be extensively used  throughout this Thesis. 
Finally, we also note that when $C\gg \{\bar n, \kappa^2/4+\Delta_a^2 \}$, we obtain $\bar n_{\text{eff}}\ll 1$, producing a cooling process of the atom since the effective temperature is close to zero. 

%
%
%
%
%

\paragraph{Effective Hamiltonian approach. }Alternatively, we could have obtained the effective description of \eqref{eq:EffectiveMasterEqPurcell} by means of the perturbative expansion presented in \eqref{eq:EffectiveHamiltonian}.
We know from the quantum trajectories formalism , presented in \refsec{Section:QuantumTraject} that the dissipative evolution of a system described by a master equation is equivalent to a description in terms of a pure wavefunction under a non-unitary evolution, consisting of the combination of random quantum jumps and the action of a non-Hermitian Hamiltonian, such that, after averaging over infinite trajectories we recover the results from a master equation in the steady-state.
Hence, since the cavity is assumed to be mostly in vacuum in the bad-cavity limit, resulting in an almost negligible probability of featuring a quantum jumps, $\delta p \propto \langle \psi |\hat a^\dagger \hat a| \psi \rangle \ll 1$, we can essentially describe the system dynamics by a non-Hermitian Hamiltonian (at zero temperature):
\begin{equation}
	\tilde  H= \hat H-i \frac{\kappa}{2} \hat a^\dagger \hat a,
\end{equation}
where $\hat H= \Delta_a \hat a^\dagger \hat a+g(\hat a^\dagger  \hat \sigma+ \hat a  \hat \sigma^\dagger )$. Now, considering the canonical basis as $|n, g/e\rangle$, with $n$ denoting the photon excitation and $g/e$ the ground/excited state of the qubit, we constraint the Hilbert space to the one-excitation manifold, $\mathcal{H}\approx \{|0,e\rangle, |1,g\rangle \}$, so that the Hamiltonian reads:
\begin{equation}
		\tilde  H= \left( \begin{array}{c|c}
			E_0 & \hat V  \\ \hline
			\hat V^\dagger&  \hat h
		\end{array} \right)= 
		\left( \begin{array}{c|c}
			0 & g  \\ \hline
			g & \Delta_a-i\frac{\kappa}{2}
		\end{array} \right).
\end{equation}
Then, applying the perturbative expansion from \eqref{eq:EffectiveHamiltonian}, we obtain the effective contributions of the lossy cavity:
\begin{equation}
	E_{\text{eff}}=\frac{g^2}{i\kappa/2 -\Delta_a}= -\frac{g^2 \Delta_a}{\Delta_a^2+\kappa^2/4}- i\frac{g^2 (\kappa/2)}{\Delta^2+\kappa^2/4},
\end{equation}
which are exactly the same terms obtained through the Nakajima-Zwanzig formalism in \eqref{eq:LambShiftPurcell} and \eqref{eq:PurcellRate}, corresponding to a Lamb shift and the Purcell rate.

\cleardoublepage
\chapter{Two-photon resonance fluorescence}\label{ch:TwoPhotonResonance}

\section{Introduction}
\begin{SCfigure}[1][b!]
	\includegraphics[width=.65\linewidth]{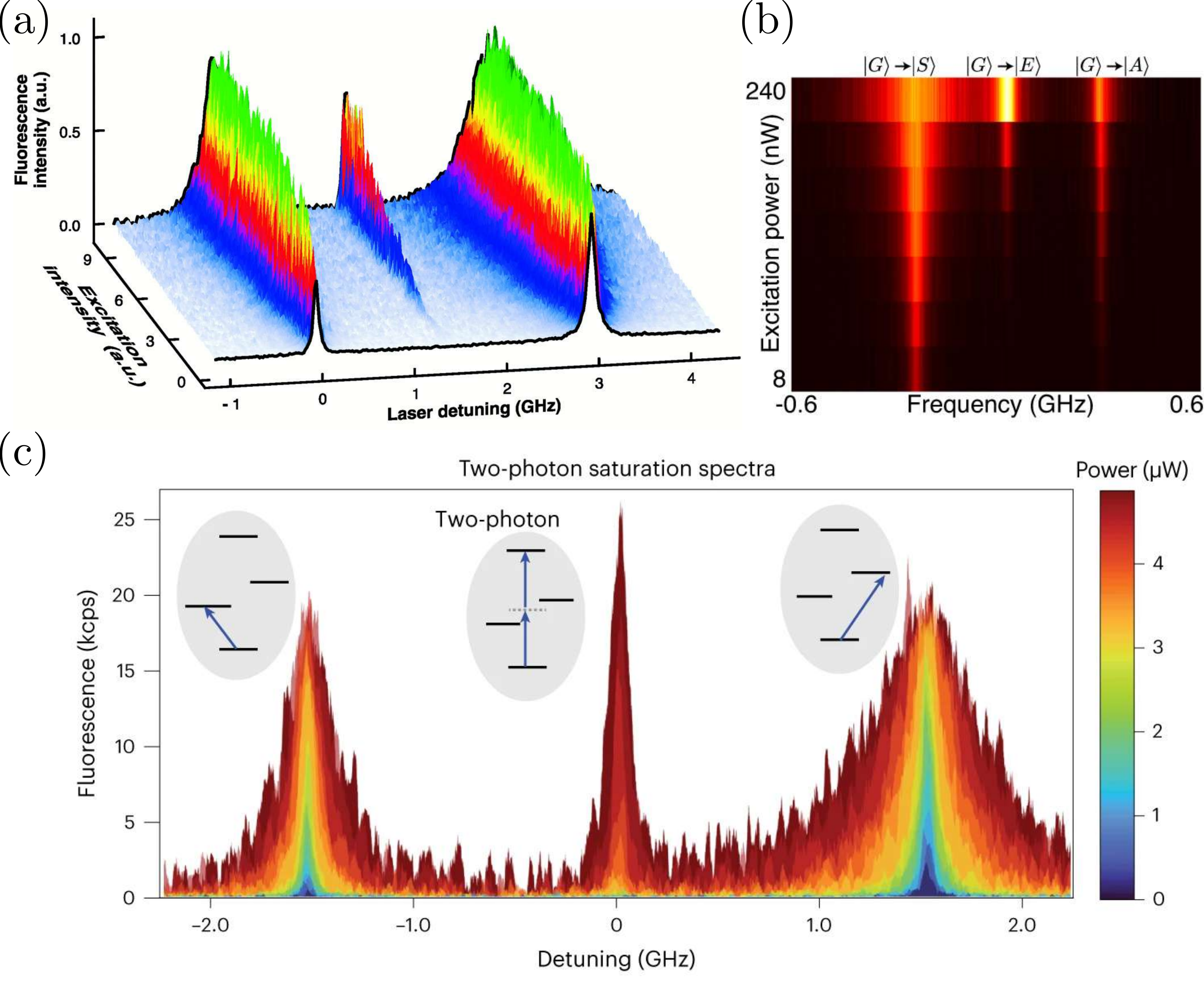}
	\captionsetup{justification=justified}
	\caption[Experiments showcasing the two-photon resonance with molecular aggregates]{
	\label{fig:TwoPhotonExperiments}
	\textbf{Experiments showcasing the two-photon resonance with molecular aggregates. }(a) Excitation spectra from terrylene molecules embedded in a p-terphenyl crystal (from \colorref{HettichNanometerResolution2002} reprinted with permission from AAAS ). (b) Fluorescence excitation spectra from a highly doped naphthalene crystal doped with DBATT molecules (extracted from Ref.~\cite{TrebbiaTailoringSuperradiant2022}). (c) Two-photon saturation spectra for DBT molecules embedded as defects in an anthracene crystal (Reproduced with permission from Springer Nature from \colorref{LangeSuperradiantSubradiant2024}).
	}
\end{SCfigure}

This Thesis is dedicated to the study of collective phenomena in quantum optics. As a starting point, we focus on the simplest realization of this paradigm: \textit{two quantum emitters interacting with a common electromagnetic environment}.  
As highlighted in the introduction, this minimal model already exhibits one of the most fundamental features of collective effects: the emergence of superradiant and subradiant emission~\cite{FicekQuantumInterference2005,ReitzCooperativeQuantum2022}. Despite the apparent simplicity of this system---which has been the subject of extensive theoretical and experimental research for configurations involving two and three quantum emitters~\cite{LehmbergRadiationAtom1970,RiosLeiteLineshapeCooperative1980,RichterPowerBroadening1982,FicekEffectInteratomic1983,PalmaPhasesensitivePopulation1989,VaradaTwophotonResonance1992,ItanoPhotonAntibunching1988,TanasStationaryTwoatom2004,BeigeTransitionAntibunching1998,LembessisTwoatomSystem2013,LakshmiEffectPairwise2014,AhmedCompetitionEffects2014,PengFilteredStrong2019,DarsheshdarPhotonphotonCorrelations2021,PengDirectionalNonclassicality2020,WangDipolecoupledEmitters2020,Juan-DelgadoTailoringStatistics2024}---, the insights extracted from it are widely applicable to a large variety of physical systems. These include coupled quantum dots~\cite{GerardotPhotonStatistics2005,ReitzensteinCoherentPhotonic2006,LauchtMutualCoupling2010,PatelTwophotonInterference2010}, trapped ions~\cite{DeVoeObservationSuperradiant1996,EschnerLightInterference2001}, Rydberg atoms~\cite{AtesAntiblockadeRydberg2007,AmthorEvidenceAntiblockade2010,PritchardCorrelatedPhoton2012}, molecular systems~\cite{HettichNanometerResolution2002,ZhangVisualizingCoherent2016,TrebbiaTailoringSuperradiant2022,LangeSuperradiantSubradiant2024}, and superconducting qubits~\cite{MlynekObservationDicke2014,LambertSuperradianceEnsemble2016}. 
Here, we focus on a particularly relevant effect that arises when interacting emitters are driven by a classical  field: the coherent, non-linear excitation of the transition from the ground state to a doubly-excited state via a \textit{two-photon resonance}, enabled by the emitter-emitter interaction~\cite{VaradaTwophotonResonance1992,HettichNanometerResolution2002,TrebbiaTailoringSuperradiant2022,LangeSuperradiantSubradiant2024}. This mechanism of two-photon excitation, which lies at the heart of important technological applications such as two-photon microscopy~\cite{ZipfelNonlinearMagic2003}, has attracted great interest for the generation of squeezing~\cite{FicekSqueezingTwoatom1994}, steady state atomic entanglement~\cite{FicekEntangledStates2002,HaakhSqueezedLight2015}, and emission of entangled photons~\cite{WangDipolecoupledEmitters2020}. 
The emergence of an extra peak in the excitation spectrum due to this two-photon resonance, which has been experimentally proven~\cite{HettichNanometerResolution2002,TrebbiaTailoringSuperradiant2022,LangeSuperradiantSubradiant2024}---as illustrated in \reffig{fig:TwoPhotonExperiments} using molecular aggregates as quantum emitters~\cite{HettichNanometerResolution2002,TrebbiaTailoringSuperradiant2022,LangeSuperradiantSubradiant2024}---, has been employed as a method to quantify dipole-dipole coupling and, indirectly, to estimate the distance between quantum emitters with nanometer resolution~\cite{HettichNanometerResolution2002}.

Despite the high fundamental and technological importance of  this mechanism, and the seemingly straightforward nature of the model that describes it, most theoretical studies have focused on the particular case of emitters with identical natural frequencies, commonly referred to as \textit{identical emitters}. In this scenario, analytical expressions for the stationary density matrix of the quantum emitters can be derived by direct diagonalization~\cite{RichterPowerBroadening1982,FicekEffectInteratomic1983,FicekEntangledStates2002,FicekQuantumInterference2005}. 
However, the situation becomes considerably more intricate when dealing with two non-identical emitters (e.g., those with different transition frequencies). For this case, only approximate solutions for steady-state populations, restricted to weak coupling and low driving strength, have been reported~\cite{VaradaTwophotonResonance1992,HaakhSqueezedLight2015}.
The challenge of dealing with non-identical emitters is particularly pronounced in solid-state emitters, where achieving identical emitters is hindered by technological limitations. In systems such as molecules or quantum dots,  this  central obstacle is commonly referred to as \textit{inhomogeneous broadening}~\cite{GrimScalableOperando2019}, which arises from variations in emitter properties such as size, strain, and composition~\cite{OBrienPhotonicQuantum2009,PatelTwophotonInterference2010,GieszCoherentManipulation2016,EvansPhotonmediatedInteractions2018,LukinIntegratedQuantum2020,LukinTwoEmitterMultimode2023}.
This situation is also relevant for related models describing phenomena such as light-harvesting~\cite{SanchezMunozPhotonCorrelation2020} and energy-transfer~\cite{LiPlasmoninducedResonance2015,ZhangVisualizingCoherent2016}. Therefore, a full theoretical description of interacting non-identical quantum emitters under coherent driving is desirable.

In this Chapter, we derive analytical expressions for the stationary density matrix of two interacting non-identical quantum emitters under coherent driving at the two-photon resonance. Our expressions are valid for an ample regime of parameters, subject only to the condition that the energy splitting between one-excitation eigenstates must be the largest energy scale in the system. This allows us to provide closed-form expressions in regimes previously inaccessible to analytical approaches~\cite{VaradaTwophotonResonance1992}, including scenarios with strong driving fields that saturate the two-photon transition and large coupling strengths between quantum emitters.
Furthermore, motivated by previous experimental and theoretical efforts based on the use of Rabi oscillation and resonance fluorescence measurements to break the diffraction limit~\cite{HettichNanometerResolution2002,ChangMeasurementSeparation2006,ShinSubopticalResolution2010,GerhardtCoherentNonlinear2010,SunSubwavelengthOptical2011,LiaoResonancefluorescencelocalizationMicroscopy2012}, we explore the potential of resonance fluorescence measurements for estimating the distance between dipoles below Abbe’s diffraction limit  within the framework of quantum parameter estimation~\cite{ParisQUANTUMESTIMATION2009,WisemanQuantumMeasurement2009,DowlingQuantumOptical2015,PetzIntroductionQuantum2011,LuisFisherInformation2012,ChaoFisherInformation2016,SafranekSimpleExpression2018,LiuQuantumFisher2020}.
We find that the maximum precision for distance estimation is achieved by driving the system at the two-photon transition in the onset of saturation.
Our results offer a pathway to advance experimental efforts towards sub-wavelength imaging of quantum emitters using resonance fluorescence.

The Chapter is organized as follows. In \refsec{section:model}, we introduce the model of the system and present two effective models that separately account for second-order and first-order processes. In \refsec{sec:2_steady_state}, we apply these results to study the steady-state observables of the light radiated by the quantum emitters. In \refsec{sec:3-spectrum}, we provide a detailed analysis of the spectrum of two-photon resonance fluorescence. Finally, in \refsec{sec:4-parameter-estimation}, we employ quantum parameter estimation theory to evaluate the potential of spectrum measurements for the estimation of the inter-emitter distance.

These results have been published in Physical Review Research~\cite{Vivas-VianaTwophotonResonance2021}.

\section{Vacuum-induced collective interactions}
\label{section:model}

We start the discussion of this Chapter by introducing a central element of this Thesis: the master equation that describes the dissipative dynamics of two nonidentical interacting quantum emitters driven by a coherent field. 
The interaction between emitters can be mediated by vacuum-induced effects~\cite{LehmbergRadiationAtom1970,RiosLeiteLineshapeCooperative1980,RichterPowerBroadening1982,FicekEffectInteratomic1983,PalmaPhasesensitivePopulation1989,VaradaTwophotonResonance1992,ItanoPhotonAntibunching1988,FicekEntangledStates2002,TanasStationaryTwoatom2004,FicekQuantumInterference2005,LembessisTwoatomSystem2013,LakshmiEffectPairwise2014,AhmedCompetitionEffects2014,PengFilteredStrong2019,DarsheshdarPhotonphotonCorrelations2021,PengDirectionalNonclassicality2020,WangDipolecoupledEmitters2020,ReitzCooperativeQuantum2022,HettichNanometerResolution2002,TrebbiaTailoringSuperradiant2022,LangeSuperradiantSubradiant2024} or through photonic structures~\cite{Martin-CanoDissipationdrivenGeneration2011,LodahlInterfacingSingle2015,ChangColloquiumQuantum2018,HaakhSqueezedLight2015,ReitzensteinCoherentPhotonic2006,LauchtMutualCoupling2010,Gonzalez-TudelaEntanglementTwo2011,MlynekObservationDicke2014,VanLooPhotonMediatedInteractions2013,ChangColloquiumQuantum2018,Miguel-TorcalInversedesignedDielectric2022,Miguel-TorcalMultiqubitQuantum2024,HallettControllingCooperative2024}. For simplicity and to provide an intuitive benchmark for the strength of the collective interaction we consider that the emitters interact via vacuum-induced effects. To obtain such a description, we extend the procedure used in \refsec{Section:SpontaneousEmission}---which computes the master equation for a single quantum emitter interacting with a vacuum electromagnetic environment---to the case of two emitters. 
While this generalization might appear straightforward, introducing two (or more) independent emitters interacting with the same environment adds two crucial ingredients to the dynamics: \textit{collective coherent and dissipative channels of interaction between the emitters}. These vacuum-induced collective phenomena will increase the complexity of the system, which we  will exploit in the subsequent chapters of this Thesis, e.g., for entanglement generation. 

\subsection{General model}
\label{sec_General_model_twoqubits}

\paragraph{Master equation. }
Our system is composed of two quantum emitters, each of them described as a two-level system (TLS) placed at position $\mathbf{r}_i$ ($i\in 1,2$), with natural frequency $\omega_i$ and dipole moment $\bm\mu_i$. The system and energy levels are sketched in \reffig{fig:SketchEmitters}~\textcolor{Maroon}{(a,b)}: each TLS spans a basis $\{|g_i\rangle,|e_i\rangle \}$, where we define the lowering operator $\hat\sigma_i \equiv |g_i\rangle\langle e_i|$, such that dipole moment reads $\bm\mu_i=\langle g_i | \hat{\bm \mu}|e_i\rangle $.  We note that each emitter preserves the properties inherited from the $\mathfrak{su}(2)$ algebra, such that
\begin{equation}
	[\hat \sigma_i^\dagger,\hat \sigma_j]=2 \hat \sigma_{z,i} \delta_{i j}, \quad 	
	[\hat \sigma_{z,i},\hat \sigma^{(\dagger)}_j]=\pm \hat \sigma_{i}^{(\dagger)}\delta_{i j}, \quad 
	\{\hat \sigma_i^\dagger,\hat \sigma_j \}= \delta_{i j},
\end{equation}
where $\hat \sigma_{z,i}\equiv 2 \hat \sigma_i^\dagger \hat \sigma_i - \mathbb{\hat I}_i$ is the $z$-Pauli matrix for the $i$th emitter.
The total basis of the composite system, $\mathcal{H}$, is defined as
\begin{equation}
	\mathcal{H}=\{|gg\rangle, |ge\rangle, |eg\rangle, |ee\rangle\},
\end{equation}
where $|gg\rangle \equiv |g_1\rangle\otimes |g_2\rangle$, and similarly for the other states. 
{
	\sidecaptionvpos{figure}{b}
	\begin{SCfigure}[1][h!]
	\includegraphics[width=.99\linewidth]{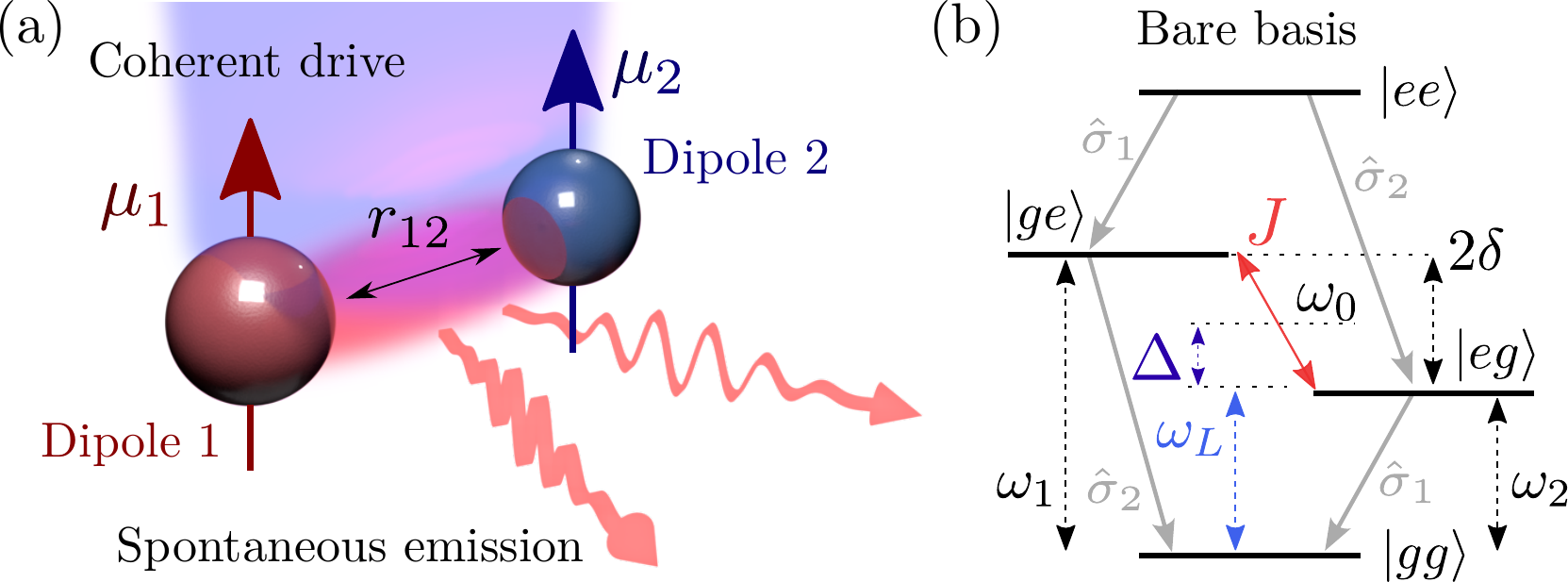}
	\captionsetup{justification=justified}
	\caption[Sketch of the system]{
		\label{fig:SketchEmitters}
		\textbf{Sketch of the system. }(a) Two interacting non-identical dipoles under coherent driving. (b) Bare basis of the system of quantum emitters. The two states of the single-excitation subspace are detuned by an energy $2\delta$, and coupled with a coupling strength $J$. 
		}
\end{SCfigure}
}

We consider a coherent coupling between both emitters with a coupling rate $J$, and a coherent laser field of frequency $\omega_L$ driving both emitter with a Rabi frequency $\Omega$. In the rotating frame of the laser, the resulting time-independent Hamiltonian is $\hat H= \hat H_\mathrm{q}+\hat H_\mathrm{d}$, where $\hat H_\mathrm{q}$ is the bare Hamiltonian of the interacting quantum emitters,
\begin{equation}
	\label{eq:HEmitters}
	\hat H_{\mathrm{q}}= ( \Delta - \delta ) \hat\sigma_1^\dagger \hat\sigma_1 + 
	( \Delta +\delta ) \hat\sigma_2^\dagger \hat \sigma_2 + J (\hat\sigma_1^\dagger \hat \sigma_2 + \text{H.c.} ), 
\end{equation}
and $\hat H_\mathrm d$ is the Hamiltonian of the coherent driving,
\begin{equation}
\hat H_\mathrm{d}= \Omega (\hat\sigma_1 + \hat\sigma_2 + \text{H.c.}),
	\label{eq:Hd}
\end{equation}
having defined the laser-qubit detuning, $\Delta$, and the qubit-qubit detuning, $\delta$, as
\begin{equation}
\Delta\equiv \frac{\omega_1 + \omega_2}{2}-\omega_L, \quad \text{and}\quad \delta\equiv \frac{\omega_2 - \omega_1}{2}.
\end{equation}

We also assume that both emitters interact with an environment, which is described as an electromagnetic field continuum in vacuum, responsible for mediating the coherent interaction between emitters and providing a decay mechanism that de-excites the quantum emitters by spontaneous emission to free-space. In the reduced Hilbert space of the emitters, this dissipative dynamics is modelled by the following master
\graffito{
	$^*$We refer the reader to Appendix~\ref{ch:Appendix_TwoPhotonResonance} for a detailed, step-by-step, derivation of the master equation. 
}
equation$^{\textcolor{Maroon}{*}}$~\cite{FicekQuantumInterference2005,BreuerTheoryOpen2007,RivasOpenQuantum2012}:
\begin{equation}
\colorboxed{Maroon}{
		\frac{d\hat\rho}{dt}=-i[\hat H,\hat \rho] +\sum_{i,j=1}^2 \frac{\gamma_{ij}}{2}\mathcal{D}[{\hat\sigma_i,\hat\sigma_j}]
	\hat\rho ,
	}
	\label{eq:LMEemitters}
\end{equation}
where we have extended the definition of the dissipator given in \eqref{eq:LindbladTerms}, such that for two system operators $\hat A, \hat B$:
\begin{subequations}
	\begin{align}
		&\mathcal{D}[\hat A,\hat B]\hat \rho \equiv 2 \hat A \hat \rho \hat B^\dagger - \{\hat B^\dagger \hat A,\hat \rho\}, \\
		&\mathcal{D}[\hat A]\hat \rho \equiv\mathcal{D}[\hat A,\hat A]\hat \rho.
	\end{align}
\end{subequations}
In \eqref{eq:LMEemitters}, $\gamma_{ii}$ is the local decay rate of spontaneous emission of the $i$th emitter, and $\gamma_{12}=\gamma_{21}$ is the dissipative coupling rate between emitters that emerges as a consequence of collective decay. Incoherent excitation by thermal photons is neglected since we consider optical transitions or cryogenic
temperatures in platforms such as superconducting circuits~\cite{GarciaRipollQuantumInformation2022}. 

In this Chapter, we choose to parametrize the coherent ($J$) and dissipative ($\gamma, \gamma_{12}$) rates using the values corresponding to point-dipoles in free space, which interact through vacuum, inducing dipole-dipole interactions~\cite{LehmbergRadiationAtom1970,RiosLeiteLineshapeCooperative1980,RichterPowerBroadening1982,FicekEffectInteratomic1983,PalmaPhasesensitivePopulation1989,VaradaTwophotonResonance1992,ItanoPhotonAntibunching1988,FicekEntangledStates2002,TanasStationaryTwoatom2004,FicekQuantumInterference2005,LembessisTwoatomSystem2013,LakshmiEffectPairwise2014,AhmedCompetitionEffects2014,PengFilteredStrong2019,DarsheshdarPhotonphotonCorrelations2021,PengDirectionalNonclassicality2020,WangDipolecoupledEmitters2020,ReitzCooperativeQuantum2022,HettichNanometerResolution2002,TrebbiaTailoringSuperradiant2022,LangeSuperradiantSubradiant2024},
since this parametrization offers an intuitive benchmark for the strength of the collective interaction.
Nonetheless, our model is entirely general, and alternative parametrizations---such as those based on couplings mediated by photonic structures---can be employed without affecting the validity of our results~\cite{Martin-CanoDissipationdrivenGeneration2011,LodahlInterfacingSingle2015,ChangColloquiumQuantum2018,HaakhSqueezedLight2015,ReitzensteinCoherentPhotonic2006,LauchtMutualCoupling2010,Gonzalez-TudelaEntanglementTwo2011,MlynekObservationDicke2014,VanLooPhotonMediatedInteractions2013,ChangColloquiumQuantum2018,Miguel-TorcalInversedesignedDielectric2022,Miguel-TorcalMultiqubitQuantum2024,HallettControllingCooperative2024}.
Therefore, within the free-space interaction parametrization, local decay rates in free space depend on the natural frequencies and dipole moments of the emitters \cite{FicekQuantumInterference2005},
\begin{equation}
\gamma_{ii}\equiv \gamma_i = \frac{\omega_i^3 \abs{\bm{\mu}_i}^2}{3\pi \epsilon_0  c^3},
	\label{eq:gamma}
\end{equation}
where $\bm{\mu}_i$ is the dipole moment of the $i$th emitter, $\epsilon_0$ is the vacuum permittivity and $c$ is the speed of light. The coherent and dissipative coupling rates also depend on the relative separation vector between emitters \cite{FicekQuantumInterference2005,ReitzCooperativeQuantum2022}, $\mathbf{r}_{12}=\mathbf{r}_1-\mathbf{r}_2$,
\begin{subequations}
	\begin{empheq}[box=\colorboxed{Maroon}]{align}
		J=
		&\frac{3}{4}\sqrt{\gamma_1 \gamma_2} \left\{ - \left[ 1- \left( \bm{\bar \mu}\cdot \mathbf{\bar r}_{12} \right)^2 \right] \frac{\cos\left( k r_{12} \right)}{kr_{12}} \right. \notag \\
		&\left.  \quad \quad \ \   + \left[ 1-3 \left( \bm{ \bar\mu}\cdot \mathbf{\bar r}_{12} \right)^2 \right] \left[  \frac{\sin\left( k r_{12} \right)}{(kr_{12})^2} +\frac{\cos\left( k r_{12} \right)}{(kr_{12})^3} \right] \right\},
		\label{eq:J}
		\\
		\gamma_{12}=
		&\frac{3}{2} \sqrt{\gamma_1 \gamma_2} \left\{ \left[ 1- \left( \bm{\bar \mu}\cdot \mathbf{\bar r}_{12} \right)^2 \right] \frac{\sin \left( k r_{12} \right)}{kr_{12}}  \right. \notag \\
		&\left.  \quad \quad \ \  +  \left[ 1-3 \left( \bm{\bar\mu}\cdot \mathbf{ \bar r}_{12} \right)^2 \right] \left[  \frac{\cos\left( k r_{12} \right)}{(kr_{12})^2} - \frac{\sin \left( k r_{12} \right)}{(kr_{12})^3} \right] \right\},
		\label{eq:gamma_12}
	\end{empheq}
\end{subequations}
where $r_{12}=|\bm r_{12}|$ is the distance between emitters, $\bm{\bar \mu_1} = \bm{\bar \mu_2} = \bm{\bar \mu}$ and $\mathbf{ \bar r}_{12}$ are unit vectors, $k=\omega_0/c$, and $\omega_0=(\omega_1+\omega_2)/2$, having assumed that the dipoles are polarized in the same direction and that $\omega_0\gg|\delta|$, which allows us to use a single value of $k$.  
Note that in \cref{eq:J,eq:gamma_12} both the product $\bm{\bar \mu}\cdot \mathbf{\bar r}_{12}$, corresponding to the relative angle between the dipole orientation and the vector connecting emitters [see \reffig{fig:RelativeOrientation}~\textcolor{Maroon}{(a)}], and the inter-emitter $r_{12}$  introduce non-trivial dependencies on the geometrical configuration of the emitters. These dependencies will be analysed in detail later in \refsec{sec:geometrical_conf}. 
{
	\sidecaptionvpos{figure}{t}
	\begin{SCfigure}[1][h!]
	\includegraphics[width=.9\columnwidth]{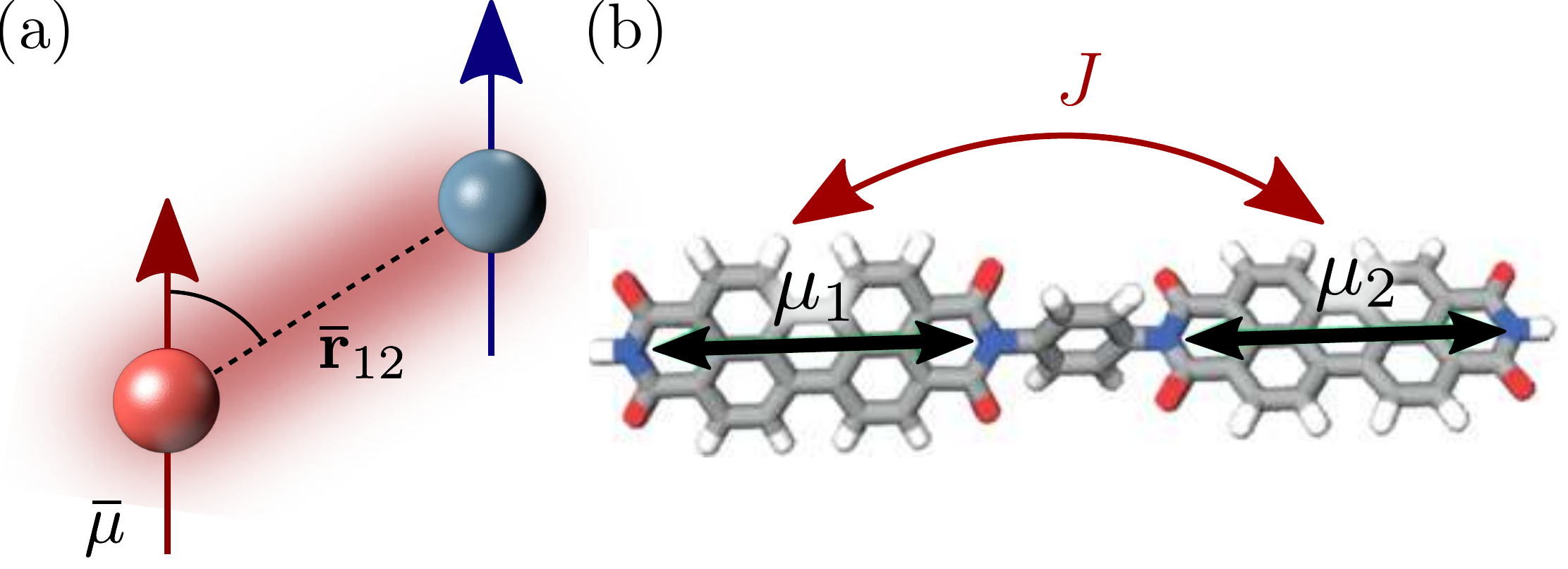}
	\captionsetup{justification=justified}
	\caption[Relative orientation of the emitters.]{\label{fig:RelativeOrientation}	\textbf{Relative orientation of the emitters. } (a) Sketch of the general geometrical configuration of dipoles. (b)  Illustration of two identical chromophores strongly interacting by an insulating bridge of length $\ll \lambda_0$ (reproduced with permission from \cite{ReitzCooperativeQuantum2022,DiehlEmergenceCoherence2014})
	 }
\end{SCfigure}}

For the sake of simplicity 
\graffito{
	$^*$When driving emitters with a coherent field, each emitter experiences a position-dependent Rabi frequency:
	$$
	\Omega(\mathbf{r}_i) = \bm \mu_i \cdot \mathbf{E}_L e^{i\mathbf{k}_L\cdot \mathbf{r}_i},
	$$
	where $\mathbf{E}_L$ is the laser field amplitude, and $\mathbf{k}_L$ its wavevector. If the dipole moments are similar and the emitters are confined to a region much smaller than the resonant wavelength ($kr_{12}$), the Rabi frequencies can be approximated as identical. 
	Alternatively, using multiple laser fields---one for each emitter---provides complete control to set their Rabi frequencies to be identical.
}
and without loss of generality, we will assume emitters with similar spontaneous decay rates and similar driving strengths for each emitter$^{\textcolor{Maroon}{*}}$,
\begin{equation}
\gamma_{ii} \approx \gamma \quad \text{and}  \quad \Omega(\mathbf{r}_1)\approx \Omega(\mathbf{r}_2)\approx\Omega.
\label{eq:DipoleAssumptions}
\end{equation}
In the parametrization based on \eqref{eq:gamma}, these assumptions remain valid even for detuned emitters ($\delta\neq 0$) provided that $\omega_0\gg \delta$, so the relative difference between decay rates is negligible, $|\gamma_{11} -\gamma_{22}|\ll1$.  
%
%
We note that all the results presented in this Chapter can be easily generalized to the case of $\gamma_1\neq \gamma_2$ and $\Omega(\mathbf{r}_1)\neq \Omega(\mathbf{r}_2)$. 
We emphasize that, even if we assume similar dipole moments, we still address the general case of nonidentical emitters that may have different natural frequencies, i.e., $\delta \neq 0$. This situation can arise, for instance, when quantum emitters with similar dipole moment (e.g., identical atoms, or molecules) have their properties modified by different local environments or inhomogeneous electric fields, such as the one created by local scanning electrodes used to produce Stark shift maps~\cite{HettichNanometerResolution2002,TrebbiaTailoringSuperradiant2022,LangeSuperradiantSubradiant2024}.

\paragraph{Eigenstructure.}
\begin{SCfigure}[1][b!]
	\includegraphics[width=1.0\linewidth]{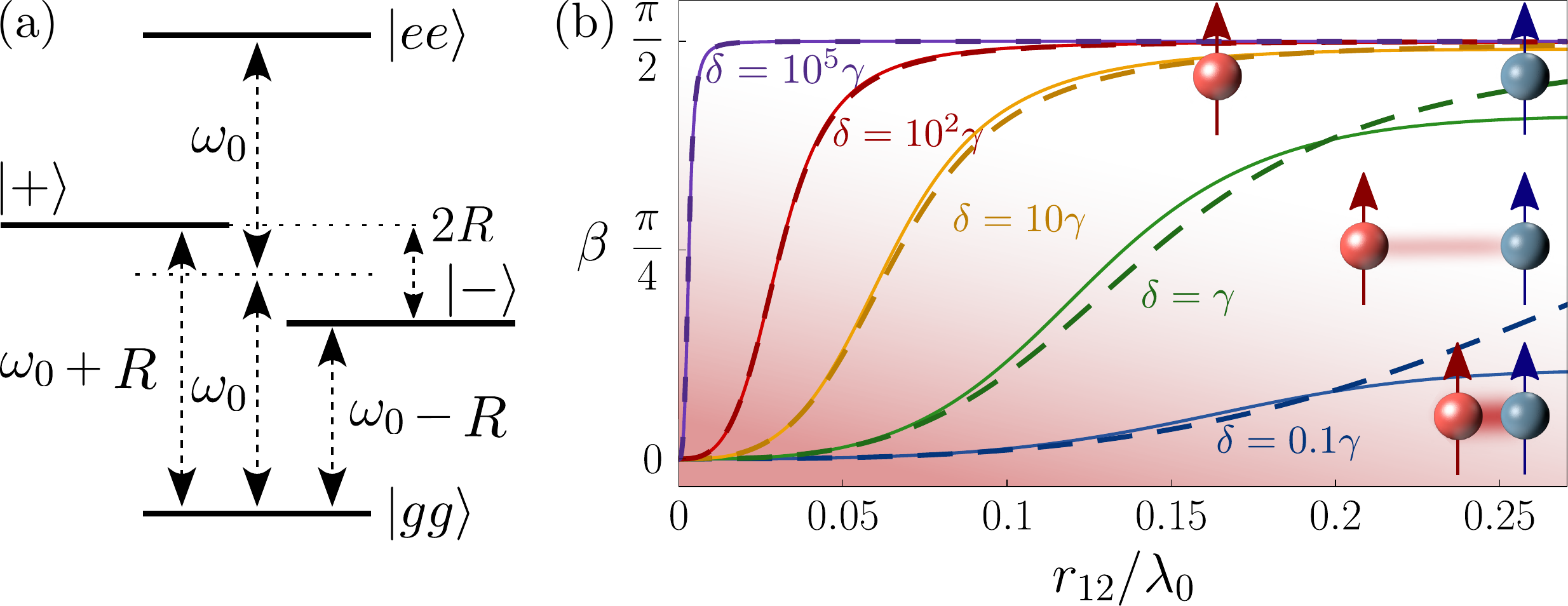}
\end{SCfigure}
The Hamiltonian of the bare, undriven emitters in \eqref{eq:HEmitters} describes a dimer yielding a diagonal basis $\{|gg\rangle, |-\rangle, |+\rangle, |ee\rangle \}$ with eigenvalues $\{0, \Delta-R,\Delta+R,2\Delta \}$ (in the rotating frame of the laser), where $\{|-\rangle, |+\rangle \}$ are the eigenstates  of the single-excitation subspace, 
\begin{empheq}[box=\colorboxed{Maroon}]{align}
	|+\rangle&= \frac{1}{\sqrt{2}}\left(\sqrt{1-\sin\beta}|eg\rangle + \sqrt{1+\sin\beta}|ge\rangle \right), 
	\label{eq:PlusEigenstate}\\ 
	|- \rangle&= \frac{1}{\sqrt{2}}\left(\sqrt{1+\sin\beta}|eg\rangle - \sqrt{1-\sin\beta}|ge\rangle \right),
	\label{eq:MinusEigenstate}
\end{empheq}
with $\beta$ denoting a mixing angle defined as~\cite{SanchezMunozPhotonCorrelation2020}
\begin{equation}
	\beta \equiv \arctan(\delta/J).
	\label{eq:betadefinition}
\end{equation}
The energies of these single-excitation eigenstates are $E_\pm= \Delta \pm R$, where we have defined the \textit{Rabi frequency of the dipole-dipole coupling} as
\begin{equation}
	R \equiv \sqrt{J^2 +\delta^2}.
\end{equation}
%
%
\graffito{\vspace{6.2cm}
	\captionof{figure}[Diagonal basis and mixing angle.]{\label{fig:Fig_DiagonalBasis_MixingAngle}	\textbf{Diagonal basis and mixing angle. }(a) Energy levels in the diagonal basis where the coupling between emitters has been diagonalized. (b) Dependence of the mixing angle, $\beta$, on the normalized inter-emitter distance, $r_{12}/\lambda_0$.  The solid lines represent numerical computations based on \eqref{eq:J}, while the dashed lines correspond to numerical results simulated using the effective coupling from \eqref{eq:DickeLimit}, obtained in the limit $r_{12}\ll \lambda_0$. }
}
This energy structure is depicted in \reffig{fig:Fig_DiagonalBasis_MixingAngle}~\textcolor{Maroon}{(a)}. 
Given the definition of the mixing angle $\beta$ in \eqref{eq:betadefinition}, we can distinguish two limit cases:
\begin{enumerate}[label=\textcolor{Maroon}{(\roman*)}]
	\item \textcolor{Maroon}{Identical emitters or strong dipole-dipole interactions. }
	The case $\beta=0$ corresponds to the limit of identical emitters ($\delta=0$) or strong dipole-dipole interactions ($J\gg \delta$). In that case, the eigenstates in the single-excitation subspace correspond to purely symmetric and antisymmetric superpositions, $|S\rangle$ and $|A\rangle$, which are the so-called superrandiant and subradiant states~\cite{ReitzCooperativeQuantum2022}, respectively,
	\begin{equation}
		\lim_{\beta \rightarrow0 }|\pm \rangle=|S/A\rangle = \frac{1}{\sqrt{2}}(|eg\rangle \pm |ge\rangle).
		\label{eq:SymAntiSym_beta0}
	\end{equation}
	\item \textcolor{Maroon}{Highly detuned emitters or disabled dipole-dipole interaction. }
	On the other hand, the limit $\beta=\pi/2$ corresponds to the case in which the interaction rate between emitters is much weaker than their detuning ($\delta \gg J$) or the dipole-dipole interaction is disabled ($J=0$). In this limit, the eigenstates tend to those of independent, non-interacting emitters,
	\begin{equation}
		\lim_{\beta \rightarrow \frac{\pi}{2}}|\pm \rangle= |ge/eg\rangle.
	\end{equation}
\end{enumerate}

In this Chapter, we focus on the case where $R$ is the largest energy scale in the system, so that $\hat H_{ \mathrm{d}}$ can be treated perturbatively with respect to $R$. This approach is different to the one taken, for instance, in \colorref{VaradaTwophotonResonance1992}, where $J$ was taken as a perturbative parameter. Our approach will allow  us to derive analytical expressions valid in more regimes, such as the one of strong dipole-dipole coupling $J\gg \delta$, which is particularly relevant for closely spaced emitters. For instance, \reffig{fig:RelativeOrientation}~\textcolor{Maroon}{(b)} illustrates a physical realization of this model based on assembled molecular dimers~\cite{DiehlEmergenceCoherence2014}, where two cromophores are strongly coupled through an insulating bridge. 
This scenario of strongly interacting quantum emitters also leads us to focus on a regime where the two resonances at $\Delta = \pm R$ are well resolved ($R\gg \gamma$), emerging as two distinct peaks in measurements such as resonance fluorescence excitation spectra~\cite{HettichNanometerResolution2002,TrebbiaTailoringSuperradiant2022,LangeSuperradiantSubradiant2024}. This is consistent with the experimental observations shown in \reffig{fig:TwoPhotonExperiments}, where the side- and central peaks are perfectly resolved. 

Since we are interested in using $R$ as our energy reference~\cite{SanchezMunozPhotonCorrelation2020}, we will reformulate the Hamiltonian parameters in terms of $R$ and $\beta$ as
\begin{equation}
\colorboxed{Maroon}{
	J=R\cos \beta \quad \text{and} \quad  \delta= R \sin \beta.
}
	\label{eq:J_delta_beta}
\end{equation}
When $J$ is determined by the dipole-dipole coupling, $\beta$ depends on both the emitter-emitter detuning $\delta$ and the distance between emitters, $k  r_{12}$. This makes $\beta$ range from $0$, at short distances, to $\pi/2$, at long distances, as we show in \reffig{fig:Fig_DiagonalBasis_MixingAngle}~\textcolor{Maroon}{(b)} for different values of $\delta$. 
In this figure, one can see that, as the detuning $\delta$ is decreased, the distance required to reach $\beta=\pi/2$ increases, until one reaches the limiting case $\delta=0$, where $\beta=0$ regardless of the distance $kr_{12}$. 
In this parametrization of $J$, the inter-emitter distance $kr_{12}$ has thus a strong impact in the structure of the eigenstates in \cref{eq:PlusEigenstate,eq:MinusEigenstate}. As we will see, this directly affects the quantum optical properties of the emitted light and the response to coherent, two-photon driving. 

\paragraph{Master equation in the eigenbasis. }
Using the eigenstructure of the bare emitters from \cref{eq:PlusEigenstate,eq:MinusEigenstate}, along with the mixing angle prescription in \cref{eq:betadefinition,eq:J_delta_beta}, we can rewrite the master equation in \eqref{eq:LMEemitters} in a more convenient form by transforming it onto the eigenbasis of the bare, undriven emitters:
\begin{equation}
	\colorboxed{Maroon}{
		\begin{aligned}[b]
			\frac{d \hat \rho}{dt}=-i[&\hat H ,\hat \rho]
			+ \sum_{i=+,-}\frac{\gamma_i}{2} \left(\mathcal{D}[|gg\rangle \langle i |]
			\hat \rho+\mathcal{D}[|i \rangle \langle ee |]
			\hat \rho\right)\\
			&+\frac{\gamma_C}{2} \left(  
			2|gg\rangle \langle -| \hat \rho |+ \rangle \langle gg| -\{ |+\rangle \langle -|,\hat \rho \} + \text{H.c.}
			\right),
		\end{aligned}
	}
	\label{eq:LMEemittersDiagonalized}
\end{equation}
where the Hamiltonians from~\cref{eq:HEmitters,eq:Hd} now reads
\begin{equation}
	\hat  H_{\mathrm q}=(\Delta+R) |+\rangle \langle +|   + (\Delta-R)  |-\rangle \langle -| ,
\end{equation}
and
\begin{equation}
\hat  H_{\mathrm d}=\sum_{i=+,-} \Omega_i \left( |gg\rangle \langle i| + |i\rangle \langle ee| + \text{H.c.}\right).
\end{equation}
In order to simplify the notation, we have defined the following \textit{excitonic parameters}:
\begin{subequations}
	\begin{align}
		\gamma_\pm &\equiv \gamma \pm \gamma_{12} \cos \beta, 
		\label{eq:ExcitonicParameters1}\\
		\gamma_C & \equiv \gamma_{12} \sin \beta, \label{eq:ExcitonicParameters2}\\
		\Omega_\pm& \equiv \Omega \sqrt{1\pm \cos \beta},\label{eq:ExcitonicParameters3}
	\end{align}
\end{subequations}
where the rates $\gamma_\pm$ describe the super- and subradiant decay through the single-excitation subspace, namely via the transitions: $|ee\rangle \rightarrow |\pm \rangle \rightarrow |gg\rangle$, while $\gamma_C$ is the rate of incoherent coupling between $|\pm \rangle$ that originates from their collective decay. The driving rates $\Omega_\pm$ are the corresponding super- and subradiant driving amplitudes. This parametrization is depicted in \reffig{fig:ExcitonicScheme}.

\addtocounter{figure}{-1}
\begin{SCfigure}[1][h!]
	\includegraphics[width=0.47\columnwidth]{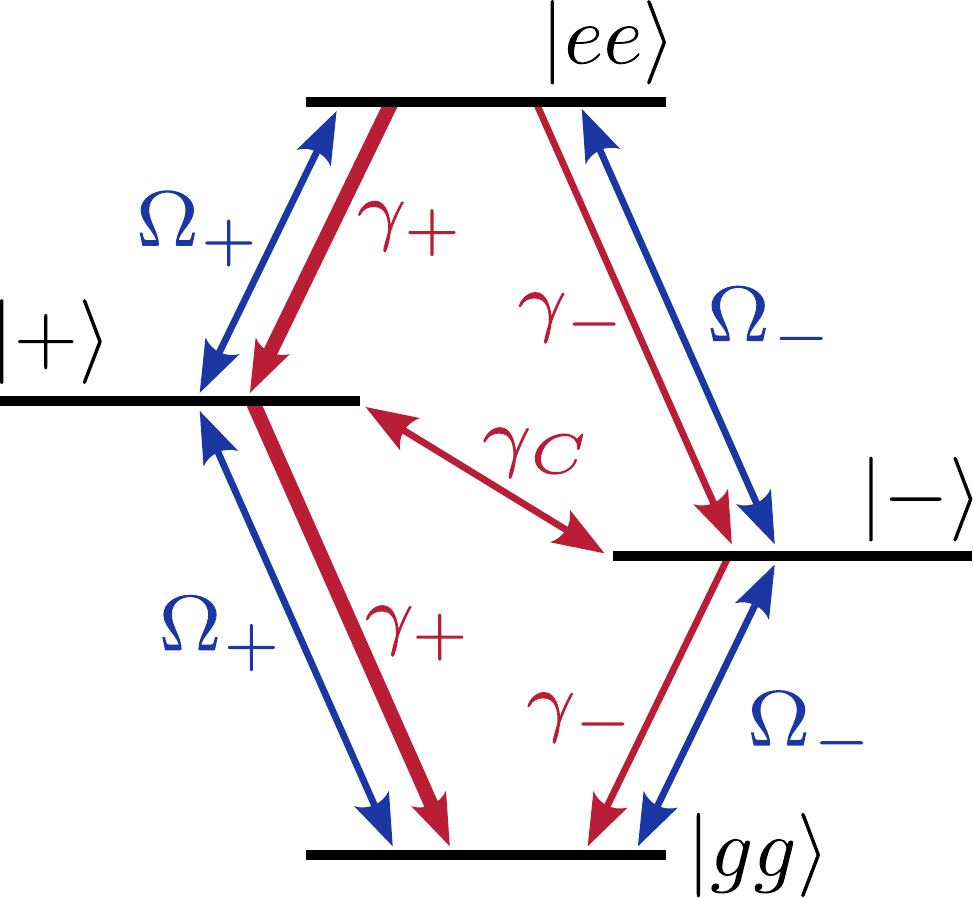}
	\captionsetup{justification=justified}
	\caption[Coherent and dissipative interactions in the excitonic basis.]{\label{fig:ExcitonicScheme}	\textbf{Coherent and dissipative interactions in the excitonic basis. } Diagram of the coherent and dissipative interactions in the excitonic basis.  Blue arrows denote coherent coupling enabled by the external drive. Red arrows denote channels of irreversible decay.}
\end{SCfigure}
In the commonly studied case of identical emitters ($\beta=0$), the symmetric and antisymmetric states exhibit distinct behaviours: 
\begin{enumerate}[label=\textcolor{Maroon}{(\roman*)}]
	\item \textcolor{Maroon}{Symmetric state, $|+\rangle=|S\rangle$. }
	 The symmetric state is characterized by superradiant properties in free space, with an enhanced coherent coupling and a superradiant decay rate, 
	\begin{equation}
		\Omega_S=\sqrt{2}\Omega\quad \text{and} \quad \gamma_S=\gamma+\gamma_{12}.
		\label{eq:Superradiant}
	\end{equation}
	\item \textcolor{Maroon}{Antisymmetric state, $|-\rangle=|A\rangle$.}   
	Conversely, the antisymmetric state becomes a dark state, decoupled from the driving field, and exhibiting a suppressed decay rate,
	\begin{equation}
	\Omega_A=0 \quad \text{and} \quad \gamma_A=\gamma-\gamma_{12}.
			\label{eq:Subradiant}
\end{equation}
\end{enumerate}
In the specific case of two closely spaced, identical emitters ($\beta=0, \gamma_{12}=\gamma$), the symmetric state acquires its maximum superradiant decay rate, $\gamma_S=2\gamma$, while the antisymmetric state becomes entirely decoupled from the system dynamics ($\Omega_A,\gamma_A,\gamma_C=0)$ and turns into a dark state whose population gets trapped~\cite{FleischhauerElectromagneticallyInduced2005,KhanCoherentPopulation2017,ZelenskyElectromagneticallyInduced2002,Jia-HuaElectromagneticallyInduced2004}. 
However, it is important to note that this idealized scenario is rather exceptional. In this Chapter, we do not consider $\beta=0$ but instead focus on $\beta\rightarrow 0^+$, where the antisymmetric state is only partially disconnected from the dynamics, enabling the emergence of interesting collective phenomena. Furthermore, we also explore the more general case where $\beta \in (0,\pi/2]$, including the possibility of non-identical emitters ($\delta\neq 0$).

\subsection{Geometrical configuration of the dipoles in free space}
\label{sec:geometrical_conf}
The collective parameters, $J$ and $\gamma_{12}$, defined in~\cref{eq:J,eq:gamma_12},  exhibit a strong dependency on the geometrical configuration of the emitters, particularly through the inter-emitter distance ($r_{12}$) and the orientation of the dipoles relative to the interatomic axis ($\bm{\bar \mu}\cdot \mathbf{\bar r}_{12}$). Consequently, variations in the geometrical configuration may imprint non-negligible effects on physical observables, such as the fluorescence spectrum or the two-photon correlation function. This is particularly evident when varying the ratio $J/\delta$, which in our parametrization, corresponds to changes in the inter-emitter distance via \eqref{eq:J}, as will be shown later in this Chapter.
\begin{SCfigure}[1][h!]
	\includegraphics[width=.65\linewidth]{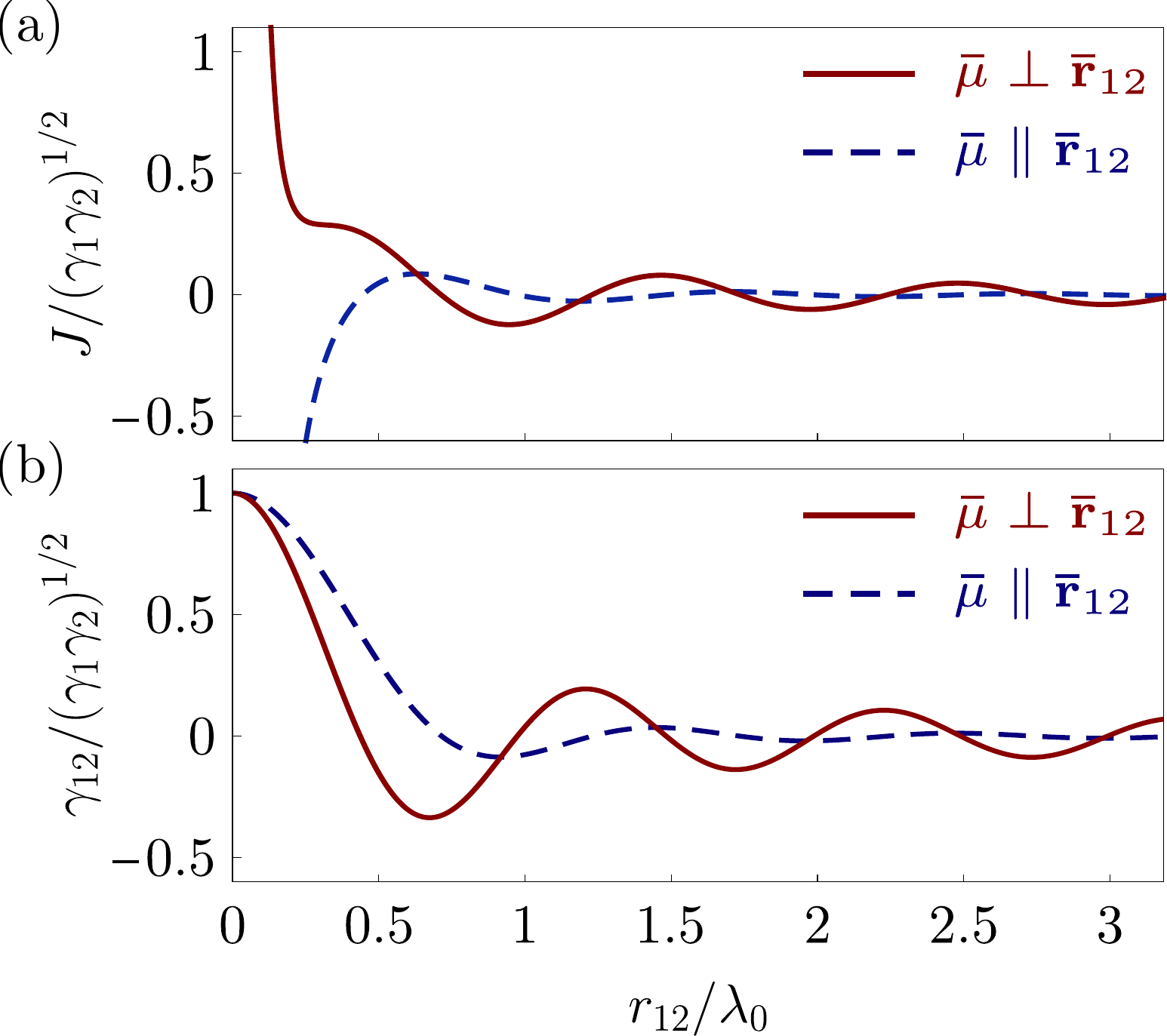}
	\captionsetup{justification=justified}
	\caption[Geometrical configuration of the dipoles in free space]{
		\label{fig:GeometricalConfig}  
\textbf{Geometrical configuration of the dipoles in free space.} (a) Collective coherent ($J$)  and (b) dissipative ($\gamma_{12}$) coupling rates versus the normalized inter-emitter distance  $r_{12}/\lambda_0$ for two different configurations: H-configuration (red solid line), and J-configuration (blue dashed line).
	}
\end{SCfigure}

\begin{SCfigure}[1][h!]
	\includegraphics[width=.7\linewidth]{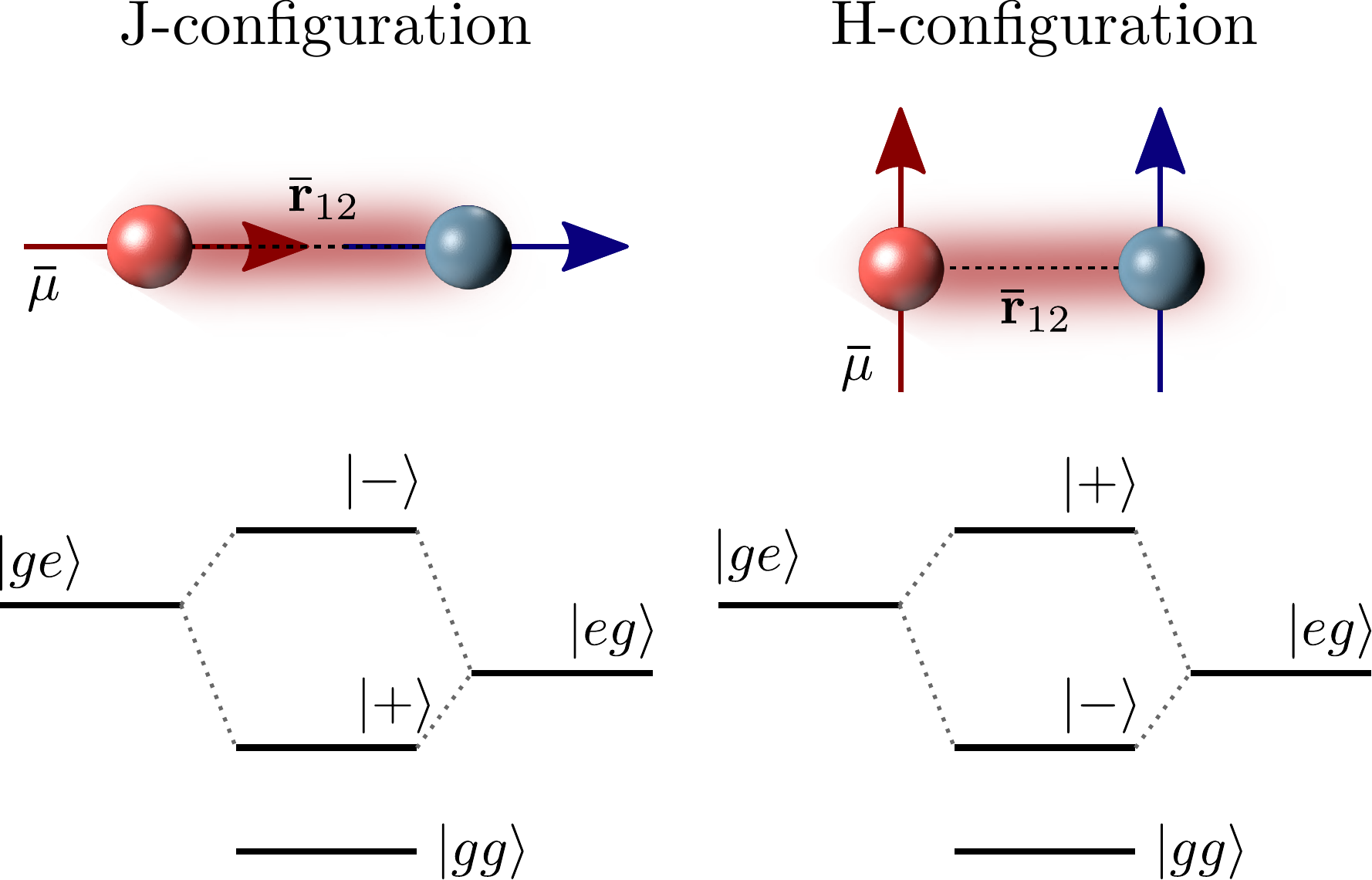}
	\captionsetup{justification=justified}
	\caption[Effect of the dye configuration in the eigenstructure]{
		\label{fig:DyeConf}  
		\textbf{Effect of the dye configuration in the eigenstructure.} 
		Energy level diagrams for J- and H-configurations at subwavelength inter-emitter distances ($r_{12}< \lambda_0$). In the J-configuration ($J<0$), the symmetric state $|+\rangle$ has a lower energy than the antisymmetric state $|-\rangle$. Conversely in the H-configuration ($J>0$), where $|+\rangle$ has a higher value than $|-\rangle$, as shown in \reffig{fig:Fig_DiagonalBasis_MixingAngle}~(a).
	}
\end{SCfigure}

In \reffig{fig:GeometricalConfig}, the geometrical dependency of the collective parameters is illustrated in terms of the normalized inter-emitter distance, $r_{12}/\lambda_0$, for two specific dipole orientations: the J-aggregate ($\bar{\bm \mu} \parallel \mathbf{\bar r}_{12}$) and H-aggregate ($\bar{\bm \mu} \perp \mathbf{\bar r}_{12}$) configurations, following the nomenclature commonly used for \textit{dyes}~\cite{HestandExpandedTheory2018}. These configurations are further detailed in \reffig{fig:DyeConf}.
The collective parameters reach their maximum values at separations much smaller than the resonant wavelength ($r_{12} \ll \lambda_0$), a regime known as \textit{the small sample model} or \textit{Dicke model}. These parameters, $J$ and $\gamma_{12}$, become less significant as the separation increases beyond half the resonant wavelength ($r_{12} \gtrsim \lambda_0/2$) and vanish at very large separations ($r_{12} \gg \lambda_0$).
 In the latter case of large separations, the system reduces to a master equation similar to \eqref{eq:MasterEqExample1}, generalized for a collection of \textit{independent emitters}.

We note that the orientation of the dipoles with respect with the interatomic distance introduces slight modifications in the values of the collective parameters in regions where $r_{12}\gtrsim \lambda_0$, as can be observed in the red solid ($\bar{\bm \mu} \perp \mathbf{\bar r}_{12}$) and blue dashed ($\bar{\bm \mu} \parallel \mathbf{\bar r}_{12}$) lines in \reffig{fig:GeometricalConfig}. 
However, for subwavelength distances, $r_{12}<\lambda_0$, while the collective dissipative coupling rate $\gamma_{12}$ does not exhibit significant changes, the magnitude and the sign of the dipole-dipole coupling are strongly influenced by the geometrical configuration [see \reffig{fig:GeometricalConfig}~\textcolor{Maroon}{(a)}]:
in the J-aggregate the dipole-dipole coupling becomes negative and in the H-aggregates positive.
This distinction affects the eigenstructure of the interacting quantum emitters [see \cref{eq:PlusEigenstate,eq:MinusEigenstate}], specifically, by interchanging the positions of the $|\pm \rangle$ eigenstates~\cite{HestandExpandedTheory2018}. The negative value of $J$ in in the J-configuration lead to a  scenario where the symmetric state has an eigenenergy $E_+=-R$ while the antisymmetric $E_-=+R$, and in a H-configuration the ordering is reversed [see \reffig{fig:DyeConf}]. 

Despite these differences, such as the interchange of locations of the one-excitation eigenstates or the sign of the incoherent coupling $\gamma_C$ via the $\beta$ parameter in \eqref{eq:ExcitonicParameters2}, the super- and subradiant nature of the symmetric and antisymmetric states remains unaffected by the dye configuration~\cite{ReitzCooperativeQuantum2022}.
In fact, we will show in \refsec{sec:2_steady_state} through analytical and numerical simulation of steady-state observables of the emitted light by the emitters, how the super- and subradiant effects are interchanged with respect the qubit-laser detuning $\Delta$.
%
%
Therefore, for the purpose of this Chapter, the choice between J- and H-orientations has no significant implications. 
%
%


Mathematically, we can summarize this discussion simply as:
\begin{enumerate}[label=\textcolor{Maroon}{(\roman*)}]
		\item \textcolor{Maroon}{Dicke model} (Small separations , $r_{1 2}\ll \lambda_0$) :
	\begin{equation}
		\colorboxed{Maroon}{\gamma_{12} \approx \sqrt{\gamma_1 \gamma_2} \quad \text{and} 	\quad 	J \approx \frac{3 \sqrt{\gamma_1 \gamma_2}}{4(k_0 r_{1 2})^3} [1-3(\bar{\bm \mu}\cdot \mathbf{\bar r}_{1 2})^2]. }
		\label{eq:DickeLimit}
	\end{equation}
	\item  \textcolor{Maroon}{Independent emitters} (Large separations, $r_{12}\gg \lambda_0$):
	\begin{equation}
	\colorboxed{Maroon}{	\gamma_{12} \approx 0 \quad \quad \quad \ \text{and} 	\quad 	J \approx 0. }
		\label{eq:IndependentEmittersLimit}
	\end{equation}
\end{enumerate}
The effective expression for $J$ in \eqref{eq:DickeLimit} provides a good approximation even for values $kr_{12}\sim 1$, as can be observed in the dashed lines in \reffig{fig:Fig_DiagonalBasis_MixingAngle}~\textcolor{Maroon}{(b)}.

Throughout this Thesis we assume an H-aggregate configuration, in which the dipole moments are perpendicular to the line that connects them, i.e.,  $\bar{\bm \mu} \perp \mathbf{\bar r}_{12}$. As we mentioned before, this choice is made out of convenience without loss of generality as the cooperative effects emerged at regions where the geometrical configuration might be relevant are not essentially affected~\cite{ReitzCooperativeQuantum2022}.  We also assume emitters with similar spontaneous decay rate and similar driving strengths for each emitter, as stated in \eqref{eq:DipoleAssumptions}. Consequently, we let the natural frequencies $\omega_i$ to be the parameter that actually determines whether the emitters are identical or not.
\subsection{Effective models}
\label{sec:effective_models}
\begin{SCfigure}[1][h!]
	\includegraphics[width=.96\columnwidth]{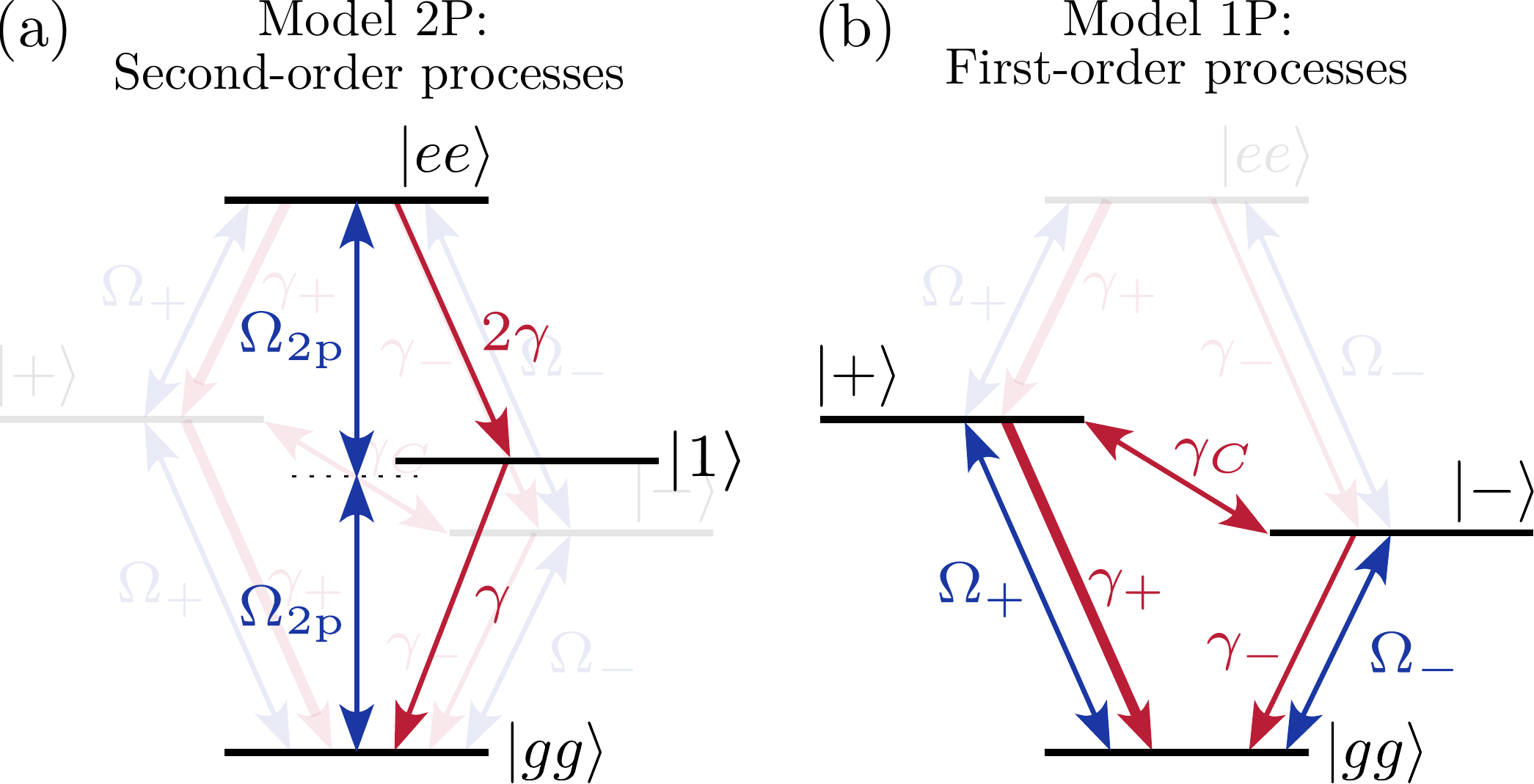}
	\captionsetup{justification=justified}
	\caption[Effective models describing 1- and 2-photon processes.]{\label{fig:EffectiveModels}	\textbf{Effective models describing 1- and 2-photon processes. } Diagrams of the two effective models used in this Chapter for describing (a) two-photon processes when the system is driven at the two-photon resonance $\Delta=0$, and (b) one-photon processes when the drive is resonant with $\Delta=\pm R$.}
\end{SCfigure}

The master equation in \eqref{eq:LMEemitters} does not yield analytical expressions in the general case of $\delta \neq 0$, which explains why the majority of literature has focused on the case $\delta=0$~\cite{RichterPowerBroadening1982,FicekEffectInteratomic1983,FicekEntangledStates2002,FicekQuantumInterference2005}. To obtain analytical expressions for the system density matrix, we will make the assumption that the dynamics is governed by two different types of processes that take place independently and that can be described by two independent effective models: 
\begin{enumerate}[label=\textcolor{Maroon}{(\roman*)}]
	\item \textcolor{Maroon}{Model 2P: Second-order processes. }The first process is the resonant, two-photon excitation that drives the $|gg\rangle \rightarrow |ee\rangle$ transition via a second-order process, and the subsequent incoherent decay towards $|gg\rangle$, passing through the single-excitation subspace.  This is described by an effective three-level cascade model, that we denote Model 2P, see \reffig{fig:EffectiveModels}~\textcolor{Maroon}{(a)}.
	\item  \textcolor{Maroon}{Model 1P: First-order processes. } The second mechanism is the excitation of the single-excitation subspace $\{|+\rangle, |-\rangle \}$ via first-order processes. This is described by a three-level Vee model that we denote Model 1P, see \reffig{fig:EffectiveModels}~\textcolor{Maroon}{(b)}, that excludes the doubly excited state $|ee\rangle$. 
\end{enumerate}

Our key assumption will be to consider that either both types of processes take place independently and very scarcely, or that the dynamics is completely dominated by one of the processes (e.g., at the two-photon resonance condition $\Delta=0$ for the second-order processes, or the one-photon resonance conditions $\Delta = \pm R$ for the first-order). In both cases, this means that the probabilities of occupation of excited states and coherences computed from each of these models contribute additively to the total density matrix. Adopting the notation $\rho_{i j}\equiv \langle i|\hat\rho|j\rangle$, we express this as
\begin{equation}
\colorboxed{Maroon}{
	\rho_{ij} \approx  \rho_{ij}^{(1)}+\rho_{ij}^{(2)},
}
	\label{eq:rho_relations}
\end{equation}
where $\rho_{ij}^{(2)}$  and $\rho_{ij}^{(1)}$ denote matrix elements computed from second-order processes (Model 2P) and first-order processes (Model 1P), respectively. In the following, we detail how these matrix elements are computed from each of these two approximated models. 
%


\paragraph{Model 2P: second-order processes. }
Model $2$P describes the nonlinear process of coherent, two-photon excitation by the driving field,  predicting sizeable probabilities for the excited states only around the two-photon resonance ($\Delta\approx 0$). 
For values of $\Delta$ close to zero, $|gg\rangle$ and $|ee\rangle$ form a \textit{quasi-degenerate subspace} in the rotating frame of the laser,  split from the states $|+\rangle$ and $|-\rangle$ by an energy difference $\pm R$---with $R\gg \Delta$ since $R$ is assumed to be the largest energy scale in our system---. 
The states $|gg\rangle$ and $|ee\rangle$ are not directly connect to first order, $\langle ee|\hat H|gg\rangle=0$, but they are coupled through second-order processes mediated by $|+\rangle$ and $|-\rangle$, as both of these states interact with $|gg\rangle$ and $|ee\rangle$ via the driving field [see \reffig{fig:TwoPhotonRabiOsc}~\textcolor{Maroon}{(a)}]. 

\begin{SCfigure}[1][h!]
	\includegraphics[width=1.\columnwidth]{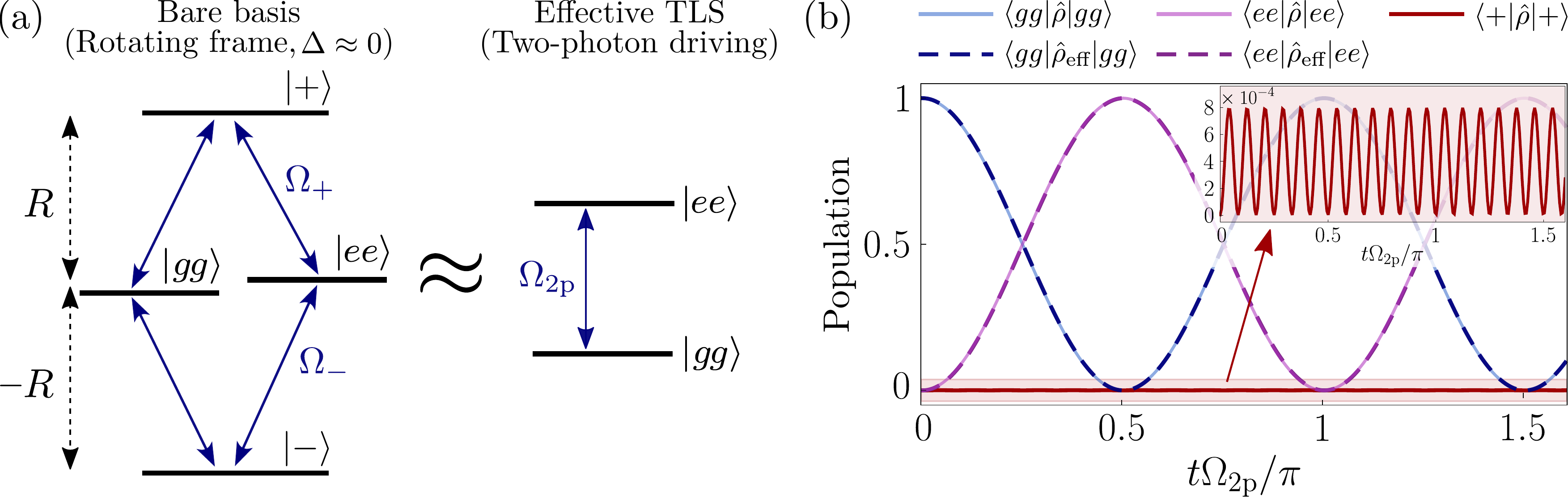}
	\captionsetup{justification=justified}
\end{SCfigure}
\graffito{\vspace{-4.7cm}
	\captionof{figure}[Two-photon Rabi oscillations.]{\label{fig:TwoPhotonRabiOsc} \textbf{Two-photon Rabi oscillations. }  In the absence of dissipation $\gamma_{ij}=0$ and driving at the two-photon resonance $\Delta=0$, the system features effective Rabi oscillations between $\langle gg| \hat \rho |gg\rangle$ and $\langle ee| \hat \rho |ee\rangle$ (a),  mediated by the single-excitation states $|\pm \rangle$ (acting as virtual states), which presents vanishingly small population $\langle +| \hat \rho |+\rangle$ (b). The actual dynamics of the two emitters (solid lines) can be effectively explained by a TLS that is resonantly driven with a driving strength $\Omega_{2\text{p}}$ (dashed lines).
	}
}

We will assume $\Omega \ll R$, allowing us to describe the effective two-photon coupling rate $\Omega_{2\mathrm p}$ between $|gg\rangle$ and $|ee\rangle$ by second-order perturbation theory, with the driving Hamiltonian $\hat H_\mathrm{d}$ from \eqref{eq:Hd} as the perturbation,
\begin{equation}
	\Omega_{2\mathrm p} \equiv -	\sum_{i=+,-}\frac{\langle ee|\hat H_\mathrm{d}|i\rangle\langle i|\hat H_\mathrm{d}|gg\rangle}{E_i} = -\frac{2\Omega^2}{R}\cos\beta.
	\label{eq:Omega_2p}
\end{equation} 
Additionally, the states $|ee\rangle$ and $|gg\rangle$ experience an effective Lamb shift, $\lambda$, which, notably, is the same for both of them and equal to
\begin{equation}
	\lambda = \lambda_j = -	\sum_{i=+,-}\frac{\langle j|\hat H_\mathrm{d}|i\rangle\langle i|\hat H_\mathrm{d}|j\rangle}{E_i}= \Omega_{2\mathrm p }, \quad \text{with}\quad j=gg,ee.
	\label{eq:lamb-shift}
\end{equation} 
The parameter $\Omega_{2\mathrm p }$ can be interpreted as a \textit{two-photon Rabi frequency}, characterized by a quadratic scaling with $\Omega$ and a strong dependence on $\beta$,  or equivalently, the ratio $J/\delta$. 
In the absence of dissipation ($\gamma, \gamma_{12}=0$) and under this particular driving regime ($\Delta=0$ and $\Omega\ll R$), the system exhibits complete Rabi oscillations between the ground and doubly-excited states, as shown in \reffig{fig:TwoPhotonRabiOsc}~\textcolor{Maroon}{(b)}. In this regime, we can also note that the symmetric $|+ \rangle$  and antisymmetric states $|-\rangle$ play the role of a virtual state~\cite{Cohen-TannoudjiAtomPhotonInteractions1998}: a state very detuned from the subspace where
the dynamics is taking place and weakly coupled to it, so
that its only effect is to provide effective energy shifts and
interactions, without even becoming populated [see inset of \reffig{fig:TwoPhotonRabiOsc}~\textcolor{Maroon}{(b)}].

A direct consequence of \eqref{eq:Omega_2p} is that, when $J=0$, there is no two-photon coupling between $|gg\rangle$ and $|ee\rangle$ due to destructive interference between the two possible pathways  mediating the interaction. This reflects the intuitive fact that, for $J=0$, the emitters are decoupled and behave as completely independent entities.
This observation explains why the emergence of optical features related to the two-photon process serves as a clear evidence of coherent coupling between quantum emitters~\cite{VaradaTwophotonResonance1992,HettichNanometerResolution2002,TrebbiaTailoringSuperradiant2022,LangeSuperradiantSubradiant2024}.

Beyond the coherent two-photon driving, the dynamics include incoherent decay from $|ee\rangle$ to $|gg\rangle$ passing through one of the single-excitation states, $|+\rangle$ or $|-\rangle$. Since this is an incoherent process that populates in an equal manner both single-excitation states, we simplify our description by considering an effective intermediate state $|1\rangle$, whose steady state population equals the combined populations of $|+\rangle$ and $|-\rangle$  [see \reffig{fig:EffectiveModels}~\textcolor{Maroon}{(a)}]. 
Since this effective state is only incoherently populated, its energy is \textit{irrelevant} for the effective description, and we set it to zero $E_{|1\rangle}=0$.

The state of the reduced subsystem $\{|gg\rangle, |ee\rangle, |1\rangle \}$ is described by a $3\times 3$ density matrix $\hat \chi^{2\mathrm p}$, whose dynamics is governed by the following master equation:
\begin{equation}
	\colorboxed{Maroon}{
		\frac{d \hat \chi^{2\mathrm p}}{dt} = -i \left[\hat H_{2\mathrm p},\hat \chi^{2\mathrm p} \right] 
		+\frac{2\gamma}{2} \mathcal{D}[| 1 \rangle \langle ee |]\hat \chi^{2\mathrm p}+ \frac{\gamma}{2} \mathcal{D}[|gg \rangle \langle 1 | ]\hat \chi^{2\mathrm p},
		}
	\label{eq:master_eq_model1}
\end{equation}
where $\hat H_{2\mathrm p}$ is the effective two-photon Hamiltonian 
\begin{multline}
	\hat H_{2\mathrm p}= (2\Delta +\Omega_{2\mathrm p})|ee\rangle\langle ee| + \Omega_{2\mathrm p}|gg\rangle\langle gg|\\ + \Omega_{2\mathrm p}(|ee\rangle\langle gg|+|gg\rangle\langle ee|).
	\label{eq:H2p}
\end{multline}
From this, we can obtain the second-order contributions to the excited-state components of the total $\hat\rho$ [see \eqref{eq:rho_relations}], establishing the following relations
\begin{subequations}
	\begin{align}
		&\rho^{(2)}_{ee,ee} \equiv \chi^{{2\mathrm p}}_{ee,ee}, \\
		&\rho^{(2)}_{+,+}  =\rho^{(2)}_{-,-}\equiv \chi^{{2\mathrm p}}_{1,1}/2, \\
		&\rho^{(2)}_{-,+} = 0.
	\end{align}
\end{subequations}
Solving for the steady-state of \eqref{eq:master_eq_model1} gives direct analytical expressions for the elements of $\hat\chi^{2\mathrm p}$, and therefore for $\hat\rho^{(2)}$. These read:
\begin{subequations}
	\begin{empheq}[box=\colorboxed{Maroon}]{align}
		\rho^{(2)}_{ee,ee} &= \frac{ \Omega^2_{2\mathrm p}}{\gamma^2 + 4\Delta^2+4\Omega^2_{2\mathrm p}},
		\label{eq:rho_ee_2PModel}
		\\
		\rho^{(2)}_{+,+} &= \rho^{(2)}_{-,-} = \rho^{(2)}_{ee,ee}.
		\label{eq:rho_SymAntisym_2PModel}
	\end{empheq}
\end{subequations}

\paragraph{Model 1P: first-order processes. }
Our second model describes the dynamics in which the single-excitation states $|+\rangle$ and $|-\rangle$ are directly excited by the driving field through a first-order process, occurring when $\Delta \approx \mp R$, respectively.
Essentially, Model 1P simplifies the system by excluding the state $|ee\rangle$ from the description, thereby neglecting the two-photon excitation mechanisms captured in Model 2P, and only allowing first-order processes to take place.
The resulting system is a three-level V-configuration comprising the basis states $\{|gg\rangle, |-\rangle, |+\rangle \}$. By incorporating now spontaneous emission in the basis of $|+\rangle$ and $|-\rangle$---as illustrated in \reffig{fig:ExcitonicScheme} and defined in \cref{eq:ExcitonicParameters1,eq:ExcitonicParameters2,eq:ExcitonicParameters3}---, we derive the following master equation for the $3\times 3$ density matrix $\hat\chi^{1\mathrm p}$ of Model 1P:
\begin{equation}
\colorboxed{Maroon}{
\begin{aligned}[b]
		\frac{d \hat\chi^{1\mathrm p}}{dt} =&
	-i \left[\hat H_{1\mathrm p},\hat\chi^{1\mathrm p}\right] +\sum_{i=+,-}\frac{\gamma_i}{2} \mathcal{D}[| g \rangle \langle i |] \hat\chi^{1\mathrm p} \\
	&+\frac{\gamma_{C}}{2} \left( 2{| g \rangle \langle - |} \hat\chi^{1\mathrm p} {{| + \rangle \langle g |}} - \left\{{{| + \rangle }} { \langle - |}, \hat\chi^{1\mathrm p}\right\}+\text{H.c.} \right),
\end{aligned}
}
	\label{eq:master_eq_model2}
\end{equation}
where $\hat H_{1\mathrm p}$ is the effective one-photon Hamiltonian 
\begin{multline}
	\hat H_{1\mathrm p} = (\Delta + R)|+\rangle \langle +|+(\Delta - R)|-\rangle \langle -|\\
	+\Omega_-(|-\rangle\langle gg| + \mathrm{H.c.})+\Omega_+(|+\rangle\langle gg| + \mathrm{H.c.}).
\end{multline}

Analogously to the model 2P, the master equation in \eqref{eq:master_eq_model2} also yields analytical solutions for the stationary state.
Hence, once we find the steady-state solution, we map the elements of the reduced density matrix $\hat\chi^{1\mathrm p}$ to the first-order contributions to the total $\hat\rho$, which we denote $\hat\rho^{(1)}$  [see \eqref{eq:rho_relations}]. For elements of the single-excitation subspace $\{|+\rangle, |-\rangle\}$, this  map is simply given by
\begin{equation}
	\rho^{(1)}_{i,j} = \chi^{1\mathrm p}_{i,j}, \quad \text{with}\quad  i,j=+,-.
\end{equation}
The resulting first-order contributions to the total steady-state density matrix are thus given by:
\begin{subequations}
	\begin{empheq}[box=\colorboxed{Maroon}]{align}
		\label{eq:rho_SymAntisym_1PModel}				
		&{\rho^{(1)}_{{+,+}/{-,-}}}\approx \frac{4 \Omega^2_\pm}{\gamma_{+/-}^2+4 (\Delta \pm R)^2+8 \Omega^2_\pm},
		\\
		&\rho_{+,-}^{(1)}=\frac{2 \Omega ^2 \sin \beta  \left(\Delta ^2-R^2-2\Omega ^2\right)}{\chi},
		\label{eq:rho_Coherences_1PModel}
	\end{empheq}
\end{subequations}
with 
\begin{multline}
	\chi\equiv2 \left[\gamma ^2 \Delta
	^2+\left(\Delta ^2+2 \Omega ^2\right)^2\right]+\gamma ^2
	R^2 \cos (2 \beta )\\
	+R^2 \left(\gamma ^2-4 \Delta ^2+8
	\Omega ^2\right)+2 R^4-4 \Delta  R \cos \beta 
	\left(\gamma ^2+4 \Omega ^2\right).
	\label{eq:chi}
\end{multline}
The expression provided in \eqref{eq:rho_SymAntisym_1PModel} is an approximation that we obtained by eliminating the antisymmetric state $|-\rangle$ from Model 1P when calculating $\rho_{+,+}^{(1)}$, and \textit{vice versa} for $\rho_{-,-}^{(1)}$ by eliminating $|+\rangle$. This effectively reduces the model to two-level system configurations, $\{|gg\rangle, |\pm\rangle\}$ when $\Delta=\mp R$, respectively, instead of the three-level system described in Model 1P. 
While this additional approximation is not strictly necessary to obtain analytical expressions from the master equation for the three-level system in \eqref{eq:master_eq_model2}, the full analytical expressions for
 $\rho_{+,+}^{(1)}$ and $\rho_{-,-}^{(1)}$  are long and cumbersome to write. Nevertheless,
 the compact and more tractable expressions derived in \cref{eq:rho_SymAntisym_1PModel} closely match the full analytical solutions using the full model 1P [not shown here], making them highly suitable for practical use and for extracting meaningful physical insights.

In principle, Model 1P ignores the excitation of the two-photon state $|ee\rangle$ via successive, first-order excitation processes. This simplification is justified since, around the region we are most interested in---the two-photon resonance $\Delta \approx 0$---, the contribution of these first-order processes to $\rho_{ee,ee}$ is  rather small compared to that of second-order processes. 
However, when one approaches the limit $\beta\approx \pi/2$, the two-photon population generated by second-order processes  tends to zero, $\rho_{ee,ee}^{(2)}\rightarrow 0$, as can be seen from \eqref{eq:rho_ee_2PModel}, and the small contribution from first-order processes may dominate and become relevant.
Even though the state $|ee\rangle$ is not included in model 1P, we can extract the first-order contribution $\rho^{(1)}_{ee,ee}$ in this regime. When $\beta  \approx \pi/2$,  the two quantum emitters are essentially decoupled, and the population $\rho_{ee,ee}$ stems from the simultaneous but independent excitation of both emitters. 
This yields factorizable correlations of the type $\langle\hat\sigma_1^\dagger \hat\sigma_2^\dagger\hat\sigma_1\hat\sigma_2\rangle =\langle\hat\sigma_1^\dagger\hat\sigma_1\rangle \langle\hat\sigma_2^\dagger\hat\sigma_2\rangle $.
Since $\rho_{ee,ee} = \langle\hat\sigma_1^\dagger \hat\sigma_2^\dagger\hat\sigma_1\hat\sigma_2\rangle$, this factorization allows us to use the expressions of the populations obtained from Model $1$P in \eqref{eq:rho_SymAntisym_1PModel}, $\langle\hat\sigma_i^\dagger\hat\sigma_i\rangle^{(1)}$, to derive the first-order contributions to the occupation 
$	\rho^{(1)}_{ee,ee}=\langle\hat\sigma_2^+\hat\sigma_2\rangle^{(1)}\langle\hat\sigma_1^\dagger\hat\sigma_1\rangle^{(1)},$
even if $|ee\rangle$ was not explicitly included in the model. For $\beta\approx \pi/2$, we have that $\langle\hat\sigma_2^\dagger\hat\sigma_2\rangle\langle\hat\sigma_1^\dagger\hat\sigma_1\rangle\approx \rho_{+,+}\rho_{-,-}$. Therefore, we will define
\begin{equation}
	\colorboxed{Maroon}{\rho_{ee,ee}^{(1)}\equiv \rho_{ee,ee}^{(1)}(\Delta \approx 0, \beta \approx \pi/2) = \rho_{+,+}^{(1)}\rho_{-,-}^{(1)}.}
	\label{eq:rho_ee_1PModel}			
\end{equation}
This expression is necessary to regularize the expected two-photon population $\rho_{ee,ee}$ in the limit of uncoupled emitters, $\beta=\pi/2$, and basically unimportant in any other case.

\paragraph{Reconstruction of the steady-state density matrix.} 
Following the scheme of \eqref{eq:rho_relations}, we can now combine the results provided by the effective models in this section and obtain an estimate of the total steady-state density matrix $\hat\rho$. The approximations used are expected to hold particularly well around the region of interest $\Delta\approx 0$, i.e., the two-photon resonance. In particular, the most relevant density matrix elements for subsequent calculations read:
\begin{subequations}
	\begin{align}
		\label{eq:rho_ee}
		\rho_{ee,ee} &\approx \rho_{ee,ee}^{(1)}+ \rho_{ee,ee}^{(2)},\\
		\label{eq:rho_SS}
		\rho_{+,+} &\approx \rho_{+,+}^{(1)}+ \rho_{+,+}^{(2)},\\
		\label{eq:rho_AA}
		\rho_{-,-} &\approx \rho_{-,-}^{(1)}+ \rho_{-,-}^{(2)},\\
		\label{eq:rho_AS}
		\rho_{+,-} &\approx \rho_{+,-}^{(1)}.
	\end{align}
\end{subequations} 
These density matrix elements will allow us to obtain physical insights on the steady-state observables of the fluorescent light emitted by the system, such as the mean intensity of the signal or the second-order correlation function. This will be extensively discussed in the next section. 

\newpage

\section{Steady state observables of the emitted light}
\label{sec:2_steady_state}

\subsection{Geometry of the emitters-detector system} 
The exploration of steady-state observables of the emitted light by the composite system is done by establishing a connection between the radiated electric field operator and the annihilation operators of the quantum emitters, as we already explained in \refsec{sec:input_output_theory}. 
For instance, if we consider the emitters to be located at the origin of coordinates and to have the same dipole moment $\bm\mu=\bm\mu_1=\bm\mu_2$---as we do throughout this Chapter---, the positive frequency part of the far-field electric field operator radiated by the two quantum emitters is given by $\hat{\mathbf{E}}^{(+)}(\mathbf{r},t)= \hat{{E}}^{(+)}(\mathbf{r},t)\mathbf{u_x}$. Here, $\mathbf{u_x}$ is a unit vector perpendicular to $\mathbf r$ and contained within the plane spanned by $\bm \mu$ and $\mathbf r$~\cite{ScullyQuantumOptics1997}, such that 
\begin{equation}
	\hat{{E}}^{(+)}(\mathbf{r},t)=\frac{\omega_0^2}{4\pi\epsilon_0 c^2 |\mathbf{r}|^2}|\bm{\mu}\times\mathbf{r}|( \hat\sigma_1 +  \hat\sigma_2)(t-|\mathbf{r}|/c).
	\label{eq:electric_field_dipoles}
\end{equation}
Nevertheless, the relation
\begin{equation}
	\colorboxed{Maroon}{\hat{{E}}^{(+)}(\mathbf{r},t)\propto ( \hat\sigma_1 +  \hat\sigma_2)(t-|\mathbf{r}|/c),}
	\label{eq:E-propto-sigma}
\end{equation}
will hold provided the dipole moments are equal and the separation between the emitters is much smaller than the resonant wavelength of the field ($k r_{12}\ll 1$)~\cite{NovotnyPrinciplesNanoOptics2012}. 
This regime is of particular relevant for our work, since a thorough understanding of the system dynamics and quantum optical properties of the emission becomes essential for inferring intrinsic properties of the system, such as the inter-emitter distance, below the diffraction limit~\cite{HettichNanometerResolution2002}, which has significant applications for microscopy and superresolution imaging~\cite{SchwartzImprovedResolution2012}. Therefore, we will assume the proportionality relation in \eqref{eq:E-propto-sigma} throughout the Chapter. 
The geometry of the emitters-detector system is illustrated in \reffig{fig:GeometryDimerDetector}. More general detection schemes can be applied to study the emitted radiation, e.g., by placing  photo-detectors at different lengths and orientation around the emitters~\cite{SanchezMunozPhotonCorrelation2020,Dura-AzorinGeometricAntibunching2024,LednevSpatiallyResolved2024}.

\addtocounter{figure}{-1}
\begin{SCfigure}[1][b!]
	\includegraphics[width=.85\linewidth]{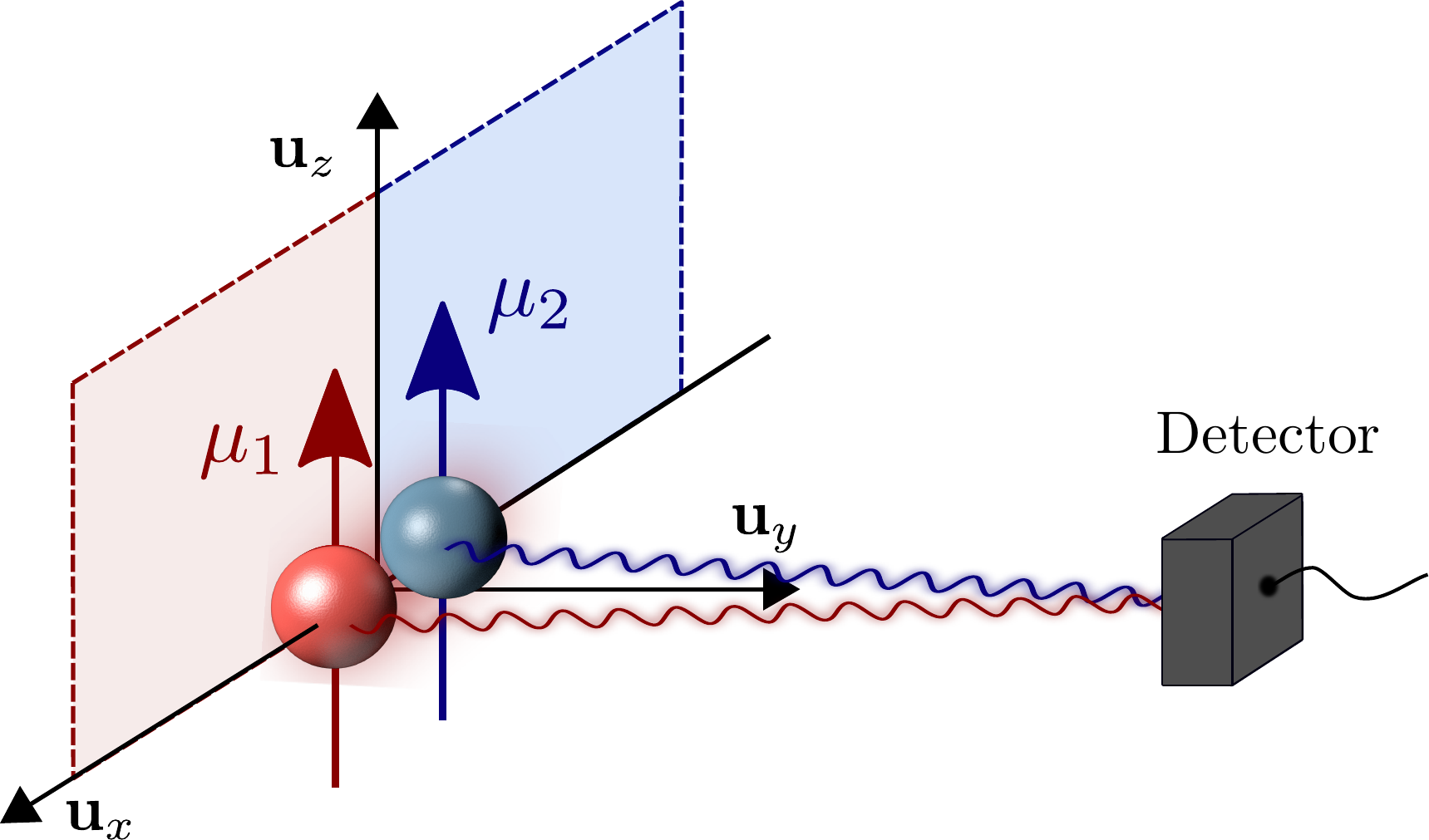}
	\captionsetup{justification=justified}
	\caption[Geometry of the emitters-detector system.]{
		\label{fig:GeometryDimerDetector}  
		\textbf{Geometry of the emitters-detector system.} 
		The emitters are positioned at the origin of the coordinate system, with both dipole moments oriented along the $z$-axis. In this configuration, a detector is placed at a distance $\mathbf{r}=|\mathbf{r}| \mathbf{u}_y$, where the detection signal is maximized
	}
\end{SCfigure}

The two main observables of the fluorescent emission that we will analyze here are mean intensity and second-order correlation function.

\subsection{Mean intensity}

\begin{SCfigure}[1][h!]
	\includegraphics[width=1.0\textwidth]{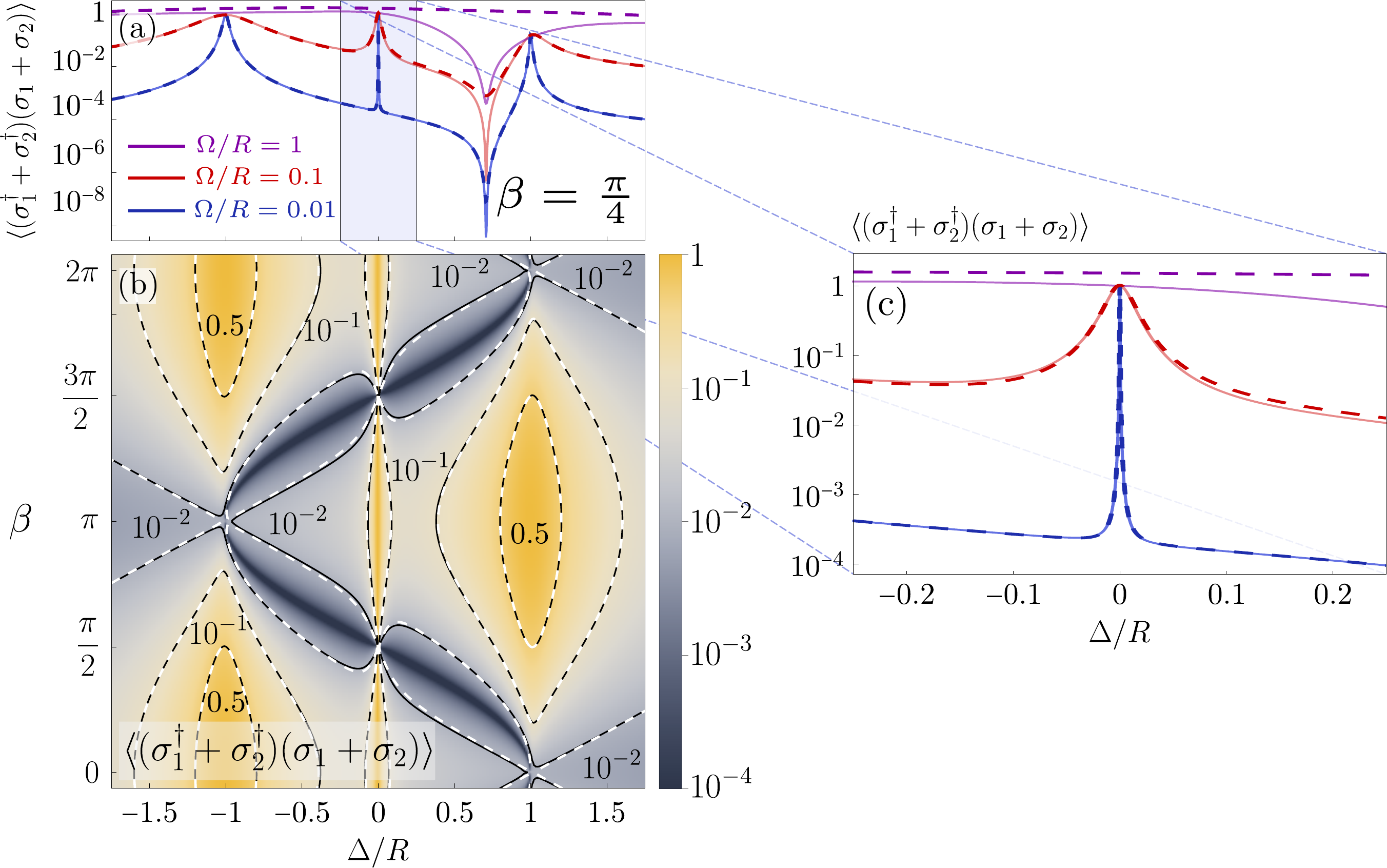}
	\captionsetup{justification=justified}
	\caption[Steady-state observables of the emitted light - Mean intensity. ]{\label{fig:EmissionProperties_MeanIntensity} \textbf{Steady-state observables of the emitted light - Mean intensity. } In all panels, dashed lines represent analytical results, solid lines are exact, numerical results.	(a-b) Intensity versus laser detuning (a) and both laser detuning and mixing angle (b). We observe a very good agreement between analytical and numerical results. Best agreement between analytical and numerical results is found around $\Delta\approx 0$ (c).
	}
\end{SCfigure}

\paragraph{Analytical results and analysis.}
A central observable of the light emitted by two interacting quantum emitters is the \textit{mean intensity of the signal}~\cite{LoudonQuantumTheory2000,ScullyQuantumOptics1997}. From the relation in \eqref{eq:E-propto-sigma}, we find that the mean intensity of the signal is given by ${\langle \hat{{E}}^{(-)}\hat{{E}}^{(+)}}\rangle \propto \langle \hat I\rangle$, where we have defined the intensity operator as
\begin{equation}
	\hat I  \equiv  (\hat\sigma_1^\dagger + \hat\sigma_2^\dagger)(\hat\sigma_1 + \hat\sigma_2).
	\label{eq:intensity_operator}
\end{equation}
We are interested in the steady-state mean values of the intensity operator, $I\equiv \langle \hat I\rangle_\text{ss}$. Then, we can express this quantity in terms of the density matrix elements
\begin{multline}
	I=	
	2\rho_{ee,ee}+\rho_{+,+}+\rho_{-,-}	\\
	+\cos \beta (\rho_{+,+}-\rho_{-,-}) 	+ 2 \sin \beta \mathrm{Re}\left[ \rho_{+,-} \right],
	\label{eq:intensity_matrix_elements}
\end{multline}
%
%
%
By combining this expression, \eqref{eq:intensity_matrix_elements}, with the analytical predictions from model $2$P in \cref{eq:rho_ee_2PModel,eq:rho_SymAntisym_2PModel}, and those from model $1$P in \cref{eq:rho_Coherences_1PModel,eq:rho_SymAntisym_1PModel} and \eqref{eq:rho_ee_1PModel}, along with the  density matrix reconstruction described in \cref{eq:rho_ee,eq:rho_SS,eq:rho_AA,eq:rho_AS}, we obtain a closed-form expression for $I$:
\begin{equation}
	\colorboxed{Maroon}{
		\begin{aligned}[b]
			I&=\frac{4\Omega_{2\mathrm p}^2}{\gamma^2+4\Delta^2+4\Omega_{2\mathrm p}^2}+\frac{4\Omega^2(R^2-\Delta^2+2\Omega^2)\sin^2 \beta}{\chi}\\
			& +\frac{4}{\xi_- \xi_+}[ 4\Omega_+^2\Omega_-^2 +\Omega_+^2(1+\cos \beta)\xi_- -\Omega_-^2(1-\cos \beta)\xi_+ ].
	\end{aligned}}
	\label{eq:SolutionIntensity}
\end{equation}
Here, $\chi$ is defined as in \eqref{eq:chi}, and, in order to ease the notation of $I$, we have identified $\xi_\pm$ with the denominators of \eqref{eq:rho_SymAntisym_1PModel}, that is,
\begin{equation}
	\xi_\pm\equiv \gamma_{+/-}^2+4 (\Delta \pm R)^2+8 \Omega^2_\pm.
\end{equation}
Our analytical results are shown in comparison with numerical results in \reffig{fig:EmissionProperties_MeanIntensity}, where panel \textcolor{Maroon}{(a)} depicts the intensity $I$ as a function of the qubit-laser detuning $\Delta$; panel \textcolor{Maroon}{(b)} displays $I$ versus both $\Delta$ and $\beta$, and panel \textcolor{Maroon}{(c)} offers a zoom of \textcolor{Maroon}{(a)} around the two-photon excitation regime, $\Delta\approx0$. 

There are three characteristic high-intensity peaks corresponding to the values $\Delta =\{-R,0,R\}$, as illustrated in  \reffig{fig:EmissionProperties_MeanIntensity}~\textcolor{Maroon}{(a)}. The two side-peaks at $\Delta=\pm R$ arise from direct, resonant excitation of the states $|\mp\rangle$, respectively. 
Focusing on the left-side peak at $\Delta=-R$,  the mean intensity exhibits a higher value and broader profile, reflecting the superradiant character of $|+\rangle$. Both the superradiant decay rate $\gamma_+=\gamma+ \gamma_{12}\cos \beta$ and the enhanced coherent excitation $\Omega_+=\Omega\sqrt{1+\cos\beta}$ amplifies the emission.
Conversely,  the right-side resonance at $\Delta=+R$ shows the opposite behaviour: a lower intensity and a narrower linewidth. This suppression stems from the subradiant nature of the antisymmetric state $|-\rangle$, featuring a reduced decay rate $\gamma_-=\gamma- \gamma_{12}\cos \beta$ and a suppressed coherent excitation $\Omega_-=\Omega\sqrt{1-\cos\beta}$. 
%
%
%

These two resonances at $\Delta=\pm R$ corresponds to first-order processes---according to our nomenclature in \refsec{sec:effective_models}---, and thus are accurately described by Model 1P. The effective expression for the intensity in \eqref{eq:intensity_matrix_elements} provides a good match with the exact, numerical calculation provided that the occupation probabilities are small enough for the simultaneous excitation probability of $\rho_{ee,ee}$ to be negligible.  
When $\beta=0$, i.e. the usually considered situation of resonant (identical) emitters, only the symmetric state $|+\rangle$ becomes significantly populated at $\Delta = -R$, while the resonance with the $|-\rangle$ state at $\Delta=R$ is suppressed. 
In contrast, at $\beta=\pi/2$, the emitters become uncoupled, allowing only classical correlations to form. 
In this situation, the mean intensity is symmetric around $\Delta=0$ and the two peaks at $\Delta=\pm R$ exhibit similar intensities, since it corresponds to the independent resonant driving of each quantum emitter.

Additionally, we also note that the peak intensity,  linewidth, and location strongly depend strongly on $\beta$, i.e., on the ratio between the dipole-dipole coupling and the relative detuning between the emitters.
As shown in \reffig{fig:EmissionProperties_MeanIntensity}~\textcolor{Maroon}{(b)},  varying the mixing angle from $0$ to $2\pi$ reveals how the underlying energy structure of the emitters imprint features in the emitted light.
Particularly, given our parametrization of the dipole-dipole interaction through vacuum in \eqref{eq:J}, this figure illustrates the effect of adopting a geometric configuration different from the H-aggregate configuration used throughout this Chapter. 
According to the definition of $\beta$ in \eqref{eq:betadefinition}, and the redefinitions of $J$ and $\delta$ in \eqref{eq:J_delta_beta}, once $\beta$ takes larger values than $\pi/2$, specifically  $\beta \in (\pi/2, 3\pi/2]$,  the coupling strength $J$ is taken to be negative, corresponding to a J-configuration [see \reffig{fig:GeometricalConfig}~\textcolor{Maroon}{(a)}].
As explained in \refsec{sec:geometrical_conf}, this choice interchanges the energy placement of the symmetric and antisymmetric states [see \reffig{fig:DyeConf}], thus explaining why the yellow (superradiant) region in \reffig{fig:EmissionProperties_MeanIntensity}\textcolor{Maroon}{(b)} is centred around $\Delta\approx -R$ for $\beta\in (0,\pi/2]$, shifts to $\Delta\approx +R$ for $\beta\in (\pi/2, 3\pi/2]$, and then returns to  $\Delta\approx -R$ for $\beta\in (3\pi/2, 2\pi]$.

The central peak at $\Delta=0$ is arguably the most relevant feature given its implications for technological applications like superresolution imaging~\cite{HettichNanometerResolution2002}. This peak emerges from the resonant two-photon excitation enabled by the coherent coupling between emitters, and thus is fully described by the contributions of second-order processes from Model 2P [see \cref{eq:rho_ee_2PModel,eq:rho_SymAntisym_2PModel}]. Our analytical solution allows us to establish that both the height and width of this peak scale as $\Omega^4 \cos^2\beta$. As expected, the peak vanishes for uncoupled emitters ($\beta=\pi/2 + k \cdot \pi  $ with $k\in \mathbb{Z}$) as a consequence of the destructive interference of the two-excitation pathways, leading to $\Omega_{2\mathrm p}=0$. 
From the analytical expression of $\rho_{ee,ee}$ in \eqref{eq:rho_ee_2PModel}, the intensity of the two-photon peak at $\Delta=0$ can be used to infer the value of $\beta$.
On the other hand, the Rabi frequency of the dipoles $R$ is readily identified from the positions of the one-photon peaks. The knowledge of these two magnitudes can then be combined to obtain information about the internal structure of the quantum emitters, specifically their natural energy detuning $2\delta$ and coherent coupling rate $J$. In turn, this information enables the estimation of quantities such as the inter-emitter distance $k r_{12}$.

To conclude our discussion of the mean intensity of the radiated field, we note that a destructive interference also emerges for values of $\beta\neq\pi/2$. This  can be observed in \reffig{fig:EmissionProperties_MeanIntensity}~\textcolor{Maroon}{(b)}, where a destructive interference dip emerges when the qubit-laser detuning is $\Delta=R\cos\beta$. 
However, unlike the particular case $\beta=\pi/2 + k \cdot \pi$ (with $k\in \mathbb{Z}$), this dip is not a destructive quantum interference between excitation pathways in the internal system dynamics; rather, it reflects an optical interference phenomenon in the radiated electric field consequence of our definition of the far-field electric field operator in \eqref{eq:electric_field_dipoles} [see \reffig{fig:GeometryDimerDetector}]. 
This interference phenomenon is captured by the cross terms appearing in \eqref{eq:intensity_matrix_elements}. 
In fact, using our Model 1P, we find that the dip occurs precisely where all the one-photon subspace contributions proportional to $\rho_{+,+}$, $\rho_{-,-}$ and $\rho_{+,-}$ in \eqref{eq:intensity_matrix_elements} cancel each other. As we will discuss below, the residual light in this regime retains a strongly two-photon character, despite the apparent destructive interference in the mean intensity.

\begin{SCfigure}[1][h!]
	\includegraphics[width=0.6\columnwidth]{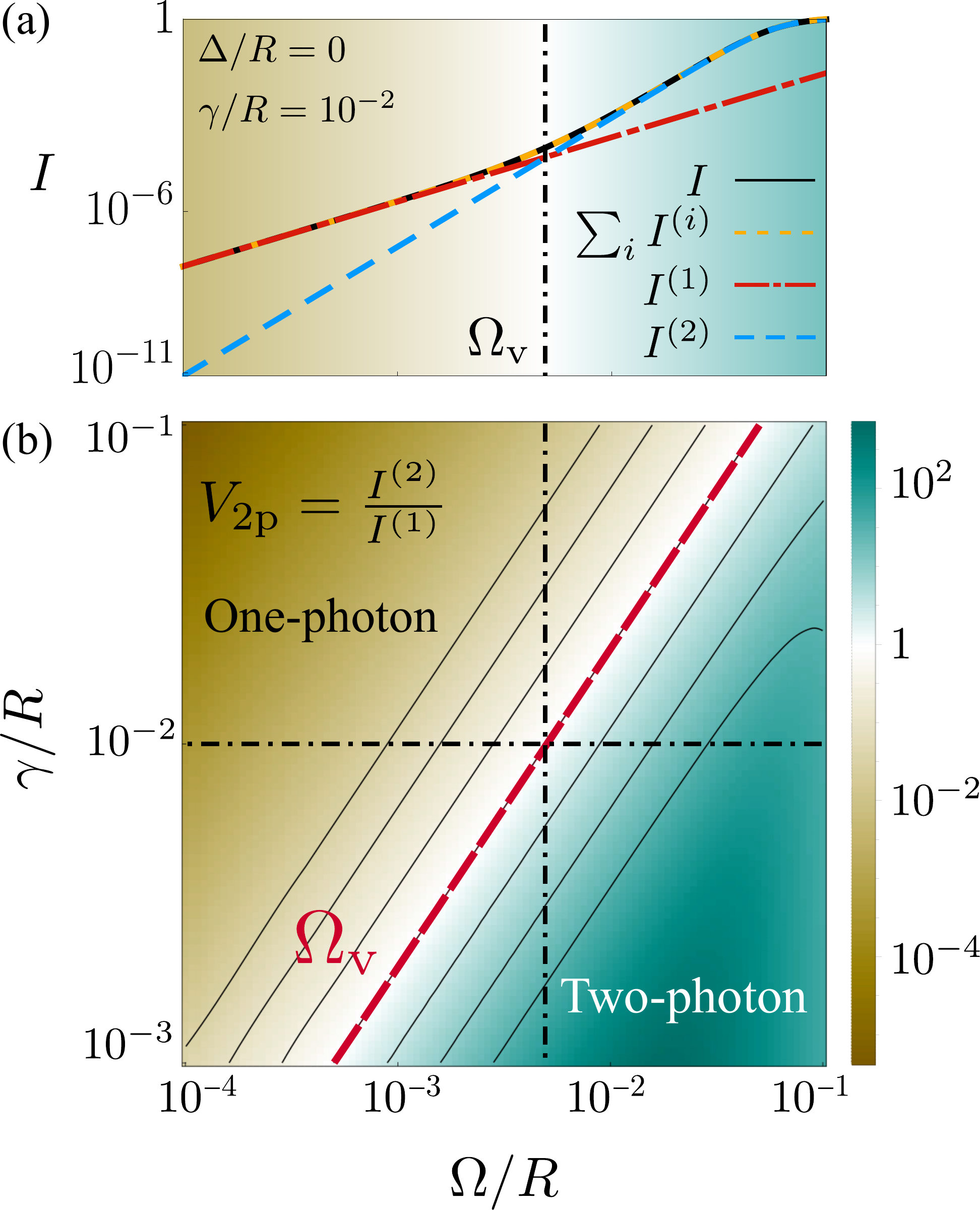}
	\captionsetup{justification=justified}
	\caption[Regimes of visibility of two-photon physics.]{\textbf{Regimes of visibility of two-photon physics.} (a) Intensity of emission $I$ computed numerically versus driving amplitude at the two-photon resonance $\Delta=0$, together with first- and second-order contributions, $I^{(1)}$ and $I^{(2)}$ obtained analytically. Their sum recovers the exact, numerical value of $I$. $\Omega_\mathrm{v}$ marks the two-photon visibility crossover where second-order contributions are larger than first-order ones, $I^{(2)}>I^{(1)}$. $\gamma/R=10^{-2}$. (b) Two photon visibility $V_{2\mathrm p}=I^{(2)}/I^{(1)}$ versus $\gamma$ and $\Omega$. Dashed-red line marks the analytical value of $\Omega_\mathrm{v}$ given by \eqref{eq:Omega_visibility}, matching perfectly the contour line $V_{2\mathrm p}=1$. Parameters: $\beta=\pi/4$, $\gamma_{12}=0.999\gamma$.}
	\label{fig:fig4-visibility}
\end{SCfigure}

\paragraph{Regimes of visibility of two-photon physics. }
Given its importance, it is desirable to determine the set of conditions under which the two-photon peak will be visible. Following our approach of separating contributions from first- and second-order processes, the total intensity can be written as $I = I^{(1)} + I^{(2)}$:
\begin{subequations}
	\begin{align}
		I^{(1)}=&\frac{4}{\xi_- \xi_+}[ 4\Omega_+^2\Omega_-^2 +\Omega_+^2(1+\cos \beta)\xi_- -\Omega_-^2(1-\cos \beta)\xi_+ ] \notag\\
		&\hspace{3.5cm} +\frac{4\Omega^2(R^2-\Delta^2+2\Omega^2)\sin^2 \beta}{\chi} .	%
		\\
		I^{(2)}=&\frac{4\Omega_{2\mathrm p}^2}{\gamma^2+4\Delta^2+4\Omega_{2\mathrm p}^2}.
	\end{align}
\end{subequations}
\newpage
The two-photon peak arises from the resonant contribution $I^{(2)}$, while the off-resonant, first-order contribution $I^{(1)}$ gives a featureless background at $\Delta=0$.
Under certain conditions, this background can be brighter and surpass the two-photon peak, effectively hiding it. In particular, since the second-order population $\rho_{ee,ee}^{(2)}$ responsible for the two-photon peak scales quartically with $\Omega$, while first-order processes yield populations  scaling quadratically with $\Omega$, there must exist a critical driving amplitude, $\Omega_{\mathrm v}$, below which first-order processes dominate [see \reffig{fig:fig4-visibility}~\textcolor{Maroon}{(a)}]. 
To determine $\Omega_\mathrm{v}$, we define a \textit{two-photon visibility}, $V_{2\mathrm p}$, as the ratio: 
\begin{equation}
	V_{2\mathrm p}\equiv \frac{I^{(2)}}{I^{(1)}},
	\label{eq:VisbilityFun}
 \end{equation}
such that the two-photon peak will be visible when $V_{2\mathrm p}>1$.  

The first- and second-order contributions of the intensity,
$I^{(1)}$ and $I^{(2)}$, can be computed by substituting the corresponding components of $\rho_{ee,ee}$, $\rho_{+,+}$, $\rho_{-,-}$ and $\rho_{+,-}$, from \cref{eq:rho_SymAntisym_1PModel,eq:rho_Coherences_1PModel,eq:rho_ee_1PModel} (Model 1P) and \cref{eq:rho_ee_2PModel,eq:rho_SymAntisym_2PModel} (Model $2$P), respectively.
Introducing these analytical results into the \textit{two-photon visibility condition} $V_{2\mathrm p}=1$ and solving it, we derive the minimum driving amplitude $\Omega_\mathrm v$ required to make two-photon effects observables.
Although the general expression for the two-photon visibility in \eqref{eq:VisbilityFun} becomes lengthy and cumbersome once the analytic expressions of the density matrix elements are included, it considerably simplifies under the regime considered in this Chapter ($\Omega, \gamma \ll R$), yielding
\begin{equation}
\colorboxed{Maroon}{V_{2\mathrm p}\approx \frac{8R^2 \Omega^2 \cos^2\beta}{\gamma^2[R^2-\Omega^2+(R^2+\Omega^2)\cos (2\beta)]}.}
	\label{eq:VisbilityFunApprox}
\end{equation}
Hence, by applying the two-photon visibility condition $	V_{2\mathrm p}=1$,  we obtain:
\begin{equation}
	\colorboxed{Maroon}{\Omega_\mathrm{v} \approx R \sqrt{\frac{2}{\tan ^2 \beta +8R^2/\gamma ^2}} \approx \frac{\gamma}{2},}
	\label{eq:Omega_visibility}
\end{equation}
where the last approximation applies provided $\gamma \ll R\tan\beta$. These results are summarized and confirmed by exact numerical calculations in \reffig{fig:fig4-visibility}. In panel~\textcolor{Maroon}{(a)}, we show that the sum of our analytical estimations of $I^{(1)}$ and $I^{(2)}$ recovers the exact value of $I$ computed numerically, which, as discussed above, features a transition from a $\propto \Omega^2$ to a $\propto \Omega^4$ scaling at $\Omega_\mathrm{v}$, which marks the onset of visibility of the two-photon peak, i.e., \textit{the emergence of features characteristic of the two-photon dynamics}. The two-photon visibility $V_{2\mathrm p}$ is shown in the full $(\gamma,\Omega)$ space in \reffig{fig:fig4-visibility}~\textcolor{Maroon}{(b)}, where the approximated expression for $\Omega_\mathrm{v}$ provided in \eqref{eq:Omega_visibility} is shown to match perfectly the condition $V_{2\mathrm p} = 1$.

\subsection{Second-order correlation function}

\begin{SCfigure}[1][h!]
	\includegraphics[width=1.0\textwidth]{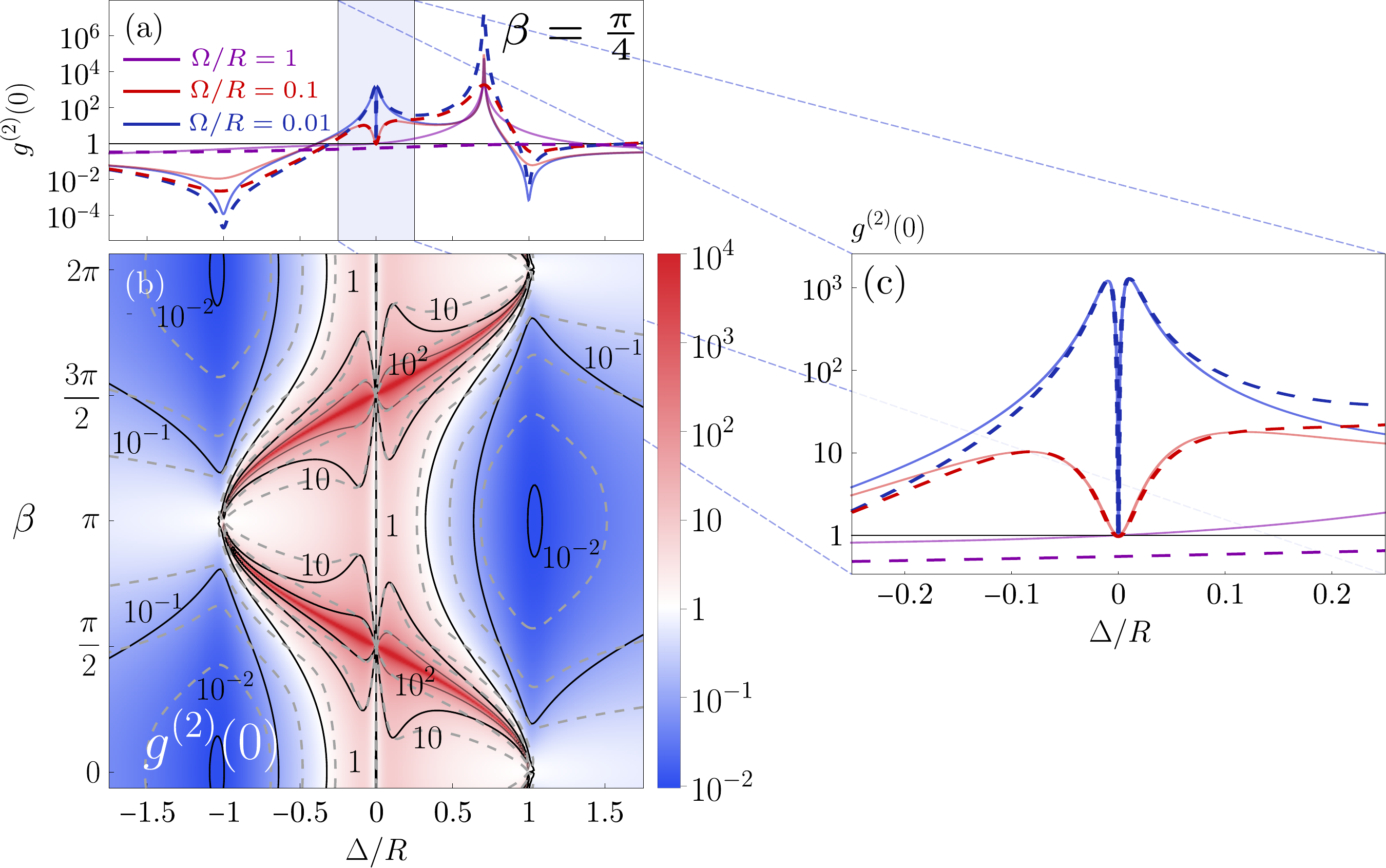}
	\captionsetup{justification=justified}
	\caption[Steady-state observables of the emitted light - Second-order correlation function. ]{\label{fig:EmissionProperties_g2} \textbf{Steady-state observables of the emitted light - Second-order correlation function. } In all panels, dashed lines represent analytical results, solid lines are exact, numerical results.(a-b) Second-order correlation function at zero delay versus laser detuning (a) and both laser detuning and mixing angle $\beta$ (b). Best agreement between analytical and numerical results is found around $\Delta\approx 0$ (c).
	}
\end{SCfigure}

\paragraph{Analytical results and analysis. }
Another central quantity in quantum optics to analyze the emitted light by a system is the \textit{two-photon correlation function}, particularly at zero-delay, defined as~\cite{GlauberQuantumTheory1963}
\begin{equation}
	g^{(2)}(0) =\frac{ {\langle {{\hat E}}^{(-)} {{\hat E}}^{(-)} {{\hat E}}^{(+)}}  {{\hat E}}^{(+)} \rangle}{\langle {{\hat E}}^{(-)} {{\hat E}}^{(+)}\rangle^2}.  
\end{equation}
Considering the relation in \eqref{eq:E-propto-sigma} and the mean intensity expression from \eqref{eq:intensity_matrix_elements}, it can be written as
\begin{equation}
\colorboxed{Maroon}{	g^{(2)}(0)= \frac{\langle ({\hat\sigma_1}^+ + {\hat\sigma_2}^+)^2 ({\hat\sigma_1} + {\hat\sigma_2})^2  \rangle}{\langle ({\hat\sigma_1}^+ + {\hat\sigma_2}^+)    ({\hat\sigma_1} + {\hat\sigma_2}) \rangle^2 } = \frac{4 \rho_{ee,ee}}{{I}^2}.}
	\label{eq:g2}
\end{equation}
This value quantifies the probability of detecting two photons simultaneously, normalized by the probability of doing so in a classical coherent field of similar intensity. As seen in \eqref{eq:g2}, in our case this is directly related to the probability of occupying the doubly excited state $|ee\rangle$. 
The analytical and numerical results are summarized in \reffig{fig:EmissionProperties_g2}, where, in analogy to \reffig{fig:EmissionProperties_MeanIntensity},  panel \textcolor{Maroon}{(a)} depicts the second-order correlation function in terms of the qubit-laser detuning $\Delta$; panel \textcolor{Maroon}{(b)} displays $g^{(2)}(0)$ versus both $\Delta$ and $\beta$, and panel \textcolor{Maroon}{(c)} offers a zoom of \textcolor{Maroon}{(a)} around the two-photon excitation regime, $\Delta\approx0$.

Our approximated analytical methods are able to describe accurately the population of the two-photon subspace---and therefore $g^{(2)}(0)$---around the two-photon resonance $\Delta\approx 0$, provided $\Omega \ll R$. This can be seen more clearly in the zoom around the two-photon resonance depicted in \reffig{fig:EmissionProperties_g2}~\textcolor{Maroon}{(c)}.
Away from the two-photon resonance, the population of the doubly excited state becomes much smaller and it is established by a more complicated mixture of one-photon and two-photon processes that our approximated models fail to capture. These small occupations of $|ee\rangle$, nevertheless, contribute very little to the actual intensity of radiation emitted, which away from $\Delta\approx 0$ is dominated by the one-photon subspace and thus is well described by our models, as could be seen in  \reffig{fig:EmissionProperties_MeanIntensity}~\textcolor{Maroon}{(b)}.
Interestingly, the maximum values of $g^{(2)}(0)$ are found at the dip of destructive interference discussed before, where the first-order contributions to the total emission interfere destructively and thus the small amount of remaining emission stems mainly from second-order processes, yielding a strong probability of detecting two photons.

At the two-photon resonance $\Delta=0$, the value of the $g^{(2)}(0)$ is sharply reduced and tends to 1 from above as $\Omega$ increases  [see \reffig{fig:EmissionProperties_g2}~\textcolor{Maroon}{(c)}]. 
The reason for this is that, as $\Omega$ increases, the light emitted is less coherent ($\langle \hat\sigma_i\rangle\rightarrow 0$), and thus the emission converges to that of two incoherent quantum emitters~\cite{SanchezMunozPhotonCorrelation2020}. The dip of $g^{(2)}$ precisely at $\Delta=0$ is a typical feature of multi-photon processes at resonance~\cite{SanchezMunozViolationClassical2014,LopezCarrenoPhotonCorrelations2017}, and it is simply a consequence of the increased intensity of emission.

\section[Two-photon resonance fluorescence spectrum]{Two-photon resonance fluorescence \protect \newline spectrum}
\label{sec:3-spectrum}
The fluorescence spectrum of the emitted radiation offers valuable insights into the energy transitions that can take place among the dressed states of the hybridized light-matter system. 
The most celebrated example of the revealing character of this type of measurement is the Mollow triplet spectrum in the emission from a two-level atom~\cite{MollowPowerSpectrum1969}---see \refsec{Chapt6_Section_Model} for detailed review of the Mollow triplet---. Its characteristic lineshape with three peaks provides key information about the structure of dressed energy levels~\cite{Cohen-TannoudjiAtomPhotonInteractions1998}, and serves as a source of strongly correlated non-classical light~\cite{Gonzalez-TudelaTwophotonSpectra2013,UlhaqCascadedSinglephoton2012,SanchezMunozViolationClassical2014,PeirisTwocolorPhoton2015,PeirisFransonInterference2017,LopezCarrenoPhotonCorrelations2017,LopezCarrenoEntanglementResonance2024,YangEntanglementPhotonic2025}. 

\begin{SCfigure}[1][h!]
	\includegraphics[width=0.9\textwidth]{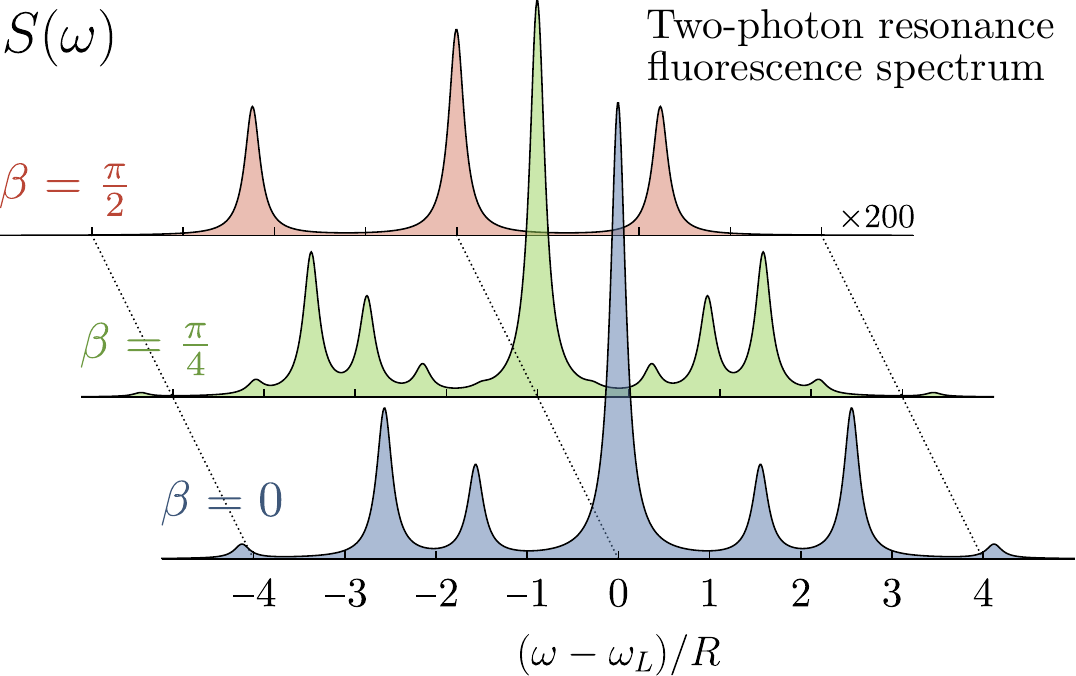}
	\captionsetup{justification=justified}
	\caption[Illustration of the two-photon resonance fluorescence spectrum. ]{\label{fig:Spectra_illustration}
		\textbf{Illustration of the two-photon resonance fluorescence spectrum.} The spectrum $S(\omega)$ is plotted for different values of the mixing angle $\beta= 0, \pi/4, \pi/2$, revealing the existence of a complex structure of sidebands that strongly depends on $\beta$.
		In the limit $\beta=\pi/2$, the spectrum simplifies to the three characteristic peaks of the Mollow triplet (scaled here by a factor of $200$ for better visibility).
		Parameters: $\Delta=0$, $\gamma=0.1R$, and $\Omega=R$.
	}
\end{SCfigure}

In general, the spectrum of emission is given by the Fourier transform of the two-time correlation function of the radiated electric field $\langle \hat E^{(-)} (t) \hat E^{(+)}(t+\tau) \rangle$---see \refsec{Section:SpectralProperties}---. Hence, using the relation in \eqref{eq:E-propto-sigma} between the radiated field and the raising/lowering operators of the quantum emitters, and disregarding global factors,  the resonance fluorescence spectrum can be expressed as %
\begin{equation}
	\colorboxed{Maroon}{
		\begin{aligned}[b]
			S(\omega) &=S(\omega; \hat\sigma_1^\dagger + \hat\sigma_2^\dagger, \hat\sigma_1 + \hat\sigma_2) \\
			& \, \lim_{t\rightarrow \infty}\frac{1}{\pi} \Re\int_0^\infty d\tau\, e^{i\omega \tau} \langle  (\hat\sigma_1^\dagger + \hat\sigma_2^\dagger) (t) (\hat\sigma_1 + \hat\sigma_2)(t+\tau) \rangle.
		\end{aligned}
	}
	\label{eq:spectrum_total}
\end{equation}
The emission spectra for different values of $\beta$ ($\beta=0,\pi/4, \pi/2$) are shown in \reffig{fig:Spectra_illustration}, for a laser detuning at  $\Delta=0$. These are examples of what we referred to as the \textit{two-photon resonance fluorescence spectra}.

The resonance fluorescence spectrum of two interacting quantum emitters can exhibit a richer structure than the Mollow triplet due to their collective behaviour. Such a multi-peaked spectrum has been reported for identical emitters~\cite{FicekEffectInteratomic1983,DarsheshdarPhotonphotonCorrelations2021}, as well as in the analogous scenario of a coherently driven three-level system at the two-photon resonance~\cite{GasparinettiTwophotonResonance2019}.
In the identical-emitter case ($\beta=0$), the result is a seven-peaked profile consisting of a central peak and six sidebands, as shown in blue in \reffig{fig:Spectra_illustration}.
Here, we reproduce that same spectrum for $\beta=0$ and extend the analysis to non-indentical emitters ($\beta>0$), where the fluorescence develops an even more intricate pattern, with up to 13 observable peaks [see green-shaded spectrum in \reffig{fig:Spectra_illustration}]. Ultimately, as $\beta$ approaches $\pi/2$---corresponding to the uncoupled emitters case---, the multi-peak spectrum converges to the three peaks characteristic of the Mollow triplet~\cite{MollowPowerSpectrum1969}, as seen in the red spectrum in \reffig{fig:Spectra_illustration}.

In the perturbative regime $\Omega \ll R$, the spectral peaks can be attributed to the hybridization of the quantum emitters with photon pairs, described by the dressed energy levels in Model 2P [see \refsec{sec:effective_models}].
By using the effective Hamiltonian from this model, we can analytically identify both the locations and physical origin of the spectral resonances in the perturbative regime ($\Omega\ll R$). Additionally, the same theoretical framework provides valuable insights into the more general case of $\Omega>R$, which we explore numerically to extend our understanding beyond strictly perturbative conditions.

\subsection{Perturbative regime}
\label{sec:perturbative_regime}
The position of all the peaks in the spectra can be obtained from the possible transitions among energy levels in the system Hamiltonian. 
Within the perturbative regime $\Omega \ll R$ and at $\Delta=0$, the ground $|gg\rangle$ and doubly-excited $|ee\rangle$ states are resonantly coupled through the two-photon excitation described by the Hamiltonian in \eqref{eq:H2p}. As a result, these two states hybridize, forming new symmetric $|S_2\rangle$  and antisymmetric $|A_2\rangle$ states, which we denote as \textit{two-photon dressed states}, 
\begin{equation}
	| S_2/A_2 \rangle \equiv \frac{1}{\sqrt{2}}\left( |gg \rangle \pm  | ee \rangle \right).
	\label{eq:two_photondressed_eig}
\end{equation} 
The resulting set of dressed eigenstates at $\Delta=0$ is $\{|+\rangle, |A_2\rangle, |S_2\rangle, |-\rangle \}$, ordered by decreasing energy, and their corresponding eigenenergies in the rotating frame of the laser are:
\begin{SCfigure}[1][b!]
	\includegraphics[width=1.\textwidth]{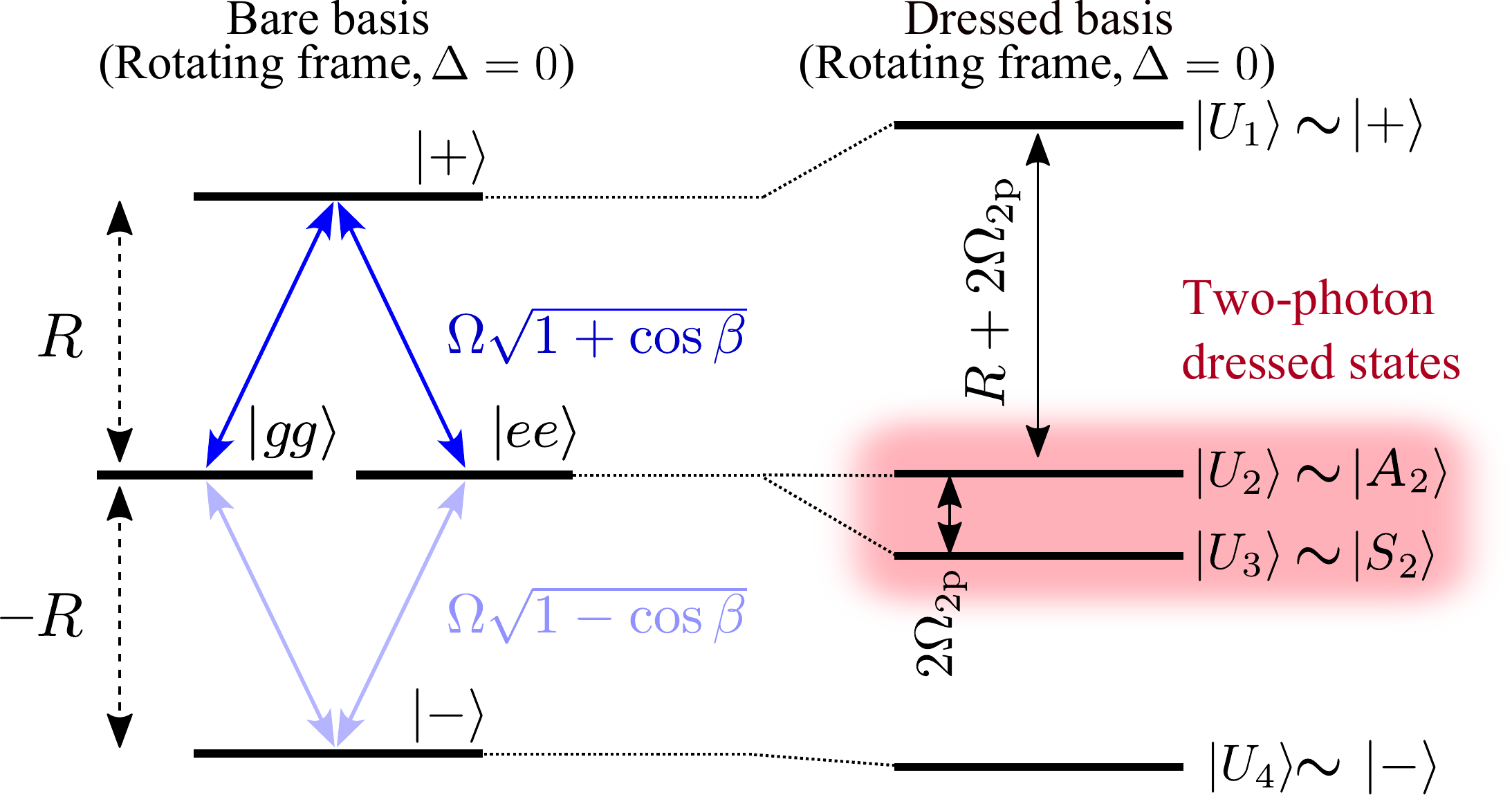}
	\captionsetup{justification=justified}
	\caption[Two-photon dressing.]{ 	\label{fig:TwoPhotonDressedStates} \textbf{Two-photon dressing.} Schemes of the two-photon dressing process: hybridization from the bare basis (at the two-photon resonance, $\Delta=0$) to the dressed basis in the limit $\beta\rightarrow0$ and $\Omega<R$. }
\end{SCfigure}
\begin{subequations}
	\label{eq_eigenenergies_twophoton}
	\begin{empheq}[box=\colorboxed{Maroon}]{align}
		&E_1 = E_{+} = R + 2(\cos\beta+1)\Omega^2/R,\\
		&E_2 = E_{A_2} = 0,\\
		&E_3 = E_{S_2} = -4\Omega^2 \cos\beta/R,\\
		&E_4 = E_{-} = -R+2(\cos\beta-1)\Omega^2/R.
	\end{empheq}
\end{subequations}
A schematic representation of this two-photon dressing mechanism is shown in \reffig{fig:TwoPhotonDressedStates}. 
In the same way that the eigenenergies of $A_{2}$ and $S_{2}$ take into account the Lamb shifts induced by the coupling to super- and subradiant states [see \eqref{eq:lamb-shift}], $E_{-}$ and $E_{+}$ also include the Lamb shifts of states $|+\rangle$ and $|-\rangle$ due to their interaction with $|ee\rangle$ and $|gg\rangle$,  given by
\begin{equation}
	\lambda_\pm = -\sum_{i=ee,gg}\frac{\langle \pm|\hat H_\mathrm{d}|i\rangle\langle i |\hat H_\mathrm{d}|\pm \rangle}{E_i-E^{(0)}_\pm}= 2(\cos\beta\pm 1)\frac{\Omega^2}{R},
\end{equation} 
where $E^{(0)}_\pm=\pm R$ are the eigenenergies of the single-excitation eigenstates in the absence of coherent driving [see \reffig{fig:Fig_DiagonalBasis_MixingAngle}].

The energy differences among these eigenvalues determine the transition frequencies that can observed as distinct peaks in the emission spectrum. 
Defining the transition energies
\begin{equation}
	\omega_{i\rightarrow j} \equiv E_i - E_j,
\end{equation}
the positions of the six positive-frequency sidebands with respect to the laser frequency can be written---in the perturbative regime---as:
\begin{subequations}
	\begin{empheq}[box=\colorboxed{Maroon}]{align}
		\label{eq:transition_omega_1}
		\omega_1 &= \omega_{1\rightarrow 4} = 2R+\frac{4\Omega^2}{R},\\
		\label{eq:transition_omega_2}
		\omega_2 &= \omega_{1\rightarrow 3} = R + \frac{2\Omega^2(3\cos\beta+1)}{R},\\
		\label{eq:transition_omega_3}
		\omega_3 &= \omega_{1\rightarrow 2} = R + \frac{2\Omega^2(\cos\beta+1)}{R},\\
		\label{eq:transition_omega_4}
		\omega_4 &= \omega_{2\rightarrow 4} = R - \frac{2\Omega^2(\cos\beta-1)}{R},\\
		\label{eq:transition_omega_5}
		\omega_5 &= \omega_{3\rightarrow 4} = R - \frac{2\Omega^2(3\cos\beta-1)}{R},\\
		\label{eq:transition_omega_6}
		\omega_6 &= \omega_{2\rightarrow 3} = \frac{4\Omega^2\cos\beta}{R}.
	\end{empheq}
\end{subequations}
The sidebands at negative frequencies stem from the reversed processes outline above, $\omega_{i\rightarrow j} = -\omega_{j\rightarrow i}$. On the other hand, the central peak at $\omega_0 = 0$ corresponds to transitions between similar states, $\omega_{i\rightarrow i}$. These relations provide the position of all the possible 13 peaks that can be observed in the fluorescence spectrum within the perturbative regime $\Omega \ll R$.

\paragraph{Visibility of the spectral sidebands. } Although the six positive sidebands described in \cref{eq:transition_omega_1,eq:transition_omega_2,eq:transition_omega_3,eq:transition_omega_4,eq:transition_omega_5,eq:transition_omega_6} characterize the two-photon resonance fluorescence spectrum, not all of them are observable for all values of $\beta$, as can be seen in \reffig{fig:Spectra_illustration}. 
For instance, in the case of coupled identical emitters $\beta=0$, only three positive sidebands are visible, yielding a total of seven peaks in agreement with previous reports~\cite{FicekEffectInteratomic1983,DarsheshdarPhotonphotonCorrelations2021}. 

\begin{SCfigure}[0.6][h!]
	\includegraphics[width=0.7\textwidth]{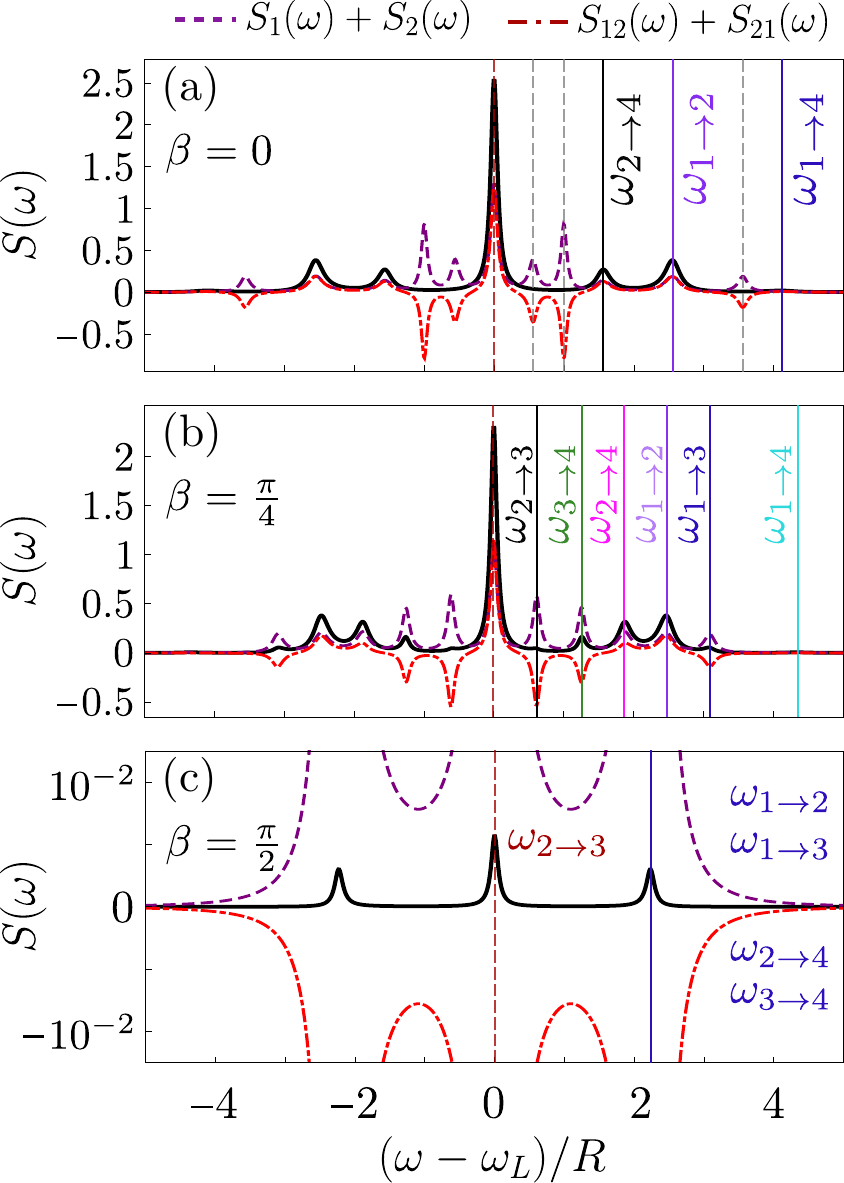}
	\captionsetup{justification=justified}
	\caption[Destructive interference in the emission spectra.]{ 	\label{fig:SpectrumInterf}
		\textbf{Destructive interference in the emission spectra.} Two-photon resonance fluorescence spectrum (black solid lines) for $\beta=0$ (a), $\beta=\pi/4$ (b), and $\beta=\pi/2$ (c).  These figures also show the contributions of the spectral expansion in \eqref{eq:spectrum_expansion} (purple dashed and red dot-dashed lines), proving the existence of peaks that are not visible due to destructive interference, but visible if only the emission from a single emitter is collected. Parameters: $\Delta=0$, $\gamma/R=0.1$ and $\Omega=R$ . }
\end{SCfigure}

In particular, the peaks that vanish are those involving transitions that start or end at the one-photon antisymmetric state $|-\rangle$. Concretely, in the perturbative regime, they correspond to the peaks $\omega_2$, $\omega_5$ and $\omega_6$ in~\cref{eq:transition_omega_1,eq:transition_omega_2,eq:transition_omega_3,eq:transition_omega_4,eq:transition_omega_5,eq:transition_omega_6}. 
The reason why these peaks are not visible stems from the destructive interference caused by the equal contributions of both emitters to the radiated electric field [see \eqref{eq:E-propto-sigma}], as illustrated in \reffig{fig:SpectrumInterf}~\textcolor{Maroon}{(a)}. 

We can gain more insights into this destructive interference phenomenon by expanding the total emission spectrum from \eqref{eq:spectrum_total}, such that
\begin{equation}
	S(\omega ) = S_1(\omega) + S_2(\omega) + S_{12}(\omega) + S_{21}(\omega),
	\label{eq:spectrum_expansion}
\end{equation}
where $S_{ij}(\omega) \equiv S(\omega; \hat\sigma_i^\dagger, \hat\sigma_j)$ and $S_i(\omega) \equiv S_{ii}(\omega)$. 
Here,  $S_1(\omega)$ and $S_2(\omega)$ describe the spectrum of emission obtained when detecting only the radiation from the first or second emitter, respectively, while $S_{12}(\omega)$ and $S_{21}(\omega)$ are interference terms arising from the superposition of both fields. 
In \reffig{fig:SpectrumInterf}~\textcolor{Maroon}{(a)}, we show that the missing peaks at $\beta=0$---all involving the state $|-\rangle$ and marked by vertical grey dashed lines---are indeed visible if $S_1(\omega)$ or $S_2(\omega)$ are measured independently (e.g., if their emission is collected locally). However, these peaks interfere destructively when the fields emitted by both QEs are superimposed, explaining the absence of these peaks in the total spectrum.
For $\beta>0$, this perfect destructive interference does not occur, and all the possible 13 peaks are visible in the  spectrum of emission [see  \reffig{fig:SpectrumInterf}~\textcolor{Maroon}{(b)}].
Finally, the opposite limit of completely decoupled non-identical emitters, $\beta=\pi/2$, yields a three-peaked structure which corresponds to the Mollow triplet of emission from the two independent emitters. In this limit, many of the transitions listed in~\cref{eq:transition_omega_1,eq:transition_omega_2,eq:transition_omega_3,eq:transition_omega_4,eq:transition_omega_5,eq:transition_omega_6} become degenerate, reducing the spectrum to only three peaks. 
The underlying reason for having fewer peak is not due to destructive interference, as in the case of $\beta=0$, but rather stems from the internal structure and available transition energies within the dressed states.

\paragraph{Emergence of two-photon sidebands.}
Away from the limit $\beta=\pi/2$, the frequencies in \cref{eq:transition_omega_1,eq:transition_omega_2,eq:transition_omega_3,eq:transition_omega_4,eq:transition_omega_5,eq:transition_omega_6}
describe transitions among dressed light-matter states, where the emitters are hybridized with photon pairs. 
This strong two-photon character is evidenced by the quadratic scaling of these frequencies with $\Omega$, in contrast to the linear scaling with $\Omega$ that one finds, e.g., for the position of the sidebands in the standard Mollow triplet~\cite{MollowPowerSpectrum1969}. 
For this reason, we refer to the sidebands described by~\cref{eq:transition_omega_1,eq:transition_omega_2,eq:transition_omega_3,eq:transition_omega_4,eq:transition_omega_5,eq:transition_omega_6} as \textit{two-photon sidebands}. 

\begin{SCfigure}[1][h!]
	\includegraphics[width=0.75\textwidth]{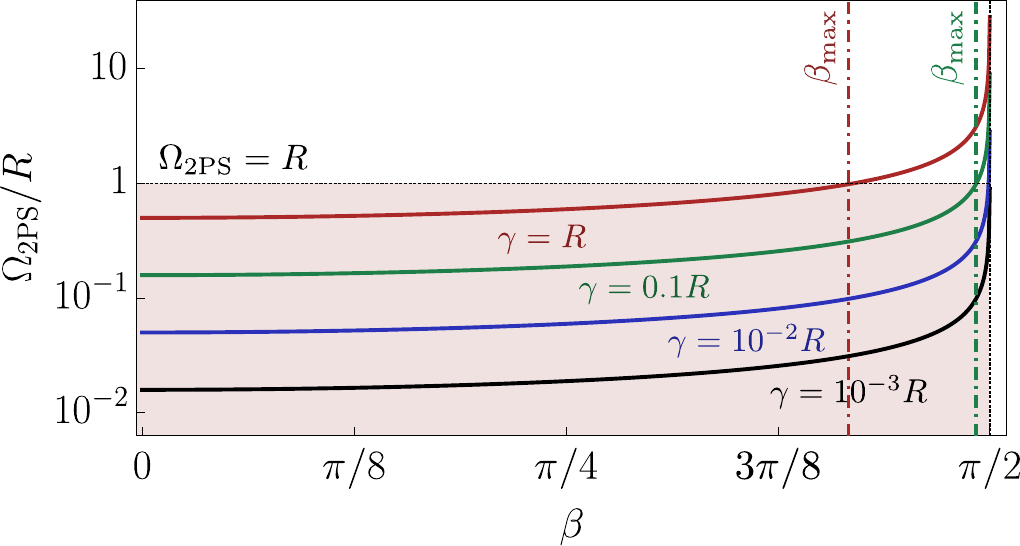}
	\captionsetup{justification=justified}
	\caption[Spectral resolution of the two-photon sidebands.]{ 	\label{fig:Omega2PS}
		\textbf{Spectral resolution of the two-photon sidebands.}
		Computation of the ratio $\Omega_{\text{2PS}}/R$ for different values of the spontaneous decay rate of the emitters $\gamma/R=10^{-3}, 10^{-2},10^{-1},1$ in terms of the mixing angle $\beta$.
		The red shaded area delimits the regime in which two-photon sidebands are well-resolved within the perturbative approach  $\Omega\ll R$. The vertical dot-dashed lines correspond to the critical mixing angle in which two-photon sidebands are no longer resolved.
	 }
\end{SCfigure}

In order to resolve two-photon sidebands, we need the energy separation
\begin{equation}
	\Delta E_{\text{2PS}} = \omega_2-\omega_3 = \omega_4-\omega_5 = 2\Omega_{2\mathrm p}
\end{equation}
to be larger than the decay rate of spontaneous emission, $\gamma$. The condition $2\Omega_{2\mathrm p}>\gamma$ defines the \textit{two-photon saturation amplitude}, $\Omega_{2\mathrm{PS}}$, which marks the onset of the regime where two-photon sidebands are resolvable:
\begin{equation}
\colorboxed{Maroon}{	\Omega_{2\mathrm{PS}} = \frac{1}{2}\sqrt{\frac{R\gamma}{\cos\beta}}.}
	\label{eq:Omega_2PS}
\end{equation}
The above condition indicates that the two-photon sidebands can always be developed within the perturbative regime, since
\begin{equation}
	\frac{\Omega_{2\text{PS}}}{R}= \frac{1}{2}\sqrt{\frac{\gamma}{R\cos\beta}} \ll 1,
\end{equation}
provided that the Rabi splitting is sufficiently large ($R\gg \gamma$) and $\beta$ is not too close to $\pi/2$ [see \reffig{fig:Omega2PS}]. As $\beta\rightarrow \pi/2$, higher values of $\Omega_{2\text{PS}}$ are required to resolved the two-photon sidebands. 
When values $\Omega_{2\mathrm{PS}}\sim R$ are reached, the perturbative approach used above breaks down:
sidebands developed purely from two-photon hybridization are no longer observable since first-order processes dominate before the former are visible. 
Setting the condition $\Omega_{2\mathrm{PS}}\approx R$, we find the critical mixing angle $\beta_{\max}$ at which two-photon sidebands are no longer visible:
\begin{equation}
	\beta_{\text{max}}\approx\arccos( \gamma/4R).
\end{equation}
For the typical parameters used in this Chapter, e.g.,  $\gamma = 10^{-3} R$, we find $\beta_\mathrm{max} \approx 0.9998 \pi/2$, implying that the two-photon sidebands remain visible for most of the range $\beta \in (0,\pi/2]$. In fact, as $\gamma/R\rightarrow 0$, $\beta_{\text{max}}$ approaches $\pi/2$. As we will see, the two-photon saturation amplitude $\Omega_{2\mathrm{PS}}$ plays a crucial role for metrological applications,  since the onset of the resolved two-photon sidebands regime marks the point of maximum sensitivity for optical estimations of the inter-emitter distance. 

\subsection{Strong driving regime}
\label{sec:StrongDriving}
\begin{SCfigure}[1][b!]
	\includegraphics[width=1.\textwidth]{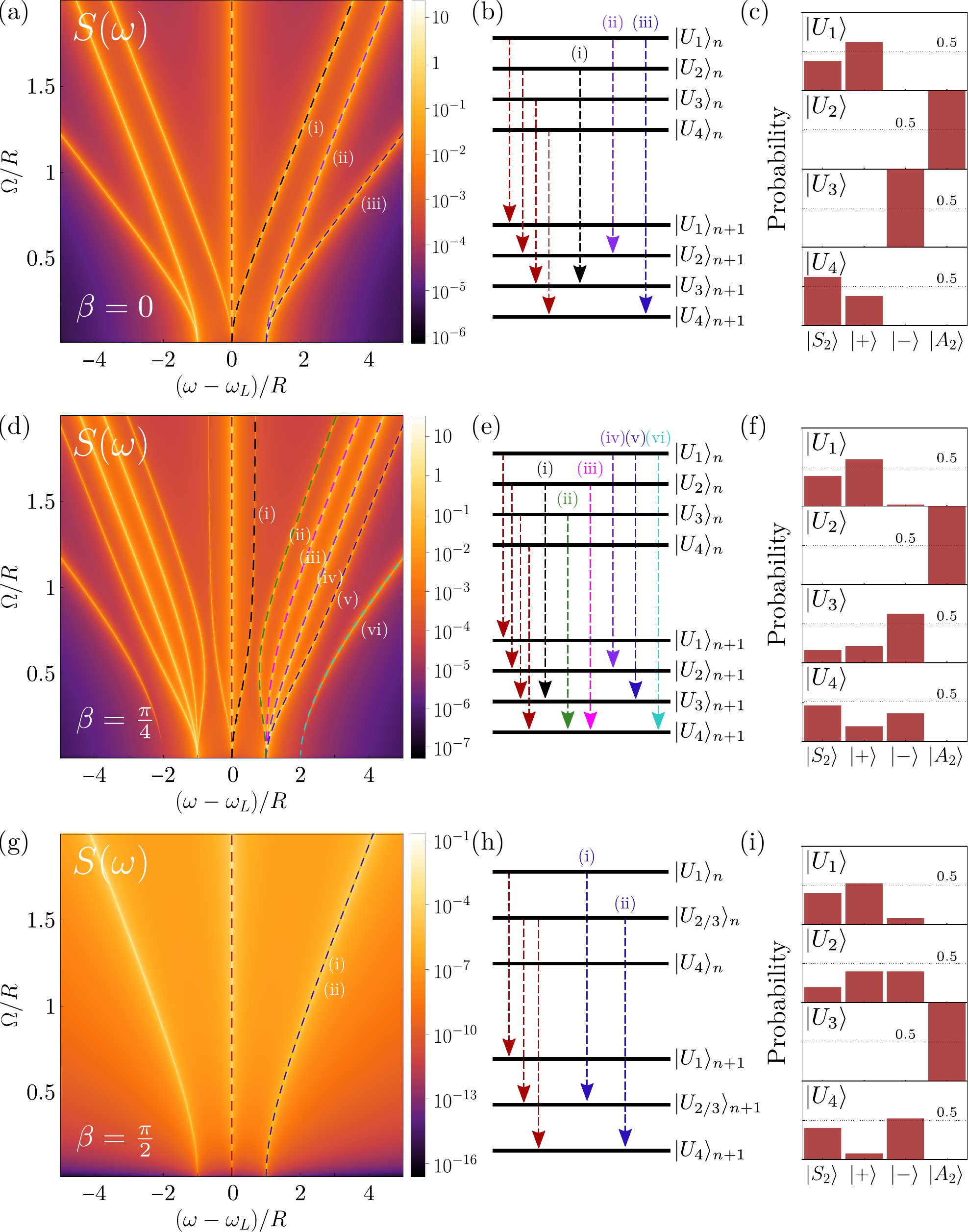}
	\captionsetup{justification=justified}
	\caption[Two-photon resonance fluorescence spectrum  in the strong driving regime]{
		\label{fig:Spectrum_vertical}
		\textbf{Two-photon resonance fluorescence spectrum  in the strong driving regime. }
		Fluorescence spectra $S(\omega)$ for $\beta=0$ (a), $\beta=\pi/4$ (d), and $\beta=\pi/2$ (g). These panels show the emergence of sidebands as $\Omega$ increases. Positive-frequency peaks are marked with dashed lines, which are identified with the corresponding transitions between ladders of eigenstates in panels (b, e, h). The structures of the eigenstates at $\Omega=R$ is depicted in panels (c, f, i). Parameters: $\gamma=10^{-3}R$ and $\gamma_{12}=0.999\gamma$.
	}
\end{SCfigure}

Beyond the perturbative regime, the set of system eigenstates is no longer given by $\{|S_2\rangle, |+\rangle, |-\rangle, |A_2\rangle \}$ but to a mixture of them due to the driving field.  Under these conditions, tractable analytical expressions for the eigenstates are only available in certain limiting cases, e.g.,  $\beta=0$ or $\beta=\pi/2$. In more general scenarios, a fully numerical approach is required to analyze the interplay between the driving and the dipole-dipole coupling.

In \reffig{fig:Spectrum_vertical}, we illustrate the effects of strong driving intensities in the fluorescence spectrum. 
The figure is organized in three rows for the particular cases $\beta = \{0,\pi/4,\pi/2\}$.
The left column depicts the spectrum $S(\omega)$ versus $\Omega$, while the middle and right columns present eigenstate information for $\Omega=R$. The middle column shows the energy transitions among dressed states, and the right column displays the structure of each eigenstate in the basis $\{|S_2\rangle, |+\rangle, |-\rangle, |A_2\rangle \}$.

Now, we analytically examine the eigenstates for the three specific 
values of $\beta$ used in \reffig{fig:Spectrum_vertical}, gaining insights into the fluorescence spectra. Throughout, we label the eigenstates  $|U_i\rangle$, with $i=1\ldots 4$, in decreasing order of  energy. 

\paragraph{$(\beta=0, \Delta=0)$. }
In this case, only the one-photon $|+\rangle$ and two-photon symmetric $|S_2\rangle$ states hybridize under the strong coherent drive. As a result, the eigenstates can be derived analytically. To make the resulting expressions more readable, we present them here in non-normalized form:
\begin{subequations}
	\begin{empheq}[box=\colorboxed{Maroon}]{align}
		|U_1 \rangle &\propto  \frac{R+\sqrt{R^2+16\Omega^2}}{4\Omega} |+ \rangle + |S_2 \rangle , \\  
		| U_2 \rangle &= |A_2 \rangle ,\\
		| U_3 \rangle &= |- \rangle ,\\		
		| U_4 \rangle &\propto \frac{R-\sqrt{R^2+16\Omega^2}}{4\Omega} |+ \rangle + |S_2 \rangle.
	\end{empheq}
\end{subequations}
In the limit $\Omega \ll R$, these equations tend to the perturbative basis used above for $\beta=0$. Their corresponding eigenenergies are
\begin{subequations}
	\begin{empheq}[box=\colorboxed{Maroon}]{align}
		E_1 &= \frac{1}{2}\left( R  + \sqrt{R^2 + 16 \Omega^2} \right), \\
		E_2 &= 0 ,\\
		E_3 &= -R ,\\		
		E_4 &= \frac{1}{2}\left( R  - \sqrt{R^2 + 16 \Omega^2} \right).
	\end{empheq}
\end{subequations}
In principle, these four eigenstates would yield up to six sidebands. However, as discussed in the perturbative analysis, any transition involving the antisymmetric state $|-\rangle$ remains undetectable due to destructive interference of the fields emitted by each emitters [see \reffig{fig:SpectrumInterf}~\textcolor{Maroon}{(a)}].
These expressions also reveal that the eigenstates $|U_3\rangle$ and $|U_4\rangle$ cross in energy when $\Omega = R/\sqrt 2$. 

\paragraph{$(\beta=\pi/4, \Delta=0)$.} This limit is not easily tractable analytically, so we limit our discussion to a description on numerical simulations. As $\beta$ increases, $|U_1\rangle$ remains predominantly as a superposition of the two symmetric states, with only a minimal component of $|-\rangle$. On the other hand, $|U_3\rangle$ and $|U_4\rangle$ become mixed, indicating that $|-\rangle$ is no longer an eigenstate of the system nor is it fully dark. This explains why the perfect destructive interference found at $\beta=0$ for transitions involving $|-\rangle$ no longer take place, resulting in 13 observables peaks.
The eigenstates in this regime can be written as:
\begin{subequations}
	\begin{empheq}[box=\colorboxed{Maroon}]{align}
		| U_1 \rangle &\approx C_{1,+} |+ \rangle +C_{1,S_2} |S_2 \rangle ,\\		
		| U_2 \rangle &= |A_2 \rangle, \\ 
		| U_3 \rangle &= C_{3,+} |+ \rangle +C_{3,-} |- \rangle +C_{3,S_2} |S_2 \rangle,\\
		| U_4 \rangle &= C_{4,+} |+ \rangle +C_{4,-} |- \rangle +C_{4,S_2} |S_2 \rangle ,
	\end{empheq}%
\end{subequations}
where the coefficients $C_{i,j}$ ($i=1,3,4$, $j=+,-,S_2)$ represent generic amplitudes, whose numerically computed values can be seen in \reffig{fig:Spectrum_vertical}~\textcolor{Maroon}{(f)}.

\paragraph{$(\beta=\pi/2, \Delta=0)$.} This is the limit of detuned, uncoupled emitters, reproducing the physics of two independent detuned Mollow triplets~\cite{MollowPowerSpectrum1969}. The non-normalized eigenstates take the following analytical form:
\begin{subequations}
	\begin{empheq}[box=\colorboxed{Maroon}]{align}
		| U_1 \rangle &\propto C_+ |+ \rangle + \frac{1}{2 C_+}|-\rangle +|S_2 \rangle ,\\		
		| U_2 \rangle &= |A_2\rangle , \\  
		| U_3 \rangle &\propto -\frac{\sqrt{2}\Omega}{R}|+ \rangle +\frac{\sqrt{2}\Omega}{R} |- \rangle + |S_2 \rangle,\\
		| U_4 \rangle &\propto C_-|+ \rangle +\frac{1}{2 C_-}|- \rangle + |S_2 \rangle .
	\end{empheq}
\end{subequations}
	with
	\begin{equation}
		C_\pm \equiv \frac{R\pm\sqrt{R^2+4\Omega^2}}{2\sqrt{2}\Omega}.
	\end{equation}
	The corresponding eigenvalues are given by
\begin{subequations}
	\begin{empheq}[box=\colorboxed{Maroon}]{align}
		E_1 &= \sqrt{R^2 + 4\Omega^2},\\
		E_2 &= 0,\\
		E_3 &= 0,\\
		E_4 &= -\sqrt{R^2+4\Omega^2}.
	\end{empheq}
\end{subequations}

\begin{SCfigure}[1][h!]
	\includegraphics[width=1.\textwidth]{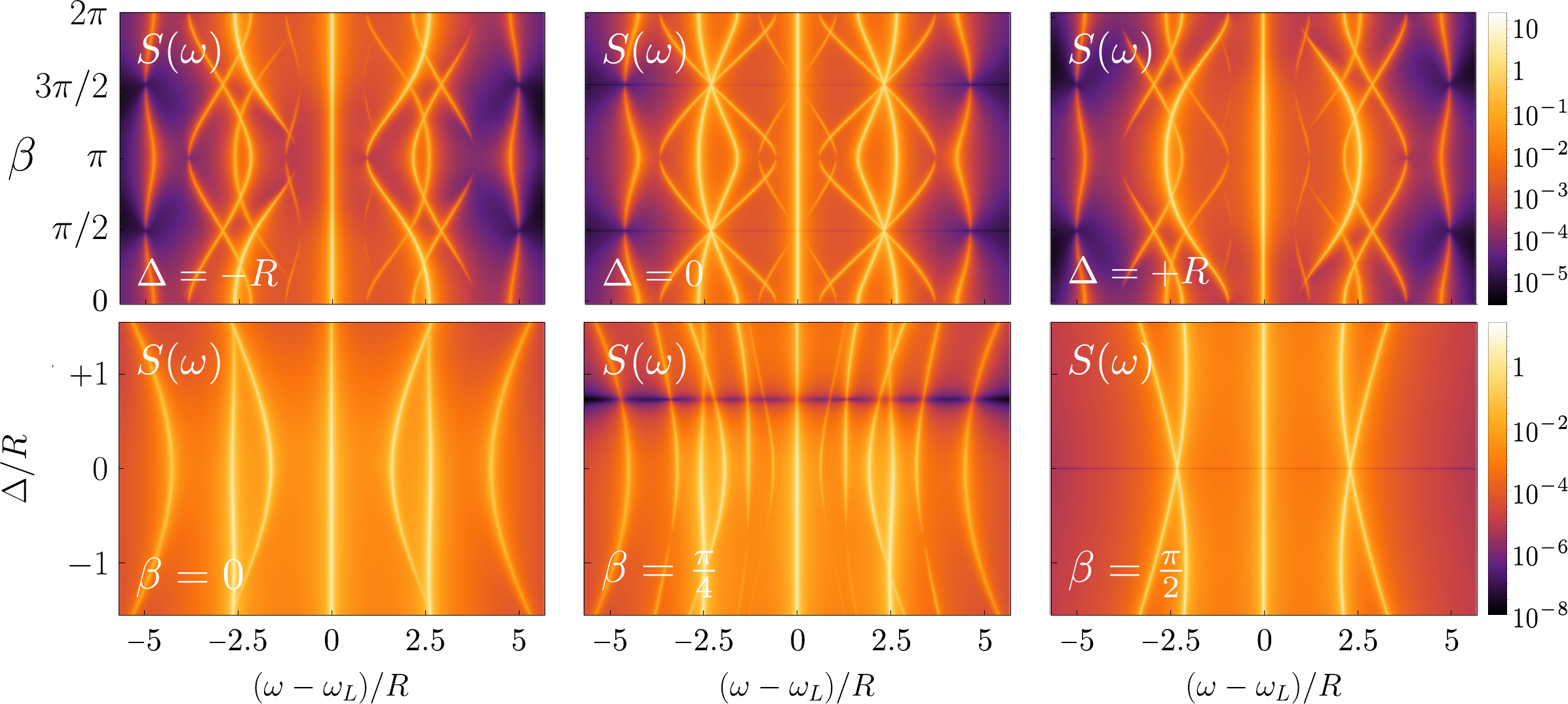}
	\captionsetup{justification=justified}
	\caption[Parameter sensitivity of the fluorescence spectrum.]{\label{fig:fig6-spectrum-Delta}
		\textbf{Parameter sensitivity of the fluorescence spectrum.} $S(\omega)$ versus $\beta$ (upper panels) and $\Delta$ (lower panels). Upper panels: from left to right, $\Delta=\{-R,0,R\}$.  Lower panels: from left to right, $\beta = \{0,\pi/4,\pi/2 \}$.
		Parameters: $\gamma/R=10^{-2}$, $\Omega/R = 1$, $\gamma_{12}=0.999\gamma$. }
\end{SCfigure}

\paragraph{Parameter sensitivity of the fluorescence spectrum. }
This analysis highlights the strong dependence of the resonance fluorescence spectrum on the ratio $J/\delta$---expressed here through the mixing angle $\beta$ [see \eqref{eq:betadefinition}]---, making it a potentially valuable source for extracting information about the parameters that characterize the system, such as the inter-emitter distance.
The continuous variation of the spectral features with $\beta$ is illustrated in  \reffig{fig:fig6-spectrum-Delta} for different values of $\Delta$. These plots not only reproduce the emergence and disappearance of peaks with $\beta$, as described earlier, but also reveal similarly complex spectral structures outside the two-photon resonance ($\Delta \neq 0 $), giving rise to distinct emission patterns that are highly sensitive to $\beta$.

\sectionmark{Quantum metrological application: estimation of ...}
\section[Quantum metrological application: estimation of inter-emitter distances]{Quantum metrological application: \protect\newline estimation of inter-emitter distances}
\sectionmark{Quantum metrological application: estimation of ...}
\label{sec:4-parameter-estimation}
The results just discussed above suggest that the measurement of emission spectra could provide valuable information for the estimation of internal parameters, such as the coherent coupling between emitters or, equivalently, their distance in real space~\cite{HettichNanometerResolution2002}. 
In this section, we address this question by establishing the metrological potential of these measurements within the framework of quantum parameter estimation~\cite{LuisFisherInformation2012,ChaoFisherInformation2016,SafranekSimpleExpression2018,DowlingQuantumOptical2015,LiuQuantumFisher2020,ParisQUANTUMESTIMATION2009,PetzIntroductionQuantum2011,WisemanQuantumMeasurement2009,Demkowicz-DobrzanskiQuantumLimits2015,PezzeQuantumMetrology2018,PolinoPhotonicQuantum2020,BarbieriOpticalQuantum2022}, specifically via the classical Fisher information.

\subsection{Metrological setup - spectral measurements }

Here, we focus on estimating the inter-emitter distance $k r_{12}$ by measuring the fluorescence spectrum described in the previous section. In order to do this, we will make a series of assumptions.

\begin{SCfigure}[1][h!]
	\includegraphics[width=1.\textwidth]{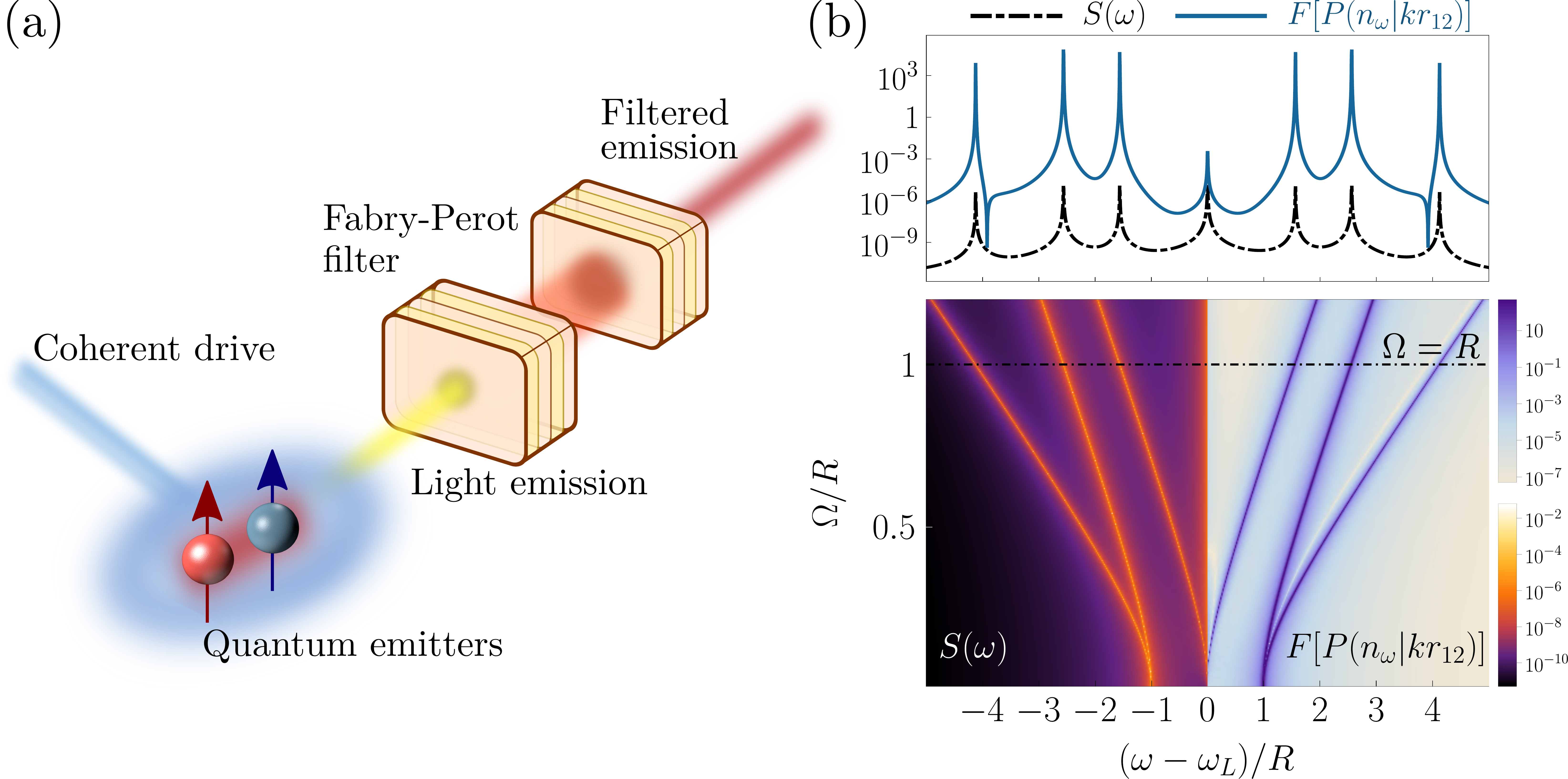}
	\captionsetup{justification=justified}
	\caption[Metrological strategy based on spectral measurements.]{\label{fig:FilteredEmission} 
		\textbf{Metrological strategy based on spectral measurements. } (a) Sketch of a Fabry-Perot cavity filtering the light emitted by the emitters. (b) Fisher information $F[P(n_\omega|kr_{12})]$ and fluorescence spectrum $S(\omega)$ versus $\Omega$ and $\omega$. The upper panel corresponds to a cut of both quantities at $\Omega=R$. Parameters: $\Gamma=\gamma$, $\delta=0$, $\Delta=0$.}
\end{SCfigure}

First, we assume that the spectrum at a frequency $\omega$ is obtained by performing a photon-counting measurement in a bosonic sensor resonant at that frequency and weakly coupled to the quantum emitters---see \refsec{sec:sensor_method}---. For instance, such a sensor could be a tunable Fabry-Perot cavity acting as a frequency filter [see \reffig{fig:FilteredEmission}~\textcolor{Maroon}{(a)}]. 
We further assume that measurements are performed at a discrete set of  $N$ frequencies $\omega_n$. 
Consequently, the corresponding POVM, $\hat \Lambda$, is given by the tensor product of photon-counting POVMS for the $N$ different sensors---as we detailed in \refsec{sec:photon_counting}--- 
\begin{equation}
	\hat \Lambda= \hat \Lambda^{(1)}_n\otimes \Lambda^{(2)}_n \otimes \ldots \otimes \Lambda^{(N)}_n, 
\end{equation}
where $\Lambda^{(i)}_n=\{| n_i\rangle\langle n_i |\}$, with $|n_i\rangle$ denoting the photon number eigenstates of the $i$th sensor. The joint photon-counting probability distribution at these frequencies is given by $P(n_1,n_2,\ldots,n_N|k r_{12})$, representing the measurement outcomes across all sensors. Assuming uncorrelated measurements at different frequencies, this distribution factorizes as:
\begin{equation}
	P(n_1,n_2,\ldots,n_N|k r_{12})=P(n_1|k r_{12})P(n_2|k r_{12})\ldots P(n_N|k r_{12}).
\end{equation}
This approximation ignores contributions to the Fisher information from correlations between photons emitted at different frequencies~\cite{DelValleTheoryFrequencyFiltered2012,Gonzalez-TudelaTwophotonSpectra2013,SanchezMunozViolationClassical2014,LopezCarrenoPhotonCorrelations2017}. 
Such an assumption is valid if the photon lifetime within the sensor is sufficiently long to erase temporal correlations or if frequencies are measured sequentially in independent experiments, e.g., by tuning the frequency of a Fabry-Perot filter [see \reffig{fig:FilteredEmission}~\textcolor{Maroon}{(a)}].
Later in \refch{chapter:Outlook}, we revisit this assumption by exploring the potential impact of quantum correlations for parameter estimation in driven-dissipative systems. Specifically, we analyse the role of frequency-resolved correlation measurements in estimating unknown atomic parameters, focusing on a single coherently driven TLS. We advance that retaining the quantum correlations of the emission  generally yield a better metrological performance, paving the way for novel quantum metrological protocols based on spectral correlations.

\paragraph{Poissonian Fisher information. }Following the same approach that is used in \colorref{DelaubertQuantumLimits2008}, 
we assume that the resulting state of the sensor is a coherent state, so that the corresponding measured photo-current displays Poissonian fluctuations---see  \refsec{Chapt6_SecMetrology} for a more detailed discussion---. 
Under this assumption, the photon-counting distribution for a sensor tuned to frequency $\omega$ is given by:
\begin{equation}
	P(n_\omega|k r_{12})=\frac{ \bar n_\omega^{n_\omega} e^{- \bar n_\omega }}{n_\omega!},
	\label{eq:p_n_coherent}
\end{equation}
where $n_\omega$ denotes the excitation number at the frequency $\omega$ and $\bar n_\omega = \eta S(\omega)$
---according to the sensor relation in \eqref{eq:spectrum_sensor}---. Here, $\eta$ is a global constant determined by the specifics of the detection scheme (e.g., detection efficiency). For simplicity, we set $\eta=1$, as it only introduces an overall factor.
Since the spectrum $S(\omega)$ is strongly dependent on the value of $\beta$, it also varies significantly with the inter-emitter distance $k r_{12}$. To emphasize this dependence, we write $S(\omega) = S(\omega,k r_{12})$.

Since different sensors are uncorrelated and their probability distributions factorize, the Fisher information associated to the measurement of the spectrum $S(\omega,k r_{12})$  is given by 
\begin{equation}
	\colorboxed{Maroon}{F =\sum_\omega F[P(n_\omega|k r_{12})]= \sum_\omega\frac{1}{S(\omega,kr_{12})}\left[\frac{\partial S(\omega,k r_{12})}{\partial k r_{12}} \right]^2.}
	\label{eq:FisherSpec}
\end{equation}
This expression represents a frequency sum of independent Poissonian Fisher information contributions---see the right panel of \reffig{fig:FilteredEmission}~\textcolor{Maroon}{(b)} for an example of a frequency-dependent FI---. 
To ensure a well-defined total Fisher information value, the number of spectral measurements must be sufficient to resolve all features in the fluorescence spectrum, such as those shown in the left panel of \reffig{fig:FilteredEmission}~\textcolor{Maroon}{(b)}.
The Fisher information in \eqref{eq:FisherSpec} provides a tool to quantify the metrological potential of fluorescence spectrum measurements. For this analysis, we incorporate a finite detector linewidth $\Gamma = \gamma$ into the spectrum~\cite{DelValleTheoryFrequencyFiltered2012} by replacing $\omega\rightarrow \omega+i\Gamma$ in \eqref{eq:spectrum_total} [see \refsec{Section:SpectralProperties}].

\subsection{Estimation of the inter-emitter distance}

In \reffig{fig:fig7-fisher}, we show the Fisher information $F$ as a function of the optically tunable parameters, $\Delta$ and $\Omega$. The results indicate that the optimal regime of operation is near the two-photon resonance  ($\Delta\approx 0$), where $F$ reaches its maximum. This is explained by the fact that the mechanism of two-photon excitation is strongly dependent on the coherent coupling between emitters, as discussed in previous sections, which is highly sensitive to the inter-emitter distance $k r_{12}$. At $\Delta=0$, there is an optimal driving amplitude $\Omega$ that maximizes $F$, thereby enhancing the precision in estimating $k r_{12}$. 
\begin{SCfigure}[1][h!]
	\includegraphics[width=1.02\textwidth]{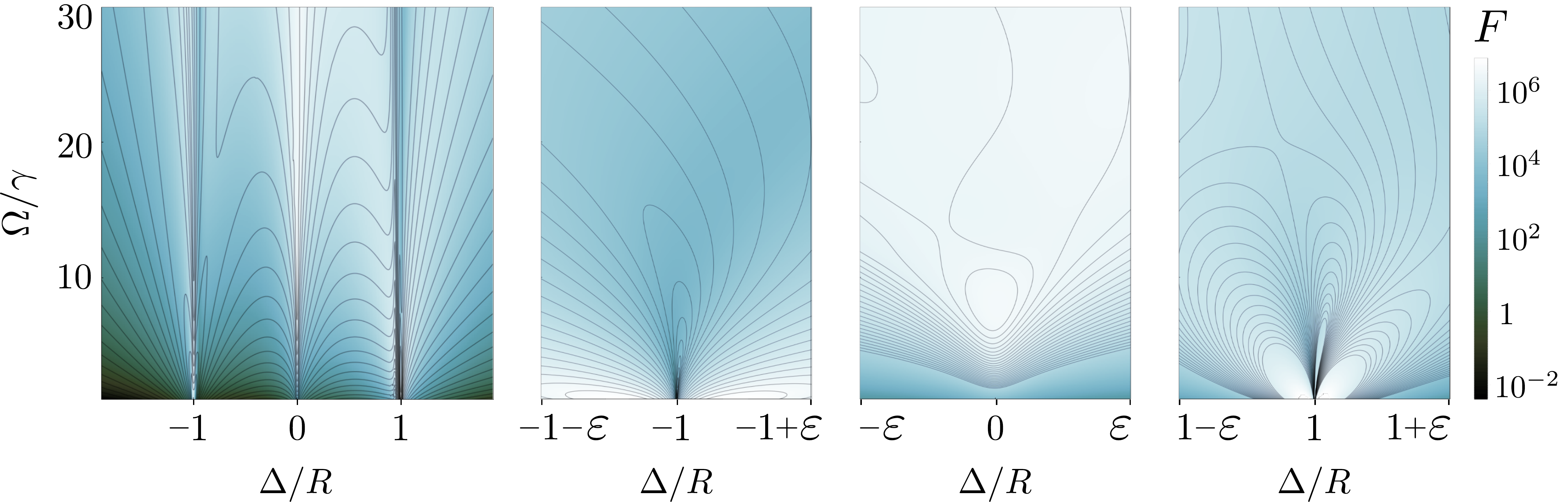}
\end{SCfigure}
\addtocounter{figure}{-1}
\graffito{\vspace{-4.8cm}
	\captionof{figure}[Fisher information of resonance fluorescence measurements for the estimation of $k r_{12}$.]{\label{fig:fig7-fisher} 
		\textbf{Fisher information of resonance fluorescence measurements for the estimation of $k r_{12}$.} Fisher information in terms of $\Omega/\gamma$ and $\Delta$, for $k r_{12}=0.17$. The three rightmost panels depict a zoom around $\Delta = \{-R, 0, R \}$, with a zoom range is $\varepsilon = 0.02 R$. We observe that the region of two-photon excitation $\Delta\approx 0 $ features the higher Fisher information.  Parameters: $\delta=50 \gamma$, $\Gamma = \gamma$, $\gamma_{12}=0.999\gamma$.}
}

\begin{SCfigure}[1][h!]
	\includegraphics[width=0.5 \textwidth]{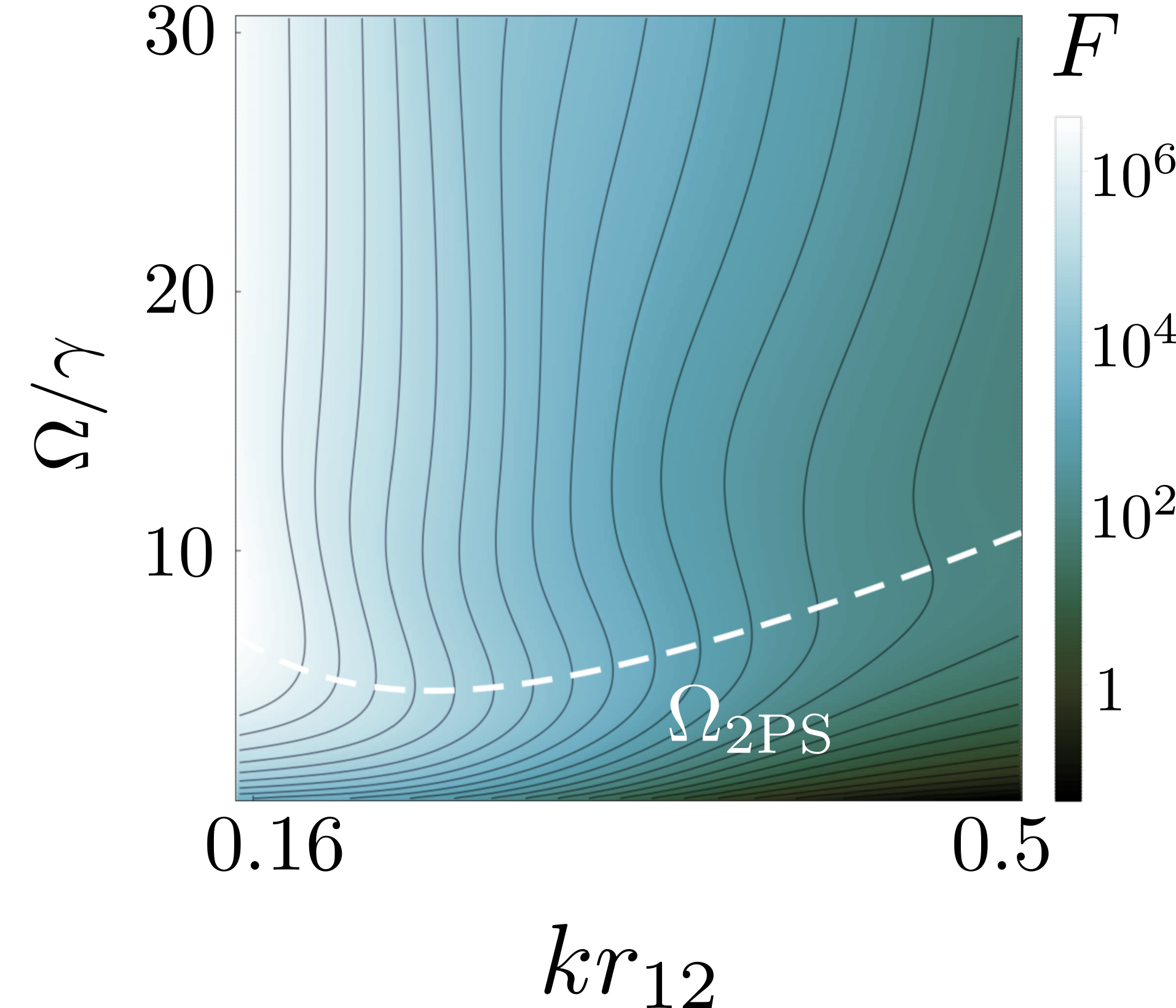}
	\captionsetup{justification=justified}
	\caption[Optimum operation regime for estimating the inter-emitter distance.]{ \label{fig:Fisher_distance}  \textbf{Optimum operation regime for estimating the inter-emitter distance. } Fisher information versus $\Omega$ and $k r_{12}$ at $\Delta=0$. The white-dashed line represents the value of two-photon saturation amplitude $\Omega_{2\text{PS}}$, which serves as a good indicator of the optimum driving amplitude that maximizes the Fisher information. Parameters: $\delta=50 \gamma$, $\Gamma = \gamma$, $\gamma_{12}=0.999\gamma$.} 
\end{SCfigure}

Motivated by the dependency of the Fisher information on the driving amplitude $\Omega$, and recalling the relevance of the two-photon saturation amplitude $\Omega_{2\text{PS}}$ to resolve two-photon sidebands, we observe that the maximum value of the Fisher information varies with the actual value of $k r_{12}$, as shown  in \reffig{fig:Fisher_distance}. This maximum is well approximated by the two-photon saturation driving amplitude $\Omega_{2\mathrm{PS}}$ that we obtained in \eqref{eq:Omega_2PS}, at which the two-photon sidebands begin to be resolved in the spectrum.
This establishes the onset of the two-photon saturation regime as the optimal operational point for precisely estimating the distance between interacting quantum emitters.   The results shown here as functions of the optically tunable parameters, $\Delta$ and $\Omega$, correspond to distinct, independent experiments. Therefore, performing a series of measurements across the $(\Omega,\Delta)$ parameter space could provide a higher precision of estimation, with a total Fisher information that would be given by the sum $F=\sum_{\Delta,\Omega}F(\Omega,\Delta)$. 
Finally, we note that variations in some assumptions considered in this text, such as the geometrical configuration of the emitters, do not significantly alter these results. 

\section{Conclusions}
In this Chapter, we have analysed in detail one of the central models of this Thesis:  two interacting, non-identical quantum emitters under coherent driving, focusing on the regime of two-photon excitation.
For the first time in the context of non-identical emitters, we derived analytical expressions for the stationary density matrix and key steady-state observables, such as the mean intensity of fluorescent emission, that are valid for an ample regime of parameters, subject only to the condition that the energy splitting between one-excitation eigenstates must be the largest energy scale in the system. These results offer valuable insights into how the emitted light depends on the coherent coupling and detuning between the emitters.
We also identified the parameter regimes where two-photon features—such as the resonance peak in the excitation spectrum and the sidebands in resonance fluorescence—dominate over first-order processes. Since these two-photon effects are highly sensitive to the coupling strength and, therefore, the inter-emitter distance, we evaluated their potential for distance estimation via the classical Fisher information. Notably, we found that the onset of two-photon effects at the resonance is the most sensitive regime for such estimations.

An intriguing question arises from the relevance of the two-photon phenomena observed in this model: 
could the emitter-emitter dynamics be effectively confined within the two-excitation manifold 
$\{|gg\rangle, |ee\rangle\}$, as in Model 2P, while neglecting the intermediate state? Such a scenario would be both fundamentally interesting and technologically relevant, as it could enable the stabilization of entanglement between these two emitters in specific parameter regimes.
In the next Chapter, we delve into this question by incorporating a photonic structure, such as a lossy single-mode cavity, to harness the two-photon effects induced by the driving for the generation of steady-state entanglement. In fact, the stabilization of stationary entanglement between emitters will be a central theme of this Thesis.
\cleardoublepage

\chapter[Unconventional mechanism  of virtual state \protect\newline  population]{Unconventional mechanism  \protect\newline 	of virtual state population}
\chaptermark{Unconventional mechanism  of virtual state population}
\label{chapter:Unconventional}

\section{Introduction}

In the previous Chapter, we introduced a model that is central to this Thesis, that is, \textit{two nonidentical interacting quantum emitters driven by a coherent field and surrounded by a vacuum electromagnetic field}, and showed its complex structure in the frequency domain through light observables, such as the second-order correlation function or the fluorescence spectrum.
Particularly, in the limit of two closely spaced, near resonant emitters ($\beta\approx 0$, $\gamma_{12}\approx \gamma$) that are perturbatively driven  ($\Omega\ll R$), the antisymmetric state $|-\rangle$ becomes a dark state,  effectively decoupling from the  dynamics.
Under these conditions, the system simplifies to an effective description involving the states $\{|+\rangle, |A_2\rangle, |S_2\rangle\}$, where $|S_2/A_2\rangle= 1/\sqrt{2}(|gg\rangle\pm|ee\rangle)$ denote \textit{two-photon dressed states}. 
These states are particularly interesting since they are maximally entangled, exhibiting a \textit{Bell-like} structure~\cite{EinsteinCanQuantumMechanical1935,BellEinsteinPodolsky1964,NielsenQuantumComputation2012,PlenioIntroductionEntanglement2007,HorodeckiQuantumEntanglement2009}.  
Entangled states  such as these play a crucial role in a range of modern quantum technological applications~\cite{OBrienPhotonicQuantum2009,AcinQuantumTechnologies2018}, including optical quantum computation~\cite{KnillSchemeEfficient2001, OBrienOpticalQuantum2007}, quantum teleportation~\cite{RenGroundsatelliteQuantum2017}, quantum metrology~\cite{GiovannettiAdvancesQuantum2011}, or quantum imaging~\cite{PittmanOpticalImaging1995, ShapiroPhysicsGhost2012, ZipfelNonlinearMagic2003}, as they are the essential resource enabling such applications.
Therefore, it would be technologically advantageous to develop a strategy for stabilizing the emitter-emitter system into  one of these entangled states, even in presence of dissipation.
In the following, we present a physical intuition that one may be tempted to follow for achieving this stabilization using reservoir engineering techniques~\cite{ChangColloquiumQuantum2018,Gonzalez-TudelaLightMatter2024}:

\emph{
Introducing a lossy cavity [see \reffig{fig:TwoPhotonDecay1}~\textcolor{Maroon}{(a)}] tuned at the two-photon resonance  would enable a fast dissipative channel from $|ee\rangle$ to $|gg\rangle$ via the Purcell effect~\cite{PurcellResonanceAbsorption1946} that would effectively bypass the one-excitation manifold $|+\rangle$.
In this particular tailored environment, we expect the one-excitation manifold to contribute as a virtual state that mediates the interaction between $|gg\rangle$ and $|ee\rangle$. As a result, the system would be described by an effective two-level system characterized by a two-photon dissipation rate  $\Gamma_{2\mathrm{p}}$ that is resonantly driven with a two-photon Rabi frequency $\Omega_{2\mathrm{p}}$, as depicted in \reffig{fig:TwoPhotonDecay1}~\textcolor{Maroon}{(b)}.
Then, under some parameter regimes, the system would be stabilized into an entangled state of the form $\sim |gg\rangle \pm |ee\rangle$ when $\Omega_{2\mathrm{p}}\approx \Gamma_{2\mathrm{p}}/2\sqrt{2}$. 
}
\begin{SCfigure}[1][h!]
	\includegraphics[width=1.\columnwidth]{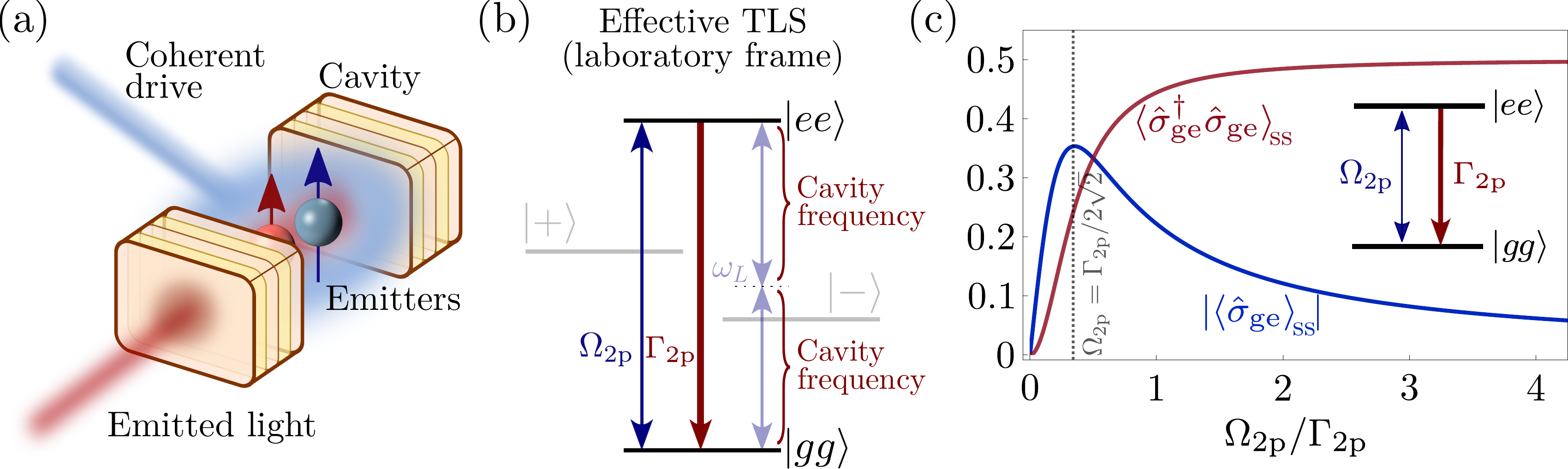}
\end{SCfigure}
\addtocounter{figure}{-1}
\graffito{\vspace{-9.5cm}
	\captionof{figure}[Effective TLS model with two-photon driving and two-photon decay rate.]{\label{fig:TwoPhotonDecay1}	\textbf{Effective TLS model with two-photon driving and two-photon decay rate. } 
		(a) Scheme of a tailored environment: two interacting TLS coherently driven and embedded in a photonic cavity. 
		(b) Scheme of the effective TLS: the states $|ee\rangle$ and $|gg\rangle$ coherently interact with a two-photon Rabi frequency $\Omega_{2\text{p}}$ and the excited state de-excites with a cavity-induced two-photon rate $\Gamma_{2\text{p}}$.
		(c) Population and absolute value of the coherence of the effective TLS in terms of the driving strength.
	}
}

This intuition is fundamentally based on the well-known dynamics of a resonantly driven TLS system, which stabilizes into a coherent superposition of $|g\rangle$ and $|e\rangle$ when the driving field is set to the value that maximizes the qubit coherence ($\Omega=\gamma/2\sqrt{2}$)~\cite{WallsQuantumOptics2008}. 
In this scenario, the system reaches a steady-state characterized by a mixed density matrix with population $\langle \hat \sigma^\dagger \hat \sigma \rangle_{\text{ss}}=1/4$ and maximum coherence $|\langle \hat \sigma \rangle_{\text{ss}}|^2=1/2\sqrt{2}$---where $\hat \sigma \equiv |g\rangle \langle e|$---, indicating the existence of a quantum superposition of both states. 
In contrast, when $\Omega\gg \gamma$, the system thermalizes into an incoherent mixture of the two states since $\langle \hat \sigma^\dagger \hat \sigma \rangle_{\text{ss}}\rightarrow 0.5$ and $|\langle \hat \sigma \rangle_{\text{ss}}|^2\rightarrow0$.
Consequently, by extending this intuition to the emitter-emitter system, the inclusion of a fast dissipative channel induced by the cavity would allow a similar description involving $|gg\rangle$ and $|ee\rangle$. 
Specifically, it would imply that the effective TLS mirrors the same behaviour [see \reffig{fig:TwoPhotonDecay1}~\textcolor{Maroon}{(c)}], replacing $\hat \sigma \rightarrow \hat \sigma_{ge} \equiv |gg\rangle \langle ee|$.
We inform the reader in advance that the following speculative picture outlined below does not work to successfully stabilize the desired entangled states. This failure arises partly because the proposed adiabatic elimination of the lossy cavity is not refined enough to faithfully describe the actual dynamics of the system. However, more critically, it is due to the emergence of an unexpected effect, which forms the central focus of this Chapter.

Let us verify how this intuition is actually incorrect. We assume that both emitters are coherently driven at the two-photon resonance ($\omega_L=\omega_0$) and embedded within a photonic structure described as a single-mode cavity with frequency $\omega_a=\omega_L$ ($\Delta_a\equiv \omega_a-\omega_L=0$), emitter-cavity coupling rate $g$, and photon leakage rate $\kappa$. The resulting qubits-cavity system is described by the following master equation in the rotating frame of the drive:
\begin{equation}
		\frac{d \hat{\rho}}{dt}=-i[ \hat H_\mathrm{q}+\hat H_{\mathrm{d}}+\hat H_\mathrm{a}, \hat{\rho}]+\sum_{i,j=1}^{2}\frac{\gamma_{ij}}{2}\mathcal{D}\left[\hat \sigma_i,\hat \sigma_j\right]  \hat{\rho}+\frac{\kappa}{2} \mathcal{D}\left[\hat a \right]  \hat{\rho},
	\label{eq:Master_eq}
\end{equation}
where $\hat H_\mathrm{q}$ and $\hat H_\mathrm{d}$ are the qubit-qubit and driving Hamiltonian---defined in \cref{eq:HEmitters,eq:Hd}---, and $\hat H_\mathrm{a}$ represents a Tavis-Cummings Hamiltonian:
\begin{equation}
	\hat H_\mathrm{a}=\Delta_a \hat a^\dagger \hat a + g\left[\hat a^\dagger (\hat \sigma_1 +\hat \sigma_2)+\text{H.c.}\right].
\end{equation}
The cavity is assumed to be in the bad-cavity limit~\cite{SavageStationaryTwolevel1988,CiracInteractionTwolevel1992,ZhouDynamicsDriven1998,Navarrete-BenllochIntroductionQuantum2022}, characterized by $\kappa\gg g \gg \gamma, \gamma_{12}$. As a result, the cavity remains predominantly in vacuum,
allowing us to effectively describe its effect by a non-Hermitian Hamiltonian, 
\begin{equation}
	\tilde{ H}=\hat H_\mathrm{q}+\hat H_\mathrm{d}+\hat H_\mathrm{a}-i\frac{\kappa}{2}\hat a^\dagger \hat a.
\end{equation}
Since the driving mechanism involves two-photon processes,
the dynamics could be constrained within the two-excitation manifold  $\{ |ee,0\rangle, |+,1\rangle, |gg,2\rangle \}$, expressed in the canonical basis $|i,n\rangle$, where $i=gg,S,ee$ denotes the excitonic levels of the emitters and $n=0,1,2$ represents the photon excitation of the cavity.
Then, the non-Hermitian Hamiltonian $\tilde H$ would read
\begin{equation}
	\tilde{ H}=\left( \begin{array}{c|c}
		E_0 & \hat V  \\ \hline
		\hat V^\dagger&  \hat h
	\end{array} \right)= 
\left( \begin{array}{c|cc}
	0 & \sqrt{2}g & 0 \\ \hline
	\sqrt{2}g &  \Delta+\Delta_a+ R - i\frac{\kappa}{2} & 2g \\ 
	0 & 2g &  2\Delta_a- i\kappa
\end{array} \right).
\label{eq:NonHermitianHamiltonian}
\end{equation}
Following the Hamiltonian perturbative expansion method outlined in \refsec{Section:QuantumTraject} and treating $\{|+,1\rangle, |gg,2\rangle\}$ as the perturbative sector, we derive the effective cavity-induced dissipative rate $\Gamma$, which modifies the dissipative properties of $|ee,0\rangle$. This is given by the relation:
\begin{equation}
	\Gamma\approx -2\text{Im}[ \hat V (E_0-\hat h)^{-1} \hat V^\dagger ] = \Gamma_{1\mathrm{p}}+\Gamma_{2\mathrm{p}},
	\label{eq:EffectiveGamma}
\end{equation}
where $E_0$, $\hat h$ and $\hat V$ are correspondingly identified from \eqref{eq:NonHermitianHamiltonian}. 
In the second equality of \eqref{eq:EffectiveGamma}, we decomposed the effective decay rate into two distinct contributions: $\Gamma_{1\mathrm{p}}\propto g^2$, corresponding to two-photon decay via one-photon processes, and $\Gamma_{2\mathrm{p}}\propto g^4$,  representing direct two-photon processes.
While a full analytical solution for both effective decay rates can be derived from \eqref{eq:EffectiveGamma}, the resulting expressions are cumbersome and not directly relevant to the current discussion---these will be presented in detail in the next Chapter---.
The two decay rates, $\Gamma_{1\mathrm{p}}$ and $\Gamma_{2\mathrm{p}}$, suggest the presence of one- and two-photon decay channels,  what bring us to intuitively propose an effective description of the reduced emitter-emitter system as an effective $\Lambda$-system [illustrated in \reffig{fig:TwoPhotonDecay2}~\textcolor{Maroon}{(a)}] governed by the master equation 
\begin{multline}
	\frac{d \hat{\rho}}{dt}=-i[ \hat H_\mathrm{q}+\hat H_{\mathrm{d}}, \hat{\rho}] +\frac{\Gamma_{2\mathrm{p}}}{2 } \mathcal{D}[|gg\rangle \langle ee|]\hat \rho\\
	+\frac{\Gamma_{1\mathrm{p}}+\gamma_+ }{2 }(\mathcal{D}[|gg\rangle \langle +|]+\mathcal{D}[|+\rangle \langle ee|])\hat \rho.
	\label{eq:Master_eq_wo_cavity}
\end{multline}
This effective model has not been obtained through a systematic adiabatic elimination but is instead phenomenologically motivated. In the next Chapter, we will employ a more sophisticated approach via the Nakajima-Zwanzig formalism~\cite{NakajimaQuantumTheory1958,ZwanzigEnsembleMethod1960,BreuerTheoryOpen2007,RivasOpenQuantum2012,Navarrete-BenllochIntroductionQuantum2022,Gonzalez-BallesteroTutorialProjector2024}, which provides deeper physical insights into the dynamics.
Nevertheless, for the purposes of this discussion, we will consider the effective master equation in \eqref{eq:Master_eq_wo_cavity} as a valid description.

\begin{SCfigure}[1][h!]
	\includegraphics[width=1.\columnwidth]{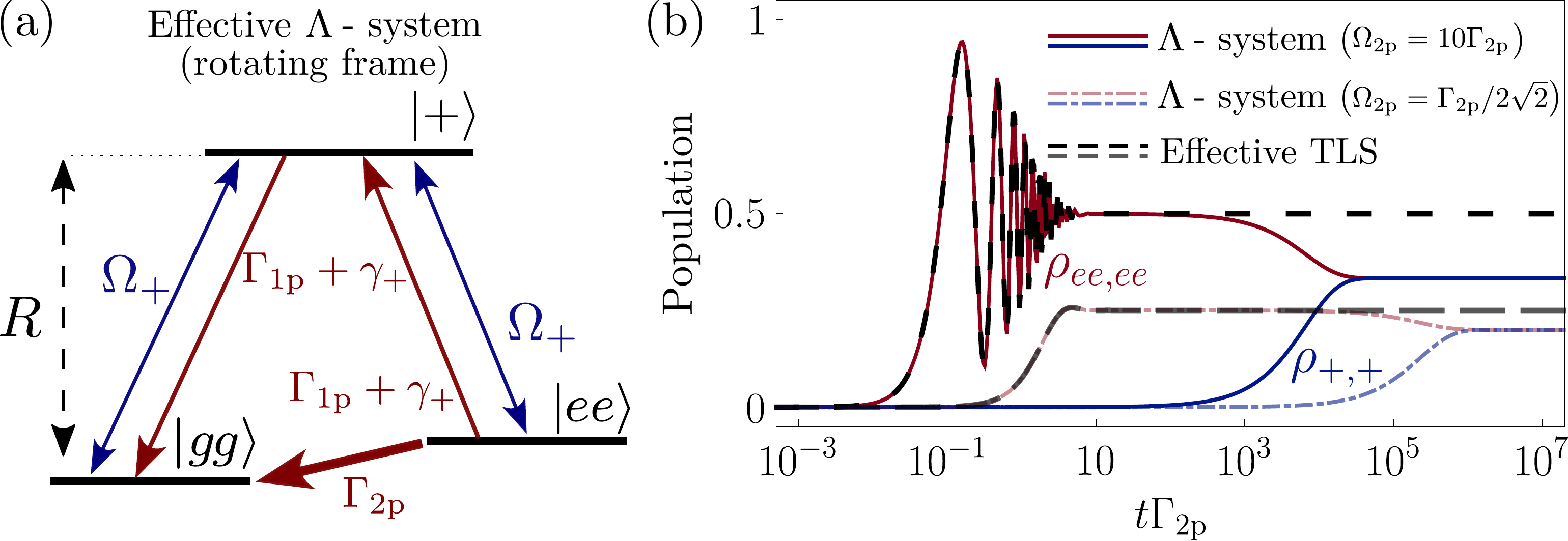}
\end{SCfigure}
\addtocounter{figure}{-1}
\graffito{\vspace{-5.cm}
	\captionof{figure}[Unconventional dynamics in the effective $\Lambda$-system.]{\label{fig:TwoPhotonDecay2}	\textbf{Unconventional dynamics in the effective $\Lambda$-system. }
	(a) Scheme of the effective $\Lambda$-system in the rotating frame of the drive. 
	%
	%
	(b) Time evolution of $\rho_{ee,ee}$ (red) and $\rho_{+,+}$ (blue). Solid and dot-dashed curves correspond to numerical simulation with \eqref{eq:Master_eq_wo_cavity} for $\Omega_{2\text{p}}=10\Gamma_{2\text{p}}$ and $\Omega_{2\text{p}}=\Gamma_{2\text{p}}/2\sqrt{2}$, respectively.  Black dashed lines represent numerical simulation with an effective TLS resonantly excited with a two-photon drive $\Omega_{2\text{p}}$ and dissipation rate $\Gamma_{2\text{p}}$. Parameters: $\Delta=0$, $\Gamma_{1\text{p}}+\gamma_+=10^{-15}R$, and $\Gamma_{2\text{p}}=10^{-5}R$.
	}
}

From the effective master equation in \eqref{eq:Master_eq_wo_cavity}, it seems reasonable---consistent  with our initial intuition---to conclude that in a regime where $\Gamma_{2\mathrm p}\gg\Gamma_{1\mathrm{p}},\gamma_+$ and $R\gg \Omega_+,\Delta$,  the dynamics would be constrained within $\{|gg\rangle, |ee\rangle \}$ through the mediation of $|+\rangle$, acting as a \textit{virtual state} since it is far-detuned from the other states---see \reffig{fig:TwoPhotonDecay2}~\textcolor{Maroon}{(a)}---. For instance, one would expect, Rabi oscillations when $\Omega_{2\mathrm{p}}>\Gamma_{2\mathrm{p}}$.
As a result, recalling the steady-state behaviour of a resonantly driven two-level system illustrated in \reffig{fig:TwoPhotonDecay1}~\textcolor{Maroon}{(c)}, the system  would relax in the long-time limit to an entangled state of the form of $\sim |gg\rangle \pm |ee\rangle$, e.g., when $\Omega_{2\mathrm{p}}=\Gamma_{2\mathrm{p}}/2\sqrt{2}$.
However, we find an unexpected result---shown in \reffig{fig:TwoPhotonDecay2}~\textcolor{Maroon}{(b)}---that defies our conventional understanding of virtual states in quantum mechanics~\cite{Cohen-TannoudjiAtomPhotonInteractions1998}.
The effective $\Lambda$-system initially exhibits the same dynamics of a resonantly driven TLS, characterized by Rabi oscillations between the states $|gg\rangle$ and $|ee\rangle$ with a two-photon Rabi frequency $\Omega_{2\mathrm{p}}$ and decay rate $\Gamma_{2\mathrm{p}}$. These oscillations would decay exponentially, stabilizing the system into a mixture of both states [black dashed lines in \reffig{fig:TwoPhotonDecay2}~\textcolor{Maroon}{(b)}].
This behaviour, however, is only metastable for a time $t\lesssim  10^3/ \Gamma_{2\mathrm{p}}$. At later times $t\gtrsim 10^3 /\Gamma_{2\mathrm{p}}$ (strong driving, $\Omega_{2\mathrm{p}}=10\Gamma_{2\mathrm{p}}$) or $t\gtrsim 10^5/\Gamma_{2\mathrm{p}}$ (weak driving, $\Omega_{2\mathrm{p}}=\Gamma_{2\mathrm{p}}/2\sqrt{2}$), the mediating character of the symmetric state breaks down, resulting in a \textit{de-virtualization} process by which it acquires a sizeable population in the long-time limit [blue curves in \reffig{fig:TwoPhotonDecay2}~\textcolor{Maroon}{(b)}].
This unexpected phenomenon demonstrates that \textit{the intuitive notion that a virtual state remains unpopulated can be false in the presence of dissipation, leading to an unconventional description of what a virtual state is in open quantum systems,  even if the virtual state is not directly coupled to any dissipative channel.
} 

In this Chapter, we delve into this unconventional mechanism by which virtual states, in the long-time limit, acquire a sizeable occupation probability comparable to that of the real states, even when dissipative processes only take place within the real subspace.
These findings are relevant for the understanding and design of dissipative processes for quantum state preparation and engineered interactions~\cite{PlenioCavitylossinducedGeneration1999,DiehlQuantumStates2008,VacantiCoolingAtoms2009,VerstraeteQuantumComputation2009,BuschProtectingSubspaces2010,BuschCoolingAtomcavity2011,LinDissipativeProduction2013,EvansPhotonmediatedInteractions2018,ChangColloquiumQuantum2018}, as well as for the generation of long-lived entanglement between emitters~\cite{LiuComparingCombining2016,Gonzalez-TudelaEntanglementTwo2011,Martin-CanoDissipationdrivenGeneration2011,Gonzalez-TudelaMesoscopicEntanglement2013,RamosQuantumSpin2014,PichlerQuantumOptics2015,HaakhSqueezedLight2015,ReitzCooperativeQuantum2022,KastoryanoDissipativePreparation2011}. In this Thesis, we explore the potential of these results for generating  entanglement between emitters by tuning the interactions in a tailored environment. Particularly, the next Chapter focuses on the case of $N$ emitters coupled to a single-mode cavity.

%
%
%
To study this \textit{de-virtualization} phenomenon, we first establish  in \refsec{sec:1} the notion of a virtual state in quantum mechanics, clarifying the standard concepts of slow---real---and fast---virtual---variables. In \refsec{sec:2}, we analyze what is arguably the simplest scenario that can be described in terms of virtual states [see {\reffig{fig:fig1_setup}~\textcolor{Maroon}{(a)}]: two quasi-resonant “real” states, effectively interacting through the mediation of a third, strongly off-resonant “virtual” state. Crucially, we enable a spontaneous decay between the real states, provided, e.g.,  by the coupling to a  Markovian environment. 
Here, we confirm the core result explained above: contrary to the familiar intuition, the situation in which the virtual state remains ``virtual''---i.e., unpopulated---is, in this case, only metastable~\cite{MacieszczakTheoryMetastability2016,MacieszczakTheoryClassical2021,BrownUnravelingMetastable2024}, and, in the long time limit, the system eventually relaxes to a stationary state where the virtual state acquires sizeable population.
In \refsec{sec:3}, we introduce a technique of \emph{hierarchical adiabatic eliminations} (HAE) to derive analytical expressions of the time-dependent elements of the system density matrix, as well as expressions for the characteristic metastability timescales.
In \refsec{sec:4}, we provide essential intuition about the underlying mechanism of population of the virtual state through the analysis of quantum trajectories~\cite{CarmichaelPhotoelectronWaiting1989,MolmerMonteCarlo1993,MolmerMonteCarlo1996,PlenioQuantumjumpApproach1998,BrunSimpleModel2002,GerryIntroductoryQuantum2004}, showing that this state gets populated due to the non-Hermitian evolution taking place between quantum jumps. This way, we establish that the process of \emph{de-virtualization} occurs through an unconventional mechanism of population entirely enabled by dissipation. 
Finally, in \refsec{sec:5} we discuss  the generality of the HAE technique and its application to other systems involving two interacting qubits.  We show that this phenomenon has strong implications for the generation of stable and metastable entanglement via dissipation~\cite{RamosQuantumSpin2014,PichlerQuantumOptics2015}.

These results have been published in Physical Review A~\cite{Vivas-VianaUnconventionalMechanism2022}.

\section{Revisiting the notion of virtual states}
\label{sec:1}
The concept of \textit{virtual state} in quantum mechanics is of paramount importance, particularly, in the context of virtual transitions between coherently unconnected states~\cite{Cohen-TannoudjiPhotonsAtoms1997,Cohen-TannoudjiAtomPhotonInteractions1998,SakuraiModernQuantum2017} or in the description of scattering processes in quantum field theory where interactions are mediated by virtual particles~\cite{RyderQuantumField1996}. For instance, \reffig{fig:SchemeVirtualState} provides a diagrammatic representation of a two-photon process involving a virtual state.

\begin{SCfigure}[1][h!]
	\includegraphics[width=0.5\columnwidth]{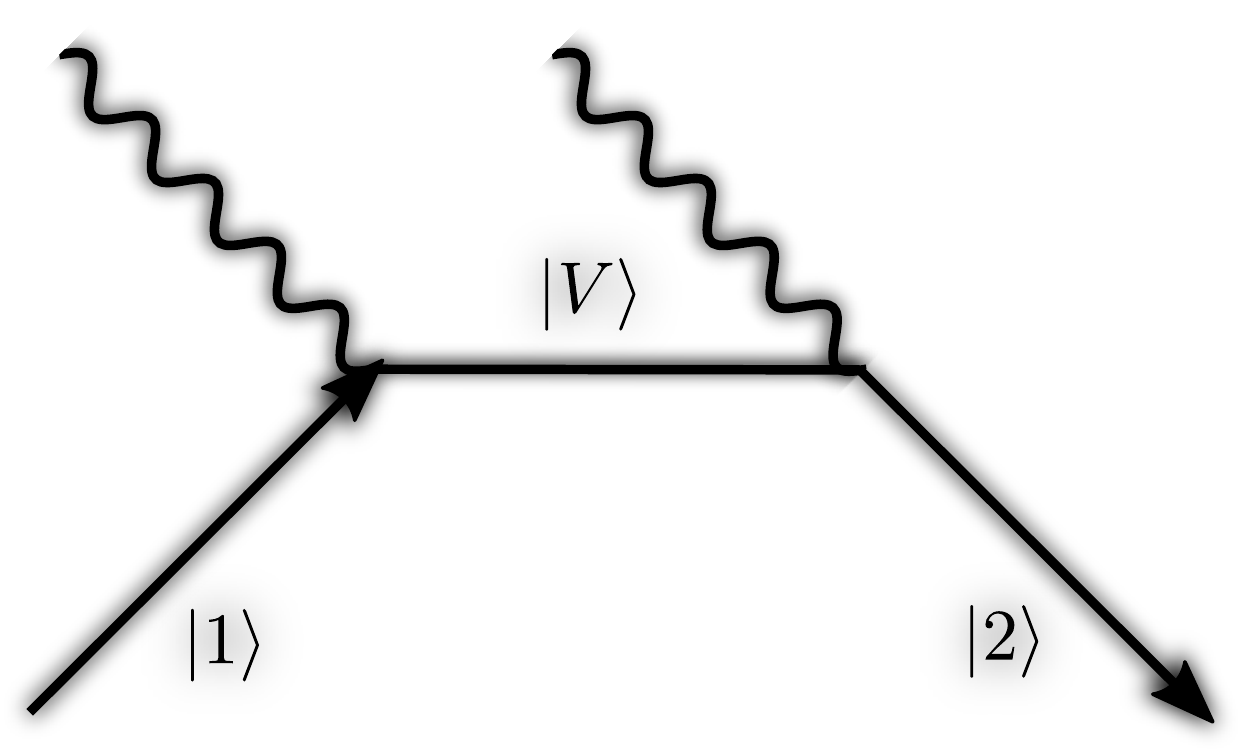}
	\captionsetup{justification=justified}
	\caption[Diagrammatic representation of a virtual process.]{\label{fig:SchemeVirtualState}	\textbf{Diagrammatic representation of a virtual process. }  Two-photon absorption process between states $|1\rangle$ and $|2\rangle$ through a virtual state $|V\rangle$. Wavy lines denote photons. }
\end{SCfigure}

In situations where strongly off-resonant ``virtual'' states mediate interactions between quasi-resonant ``real'' states, an adiabatic elimination over the fast degrees of freedom---the virtual states---provides an effective, dimensionally reduced description of the slow degrees of freedom—the real states---.
 This technique of adiabatic elimination, which can be formulated in several alternatives ways---e.g.,  standard second-order perturbation theory~\cite{SakuraiModernQuantum2017}, the Schrieffer-Wolff transformation~\cite{Cohen-TannoudjiAtomPhotonInteractions1998}, or the Nakajima-Zwanzig approach~\cite{NakajimaQuantumTheory1958,ZwanzigEnsembleMethod1960,BreuerTheoryOpen2007,RivasOpenQuantum2012,Navarrete-BenllochIntroductionQuantum2022,Gonzalez-BallesteroTutorialProjector2024}---is ubiquitous in the description and design of quantum phenomena. Examples of its application are found, e.g., in quantum optics and atomic physics~\cite{GaubatzPopulationTransfer1990,BergmannCoherentPopulation1998,LutkenhausMimickingSqueezedbath1998,WarszawskiAdiabaticElimination2000,BrionAdiabaticElimination2007,DimerProposedRealization2007,BurgarthGeneralizedAdiabatic2019,GamelTimeaveragedQuantum2010,DamanetAtomonlyDescriptions2019,KaufmanAdiabaticElimination2020,BurgarthEternalAdiabaticity2021}, as well as in the exploration of exotic dynamics in the ultrastrong coupling regime of cavity QED~\cite{GarzianoMultiphotonQuantum2015,GarzianoOnePhoton2016,StassiQuantumNonlinear2017}. 
 In fact, a significant effort has been devoted to establish the mathematical foundations of adiabatic elimination~\cite{ComparatGeneralConditions2009,MirrahimiSingularPerturbations2009,PaulischAdiabaticElimination2014} and its extension to dissipative contexts, enabling its application in open quantum systems~\cite{SantosGeneralizedTransitionless2021,ReiterEffectiveOperator2012,AzouitAdiabaticElimination2016,Finkelstein-ShapiroAdiabaticElimination2020}. 

The notion of virtual states can be unambiguously invoked in a Hamiltonian context when the eigenenergies of a bare Hamiltonian $\hat H_0$ are grouped into clusters that are energetically well-separated from each other~\cite{Cohen-TannoudjiAtomPhotonInteractions1998}. 
If we denote these levels as $E_{i,\alpha}$, where the index $i$ labels states within an energy manifold $\mathcal{E}_\alpha^0$, and $\alpha$ labels the different manifolds, such a condition reads
\begin{equation}
	\colorboxed{Maroon}{
	|E_{i,\alpha} - E_{j,\alpha}| \ll |E_{i,\alpha} - E_{j,\beta}|\quad \text{with} \quad  \alpha\neq \beta .
	}
		\label{eq:VirtualityCond1}	
\end{equation}
This condition explicitly reveals the existence of two types of physical degrees of freedom:
\begin{enumerate}[label=\textcolor{Maroon}{(\roman*)}]
\item \textcolor{Maroon}{Slow degrees of freedom:} Differences in the quantum number $i$, i.e., interactions within the same manifold $|E_{i,\alpha}-E_{j,\alpha}|$, lead to small energy shifts. 
\item \textcolor{Maroon}{Fast degrees of freedom:} Differences in the quantum number $\alpha$, i.e., interactions between different manifolds $|E_{i,\alpha}-E_{j,\beta}|$, lead to larger energy changes.
\end{enumerate}

Given a manifold $\alpha$, the eigenstates $|i,\alpha\rangle$ forming it span a Hilbert subspace denoted as $\mathcal H_\alpha$, such that
\begin{equation}
\hat H_0 |i,\alpha\rangle = E_{i,\alpha} |i,\alpha\rangle
\end{equation}
where $P_\alpha=\sum_i |i,\alpha\rangle \langle i, \alpha|$ is the projector onto the manifold $\mathcal{E}_\alpha^0$.
Now, let us consider that the Hamiltonian contains a small perturbation term $\hat V$, 
\begin{equation}
	\hat H= \hat H_0+\hat V,
\end{equation}
yielding a coupling rate between manifolds that is much smaller than their energy difference,
\begin{equation}
\colorboxed{Maroon}{
	|\langle i,\alpha| \hat V |j,\beta\rangle |	 \ll |E_{i,\alpha}-E_{j,\beta}| \quad \text{with} \quad  \alpha\neq \beta .	
	}
	\label{eq:VirtualityCond2}
\end{equation} 
Under this condition, the resulting eigenstate energies remain clustered in manifold structures $\mathcal{E}_{\alpha},\mathcal{E}_{\beta}, \ldots$, which are well separated from each other [see \reffig{fig:EnergyManifolds}]. In this case,  second-order perturbation theory allows the slow dynamics to be described in terms of an effective block-diagonal Hamiltonian that does not couple different manifolds. 
This implies that the evolution of any initial state within a given subspace, that we denote the ``real'' subspace $\mathcal H_R$, evolves effectively constrained within that subspace. The rest of manifolds, which are strongly out-of-resonance, spans the subspace of ``virtual'' states $\mathcal H_V$. These virtual states remain effectively unpopulated at all times---hence their name---but contribute to the dynamics in $\mathcal H_R$ by inducing effective energy shifts and interactions among the ``real'' states.
\begin{SCfigure}[1][h!]
	\includegraphics[width=0.46\columnwidth]{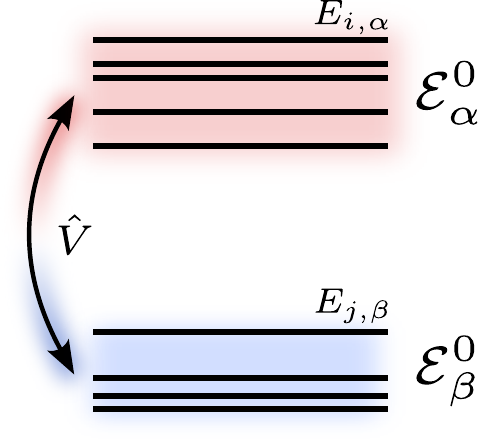}
	\captionsetup{justification=justified}
	\caption[Energy manifolds.]{\label{fig:EnergyManifolds}	\textbf{Energy Manifolds. } Diagrammatic representation of energy levels $E_{i,\alpha}$ grouped into energy manifolds $\mathcal{E}_{\alpha}^0,\mathcal{E}_{\beta}^0, \ldots$ that are well separated from each other. The subscript $i$ denotes the different energy levels within the same manifold. $\hat V$ is a coupling operator that perturbatively connects elements from different manifolds. }
\end{SCfigure}

This discussion establishes the definition of virtual state that we consider here~\cite{Cohen-TannoudjiAtomPhotonInteractions1998}: \textit{a state far-detuned from the subspace where the dynamics is taking place and weakly coupled to it, so that its only effect is to provide effective energy shifts and interactions, without ever becoming populated}.

\section{Minimal model}
\label{sec:2}
\subsection{Definitions} 

In order to unravel the physics behind the unconventional behaviour observed in \reffig{fig:TwoPhotonDecay2}~\textcolor{Maroon}{(b)}, where the symmetric state---acting as a virtual state---develops a sizeable population in the long-time limit,  we consider a simplified three-level system configuration.
Following the nomenclature on \textit{real} and \textit{virtual} states introduced in the previous section, the Hilbert space of this model spans a basis 
\begin{equation}
	\mathcal{H}=\left\{|1\rangle , |2\rangle, |V\rangle \right\},
	\label{eq:HilbertBasis}
\end{equation}
where the states $|1\rangle$ and $|2\rangle$ represent two \textit{real} states, and $|V\rangle$ will play the role of a \textit{virtual} state, being strongly detuned from $|1\rangle$ and $|2\rangle$. 	Given this basis, we define the following lowering operators as 
\begin{equation}
\hat \sigma_{i,j}\equiv |i\rangle \langle j|\quad \text{with}\quad (i,j ={1,2,V}).
\end{equation}
In this configuration, the real states $|1\rangle$ and $|2\rangle$ are coupled to $|V\rangle$ with a coupling rate $\Omega$. Additionally, there is an irreversible decay process within the real subspace, where state $|2\rangle$ decays towards $|1\rangle$ with a decay rate $\Gamma$. 
A diagrammatic illustration of the system is provided in \reffig{fig:fig1_setup}.

By comparing this specific $\Lambda$-model  to the system introduced earlier in this Chapter [see \reffig{fig:TwoPhotonDecay2}~\textcolor{Maroon}{(b)}], we observe the following correspondences: 
the ground state $|gg\rangle$ corresponds to the real state $|1\rangle$, 
the doubly-excited state $|ee\rangle$ maps to the excited real state $|2\rangle$, 
and the symmetric single-excitation state $|+\rangle$---detuned from the two-photon transition energy due to the interaction between qubits---represents the virtual state $|V\rangle$. 
The two-photon decay channel could be enabled, for instance, by a cavity in resonance with the two-photon transition~\cite{DelValleTwophotonLasing2010,OtaSpontaneousTwoPhoton2011}. 
A transformation to the rotating frame of the drive directly leads to the configuration depicted in \reffig{fig:fig1_setup}, where $\gamma_+$ is effectively set to zero, assuming it is negligible compared to the two-photon decay rate.
\begin{SCfigure}[1][h!]
	\includegraphics[width=0.54\columnwidth]{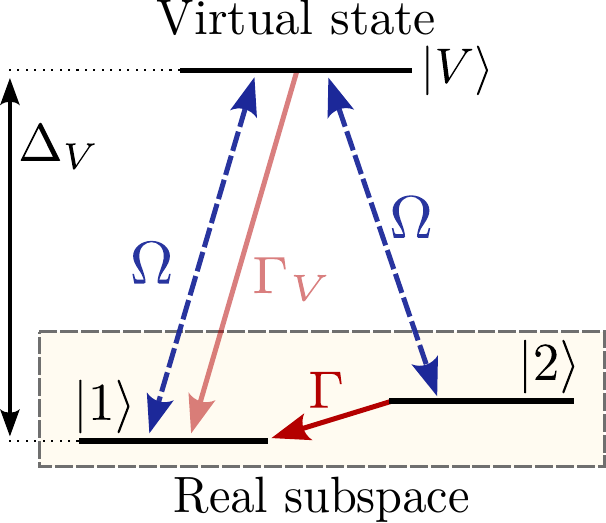}
	\captionsetup{justification=justified}
	\caption[Scheme of the $\Lambda$-system.]{\label{fig:fig1_setup}	\textbf{Scheme of the $\Lambda$-system. }  Two quasi-resonant ``real'' states, interacting via a third, strongly off-resonant ``virtual'' state. There is spontaneous decay between the real states, from $|2\rangle$ towards $|1\rangle$ with a rate $\Gamma$. There may be an additional decay channel from the virtual state $|V\rangle$ towards $|1\rangle$ with a rate $\Gamma_V$.}
\end{SCfigure}
%
%


	%
	%
	%
	%
The resulting time-independent Hamiltonian is expressed as $\hat H=\hat H_0+\hat H_d$, where $\hat H_0$ is the bare Hamiltonian,
\begin{equation}
		\hat H_0=\Delta_1 |1\rangle \langle 1  |+ \Delta_2 |2\rangle \langle 2 | + \Delta_V |V\rangle \langle V|, 
	\label{eq:Hamiltonian_0}
\end{equation}
and $\hat H_d$ is the interaction Hamiltonian
\begin{equation}
	\hat H_d=\Omega \left(\hat \sigma_{1,V}+\hat \sigma_{2,V}+ \text{H.c}  \right), 
\label{eq:Hamiltonian_d}
\end{equation}
where $\Delta_i$ ($i=1,2,V$) stand for the free energy parameters and $\Omega$ denotes the coupling rate. To satisfy the virtuality conditions given by \cref{eq:VirtualityCond1,eq:VirtualityCond2}, we will assume that
\begin{equation}
\Delta_V\gg \Delta_2,\Delta_1,\Omega, \quad \text{and}\quad \Delta_2\approx \Delta_1 \approx 0.
\end{equation}
We finally consider that the evolution of the system is governed by a Lindblad master equation~\cite{BreuerTheoryOpen2007,RivasOpenQuantum2012},
\begin{equation}
	\colorboxed{Maroon}{	\frac{d \hat\rho}{dt}= -i[\hat H, \hat \rho]+\frac{\Gamma}{2} \mathcal{D}[\hat \sigma_{12}]\hat \rho  + \frac{\Gamma_V}{2} \mathcal{D}[\hat \sigma_{1V}]\hat \rho ,}
	\label{eq_Master}
\end{equation}
where the Lindblad terms describe processes of spontaneous decay.
Unless otherwise specified, we will assume $\Gamma_V=0$ throughout the Chapter, meaning there is only one process of spontaneous decay, from $|2\rangle$ to $|1\rangle$. The case where $\Gamma_V\neq 0$ will be also discussed to test the robustness of the effect against additional dissipation channels.

\vspace{-3mm}
\subsection{Dynamics and metastability}
\vspace{-3mm}
The dynamics of the system can be straightforwardly studied by numerically solving the master equation in \eqref{eq_Master}, e.g., in QuTiP~\cite{JohanssonQuTiPOpensource2012,JohanssonQuTiPPython2013,LambertQuTiP52024}. However, to gain deeper insights into the internal structure of the density matrix, which is otherwise inaccessible, it is advantageous to perform its spectral decomposition [see \refsec{sec:LiouvProp}]. As we will elaborate later in this section, the spectral decomposition of the density matrix enables a detailed analysis of the different spectral contributions to the dynamics and facilitates the identification of metastable regimes by analysing the eigenvalues of the Liouvillian~\cite{MacieszczakTheoryMetastability2016,MacieszczakTheoryClassical2021,BrownUnravelingMetastable2024}. Then, the general solution for $\rho(t)$ reads
\\[-1em]
\begin{equation}
	\hat \rho(t)= \hat \rho^{\mathrm{ss}}+\sum_{\mu=2}^{9}e^{\Lambda_\mu t} \text{Tr}[ \hat \Lambda_\mu^L \hat \rho(0)] \hat \Lambda_\mu^R,
	\label{eq:spectralDecomp}
\end{equation}
\\[-1em]
where $\hat \Lambda_\mu^{L/R}$ are the left and right eigenmatrices associated with the eigenvalue $\Lambda_\mu$, and $ \hat \rho^{\mathrm{ss}}$ is the steady-state density matrix. The steady-state can be analytically determined by identifying the nullspace of the Liouvillian [see \refsec{sec:LiouvProp}], such that, by recalling the order of the basis $\left\{|1\rangle , |2\rangle, |V\rangle \right\}$,
\begin{equation}
	\colorboxed{Maroon}{
		\text{ \fontsize{7.0}{10}\selectfont $
		\hat \rho^{\mathrm{ss}}=
	\begin{pmatrix}
		\frac{\Gamma ^2 \Omega ^2+\Gamma ^2 \Delta _V^2+4 \Omega
			^4}{2 \Gamma ^2 \Omega ^2+\Gamma ^2 \Delta _V^2+12
			\Omega ^4} & -\frac{2 i
			\Gamma  \Omega ^2 \Delta _V}{2 \Gamma ^2 \Omega
			^2+\Gamma ^2 \Delta _V^2+12 \Omega ^4} & -\frac{\Gamma  \Omega  \left(\Gamma  \Delta
			_V-2 i \Omega ^2\right)}{2 \Omega ^2 \left(\Gamma ^2+6
			\Omega ^2\right)+\Gamma ^2 \Delta _V^2} \\
			\frac{2 i \Gamma  \Omega ^2 \Delta _V}{2 \Gamma ^2 \Omega
				^2+\Gamma ^2 \Delta _V^2+12 \Omega ^4} 
			& 
				\frac{4 \Omega ^4}{2 \Gamma
				^2 \Omega ^2+\Gamma ^2 \Delta _V^2+12 \Omega ^4}
			&- \frac{2 i \Gamma  \Omega ^3}{2 \Gamma ^2 \Omega
			^2+\Gamma ^2 \Delta _V^2+12 \Omega ^4} \\
	-\frac{\Gamma  \Omega  \left(\Gamma  \Delta _V+2 i \Omega
	^2\right)}{2 \Omega ^2 \left(\Gamma ^2+6 \Omega
	^2\right)+\Gamma ^2 \Delta _V^2} 			
			& \frac{2 i
			\Gamma  \Omega ^3}{2 \Gamma ^2 \Omega ^2+\Gamma ^2
			\Delta _V^2+12 \Omega ^4} & 		
			\frac{\Omega ^2
				\left(\Gamma ^2+4 \Omega ^2\right)}{2 \Omega ^2
				\left(\Gamma ^2+6 \Omega ^2\right)+\Gamma ^2 \Delta
				_V^2} 
	\end{pmatrix}.
	$}
	}
	\label{eq:rho_ss}
\end{equation}
In \reffig{fig:Fig5_Dynamics}~\textcolor{Maroon}{(a)}, we show the occupation probability of the excited state, $\rho_{2,2}\equiv \langle 2|\hat\rho|2\rangle$ and the virtual state $\rho_{V,V} \equiv \langle V|\hat\rho| V\rangle$ in terms of the normalized time $t\Gamma$. The figure clearly reveals the existence of two distinct relaxation timescales:
\begin{enumerate}[label=\textcolor{Maroon}{(\roman*)}]
	\item  	\textcolor{Maroon}{First relaxation timescale. } 
	%
	Initially,  the system behaves according to the standard intuition on virtual states: $|V\rangle$ remains unpopulated, mediating the interaction between $|1\rangle$ and $|2\rangle$. This interaction gives rise to coherent Rabi oscillations between these two states with a two-photon Rabi frequency $\Omega_{2\mathrm p} =  \Omega^2/\Delta_V$, damped by spontaneous emission at a rate $\Gamma$, eventually leading to a stationary state at a timescale $t \sim 1/\Gamma$. 
	This situation can be effectively described in terms of a coherently driven two-level system spanned by $|1\rangle$ and $|2\rangle$.
	\item \textcolor{Maroon}{Second relaxation timescale. } 
	The stationary state from the first timescale is, however, metastable. In a much longer timescale, $\rho_{V,V}$ develops a population comparable to $\rho_{2,2}$, indicating a breakdown of the initial virtual state behaviour. Note that for this particular choice of parameters, the second relaxation timescale occurs at $t\sim 10^4/\Gamma$.
\end{enumerate}
\begin{SCfigure}[1][h!]
	\includegraphics[width=1.\columnwidth]{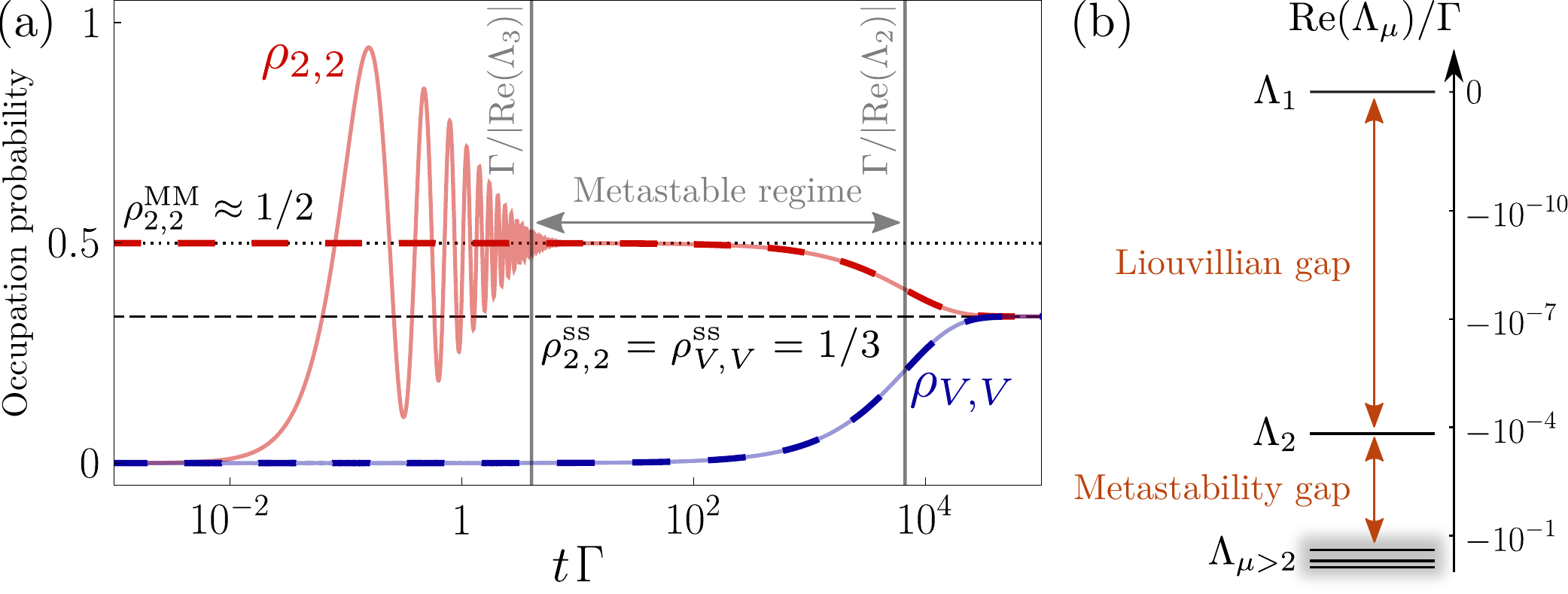}
	\captionsetup{justification=justified}
\end{SCfigure}
\addtocounter{figure}{-1}
\graffito{\vspace{-5.5cm}
	\captionof{figure}[Dynamics and metastability. ]{\label{fig:Fig5_Dynamics} \textbf{Dynamics and metastability. }
		(a) Dynamics of $\rho_{2,2}(t)$ and $\rho_{V,V}(t)$. Solid lines are numerical calculations using \eqref{eq:spectralDecomp},  and dashed lines are numerical calculation from the truncated spectral decomposition up to $m=2$ from \eqref{eq:spectralDecompApprox}. 
		%
		%
		(b) Liouvillian spectrum illustrating the defining feature of metastability in open quantum systems: a spectral gap between  $\Lambda_2$ and  $\Lambda_{\mu>2}$.
		Parameters: $\Omega/\Delta_V=0.01$, $\Gamma/\Delta_V=10^{-5}$. The initial state was set to be $\hat \rho(0)=|1\rangle \langle 1|$.
	}
}
Clear evidences of this metastable behaviour in open quantum systems can be found in the spectrum of the Liouvillian superoperator $ \mathcal{\hat L}$~\cite{MacieszczakTheoryMetastability2016,MacieszczakTheoryClassical2021,BrownUnravelingMetastable2024}. 
All these eigenvalues  $\{\Lambda_\mu,\  \mu=1,2,\ldots \}$---ordered by its real values, so that $\Re(\Lambda_\mu)\geq \Re(\Lambda_{\mu+1})$---have a negative real part, except for the eigenvalue with the largest real part, $\Lambda_1=0$, whose eigenstate corresponds to the steady-state of the system [see \refsec{sec:LiouvProp}].  
The second largest real value of the Liouvillian spectrum, $\Re(\Lambda_2)$, is the Liouvillian gap~\cite{EvansIrreducibleQuantum1977,EvansGeneratorsPositive1979,KesslerDissipativePhase2012}, which determines the relaxation time necessary to reach the steady state, $\tau_2 = 1/|\Re(\Lambda_2)|$.

Following the \textit{spectral theory of metastability}~\cite{MacieszczakTheoryMetastability2016,MacieszczakTheoryClassical2021,BrownUnravelingMetastable2024},  a large separation in the real part of the spectrum between the $m$th Liouvillian eigenvalue and the $(m+1)$th gives rise to a metastable behaviour in the system dynamics. This conditions is expressed as
\begin{equation}
	\label{eq:Metacond}
\colorboxed{Maroon}{\Re(\Lambda_{m})\gg \Re(\Lambda_{m+1}) \quad  \quad \text{(Metastability condition)}.}
\end{equation}
At times beyond the first relaxation, $t\gtrsim1/|\text{Re}(\Lambda_m)|$, terms $e^{t \Lambda_\mu}\approx0$ for $\mu \geq m+1$ can be neglected, as they become negligible  these timescales. As a consequence, the formal expression for the time-dependent density matrix in \eqref{eq:spectralDecomp} can be restricted to the $(m-1)$-dimensional metastable manifold (MM)~\cite{MacieszczakTheoryClassical2021}---characterized by the bounded coefficients $\{ \text{Tr}[ \hat \Lambda_\mu^L \hat \rho(0)],\  \mu=2,\ldots, m\}$---, allowing the density matrix to be approximated as
\begin{equation}
\hat \rho^{\text{MM}}(t)\approx \hat \rho^{\mathrm{ss}}+\sum_{\mu=2}^m e^{\Lambda_\mu t} \text{Tr} [ \hat \Lambda_\mu^L \hat \rho(0) ] \hat \Lambda_\mu^R.
	\label{eq:spectralDecompApprox0}
\end{equation}
Additionally, for large spectral separations, we can approximate $e^{t \Lambda_\mu}\approx1$ for $\mu \leq m$~\cite{MacieszczakTheoryClassical2021,BellomoQuantumSynchronization2017}, yielding an expression that describes the metastable regime, during which the system state is approximately stationary:
\begin{equation}
\colorboxed{Maroon}{
	\hat \rho^{\text{MM}}\approx \hat \rho^{\mathrm{ss}}+\sum_{\mu=2}^m \text{Tr}[ \hat \Lambda_\mu^L \hat \rho(0)] \hat \Lambda_\mu^R
	\quad \quad \text{(Metastable density matrix).}
}
\label{eq:spectralDecompApprox}
\end{equation}
%
%
%
%

In our particular case, there is only one metastable state rather than a manifold of metastabilities: $\Lambda_2$ is well separated from the rest of eigenvalues by a second gap, such that $
\Re(\Lambda_2)\gg \Re(\Lambda_3)$.
The system relaxes to a metastable state in a timescale
\begin{equation}
	\tau_3 = 1/|\Re(\Lambda_3)| \quad \quad \text{(Metastability timescale),}
\end{equation}
 as described by \eqref{eq:spectralDecompApprox} with $m=2$. Particularly, for the particular choice of parameters used in \reffig{fig:Fig5_Dynamics}, the state $|2\rangle$ acquires a metastable population given by
\begin{equation}
	\rho_{2,2}^{\text{MM}}=\langle 2| \hat \rho^{\text{MM}}|2\rangle = \hat \rho_{2,2}^{\mathrm{ss}}+ \text{Tr}[ \hat \Lambda_2^L \hat \rho(0)]  \langle 2| \hat \Lambda_2^R|2\rangle \approx \frac{1}{2},
\end{equation}
where $\hat \rho_{2,2}^{\mathrm{ss}}$ is obtained from \eqref{eq:rho_ss}, while $\hat \Lambda_2^{L/R}$ must be computed numerically.
Eventually, the metastable state evolves into the actual steady-state in a time $\tau_2 \gg \tau_3$,
\begin{equation}
	\tau_2 = 1/|\Re(\Lambda_2)| \quad \quad \text{(Relaxation timescale).}
\end{equation}
As shown in \reffig{fig:Fig5_Dynamics}~\textcolor{Maroon}{(a)}, the approximated time-dependent density matrix in \eqref{eq:spectralDecompApprox0} (red and blue dashed lines) and the metastable density matrix in \eqref{eq:spectralDecompApprox} (black dotted line), with $m=2$, perfectly match the dynamics of $\rho_{2,2}$ and $\rho_{V,V}$ after the first relaxation at $t\sim 1/|\text{Re}(\Lambda_3)|$.

The system under consideration exhibits precisely the clustering of eigenvalues characteristic of metastability, according to \eqref{eq:Metacond}. This spectral structure is evident in \reffig{fig:Fig5_Dynamics}~\textcolor{Maroon}{(b)}, where we confirm that $\tau_3 \sim 1/\Gamma$ and $\tau_2 \sim 10^4/\Gamma$, and thus the existence of a metastability gap. 
In fact, considering the analytical expression for the stationary density matrix in \eqref{eq:rho_ss}, we observe that in the limit  $\Omega^2 \gg \Gamma \Delta_V$, the virtual state population yields a sizeable population
\begin{equation}
		\rho_{V,V}^{\mathrm{ss}}=\frac{\Omega ^2 \left(\Gamma ^2+4 \Omega ^2\right)}{2 \Omega ^2 \left(\Gamma ^2+6 \Omega ^2\right)+\Gamma ^2 \Delta _V^2} \xrightarrow[\Omega^2 \gg \Gamma \Delta_V]{} \frac{1}{3},
\end{equation}
%
clearly establishing that the virtual state will get populated in the long time limit. 


\section{Hierarchical adiabatic elimination (HAE)}
	\label{sec:3}
In order to have an estimate of the survival time of the metastable state, it is desirable to obtain an analytical expression of $\Lambda_2$, i.e., the Liouvillian gap. Despite the reduced dimensionality of the system, a direct analytical solution for the time evolution of the density matrix through the diagonalization of the Liouvillian superoperator $\mathcal{\hat L}$ is not readily achievable. However, as discussed in the previous section, the system exhibits a clear hierarchy of timescales [see \reffig{fig:Fig5_Dynamics}~\textcolor{Maroon}{(a)}], suggesting that a series of adiabatic eliminations could be employed:
\begin{enumerate}[label=\textcolor{Maroon}{(\roman*)}]
	\item \textcolor{Maroon}{First adiabatic elimination.}  
	The shortest timescale is clearly governed by Hamiltonian dynamics, evidenced by a fast oscillatory evolution of the density matrix elements. This oscillatory dynamics stabilizes into a metastable state in a timescale of the order $\tau_3 \sim 1/\Gamma$. During this regime, the states $|1\rangle$ and $|2\rangle$ span the ``real'' subspace $\mathcal{H}_R$---as defined in \refsec{sec:1},  where real states correspond to slow variables, and virtual states to fast variables---while the state $|V\rangle$ plays the role of a fast variable: $|V\rangle$ mediates effective interactions within the real subspace, i.e., it plays the role of a virtual state that can be eliminated within a purely Hamiltonian evolution. 
	This forms the basis for the first adiabatic elimination.
	\item \textcolor{Maroon}{Second adiabatic elimination.} 
	The longest timescale is characterized by the very slow evolution of $|V\rangle$, governed by the actual relaxation timescale $\tau_2\sim 1/|\text{Re}(\Lambda_2)|$. 
	For this variable evolving slowly within this very long characteristic timescale, the relaxation of the ``real'' states---i.e., $|1\rangle$ and $|2\rangle$---occurs almost instantaneously ($1/\Gamma$).
	As a result, the roles of  ``slow'' and ``fast'' variables from the first adiabatic elimination are effectively reversed: 
	 one can treat the ``real'' variables as the ``virtual'' variables in a dissipative sense, i.e., they relax quickly into a time-dependent quasi-steady state that follows the slow evolution of $\rho_{V,V}$. 
	 This forms the basis for the second adiabatic elimination.
\end{enumerate}
It is important to note that a direct application of standard adiabatic elimination techniques in dissipative context, e.g., the projection-operator method~\cite{BreuerTheoryOpen2007,RivasOpenQuantum2012,Finkelstein-ShapiroAdiabaticElimination2020}, would eliminate the virtual subspace entirely, failing to capture the mechanism that populates this state in the long-time limit. 
Instead, more detailed techniques must be applied, such as artificially extending the Hilbert space~\cite{Gonzalez-BallesteroTutorialProjector2024}.
Here, we introduce a novel two-step approach that we term \textit{hierarchical adiabatic elimination} (HAE). This method involves performing two consecutive adiabatic eliminations to derive an effective description of the system dynamics and, when possible, tractable analytical expressions for the time-dependent density matrix and the relaxation timescale.
We explore the validity of the results obtained using the HAE method across the parameter space, showing excellent agreement with numerical simulations. Finally, we highlight the generality of this approach by extending it to the arbitrary case of  $N$ quasi-resonant real states and $M$ off-resonant virtual states.

\subsection{First adiabatic elimination}
Our starting point is the set of differential equations describing the evolution of the elements of the total density matrix, obtained from \eqref{eq_Master} as
	\begin{subequations}
		\begin{align}
			\label{eq:drho_VV}
			\dot \rho_{V,V}&=
			2 \Omega \Im\left[ \rho_{1,V}-\rho_{V,2}\right], \\
			\label{eq:drho_22}
			\dot \rho_{2,2}&=-\Gamma \rho_{2,2}+2\Omega \Im \left[\rho_{V,2}\right], \\
			\label{eq:drho_1V}
			\dot \rho_{1,V}&=i \Delta_V \rho_{1,V}
			+i\Omega \left[1+\rho_{1,2}-2\rho_{V,V}-\rho_{2,2}\right], \\
			\label{eq:drho_12}
			\dot \rho_{1,2}&=-\frac{\Gamma}{2}\rho_{1,2}+i\Omega \left[\rho_{1,V}-\rho_{V,2}\right], \\
			\label{eq:drho_V2}
			\dot \rho_{V,2}&=-\left(i \Delta_V+\frac{\Gamma}{2}\right)\rho_{V,2}			-i\Omega \left[\rho_{1,2}+\rho_{2,2}-\rho_{V,V}\right].
		\end{align}
	\end{subequations}
The first stage of the dynamics can be completely described within the real Hilbert subspace, $\mathcal{H}_{R}=\left\{\ket{1},\ket{2}\right\}$, after performing an adiabatic elimination of the virtual state $\ket{V}$. More specifically, this adiabatic elimination consists in setting
\begin{equation}
	\dot{\rho}_{1,V}=\dot{\rho}_{V,2}=0.
	\label{eq:FirstAE}
\end{equation}

The is done under the assumption that the energy difference between the real and virtual subspaces is much larger than the coupling rate, i.e., $|\Delta_V - \Delta_i | \gg \Omega$ for $i=1,2$. 
Thus, by solving  \cref{eq:drho_1V,eq:drho_V2} for the virtual coherence terms, we obtain their steady-state values
\begin{equation}
	\{\rho_{1,V}(t),\rho_{V,2}(t) \}\rightarrow \{\rho_{1,V}^{\mathrm{ss}},\rho_{V,2}^{\mathrm{ss}}\} ,
\end{equation}
such that the dynamics of the real subspace gets effectively described by the following differential equations:
\begin{subequations}
	\begin{empheq}{align}
		&\text{ \fontsize{9.5}{10}\selectfont $\dot \rho_{1,2}\approx\left[-\frac{\Gamma}{2}+\frac{\Gamma \Omega^2}{(i\Gamma -2\Delta_V)\Delta_V}\right]\rho_{1,2}(t)-\frac{i\Omega^2}{\Delta_V}$}\notag\\
		&\quad  \quad \text{ \fontsize{9.5}{10}\selectfont $+\left(-\frac{2\Omega^2}{\Gamma+2i\Delta_V}+\frac{i\Omega^2}{\Delta_V}\right)\rho_{2,2}(t)+\left( \frac{2\Omega^2}{\Gamma+2i\Delta_V}+\frac{2i\Omega^2}{\Delta_V} \right)\rho_{V,V}(t),$} 	\label{eq:FirstAdiabatic1} \\
		&\text{ \fontsize{9.5}{10}\selectfont $\dot \rho_{2,2}\approx\left(-\Gamma -\frac{4 \Gamma \Omega^2 }{\Gamma^2+4\Delta_V^2} \right) \rho_{2,2}(t)-\frac{2\Omega^2}{\Gamma+2 i \Delta_V}\rho_{1,2}(t)$} \notag \\
		&\quad \quad\quad \quad \quad \quad \quad \quad\quad \quad \quad \quad \ \text{ \fontsize{9.5}{10}\selectfont $-\frac{2\Omega^2}{\Gamma-2 i \Delta_V}\rho_{2,1}(t)+\frac{4\Gamma \Omega^2}{\Gamma^2+4\Delta_V^2}\rho_{V,V}(t).$}
		\label{eq:FirstAdiabatic2}
	\end{empheq}	
\end{subequations}
In the limit $\Delta_V\gg \Gamma$, we can simplify \cref{eq:FirstAdiabatic1,eq:FirstAdiabatic2} and obtain a more familiar set of equations:
\begin{subequations}
	\begin{empheq}[box=\colorboxed{Maroon}]{align}
		\label{eq:adiabatic-1st-a}
		\dot \rho_{1,2}&\approx-\frac{\Gamma}{2} \rho_{1,2}(t)+i\Omega_{2\mathrm p} [2\rho_{2,2}(t)+\rho_{V,V}(t)-1], \\
		\dot \rho_{2,2}&\approx-\Gamma \rho_{2,2}(t)-2\Omega_{2\mathrm p}\Im [\rho_{1,2}(t)]+\frac{\Gamma \Omega_{2\mathrm p}}{\Delta_V}\rho_{V,V}(t).
		\label{eq:adiabatic-1st-b}
	\end{empheq}
\end{subequations}
These equations resemble those that would correspond to a two-level system resonantly driven with a \textit{two-photon Rabi frequency}, $\Omega_{2\mathrm p }\equiv \Omega^2/\Delta_V$, and a decay rate $\Gamma$. The only addition is an extra term related to the virtual state population $\rho_{V,V}$ which can be treated as a time-independent parameter with a fixed value, e.g., $\rho_{V,V}=0$, since it evolves in a much slower timescale than $\rho_{1,2}$ and $\rho_{2,2}$.  One can clearly see that, for these equations, the relaxation time towards a stationary state occurs in a timescale $\sim 1/\Gamma$. 

The time-dependent analytical solutions of \cref{eq:adiabatic-1st-a,eq:adiabatic-1st-b} are well known and can be found in any quantum optics textbook~\cite{Cohen-TannoudjiAtomPhotonInteractions1998,WallsQuantumOptics2008,GardinerQuantumNoise2004}:
\begin{subequations}
	\begin{empheq}[box=\colorboxed{Maroon}]{align}
		\label{eq:rho_12t_meta}
		&\text{ \fontsize{9.5}{10}\selectfont $	\rho_{1,2}(t)\approx  \frac{-2 i \Omega_{2\mathrm p} \Gamma}{\Gamma^2+8\Omega_{2\mathrm p}^2}\bigg\{1-e^{-3\Gamma t/4} \bigg[ \cosh(\kappa t)  +\left(\frac{\kappa}{\Gamma}+\frac{3\Gamma}{16\kappa}\right)\sinh(\kappa t) \bigg]  \bigg\} $}
		, \\
		\label{eq:rho_22t_meta}
		&\text{ \fontsize{9.5}{10}\selectfont $\rho_{2,2}(t)\approx\frac{4\Omega_{2\mathrm p}^2}{\Gamma^2+8\Omega_{2\mathrm p}^2} 	\bigg\{1-e^{-3\Gamma t/4}\left[\cosh(\kappa t)+\frac{3\Gamma}{4\kappa}\sinh(\kappa t) \right]  \bigg\}, $}
	\end{empheq}
\end{subequations}
where $\kappa\equiv \frac{1}{2}\sqrt{\frac{\Gamma^2}{4}-16\Omega_{2\mathrm p}^2}$. 
Such a two-level system dynamics describes accurately the short-timescale oscillatory dynamics of the system, as will be shown later in \reffig{fig:Fig5_Dynamics_HAE}, where the initial state was set to be $\hat \rho(0)=|1\rangle \langle 1|$.
From these results, it becomes evident that to observe these oscillations, the system must be strongly driven, with $\Omega_{2\mathrm p} \gg \Gamma$; otherwise, the system is overdamped and will basically remain in the ground state $|1\rangle$. 

\subsection{Second adiabatic elimination}

While the usual approach when eliminating a virtual state is to indeed  assume $\rho_{V,V}=0$ for all times, we have already seen that this approach is eventually bound to fail since the virtual state develops a sizeable population within a characteristic timescale $\tau_2\gg 1/\Gamma$.
Crucially, this timescale is orders of magnitude longer than the relaxation time of \cref{eq:adiabatic-1st-a,eq:adiabatic-1st-b}, suggesting that we can make a second adiabatic elimination based on this separation of timescales.

In much longer timescales than $1/\Gamma$, the terms $\rho_{1,2}$ and $\rho_{2,2}$ can be considered as ``fast'' variables since they relax to a steady-state in a very short time in comparison with the relaxation timescale of $\rho_{V,V}$.
From \cref{eq:FirstAdiabatic1,eq:FirstAdiabatic2}, it is clear that, if we assume that $\rho_{V,V}$ will be virtually unchanged in a timescale $\sim 1/\Gamma$, we could take it as a time-independent parameter and obtain a stationary solution for $\rho_{2,2}$ and $\rho_{1,2}$ that depends on the fixed value of $\rho_{V,V}$. This \textit{quasi-steady state} will \textit{adiabatically follow any slow change} of $\rho_{V,V}$ taking place in a much longer characteristic timescale. The expression of this $\rho_{V,V}$-dependent steady state can be derived by solving a linear system of equations of the form $\hat M.\vec{\rho}+\vec{b}=0$ for the vector $\vec{\rho}=(\rho_{2,2}^{\mathrm{ss}},\rho_{2,1}^{\mathrm{ss}},\rho_{1,2}^{\mathrm{ss}})^T$, where $\hat M$ is given by
\begin{equation}
	\hat M=	\begin{bmatrix}
		-\Gamma\left(1+\frac{4 \Omega^2}{\Gamma ^2+4 \Delta_V
			^2}\right) & 
		-\frac{2 \Omega ^2}{\Gamma -2i \Delta_V } &
		-\frac{2 \Omega ^2}{\Gamma +2i \Delta_V }  \\
		-\frac{2\Omega^2}{\Gamma -2 i\Delta_V }-i\Omega_{2\mathrm p}
		&  
		-\Gamma\left(\frac{1}{2} -\frac{i\Omega_{2\mathrm p}}{\Gamma- 2i \Delta_V} \right)
		& 0 \\
		-\frac{2\Omega^2}{\Gamma +2 i\Delta_V }+i\Omega_{2\mathrm p} & 0 &   
		-\Gamma \left( \frac{1}{2}-\frac{\Gamma \Omega_{2\mathrm p}}{		i \Gamma -2 \Delta_V }\right)
	\end{bmatrix},
\end{equation}
and $\vec b$ reads
\begin{equation}
		\vec{b}=
	\begin{pmatrix}
		\frac{4 \Gamma  \Omega ^2}{\Gamma ^2+4\Delta_V ^2} \\ 
		\frac{2\Omega ^2}{\Gamma -2 i\Delta_V }-2i\Omega_{2\mathrm p}\\ 
		\frac{2\Omega ^2}{\Gamma +2 i\Delta_V }+2i\Omega_{2\mathrm p}
	\end{pmatrix}	\rho _{V,V}(t)+
	\begin{pmatrix}
		0 \\ 
		i\Omega_{2\mathrm p} \\ 
		-i\Omega_{2\mathrm p}
	\end{pmatrix}.
\end{equation}
By solving this linear system, we obtain a set of equations for the quasi-stationary values of $\rho_{2,2}$ and $\rho_{1,2}$ that depend on the population of the virtual state at any given time,
\begin{subequations}
	\begin{empheq}[box=\colorboxed{Maroon}]{align}
		& \text{ \fontsize{9.0}{10}\selectfont $\rho_{2,2}^{\mathrm{ss}}[\rho_{V,V}(t)]=\frac{16\Omega^4(\Delta_V^2
				+\Omega^2)}{\chi}+\frac{4\left[\Gamma^2\Delta_V^2\Omega^2-4\Omega^4(\Delta_V^2+\Omega^2)\right]}{\chi}\rho_{V,V}(t),$ } 
		\label{eq:rho22solution} \\
		&\text{ \fontsize{9.0}{10}\selectfont $\rho_{1,2}^{\mathrm{ss}}[\rho_{V,V}(t)]=-\frac{2 \Omega ^2 \left[i
				\Gamma  \Delta_V 
				\left(\Gamma ^2+4 \Delta_V
				^2\right)+2 \Gamma  \Omega
				^2 (\Gamma +4 i \Delta_V )+8
				\Omega ^4\right]}{\chi} $}\notag \\
		&\quad \ \ \text{ \fontsize{9.0}{10}\selectfont $+ \frac{4 \Omega ^2  \left(2 \Gamma  \Omega ^2 (\Gamma +4 i
				\Delta_V )+\Gamma  \Delta_V  (\Gamma +i \Delta_V ) (2 \Delta +i \Gamma )+12
				\Omega ^4\right)}{\chi}\rho_{V,V}(t)$},
		\label{eq:rho12solution}
	\end{empheq}
\end{subequations}
where $\chi$ is defined as
\begin{equation}
			\chi \equiv \Gamma^4 \Delta_V^2 + 32\Omega^2(\Delta_V^2 + \Omega^2) + 4\Gamma^2(\Delta_V^4 + 3\Delta_V^2\Omega^2 + \Omega^4).
\end{equation}

From the first adiabatic elimination conditions [$\dot\rho_{V,i}=0$, \eqref{eq:FirstAE}] and the dynamical equation for the virtual population in \eqref{eq:drho_VV}, we can obtain a differential equation for $\rho_{V,V}$, which is a function of itself and the real-subspace elements, i.e.,  $\dot\rho_{V,V}(t) = f[\rho_{V,V}(t);\rho_{1,2}(t);\rho_{2,2}(t)]$:
\begin{multline}
	\dot \rho_{V,V}(t) \approx-\frac{4\Gamma \Omega^2}{\Gamma^2 + 4\Delta_V^2} \rho_{V,V}(t) +
\frac{\Omega_{2\mathrm p}\Gamma(i\Gamma+2\Delta_V)}{\Gamma^2 + 4\Delta_V^2} \rho_{1,2}(t) \\
  + \frac{\Omega_{2\mathrm p}\Gamma(-i\Gamma+2\Delta_V)}{\Gamma^2 + 4\Delta_V^2} \rho_{2,1}(t) + \frac{4\Gamma \Omega^2}{\Gamma^2 + 4\Delta^2}\rho_{2,2}(t).
\end{multline}
The second adiabatic elimination consists  in substituting $\rho_{1,2}(t)$ and  $\rho_{2,2}(t)$ in that equation by their corresponding steady state solutions of $\rho_{1,2}^{\mathrm{ss}}[\rho_{V,V}(t)]$ and $\rho_{2,2}^{\mathrm{ss}}[\rho_{V,V}(t)]$, \cref{eq:rho12solution,eq:rho22solution}, which are obtained for a given $\rho_{V,V}=\rho_{V,V}(t)$. This yields a dynamical equation that only depends on $\rho_{V,V}(t)$, that is,
\begin{equation}
	 \dot\rho_{V,V}(t) = f[\rho_{V,V}(t);\rho_{1,2}^{\mathrm{ss}}[\rho_{V,V}(t)];\rho_{2,2}^{\mathrm{ss}}[\rho_{V,V}(t)]].
\end{equation}
In this approximation, $\rho_{1,2}$ and $\rho_{2,2}$ act as fast variables that relax into a time-dependent stationary state that follows the slow evolution of $\rho_{V,V}$. Namely, the differential equation for the virtual state population becomes
\begin{equation}
\colorboxed{Maroon}{
\begin{aligned}[b]
	\dot{\rho}_{V,V}(t)= &\frac{4\Gamma \Omega^4(\Gamma^2+4\Omega^2)}{\chi} \\ &\quad \quad -\frac{4\left[12\Gamma\Omega^6+ 
		\Gamma^3\Omega^2(\Delta_V^2+2\Omega^2)\right]}{\chi}\rho_{V,V}(t).
\end{aligned}
}
\end{equation}
Solving this differential equation, we obtain an analytical expression for the evolution of $\rho_{V,V}(t)$
\begin{equation}
	\rho_{V,V}(t)=\rho_{V,V}^{\mathrm{ss}}\left(1-e^{-\Gamma_c t}\right),
\end{equation}
where $\rho_{V,V}^{\mathrm{ss}}$ stands for the steady-state value, and $\Gamma_c$ stands for the relaxation rate,
\begin{equation}
\colorboxed{Maroon}{
	\Gamma_c= \frac{4\left[12\Gamma\Omega^6+	\Gamma^3\Omega^2(\Delta_V^2+2\Omega^2)\right]}{\Gamma^4 \Delta_V^2 + 32\Omega^4(\Delta_V^2 + \Omega^2) + 4\Gamma^2(\Delta_V^4 + 3\Delta_V^2\Omega^2 + \Omega^4)}.
	}
	\label{eq:Gamma_c}
\end{equation}
This expression for the relaxation rate can be further simplified under the assumptions:
 \textcolor{Maroon}{(i)} ${\Delta_V \gg \Omega, \Gamma}$ and \textcolor{Maroon}{(ii)} the effective strong driving regime ${\Omega_{2\mathrm p} \gg \Gamma}$, such that
 \begin{equation}
\Gamma_c^{\text{approx}}
 			\approx \frac{3 \Gamma \Omega^2}{2\Delta_V^2}.
 	\label{eq:Gamma_c_approx}
 \end{equation}
The desired expression for the survival time of the metastable regime is then given by the inverse of \cref{eq:Gamma_c,eq:Gamma_c_approx}, and thus, must correspond to the Liouvillian gap, 
\begin{equation}
	\colorboxed{Maroon}{
	\Gamma_c = |\Re(\Lambda_2)| \quad \quad \text{(Liouvillian gap)}.
	}
\end{equation}
\begin{SCfigure}[1][b!]
	\includegraphics[width=0.65\columnwidth]{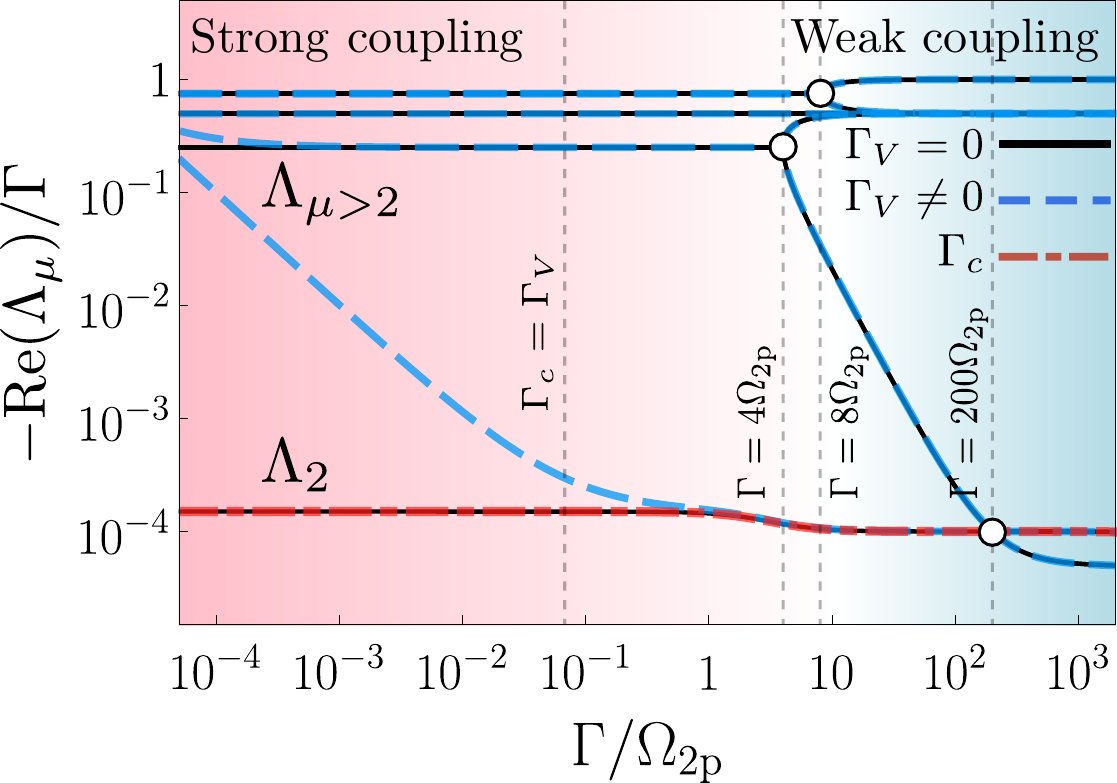}
	\captionsetup{justification=justified}
	\caption[Liouvillian spectrum and effective relaxation rate $\Gamma_c$.]{\label{fig:LiouvillianEigen}	\textbf{Liouvillian spectrum and effective relaxation rate $\Gamma_c$. } Liouvillian eigenvalues in terms of $\Gamma$	for $\Gamma_V=0$ (solid black lines), and $\Gamma_V=10^{-5}\Omega_{2\mathrm p }$ (blue dashed lines). Our analytical prediction for the value of the Liouvillian gap $\Lambda_2$, i.e.,  $\Gamma_c$ from \eqref{eq:Gamma_c}, is depicted by a red dashed line. If $\Gamma_V\neq0$, it is seen that the metastability disappears when $\Gamma_c < \Gamma_V$ as the metastability gap closes.
		Parameters:  $\Omega/\Delta_V=0.01$.
	}
\end{SCfigure}
We have confirmed that this is indeed the case in \reffig{fig:LiouvillianEigen}, which depicts the real part of the Liouvillian spectrum, $\text{Re}(\Lambda_\mu)$, as a function of $\Gamma$.
The lower part of the figure illustrates a perfect agreement between our analytical  expression for $\Gamma_c$ [\eqref{eq:Gamma_c},  red dot-dashed line]  and $\Lambda_2$ (black solid line) in both the weak ($\Gamma\ll \Omega_{2\mathrm{p}}$) and strong ($\Gamma\gg \Omega_{2\mathrm{p}}$) driving regimes.
Additionally, the case $\Gamma_V\neq0$---i.e., the non-zero decay from the virtual state---is also considered in this figure (blue dashed lines), showing that the metastability disappears when $\Gamma_c \lesssim \Gamma_V$. In this scenario, the dissipation from $|V\rangle$ dominates the dynamics, reducing the system to the simpler behaviour of a coherently driven two-level system.
Finally, the white circles correspond to \textit{exceptional points} (EP), which represent bifurcations, convergence, or crossings of the Liouvillian eigenvalues. EPs are associated to the emergence of exotic phenomena, such as quantum phase transitions, characterization of topological phases, or novel enhanced sensing~\cite{MullerExceptionalPoints2008,OzdemirParityTime2019,MiriExceptionalPoints2019,MingantiQuantumExceptional2019,RoccatiNonHermitianPhysics2022}.

\paragraph{Summary of analytic expressions for the time-dependent  density matrix elements. }
Once we know the analytic expression for the time-dependent virtual state population, $\rho_{V,V}(t)$, the remaining equations are easily computed:
		%
		%
		%
		%
		%
		%
		%
		%
		%
		%
%
%
\begin{equation}
	\colorboxed{Maroon}{
	\begin{aligned}[b]
	&\text{ \fontsize{6.5}{10}\selectfont $\rho_{V,V}(t)\approx \rho_{V,V}^{\mathrm{ss}} \left[1-e^{-\Gamma_c t}\right] ,$} \\
	&\text{ \fontsize{6.5}{10}\selectfont $\rho_{2,2}(t)\approx \rho_{2,2}^{\mathrm{ss}} \left[1+\frac{4 \Omega ^4 \left(\Gamma ^2+4 \Delta_V ^2\right)-\Gamma ^4 \Delta_V
		^2+16 \Omega ^6}{\Gamma^4 \Delta_V^2 + 32\Omega^2(\Delta_V^2 + \Omega^2) + 4\Gamma^2(\Delta_V^4 + 3\Delta_V^2\Omega^2 + \Omega^4)}e^{-\Gamma_c t}\right],$}    \\
	&\text{ \fontsize{6.5}{10}\selectfont $\rho_{1,2}(t)\approx \rho_{1,2}^{\mathrm{ss}}  \left[1-\frac{2 i \Omega ^2 \left(\Gamma ^2+4 \Omega ^2\right) \left(2 \Gamma 
		\Omega ^2 (\Gamma +4 i \Delta_V )+\Gamma  \Delta_V  (\Gamma +i \Delta_V )
		(2 \Delta_V +i \Gamma )+12 \Omega ^4\right)}{\Gamma  \Delta_V  [\Gamma^4 \Delta_V^2 + 32\Omega^2(\Delta_V^2 + \Omega^2) + 4\Gamma^2(\Delta_V^4 + 3\Delta_V^2\Omega^2 + \Omega^4)]}e^{-\Gamma_c t}\right] ,$}  \\
	&\text{ \fontsize{6.5}{10}\selectfont $ \rho_{1,V}(t)\approx \rho_{1,V}^{\mathrm{ss}}  \left[1-\frac{2 \Omega ^2 \left(\Gamma ^2+4 \Omega ^2\right) \left(-2 i \Gamma 
		\Omega ^2 \left(\Gamma ^2+6 i \Gamma  \Delta_V +2 \Delta_V
		^2\right)+\Gamma ^2 \Delta_V  \left(\Gamma ^2+4 \Delta_V ^2\right)+8
		\Omega ^4 (3 \Delta_V -2 i \Gamma )\right)}{\Gamma  \left(\Gamma 
		\Delta_V -2 i \Omega ^2\right) [\Gamma^4 \Delta_V^2 + 32\Omega^2(\Delta_V^2 + \Omega^2) + 4\Gamma^2(\Delta_V^4 + 3\Delta_V^2\Omega^2 + \Omega^4)] }e^{-\Gamma_c t}\right]$} ,  \\
	& \text{ \fontsize{6.5}{10}\selectfont $\rho_{2,V}(t)\approx \rho_{2,V}^{\mathrm{ss}}  \left[1-\frac{\left(\Gamma ^2+4 \Omega ^2\right) \left(\Gamma ^2 \Delta_V ^2
		(\Gamma +2 i \Delta_V )-4 \Omega ^4 (\Gamma -6 i \Delta_V )+4 i \Gamma 
		\Delta_V  \Omega ^2 (\Gamma +i \Delta_V )\right)}{
			\Gamma[\Gamma^4 \Delta_V^2 + 32\Omega^2(\Delta_V^2 + \Omega^2) + 4\Gamma^2(\Delta_V^4 + 3\Delta_V^2\Omega^2 + \Omega^4)] 
	}e^{-\Gamma_c t}\right]$}, 
	\end{aligned}
}
\label{eq:rho_HAE}
\end{equation}
where $\rho_{i,j}^{\mathrm{ss}}$ $(i,j=1,2,V)$ are the steady state density matrix elements from \eqref{eq:rho_ss}.  The stationary values of $\rho_{1,2}$ and $\rho_{2,2}$ reached in the metastable regime correspond to  the long-time limit of \cref{eq:rho_22t_meta,eq:rho_12t_meta},
\begin{subequations}
	\label{eq:TwoPhotonSS}
	\begin{empheq}[box=\colorboxed{Maroon}]{align}
	&\rho_{1,2}^{\text{MM}}=\frac{-2i\Gamma \Omega_{2\mathrm p}}{\Gamma^2+8\Omega_{2\mathrm p}^2} ,
	\label{eq:rho_12_metaSS}\\
	& \rho_{2,2}^{\text{MM}}=\frac{4\Omega_{2\mathrm p}^2}{\Gamma^2+8\Omega_{2\mathrm p}^2} .
		\label{eq:rho_22_metaSS}
\end{empheq}
\end{subequations}

In \reffig{fig:Fig5_Dynamics_HAE}, we observe a perfect match between the analytical expressions derived from the HAE (dashed lines) for the time-dependent density matrix elements and the numerical simulations (solid lines) from direct simulation of the master equation in \eqref{eq_Master}. Specifically, the figure illustrates the time evolution of $\rho_{2,2}(t)$ and $\rho_{V,V}(t)$.
Furthermore,  the effective description of the Liouvillian gap, $\Gamma_c$, using either the general expression [\eqref{eq:Gamma_c}] or its approximated version [\eqref{eq:Gamma_c_approx}], accurately captures the relaxation timescale leading to the stationary state. 

\begin{SCfigure}[1][h!]
	\includegraphics[width=0.75\columnwidth]{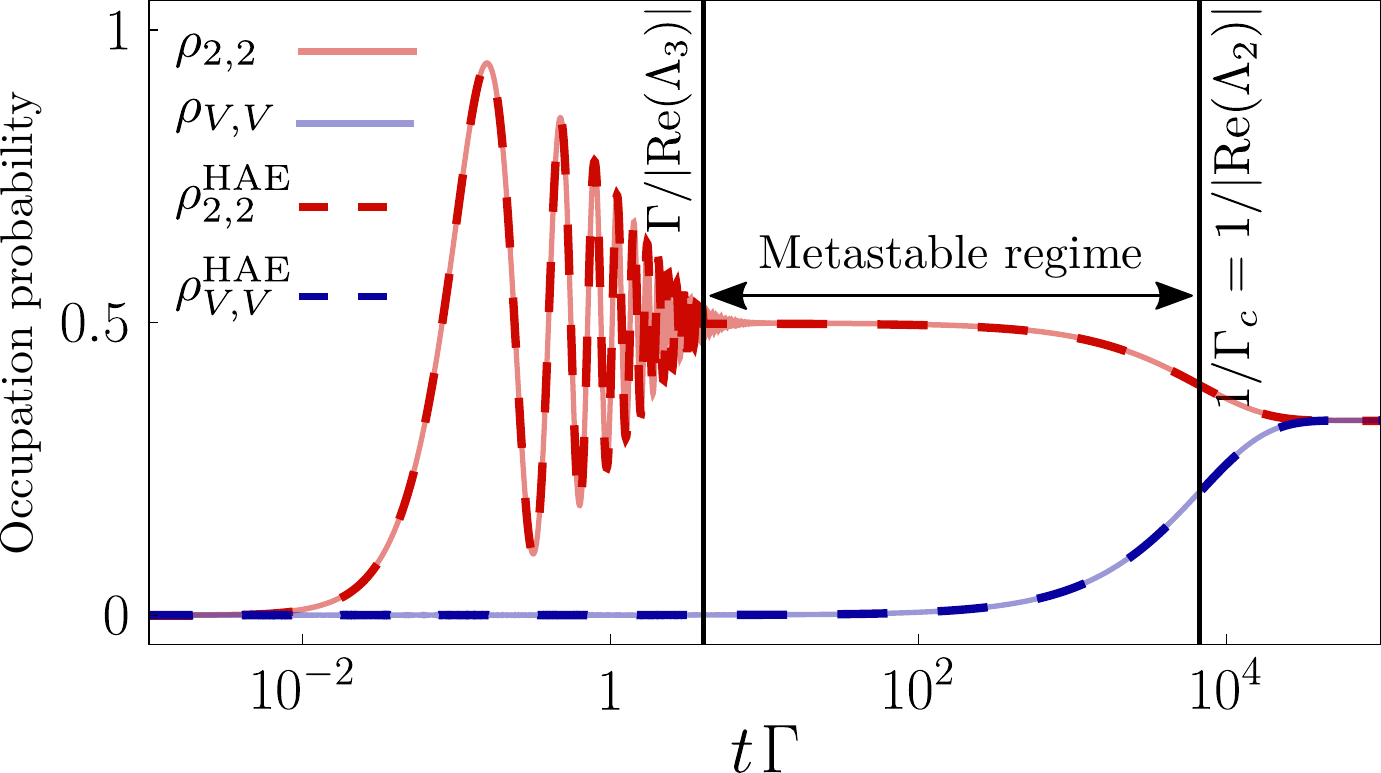}
	\captionsetup{justification=justified}
	\caption[Application of the HAE to the system dynamics]{
	\label{fig:Fig5_Dynamics_HAE} \textbf{Application of the HAE to the system dynamics. }
	Dynamics of $\rho_{2,2}(t)$ and $\rho_{V,V}(t)$. Solid lines are exact numerical calculations from \eqref{eq_Master},  and dashed lines are analytical predictions from the hierarchical adiabatic elimination (HAE) from \cref{eq:rho_12t_meta,eq:rho_22t_meta,eq:rho_HAE}. 	
	Parameters: $\Omega/\Delta_V=0.01$, $\Gamma/\Delta_V=10^{-5}$. The initial state was set to be $\hat \rho(0)=|1\rangle \langle 1|$.
	}
\end{SCfigure}

\subsection{Disrupting the unconventional mechanism}

The unconventional mechanism of virtual state population can be entirely disrupted  by introducing additional dissipative channels involving the virtual state, as we observed in \reffig{fig:LiouvillianEigen}. 
To explore this scenario, we set $\Gamma_V\neq 0$ in the master equation [see \eqref{eq_Master}], thereby including a channel of spontaneous emission from the virtual state to the ground state, as illustrated in \reffig{fig:fig1_setup}. 
The ratio of decay rates, $\Gamma_V/\Gamma_c$, will determine whether the unconventional mechanism of virtual state population is enabled or suppressed:

\begin{SCfigure}[1][b!]
	\includegraphics[width=.99\columnwidth]{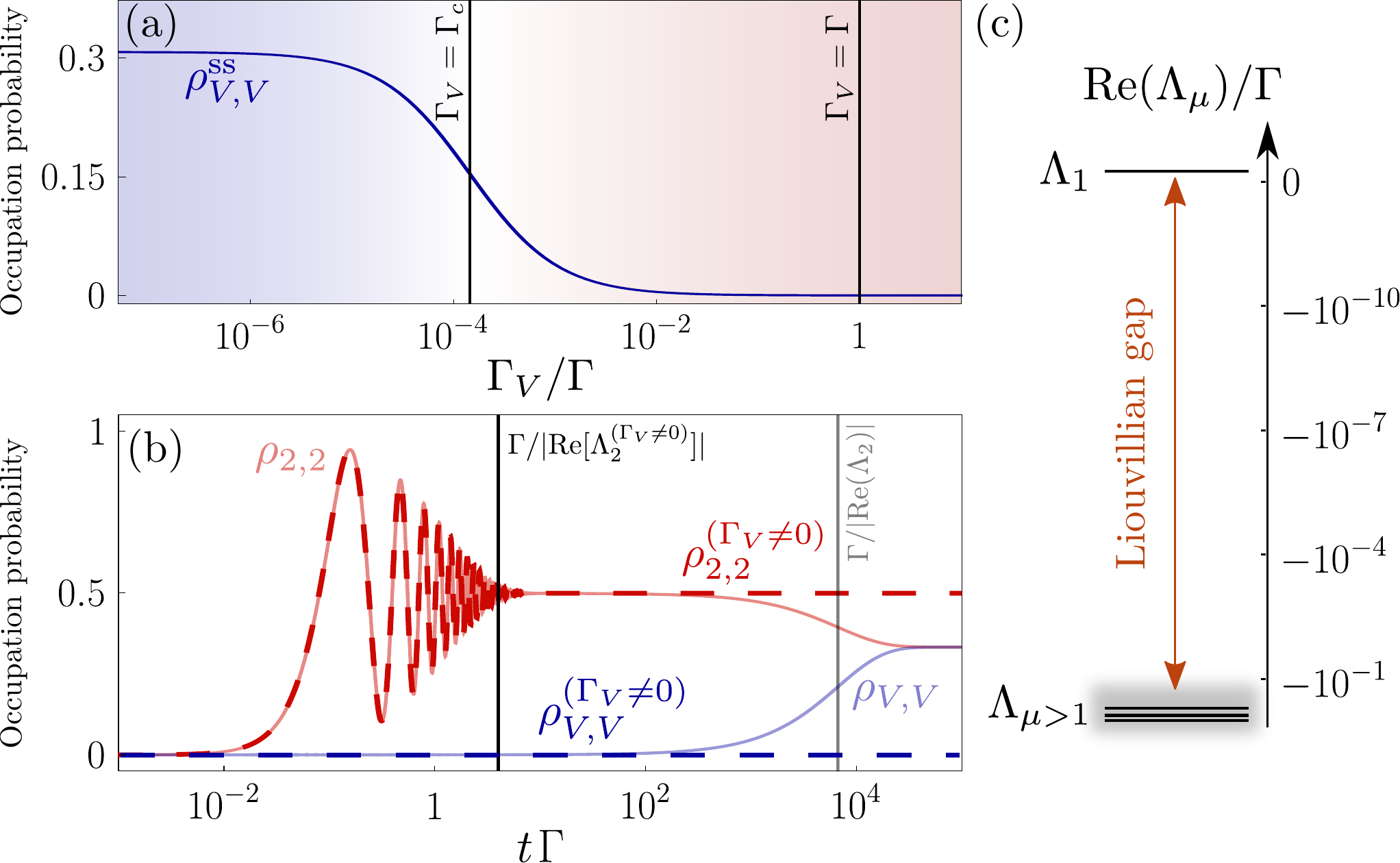}
\end{SCfigure}
\addtocounter{figure}{-1}
\graffito{\vspace{8.0cm}
	\captionof{figure}[Disruption of the metastability. ]{ \label{fig:Fig5_Dynamics_Gamma_V} \textbf{Disruption of the metastability. }
		(a) Steady-state population of the virtual state, $\rho_{V,V}^{\text{ss}}$, in terms of the additional dissipative rate $\Gamma_V$. When $\Gamma_V >\Gamma$, $\rho_{V,V}^{\text{ss}}$ reduces its population, eventually vanishing for $\Gamma_V\gg \Gamma$.
		(b) Dynamics of $\rho_{2,2}(t)$ and $\rho_{V,V}(t)$,  showing exact numerical simulations from \eqref{eq_Master} for both ($\Gamma_V=0$) and dashed ($\Gamma_V\neq0$).
		(c) Liouvillian spectrum for $\Gamma_V=\Gamma$. The closure of the metastability gap leaves the spectrum characterized only by the Liouvillian gap.
		Parameters: (a-c) $\Omega/\Delta_V=0.01$, and $\Gamma/\Delta_V=10^{-5}$; (b, c)  $\Gamma_V=\Gamma$ . The initial state was set to be $\hat \rho(0)=|1\rangle \langle 1|$.
	}
}
\begin{enumerate}[label=\textcolor{Maroon}{(\roman*)}]
	\item When $\Gamma_V \ll \Gamma_c$, the additional dissipative channel does not significantly affect the mechanism enabling virtual state population in the steady state.
	As a result, the virtual state retains a sizeable population, with $\rho_{2,2}^{ss}\approx1/3$, as shown in the left region of \reffig{fig:Fig5_Dynamics_Gamma_V}~\textcolor{Maroon}{(a)}.
	\item  When $\Gamma_V\approx \Gamma_c$, the metastability gap begins to close---as seen in \reffig{fig:LiouvillianEigen}---, resulting in a reduction of the virtual state population  due to the competition of dissipative effects.
	\item When $\Gamma_V \gg \Gamma_c$, dissipation from $|V\rangle$ outcompetes the mechanism of population, leading to a vanishing virtual state population [see right region of \reffig{fig:Fig5_Dynamics_Gamma_V}~\textcolor{Maroon}{(a)}].
	As a result, the system dynamics simplify to those of a driven two-level system, as shown in \reffig{fig:Fig5_Dynamics_Gamma_V}~\textcolor{Maroon}{(b)}.
	In this regime, there is no longer a Liouvillian eigenvalue corresponding to a metastable state as the Liouvillian gap is pulled towards values $\Lambda_2\sim \Gamma$. Consequently, the metastability gap closes, leaving the Liouvillian spectrum characterized only  by the Liouvillian gap [see \reffig{fig:Fig5_Dynamics_Gamma_V}~\textcolor{Maroon}{(c)}]. 
\end{enumerate}
The analytical expression of $\Gamma_c$ provides a critical threshold for $\Gamma_V$, establishing the minimum value required to suppress the unconventional virtual state population. This insight is particularly relevant in scenarios where such effects are undesirable, such as in protocols for dissipative cooling into entangled states~\cite{VacantiCoolingAtoms2009,BuschProtectingSubspaces2010,KastoryanoDissipativePreparation2011,BuschCoolingAtomcavity2011,LinDissipativeProduction2013}.

\subsection[Validity of the analytical expression for the \protect\newline Liouvillian gap]{Validity of the analytical expression for the Liouvillian gap}
\begin{SCfigure}[1][b!]
	\includegraphics[width=0.64\columnwidth]{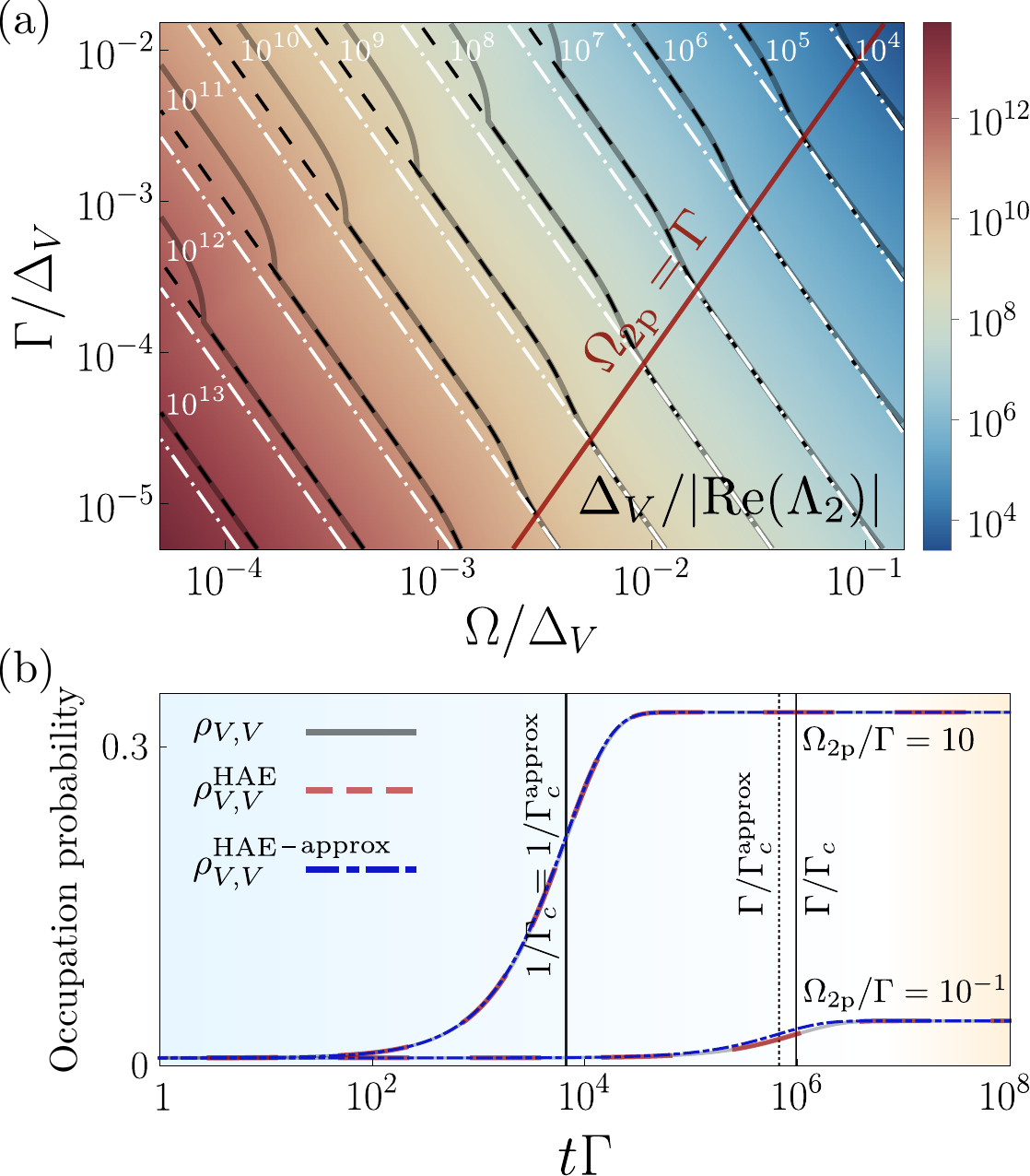}
	\captionsetup{justification=justified}
	\caption[Validity of the analytical expression for the Liouvillian gap.]{\label{fig:accuracy}	\textbf{Validity of the analytical expression for the Liouvillian gap. } (a) Liouvillian gap in terms of $\Omega/\Delta_V$ and $\Gamma/\Delta_V$. Solid-black lines are exact numerical computations, dashed black and dot-dashed white lines are analytical predictions from the general relaxation rate, \eqref{eq:Gamma_c}, and its approximated expression, \eqref{eq:Gamma_c_approx}, respectively. The solid red line divides the figure into the weak (left) and strong (right) driving regimes. 
	(b) Dynamics of $\rho_{V,V}$ at two different regimes: strong driving $\Omega_{2\text p}/\Gamma=10$  and weak driving $\Omega_{2\text p}/\Gamma=10^{-1}$. Solid lines correspond to exact numerical calculations, dashed and dot-dashed lines are analytical predictions from the HAE considering the complete relaxation rate, $\rho_{V,V}^{\text{HAE}}$, and the approximated relaxation rate, $\rho_{V,V}^{\text{HAE-approx}}$, respectively.
		Parameters: (b) $\Gamma/\Delta_V=10^{-5}$; strong driving, $\Omega/\Delta_V=10^{-2}$; weak driving, $\Omega/\Delta_V=10^{-3}$.}
\end{SCfigure}
In \reffig{fig:LiouvillianEigen} and \reffig{fig:Fig5_Dynamics_HAE}, we used the general expression for $\Gamma_c$ in \eqref{eq:Gamma_c} to compute the Liouvillian gap and the analytical curves for the population dynamics, respectively, instead of its approximated form in \eqref{eq:Gamma_c_approx}.
It is therefore useful to offer a comparison between both expressions with the numerical computation of the Liouvillian gap, and to clarify their respective regimes of validity, particularly when the condition $\Omega_{2\text{p}}\gg \Gamma$---assumed for the derivation of the simplified expression for $\Gamma_c$---no longer holds.

To test the validity of these expressions, in \reffig{fig:accuracy}~\textcolor{Maroon}{(a)} we show an exact numerical computation (grey solid lines) of the Liouvillian gap versus $\Gamma$ and $\Omega$, compared to the analytical estimations of $\Gamma_c$ [\eqref{eq:Gamma_c}, black dashed lines] and $\Gamma_c^{\text{approx}}$ [\eqref{eq:Gamma_c_approx}, white dot-dashed lines].
This figure confirms that both expressions accurately describe the Liouvillian gap when the condition $\Omega_{2\text{p}}\gg \Gamma$ is satisfied, as can be seen in the lower-right region of \reffig{fig:accuracy}~\textcolor{Maroon}{(a)}. 
On the other hand, when $\Omega_{2\text{p}}< \Gamma$, the simplified expression  $\Gamma_c^{\text{approx}}$ deviates slightly from the exact value. 
The eigenvalue crossing that occurs when $\Gamma\gg\Omega_{2\mathrm p}$---as shown in \reffig{fig:LiouvillianEigen}---manifests as a kink in the Liouvillian gap,  which is not captured by any of the analytical expressions obtained from the HAE. 
However, this overdamped regime is not expected to produce any particularly interesting dynamics, as the system predominantly remains in the ground state  $|1\rangle$.
We note that, even in the regime $\Omega_{2\mathrm p} < \Gamma$, the simplified expression $\Gamma_c^{\text{approx}}$ still provides a very good approximation of the relaxation timescale toward the steady state. This is illustrated in \reffig{fig:accuracy}~\textcolor{Maroon}{(b)}, where the population dynamics of the virtual state is computed for two driving regimes: weak ($\Omega_{2\mathrm p}<\Gamma$) and strong ($\Omega_{2\mathrm p}>\Gamma$) driving. Here, it can be seen that, although there is a small discrepancy at the onset of the relaxation stage, the approximated expression for $\Gamma_c$ provide a good description of the system.

	%

\subsection{Generalization of the HAE to arbitrary systems }
The \textit{hierarchical adiabatic elimination}  method was initially introduced for a simplified system in a three-state Hilbert space, where two states are real and one is virtual. This represents the most basic case where this method can be applied. 
In what follows, we discuss the implications of applying this technique in systems of arbitrary dimension, focusing in particular on the computational cost associated with its implementation.

Let us assume we aim to describe a general system composed of $N$ quasi-resonant real states and $M$ off-resonant virtual states, as it is sketched in \reffig{fig:figApendix_generalization}. These states span a Hilbert space
\begin{equation}
	\mathcal{H}=\{|i\rangle\}\quad \text{with}\quad i=1,\ldots, N+M,
\end{equation}
where we assume the indices are organized such that the first $N$ states correspond to the real states, and the last $M$ states represent the virtual ones. To explicitly denote a virtual state index, we adopt the notation:
\begin{equation}
|V_j\rangle \equiv |N+j\rangle\quad  \text{with}\quad j=1,\ldots, M.
\end{equation}
The formal computation of the dynamics for this general model requires solving $(N+M)^2-1$ coupled differential equations for all density matrix elements $\rho_{i,j}=\langle i|\hat\rho|j\rangle$, excluding one diagonal element determined by the normalization condition, $\text{Tr}[\hat\rho]=1$. 
To address the complexity of such a system, we outline the steps involved in solving the dynamics using the hierarchical adiabatic elimination (HAE) method.

\begin{SCfigure}[1][h!]
	\includegraphics[width=1.\textwidth]{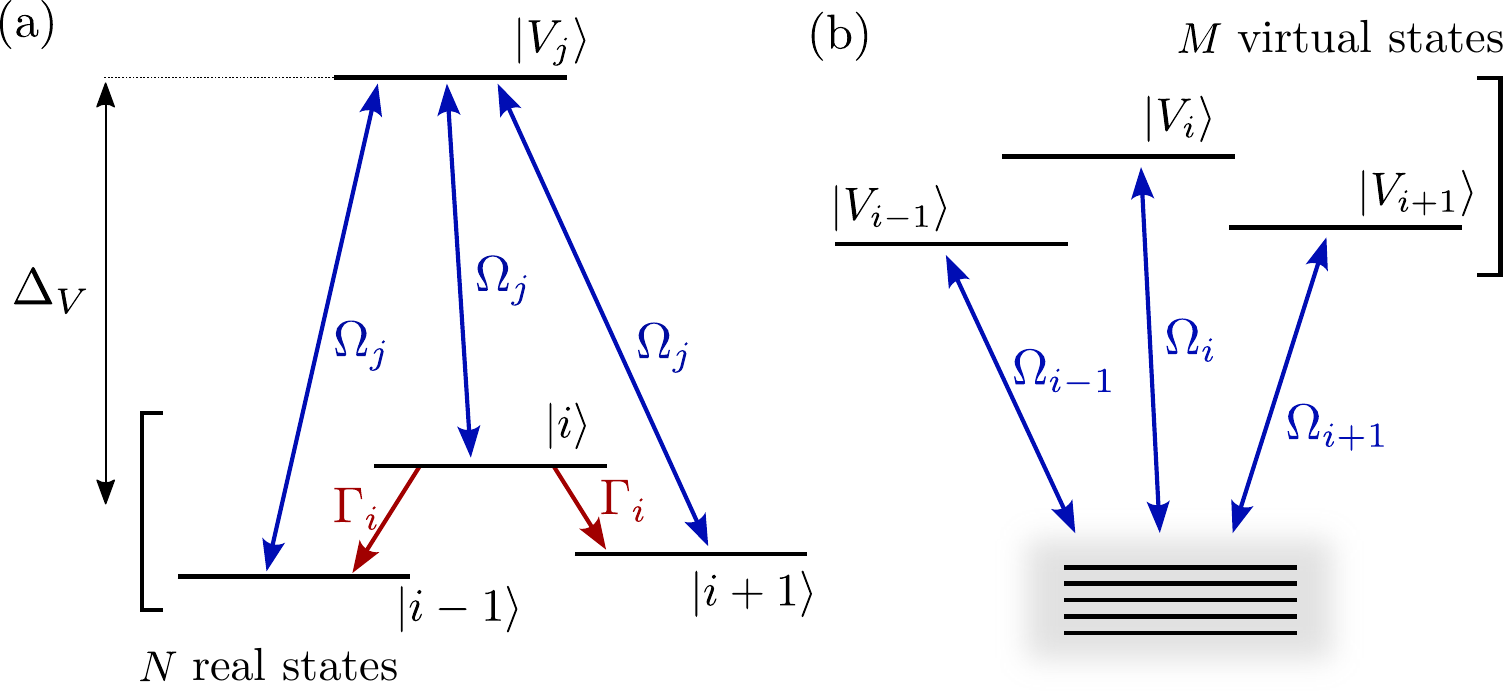}
	\captionsetup{justification=justified}
	\caption[Generalization of the HAE method to arbitrary systems]{	\label{fig:figApendix_generalization} \textbf{Generalization of the HAE method to arbitrary systems. }
		(a) Scheme of $N$ real states in a quasi-resonant energy manifold interacting with a single virtual state. (b) Scheme of $N$ real states in a quasi-resonant energy manifold interacting with $M$ virtual states. 
	}
\end{SCfigure}

\paragraph{First adiabatic elimination. }
In the first stage of the HAE, we set the derivative of real-virtual coherences to zero,
\begin{equation}
\dot{\rho}_{i,V_j}=0, \quad \text{where}\quad i=1,\ldots, N,\quad \text{and} \quad j=1,\ldots,M.
	\label{eq:eq-coherences}
\end{equation}
This results in a total of $N_c = 2M\times N$ equations. Hence, the first adiabatic elimination involves solving these $N_c$  equations, \eqref{eq:eq-coherences}, for the corresponding $N_c$ variables. 
The solutions are expressed in terms of the remaining $N^2 + M^2 -1$ variables, which include all populations and coherences within the real and virtual subspaces, reduced by one due to the normalization condition.

\paragraph{Second adiabatic elimination.} 
The second step of the adiabatic elimination involves solving for the steady-state dynamics within the real sector, treating the $M^2$ variables of the virtual sector as fixed, time-independent parameters.
Therefore, one has to solve a linear system of equations of the form 
\begin{equation}
	\hat W.\vec\rho+\vec b=0,
\end{equation}
where $\hat W$ is a matrix of dimension $(N^2-1)\times(N^2-1)$, while  $\vec \rho$ and $\vec b$ are vectors of dimension $(N^2-1)$. Here, $\vec\rho$ represents the flattened density matrix of the real sector, given by
\begin{equation}
	\vec{\rho}= (\rho_{1,1},\, \rho_{1,2},\ldots,\, \rho_{N,N-1})^\text{T},
\end{equation}
where the element $\rho_{N,N}$ is not included since it is removed by applying the normalization condition. Similarly, 
\begin{equation}
	\vec{b}=\hat \Lambda \vec{\rho}_{V}+\vec{\mu},
\end{equation}
where $\hat\Lambda$ is a $(N^2-1)\times M^2$ matrix, $\vec\rho_{V}$ is the vectorial representation of the $M^2$-dimensional density matrix of the virtual sector,  and $\vec\mu$ is a  vector of dimension $(N^2-1)$.  
The stationary solution of this coupled system will take the functional form
\begin{equation}
	\vec{\rho}_{\mathrm{ss}}=\vec{\rho}_{\mathrm{ss}}\left[\vec{\rho}_{V}(t)\right].
\end{equation} 

The solution of this linear system of $N^2 -1$ equations allows us to write the set of differential equations that governs the dynamics of $\vec{\rho}_{V}(t)$ with the following functional structure, 
\begin{equation}
		\dot{\vec \rho}_{V}=f\left[\vec \rho_{V},\vec \rho_{\mathrm{ss}}\left[\vec \rho_{V}(t)\right]\right].
\end{equation}
This can be expressed as a standard matrix differential equation of the form $\dot{\vec{\rho}}_{V}=\hat A\vec{\rho}_{V}(t)+\vec{B}$, where $\hat A$ is a matrix of dimension $M^2 \times M^2 $ and $\vec B$ a vector of dimension $M^2$.

\paragraph{Summary.}
In summary, given $N$ real states and $M$ virtual states, the method of HAE allows us to write an effective, linear differential equation for the $M^2$ slow degrees of freedom---the virtual sector---after performing the following steps:
\begin{enumerate}[label=\textcolor{Maroon}{(\roman*)}]
	\item \textcolor{Maroon}{First Adiabatic Elimination:}
	Solving a linear system of $2M \times N$ equations.
	\item \textcolor{Maroon}{Second Adiabatic Elimination:} Solving a second linear system of $N^2 -1$ equations.
\end{enumerate} 
The availability of tractable closed-form expressions for the final differential equation and the associated relaxation rates strongly depends on the complexity of the system and may not always be guaranteed. 
However, it is important to emphasize that even when closed-form solutions are challenging to derive, the HAE method can be computationally advantageous. Specifically, it reduces the computational cost of numerically determining the Liouvillian gap from diagonalizing an  $(N+M)^2\times(N+M)^2$ matrix to diagonalizing a $M^2 \times M^2 $ matrix.

\section[A quantum trajectory perspective on virtual state population]{A quantum trajectory perspective on \protect\newline virtual state population}
	\label{sec:4}
\begin{SCfigure}[1][b!]
	\includegraphics[width=0.71\textwidth]{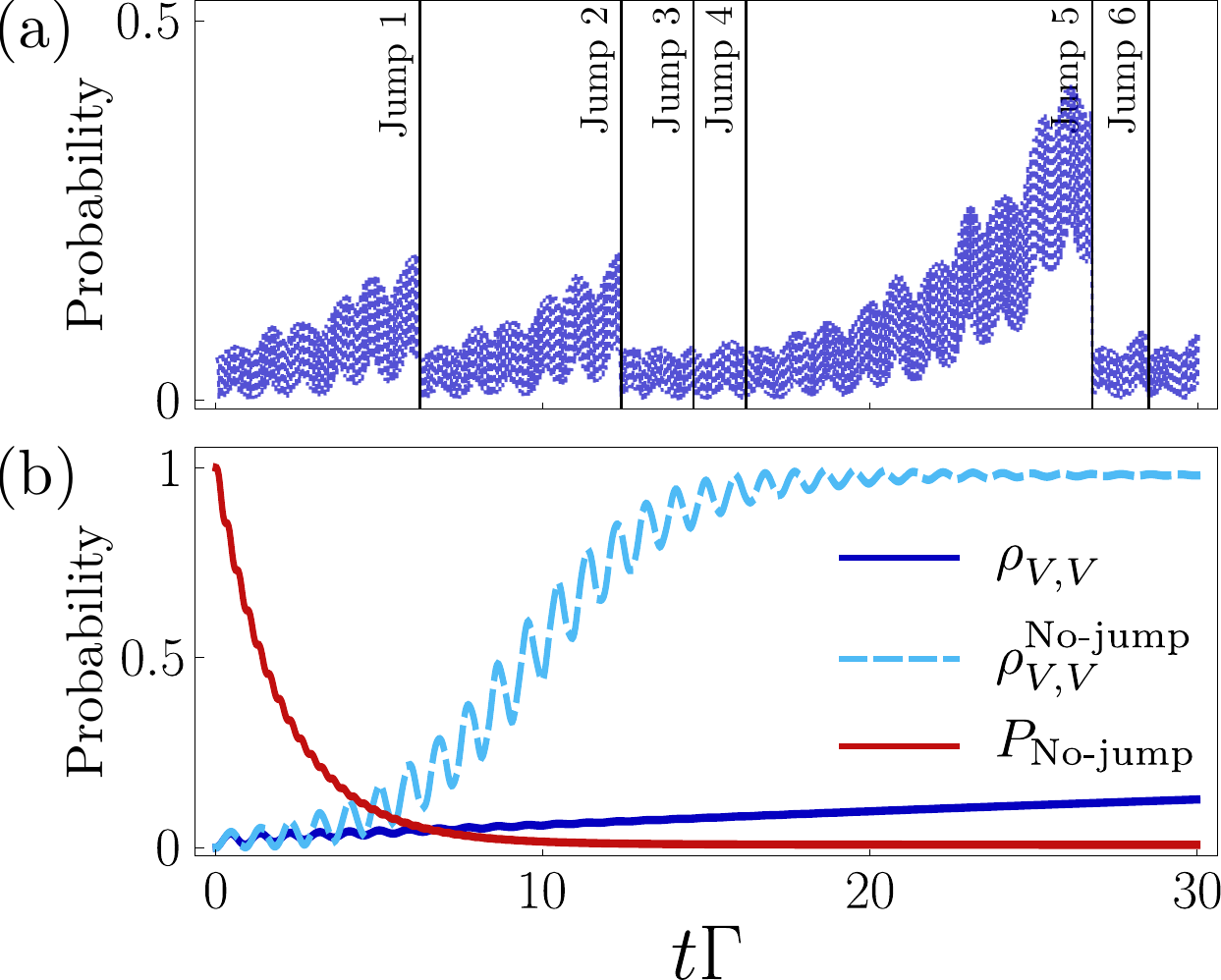}
	\captionsetup{justification=justified}
	\caption[Analysis of the unconventional mechanism of virtual state population at the single trajectory level.]{		\label{fig:fig3_nojump} \textbf{Analysis of the unconventional mechanism of virtual state population at the single trajectory level.} (a) Population of the virtual state through non-Hermitian evolution between quantum jumps. (b)  Conditional evolution when no jumps take place. Red line: probability of no-jump. Blue dashed line: population of the virtual state conditioned to no jumps, showing that, in this case, it reaches its maximum possible value. Blue straight line: population of the virtual state for the general evolution. 
		Parameters: $\Omega/\Delta_V=0.1,\ \Gamma/\Delta_V=10^{-3},\  \Delta_2=\Delta_1=0.$}
	
\end{SCfigure} 
\subsection{Mechanism of population at the single trajectory level}
We will now provide some insights into the dissipative mechanism responsible for populating the virtual state in the long-time limit.
To achieve this, we perform an analysis of the system from the perspective of quantum trajectories, employing the method of quantum jumps~\cite{CarmichaelPhotoelectronWaiting1989,MolmerMonteCarlo1993,MolmerMonteCarlo1996,PlenioQuantumjumpApproach1998,BrunSimpleModel2002,GerryIntroductoryQuantum2004}, already introduced in \refsec{Section:QuantumTraject}.
To carry out this analysis, we construct the following \textit{non-Hermitian Hamiltonian}:
\begin{equation}
	\tilde H=\hat H-i \frac{\Gamma}{2} \hat \sigma_{12}^\dagger \hat \sigma_{12},
\end{equation}
where $\hat H=\hat H_0+\hat H_\mathrm{d}$ is the Hamiltonian introduced in \refsec{sec:2}, specifically in \cref{eq:Hamiltonian_0,eq:Hamiltonian_d}, respectively. Here we assume that the system only experiences spontaneous decay within the real subspace, i.e., $|2\rangle \rightarrow |1\rangle$.
Therefore, the \textit{gedanken measurement process} of quantum trajectories reduces to:
\begin{subequations}
\begin{align}
		&|\psi (t+\delta t) \rangle_\text{emit} = \frac{ \hat \sigma_{12} |\psi (t) \rangle}{(\delta p/\delta t )^{1/2}}, \quad \text{if}\quad r < \delta p, \\
	&|\psi (t+\delta t) \rangle_\text{no-emit} \approx 
	\frac{(\mathbb{I}-i\delta t \tilde H) |\psi(t)\rangle}{(1-\delta p)^{1/2} }, \quad \text{if}\quad r > \delta p.
\end{align}
\end{subequations}
where $r$ is a random number uniformly distributed between $0$ and $1$, and $\delta p$---i.e., the probability that during the evolution from $t$ to $t+\delta t$ the system undergoes a quantum jump---is simply given by
\begin{equation}
	\delta p= i \delta t \Gamma \langle \psi(t)| \hat \sigma_{12}^\dagger \hat \sigma_{12}|\psi(t)\rangle.
\end{equation}

Inspection of individual trajectories reveals that \textit{the virtual state gets populated through the non-Hermitian evolution between quantum jumps}, and that the effect of a jump is, in fact, to strongly decrease the population of the virtual state. Such a behaviour is illustrated in \reffig{fig:fig3_nojump}~\textcolor{Maroon}{(a)}.

This phenomenon can be understood by considering the information about the system leaked to the environment during a time interval with no jumps~\cite{HarocheExploringQuantum2006}. 
When no jump occurs, the system is more likely to be in a state that cannot emit, i.e., either $|1\rangle$ or $|V\rangle$. However, since $|1\rangle$ is resonantly coupled to $|2\rangle$ through second-order in perturbation theory, a system in $|1\rangle$ will eventually evolve into $|2\rangle$ and lead to a jump. In other words, as a period without a jump becomes longer, $|V\rangle$ becomes the most likely state, and the system is updated accordingly, increasing its population.   
This purely dissipative mechanism will slowly accumulate over time, explaining the population build-up of the virtual state in our system.

\subsection{Conditional evolution without jumping}
This intuition is further confirmed by computing the conditional density matrix for the particular trajectory in which no jumps occur at all,  as shown in \reffig{fig:fig3_nojump}~\textcolor{Maroon}{(b)}. 
The resulting conditional dynamics is governed by 
\begin{equation}
	\frac{d \hat\rho^{\text{No-jump}}}{dt}= -i[\hat H, \hat \rho^{\text{No-jump}}] -\frac{\Gamma}{2} \left(\hat \sigma_{12}^\dagger \hat \sigma_{12} \hat \rho^{\text{No-jump}} + \hat \rho^{\text{No-jump}}\hat \sigma_{12}^\dagger \hat \sigma_{12} \right),
\label{eq_Master_NoJump}
\end{equation}
which is similar to \eqref{eq_Master} but having neglected the jump term, $2\hat \sigma_{12}^\dagger\rho \hat \sigma_{12}$.  The dynamics described by this master  equation yields an unnormalized density matrix $ \hat\rho^{\text{No-jump}}(t)$ conditioned to that no jump occurs during a time $t$, with its trace determining the probability of occurrence,
\begin{equation}
	P_{\text{No-jump}}(t)=\text{Tr}[ \hat\rho^{\text{No-jump}}(t)].
\end{equation}
Therefore, to make $\hat\rho^{\text{No-jump}}(t)$ suitable for further comparison with the full density matrix $\hat \rho(t)$ from \eqref{eq_Master}, we must re-renormalize the conditional density matrix as
\begin{equation}
\hat\rho^{\text{No-jump}}(t) \longrightarrow \frac{
		\hat\rho^{\text{No-jump}}(t)
		}{
		\text{Tr}[\hat\rho^{\text{No-jump}}(t)]
		}.
\end{equation}

In our particular scenario, the population of the virtual state saturates to its maximum possible value, $\rho_{V,V}^{\text{No-jump}}\approx 1$, while the probability of \textit{no jumping} decreases over time, $P_{\text{No-jump}}\rightarrow0$, as it is expected. This behaviour---illustrated in \reffig{fig:fig3_nojump}~\textcolor{Maroon}{(b)}---confirms that \textit{the non-Hermitian evolution in the absence of jumps is the mechanism responsible for populating the virtual state.}

\section[Application of the HAE: dissipative generation of entanglement]{Application of the HAE: \protect\newline  dissipative generation of entanglement }
	\label{sec:5}
	The HAE method introduced in this Chapter offers a powerful tool for gaining analytical insights into metastable dynamics in open quantum systems.  This may have significant implications for quantum technological applications, particularly in the context of \textit{entanglement generation in dissipative scenarios}. To illustrate this, we now apply the HAE to analytically describe the generation of entanglement in two different systems featuring metastability.

	\subsection{Example 1: Interacting emitters embedded in a cavity}
			\begin{SCfigure}[1][b!]
		\captionsetup{justification=justified}
		\includegraphics[width=0.69\textwidth]{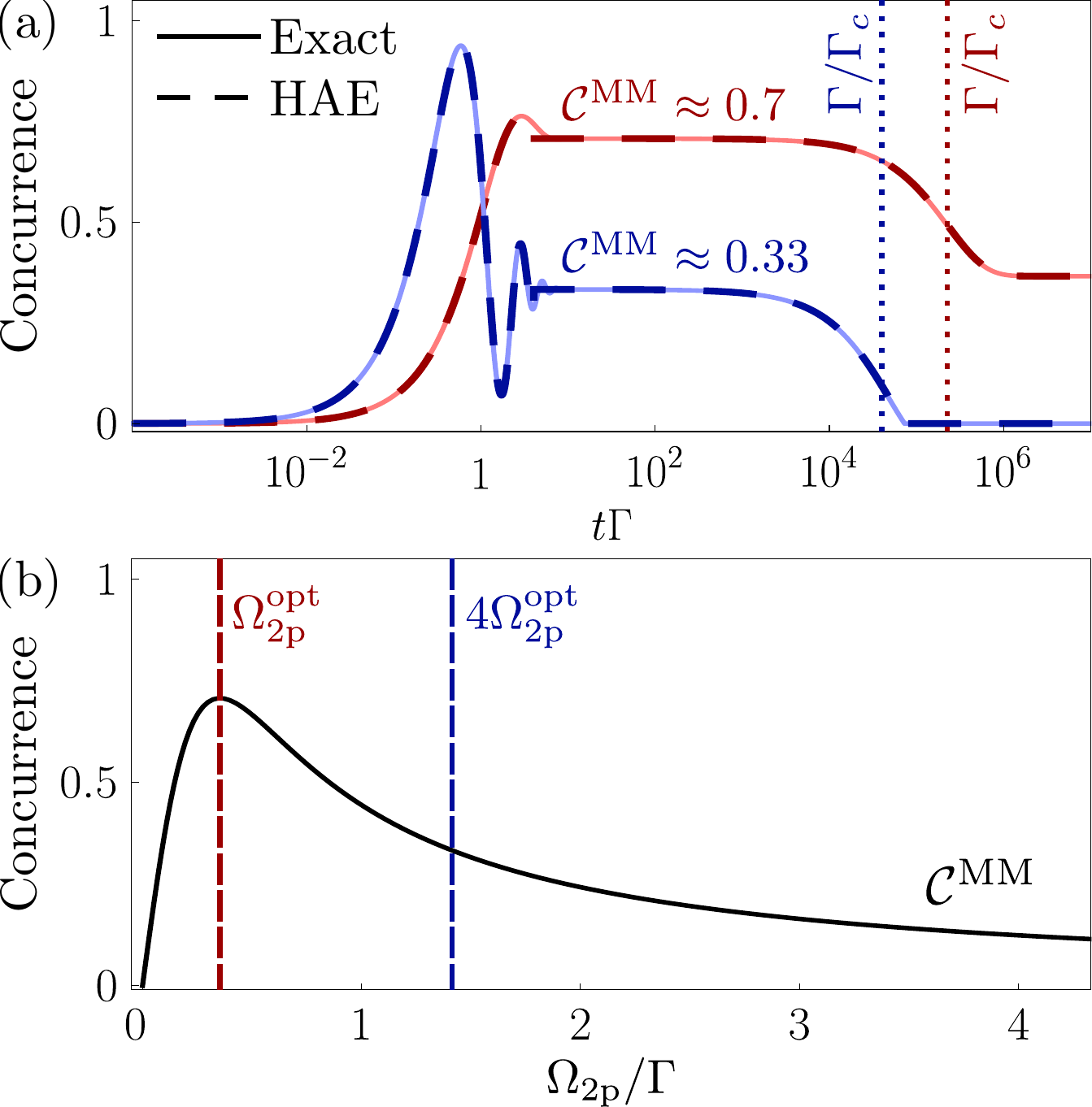}
		\caption[Example 1: generation of metastable and stationary entanglement between two coupled emitters embedded in a cavity.]{\label{fig:fig4_ConcurrencePlot_Example1}
			\textbf{Example 1: generation of metastable and stationary entanglement between two coupled qubits embedded in a cavity.}
			(a) Formation of metastable entanglement in the $\Lambda$-system sketched in Fig.~\ref{fig:TwoPhotonDecay2}~(a) for two driving strengths: $\Omega_{2\text{p}}=\Omega_{2\text{p}}^{\text{opt}}$ (red) and $\Omega_{2\text{p}}=4\Omega_{2\text{p}}^{\text{opt}}$ (blue). Solid lines correspond to exact numerical simulations and dashed lines are analytical predictions from the HAE results.
			The survival time of entanglement is given by $\tau_2 \sim 1/\Gamma_c$. 			 
			Parameters: $\Gamma/\Delta_V=10^{-5},\  \Delta_2/\Delta_V=0.$ 
			(b) 
			Metastable concurrence from \eqref{eq:Meta_Concurrence} in terms of the two-photon driving strength $\Omega_{2\text{p}}$.
		}
	\end{SCfigure}
	We first consider the phenomenologically proposed system introduced at the beginning of this Chapter: two interacting two-level systems coherently driven at the two-photon resonance, with a cavity-induced decay channel from $|ee\rangle$ towards $|gg\rangle$ [see  \reffig{fig:TwoPhotonDecay2}~\textcolor{Maroon}{(a)}]. 
	As previously discussed, this system could be mapped onto the $\Lambda$-model studied so far in this Chapter, with the antisymmetric state $|-\rangle$ considered as a dark state that is completely decoupled from the dynamics and exhibits zero occupation probability$^{\textcolor{Maroon}{*}}$~\cite{FicekQuantumInterference2005}.
	\graffito{
	$^*$This behavior was thoroughly analyzed in the previous Chapter. In particular, \refsec{sec_General_model_twoqubits} provides a detailed discussion of the properties of the emitter-emitter system. As shown in \eqref{eq:Subradiant}, the antisymmetric state can evolve into a dark state, becoming decoupled from both coherent and dissipative interactions.
	}
	The mapping between the $\Lambda$-model and the two interacting emitters model allows us to use the density matrix elements derived earlier [see \cref{eq:rho_HAE,eq:rho_12_metaSS,eq:rho_22_metaSS}] to estimate the degree of entanglement between the two qubits.
	This can be done via the concurrence $\mathcal{C}$---already introduced in \refsec{sec:Conc_and_Negat}---, which is a widely used measure for witnessing  entanglement in bipartite systems~\cite{WoottersEntanglementFormation1998,WoottersEntanglementFormation2001,PlenioIntroductionEntanglement2007,HorodeckiQuantumEntanglement2009}. We recall that $\mathcal{C}=0$ corresponds to separable states (e.g., vacuum or thermal) and $\mathcal{C}=1$ indicates maximally entangled state (i.e., Bell states). 
	 
	In  \reffig{fig:fig4_ConcurrencePlot_Example1}~\textcolor{Maroon}{(a)}, we evidence the formation and stabilization of entanglement at short timescales $t\sim 1/\Gamma$, due to the coherence built between the states $|ee\rangle$ and $|gg\rangle$ via the two-photon drive.  
	Notably, \textit{this entanglement is long-lived, but metastable}, reaching values of $\mathcal{C}^{\text{MM}}\approx 0.7$ (in red) and $\mathcal{C}^{\text{MM}}\approx 0.33$ (in blue) for the two different driving strengths used in the figure. The survival time of the entanglement is determined by the relaxation rate derived in \eqref{eq:Gamma_c} via the HAE
	\begin{equation}
		\colorboxed{Maroon}{	
			\tau_{2}\sim 1/\Gamma_c \quad \quad\text{(entanglement survival time)}.
		}
	\end{equation}
	In the long-time limit, the unconventional population of the virtual state---in this case, the symmetric state $|+\rangle$---reduces, or even destroys, the entanglement depending on the driving strength.

	Since we have derived analytical expressions for the time-dependent density matrix elements over the entire evolution [\cref{eq:rho_12t_meta,eq:rho_22t_meta,eq:rho_HAE}], we can accurately reproduce the numerical simulations of the concurrence, as shown by the dashed curves in \reffig{fig:fig4_ConcurrencePlot_Example1}~\textcolor{Maroon}{(a)}.
	While a general time-dependent expression for the concurrence $\mathcal{C}(t)$ can be obtained from these results, such an expression would be cumbersome and untractable. However, the most interesting regime for entanglement generation occurs during the metastable stage.
	By using the analytical expressions for the density matrix during this regime [see \cref{eq:rho_12_metaSS,eq:rho_22_metaSS}], we derive a tractable expression for the metastable concurrence,
	\begin{equation}
		\mathcal{C}^{\text{MM}}\approx \text{Max}\left[0, 
		\frac{2\sqrt{2} \Omega_{2\mathrm{p}}}{\Gamma^2+8\Omega_{2\mathrm{p}}^2}
		(\tilde \Omega_+-\tilde \Omega_-) \right],
		\label{eq:Meta_Concurrence}
	\end{equation} 
	having defined  
	\begin{equation}
		\tilde \Omega_\pm \equiv 	\sqrt{2\Omega_{2\mathrm{p}}^2 + \Gamma\left(\Gamma+\sqrt{\Gamma^2\pm\Omega_{2\mathrm{p}}^2}\right)} .
	\end{equation}
	From this expression, which is illustrated in \reffig{fig:fig4_ConcurrencePlot_Example1}~\textcolor{Maroon}{(b)}, we can determine the optimal driving strength, $\Omega_{2\mathrm{p}}^{\text{opt}}$, that maximizes the generation of entanglement during the metastable regime,
$
		\Omega_{2\mathrm{p}}^{\text{opt}}=\Gamma/2\sqrt{2}.
$
	This value corresponds to the optimal driving strength maximizing the coherence in a resonantly driven qubit, as we discussed in the introduction of this Chapter in \reffig{fig:TwoPhotonDecay1}~\textcolor{Maroon}{(a)}.
	Inserting  $\Omega_{2\mathrm{p}}^{\text{opt}}$ into \eqref{eq:Meta_Concurrence}, the maximum metastable concurrence reaches a value of $\mathcal{C}_{\text{max}}^{\text{MM}}=1/\sqrt{2}\approx 0.7$, as it illustrated in  \reffig{fig:fig4_ConcurrencePlot_Example1}~\textcolor{Maroon}{(a, b)}.
	Note that in \reffig{fig:fig4_ConcurrencePlot_Example1}~\textcolor{Maroon}{(a)}, the chosen driving strengths are $\Omega_{2\mathrm{p}}=\Omega_{2\mathrm{p}}^{\text{opt}}$ (in red) and $\Omega_{2\mathrm{p}}=4\Omega_{2\mathrm{p}}^{\text{opt}}$ (in blue).

	\subsection{Example 2: Emitters coupled to a chiral waveguide }

	Next, we consider another two-qubit system originally introduced in \colorref{RamosQuantumSpin2014,PichlerQuantumOptics2015} in the context of chiral waveguides. In this setup, the qubits experience collective decay with rate $\Gamma$, inducing transitions $|ee\rangle \rightarrow |S\rangle$ and $|S\rangle \rightarrow |gg\rangle$. Additionally, both qubits are coherently driven with the same Rabi frequency $\Omega$, but each of them is detuned from the drive frequency by an absolute value $\delta$ with opposite sign---in fact, this is an alternative way of establishing that the system is driven at the two-photon resonance---.
	The dynamics of this model gets described by the master equation,
	\begin{equation}
		\frac{d \hat \rho }{d t}= -i[\hat H, \hat\rho ]+ \frac{\Gamma}{2}\left(\mathcal{D}[|gg\rangle \langle S|]+\mathcal{D}[|S\rangle \langle ee|]\right)\hat \rho,
	\end{equation}
	where the Hamiltonian reads
	\begin{equation}
		\hat H = \sqrt{2}\Omega (|S\rangle\langle gg| + |ee\rangle \langle S| ) + (\delta-i\Delta\gamma/2)|A\rangle\langle S| + \mathrm{H.c.},
		\label{eq:H-Ramos}
	\end{equation}
	with $\Gamma = 2(\gamma_R + \gamma_L)$ and $\Delta\gamma \equiv \gamma_R-\gamma_L$, where $\gamma_R$ and $\gamma_L$ describe decay into right and left propagating modes, respectively. The corresponding decay channels are sketched in the inset of \reffig{fig:fig4_ConcurrencePlot_Example2}~\textcolor{Maroon}{(a)}. We observe that the state $|A\rangle$ is not coupled to any dissipative channel but just connected with the state $|S\rangle$ with a rate $\delta-i\Delta \gamma/2$.
	This configuration was shown to stabilize in the long time limit to the fully-entangled dark state $|A\rangle$ provided $\Omega \gg \Delta\gamma, \delta$ \cite{RamosQuantumSpin2014,PichlerQuantumOptics2015}, as we explicitly  show in \reffig{fig:fig4_ConcurrencePlot_Example2}~\textcolor{Maroon}{(a)} by computing the population  dynamics of the symmetric and antisymmetric states.

	The application of the HAE to describe this system follows exactly the same reasoning detailed in \refsec{sec:3}, with $|A\rangle$ playing the role of the ``virtual'' state that gets populated over time. Indeed, one can observe that, if one diagonalizes the driving term proportional to $\Omega$ in \eqref{eq:H-Ramos}, only two of the three resulting eigenstates  are coherently coupled to $|A\rangle$, and both are detuned from it by values $\pm 2\sqrt{2}\Omega$, so that the condition $\Omega \gg \Delta\gamma, \delta$ implies that, at a Hamiltonian level, $|A\rangle$ is expected to play the role of a virtual state. Consequently, \textit{the unconventional mechanism of virtual state population is the responsible for the formation of entanglement in the long time limit}. 

\begin{SCfigure}[1][b!]
	\captionsetup{justification=justified}
	\includegraphics[width=0.69\textwidth]{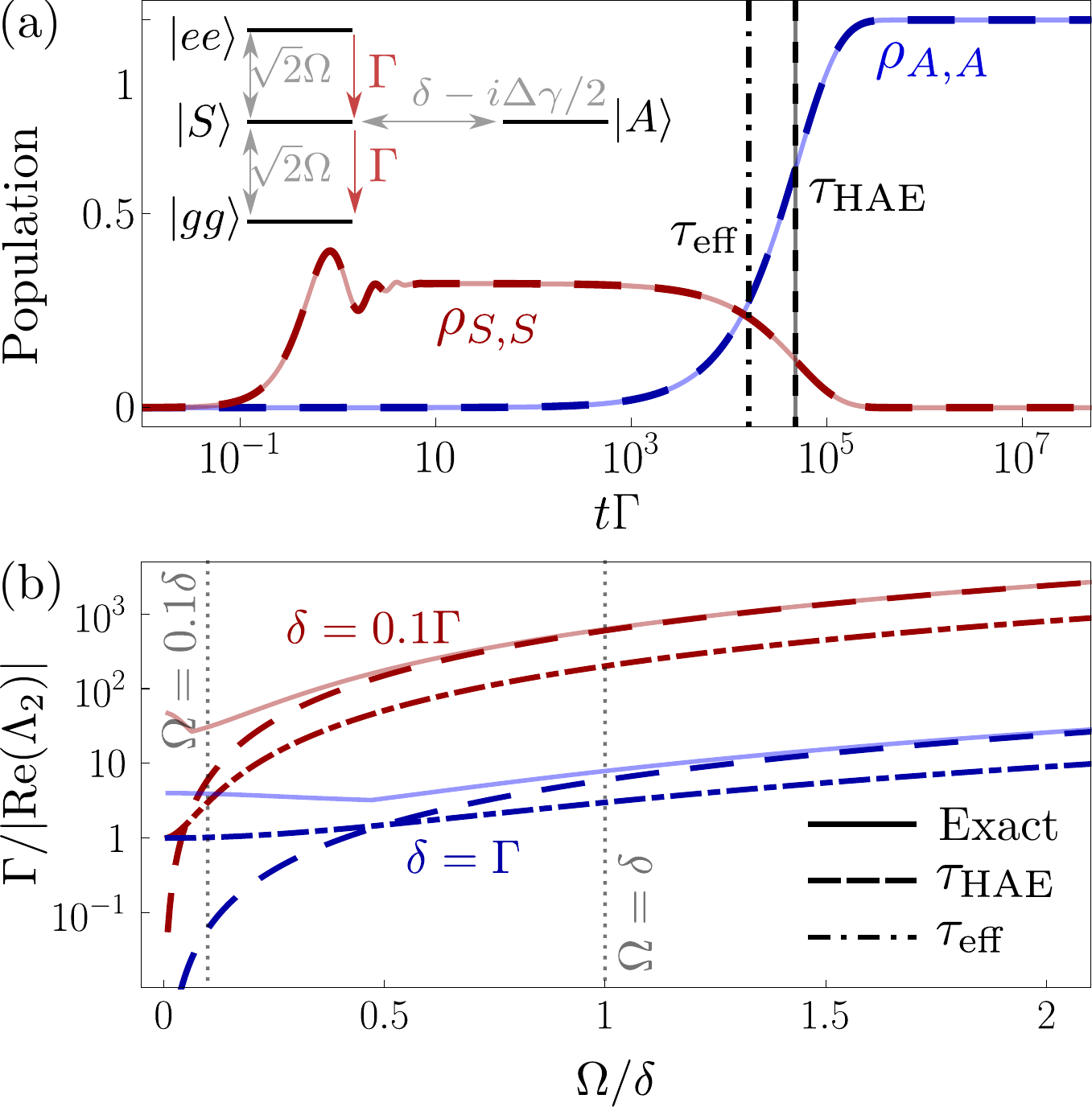}
	\caption[Example 2: generation of metastable and stationary entanglement between two emitters coupled to a chiral waveguide]{\label{fig:fig4_ConcurrencePlot_Example2}
		\textbf{Example 2: generation of metastable and stationary entanglement between two emitters coupled to a chiral waveguide.} 
		(a) Stabilization of the entangled antisymmetric state in the two-qubit system sketched in the inset. The evolution is perfectly described by the HAE. Parameters: $\Omega/\Gamma=1$, $\delta/\Gamma=0.01$, $\Delta\gamma/\Gamma=0.01$. 
		(b) Comparison of the exact numerical relaxation timescale with the effective timescale derived from the HAE in \eqref{eq:Tomas_HAE} and from Refs.~\cite{RamosQuantumSpin2014,PichlerQuantumOptics2015} in \eqref{eq:Tomas}.
		 }
\end{SCfigure}
	
	By applying the HAE technique, we can determine the timescale for the formation of entanglement, which is given by
	\begin{equation}
	\colorboxed{Maroon}{		\tau_{\text{HAE}} \approx \frac{24\Omega^2}{\Gamma(4\delta^2+\Delta\gamma^2)}.}
	\label{eq:Tomas_HAE}
	\end{equation}
	The analytical results obtained from the HAE method [not shown here] perfectly match the exact numerical calculations, as shown in \reffig{fig:fig4_ConcurrencePlot_Example2}~\textcolor{Maroon}{(a)}. 
	We emphasize that the present effective relaxation rate $\tau_{\text{HAE}}$ provides a different and more accurate prediction that the one presented in the original work~\cite{RamosQuantumSpin2014,PichlerQuantumOptics2015}, given by
		\begin{equation}
		\tau_{\text{eff}}=\frac{2(\gamma_L+\gamma_R)(\Delta \gamma^2/4+\delta^2)}{\Delta\gamma^2/4+\delta^2+2\Omega^2}.
		\label{eq:Tomas}
	\end{equation}	
	As shown in \reffig{fig:fig4_ConcurrencePlot_Example2}~\textcolor{Maroon}{(a)},  the difference between these two expressions is evident: $\tau_{\text{HAE}}$ accurately matches the actual relaxation timescale, whereas $\tau_{\text{eff}}$ exhibits a slight deviation.
	We further explored the validity of both relaxation timescales in \reffig{fig:fig4_ConcurrencePlot_Example2}~\textcolor{Maroon}{(b)}, where the relaxation timescale $1/|\text{Re}(\Lambda_2)|$ is computed in terms of the driving strength for two detuning: $\delta=0.1\Gamma$ (in red) and $\delta=\Gamma$ (in blue). 
	The figure shows that when the condition $\Omega\gg \delta$ is satisfied, the effective relaxation timescale derived from the HAE [\eqref{eq:Tomas_HAE}] perfectly matches the actual value. On the other hand, as already observed in \reffig{fig:fig4_ConcurrencePlot_Example2}~\textcolor{Maroon}{(a)}, the effective expression $\tau_{\text{eff}}$ [\eqref{eq:Tomas}] derived in \colorref{RamosQuantumSpin2014,PichlerQuantumOptics2015} slightly deviates from the exact numerical results. 
	
	Finally, We note that in the limit $(\delta,\Delta\gamma)\rightarrow 0$, the effective relaxation timescale $\tau_{\text{HAE}}\rightarrow \infty$, implying that the metastable state becomes the steady state, as reported, e.g., in \colorref{Gonzalez-TudelaEntanglementTwo2011}.

	\section{Conclusions}
	We have shown that, in open quantum systems, off-resonant virtual states may exhibit an \textit{unconventional} behaviour, where they become populated in the long-time limit even without being connected to any dissipative channel. This implies that the regime in which the virtual state remains unpopulated is merely \textit{metastable}.
	In order to unveil this unconventional dynamics of virtual states, we introduced a novel method based on \textit{hierarchical adiabatic eliminations} (HAE). This method approximates the  dynamics of the system and provides analytical expressions of relevant quantities: \textcolor{Maroon}{(i)} the lifetime of the metastable state, corresponding to the Liouvillian gap, $\Gamma_c\sim 1/|\text{Re}(\Lambda_2)|$, and \textcolor{Maroon}{(ii)} the time-dependent density matrix elements $\rho_{ij}(t)=\langle i | \hat \rho(t)|j\rangle$. Notably, this method can be formally applied to a wide range of systems where real and virtual states can be clearly distinguished,  reducing thus the computational complexity of the problem.
	Particularly, our method  is well-suited for analysing metastable open quantum systems, such as qubits embedded in photonic environments, offering valuable insights into collective phenomena, e.g., the dissipative stabilization of entangled states.
	
	The examples presented in \refsec{sec:5} illustrate the potential of harnessing collective effects, both coherent (e.g., two-photon driving) or dissipative (e.g., collective decay), to generate metastable and stationary entanglement between qubits.  
	Motivated by these findings, in the next Chapter, we will study in detail the cavity QED system proposed in \eqref{eq:Master_eq}, highlighting the rich interplay between the cavity-enabled losses and the dressed energy structure of the emitters. 

\cleardoublepage
\chapter{Dissipative generation of steady-state entanglement}\label{chapter:Entanglement}

\section{Introduction }
\begin{SCfigure}[1][b!]
	\includegraphics[width=1.\columnwidth]{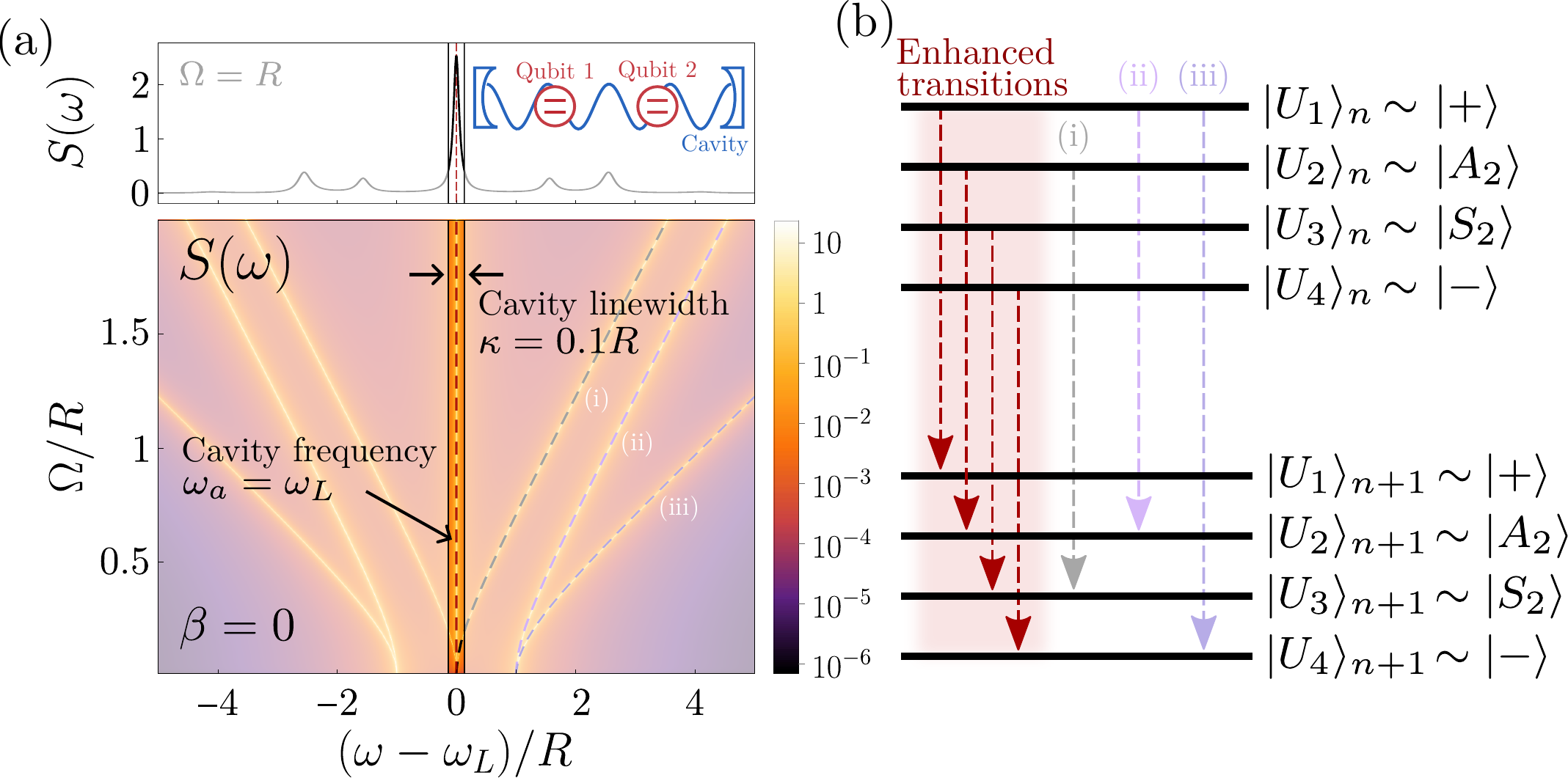}
\end{SCfigure}
\addtocounter{figure}{-1}
\graffito{\vspace{10cm}
	\captionof{figure}[Selective enhancement of energy transitions through a narrow-band cavity.]{\label{fig:Fig1_FailureTwoPhotonDecay}	
		\textbf{Selective enhancement of energy transitions through a narrow-band cavity. } 
		(a) Two-photon resonance fluorescence spectrum of two nonidentical emitters when $\beta=0$: in the frequency domain (upper panel) and driving strength (lower panel). A lossy single-mode cavity in resonance with a concrete frequency (e.g., $\omega_a=\omega_L$) and within its frequency linewidth ($\sim 2\kappa$) produces a Purcell-enhancement of the dressed-energy transitions, such that non-resonant transitions are quenched (b).
	}
}

In the previous Chapter, we took  a brief \textit{detour} from our original idea on harnessing the two-photon phenomena presented in \refch{ch:TwoPhotonResonance},  unveiling an unconventional mechanism by which virtual states \textit{de-virtualize} in dissipative scenarios.
In this Chapter, we return to the concept of generating entanglement between two nonindentical, coherently driven quantum emitters when a photonic structure~\cite{ChangColloquiumQuantum2018,Gonzalez-TudelaLightMatter2024}, such as a single-mode cavity, is included in the description.
%
%
Specifically, we follow up the discussion about the possibility of stabilizing two-photon dressed states, $|S_2/A_2\rangle=1/\sqrt{2}(|gg\rangle \pm |ee\rangle)$,
by properly tailoring the photonic structure.
We remind the reader that these two-photon dressed states are two of the four eigenstates of the emitter-emitter system,  emerging in the limit of two closely spaced, near resonant emitters ($\beta\approx 0,\  \gamma_{12}\approx\gamma$) that are perturbatively driven at the two-photon resonance ($\Omega\ll R, \ \Delta=0$). Under these conditions, the system eigenstates, $\{|U_i\rangle\}$, with $i=1,\ldots,4$, and ordered in decreasing energy, take the form $\{ |+\rangle, |A_2\rangle, |S_2\rangle, |-\rangle \}$, where $|\pm\rangle\approx |S/A\rangle=1/\sqrt{2}(|eg\rangle \pm |ge\rangle)$. For a detailed analysis of the emitter-emitter system, see \refsec{sec:perturbative_regime}.

In a regime where the Rabi frequency of the dipoles, $R$, is the dominant energy scale of the system, a single-mode cavity is able to resolve the different energy transitions between eigenstates of the interacting emitters. This is possible because the cavity linewidth ($\kappa$) is much narrower than the different energy transitions occurring within the emitters ($\Delta E$), such that $\kappa \ll \Delta E\propto R$.  
As a consequence of considering a narrow-band cavity, it can selectively enhance any of these transitions via the Purcell effect~\cite{PurcellResonanceAbsorption1946}, while suppressing non-resonant transitions.  
We illustrate this frequency-resolving effect in \reffig{fig:Fig1_FailureTwoPhotonDecay}~\textcolor{Maroon}{(a)}, where a narrow  window overlapping the fluorescence spectrum of the two emitters depicts the resolution of the cavity.
\begin{SCfigure}[1][b!]
	\includegraphics[width=0.75\columnwidth]{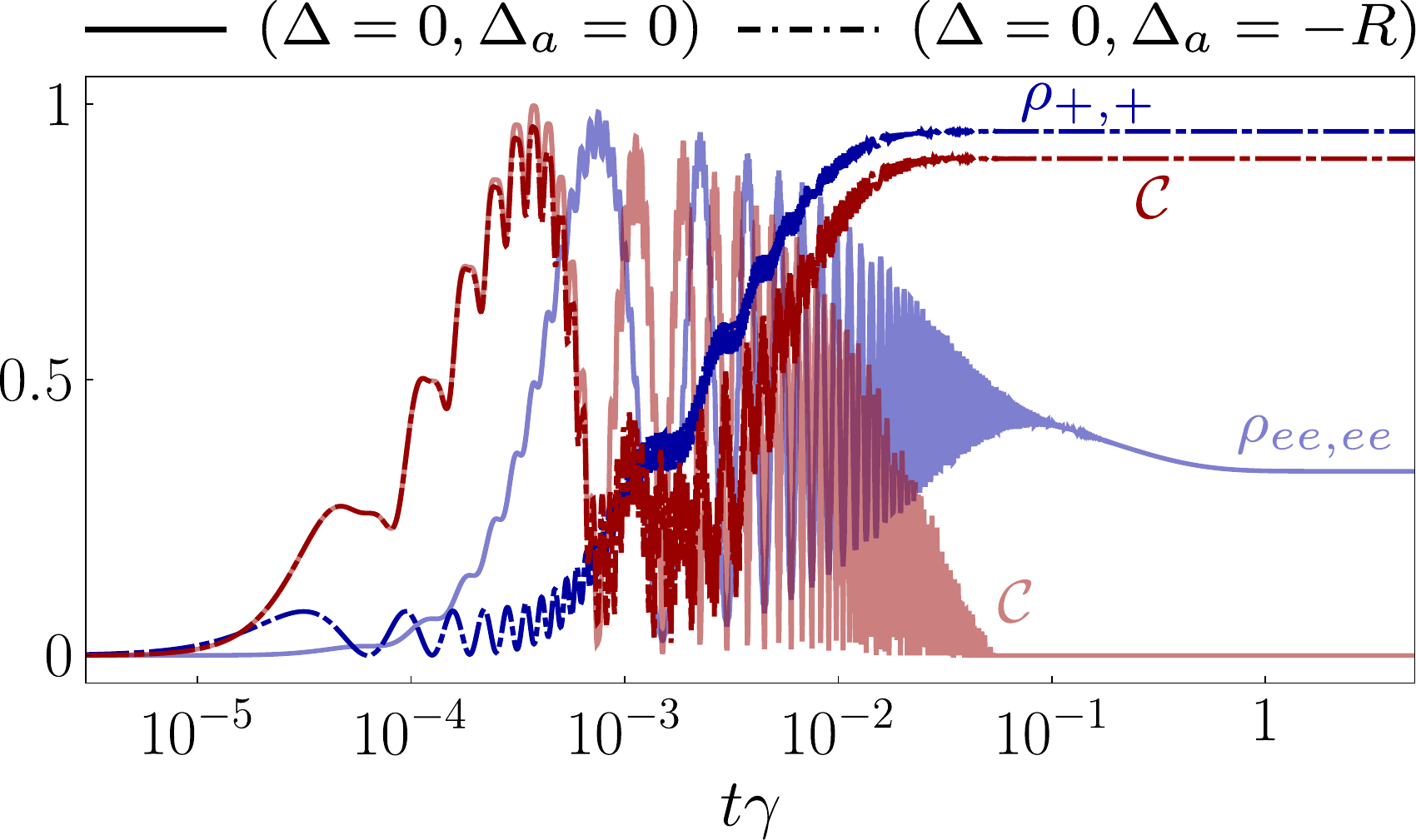}
	\captionsetup{justification=justified}
	\caption[Tuning the cavity at different frequencies.]{ \label{fig:Fig1_FailureTwoPhotonDecay1}	\textbf{Tuning the cavity at different frequencies. } Time evolution of the doubly-excited state ($\rho_{ee,ee}$, blue solid curve), symmetric state ($\rho_{+,+}$, blue dot-dashed curve) and concurrence ($\mathcal{C}$, red curves). Solid lines correspond to $\Delta_a=0$, and dot-dashed lines correspond to $\Delta_a=-R$. Parameters: $r_{12}=2.5$ nm, $k=2\pi/780\ \text{nm}^{-1}$, $J=9.18\times 10^4\gamma$, $\gamma_{12}=0.999\gamma$, $\delta=10^{-2}J$, $R=9.18\times 10^4\gamma$, $\Omega=10^4\gamma$, $\kappa=10^4\gamma$, and $g=10^{-1}\kappa$. Initial state: $\hat \rho(0)=|gg,0\rangle \langle gg,0|$.  }
\end{SCfigure}

Building on this essential phenomenon of the cavity resolving the eigenstates of the emitters, we observe that tuning the cavity frequency in resonance with the laser---that is, the two-photon resonance---, the cavity Purcell-enhances the desired transitions for stabilizing a two-photon dressed state, $|U_2/U_3\rangle_{n}\rightarrow |U_2/U_3\rangle_{n+1}$, where  $|U_{2/3}\rangle \sim |A_2/S_2\rangle$, as we firstly speculated in the previous Chapter. However, it also enhances the transitions $|U_1/U_4\rangle_{n}\rightarrow |U_1/U_4\rangle_{n+1}$, where $|U_{1/4}\rangle \sim |\pm\rangle$ [see \reffig{fig:Fig1_FailureTwoPhotonDecay}~\textcolor{Maroon}{(b)}]. 
As a result of this collective enhancement of the eigenenergy transitions, the system is prevented from being stabilized into an entangled state. This behaviour is evident in \reffig{fig:Fig1_FailureTwoPhotonDecay1}, where we show the time evolution of the doubly-excited state population  (blue  solid curve) and concurrence (red solid curve). The figure illustrates that, in the long-time limit, the system thermalizes, causing the concurrence to vanish at times $t\gtrsim 10^{-1}/ \gamma$ (for the specific set of parameters used in this figure):
\begin{equation}
	\hat \rho_{\mathrm{ss}}= \frac{1}{4} \mathbb{I}_{4} \quad  \text{and} \quad \mathcal{C}=0 \quad \text{(Resonant cavity, $\Delta_a=0$)}.
\end{equation}
Nevertheless, we can ask ourselves the following question:
\emph{Can we exploit the  eigenenergy structure of the emitters and stabilize other eigenstates, such as $|\pm \rangle$, by tuning the frequency of the cavity at different energy transitions?}

Interestingly, when we tune the cavity in resonance with the symmetric transition at $\Delta_a\approx-R$, the symmetric state $|+\rangle$ nearly achieves saturation in the long-time limit, accompanied with an almost saturated concurrence, as shown by the dot-dashed curves in \reffig{fig:Fig1_FailureTwoPhotonDecay1},
\begin{equation}
	\hat \rho_{\mathrm{ss}} \approx |+\rangle \langle +|\quad  \text{and} \quad  \mathcal{C}\approx 0.9 \quad \text{(Tuned cavity, $\Delta_a=-R$)}.
\end{equation}
Thus, we can affirmatively answer the previous question: it is indeed possible to stabilize the system into a maximally entangled state by \textit{frequency-resolving} the eigenenergy structure of the emitters.  
This constitutes one of the main results of this Chapter, which we refer to as the \textit{frequency-resolved Purcell effect}.
This finding raises further questions, such as: 
\emph{What is the physical mechanism underlying this effect? Can it be extended to other resonances? What is the range of applicability? Is it physically feasible within state-of-the-art technological platforms? Or, can it be generalize to $N>2$ interacting qubit systems? }

We dedicate this Chapter to answer these questions and others that will rise throughout the analysis. To this end, we perform an extensive and detailed study of the system originally proposed in the previous Chapter---a collection of interacting quantum emitters coupled to a photonic structure described as a single-mode cavity in the bad-cavity limit---for the generation of long-lived entanglement between emitters. By exploring the parameter space of the composite system,
we realize that the frequency-resolved mechanism for generating stationary entanglement is just one among others: \textit{the system exhibits a landscape of five different driven-dissipative mechanisms to generate long-lived of both stable and metastable entanglement among the emitters, which are tunable and robust against decoherence}. 
The structure of this Chapter is as follows. In \refsec{sec:model}, we revisit the model of the emitters-cavity system
and introduce an effective description when the cavity is traced out. Here, we explicitly show the failure of the adiabatic elimination performed in the previous Chapter, and show why a more refined approach---via the Nakajima-Zwanzig formalism~\cite{NakajimaQuantumTheory1958,ZwanzigEnsembleMethod1960,BreuerTheoryOpen2007,RivasOpenQuantum2012,Navarrete-BenllochIntroductionQuantum2022,Gonzalez-BallesteroTutorialProjector2024}---is needed to preserve all the physical information.
Then, in \refsec{sec:Freq_Resolved_Mech}, we provide an overview of the different mechanisms for entanglement generation within the parameter space, providing both novel physical insights and analytical descriptions in the subsequent sections dedicated to these mechanisms. 
In \refsec{sec:FreqResolvedMech}, we describe the \textit{frequency-resolved Purcell effect} for the dissipative generation of steady-state entanglement, by which the cavity selectively enhances a certain set of  eigenenergy transitions to stabilize the emitters into the superradiant or subradiant eigenstates. We show this driven-dissipative strategy can be implemented using either a coherent or incoherent drive for the case of $N=2$ emitters, and that, in the particular case of incoherent excitation, this strategy can be scaled up to $N$ interacting emitters. 
%
%
Additionally, we address two of the primary challenges of quantum entanglement: its detection via the quantum and classical properties of the emitted light and its robustness against additional decoherence channels. 
In \refsec{sec:MechanismII}, we discuss the case in which two emitters are strongly detuned and have negligible coherent interaction in the absence of the cavity. In this regime, as originally reported in \colorref{RamosQuantumSpin2014,PichlerQuantumOptics2015}, the combination of driving at the two-photon resonance and collective decay induced by a cavity or a waveguide can stabilize the system into the antisymmetric state. Here, we contribute to the understanding of this effect from the physical insights inherited from the previous Chapter.
%
%
In \refsec{sec:MechanismIII} and \refsec{sec:MechanismIV}, we discuss the last two mechanisms for generating stationary entanglement. The first mechanism stabilizes the emitters into a mixed state of two-photon dressed states via either two-photon driving and spontaneous emission or through stimulated, collective emission through the cavity. The second mechanism stabilizes the system into a mixture of the ground and the antisymmetric states when the main decay channel is mediated by the cavity.
Finally, in \refsec{sec:MechanismV}, we describe a mechanism for generating metastable entanglement.

These results have been published in Physical Review Letters~\cite{Vivas-VianaFrequencyResolvedPurcell2024} and Physical Review Research~\cite{Vivas-VianaDissipativeStabilization2024}.

\section{Revisiting the two emitter model}
\label{sec:model}

\subsection{Model: two non-identical interacting emitters and a cavity}
In this Chapter, we extend the theoretical model of two interacting emitters introduced in \refch{ch:TwoPhotonResonance} [see \eqref{eq:LMEemitters}]:
two nonidentical interacting quantum emitters coherently driven and embedded in a photonic structure~\cite{ChangColloquiumQuantum2018,Gonzalez-TudelaLightMatter2024}, described as a single-mode cavity. 
The present configuration is illustrated in \reffig{fig:Scheme_Blender}.   
Additionally, we allow the emitters to be incoherently excited, e.g., as a result of an off-resonant coherent drive that excites higher energy levels, which then subsequently decay into the target state$^{\textcolor{Maroon}{*}}$.
\graffito{
$^*$We refer the reader to \refap{Appendix:IncoherentPump} for a detailed derivation of how incoherent driving effectively emerges in a specific example involving a three-level system.
}
Despite the apparent simplicity of the present model, the interplay of these elements---a collection of interacting quantum emitters, a coherent and incoherent drive, and a single-mode cavity---introduces a rich complexity, which, when properly tuned, can be harnessed for the generation of steady-state entanglement, as we will see throughout this Chapter.

\begin{SCfigure}[1][b!]
	\includegraphics[width=0.85\columnwidth]{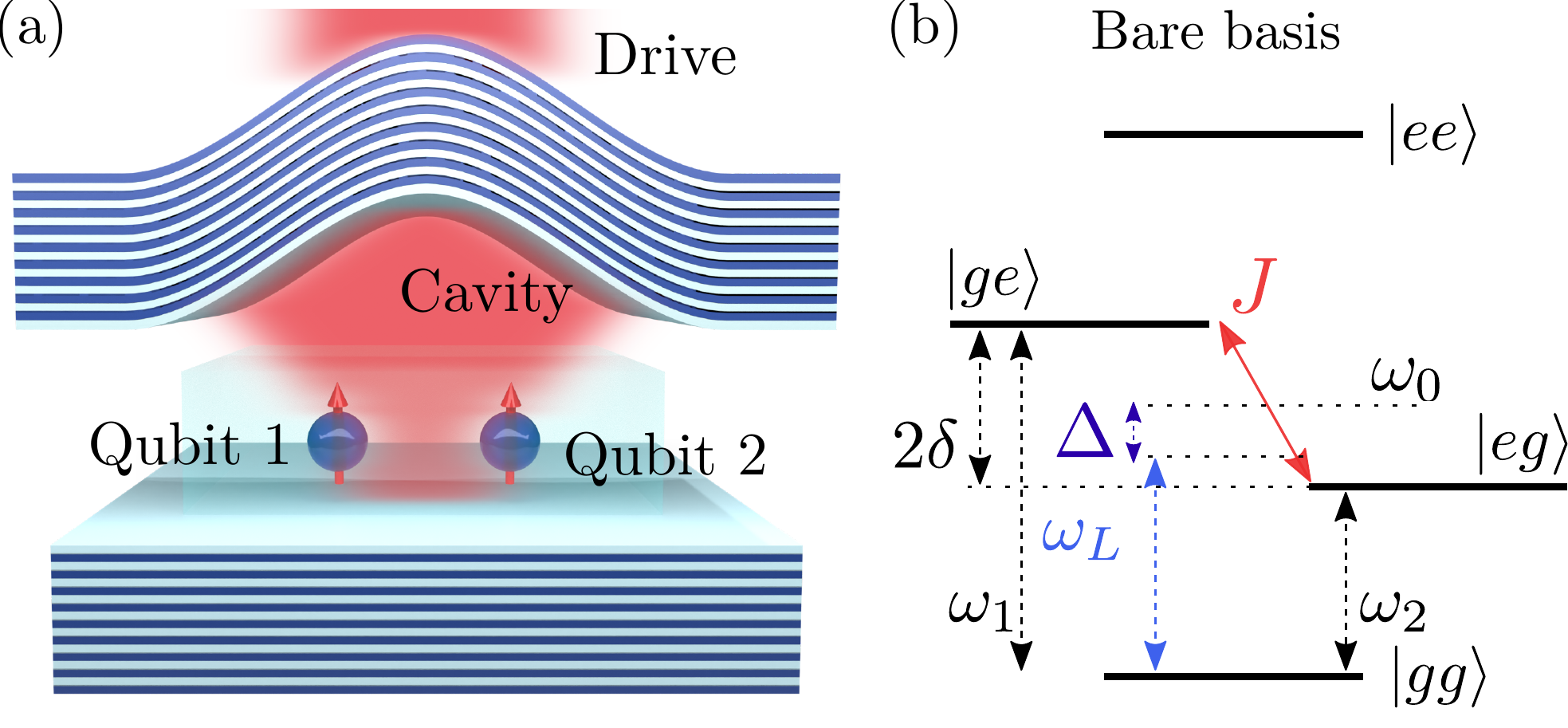}
	\captionsetup{justification=justified}
	\caption[Sketch of the emitters-cavity system.]{ \label{fig:Scheme_Blender}	\textbf{Sketch of the emitters-cavity system. } 
	(a) Two interacting nonidentical QEs under coherent drive and coupled to a single mode cavity. The cavity is illustrated as Fabry-Perot hybrid resonator~\cite{ShlesingerHybridCavityantenna2023}. (b) Energy 	diagram of the QEs in the bare basis. The states of the one-photon manifold are detuned by an energy $2\delta$ and coupled with a coupling strength $J$. The system is driven with a laser with frequency $\omega_L$
	}
	%
\end{SCfigure}

\paragraph{Master equation.} Our system is composed of two nonidentical interacting quantum emitters, both coupled to a single mode cavity, and driven by a coherent source of light. 
As usually assumed in this Thesis, we consider the quantum emitters as two-level systems, in agreement with state-of-the-art experiments using coherently-driven solid-state QEs well described by TLS, including QDs~\cite{FischerSelfhomodyneMeasurement2016}, molecules~\cite{WriggeEfficientCoupling2008,GerhardtCoherentState2009}, and NV centers~\cite{PigeauObservationPhononic2015,JoasQuantumSensing2017}. 
The cavity is modelled by a single bosonic mode with frequency $\omega_a$, annihilation operator $\hat a$, and emitter-photon coupling $g$. We recall the theoretical definitions introduced in \refsec{section:model} to describe the emitter-emitter system:
\begin{enumerate}[label=\textcolor{Maroon}{(\roman*)}]
	\item  The two natural frequencies of the emitters are given by $\omega_1\equiv \omega_0-\delta$ and $\omega_2\equiv \omega_0+\delta$, such that the emitters have an average frequency of $\omega_0=(\omega_1+\omega_2)/2$, with a relative energy detuning of $2\delta$.
	\item  The classical source of light is described by a laser field of frequency $\omega_L$, driving both emitters with a Rabi frequency $\Omega$.
\end{enumerate}
We can write, in the rotating frame of the laser and under the rotating wave approximation, the time-independent Hamiltonian $\hat H= \hat H_{\mathrm{q}} +\hat H_{\mathrm{a}} +\hat H_{\mathrm{d}}$, where $\hat H_{\mathrm{q}}$ is the bare Hamiltonian of the interacting emitters
\begin{equation}
		\hat H_{\text{q}}=(\Delta -\delta)\hat \sigma_1^\dagger \hat \sigma_1 +(\Delta +\delta)\hat \sigma_2^\dagger \hat \sigma_2 + J(\hat \sigma_1^\dagger  \hat \sigma_2+\text{H.c.}),
	\label{eq: H_emitters}
\end{equation}
$\hat H_\mathrm{a}$ is the Hamiltonian that involves the cavity
\begin{equation}
		\hat H_\mathrm{a}=\Delta_a \hat a^\dagger \hat a+g\left[\hat a^\dagger (\hat \sigma_1 +\hat \sigma_2)+\text{H.c.}\right],
	\label{eq:H_cavity}
\end{equation}
and $\hat H_\mathrm{d}$ is the Hamiltonian of the coherent drive,
\begin{equation}
		\hat H_\mathrm{d}=\Omega (\hat \sigma_1+\hat \sigma_2+ \text{H.c.}),
	\label{eq: H_laser}
\end{equation}
where we define the laser-qubit detuning, $\Delta$, and the laser-cavity detuning, $\Delta_a$, as
\begin{equation}
	\Delta\equiv \omega_0-\omega_L, \quad \text{and} \quad \Delta_a\equiv \omega_a-\omega_L.
\end{equation}

We also assume that the total system interacts with an environment that introduces incoherent processes by which the emitters are deexcited and photons leak out of the cavity. In the Markovian regime, the dissipative dynamics of the reduced qubits-cavity density matrix is described by the master equation~\cite{GardinerQuantumNoise2004,FicekQuantumInterference2005,BreuerTheoryOpen2007},
\begin{SCfigure}[1][b!]
	\includegraphics[width=0.9\columnwidth]{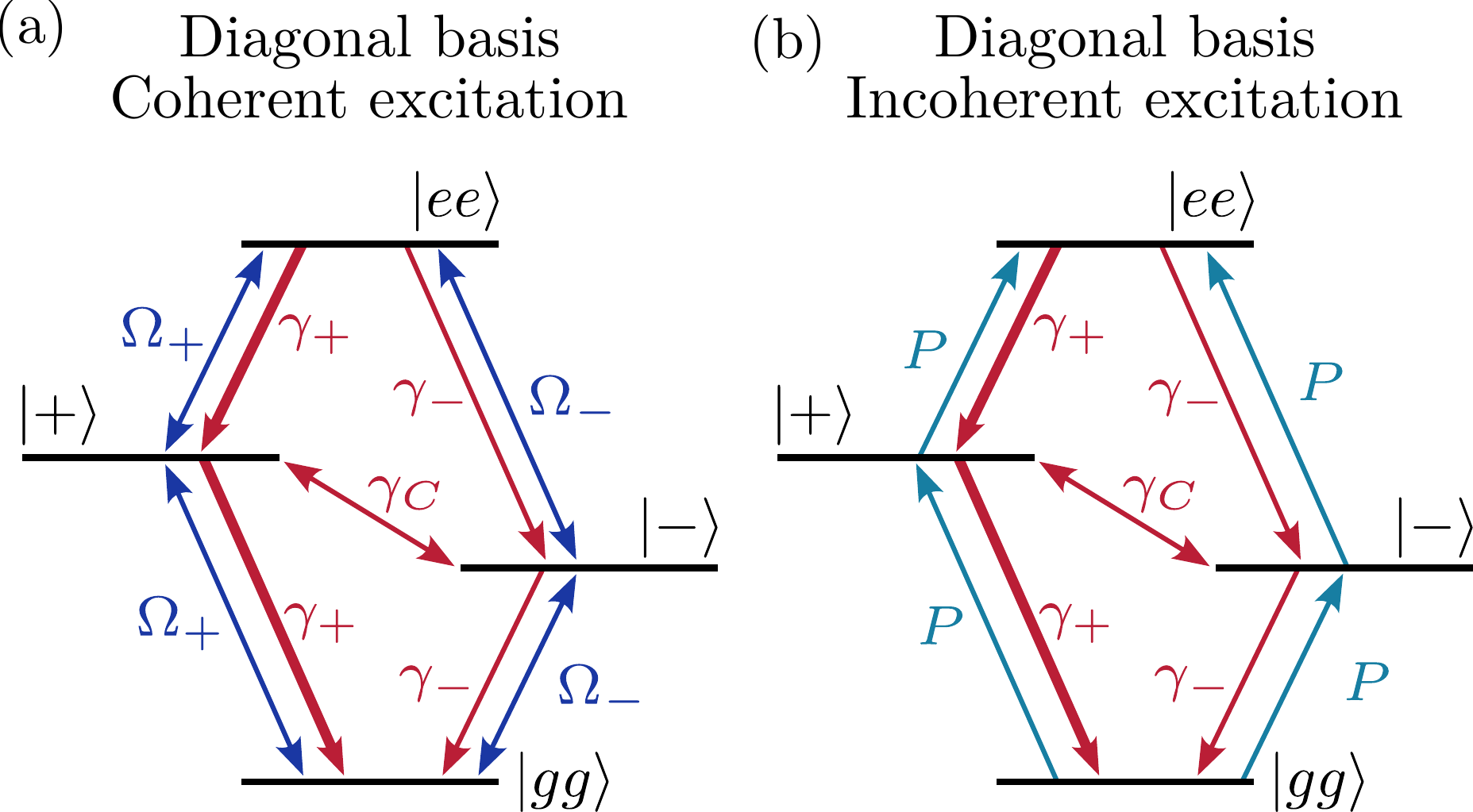}
	\captionsetup{justification=justified}
	\caption[Energy diagrams for the two pumping schemes.]{ \label{fig:PumpingSchemes}	\textbf{Energy diagrams for the two pumping schemes. }  Energy diagrams of the coherent and incoherent processes in the diagonal basis when coherent (a) and incoherent (b) channels of excitation are considered. Notice that blue double-ended arrows indicate coherent processes, whereas red (local and collective decay) and blue (incoherent pumping) arrows represent dissipative process.
	}
\end{SCfigure}
\begin{equation}
	\colorboxed{Maroon}{
		\frac{d \hat{\rho}}{dt}=-i[ \hat H, \hat{\rho}]+\sum_{i,j=1}^{2}\frac{\gamma_{ij}}{2}\mathcal{D}\left[\hat \sigma_i,\hat \sigma_j\right]  \hat{\rho}+\frac{P_i}{2}\mathcal{D}[\hat \sigma_i^\dagger]\hat \rho+\frac{\kappa}{2} \mathcal{D}\left[\hat a \right]  \hat{\rho}.
	}
	\label{eq:full_master_eq}
\end{equation}
Here, $\gamma_{ii}=\gamma$ is the local decay rate of spontaneous emission of the $i$th emitters, $\gamma_{12}=\gamma_{21}$ is the dissipative coupling rate between emitters resulting from the coupling to a common environment~\cite{FicekQuantumInterference2005}, $P_i=P$ is the local rate of incoherent pumping, and $\kappa$ is the photon leakage rate. 
As considered in previous Chapters, incoherent excitation by thermal photons is neglected since we consider optical transitions or cryogenic temperatures in platforms such as superconducting circuits~\cite{GarciaRipollQuantumInformation2022}. 

In this case,  the emitters can be excited either via the coherent field [\eqref{eq: H_laser}] or via single-photon incoherent excitations [\eqref{eq:full_master_eq}].
%
%
%
In order to provide a clear description of the different mechanisms of entanglement generation, we simplify the discussion by independently considering the following two pumping schemes:
\begin{enumerate}[label=\textcolor{Maroon}{(\roman*)}]
	\item \textcolor{Maroon}{Coherent excitation:} $\Omega\neq0 ,\ P=0 $.
	\item \textcolor{Maroon}{Incoherent excitation:} $\Omega=0 ,\ P\neq0 $.
\end{enumerate}
These two scenarios are depicted in \reffig{fig:PumpingSchemes}~\textcolor{Maroon}{(a, b)}, respectively. We note that in the case of incoherent excitation, since the coherent drive is disabled, the energy reference fixing the rotating frame is chosen to be the average qubit frequency, $\omega_0$. Consequently, the cavity detuning term reads
\begin{equation}
	\Delta_a\equiv \omega_a-\omega_0, \quad \text{(Incoherent excitation)}
	\label{eq:Incoh_Cond}
\end{equation}
and the qubit Hamiltonian simplifies to
\begin{equation}
			\hat H_{\text{q}}= -\delta\hat \sigma_1^\dagger \hat \sigma_1 +\delta \hat \sigma_2^\dagger \hat \sigma_2 + J(\hat \sigma_1^\dagger  \hat \sigma_2+\text{H.c}) \quad \text{(Incoherent excitation)}.
\end{equation}

\paragraph{Validity of the model.} The  model presented in \eqref{eq:full_master_eq}, and the results derived from it, apply to any type of interaction between the emitters, as long as such general interaction admits a description in terms of a Tavis-Cummings model$^{\textcolor{Maroon}{*}}$. 
\graffito{
$^*$In \refap{sec:SM_Validity_RWA}, we prove the validity of the different RWAs applied to the interaction terms in the total Hamiltonian: the coherent excitation and the emitters-cavity interaction. 
}
Nevertheless, in order to provide an intuitive reference of the strength of the coherent ($J$) and dissipative rates ($\gamma$, $\gamma_{12}$), we will parametrize these terms as the values corresponding to point-dipoles in free space, which interact through vacuum, inducing dipole-dipole interactions~\cite{CarmichaelStatisticalMethods1999,FicekQuantumInterference2005,HettichNanometerResolution2002,TrebbiaTailoringSuperradiant2022,LangeSuperradiantSubradiant2024}. Therefore, these terms would be described by \cref{eq:J,eq:gamma_12}, exhibiting a behaviour in terms of the inter-emitter distance and the geometrical configuration of the dipoles already discussed in \refsec{sec:geometrical_conf}.
We stress that this parametrization serves only as a reference, and that values of $J$ that may seem unrealistic under this parametrization could nevertheless emerge from other scenarios also described by our model, such as when the coherent interaction is mediated by engineered photonic structures~\cite{Gonzalez-TudelaEntanglementTwo2011,Martin-CanoDissipationdrivenGeneration2011,ChangColloquiumQuantum2018,Miguel-TorcalInversedesignedDielectric2022,Miguel-TorcalMultiqubitQuantum2024}.

We finally note that all the results presented in this Chapter are valid in a general case where $\gamma_1\neq \gamma_2$ as long as $|\gamma_1-\gamma_2|\ll (\gamma_1+\gamma_2),\gamma_{12}$---see \refap{Appendix:diff_gamma}---.
Also, we assume an H-aggregate configuration, in which the dipole moments are perpendicular to the line that connects them, i.e.,  $\bm{\bar \mu }\perp \bm{\bar r_{12}}$.
As we mentioned in \refsec{sec:geometrical_conf}, this choice is made out of convenience without loss of generality as the cooperative effects emerged at regions where the geometrical configuration might be relevant are not essentially affected~\cite{ReitzCooperativeQuantum2022}. 

\subsection{Adiabatic elimination of the cavity}

In this Chapter, we explore the applications of lossy cavities in the design of dissipative phenomena by means of the Purcell effect~\cite{PurcellResonanceAbsorption1946}. 
We specifically consider the cavity operating in the weak coupling regime, where the losses through the cavity significantly exceed the coherent interaction between the cavity and the emitters ($\kappa > g$). In particular, we restrict the cavity to be in the bad-cavity limit~\cite{SavageStationaryTwolevel1988,CiracInteractionTwolevel1992,ZhouDynamicsDriven1998,Navarrete-BenllochIntroductionQuantum2022}, a regime characterized by
\begin{equation}
	\colorboxed{Maroon}{
		\kappa \gg g \gg \gamma, \gamma_{12} \quad \text{(Bad cavity limit)}.
	}
\end{equation}
In this limit,  we can perform an adiabatic elimination of the cavity degrees of freedom and derive an effective model for the reduced two-emitter system, similar to the approach followed for a single emitter coupled to a lossy cavity in \refsec{sec:AE_singleEmitter}. 
Notably, even though the emitters are weakly coupled to the cavity  ($g<\kappa$), the cooperativity---already introduced in \eqref{eq:coop_intro}---,
\begin{equation}
	\colorboxed{Maroon}{
		C \equiv \frac{4g^2}{\kappa \gamma}, \quad \quad \text{(Cooperativity)}
	}
	\label{eq:Coop}
\end{equation}
might exhibit values much larger than $1$ as long as the emitters decay rates are sufficiently suppressed ($\kappa \gg \gamma$) and $g$ is non-negligible, satisfying the relation $\kappa \gamma\ll 4g^2$.  In such a regime, the losses in the system are predominantly channelled through the cavity mode rather than through spontaneous emission, effectively enhancing the cooperative effects between the emitters and the cavity~\cite{PlankensteinerEnhancedCollective2019,ReitzCooperativeQuantum2022}.
As we will see in this Chapter, the weak coupling regime ($\kappa\gg g$), combined with the bad-cavity limit $C\gg1$, are essential requirements for achieving near-maximal, long-lived entanglement between the emitters.

\paragraph{Failure of the Hamiltonian adiabatic elimination. }In  \refch{chapter:Unconventional}, we introduced an effective description of this system by performing a Hamiltonian adiabatic elimination, where the cavity contributions were treated as a non-Hermitian Hamiltonian---see \refsec{Section:QuantumTraject}---. As a result, we phenomenologically proposed an effective $\Lambda$-system within the subspace $\{|gg\rangle, |+\rangle, |ee\rangle \}$,  
\begin{multline}
	\frac{d \hat{\rho}}{dt}=-i[ \hat H_\mathrm{q}+\hat H_{\mathrm{d}}, \hat{\rho}] +\frac{\Gamma_{2\mathrm{p}}}{2 } \mathcal{D}[|gg\rangle \langle ee|]\hat \rho\\
	+\frac{\Gamma_{1\mathrm{p}}+\gamma_+ }{2 }(\mathcal{D}[|gg\rangle \langle +|]+\mathcal{D}[|+\rangle \langle ee|])\hat \rho.
\end{multline}
exhibiting two additional Lindblad terms with dissipative rates $\Gamma_{1\mathrm{p}}$ and $\Gamma_{2\mathrm{p}}$, derived from \eqref{eq:Master_eq_wo_cavity}:
\begin{align}
	\Gamma_{2\mathrm{p}}&=\frac{64 g^4 \kappa}{\chi}, \\
	\Gamma_{1\mathrm{p}}&=\frac{8g^2 \kappa [4(\Delta-\Delta_a)^2+\kappa^2]}
	{ \chi		}
\end{align}
with $\chi$ given by
\begin{multline}
	\chi=64 g^4 + 16 g^2 [ 4(\Delta-\Delta_a)(R-\Delta+\Delta_a)+\kappa^2 ] \\
	+[4(\Delta-\Delta_a)^2+\kappa^2][4(R-\Delta+\Delta_a)^2+\kappa^2].
\end{multline}

Nevertheless, as already anticipated in the previous Chapter, that adiabatic elimination does not fully capture the physics of the full model from \eqref{eq:full_master_eq}. The failure of this approach is evident in \reffig{fig:ComparisonAE}, where the time evolution of the symmetric state population $\rho_{+,+}$ is shown for two different laser-cavity detunings, $\Delta_a=\{0,-R\}$.
The Hamiltonian adiabatic elimination (represented by dot-dashed lines) do not reproduce the actual system dynamics (denoted by solid lines) at any point in time. 
So, \textit{what is it happening?} \textit{Why does this adiabatic elimination fail to reproduce the system dynamics, even though we are in the bad-cavity limit, where an effective description of the emitters without the cavity should theoretically be valid?} 

We might think the issue lies in restricting the Hilbert space to the two-excitation manifold $\{|ee,0\rangle, |+,1\rangle, |gg,2\rangle\}$, and that extending it to include higher and lower excitation states could solve the problem. However, pursuing this approach leads to the same outcome: the adiabatic elimination still fails to fully capture the system dynamics. This suggests that a more sophisticated adiabatic elimination method is needed.
{
	\sidecaptionvpos{figure}{t}
\begin{SCfigure}[1][t!]
	\includegraphics[width=0.65\columnwidth]{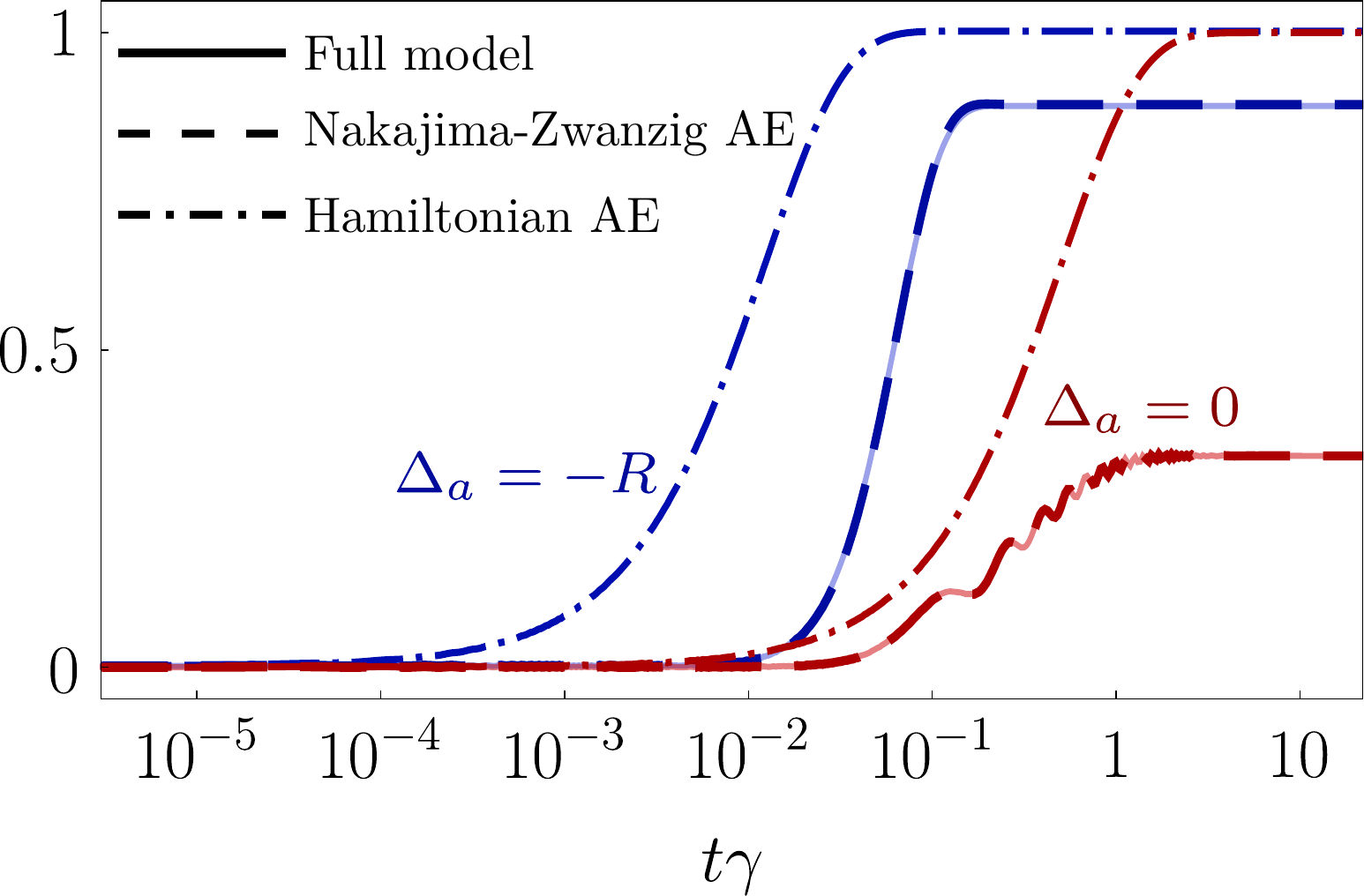}
	\captionsetup{justification=justified}
	\caption[Approaches to adiabatically eliminate the cavity]{ \label{fig:ComparisonAE}	\textbf{Approaches to adiabatically eliminate the cavity. }  Time evolution of $\rho_{+,+}$ when $\Delta_a=0$ (in red) and $\Delta_a=-R$ (in blue). Solid lines correspond to numerical simulation from the full model in \eqref{eq:full_master_eq}, dot-dashed lines correspond to numerical simulation from the Hamiltonian AE in \eqref{eq:Master_eq_wo_cavity}, and dashed lines correspond to numerical simulation from the Nakajima-Zwanzig AE in \eqref{eq:Nakajima}. Parameters: $r_{12}=2.5$ nm, $k=2\pi/780\ \text{nm}^{-1}$, $J=9.18\times 10^4\gamma$, $\gamma_{12}=0.999\gamma$, $\Delta=0$, $\delta=0$, $R=9.18\times 10^4\gamma$, $\Omega=10^3\gamma$, $\kappa=10^3\gamma$, and $g=10^{-1}\kappa$. 
	}
\end{SCfigure}}

\paragraph{The Nakajima-Zwanzig approach. }To overcome this issue, we consider the application of the Nakajima-Zwanzig formalism~\cite{NakajimaQuantumTheory1958,ZwanzigEnsembleMethod1960,BreuerTheoryOpen2007,RivasOpenQuantum2012,Navarrete-BenllochIntroductionQuantum2022,Gonzalez-BallesteroTutorialProjector2024}.
 This systematic approach---introduced in \refsec{sec_Nakajima}---allows us to formally eliminate the degrees of freedom of the lossy cavity by identifying the relevant contribution of the system dynamics when $1/\kappa$ is the shortest dissipative timescale. In this case,  the action of the emitter-laser Hamiltonian, $\hat H_{\mathrm{q}}+\hat H_{\mathrm{d}}$. 
Then, following the Nakajima-Zwanzing formalism, we derive a master equation for the emitters, where the effect of the cavity takes the form of Bloch-Redfield terms$^{\textcolor{Maroon}{*}}$ \cite{WhitneyStayingPositive2008,JeskeBlochRedfieldEquations2015,FernandezDeLaPradillaRecoveringAccurate2024}, %
\graffito{
The details of the step-by-step derivation can be found in \refap{appendix:AdiabaticElimination}.
}
\begin{equation}
	\colorboxed{Maroon}{
		\begin{aligned}[b]
			\frac{d \hat{\rho}}{dt}=-i[&\hat{H}_q+\hat{H}_d,\hat{\rho}]+\sum_{i,j=1}^2\frac{\gamma_{ij}}{2}\mathcal{L}[\hat{\sigma}_i,\hat{\sigma}_j]\hat{\rho}+\frac{P_i}{2}\mathcal{D}[\hat \sigma_i^\dagger]\hat \rho
			\\ &\ + \sum_{i,j,m,n=1}^4 \left( \frac{g_{ij}g_{mn}^* }{\kappa/2+i(\Delta_a-\omega_{ij})}\left[\hat \sigma_{ij}\hat \rho,\hat \sigma_{mn}^\dagger\right] + \text{H.c.} \right).
		\end{aligned}
	}
	\label{eq:Nakajima}
\end{equation}
Here,  we defined
\begin{subequations}
	\begin{empheq}[box=\colorboxed{Maroon}]{align}
	\hat\sigma_{ij}&\equiv |j\rangle \langle i |, \quad \quad\quad \quad \hspace{0.15cm} \text{(dressed qubit-laser ladder operators)} \\
	g_{ij}&\equiv g\langle j| \hat \sigma_1 +\hat \sigma_2 |i\rangle, \hspace{0.3cm} \text{(effective cavity coupling rate)}\\
	\omega_{ij}&\equiv\lambda_i-\lambda_j, \quad \quad \quad \ \ \text{(dressed qubit-laser transition energy)}
\end{empheq}
\end{subequations}
where $|i\rangle$ and $\lambda_i$, $i=1,\ldots,4$, are the eigenvectors and eigenvalues of the dressed qubit-laser system, extensively studied in \refch{ch:TwoPhotonResonance}.
Note that in the incoherent excitation scheme ($\Omega=0,\ P\neq 0$), the system eigenvectors $|i\rangle$ reduce to the qubit-qubit eigenstates, labelled as
\begin{equation}
	|1\rangle =|gg\rangle; \quad 	|2\rangle =|+\rangle; \quad	|3\rangle =|-\rangle;  \quad	|4\rangle =|ee\rangle.
\end{equation}

We test the resulting adiabatic elimination in \eqref{eq:Nakajima} by computing the time evolution of the symmetric state population, shown as dashed lines in \reffig{fig:ComparisonAE}. Unlike the previous Hamiltonian adiabatic elimination from \eqref{eq:Master_eq_wo_cavity}, this effective description accurately captures all the physical features of the full model, such as the dressing of the qubits by the drive.
In addition to providing a more efficient computational method since the cavity degrees of freedom have been traced out, this effective master equation will be particularly useful for analysing the underlying physical processes involved in the different mechanisms of entanglement generation, which we will discuss in this Chapter. Notably, the key features of this effective master equation are: 
\begin{enumerate}[label=\textcolor{Maroon}{(\roman*)}]
\item its dependence on the dressed qubit-laser structure through $\hat{\sigma}_{ij}$, $g_{ij}$, and $\omega_{ij}$,
\item its dependence on the cavity frequency via $\Delta_a$.
\end{enumerate}
The first of these features, the dependence on the dressed qubit-laser structure, was not accounted for in the previous phenomenological approach considered in \eqref{eq:Master_eq_wo_cavity}. As we will see, this feature is of vital importance in the mechanism of entanglement generation.

We finally note that a characteristic quantity emerging in several limits of \eqref{eq:Nakajima} is the effective decay rate $ \Gamma_P$, induced in the qubits by the coupling to the lossy cavity. This is given by the standard expression of the Purcell rate~\cite{PurcellResonanceAbsorption1946,KavokinMicrocavities2017}---already introduced in \eqref{eq:Purcell_intro}---,
\begin{equation}
	\colorboxed{Maroon}{
	\Gamma_P \equiv \frac{4g^2}{\kappa}, \quad  \quad \text{(Purcell rate)}.
	}
	\label{eq:purcell-rate}
\end{equation}
Given the definition of cooperativity in \eqref{eq:Coop} and the Purcell rate in \eqref{eq:purcell-rate}, it is now more evident that when $C=\Gamma_P/\gamma>1$, the losses in the system are predominantly channelled through the cavity, as stated earlier.
Both quantities, the cooperativity and the Purcell rate, will be extensively used throughout this Chapter.

\section{Overview of entanglement mechanisms in parameter space}
\label{sec:Freq_Resolved_Mech}
%
%
%
\subsection{Dependence with laser frequency}
\label{sec:laser-freq}

In \reffig{fig:Fig1_FailureTwoPhotonDecay1}, we observed that when the qubits are driven at the two-photon resonance ($\Delta=0$) and the cavity---in the bad-cavity limit, and  \textit{resolving} the emitters eigenstructure---is tuned in resonance with the symmetric state, the emitter-emitter system stabilizes into an almost maximally symmetric state, achieving near-maximum concurrence $\mathcal{C}\approx 0.9$. However, when $\Delta_a=0$, the entanglement completely vanishes in the long-time limit. 
This frequency dependency through $\Delta$ and $\Delta_a$ suggests the presence of a frequency structure underlying the generation of entanglement between emitters.
Therefore, as a first step of our analysis,  we study the impact of the choice of laser frequency in the generation of entanglement. 

\begin{SCfigure}[1][b!]
	\includegraphics[width=0.9\columnwidth]{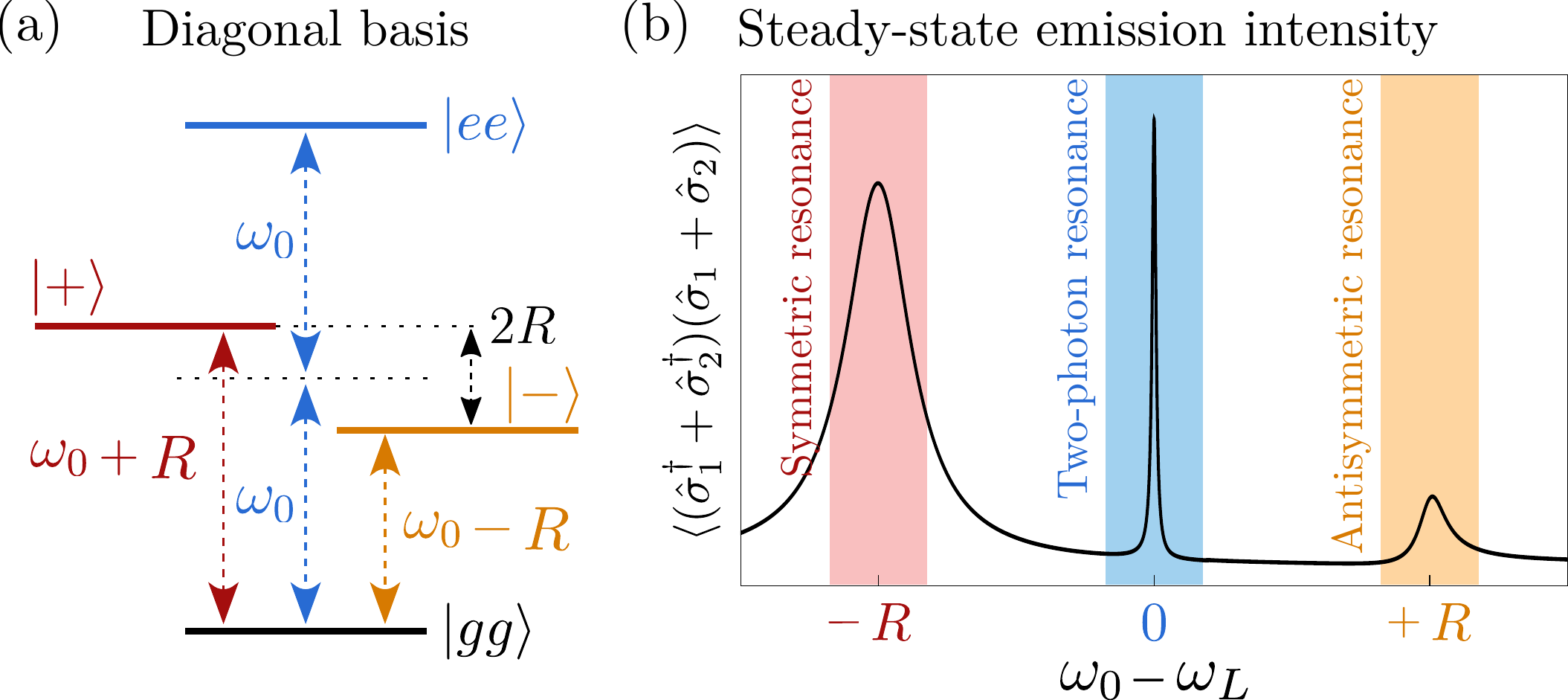}
	\captionsetup{justification=justified}
	\caption[Eigenstructure and laser resonances.]{\label{fig:Fig_Eigenresonances} \textbf{Eigenstructure and laser resonances. }(a) Energy levels in the excitonic basis where the coupling between emitters has been diagonalized. (b) Steady-state emission intensity as the frequency of the laser
		is varied. 
	}
\end{SCfigure}

To address this question, we first recall the eigenenergy structure of the bare, undriven emitters$^{\textcolor{Maroon}{*}}$ $\{|gg\rangle, |+\rangle, |-\rangle,|ee\rangle\}$, 
\graffito{
$^*$A detailed discussion of the emitter eigenstructure and its geometrical properties can be found in \refsec{sec:geometrical_conf}.
}
where $|\pm \rangle$ are the eigenstates of the single-excitation subspace,
\begin{equation}
	|\pm \rangle=\frac{1}{\sqrt{2}}(\sqrt{1\mp \sin \beta}|eg\rangle \pm \sqrt{1\pm \sin \beta}|ge\rangle),
\end{equation}
with $\beta\equiv\arctan(\delta/J)$ and eigenenergies $E_\pm=\Delta \pm R$, where $R=\sqrt{J^2+\delta^2}$. 
The energy structure is shown in \reffig{fig:Fig_Eigenresonances}~\textcolor{Maroon}{(a)}.
This configuration is evidenced in observables such as the stationary emission intensity, $\langle (\hat \sigma_1^\dagger+ \hat \sigma_2^\dagger)(\hat \sigma_1+ \hat \sigma_2)\rangle$, revealing a three-peak structure in the laser frequency domain, corresponding to the excitonic resonances of the dimer at $\Delta = \{-R, 0, R\}$, as shown in \reffig{fig:Fig_Eigenresonances}~\textcolor{Maroon}{(b)}. From left to right, these resonances correspond to the symmetric ($\Delta=-R$), two-photon ($\Delta=0$) and antisymmetric ($\Delta=R$) resonances, as we explained in \refch{ch:TwoPhotonResonance}.
Considering this structure, we explore the stationary concurrence $\mathcal{C}$ of the qubits in the frequency domain in terms of the laser-qubit detuning $\Delta$ and laser-cavity detuning $\Delta_a$, as illustrated in \reffig{fig:Fig3_ConcurrenceQubitLaserDetuning}. 
In this figure, we observe how the stationary value of $\mathcal C$ features clear resonant patterns at these specific system resonances. In \reffig{fig:Fig3_ConcurrenceQubitLaserDetuning}~\textcolor{Maroon}{(a)}, we present a cut of the concurrence (solid black curve) versus $\Delta$ while fixing $\Delta_a=-R$, exhibiting three peaks centred at the three excitonic resonaces emerging from the dimer. 

\begin{SCfigure}[0.9][b!]
	\includegraphics[width=0.6\columnwidth]{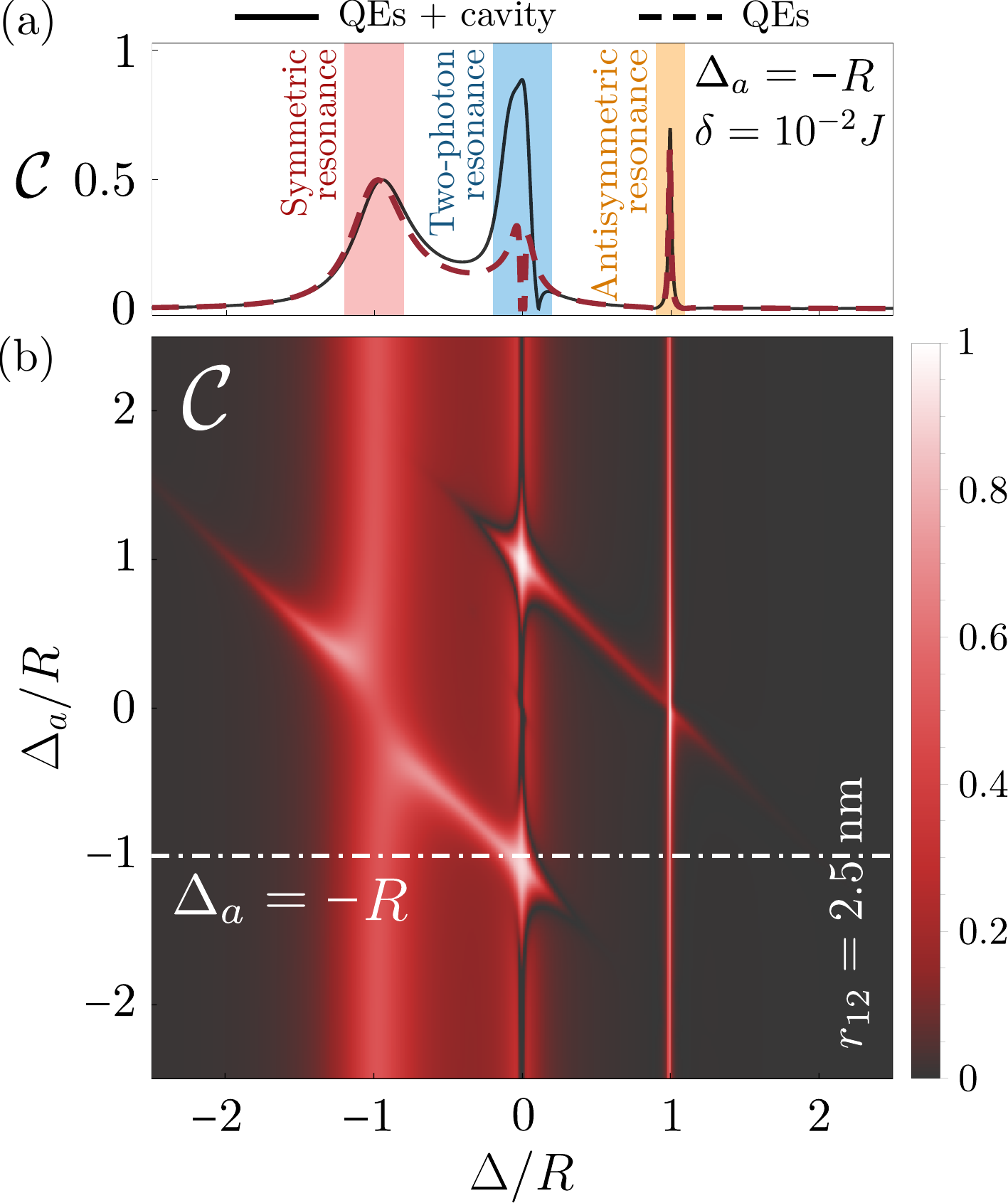}
	\captionsetup{justification=justified}
	\caption[Dependence on both the qubit-laser $\Delta$ and cavity-laser $\Delta_a$ detunings for entanglement generation.]{\label{fig:Fig3_ConcurrenceQubitLaserDetuning} \textbf{Dependence on both the qubit-laser $\Delta$ and cavity-laser $\Delta_a$ detunings for entanglement generation. } Steady-state concurrence versus (a) laser-qubit detuning $\Delta$ 
		and (b) both $\Delta$ and laser-cavity detuning $\Delta_a$ (b). In (a), the laser-cavity detuning is fixed at $\Delta_a=-R$ [dot-dashed line in (b)] The solid black line corresponds to numerical computation with the emitters-cavity model, and the dashed red line corresponds to numerical computation with just the emitters model ($g=0, \kappa=0$). 
		Parameters (a-b): $r_{12}=2.5\ \text{nm}$, $k=2\pi/780\ \text{nm}^{-1}$, $J=9.18\times 10^4\gamma$, $\gamma_{12}=0.999\gamma$, $\delta=10^{-2}J$, $R=9.18\times 10^4\gamma$, $\Omega=10^4 \gamma$, $\kappa=10^4 \gamma$, $g=10^{-1}\kappa$. }
\end{SCfigure}

\paragraph{One-photon resonances.} The leftmost and rightmost peaks, that we label as ``symmetric'' and ``antisymmetric'', are explained by the fact that, at laser-qubit detunings $\Delta=\pm R$, the laser is resonant with the transition between the ground state and one of the one-excitation eigenstates [see \reffig{fig:Fig_Eigenresonances}~\textcolor{Maroon}{(a)}], which for strongly coupled qubits correspond to the superradiant and subradiant
states $|\pm\rangle \approx |S/A\rangle$, since $\beta\ll 0$ when $J\gg \delta$. 
In this case of direct excitation of the single-excitation manifold, the dynamics  gets approximately confined within a reduced subspace involving only the ground state and the symmetric, or the antisymmetric, state
\begin{subequations}
\begin{align}
	&\{|gg\rangle, |S\rangle \} \quad \text{when}\quad \Delta=-R, \\
	&\{|gg\rangle, |A\rangle \} \quad \text{when}\quad \Delta=+R.
\end{align}
\end{subequations}
Since in the symmetric and the antisymmetric resonances the qubits are excited only by populating either the state $|S\rangle$ or $|A\rangle$, respectively, and these are maximally entangled states, the finite population of these states leads to high values of the stationary concurrence---see \refap{appendix:B} for a detailed analysis---. These two resonances feature very different widths due to their respective superradiant and subradiant nature \cite{FicekQuantumInterference2005}, as we could already observe in \refch{ch:TwoPhotonResonance}.

Here, the cavity does not play a significant role, as shown in \reffig{fig:Fig3_ConcurrenceQubitLaserDetuning}~\textcolor{Maroon}{(a)}, where the red dashed curve represents the concurrence of the qubits without the influence of the cavity [$\kappa, g=0$ in \eqref{eq:full_master_eq}]. From this plot, we highlight the fact that the symmetric and antisymmetric regions already exhibit stationary entanglement without considering the effect of the cavity. Hence,
while the cavity slightly modifies the effective decay rates via Purcell effect, it is not responsible of that particular mechanism of entanglement generation. 
This type of direct resonant excitation of the subradiant and superradiant states has been reported experimentally, e.g., in \colorrefs{TrebbiaTailoringSuperradiant2022,LangeSuperradiantSubradiant2024}.

\paragraph{Two-photon resonance.} 
On the other hand, the highest value of the concurrence is obtained when the system is driven at the two-photon resonance, i.e., $\Delta=0$, with the concurrence reaching values close to the maximum achievable entanglement $\mathcal{C}\approx 1$ when the cavity frequency is tuned to one of the two resonant frequencies, $\Delta_a \approx \pm R$. 
These two resonance conditions correspond to the stabilization of the states $|A\rangle$ and $|S\rangle$, respectively. 
Notably, the stabilization time---the time the system takes to reach the steady-state, characterized by the Liouvillian gap---is several orders of magnitude faster when exciting the two-photon resonance compared to the one-photon resonances. This  behaviour is clearly illustrated in \reffig{fig:Fig3_ConcurrenceTimeQubitLaserDetuning}, where we show the time evolution of the symmetric and antisymmetric state populations when exciting the one-photon resonances $(\Delta=\pm R)$ and the two-photon resonance $(\Delta=0,\ \Delta_a =\pm R)$. In the latter case, the stabilization of $\rho_{S,S}$ (red solid line) and $\rho_{A,A}$ (blue solid line) is orders of magnitude faster than their one-photon excitation counterparts, represented by the red dashed and blue dot-dashed lines, respectively.
Moreover, the entanglement created at the two-photon resonance is the only one that survives as the distance between emitters increases and the one-excitation eigenstates lose their superradiant/subradiant character
$|\pm\rangle \neq |S/A\rangle$---see \refap{appendix:B} for a detailed analysis---. All these features justify that, in this Chapter, we focus our attention on the regime of resonant two-photon excitation $\Delta=0$.
\begin{SCfigure}[1][h!]
	\includegraphics[width=0.75\columnwidth]{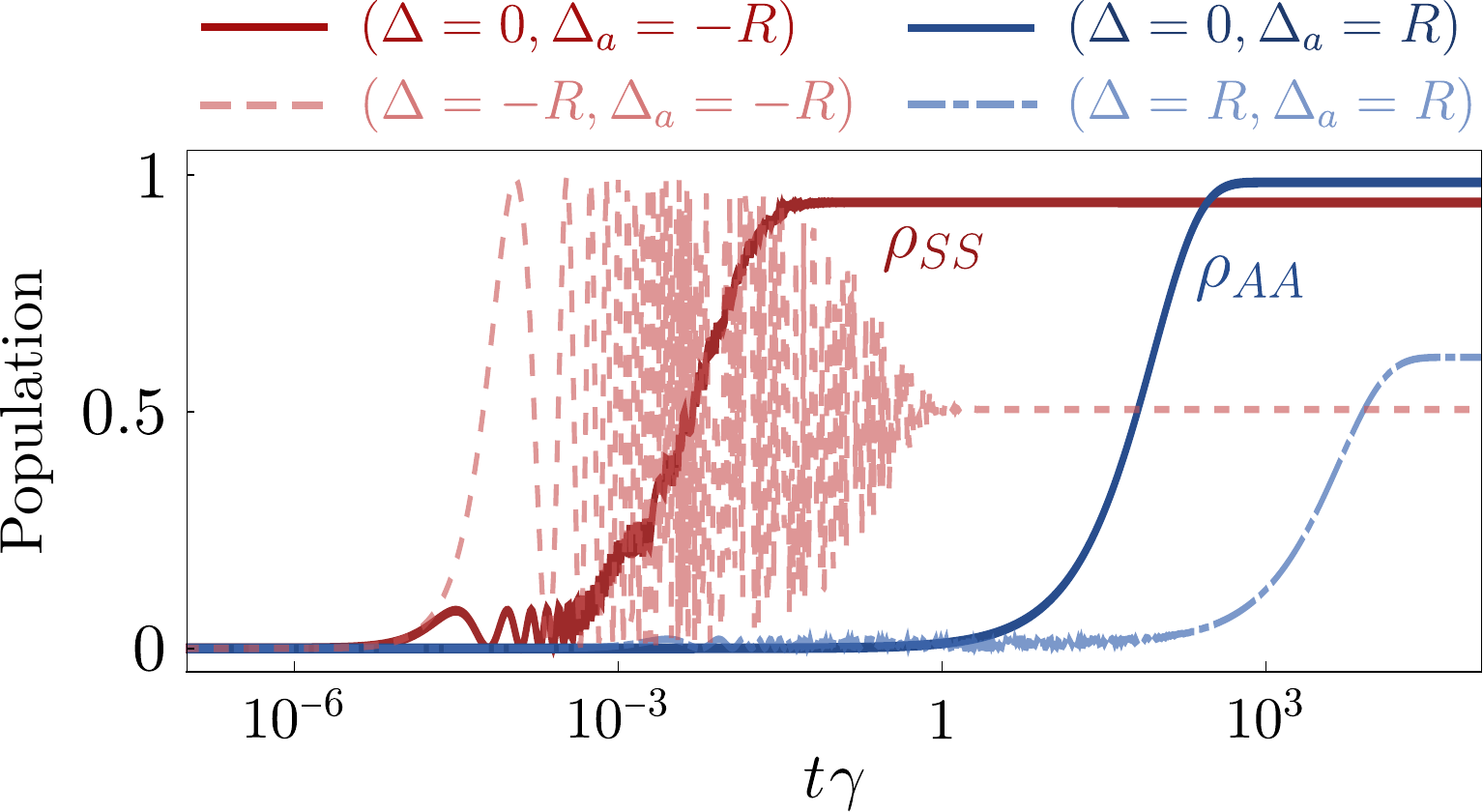}
	\captionsetup{justification=justified}
	\caption[Time evolution of symmetric and antisymmetric populations.]{\label{fig:Fig3_ConcurrenceTimeQubitLaserDetuning} \textbf{Time evolution of symmetric and antisymmetric populations.} Time evolution of the symmetric ($\rho_{SS}$, red curves) and antisymmetric ($\rho_{AA}$, blue curves) populations for different laser-qubit and laser-cavity resonances. Solid lines correspond to two-photon resonance condition ($\Delta=0$), and dot-dashed and dashed lines correspond to one-photon resonance  condition ($\Delta=\pm R$), respectively.
		Parameters: $r_{12}=2.5\ \text{nm}$, $J=9.18\times 10^4\gamma$, $\gamma_{12}=0.999\gamma$, $\delta=10^{-2}J$, $R=9.18\times 10^4\gamma$, $\Omega=10^4 \gamma$, $\kappa=10^4 \gamma$. }
	
\end{SCfigure}

We highlight that, in contrast to the symmetric and antisymmetric resonances, the dissipative generation of entanglement  at the two-photon resonance is enabled by the cavity, as we observe in \reffig{fig:Fig3_ConcurrenceQubitLaserDetuning}~\textcolor{Maroon}{(a)} around $\Delta\approx0$ (blue region). In the absence of the cavity, the concurrence exhibits a sharp dip at the two-photon resonance ($\Delta=0$), with $\mathcal{C}=0$ (dashed red curve), and only when the cavity is included (solid black curve), nearly maximal entanglement is generated.   
This cavity-enabled process is discussed in detail in the following sections.
 
We additionally note that similar cavity-enabled mechanisms of entanglement stabilization in systems under two-photon excitation have been proposed in biexciton-exciton cascades for the case of single QDs~\cite{SeidelmannDifferentTypes2021,
	SeidelmannTimedependentSwitching2021,
	SeidelmannPhononinducedTransition2023}.

\subsection{Dependence with other parameters}
\label{sec:other_mechanisms}

As we anticipated in the introduction of this Chapter, the model considered here [\eqref{eq:full_master_eq}]
exhibits a rich complexity that can be harnessed for generating entanglement among the emitters by properly tuning the system parameters. 
Specifically, we identify five different driven-dissipative mechanisms capable of generating long-lived of both stable and metastable entanglement among the emitters, which are tunable and robust against decoherence. 
%
%
A systematic exploration of the parameter space reveals a rich landscape of high stationary entanglement regions, each arising from a different physical process.
In this section, we provide an overview of the stationary mechanisms of entanglement generation and their dependence on the system parameters. These are illustrated in \reffig{fig:Fig9_EntanglementRegimes_Coherent} for the case of coherent excitation and in \reffig{fig:Fig9_EntanglementRegimes_Incoherent} for the case of incoherent excitation.

	\addtocounter{figure}{-1}
\begin{SCfigure}[1][b!]
	\includegraphics[width=1.0\columnwidth]{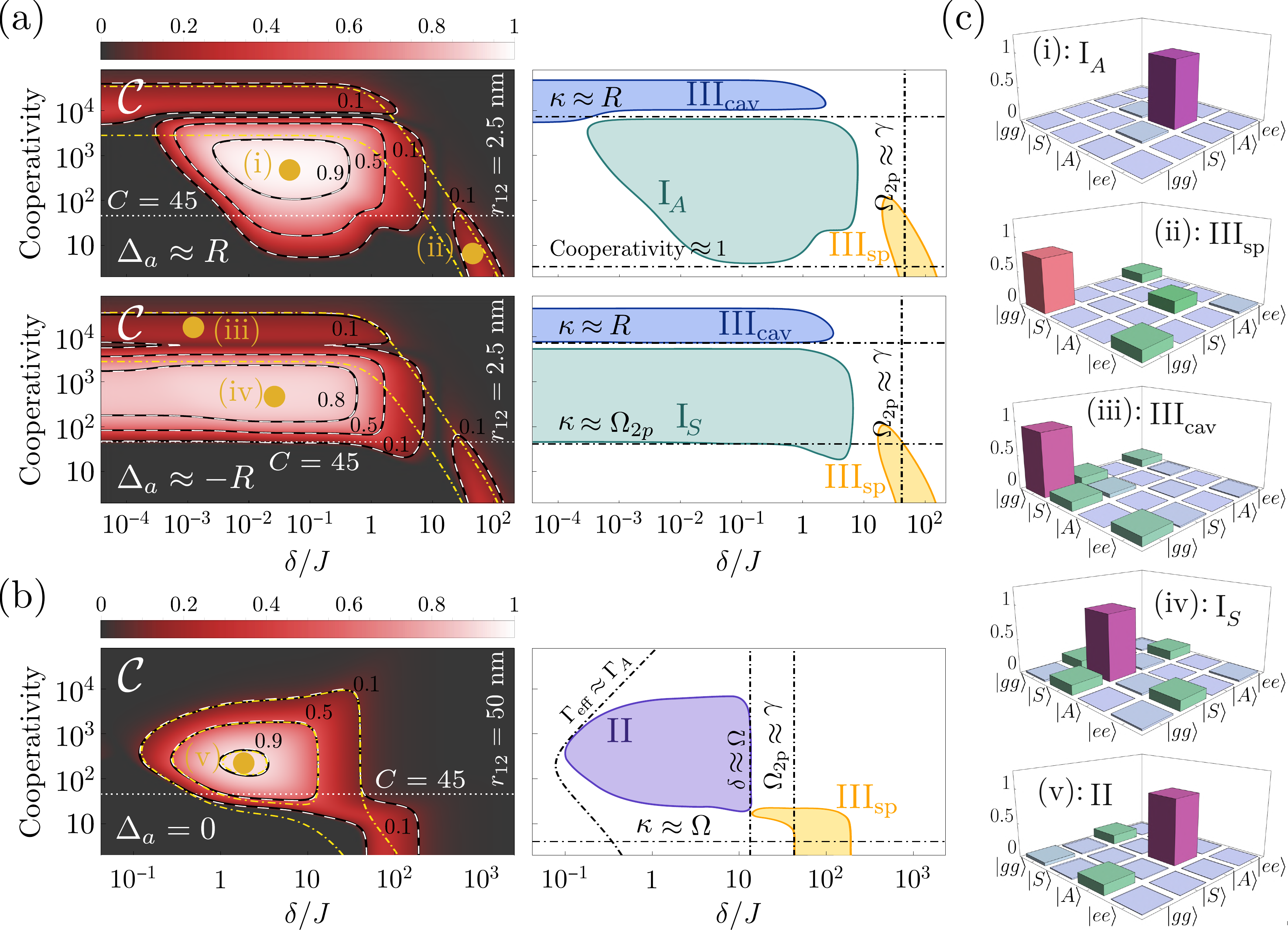}
	\captionsetup{justification=justified}
\end{SCfigure}
\graffito{\vspace{1.6cm}
	\captionof{figure}[Entanglement regimes under coherent excitation.]{\label{fig:Fig9_EntanglementRegimes_Coherent}	\textbf{Entanglement regimes under coherent excitation. } Stationary concurrence [left column, (a,b)] and diagram of entanglement mechanisms [left column, (a,b)]  versus cooperativity and qubit-qubit detuning for three different configurations: (a)  $\Delta_a \approx R$ (first row) and $\Delta_a\approx -R$ (second row) transitions, with $r_{12}=2.5\ \text{nm}$; (b) $\Delta_a=0$, with $r_{12}=50\ \text{nm}$. Solid contour lines correspond to exact calculations from the full model, white dashed lines are numerical predictions from the adiabatic model in \eqref{eq:Nakajima}, and yellow dot-dashed lines are numerical predictions from the approximated collective model in \eqref{eq:CollectivePurcell}. (c) Absolute value of the steady-state density matrix corresponding to points (i)-(v) in panels (a,b).
		Parameters: (a) $J=9.18\times 10^4\gamma$, $\gamma_{12}=0.999\gamma$, $\Omega=10^4 \gamma$; (b) $J=10.65 \gamma$, $\gamma_{12}=0.967\gamma$, $\Omega=10^2 \gamma$.
	}
}	

{
	\sidecaptionvpos{figure}{t}
\begin{SCfigure}[1][t!]
	\includegraphics[width=0.85\columnwidth]{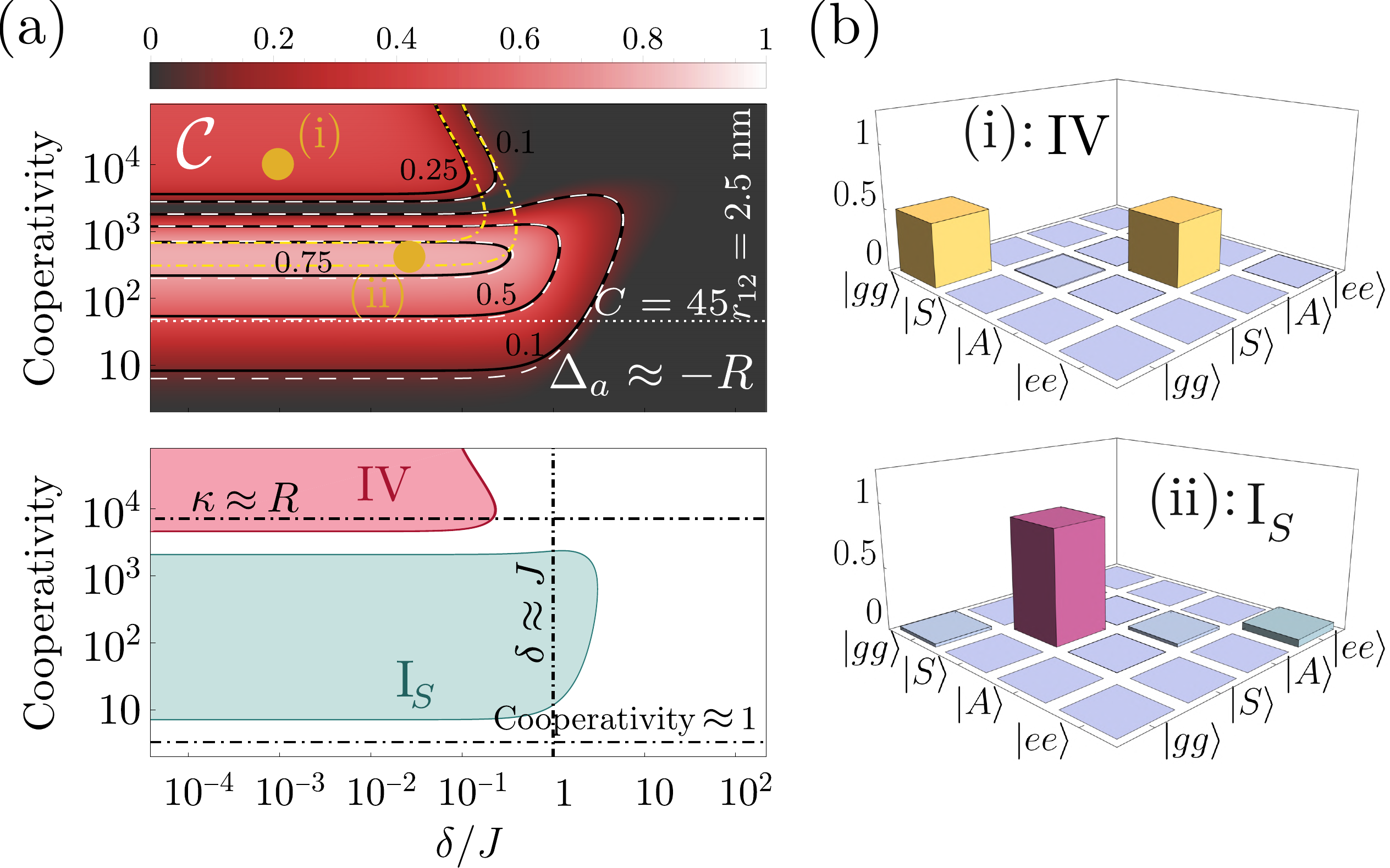}
	\captionsetup{justification=justified}
	\caption[Entanglement regimes under incoherent excitation.]{\label{fig:Fig9_EntanglementRegimes_Incoherent}	\textbf{Entanglement regimes under incoherent excitation. } (a) Stationary concurrence (top row) and diagram of entanglement mechanisms (low row) versus cooperativity and qubit-qubit detuning when $\Delta_a \approx -R$ and $r_{12}=2.5\ \text{nm}$.  (b) Absolute value of the steady-state density matrix corresponding to points (i,ii) in panel (a).
		Parameters: $J=9.18\times 10^4\gamma$, $\gamma_{12}=0.999\gamma$, $P=50\gamma$.
	}
\end{SCfigure}}

\begin{SCtable}[1.][b!]
	\centering
	\captionsetup{justification=justified}
	\caption[Overview of entanglement generation mechanisms]{\textbf{Overview of entanglement generation mechanisms.} 
		 Summary of the different mechanisms of entanglement generation, including the conditions that activate them and the specific entangled states they produce.
		 For mechanisms involving coherent excitation, the two-photon resonance condition ($\Delta=0$) is assumed throughout.
		 In Mech. II, the entangled state $|W_N\rangle$ is defined in \eqref{eq:W_state} while in Mechs. III$_{\text{sp}}$ and III$_{\text{cav}}$, the Bell state $|\psi_{\text{Bell}}\rangle$ is defined in \eqref{eq:Bell_state}.}
	\label{tab:EntanglementTab}
	\resizebox{1.0\textwidth}{!}{%
		\begin{tabular}{lllll}
			\toprule
			\toprule
			\multicolumn{2}{l}{\textbf{\textcolor{Maroon}{Mechanism}}}  & \multicolumn{2}{l}{\textbf{\textcolor{Maroon}{Conditions}}}  & \textbf{\textcolor{Maroon}{Entangled states}} \\
			\midrule
			\multicolumn{2}{l}{{\textbf{\textcolor{Maroon}{I. Frequency-Resolved}}}} & \multicolumn{2}{l}{ \textcolor{Maroon}{Coherent Excitation ($N=2$)}			}			      & $\hat  \rho_{\text{I}_A}\approx |A\rangle \langle A|$ \\
			\multicolumn{2}{l}{{\hspace{0.328cm}\textbf{\textcolor{Maroon}{Purcell enhancement}}}} & \multicolumn{2}{l}{(i) $\text{Cooperativity}>1$}						      & $ \hat  \rho_{\text{I}_S}\approx |S\rangle \langle S|$ \\
			\multicolumn{2}{l}{}                               				& \multicolumn{2}{l}{(ii) $R\gg \kappa, \delta,\Omega$ }							  &   \\
			\multicolumn{2}{l}{}                               				& \multicolumn{2}{l}{(iii) $\Omega_{\text{2p}}(\beta) \lesssim \kappa \lesssim J$ }  &  \\	  
			\multicolumn{2}{l}{}                                			& \multicolumn{2}{l}{
				(iv) $\Delta_a\approx R$, $\delta \neq 0$,$\gamma_{12}\neq \gamma$, }						  &  \\	     
				\multicolumn{2}{l}{}                                			& \multicolumn{2}{l}{
		 \hspace{0.55cm} and $\Gamma_{\mathrm I,A}(\beta)>\gamma_-$ (Mech. $\text{I}_A$)}						  &  \\	     
			\multicolumn{2}{l}{}                                			& \multicolumn{2}{l}{
				(iv) $\Delta_a\approx -R$, $\kappa\gg \Omega_{\text{2p}}$, }		          &  \\	    
				\multicolumn{2}{l}{}                                			& \multicolumn{2}{l}{
				 \hspace{0.55cm}  and $P_S> \gamma_{S}$ (Mech. $\text{I}_S$)}		          &  \\	    
			\multicolumn{2}{l}{} & \multicolumn{2}{l}{\textcolor{Maroon}{Incoherent Excitation ($N\geq 2$)}}						      &   $ \hat  \rho_{\text{I}_S}\approx |W_N\rangle \langle W_N|$ \\ 
			\multicolumn{2}{l}{}                                			& \multicolumn{2}{l}{(i) $\text{Cooperativity}>1$	   }       &  \\	    
			\multicolumn{2}{l}{}                               				& \multicolumn{2}{l}{(ii) $R\gg \kappa, \delta$ }							  &   \\
			\multicolumn{2}{l}{}                                			& \multicolumn{2}{l}{(iii) $\Delta_a\approx -R$, $\kappa\gg \Omega_{\text{2p}}$, }		          &  \\	    
			\multicolumn{2}{l}{}                                			& \multicolumn{2}{l}{
			 \hspace{0.55cm}	 and $P_S> \gamma_{S}$ (Mech. $\text{I}_S$)}		          &  \\	    
			\midrule
			\multicolumn{2}{l}{{\textbf{\textcolor{Maroon}{II. Collective}}}}         & \multicolumn{2}{l}{(i)  $\text{Cooperativity}>1$}   &  $\hat  \rho_{\text{II}}\approx  |A\rangle \langle A|$    \\
			\multicolumn{2}{l}{ \hspace{0.463cm}\textbf{\textcolor{Maroon}{Purcell enhancement}} }                                             & \multicolumn{2}{l}{(ii) $\kappa \gg R, \Omega $} &             						        \\
			\multicolumn{2}{l}{}                                             & \multicolumn{2}{l}{(iii) $\Omega \gg \delta$} &              							    \\
			\multicolumn{2}{l}{}                                             & \multicolumn{2}{l}{(iv) $\Gamma_{\text{eff}} \gg \Gamma_A$} &              				    \\ 	
			\midrule
			\multicolumn{2}{l}{{\textbf{\textcolor{Maroon}{III. Two-photon}}}}              & \multicolumn{2}{l}{\textcolor{Maroon}{Mechanism $\text{III}_{\text{sp}}$}} &     $\hat  \rho_{\text{III}_{\text{sp}}}\approx (1-\varepsilon) \hat  \rho_{\text{mix}}$              \\ 			
			\multicolumn{2}{l}{{ \hspace{0.493cm}\textbf{\textcolor{Maroon}{resonance fluorescence}}}}  												  & \multicolumn{2}{l}{(i)  $\delta \gtrsim J$}  					 &   \hspace{1.45cm}  $+\varepsilon |\psi_{\text{Bell}} \rangle \langle \psi_{\text{Bell}}|$   \\
			\multicolumn{2}{l}{}                                           			  & \multicolumn{2}{l}{(ii) $\kappa < J$}   						 &            \\
			\multicolumn{2}{l}{}                                           			  & \multicolumn{2}{l}{(iii) $\Omega_{\text{2p}} \approx \gamma$}   &              \\
			\multicolumn{2}{l}{}                                            		  & \multicolumn{2}{l}{\textcolor{Maroon}{Mechanism $\text{III}_{\text{cav}}$}} &      $ \hat \rho_{\text{III}_{\text{cav}}}\approx (1-\varepsilon) \hat \rho_{\text{mix}}$              \\
			\multicolumn{2}{l}{{}}    												  & \multicolumn{2}{l}{(i)   $\text{Cooperativity}>1$}   &     \hspace{1.45cm} $+\varepsilon |\psi_{\text{Bell}} \rangle \langle \psi_{\text{Bell}}|$  \\
			\multicolumn{2}{l}{}                                           			  & \multicolumn{2}{l}{(ii)  $\kappa\gtrsim J,\delta, \Omega$} &                \\
			\multicolumn{2}{l}{}                                            		  & \multicolumn{2}{l}{(iii) $\Gamma_P >\Omega $ when $\Omega \gtrsim J$} &                 \\
			\midrule
			\multicolumn{2}{l}{{\textbf{\textcolor{Maroon}{IV. Incoherently driven collective}}}}         & \multicolumn{2}{l}{(i)  $\text{Cooperativity}>1$}   &  $\hat  \rho_{\text{IV}}\approx  \frac{1}{2}( |gg\rangle \langle gg|+ |A\rangle \langle A|)$    \\
			\multicolumn{2}{l}{ \hspace{0.47cm}\textbf{\textcolor{Maroon}{Purcell enhancement}} }                                             & \multicolumn{2}{l}{(ii) $\kappa \gg R $} &             						        \\
			\multicolumn{2}{l}{}                                             & \multicolumn{2}{l}{(iii) $J \gg \delta$} &              							    \\
			\multicolumn{2}{l}{}                                             & \multicolumn{2}{l}{(iv) $\Gamma_P \gg P \gg \gamma$} &              				    \\ 	
			\midrule
			\multicolumn{2}{l}{{\textbf{\textcolor{Maroon}{V. Metastable two-photon}}}}        & \multicolumn{2}{l}{(i)  $\text{Cooperativity}>1$  }   							& $\hat \rho_{\text{V}}\approx (1-\varepsilon) |S\rangle\langle S| $    \\
			\multicolumn{2}{l}{ \hspace{0.45cm}\textbf{\textcolor{Maroon}{entanglement}}}                                         	    & \multicolumn{2}{l}{(ii)  $ \kappa \lesssim J, \Omega_{\text{2p}}(\beta)$ }    &     \hspace{1.45cm}     $+\varepsilon  |S_2/A_2\rangle\langle S_2/A_2|$    \\
			\multicolumn{2}{l}{}                                         	    & \multicolumn{2}{l}{(iii)  $ R\gg \Omega$ }    &           \\
			\multicolumn{2}{l}{}                                         	    & \multicolumn{2}{l}{(iv)  $ \Delta_a\approx \pm 2 \Omega_{2\mathrm{p}}$ }    &           \\
			\bottomrule
			\bottomrule
		\end{tabular}%
	}
\end{SCtable}

In the left panels of \reffig{fig:Fig9_EntanglementRegimes_Coherent}~\textcolor{Maroon}{(a,b)} and the top panel of \reffig{fig:Fig9_EntanglementRegimes_Incoherent}~\textcolor{Maroon}{(a)}, we plot the stationary concurrence $\mathcal{C}$ in terms of the qubit-qubit detuning $\delta$ and the cavity cooperativity $C$ (while fixing $g = 0.1\kappa$).
The panels in these figures correspond to different choices of the cavity-laser detuning $\Delta_a$ and the inter-emitter distance $r_{12}$. 
%
%
%
To provide a realistic benchmark, a horizontal white dotted line at a realistic value of $C = 45$~\cite{PschererSingleMoleculeVacuum2021} has been added as a reference, illustrating that our results are feasible with state-of-the-art solid-state platforms. 
The solid contour lines represent numerical results obtained using the full model from \eqref{eq:full_master_eq}, while the white dashed contour lines represent results from the effective model in \eqref{eq:Nakajima}, which confirm its validity across all parameter regimes explored in this work.
In the right panels of \reffig{fig:Fig9_EntanglementRegimes_Coherent}~\textcolor{Maroon}{(a,b)} and the bottom panel of \reffig{fig:Fig9_EntanglementRegimes_Incoherent}~\textcolor{Maroon}{(a)}, we identify regions of the parameter space corresponding to four different mechanisms of entanglement generation.

Additionally, \reffig{fig:Fig9_EntanglementRegimes_Coherent}~\textcolor{Maroon}{(c)} and \reffig{fig:Fig9_EntanglementRegimes_Incoherent}~\textcolor{Maroon}{(b)} illustrate the types of entangled states achieved in the long-time limit, depending on the underlying mechanism of entanglement generation. 
To clarify the analysis, we have included yellow markers in the concurrence plots that  unambiguously point the mechanism and the corresponding entangled state.
From these plots, we note that the different mechanisms stabilize the system into the antisymmetric state [points (i, v)], the symmetric state [point (iv)], or non-trivial superpositions of the ground and doubly-excited states [points (ii, iii)].

We discuss these emergent mechanisms, along with the metastable mechanism of entanglement generation [not shown in these figures], in greater detail in the following sections. 
To provide a comprehensive and concise overview of all the mechanisms identified within the parameter space, \textcolor{Maroon}{Table}~\ref{tab:EntanglementTab} summarizes the results presented in this Chapter, including their corresponding conditions and the resulting entangled steady states.

\section{Mechanism I: Frequency-Resolved Purcell enhancement}
\label{sec:FreqResolvedMech}
\subsection{Identification of the mechanism}

The first mechanism we present, referred to as Mechanism I, 
corresponds to the \textit{frequency-resolved Purcell effect}. It is characteristic of situations in which the emitters have a strong interaction in the absence of the cavity, e.g., via dipole-dipole interaction, forming a dimer. This occurs if $J \gg \delta$, so that $\beta \ll 1 $ and the one-excitation eigenstates are close to the purely symmetric and antisymmetric superpositions $|S\rangle$ and $|A\rangle$, i.e., the superradiant and subradiant states, respectively, 
\begin{equation}
	|\pm \rangle \approx |S/A\rangle = \frac{1}{\sqrt{2}}(|eg\rangle \pm |ge\rangle) \quad \text{when}\quad \beta\ll 1.
\end{equation}
This scenario can be observed, for instance, in closely spaced molecules~\cite{HettichNanometerResolution2002, TrebbiaTailoringSuperradiant2022, LangeSuperradiantSubradiant2024} or molecular dimers assembled via insulating bridges~\cite{DiehlEmergenceCoherence2014}, where distances of just a few nanometers lead to strong dipole-dipole coupling.
Other possible candidates are quantum dot molecules~\cite{KrennerDirectObservation2005,ReitzensteinCoherentPhotonic2006,ArdeltOpticalControl2016}. 
In order to illustrate the effect of strongly interacting emitters, we consider values of $J\sim 10^5 \gamma$ and $\gamma_{12}\approx \gamma$, corresponding to those obtained from the dipole-dipole interaction between two emitters in an H-aggregate configuration in free space, separated by $r_{12} = 2.5$ nm, with an emission wavelength of $\lambda_{0} = 780$ nm [see \refsec{sec:geometrical_conf}]. 
While such subnanometer interemitter distance might arguably require a more detailed description of the interaction between emitters beyond a point-dipole approximation~\cite{CzikklelyExtendedDipole1970}, these large values of $J$ are purposefully selected to unambiguously illustrate the effect. Nonetheless, we note that more modest values of $J\sim 10^3\gamma$, compatible with distances larger than $10$ nm in free space, can still generate entanglement via this mechanism, as shown in \refap{sec:SM_Validity}. 
We note that the first evidence of entanglement at the two-photon resonance that we presented in \reffig{fig:Fig3_ConcurrenceQubitLaserDetuning}, \reffig{fig:Fig3_ConcurrenceTimeQubitLaserDetuning}, and \reffig{fig:Fig9_EntanglementRegimes_Coherent}~\textcolor{Maroon}{(a)}  also belongs to this type of mechanism.

\begin{SCfigure}[1][b!]
	\includegraphics[width=0.75\columnwidth]{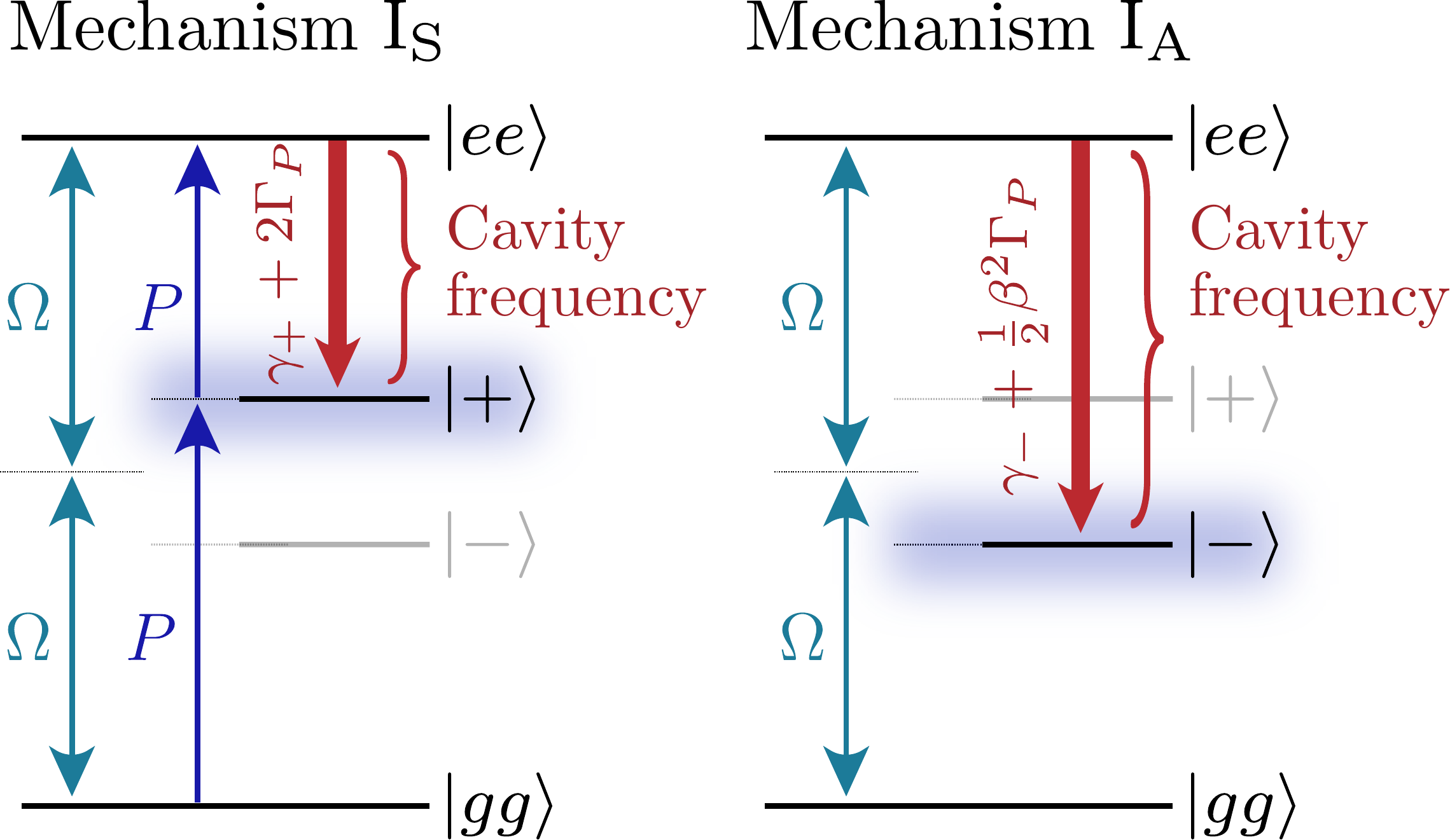}
	\captionsetup{justification=justified}
	\caption[Schematic representations of mechanism I$_S$ and I$_A$. ]{
		\label{fig:Fig6_DiagramPurcellMechanisms}
		\textbf{Schematic representations of mechanism I$_S$ and I$_A$. }Diagrams of Mechanism I for the stabilization of $|+\rangle$ and $|-\rangle$. 
		Mechanism I$_S$ is activated either by coherently exciting the emitters at the two-photon resonance or through incoherent excitation, whereas Mechanism I$_A$	is enabled exclusively via two-photon driving.
	}
\end{SCfigure}

\subsection{Origin of the mechanism}

Mechanism I relies on the cavity providing a Purcell enhancement of the decay from the doubly excited state $|ee\rangle$ to either the superradiant state $|S\rangle$ or the subradiant state $|A\rangle$, both of which are maximally entangled states. This enhancement is activated when the cavity frequency is close to one of the two main transition energies in the dimer, $\Delta_a \approx \pm R$. Specifically, when $\Delta_a = R$, the cavity enhances resonantly the transition $|ee\rangle \rightarrow |A\rangle$, and when $\Delta_a = -R$, it enhances the transition $|ee\rangle \rightarrow |S\rangle$.
Crucially, by combining the direct two-photon excitation of the doubly-excited state---in the case of coherent excitation---, or a sufficiently large pumping rate $P\gg \gamma$---in the incoherent excitation case---, and  the subsequent cavity-enhanced decay towards either $|S\rangle$ or $|A\rangle$, a stationary occupation probability of these states close to one can be reached. We label each of these two possibilities as Mechanism I$_S$ and I$_A$, depending on whether the cavity enhances the decay towards $|S\rangle$ or $|A\rangle$, respectively:
\begin{subequations}
	\begin{align}
		&\text{Mechanism I$_A$ ($\Delta_a\approx R,\ |ee\rangle \rightarrow |A\rangle$):} \quad \ \ \  \hat \rho_{\mathrm{ss}}\approx \rho_{AA}, \\
		&\text{Mechanism I$_S$  \ ($\Delta_a\approx- R,\ |ee\rangle \rightarrow |S\rangle$):} \quad   \hat \rho_{\mathrm{ss}}\approx \rho_{SS}.
	\end{align}
\end{subequations}
These two mechanisms are schematically depicted in  \reffig{fig:Fig6_DiagramPurcellMechanisms} for both coherent and incoherent excitation.

By using the Bloch-Redfield master equation from \eqref{eq:Nakajima}, we have demonstrated---see \refap{appendix:MechanismI}---that, when resonances activating either Mechanism I$_A$ or Mechanism I$_S$ are selected, the effective contribution of the cavity to the dynamics of the emitters can described by a single Lindblad term 
\begin{equation}
	\colorboxed{Maroon}{
		\mathcal L_\mathrm{eff}[\hat\rho] = \Gamma_P\mathcal D[\hat\xi_{A/S}]\hat\rho,
	}
	\label{eq:SingleLindbladEff}
\end{equation}
with jump operators
\begin{subequations}
	\begin{align}
		\hat\xi_A &\equiv |gg\rangle\langle + | - \frac{\beta}{2}|-\rangle \langle ee|, \quad \  \ \  \text{(Mechanism I$_A$)}, 
		\label{eq:xi_A_main}\\
		\hat\xi_S &\equiv -|+\rangle \langle ee| + \frac{\beta}{2}|gg\rangle\langle -|, \quad \text{(Mechanism I$_A$)} .
		\label{eq:xi_S_main} 
	\end{align}
\end{subequations}
This expression for the effective dynamics [\eqref{eq:SingleLindbladEff}], derived in \refap{appendix:MechanismI}, provides analytical insights into the process of entanglement stabilization. 

\begin{SCfigure}[1][b!]
	\includegraphics[width=0.8\columnwidth]{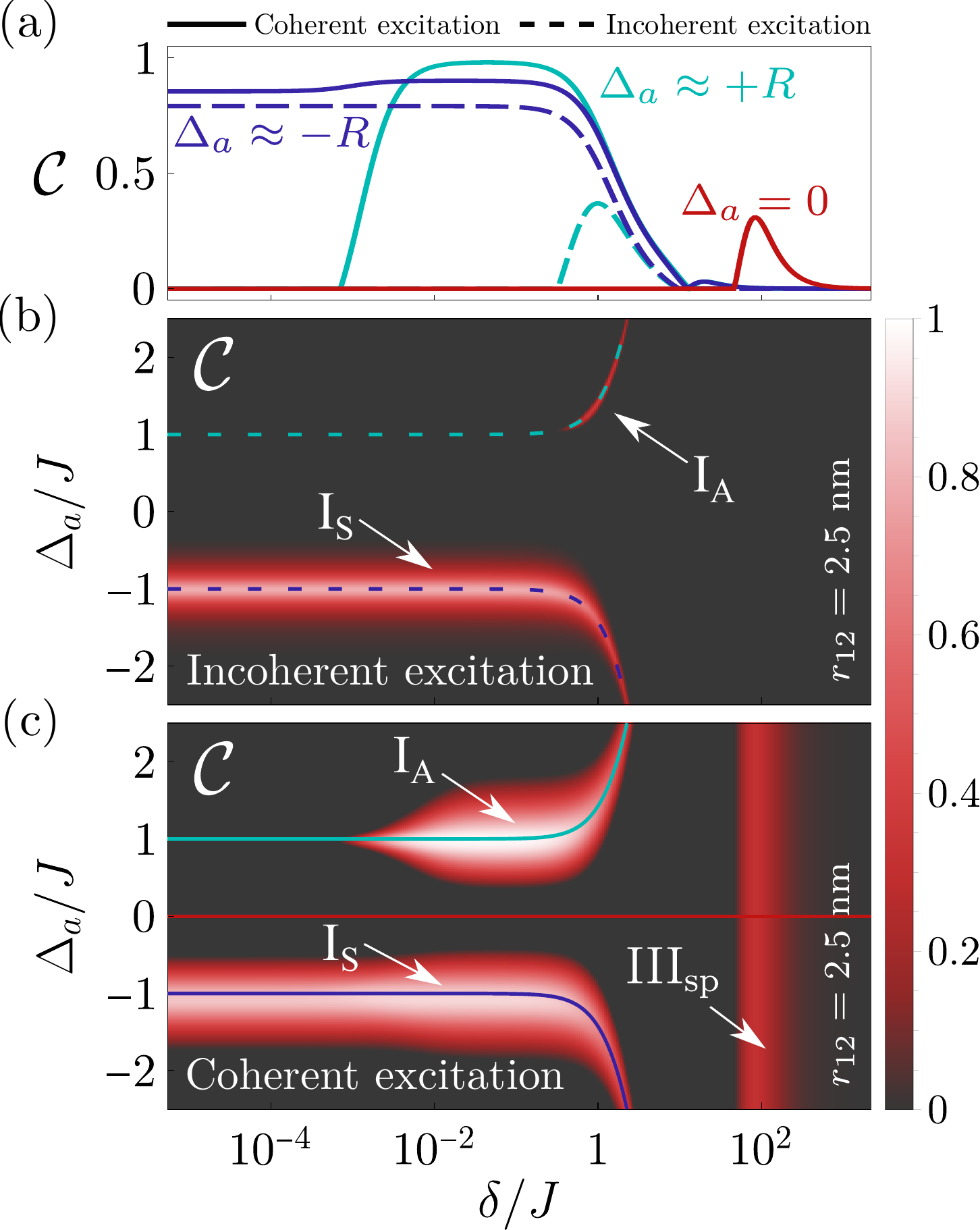}
	\captionsetup{justification=justified}
	\caption[Dependence on the cavity-laser detuning $\Delta_a$ for entanglement generation]{
		\label{fig:Fig4_ConcurrenceQubitCavityDetuning}
		\textbf{Dependence on the cavity-laser detuning $\Delta_a$ for entanglement generation.}
		(a)
		Steady-state concurrence versus laser-cavity detuning $\Delta_a$ and qubit-qubit detuning $\delta$ when the two-qubit system is driven at the two-photon resonance $\Delta=0$. Top panel corresponds to a cut of the concurrence when the cavity is in resonance with some frequencies: antisymmetric transition (blue) and the two-photon resonance (red). Solid lines correspond to numerical computations, blue dashed ($\Delta_a\approx R$) and red dot-dashed ($\Delta_a=0$) lines are numerical predictions from the adiabatic model in Eq.~\eqref{eq:Nakajima}.
		Parameters (b): $r_{12}=2.5\ \text{nm}$, $k=2\pi/780\ \text{nm}^{-1}$, $J=9.18\times 10^4\gamma$, $\gamma_{12}=0.999\gamma$, $\Delta=0$, $\Omega=10^4 \gamma$, $\kappa=10^4 \gamma$, $g=10^{-1}\kappa$. 
	}
\end{SCfigure}

\subsection{Mechanism I$_A$ } 
The combination of the effective jump operator $\hat \xi_A$ and the coherent drive at the two-photon resonance eventually stabilizes the subradiant state $|-\rangle \approx |A\rangle$. This can be intuitively understood in the following way: firstly, the two-photon excitation populates the state $|ee\rangle$; then, the term $\frac{\beta}{2} |-\rangle \langle ee|$ in the jump operator $\hat \xi_A$ [\eqref{eq:xi_A_main}] takes the system from $|ee\rangle$ into $|-\rangle$, as it is schematically illustrated in \reffig{fig:Fig6_DiagramPurcellMechanisms}. 

From this effective description in terms of \eqref{eq:SingleLindbladEff}, we find that the steady state population of the entangled subradiant state $|-\rangle \approx |A\rangle$ via Mechanism I$_A$ is given by---see \refap{appendix:DeltapR_Coh}---
\begin{equation}
		\colorboxed{Maroon}{
			\rho^{\mathrm{Coh}}_{A,\mathrm{ss}} = \frac{\Gamma_{\mathrm I,A}}{\Gamma_{\mathrm I,A}+\gamma_-}, \quad \text{(Coherent excitation)}
		}
\end{equation}
where we identify $\Gamma_{\mathrm{I},A}\equiv \beta^2\Gamma_P/2$
as the effective decay rate from $|ee\rangle$ to $|-\rangle$ provided by the cavity, while $\gamma_- = \gamma - \gamma_{12}\cos\beta$ is the subradiant decay rate$^{\textcolor{Maroon}{*}}$.
\graffito{
Note that the definition of the super- and subradiant decay rates, $\gamma_\pm$, were already introduced in \refch{ch:TwoPhotonResonance}, in \eqref{eq:ExcitonicParameters1}.
}
 This mechanism becomes efficient when $\Gamma_{\mathrm{I},A}\gg \gamma_-$, which for $\beta \ll 1$ sets the condition
\begin{equation}
	\beta^2 > \frac{2}{C}\left(1-\frac{\gamma_{12}}{\gamma}\right).
	\label{eq:beta_min_IA}
\end{equation}
Notice that the presence of the factor $\beta$ in the definition of $\Gamma_{\mathrm{I},A}$ indicates that a certain non-zero detuning $\delta$ is required to ensure that $\Gamma_{\mathrm{I},A}\neq 0$, enabling the stabilization of the state $|-\rangle$ in the long-time limit. 
The finite detuning prevents the state $|-\rangle$ from being entirely subradiant and decoupled from the cavity. 
This minimum required value of $\beta$ is illustrated in \reffig{fig:Fig4_ConcurrenceQubitCavityDetuning}~\textcolor{Maroon}{(a, c)}, where a qubit-qubit detuning value of $\delta\gtrsim 10^{-2}J$ is needed to enable the generation of entanglement.
From this effective description in terms of \eqref{eq:SingleLindbladEff}, we also obtain that this state is stabilized in a timescale which, in the limit $\Gamma_{\mathrm I,A}\gg \gamma_-$ is given by---see \refap{appendix:DeltapR_Coh}--- 
\begin{equation}
	\colorboxed{Maroon}{
	\tau^{\text{Coh}}_{A}\approx \frac{2}{\Gamma_{\mathrm I,A}} = \frac{4}{\beta^2 \Gamma_P}, \quad \text{(Coherent excitation)}.
	}
	\label{eq:tau_IA}
\end{equation}

We note that mechanism I$_A$ is highly inefficient when the emitters are incoherently excited.
This inefficiency arises from the depletion of the antisymmetric state due to incoherent repumping: 
any population in the antisymmetric state is repumped by the incoherent excitation to the excited state and subsequently redirected to the symmetric dissipation channel, preventing the stabilization of the antisymmetric state in the long-time limit.
This behaviour is shown in \reffig{fig:Fig4_ConcurrenceQubitCavityDetuning}~\textcolor{Maroon}{(a, b)}, where entanglement is suppressed within the antisymmetric region, $\Delta_a\approx R$, and only a small amount of concurrence, $\mathcal{C}\lesssim 0.3$, appears when $\delta \approx J$, and completely vanishes when $\delta \gg J$. Therefore, in practice, mechanism I$_A$ is only enabled via coherent excitation.

\subsection{Mechanism I$_S$ }
For the symmetric resonance, $\Delta_a\approx -R$, we follow the same procedure as in the antisymmetric case, with the addition of considering both the  coherent or incoherent excitation schemes. 
In this case, the combination of the effective jump operator $\hat \xi_S$ [\eqref{eq:xi_S_main}] and the excitation of the highest excited state $|ee\rangle$ via two-photon excitation or subsequent incoherent pumping, stabilizes the system into the superradiant state $|+\rangle \approx |S\rangle$. Intuitively, the stabilization  process is analogous to the subradiant case, with the primary difference being the decay from the doubly excited state, which is enhanced in the supperradiant case and given by $\Gamma_{\mathrm{I},S}=2\Gamma_P$.

\paragraph{Coherent excitation. }When the emitters are coherently driven at the two-photon resonance, we find that the steady-state population of the entangled superradiant state $|+\rangle \approx |S\rangle$ formed via Mechanism I$_S$ is given by---see \refap{appendix:DeltamR_Coh}---
\begin{equation}
\colorboxed{Maroon}{
	\rho^{\mathrm{Coh}}_{S,\mathrm{ss}} = \frac{1}{1 + \gamma_+\left(P_S^{-1}+\Gamma_P^{-1}\right)}, \quad \text{(Coherent excitation)}
}
\end{equation}
where we defined the effective pumping rate $P_S \equiv 2\Omega_\mathrm{2p}^2/\Gamma_P$ and the superradiant decay rate $\gamma_+\equiv \gamma +\gamma_{12}\cos\beta$. Intuitively, we  may understand the stabilization process as an effective incoherent pumping of the state $|S\rangle$ from $|gg\rangle$ with a rate $P_S$. 
In this case, the mechanism will be efficient when $P_S> \gamma_+$, which sets the condition
\begin{equation}
	\left(\frac{\Omega_\mathrm{2p}}{\gamma}\right)^2 > C.
\end{equation}
Here, the timescale of stabilization in the efficient regime $P_S\gg \gamma_+$ is directly given by---see \refap{appendix:DeltamR_Coh}---
\begin{equation}
\colorboxed{Maroon}{
\tau^{\text{Coh}}_{S} =  \frac{2/\Gamma_P}{1 - \text{Re}\sqrt{1-(2\Omega_\mathrm{2p}/\Gamma_P)^2}}, \quad \text{(Coherent excitation)}
}	
	\label{eq:tau_IS}
\end{equation}
where $\text{Re}(*)$ denotes real part.
Notice the two significant limits of this equation: \textcolor{Maroon}{(i)} when $\Gamma_P < 2\Omega_\mathrm{2p}$, we have $\tau^{\text{Coh}}_{S} \approx  2/\Gamma_P$, \textcolor{Maroon}{(ii)} while when $\Gamma_P \gg \Omega_\mathrm{2p}$, one obtains $\tau^{\text{Coh}}_{S}\approx \Gamma_P/\Omega_\mathrm{2p}^2$ after Taylor-expanding the square-root. 

\paragraph{Incoherent excitation. }On the other hand, when the emitters are incoherently driven, we find that the steady-state population of the entangled superradiant state $|+\rangle \approx |S\rangle$ can be directly derived from the diagonalization of \eqref{eq:Nakajima}---see \refap{sec:SM_AnalyticalDensityMatrix}---
\begin{equation}
\colorboxed{Maroon}{
	\rho^{\text{Incoh}}_{S,\text{ss}}\approx \left[1+ \frac{P}{2 \Gamma_{P}}+\frac{\gamma_+}{P}+\frac{1}{8}\frac{\Gamma_P}{P}\left(\frac{\kappa}{J}\right)^2\right]^{-1},
	\quad \text{(Incoh. excitation)}
}
	\label{eq:rhoS_purcell}
\end{equation}
where we have assumed that $C\gg 1$. From this expression, we note that the mechanism is efficient when  $P\gg \gamma$ and $\Gamma_P\gg P$, which imply $\kappa \gg P$. These two conditions are easily understood in the following way:  the pumping rate must be faster than the natural emitter lifetime but slower than the cavity-induced decay rate, since, if not, the system would be stabilized into the  doubly-excited state $|ee\rangle$. Here, the stabilization timescale $\tau^{\text{Incoh}}_S$ is expressed as---see  \refap{sec:SM_AnalyticalDensityMatrix}---
\begin{equation}
\colorboxed{Maroon}{
	\tau_S^{\text{Incoh}}  \approx \frac{1}{P+\Gamma_{P}-\sqrt{\Gamma_{P}^2+P\gamma_+}},
	\quad \text{(Incoherent excitation)}.
}
\label{eq:tau_IS_incoh}
\end{equation}

\subsection{Summary of conditions}
\label{sec:Conditions}

There are three primary conditions for Mechanism I to occur:
\begin{enumerate}[label=\textcolor{Maroon}{(\roman*)}]
	\item  The cavity must be able to resolve the main transitions in the dimer taking place at $\omega_0 \pm R$ in order to enhance them selectively. This requires $\kappa \ll R$.
	\item To ensure that the cavity significantly enhances transition rates with respect to spontaneous emission, the cooperativity parameter should be greater than one, i.e., $C>1$.
	\item  The dimer must be strongly coupled, i.e., $J\gg\delta$. 
\end{enumerate}
This list suggests a hierarchy of rates that must be respected for it to occur:
\begin{equation}
	\colorboxed{Maroon}{
	J\gg \kappa \gg \Gamma_P \gg \gamma,
	}
\end{equation}
which ensures a frequency-dependent Purcell effect ($	J\gg \kappa \gg \Gamma_P$) with high cooperativity ($\Gamma_P\gg \gamma$).
Furthermore, there are specific requirements for Mechanism I$_A$ and Mechanism I$_S$ in addition to the previously mentioned conditions depending on the driving mechanism that ensure the target entangled state is efficiently populated:
\begin{enumerate}[label=\textcolor{Maroon}{(\roman*)}]
	\item \textcolor{Maroon}{Mechanism I$_A$ (Coherent excitation): }  A finite detuning $\delta$ is required to break perfect subradiance and fulfill the condition $\Gamma_{\mathrm I, A}>\gamma_-$, which sets the minimum for $\beta$,
	\begin{equation}
		\colorboxed{Maroon}{
		\beta^2 > \frac{2}{C}\left(1-\frac{\gamma_{12}}{\gamma}\right).
		}
	\end{equation}
	For instance, this condition can be clearly observed in the upper panel of  \reffig{fig:Fig9_EntanglementRegimes_Coherent}~\textcolor{Maroon}{(a)}, where a minimum  value of $\delta$ is required  to prevent the state $|-\rangle$ from being entirely subradiant and decoupled from the cavity.
	\item \textcolor{Maroon}{Mechanism I$_S$ (Coherent excitation): }
	The cavity should be not able to resolve the fine dressed-state structure of the system, i.e., $\kappa \gg \Omega_{\mathrm{2p}}$, to ensure that all possible transitions towards $|+\rangle$ are enhanced by the cavity and the transitions towards $|-\rangle$ are subsequently quenched, since otherwise $|-\rangle$ would acquire a non-negligible population due to its long lifetime:
	\begin{equation}
		\colorboxed{Maroon}{
		P_S\gg \gamma,\quad \text{and}\quad \kappa\gg \Omega_{2\mathrm{p}}.
		}
	\end{equation}
	This is in stark contrast to Mechanism I$_A$, which benefits from the subradiant nature of $|-\rangle$. Even if the cavity is able to resolve the energy levels dressed by the laser$^{\textcolor{Maroon}{*}}$,
	\graffito{
In the perturbative regime $\Omega\ll R$, the separation between sidebands at $\omega_0\pm R$ is given by $\Omega_\text{2p}$, since, from the expression of the eigenvalues presented in \refch{ch:TwoPhotonResonance} [\cref{eq:transition_omega_1,eq:transition_omega_2,eq:transition_omega_3,eq:transition_omega_4,eq:transition_omega_5,eq:transition_omega_6}], we find:
$$
		\begin{aligned}
			\omega_{12} &= -\omega_{21} = R + 2\Omega_\text{2p},\\
			\omega_{13}&= -\omega_{31} = R + 4\Omega_\text{2p},\\
			\omega_{24} &= -\omega_{42} = R, \\
			\omega_{34} &= -\omega_{43} = R - 2\Omega_\text{2p}.
		\end{aligned}
$$
Hence, in order to resolve the two-photon sidebands, the cavity linewidth must be $\kappa\ll \Omega_{2\text{p}}$.
	}
	 split by $\Omega_\text{2p}$, the entangled state $|-\rangle \approx |A\rangle$ will be stabilized as long as one of the transitions $\omega_{24}: |A_2\rangle \rightarrow |A\rangle$ or $\omega_{34}: |S_2\rangle \rightarrow |A\rangle$ are selectively enhanced. 
	 In contrast, stabilizing $|S\rangle$ requires that the cavity enhances both the transitions $\omega_{12}: |A_2\rangle\rightarrow |S\rangle$ and $\omega_{31}: |S_2\rangle \rightarrow|S\rangle$.
	This conclusion is supported by the lack of well-resolved peaks in the entanglement corresponding to these two transitions in \reffig{fig:FigAppendix_ConcurrenceKappaLessOmega2p}, observing instead broader peak that encompasses both of them and that is much weaker than in the antisymmetric case. This indicates that both transitions must be enhanced to observe any effect, which necessarily requires $\kappa\gg \Omega_{2\mathrm{p}}$.
	\item \textcolor{Maroon}{Mechanism I$_S$ (Incoherent excitation):} A pumping rate faster than the natural emitter lifetime, but slower than the cavity-induced decay rate, since, if not, the system would be stabilized into $|ee\rangle$,
	\begin{equation}
		\colorboxed{Maroon}{
				\Gamma_P \gg P \gg \gamma.
		}
	\end{equation}
\end{enumerate}

Note that the conditions that activate mechanism I$_A$ and I$_S$ are perfectly illustrated in the green regions of the phase diagrams in  \reffig{fig:Fig9_EntanglementRegimes_Coherent}~\textcolor{Maroon}{(a)} and \reffig{fig:Fig9_EntanglementRegimes_Incoherent}~\textcolor{Maroon}{(a)}.

{
	\sidecaptionvpos{figure}{t}
\begin{SCfigure}[1][t!]
	\includegraphics[width=0.65\textwidth]{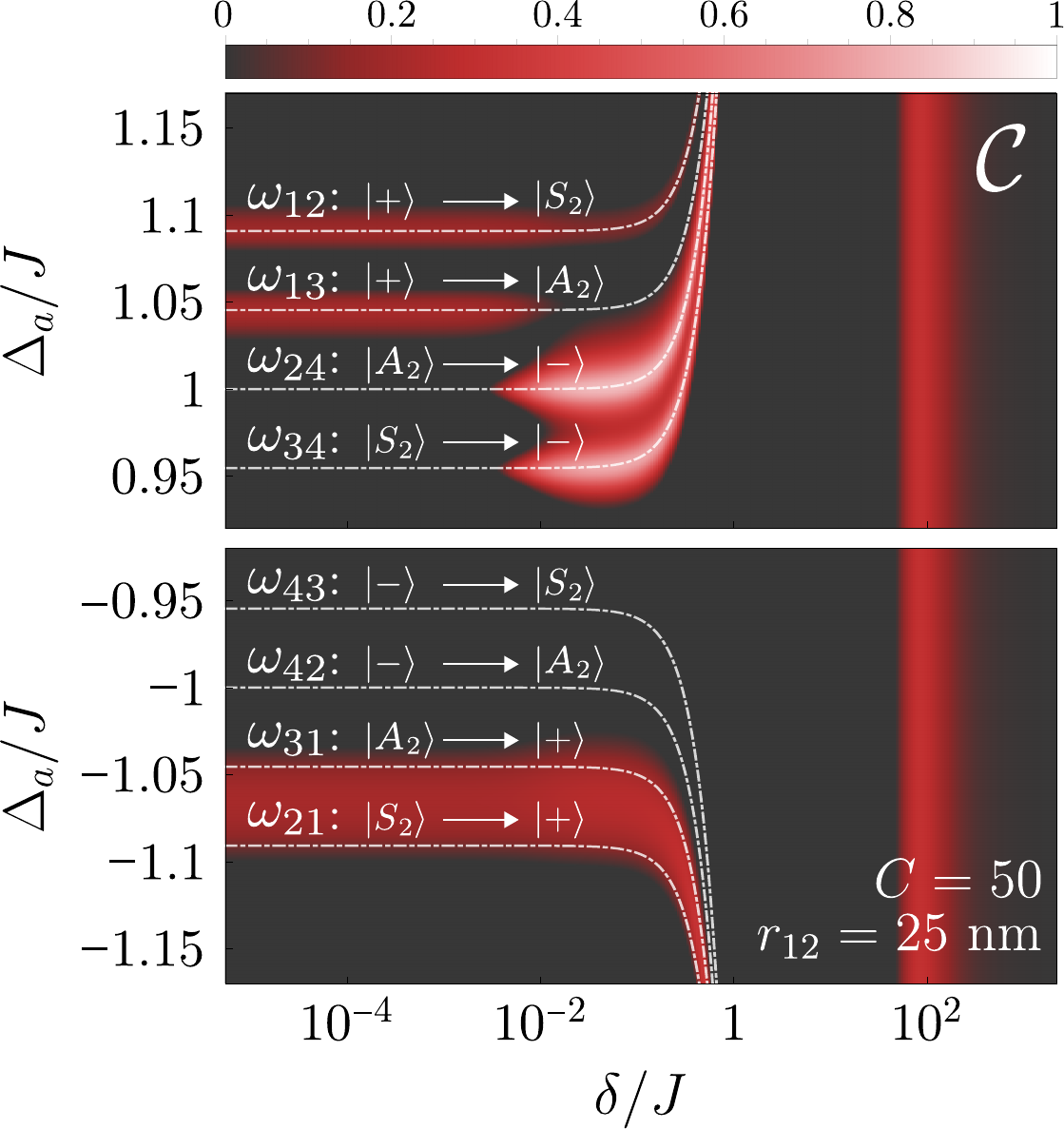}
	\captionsetup{justification=justified}
	\caption[Formation of entanglement in the resolved two-photon sideband regime $\kappa < \Omega_\text{2p}$.]{\textbf{Formation of entanglement in the resolved two-photon sideband regime $\kappa < \Omega_\text{2p}$}. Compared to the results shown in \reffig{fig:Fig4_ConcurrenceQubitCavityDetuning}, here the cavity is able to resolve the laser-dressed state structure of the driven dimer. 
	The asymmetric behaviour in this figure is interpreted as a result of the mechanism being activated when $J\gg\delta$, which is a limit where the target states are almost perfectly subradiant $|-\rangle \approx |A\rangle$ and super-radiant $|+\rangle \approx |S\rangle$ states, therefore having significantly different lifetimes. Due to the extremely long lifetime of the antisymmetric state $|A\rangle$, even a small enhancement of the dissipative transitions towards this state (which the cavity can provide when $C\gg 1$) contributes to a significant increase in the stationary population of this state.
		Parameters: $r_{12}=2.5\ \text{nm}$, $k=2\pi/780\ \text{nm}^{-1}$, $J=9.18\times 10^4 \gamma$, $\gamma_{12}=0.999\gamma$, $\Delta=0$, $\Omega=10^4 \gamma$, $\kappa=1.25\times10^3 \gamma$, $g=10^{-1}\kappa$.}
	\label{fig:FigAppendix_ConcurrenceKappaLessOmega2p}
\end{SCfigure}}

\subsection{Relaxation timescale}

The relaxation time into the entangled state is another important figure of merit, as it allows one to assess the resilience of the process against extra dephasing mechanisms by comparing the stabilization timescale with the decoherence timescale. 
In this regard, it is clear from the different expressions for the relaxation timescales $\tau^{\text{Coh}}_{A/S}$ in \eqref{eq:tau_IA} and \eqref{eq:tau_IS} for coherent excitation, respectively, and $\tau^{\text{Incoh}}_{S}$ in  \eqref{eq:tau_IS_incoh} for incoherent excitation, that 
the choice between the symmetric or antisymmetric transition has a strong impact on the timescale of stabilization, a difference that stems from the superradiant and subradiant nature of the corresponding target states. This is explicitly shown in \reffig{fig:Fig5_RelaxationRates}~\textcolor{Maroon}{(a)}, where the time evolution of the symmetric and antisymmetric state population is depicted for three possible scenarios: $\rho_{SS}$ when $\Delta_a\approx -R$ for both coherent and incoherent excitation, and $\rho_{AA}$ when $\Delta_a\approx R$ for coherent excitation. 

From \reffig{fig:Fig5_RelaxationRates}~\textcolor{Maroon}{(a)}, we confirm the significant difference in the relaxation timescale between stabilizing the antisymmetric state via coherent excitation and the symmetric state, which exhibits a similar relaxation timescales either considering coherent or incoherent excitation. 
For the shake of concreteness, let us compare the coherent excitation only, as the difference in the relaxation timescale of the superradiant state either using coherent or incoherent excitation is not dramatic. In particular, the ratio between both quantities will be given by 
{
	\sidecaptionvpos{figure}{t}
	\begin{SCfigure}[1][t!]
	\includegraphics[width=0.8\columnwidth]{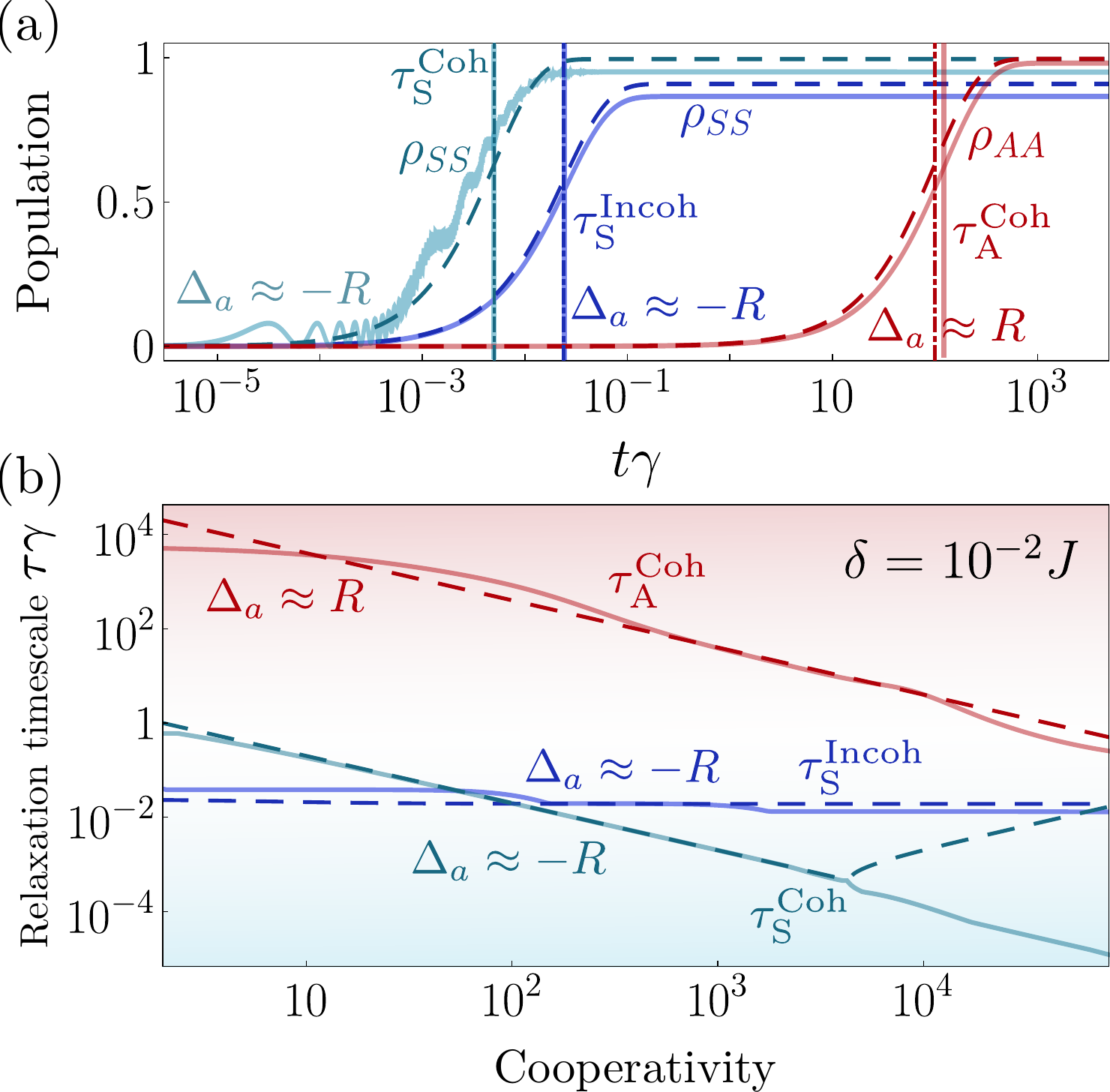}
	\captionsetup{justification=justified}
	\caption[Validity of the analytical expressions for the time-dependent density matrix elements and relaxation timescales.]{\label{fig:Fig5_RelaxationRates}
		\textbf{Validity of the analytical expressions for the time-dependent density matrix elements and relaxation timescales. }
		(a) Time evolution of the density matrix elements when the cavity is in resonance with the antisymmetric transition $\Delta_a=\omega_{34}\approx R$ in blue ($\rho_{AA}\equiv \langle A | \hat  \rho |A\rangle$, antisymmetric state) and the symmetric transition $\Delta_a=\omega_{21} \approx -R$ in red ($\rho_{SS}\equiv \langle S | \hat  \rho |S\rangle$, symmetric state).  
		(b) Relaxation time towards the entangled steady-state versus the cooperativity when the cavity is in resonance with the antisymmetric transition $\Delta_a=\omega_{34}\approx R$ (blue) and the symmetric transition $\Delta_a=\omega_{21} \approx -R$ (red). In (a,b), blue dashed and red dot-dashed curves correspond to analytical predictions.
		Parameters (a-b): $r_{12}=2.5\ \text{nm}$, $k=2\pi/780\ \text{nm}^{-1}$, $J=9.18\times 10^4\gamma$, $\gamma_{12}=0.999\gamma$, $\delta=10^{-2}J$, $\Delta=0$, $\Omega=10^4 \gamma$, $g=10^{-1}\kappa$. 
	}
\end{SCfigure}}
\begin{equation}
	r_\tau \equiv \frac{\tau^{\text{Coh}}_{\mathrm S}}{\tau^{\text{Coh}}_{\mathrm A}}= 
	\frac{\beta^2/2}{1 - \text{Re} \sqrt{1-(2\Omega_\mathrm{2p}/\Gamma_P)^2}},
\end{equation}
which, for the particular set of parameters chosen in \reffig{fig:Fig5_RelaxationRates}~\textcolor{Maroon}{(a)}, yields $r_{\tau} = \beta^2 \sim 10^{-4}$. This implies that 
the relaxation towards the symmetric state is four orders of magnitude faster than the relaxation towards the antisymmetric state. This observation is further supported by \reffig{fig:Fig5_RelaxationRates}~\textcolor{Maroon}{(b)}, where we evaluate the relaxation timescales $\tau^{\text{Coh}}_\mathrm{A} $ and $\tau^{\text{Coh/Incoh}}_\mathrm{S}$  with respect to the cooperativity $C$, confirming a ratio $\propto 10^{-4}$ between them. From this figure, we confirm that the mechanism of stabilization is most efficient when $\Gamma_P$ is the dominant dissipative rate of the system. 
Additionally, we note that our analytical calculations (depicted as dashed lines) show a good agreement with the corresponding timescales computed numerically through direct diagonalization of the Liouvillian. We only observe a discrepancy for $\tau^{\text{Coh}}_{S}$ at large values of $C>10^3$, where, nevertheless, this mechanism is no longer activated, since $\kappa > R$.

\subsection{Tunability and optical signatures of entanglement}
\label{sec:observation}

In this section, we discuss a possible measurement scheme of the optical properties of the light emitted by the system that signal the formation of entanglement. Furthermore, we discuss how changing the strength of the excitation can be used to optically tune the system in and out of the resonant entanglement condition necessary for the activation of Mechanism I.

\paragraph{Impact of the coherent fraction of light. }
A first element to discuss before describing experimental measurements is the impact of the coherent fraction of light generated by the 
strong driving field, which may hinder the measurements of interest. We note that our model describes a configuration in which the two qubits are coherently and incoherently driven, but the cavity is not.
This situation may be challenging to realize in an experimental configuration in which the emitters are embedded into the cavity, which often forces a simultaneous excitation of both systems. However, we note that the simultaneous driving of both the cavity and the emitters can also be described by our model, after applying a displacement transformation into the cavity that removes its coherent fraction---see \refap{Appendix:CoherentDisp}---. 

In other words, our model will always describe the quantum fluctuations of the cavity field, on top of any coherent fraction that may be developed because of the driving, which we disregarded here. In the language of input-output theory~\cite{GardinerQuantumNoise2004}, the output operator will be given by
\begin{equation}
	\hat a_{\text{out}}(t) =\sqrt{\kappa}  \hat a(t) + \alpha e^{-i\omega_\mathrm L t},
\end{equation}
where $\alpha$ corresponds to the classical, coherent fraction of the output field coming from the input laser plus the coherent contribution of the cavity field. 
Since the protocols we propose require rather strong driving fields, we will normally find that $|\alpha|\gg \sqrt{\langle\hat a^\dagger \hat a \rangle}$, which means that this bright coherent signal will hinder the measurement of quantum fluctuations unless it is rejected.
This is a common problem in the study of resonance fluorescence in solid-state quantum optics, where the challenge has been successfully addressed via, e.g., background-suppression through orthogonal excitation-detection geometries~\cite{FlaggResonantlyDriven2009},
cross-polarization schemes~\cite{SomaschiNearoptimalSinglephoton2016} or self-homodyning of the signal, as illustrated in \reffig{fig:Fig_SelfHomodyning}, where the coherent component is eliminated through the interference with the input field~\cite{FischerSelfhomodyneMeasurement2016,FischerOnChipArchitecture2017,HanschkeOriginAntibunching2020}.
In the following, we assume that one of these approaches has been performed, and describe the direct measurement of quantum fluctuations given by an output operator without a coherent fraction, $\hat a_\mathrm{out} = \sqrt\kappa \hat a$.
	\addtocounter{figure}{-1}
\begin{SCfigure}[1][h!]
	\includegraphics[width=1.\columnwidth]{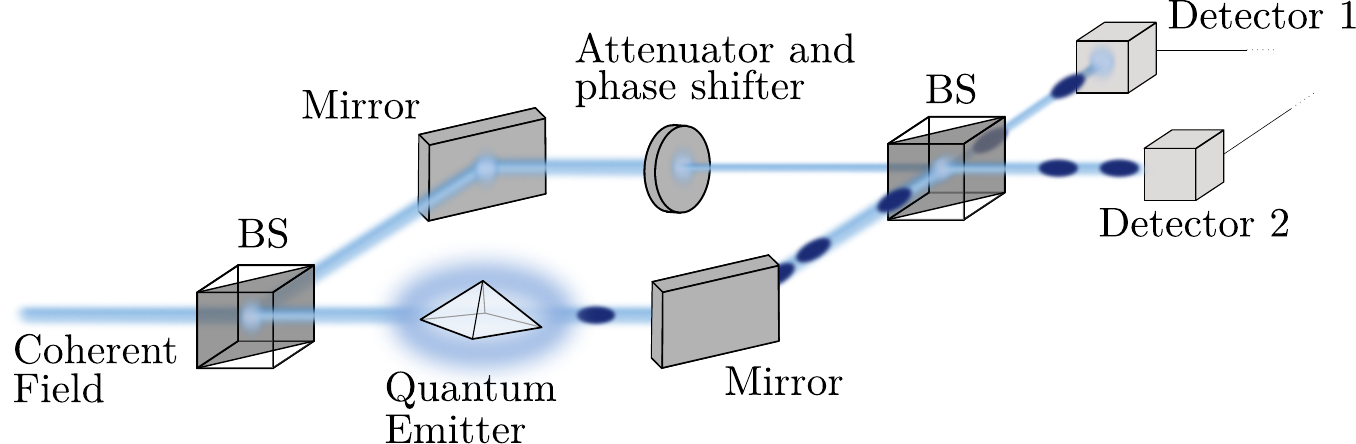}
	\captionsetup{justification=justified}
\end{SCfigure}
\graffito{\vspace{-5.0cm}
	\captionof{figure}[Schematic representation of the self-homodyning process.]{	\label{fig:Fig_SelfHomodyning}
		\textbf{Schematic representation of the self-homodyning process.} The coherent field used to drive the quantum emitter is recombined with the output field at the second beam splitter (BS), producing destructive interference that removes any coherent contribution, leaving only the quantum fluctuations of the light. The dark blue spots represents single photons emitted by the quantum emitter.
	}
}

\paragraph{Signalling the generation entanglement through light emission properties. }
In this section, we employ the first- and second-order correlation functions to signal the formation of entanglement between emitters. Particularly, we establish a correspondence between regions of high stationary entanglement and certain properties of the light, such as subradiant emission or photon bunching, which are characteristic of collective dynamics.

In order to witness the generation of entanglement between quantum emitters, we study the optical properties of the emitted light by the cavity in terms of optically tunable parameters that are, in principle, accessible in a real experiment.
In \reffig{fig:Fig10_Observability}, we show plots of the concurrence $\mathcal{C}$ and the corresponding properties of the radiated field as a function of the cavity detuning $\Delta_a$ and the pumping rate $P$ (left-column) or driving amplitude $\Omega$ (right column). The properties of the emission that we show are the emission intensity, given by 
$
I\propto \langle \hat a^\dagger \hat a \rangle,
$
and photon statistics, through the zero-delay second-order correlation function 
$
	g^{(2)}(0) \equiv \langle {\hat a^\dagger}{\hat a^\dagger} \hat a \hat a\rangle/\langle \hat a^\dagger \hat a \rangle^2,
$
which has been used in the study of cooperative emission in pairs of quantum dots~\cite{SipahigilIntegratedDiamond2016,KimSuperRadiantEmission2018,KoongCoherenceCooperative2022,CygorekSignaturesCooperative2023}.
We observe that the resonances giving rise to the stabilization of entanglement correlate with strong features in the properties of the emission, which depend on whether the emitters are driven coherently or incoherently:
\begin{enumerate}[label=\textcolor{Maroon}{(\roman*)}]
	\item \textcolor{Maroon}{Antisymmetric transition and coherent excitation. } For the antisymmetric transition ($\Delta_a\approx R$), a high value of the concurrence correlates with a dip in the emission intensity and bunching in the photon statistics. 
	Here, the single-photon emission is strongly suppressed ($\langle \hat a^\dagger \hat a\rangle \ll 1$) due to the subradiant nature of the stabilized state $|-\rangle \approx |A\rangle$, as we explained in \refsec{sec:2_steady_state} in the absence of the single-mode cavity.
	 Consequently,  the small amount of emission observed mostly comes from two-photon emission, yielding a high probability of detecting two photon simultaneously. 
	\item \textcolor{Maroon}{Symmetric transition and both coherent and incoherent excitation. }  In the symmetric transition ($\Delta_a\approx -R$), the stabilization of a superradiant state $|+\rangle \approx |S\rangle$ corresponds to an increase in emission intensity, with a broader profile in the case of incoherent excitation.
	Photon correlations in this regime exhibit antibunching, as the emission is primarily dominated by single-photon processes.
\end{enumerate}
We note that, given the superradiant and subradiant nature of these states, other methods such as emission dynamics measurements of the lifetime~\cite{TiranovCollectiveSuper2023} could also provide reliable evidence of their stabilization, as we will discuss in the following mechanism of entanglement generation. 

\reffig{fig:Fig10_Observability} also offers additional valuable insights into the optical tunability of entanglement. 
When exciting coherently, we observe that the reported optical features persist even in the non-perturbative regime $\Omega \gg R$, where the resonant conditions $\Delta_a \approx \pm R$ do not longer hold since the eigenstates are now strongly dressed by the drive and their frequencies are subsequently shifted---see \refsec{sec:StrongDriving}---.
These results demonstrate that strong hybridization with the drive does not inhibit entanglement generation. Instead, it provides a mechanism to optically tune the cavity in or out of resonance with the energy transitions associated with these dressed states by varying the driving intensity $\Omega$ while keeping the cavity detuning $\Delta_a$ fixed.
This tunability is clearly illustrated in the top panels of the right column of \reffig{fig:Fig10_Observability}, where a cut of $\mathcal{C}$, $\langle \hat a^\dagger \hat a \rangle$ and $g^{(2)}(0)$ are depicted versus the drive amplitude $\Omega$ for a fixed $\Delta_a$. The concurrence [\textcolor{Maroon}{panel (a)}] features a maximum when the corresponding dressed-state transition is placed in resonance with the cavity, correlating perfectly with a dip in the emission [\textcolor{Maroon}{panel (b)}]  and a peak in the statistics [\textcolor{Maroon}{panel (c)}] .
Finally, we note that for sufficiently high incoherent pumping rate $P\gg \Gamma_P$, the generation of entanglement is disabled as the emitters stabilize in $|ee\rangle$. 
{
	\sidecaptionvpos{figure}{t}
	\begin{SCfigure}[1][t!]
	\includegraphics[width=1.\columnwidth]{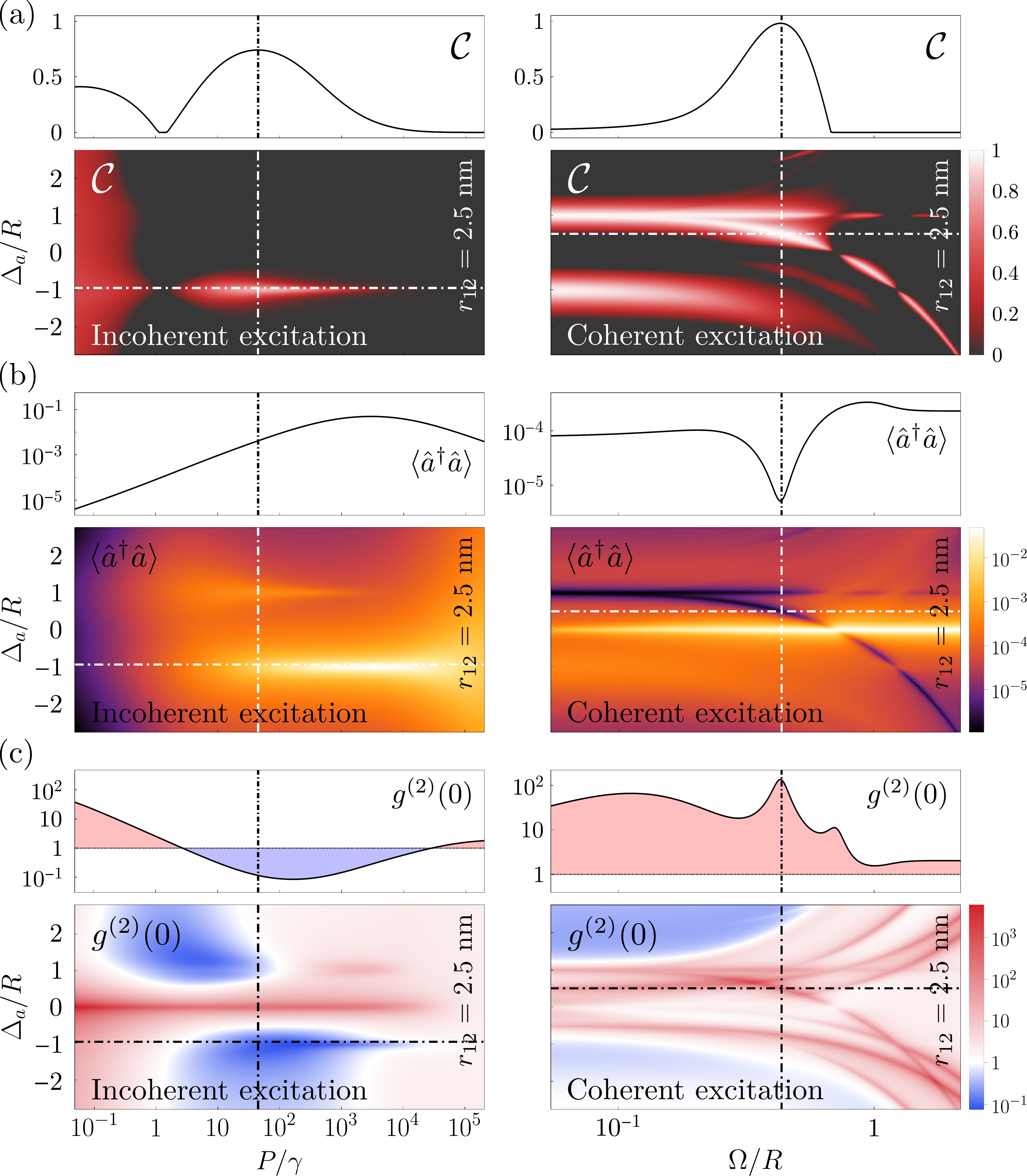}
	\captionsetup{justification=justified}
	\caption[Entanglement detection and control for Mechanism I.]{\label{fig:Fig10_Observability}
		\textbf{Entanglement detection and control for Mechanism I.}
		 Stationary concurrence and optical properties of the emitted light by the cavity.
		(a), (b), (c) Steady-state concurrence, transmission intensity and photon correlations versus the Rabi frequency of the laser $\Omega$ and the laser-cavity detuning $\Delta_a$. Top panels show a cut along $\Omega$, marked by the horizontal white dashed lines in the density plots. Parameters:  $r=2.5$ nm, $k=2\pi/780\ \text{nm}^{-1}$, $J=9.18\times 10^4\gamma$, $\gamma_{12}=0.999\gamma$, $\delta=10^{-2}J$, $R=9.18\times 10^4\gamma$, $\Delta=0$, $\kappa =10^4 \gamma $, $g=10^{-1} \kappa$.
	}
\end{SCfigure}}

\paragraph{Validity in state-of-the-art open cavities. }
As shown in \reffig{fig:Fig10_Observability}, Mechanism I is optically tunable via the pump intensity. To explore how this tunability extends across a range of technological platforms---commonly characterized by the cooperativity---we compute the concurrence in terms of the driving intensity ($P$ or $\Omega$) and the cooperativity ($C$) in \reffig{fig:Fig3_Tunability}.

The regions of high entanglement  are enclosed within the white lines that delineate the set of conditions outlined in \refsec{sec:Conditions}, confirming their validity for both pumping mechanisms. Notably, our results show that one can observe high values of the concurrence in the range of
cooperativities $C\sim 10-10^2$, compatible with cavity QED platforms based on quantum
emitters coupled to single-mode cavities. 
Experimental values of the cooperativity have been reported in
the range of $C\sim 1-40$ for color centres~\cite{EvansPhotonmediatedInteractions2018,LukinTwoEmitterMultimode2023}, quantum dots~\cite{SomaschiNearoptimalSinglephoton2016,TommBrightFast2021}, and molecules~\cite{GerhardtCoherentState2009,WangCoherentCoupling2017,WangTurningMolecule2019,PschererSingleMoleculeVacuum2021}. Additionally,
situations in which the emitters are coupled to plasmonic nanoantennas have reported an enhancement of the
emission rate of the order of  $\sim 10^3 - 10^6$~\cite{AkselrodProbingMechanisms2014,ChikkaraddySinglemoleculeStrong2016,BaumbergExtremeNanophotonics2019}. Novel approaches like inverse design~\cite{MoleskyInverseDesign2018,Miguel-TorcalInversedesignedDielectric2022,Miguel-TorcalMultiqubitQuantum2024} are also opening
new possibilities for the development of photonic environments for the generation of entanglement.
See \textcolor{Maroon}{Table}~\ref{tab:CoopPlatforms} for a summary of achievable cooperativities. 

{
	\sidecaptionvpos{table}{t}
	\begin{SCtable}[0.57][t!]
	\centering
	\begin{tabular}{p{6.5cm} p{2.5cm}}
		\toprule
		\toprule
		\textbf{Platform} & \textbf{Cooperativity}\\
		\toprule
		Color centres~\cite{EvansPhotonmediatedInteractions2018,LukinTwoEmitterMultimode2023} & $\sim 1- 20$ \\
		Quantum dots~\cite{SomaschiNearoptimalSinglephoton2016,TommBrightFast2021}  & $\sim 10-40$ \\
		Molecules~\cite{GerhardtCoherentState2009,WangCoherentCoupling2017,WangTurningMolecule2019,PschererSingleMoleculeVacuum2021}     & $\sim 1- 40$ \\
		Plasmonic nanoantennas~\cite{AkselrodProbingMechanisms2014,ChikkaraddySinglemoleculeStrong2016,BaumbergExtremeNanophotonics2019} & $\sim 10^3-10^6$\\
		\bottomrule
		\bottomrule
	\end{tabular}
	 \captionsetup{justification=justified} 
	\caption[Achievable cooperativities in state-of-the-art solid-state platforms]{\textbf{Achievable cooperativities in state-of-the-art solid-state platforms.}  }
	\label{tab:CoopPlatforms}
\end{SCtable}}
{
	\sidecaptionvpos{figure}{b}
	\begin{SCfigure}[0.7][b!]
		\includegraphics[width=.67\textwidth]{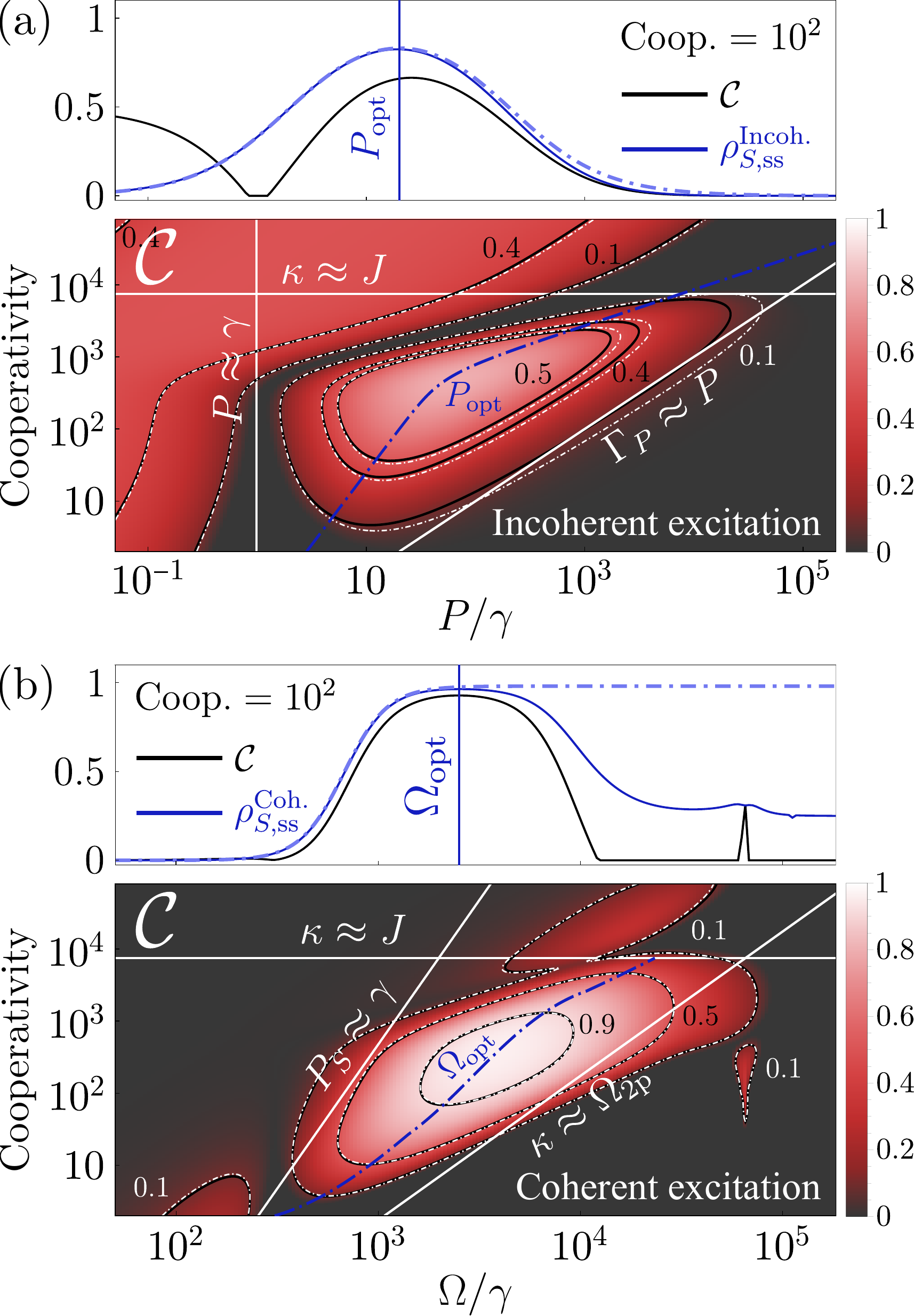}
		\captionsetup{justification=justified}
		\caption[Tunability of the maximum achievable concurrence.]{
			\label{fig:Fig3_Tunability}
			\textbf{Tunability of the maximum achievable concurrence.}
			Stationary concurrence versus cooperativity $C$ and driving intensity: (a) incoherent excitation, (b) coherent excitation, for $N=2$ and  $\Delta_a\approx -J$. 
			White straight lines mark the conditions outlined in the text and bound the regions of formation of entanglement. Blue dot-dashed lines indicate the optimal driving strength that maximizes the concurrence.
			Black contours correspond to numerical calculations with a full model from \eqref{eq:full_master_eq} that perfectly match thee grey-dashed contours, obtained from the adiabatic model in \eqref{eq:Nakajima}.
			Top panels correspond to a cut of the concurrence (black) for a fixed cooperativity: $C=100$. Exact (blue) and analytical (dot-dashed blue) predictions of the population of $|S\rangle$ are depicted.
		}
\end{SCfigure}}

To characterize the maximum entanglement achievable in a given platform, a key parameter is the optimal pumping rate. For the case of incoherent excitation, we find the $P_{\text{opt}}$ that maximizes $\rho_{S,\text{ss}}^{\text{Incoh}}$ in \eqref{eq:rhoS_purcell},
\begin{equation}
\colorboxed{Maroon}{
P_\text{opt}=\frac{\Gamma_P}{2}\sqrt{\left(\frac{\kappa}{J}\right)^2+ \frac{16}{C}},
}
\end{equation}
whose accuracy is confirmed in \reffig{fig:Fig3_Tunability}~\textcolor{Maroon}{(a)}---see \refap{sec:SM_AnalyticalDensityMatrix} for detailed derivation---.
In contrast, under coherent excitation, the optimum value of the coherent drive cannot be computed using our analytical expressions since these were derived for perturbative driving regimes. Beyond this limit, $\Omega \gg R$, there are no general closed-form expressions for the stationary density matrix elements. Therefore, we determine numerically the optimum value for the coherent drive, $\Omega_{\text{opt}}$, as shown in \reffig{fig:Fig3_Tunability}~\textcolor{Maroon}{(b)}.

\subsection{Effect of additional decoherence channels}
\label{sec:decoherence}
So far we have considered a general situation in which the system undergoes decoherence due to its interaction with a vacuum electromagnetic bath, giving rise to both local and collective spontaneous decay and cavity leakage~\cite{FicekQuantumInterference2005,CarmichaelStatisticalMethods1999}. 
However, considerations of more specific platforms, such as molecules or semiconductor quantum dots, may require to account for additional decoherence channels~\cite{ShammahOpenQuantum2018}. 

In order to test the robustness of our mechanism for generating high stationary entanglement, we consider individually the impact of three distinct scenarios, each involving a different type of decoherence:
\begin{enumerate}[label=\textcolor{Maroon}{(\roman*)}]
	\item \textcolor{Maroon}{Extra spontaneous decay. }
	The presence of additional decay channels can make the value of the local spontaneous decay rate $\gamma$ to be larger than the value predicted by the parametrization used in this Chapter, given by \eqref{eq:gamma}. This may occur, e.g.,  due to the coupling to phonon baths or to lossy nanophotonic structures with specific geometries that cannot be captured by the coupling to a single, lossy bosonic mode described here, yielding additional spontaneous channels beyond the zero-phonon line~\cite{HaakhSqueezedLight2015, MlynekObservationDicke2014}. 
	\item \textcolor{Maroon}{Local pure dephasing. }
	Local pure dephasing, i.e., dissipative processes that destroy coherence terms in the density matrix of each qubit independently, arises in numerous systems. For instance, this effect is prevalent in semiconductor quantum dots~\cite{KrummheuerTheoryPure2002} or in
	molecular systems where intra-molecular vibrations and crystal motion are taken into account~\cite{delPinoQuantumTheory2015,ReitzMoleculephotonInteractions2020}.
	\item \textcolor{Maroon}{Collective pure dephasing. }
	Collective dephasing, i.e., dissipative processes that destroy coherence terms in the density matrix of qubits collectively, may arise in different scenarios, such as color centers devices interacting with photonic baths~\cite{PrasannaVenkateshCooperativeEffects2018}, or in several qubits interacting with common dissipative bath~\cite{WangDissipationDecoherence2015}.
\end{enumerate}
These additional decoherence channels translate into three extra Lindblad terms in the master equation:
\begin{multline}
	\frac{d \hat \rho}{d t}=\mathcal{L}[\hat \rho]
	+
\underbrace{\frac{\Gamma}{2}(\mathcal{D}[\hat \sigma_1]\hat \rho+\mathcal{D}[\hat \sigma_2]\hat \rho)}_{\substack{\text{Extra spontaneous} \\ \text{decay},\  \mathcal{L}_{\Gamma}[\hat \rho]}}
\\
+
\underbrace{\frac{\gamma_\phi}{2}(\mathcal{D}[\hat \sigma_{z,1} ]\hat \rho+\mathcal{D}[\hat \sigma_{z,2} ]\hat \rho)}_{\text{Local pure dephasing},\ \mathcal{L}_{\gamma_\phi} [\hat \rho]}
+
\underbrace{\frac{\Gamma_\phi}{2}\mathcal{D}[\hat \sigma_{z,1}+\hat \sigma_{z,2} ]\hat \rho}_{\substack{\text{Collective pure} \\ \text{dephasing},\  \mathcal{L}_{\Gamma_\phi}[\hat \rho]}},
\end{multline}
where $\hat \sigma_{z,i}\equiv 2\hat \sigma_i^\dagger \hat \sigma_i-\mathbb{I}$ is the z-component of the Pauli matrices, and $\mathcal{L}[\hat \rho]$ denotes the original Liouvillian from \eqref{eq:full_master_eq},
\begin{equation}
	\mathcal{L}[\hat \rho]\equiv -i [\hat H,\hat  \rho] 
	+\frac{\kappa}{2}\mathcal{D}[\hat a]\hat \rho 
	+ \sum_{i,j=1}^{N=2}\frac{\gamma_{ij}}{2}\mathcal{D}[\hat \sigma_i,\hat \sigma_j]\hat \rho +
	\frac{P_i}{2}\mathcal{D}[\hat \sigma_i^\dagger ]\hat \rho.
\end{equation}

We will exhibit the effects of the additional decoherence channels independently, resulting in a master equation
\begin{equation}
\frac{d\hat \rho}{dt}=\mathcal{L}[\hat \rho]+\mathcal{L}_{\gamma_\text{d}}[\hat \rho], \quad \text{with}\quad \gamma_\text{d}=\{\Gamma, \gamma_\phi, \Gamma_\phi\}.
\label{eq:Additional_Decoh}
\end{equation}
\reffig{fig:Fig11_EntanglementProtection}~\textcolor{Maroon}{(a-c)} shows the concurrence versus the cooperativity and the additional decoherence rates for both incoherent (left panels) and coherent (right panels) excitation schemes, considering an inter-emitter distance of $r_{12}=2.5\ \text{nm}$ and the cavity in the 
\graffito{
	$^*$When the cavity is placed in the antisymmetric resonance, $\Delta_a=R$,  the concurrence is significantly degraded since the condition for $\Gamma_{\text{I}, A}=\beta^2 \Gamma_P/2$ is quickly compromised---not shown here, see \colorref{Vivas-VianaDissipativeStabilization2024}---. For instance, in the case of extra spontaneous decay, this is due to the fact that $\gamma_-$ is now increased by $\Gamma$, making the state $|-\rangle$ no longer subradiant, while it remains still weakly coupled to the cavity, with a rate which is proportional to $\beta \ll 1$. The enhancement provided by the cavity is then lost.
}
symmetric resonance$^{\textcolor{Maroon}{*}}$ $\Delta_a\approx -R$.

{
	\sidecaptionvpos{figure}{t}
\begin{SCfigure}[1][t!]
	\includegraphics[width=1.0\textwidth]{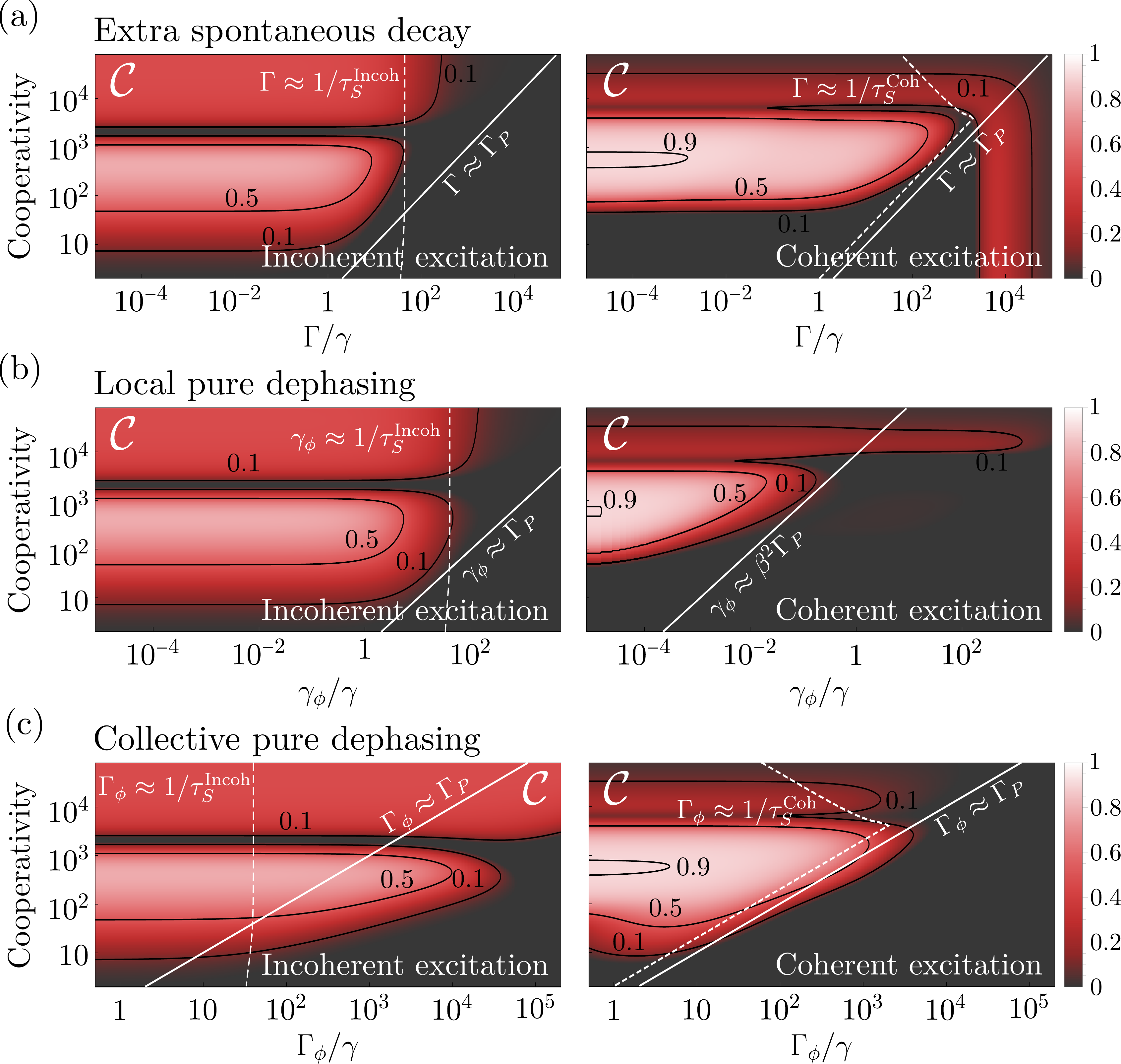}
	\captionsetup{justification=justified}
	\caption[Robustness test of the mechanisms of entanglement generation]{\label{fig:Fig11_EntanglementProtection}
		\textbf{Robustness test of the mechanisms of entanglement generation. }
		Stationary concurrence for both excitation schemes incoherent (left column) and coherent (right column). The considered additional channels are: (a) extra spontaneous decay; (b) local pure dephasing; and (c) collective pure dephasing. Each panel depicts the concurrence as a function of the cavity cooperativity $C$ and the corresponding extra decoherence rate.
		Parameters: 
		$J=9.18\times 10^4\gamma,\ \gamma_{12}=0.999\gamma,\ \delta=10^{-2}J,\ \ g=10^{-1}\kappa,\ \Delta_a\approx -R,\ \Delta=0,\ P=40\gamma,\ \Omega=10^4 \gamma$
	}
	
\end{SCfigure}}

\vspace{-2mm}

\paragraph{General conditions. }

The frequency-resolved Purcell effect is efficient when the Purcell rate $\Gamma_P$ is the dominant dissipative rate. 
Consequently, for the mechanism to be robust against additional decoherence, the Purcell rate should exceed the corresponding decoherence rates, i.e., 
\begin{equation}
	\Gamma_P\gg \gamma_{\text{d}},  \quad \text{with}\quad \gamma_\text{d}=\{\Gamma, \gamma_\phi, \Gamma_\phi\}.
	\label{eq:decoh_cond1}
\end{equation}
This condition is illustrated in \reffig{fig:Fig11_EntanglementProtection}~\textcolor{Maroon}{(a-c)} by a solid-white line. This line bounds the region of entanglement in most cases, with some exceptions that we discuss below.
Additionally, for entanglement to survive in the long-time limit, the stabilization timescale $\tau_S$ must also be faster than the additional decoherence rates. This translates into the condition
\begin{equation}
\colorboxed{Maroon}{
\tau_S\ll \gamma_{\text{d}}^{-1},  \quad \text{with}\quad \gamma_\text{d}=\{\Gamma, \gamma_\phi, \Gamma_\phi\},
}
	\label{eq:decoh_cond2}
\end{equation}
depicted in \reffig{fig:Fig11_EntanglementProtection}~\textcolor{Maroon}{(a-c)} by dashed white lines. 
\newpage

In summary, we require:%
\begin{enumerate}[label=\textcolor{Maroon}{(\roman*)}]
	\item $\Gamma_P \gg \gamma_{\text{d}} $: The Purcell rate must dominate over all decoherence rates.
	\item $\tau_S \ll \gamma_{\text{d}}^{-1}$: The stabilization timescale must be faster than the decoherence.
\end{enumerate}
There are two exceptions to these general observations:

\begin{enumerate}[label=\textcolor{Maroon}{(\roman*)}]
\item  \textcolor{Maroon}{Local pure dephasing with coherent excitation.}
This first exception arises from the nature of the subradiant state, $|-\rangle$. As proven in \refap{appendix:MechanismI}, the effective Lindblad term is given by a jump operator $\hat \xi_S=-|+\rangle \langle ee | +\beta/2|gg\rangle \langle -|$, resulting in
\begin{multline}
	\mathcal{L}_{\text{cav}}^{\text{eff}}=\Gamma_P\mathcal{D}[\hat \xi_S]\hat \rho \approx \Gamma_P \mathcal{D}[|+\rangle \langle ee| ]\hat \rho \\
	 + \frac{(\beta^2/4) \Gamma_P}{2} \mathcal{D}[|gg\rangle \langle - |]\hat \rho,
\end{multline}
stabilizing the system in the symmetric state $|+\rangle \langle +|$. However, local pure dephasing tends to mix the states $|+\rangle$ and $|-\rangle$ by transferring population from one to the other, as can be seen in the following terms of the corresponding rate equations
\begin{subequations}
		\label{eq:rateLocalDeph}
\begin{align}
		\dot{\rho}_{+,+}&\propto-\left(\gamma_+ +2\Gamma_P+2\gamma_\phi \right)\rho_{-,-}+ 2\gamma_\phi \rho_{-,-}, \\
	\dot{\rho}_{-,-}&\propto-\left(\gamma_- + \beta^2/4 \Gamma_P+2\gamma_\phi \right)\rho_{-,-}+ 2\gamma_\phi \rho_{+,+}.
\end{align}
\end{subequations}
Here, we can observed that $\rho_{+,+}$ is fed by the population $\rho_{-,-}$ with a rate $\propto \gamma_\phi$, and vice versa.  Therefore, in order for entanglement emerging from the population of $|+\rangle \approx |S\rangle$ to survive, the subradiant state $|-\rangle$ must be depleted faster than it is populated, which sets the condition 
\begin{equation}
\colorboxed{Maroon}{
	\beta^2 \Gamma_P \gg \gamma_\phi.
}
	\label{eq:condition_puredeph}
\end{equation}
When this condition does not hold, the mechanism of entanglement generation breaks down since the population transfer between the super- and subradiant states occur with a rate $\propto \gamma_\phi$, destroying any possible entanglement.
Notably, the generation of entanglement exhibits greater robustness against local pure dephasing under incoherent excitation. In contrast to the coherent excitation scenario, once the subradiant state acquires population due to the population transfer induced by this decoherence term, the incoherent excitation is able to deplete this state by repumping towards the doubly-excited state, leading it back into the symmetric state by means of the frequency-resolved Purcell effect.

If the condition in \eqref{eq:condition_puredeph} does not hold, but the formation of the entangled state is faster than the decoherence timescale, 
\begin{equation}
	\tau_S\ll \gamma_\phi^{-1},
\end{equation}
we can still observe the formation of entanglement as a metastable state that survives for a time $ \gamma_\phi^{-1}$---not shown here---.
\item \textcolor{Maroon}{Collective pure dephasing and incoherent excitation. }
The second exception is the specific case of collective pure dephasing and incoherent excitation, where the stationary entanglement is maintained even when the conditions \textcolor{Maroon}{(i-ii)} do not hold: $\Gamma_P\gg \gamma_{\text{d}}$ and $\tau_S \ll \gamma_{\text{d}}^-1$, respectively. This type of decoherence essentially degrades the coherence between the ground state $|gg\rangle$ and the doubly-excited state $|ee\rangle$. Contrary to the coherent excitation scheme, in the case of incoherent excitation, population transfer among states occurs via incoherent one-photon processes, and does not rely on any type of coherence between $|gg\rangle$ and $|ee\rangle$. Therefore, the mechanism of entanglement generation via incoherent excitation proves to be highly robust against collective pure dephasing, maintaining entanglement even at very high values of this decoherence term ($\Gamma_\phi \sim 10^5 \gamma$). This type of dephasing only becomes detrimental when its rate becomes comparable to the coherent coupling, $J\approx \Gamma_\phi$, since then the relevant transitions are no longer resolved due to the dephasing-induced broadening.
\end{enumerate}

\subsection{Generalization to $N$ emitters}

So far, we have explored Mechanism I in the case of two interacting emitters under both coherent and incoherent excitation. In particular,  \reffig{fig:Fig4_ConcurrenceQubitCavityDetuning}, \reffig{fig:Fig10_Observability}, or
\reffig{fig:Fig3_Tunability}, suggest that, for a fixed set of cavity parameters, coherent driving yields higher values of entanglement than the incoherent case. 
However, incoherent excitation offers a straightforward path for extending this scheme to cases where $N>2$, due to the simplicity of its excitation mechanism---subsequent single-excitation through the energy levels---. 
Note that for Mechanism I to operate under coherent driving, the emitters must be driven at the two-photon resonance to enable the direct excitation of the doubly-excited state via a perturbative two-photon process---see \refch{ch:TwoPhotonResonance}---. In a more general scenario with $N>2$ emitters, we would need to engineer conditions where, e.g, a perturbative $N$-photon process excites the highest excited state. While theoretically possible, achieving such a setup experimentally would be challenging; indeed, it is worth noting that just a few observations of two-photon processes in molecular aggregates have been reported~\cite{HettichNanometerResolution2002,TrebbiaTailoringSuperradiant2022,LangeSuperradiantSubradiant2024}.

\begin{SCfigure}[1][b!]
	\includegraphics[width=0.6\textwidth]{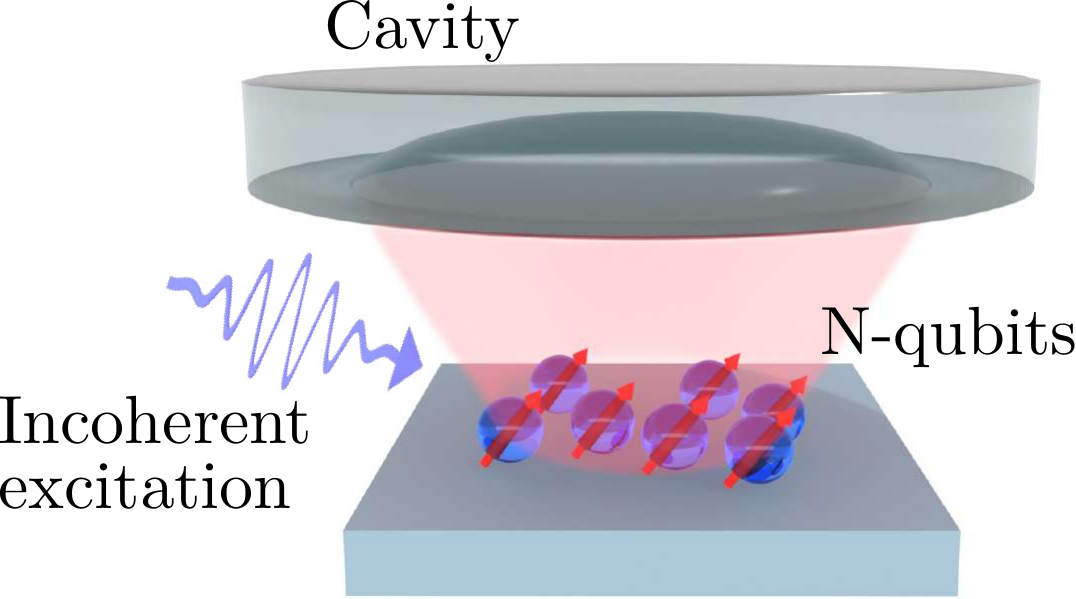}
	\captionsetup{justification=justified}
	\caption[Sketch of the setup of entanglement generation for $N>2$.]{\label{fig:Fig4_Generalization}
		\textbf{Sketch of the setup of entanglement generation for $N>2$. }
		Illustration of $N$ all-to-all interacting quantum emitters, incoherently excited and coupled to a single-mode cavity. The cavity is illustrated as a Fabry-Perot hybrid resonator with one of the mirrors of the cavity exhibiting a parabolic geometric that benefits the confinement of light~\cite{ShlesingerHybridCavityantenna2023}.
	}
\end{SCfigure}

\paragraph{Dicke model. }
As an illustrative example, we consider $N$ degenerate emitters with transition frequency $\omega_q$, and an all-to-all interaction Hamiltonian, as could be provided by the coupling to a common photonic environment~\cite{GoldsteinDipoledipoleInteraction1997,EvansPhotonmediatedInteractions2018,ChangColloquiumQuantum2018}. The system is depicted in \reffig{fig:Fig4_Generalization}.
The $N$ TLSs tensor product space of dimension $2^N$  can be expressed as a direct sum of spaces:
\begin{equation}
		\mathcal{H}_{\text{N-TLS}}=\frac{1}{2} \otimes \frac{1}{2} \otimes \ldots \otimes \frac{1}{2}= 
		\left\{
		\begin{aligned}
			&0 \oplus 1 \oplus \ldots \oplus \frac{N}{2} \quad \  \text{(even $N$)} \\
			&\frac{1}{2} \oplus \frac{3}{2} \oplus \ldots \oplus \frac{N}{2} \quad \text{(odd $N$)}
		\end{aligned}
		  \right.
\end{equation}
%
		%
		%
		%
		%
		%
		%
%
%
allowing the description of the composite system in terms of a \textit{big spin}, with collective spin operators:
\begin{equation}
\hat S_z=\sum_j^N \frac{\hat \sigma_{z,j}}{2}, \quad \hat S^- = \sum_j^N \hat \sigma_j,\quad \text{and}\quad  \mathbf{\hat S}=  \hat S_z +\frac{ (\hat S^+ \hat S^-+\hat S^- \hat S^+)}{2},
\end{equation}
where $S^+=(S^-)^\dagger=\sum_j^N \hat \sigma_j^\dagger$ and $ \mathbf{\hat S}$ denoting the total spin---the \textit{casimir} of the group---. This collective representation of the $N$ emitters forms an effective $\mathfrak{su}(N)$ algebra via the operator set $\{\mathbf{\hat S},\hat S_z,\hat S^\pm  \}$, whose natural collective basis is known as the \textit{Dicke basis}~\cite{DickeCoherenceSpontaneous1954, ShammahOpenQuantum2018,ReitzCooperativeQuantum2022}, or \textit{Dicke states}, indexed by two quantum numbers  $|j,m\rangle$:
\begin{SCfigure}[1][b!]
	\includegraphics[width=0.95\textwidth]{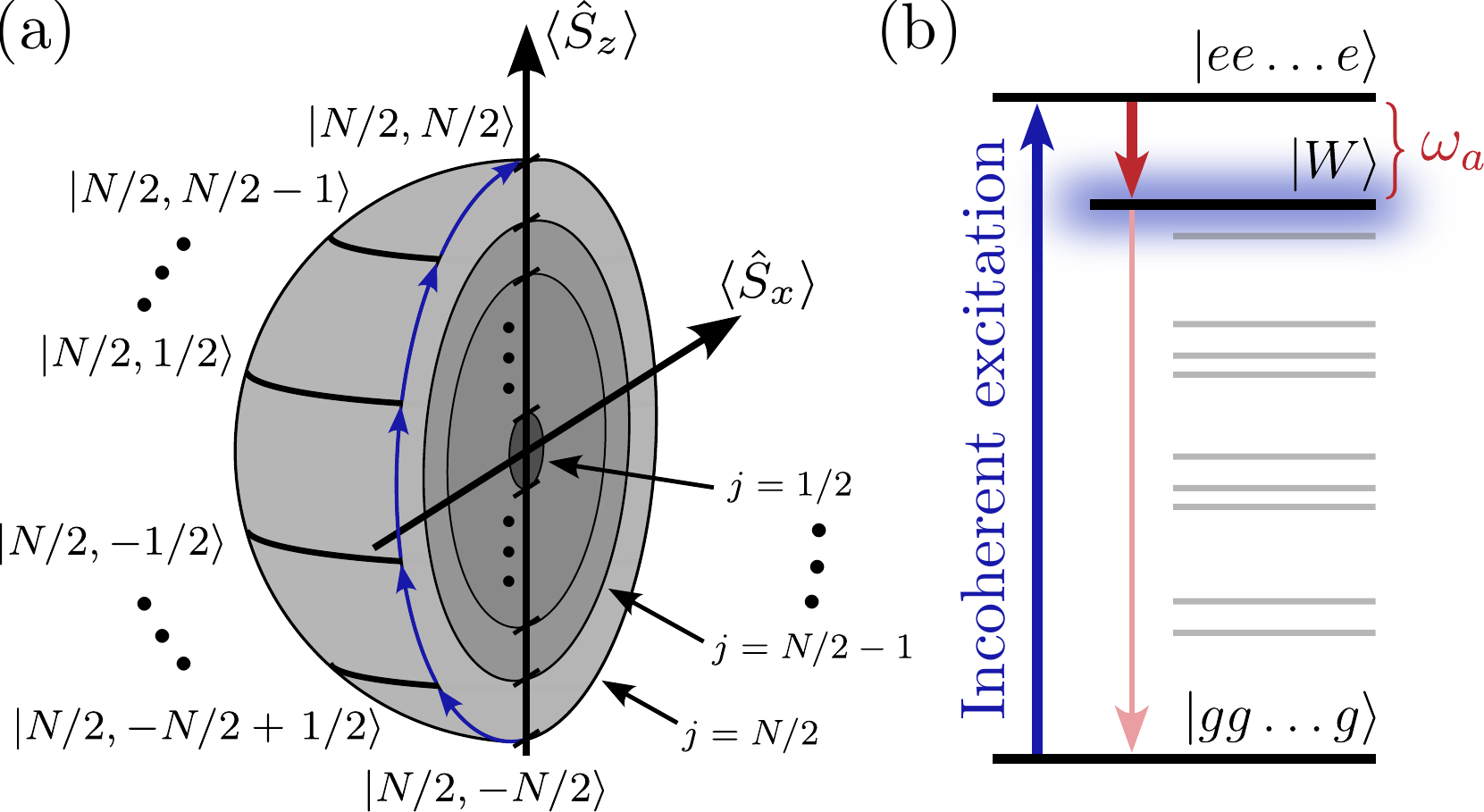}
	\captionsetup{justification=justified}
	\caption[Illustration of the Dicke ladder and stabilization of $W$ states.]{\label{fig:DickeLadder}
		\textbf{Illustration of the Dicke ladder and stabilization of $W$ states.}
		(a) Collective-emitter Bloch sphere for an odd number of spins showing an onion-like structure. The lines of latitude (blue lines) represents Dicke states $|j,m\rangle$ at a fixed $j$. The concentric spheres represents the antisymmetric subspaces of $j\leq N/2$.   (b) Enhanced dissipative transition in the Dicke ladder.
	}
\end{SCfigure}
\begin{subequations}
	\begin{align}
		&\mathbf{\hat S} |j,m\rangle= j(j+1) |j,m\rangle, \\
		&\hat S_z |j,m\rangle =m |j,m\rangle, \\
		&\hat S^\pm|j,m\rangle= \sqrt{(j\mp m)(j\pm m+1)}|j,m\pm 1\rangle,
	\end{align}
\end{subequations}
where $j\leq \frac{N}{2}$ with $ j_{\text{min}}=0$ (even $N$) or $ j_{\text{min}}=1/2$ (odd $N$), and $|m|\leq j $. 
See \reffig{fig:DickeLadder}~\textcolor{Maroon}{(a)} for a schematic representation of the Dicke basis, forming a \textit{onion-like} structure~\cite{ReitzCooperativeQuantum2022}, where the external sphere represents the highest spin subspace $j=N/2$, while the concentric spheres represent the antisymmetric subspaces of $j\leq N/2$ . 
Each Dicke state $|j,m\rangle$ exhibits a degeneracy with respect to an irreducible representation in the uncoupled TLS basis given by:
\begin{equation}
	d_N^j= (2j+1)\frac{N!}{(N/2+j+1)!(N/2-j)!}.
\end{equation}

Therefore, the $N$ identical all-to-all interacting emitter system, depicted in \reffig{fig:Fig4_Generalization}, is described by a Hamiltonian
\begin{equation}
\hat H^N_q = \frac{1}{2}\omega_q \hat S_z + J \hat S^+\hat S^-= \frac{1}{2}(\omega_q+J) \hat S_z + J \sum_{i\neq j}^N\hat \sigma_i^\dagger \hat \sigma_j + \text{constant},
\label{eq:H_q_N_Emitters}
\end{equation}
where counter-rotating terms have been neglected based on a RWA that assumes $J\ll \omega_q$. 
For $N=2$ emitters, the results of the previous sections are recovered by setting $\omega_0\equiv \omega_q+J$, i.e., absorbing the shift $J$ into the definition of the atomic energy. 
%

As indicated  in \eqref{eq:Incoh_Cond} in \refsec{sec:model}, we will study the system in the rotating frame of the emitters, so that \eqref{eq:H_q_N_Emitters} simply reads 
\begin{equation}
	\hat H^N_q =  J \hat S^+\hat S^-,
\end{equation}
with eigenvalues
\begin{equation}
\lambda_{j,m} = J[j(j+1) - m(m-1)].
\label{eq:Eigenvals_Dicke}
\end{equation}
Additionally, we assume a description of the cavity with a coupling to the $N$ emitters of the form
\begin{equation}
	\hat H^N_a=\tilde \Delta_a \hat a^\dagger \hat a+ g \left(\hat a^\dagger \hat S^- + \hat a \hat S^+\right),
\end{equation}
where we have defined a new cavity detuning 
\begin{equation}
	\tilde \Delta_a \equiv \omega_a-\omega_q,
\end{equation}
using a tilde to differentiate it from the detuning $\Delta_a=\omega_a-\omega_0$ in \eqref{eq:Incoh_Cond} for the case of $N=2$. This distinction, together with the relation $\omega_0=\omega_q+J$ laid out before, makes the results consistent and unambiguous.
Finally, assuming the same type of dissipators considered before, the generalized master equation to $N$ emitters reads
\begin{equation}
\colorboxed{Maroon}{	\frac{d \hat \rho}{dt}=-i[\hat H^N, \hat \rho]+\frac{\kappa}{2}\mathcal{D}[\hat a]\hat \rho +\sum_{i,j}^N \frac{\gamma_{ij}}{2}\mathcal{D}[\hat \sigma_i, \hat \sigma_j]\hat \rho + \sum_i^N \frac{P_i}{2} \mathcal{D}[\hat \sigma_i^\dagger]\hat \rho,}
\end{equation}
with $\hat H^N=\hat H^N_q+\hat H^N_a$, and having defined $\gamma_{ii}=\gamma$, $P_i=P$, and $\gamma_{ij}=\gamma_{ji}$ (with $i\neq j$) for all $(i,j)=1,\ldots, N$. Then, we denote the collective dissipative coupling simply as $\gamma_{ij} \rightarrow \gamma_{\text{col}}$.

\begin{SCfigure}[0.75][b!]
	\includegraphics[width=0.65\textwidth]{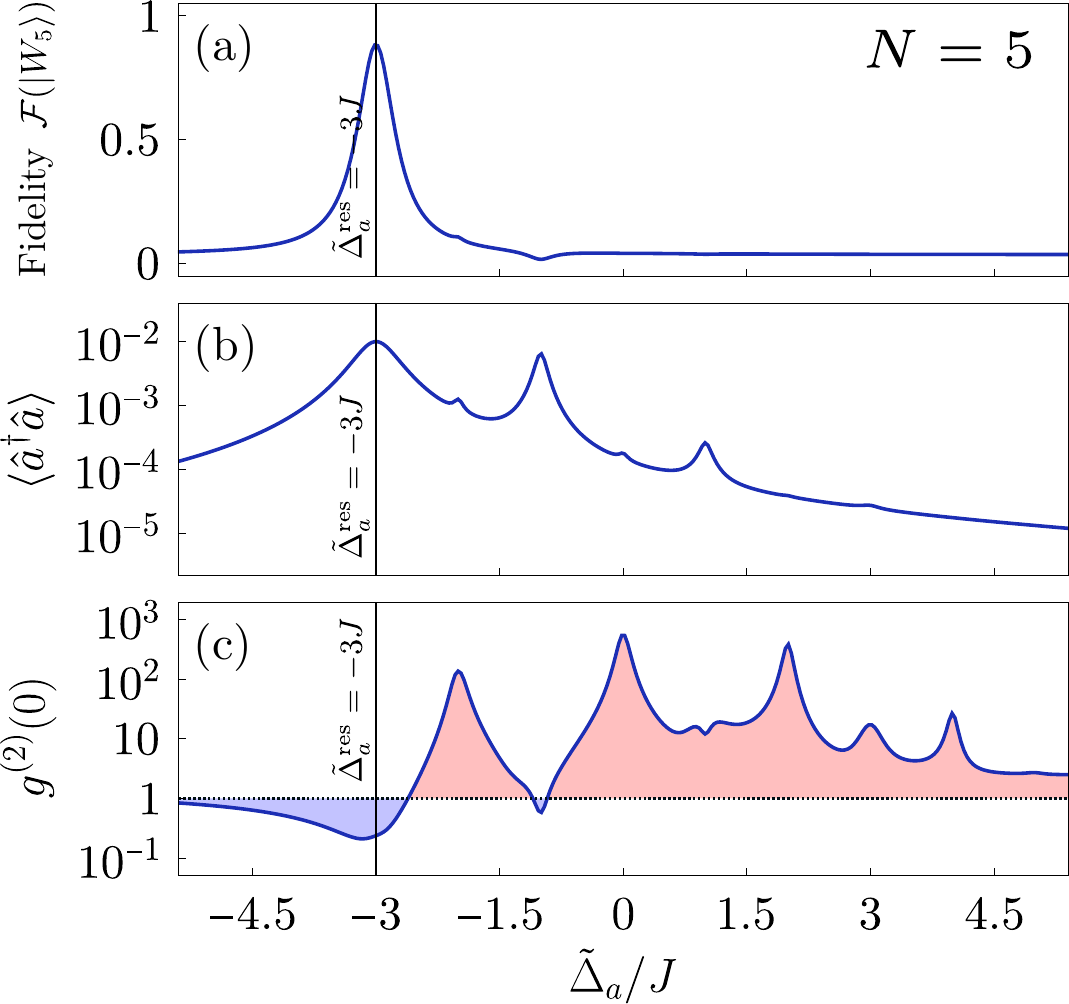}
	\captionsetup{justification=justified}
	\caption[Entanglement detection and control for Mechanism I for $N>2$ emitters]{
		\label{fig:Fig_Observability_N5_Emitters}
		\textbf{	Entanglement detection and control for Mechanism I for $N>2$ emitters.}
		Stationary concurrence and optical properties of the emitted light by the cavity.
		(a) Steady-state fidelity to the entangled state $|W_5\rangle$, (b) transmission intensity and  (c) two-photon correlations in terms of the laser-cavity detuning  $\tilde \Delta_a$.
		Parameters: $J=10^5 \gamma$, $\gamma_{\text{col}}=0.999\gamma$,  $g=10^{-1}\kappa$, $P_{\text{opt}}=132.3 \gamma$, $C=496.84$. 
	}
\end{SCfigure}
\paragraph{Generation of $W$ states. }
As for $N=2$, qubit-qubit interactions provide a non-equidistant set of eigenvalues, enabling the tuning of the cavity to enhance specific transitions within the ladder, energetically separated from each other by values $\Delta \lambda = \lambda_{j,m}- \lambda_{j',m'}   \propto J$, as can be seen from the eigenvalue relation in \eqref{eq:Eigenvals_Dicke}.

Following our previous discussion in \refsec{sec:FreqResolvedMech}, a strong pump $P$ will tend to populate the highest excited state,
\begin{equation}
	|N/2,N/2\rangle = |ee\ldots e\rangle,
\end{equation}
allowing the cavity to successively enhance the transition from this state to 
the first \textit{symmetric Dicke state}~\cite{DickeCoherenceSpontaneous1954,ShammahOpenQuantum2018} with $j=N/2$ [see \reffig{fig:DickeLadder}~\textcolor{Maroon}{(b)}], corresponding to a maximally symmetric superposition of a state with $N-1$ excited TLSs. In other words, a symmetric superposition
of all the one-deexcitation states. This kind of state is commonly known as $W$ state~\cite{DurThreeQubits2000}, and it is given by
\begin{equation}
	|W_N\rangle \equiv |N/2,N/2-1\rangle =\frac{1}{\sqrt{N}}(|ge\ldots e\rangle + |eg\ldots e\rangle)+\ldots + |ee\ldots g\rangle).
	\label{eq:W_state}
\end{equation} 
This specific transition is selected by setting the cavity detuning at the energy difference between $	|N/2,N/2\rangle$ and $	|N/2,N/2-1\rangle$. Therefore, using the eigenvalue relation in \eqref{eq:Eigenvals_Dicke}, we find that the resonance occurs at
\begin{equation}
\colorboxed{Maroon}{
	\tilde\Delta_a^{\text{res}} \equiv \lambda_{\frac{N}{2},\frac{N}{2}}-\lambda_{\frac{N}{2},\frac{N}{2}-1}=  J(2-N).
}
\end{equation}

In \textcolor{Maroon}{Figure}~\ref{fig:Fig_Observability_N5_Emitters}~\textcolor{Maroon}{(a)}, we show the degree of entanglement among the emitters via the fidelity between the steady-state and the $W$ state, defined as 
\begin{equation}
	\mathcal{F}(|W_N\rangle)\equiv \langle W_N| \hat \rho |W_N\rangle,
\end{equation}\textbf{}
for $N=5$ versus cavity detuning, confirming that high fidelity values are achieved at resonance. 
Notably, the activation of this mechanism translates into measurable properties of the emission by the cavity. In \reffig{fig:Fig_Observability_N5_Emitters}, we observe that a region of high fidelity---i.e., high degree of entanglement---correlates with a peak in the emission intensity [\hyperref[fig:Fig_Observability_N5_Emitters]{panel (b)}], as well as an antibunching resonance in the photon statistics [\hyperref[fig:Fig_Observability_N5_Emitters]{panel (c)}] at the resonant frequency $\tilde \Delta_a=\tilde{\Delta}_a^{\text{res}}\approx -3J$ for the case of $N=5$ emitters. It is noteworthy that, recalling the emission properties for the case of $N=2$ emitters shown in the left panels in \reffig{fig:Fig10_Observability}, the antibunching resonance remains even as the number of emitters is increased, in contrast with the general tendency of light emitted from an ensemble to thermalize. These features suggest a way to evidence the formation of entanglement.

\paragraph{Post-selection. }
Interestingly, the results discussed so far in this Chapter for Mechanism I can be used in a projective preparation scheme based on post-selection~\cite{NielsenQuantumComputation2012,HaasEntangledStates2014}. 
In particular, we consider a measurement scheme for the detection of entanglement based on the post-selection of the state heralded by the detection of a photon in the steady-state. The conditional state upon detection is expressed as
\begin{equation}
\hat\rho_c = \frac{
\hat a \hat \rho \hat a^\dagger
}{
\text{Tr}[\hat a \hat \rho \hat a^\dagger]
}
, 
\end{equation}
so that the heralded fidelity is labelled by
\begin{equation}
\mathcal{F}^\text{H}(|W_N\rangle)\equiv \langle W_N|\hat \rho_c| W_N\rangle.
\end{equation}
This conditioned quantum state $\hat \rho_c$ always shows a significant increase in the fidelity [$\mathcal{F}^\text{H}(|W_N\rangle)$] with respect to the non-heralded case [$\mathcal{F}(|W_N\rangle)$], highlighting the relevance of using more refined measurement schemes to improve the detection of entanglement. 
\reffig{fig:Fidelity_Heralded_1D}~\textcolor{Maroon}{(a)} illustrates the fidelity (solid curves) to the $W$ state for $N=5$ emitters in terms of the cavity detuning $\tilde \Delta_a$ for two different values of $J=\{10^3 \gamma, 10^5\gamma\}$.
A reduction of two orders of magnitude in $J$ results in a moderate decrease in fidelity, from approximately $0.9$ to about $0.6$. Nevertheless, when we post-select the state, the heralded fidelity to the entangled state $|W_5\rangle$ increases notably, almost saturating for the case $J=10^5\gamma$, as can be observed via the dashed curves  \reffig{fig:Fidelity_Heralded_1D}~\textcolor{Maroon}{(a)}.
As noted earlier in \reffig{fig:Fig_Observability_N5_Emitters}, regions with high fidelity values correlate with measurable properties of the cavity emission. This is further illustrated in \reffig{fig:Fidelity_Heralded_1D}~\textcolor{Maroon}{(b)},
which shows that the resonant fidelity profile aligns with a corresponding profile in the cavity population.
A more systematic analysis of how the efficiency of the mechanism depends on $J$ is provided in \refap{sec:SM_Validity}.

\begin{SCfigure}[0.5][h!]
	\includegraphics[width=0.79\textwidth]{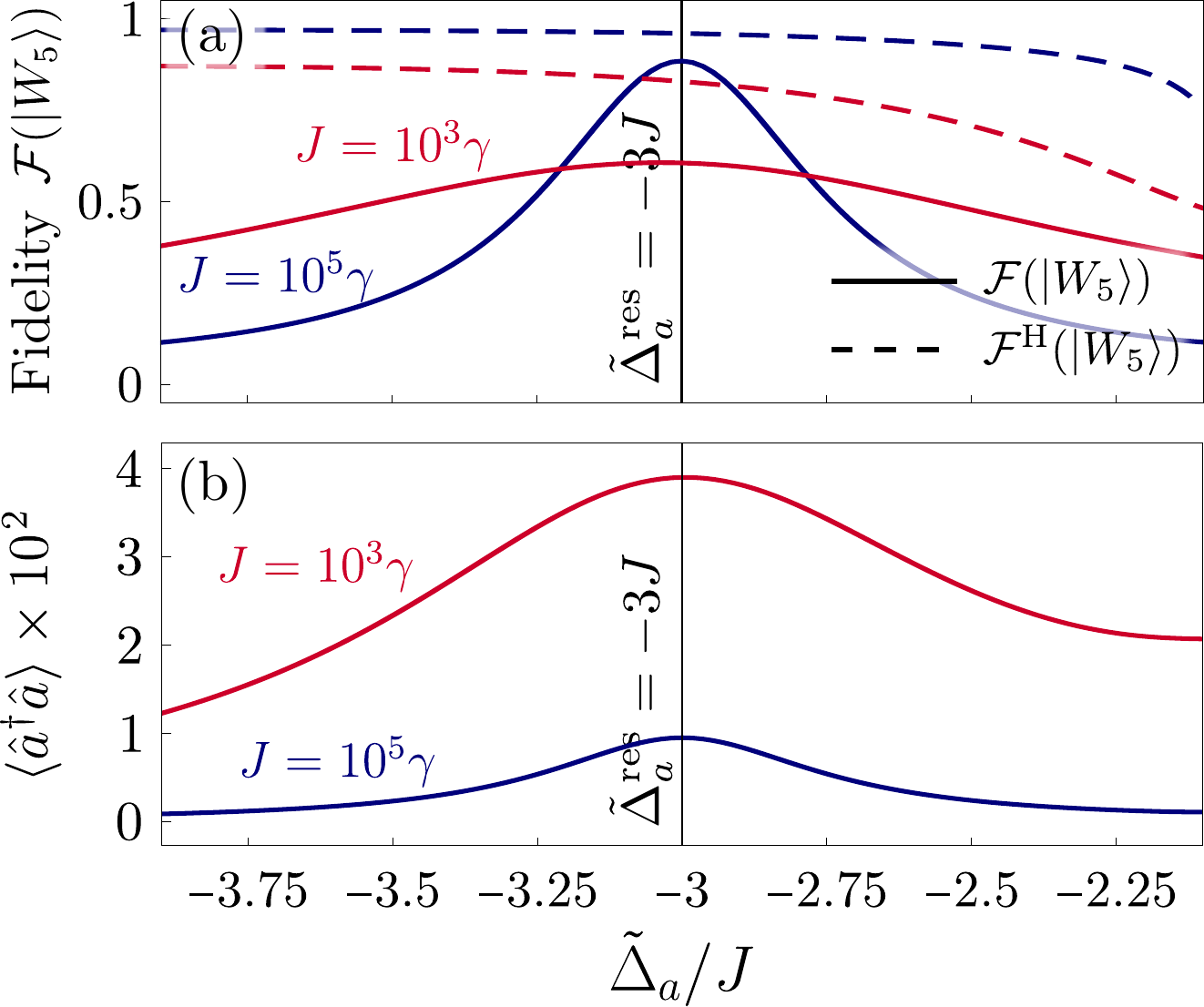}
	\captionsetup{justification=justified}
	\caption[Impact of the dipole-dipole coupling in the fidelity and emission intensity for $N=5$ emitters.]{\label{fig:Fidelity_Heralded_1D}
		\textbf{Impact of the dipole-dipole coupling in the fidelity and emission intensity for $N=5$ emitters.}
		(a) Fidelity with the $W$ state versus cavity detuning for $N=5$ for two values of $J$. 
		Dashed lines show the fidelities for a post-selected state, heralded by the detection of a photon from the cavity.
		The corresponding cavity population in panel (b). 
		Parameters: $\gamma_{\text{col}}=0.999\gamma$, $J=10^5 \gamma$, $g=10^{-1}\kappa$, $\tilde \Delta_a=J(2-N)$, $N=25$.
	}
\end{SCfigure}

\newpage

Post-selecting the quantum state not only offer higher values of the fidelity, but  also relaxes  theoretical requirements for the Mechanism I to be enabled, such as the necessity of high cooperativity or high coherent coupling among the emitters. 
\reffig{fig:Fidelity_Heralded} demonstrates how high values of the fidelity---consistent with the patterns previously discussed in \reffig{fig:Fig3_Tunability}~\textcolor{Maroon}{(a)}---can be achieved for a number of emitters as high as $N=25$ both using non-heralded [\hyperref[fig:Fidelity_Heralded]{panel (a)}, $\mathcal{F}(|W_{25}\rangle)$] and post-selected [\hyperref[fig:Fidelity_Heralded]{panel (b)}, $\mathcal{F}^\text{H}(|W_{25}\rangle)$] detection schemes. 
These two panels highlight the significant advantage of using post-selection measurement techniques to achieve higher fidelity in wider ranges of the parametric space. 
Finally, in \refap{sec:SM_AddingNEmitters}, we present a more systematic
analysis of how the efficiency of the mechanism depends on the number of emitters $N$,  the interaction strength among emitters
$J$, and the post-selection scheme. 
All these simulations were done using the PIQS library in QuTiP, which offer an efficient computational method to simulate Dicke systems with $N\gg 1$ emitters  ~\cite{ShammahOpenQuantum2018,JohanssonQuTiPOpensource2012,JohanssonQuTiPPython2013,LambertQuTiP52024}.

\begin{SCfigure}[0.75][h!]
	\includegraphics[width=0.65\textwidth]{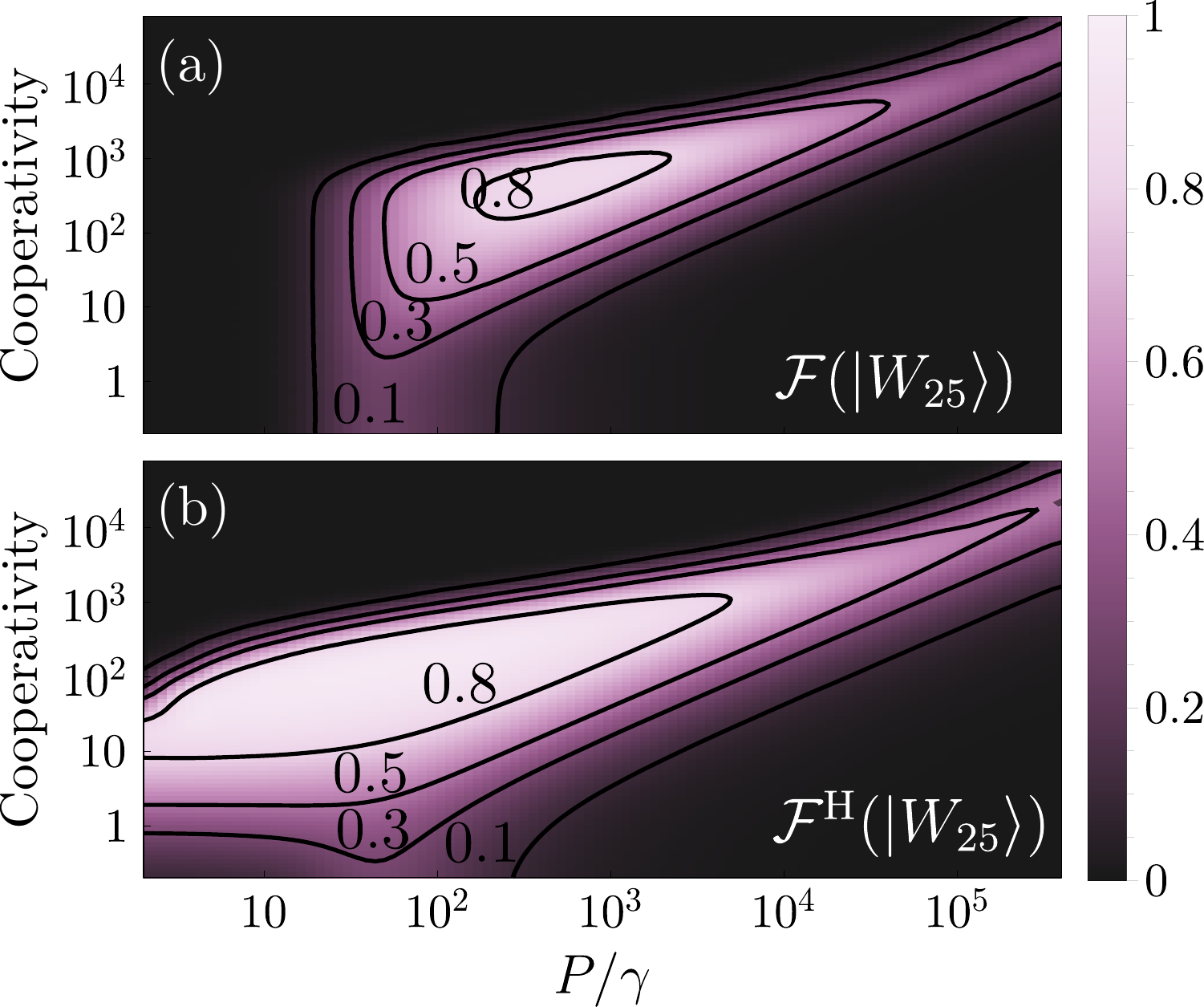}
	\captionsetup{justification=justified}
	\caption[Non-heralded and post-selected fidelities for $N=25$ emitters.]{\label{fig:Fidelity_Heralded}
		\textbf{Non-heralded and post-selected fidelities for $N=25$ emitters.}
		Comparison of fidelities between non- (a, $\mathcal{F}(|W_{25}\rangle)$) and  post-selection (b, $\mathcal{F}^\text{H}(|W_{25}\rangle)$ ) measurement schemes for detecting $W$ states. In both plots, the fidelity to the $W$ state for $N=25$ emitters is depicted versus cavity cooperativty and incoherent pumping rate.
		Parameters: $\gamma_{\text{col}}=0.999\gamma$, $J=10^5 \gamma$, $g=10^{-1}\kappa$, $\tilde \Delta_a=J(2-N)$, $N=25$.
	}
\end{SCfigure}

\section[Mechanism II. Collective Purcell enhancement]{Mechanism II. Collective Purcell \protect\newline enhancement}
\label{sec:MechanismII}

\subsection{Identification of the mechanism}

The next mechanism we consider, that we refer to as Mechanism II, occurs when the cavity is not able to resolve any of the transitions taking place within the dimer, i.e., when
\begin{equation}
	\vspace{-1mm}
\kappa \gg R, \Omega,
\vspace{-1mm}
\end{equation}
including possible dressed-state transitions when $\Omega\gg R$. Importantly, this mechanism does not require any coherent coupling between the QEs. In other words, even if $J\ll \delta$ so that $R\approx \delta$, the process will be activated provided that $\delta < \kappa$. Hence, this mechanism is particularly relevant in scenarios involving weakly interacting non-identical emitters, which commonly arise in solid-state quantum systems, such as quantum dots or molecules.
In these systems, precise control over the emitter positioning is often challenging, resulting in emitters with different frequencies that are too separated to present any significant interaction~\cite{ToninelliSingleOrganic2021}. In this limit, the main effect brought in by the cavity is a Purcell enhancement of the collective decay along the symmetric state $|S\rangle$ which, in combination with the drive, results in a stabilization of the entangled antisymmetric state $|A\rangle$ via a mechanism that we will discuss in detail below.
We note that this effect has been previously described in the context of waveguide QED~\cite{RamosQuantumSpin2014,PichlerQuantumOptics2015} and cavity QED~\cite{OliveiraSteadystateEntanglement2023}. Here, we offer a comprehensive explanation of the underlying mechanism, providing analytical expressions of the regions in parameter phase space where it occurs, the steady-state populations, and the timescale of formation of entanglement. 

The generation of entanglement via this mechanism was already evidenced in \reffig{fig:Fig9_EntanglementRegimes_Coherent}~\textcolor{Maroon}{(b)}, where the steady-state concurrence $\mathcal C$ is evaluated in the parameter phase space spanned by the cavity cooperativity $C$ and $\delta$ for two emitters spatially separated by $r_{12}=50$ nm. This separation is reflected in a small dipole-dipole coupling $J\approx 10\gamma$, meaning that most of the phase diagram corresponds to the situation of non-interacting emitters, with $\delta \gg J$. We observe ample regions in this parameter phase space where the system reaches very high values of the concurrence, close to the maximum possible value $\mathcal  C = 1$. These regions are shaded in purple in the phase diagram at the right column of  \reffig{fig:Fig9_EntanglementRegimes_Coherent}~\textcolor{Maroon}{(b)}, and labelled as II. High concurrence values correspond to sizeable population of the state $|A\rangle$, as shown in \reffig{fig:Fig9_EntanglementRegimes_Coherent}~\textcolor{Maroon}{(c)}.

\subsection{Origin of the mechanism}

The origin of this mechanism resides in the collective decay induced by the cavity, which can be easily obtained by analyzing the Bloch-Redfield master equation \eqref{eq:Nakajima}. One can see that, in the limit
$\kappa \gg R$, all the denominators in the last line of that equation are dominated by $\kappa$, resulting in an effective master equation with a term of collective decay---details are provided in \refap{appendix:AdiabaticElimination}---
\
\begin{equation}
\colorboxed{Maroon}{
	\frac{d \hat{\rho}}{dt}=-i[\hat{H}_q+\hat{H}_d,\hat{\rho}]+\sum_{i,j=1}^2\frac{\gamma_{ij}}{2}\mathcal{D}[\hat{\sigma}_i,\hat{\sigma}_j]\hat{\rho}
+\frac{\Gamma_P}{2}\mathcal{D}[\hat \sigma_1+\hat \sigma_2 ]\hat{\rho} ,
}
	\label{eq:CollectivePurcell}
\end{equation}
where $\Gamma_P$ is the standard Purcell rate, defined in \eqref{eq:purcell-rate}.  The yellow dot-dashed contour lines in \reffig{fig:Fig9_EntanglementRegimes_Coherent}~\textcolor{Maroon}{(b)} correspond to numerical calculations with the reduced effective model from \eqref{eq:CollectivePurcell}. These contours provide a clear indication of the regimes of validity of that equation, which effectively describes the system when the cavity cannot resolve any transitions occurring within the emitters ($\kappa \gg R, \Omega$).
\graffito{
	$^*$Note that in \eqref{eq:LMEemittersDiagonalized}, we already presented the master equation for the emitter-emitter system in the general excitonic basis. In this case, we have included the cavity-induced Purcell rate $\Gamma_P$, which only affects the symmetric channel.}
Writing the master equation from \eqref{eq:CollectivePurcell} in the completely symmetric and antisymmetric basis$^{\textcolor{Maroon}{*}}$
 $\{|gg\rangle, |S\rangle,|A\rangle,|ee\rangle\}$, 
\begin{multline}
	\frac{d \hat \rho}{dt}=-i[\hat H ,\hat \rho]
	+ \frac{\Gamma_S}{2} \left(\mathcal{D}[|gg\rangle \langle S |]
	\hat \rho+\mathcal{D}[|S \rangle \langle ee |]
	\hat \rho\right)\\
	+\frac{\Gamma_A}{2} \left(\mathcal{D}[|gg\rangle \langle A |]
	\hat \rho+\mathcal{D}[|A \rangle \langle ee |]
	\hat \rho\right),
	\label{eq:CollectivePurcell_Exc}
\end{multline}
we find that the spontaneous decay in the system can be expressed in terms of a decay through:
\begin{enumerate}[label=\textcolor{Maroon}{(\roman*)}]
	\item \textcolor{Maroon}{Antisymmetric channel:} 
	\begin{equation}
|ee\rangle \rightarrow |A\rangle \rightarrow |gg\rangle \quad  \text{with a rate} \quad \Gamma_A\equiv \gamma -\gamma_{12},
	\end{equation}
	\item  \textcolor{Maroon}{Symmetric channel:} 
		\begin{equation}
		|ee\rangle \rightarrow |S\rangle \rightarrow |gg\rangle \quad  \text{with a rate} \quad	\Gamma_S\equiv \gamma +\gamma_{12}+2\Gamma_P.
			\label{eq:Gamma_S}
	\end{equation}
\end{enumerate}
Crucially, the symmetric channel, is enhanced by the Purcell rate $\Gamma_P$ brought in by the cavity. In this basis, the Hamiltonian---with the driving at the two-photon resonance---is written in the rotating frame of the drive as 
\begin{multline}
\hat H = \sqrt 2\Omega(|gg\rangle \langle S| + |S\rangle \langle gg|+|S\rangle \langle ee| + |ee\rangle \langle S| )\\
 - \delta (|S\rangle \langle A| + |A\rangle \langle S|).
\end{multline}
The resulting scheme of energy levels with the interactions and incoherent processes associated to them are depicted in \reffig{fig:Fig_EnergyDiagramPurcell}~\textcolor{Maroon}{(a)}.

To understand this mechanism, it is useful to diagonalize the driving term in the Hamiltonian. In the completely symmetric and antisymmetric basis, the coherent drive couples the states $|gg\rangle$, $|ee\rangle$, and $|S\rangle$, while $|A\rangle$ becomes coherently decoupled. 
The resulting eigenstates can be expressed in terms of the two-photon dressed-states---$|S_2/A_2\rangle=1/\sqrt{2}(|gg\rangle \pm | ee \rangle)$---as
\begin{equation}
|\phi_\pm\rangle = \frac{1}{\sqrt{2}}(|S\rangle \pm |S_2\rangle),\quad  \text{and} \quad |\phi_0\rangle=|A_2\rangle.
\end{equation}
with eigenenergies 
\begin{equation}
\lambda_\pm=\pm 2\Omega, \quad \text{and}\quad  \lambda_0=0.
\end{equation}

The resulting energy level structure in the diagonal basis is depicted in \reffig{fig:Fig_EnergyDiagramPurcell}~\textcolor{Maroon}{(b)}. The present description of the system features several characteristics that, combined, lead to the stabilization of state $|A\rangle$:
\begin{enumerate}[label=\textcolor{Maroon}{(\roman*)}]
	\item $|A\rangle$ is coherently coupled with a rate $\delta/\sqrt 2$ to the states $|\phi_\pm\rangle$, which are detuned from $|A\rangle$ by an energy difference of $\pm 2\Omega$ due to the dressing by the drive. In the regime where $\Omega\gg \delta$, this coupling is substantially off-resonant, and $|A\rangle$ only affects $|\phi_\pm\rangle$ perturbatively, acting as a virtual state~\cite{Cohen-TannoudjiAtomPhotonInteractions1998}. 
	\item Losses in the system predominantly induce transitions within the manifold $\{ |\phi_\pm\rangle, |\phi_0\rangle \}$, characterized by rates given by $\Gamma_S$, which are significantly larger than the characteristic rate $\Gamma_A$ associated with the incoherent coupling between this manifold and the state $|A\rangle$. Note that in the basis  $\{|gg\rangle, |S\rangle,|A\rangle,|ee\rangle\}$, there is not collective incoherent interaction since $\gamma_C=0$ [see \eqref{eq:ExcitonicParameters2}].
\end{enumerate}

In \refch{chapter:Unconventional}, we demonstrated  that a state satisfying these two conditions---\textcolor{Maroon}{(i)} being strongly off-resonant and \textcolor{Maroon}{(ii)} having negligible dissipation---will develop a sizable population in the long-time limit. We termed this effect \textit{``unconventional population of virtual states''}, since it arguably challenges the common intuition that a virtual state---which is both out of resonance from the states involved in the dynamics and uncoupled to any incoherent channel---would remain unpopulated at all times. In the current system, this is the effect behind the formation of entanglement via the Mechanism II.
In \refch{chapter:Unconventional}, we also showed that the virtual state slowly accumulates population through the non-Hermitian evolution between quantum jumps, due to the fact that, the longer a period without quantum jumps, the more likely it is that the state is to be found in $|A\rangle$. This information provided by the absence of jumps reflects back on the state, and this effect accumulates over time eventually leading to a sizeable population of the ``virtual'' state.

\begin{SCfigure}[1][t!]
	\includegraphics[width=1.0\textwidth]{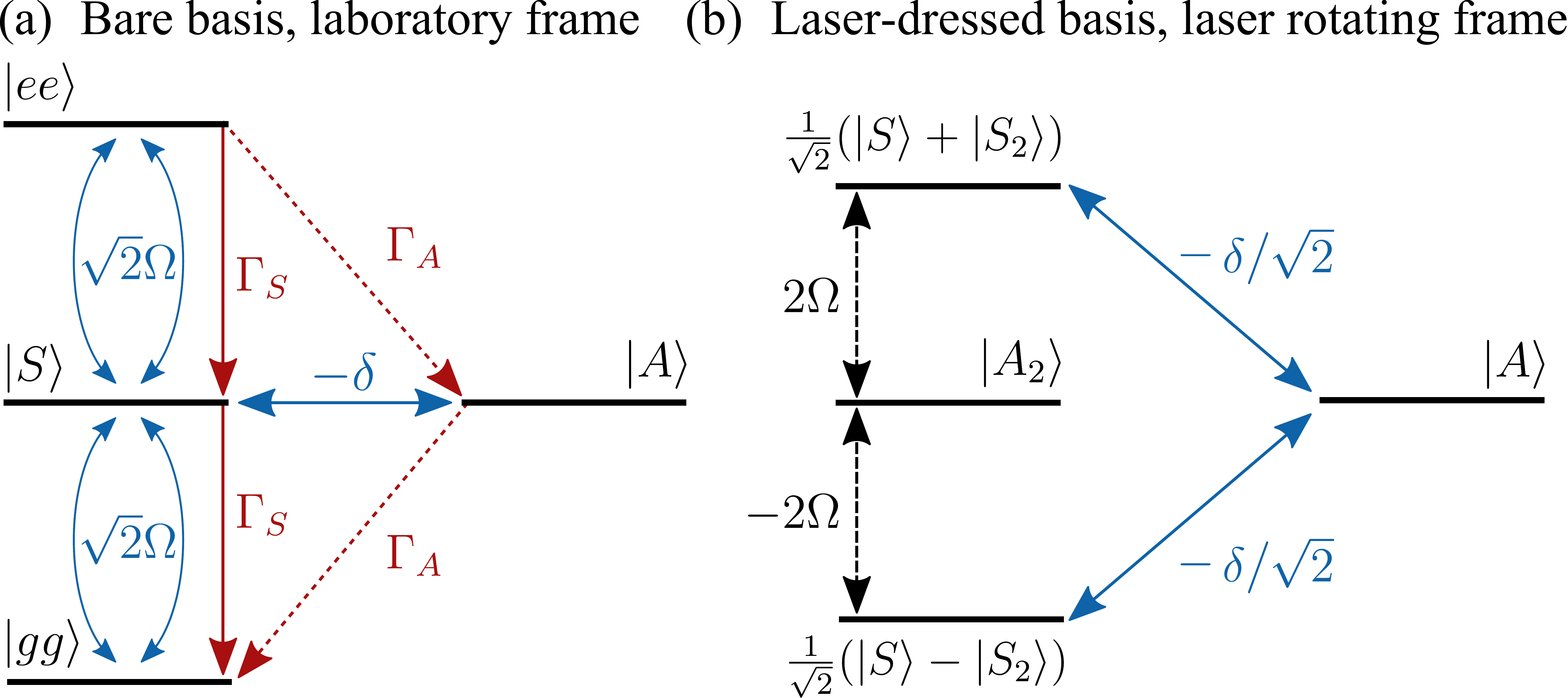}
\end{SCfigure}
\graffito{\vspace{-13.2cm}
	\captionof{figure}[Schematic representation of Mechanism II.]{	\label{fig:Fig_EnergyDiagramPurcell}
		\textbf{Schematic representation of Mechanism II. }
		(a) Scheme of the qubit-laser system in the regime where Mechanism II is activated ($J\ll \delta$). Blue and red lines correspond to coherent and incoherent interactions, respectively. 
		The antisymmetric state is only coupled coherently via the qubit-qubit detuning $\delta$ and dissipatively connected with reduced decay rate $\Gamma_A\ll \Gamma_S$. (b) In the basis dressed by the laser, it becomes clear that $|A\rangle$ is detuned from the states to which it is coupled by an energy difference $2\Omega$, contributing only perturbatively to the Hamiltonian dynamics of those states as a virtual state, provided that $\Omega \gg \delta$. Incoherent processes are no longer shown. 
	}
}

Under these conditions, the system becomes metastable and develops two characteristic, dissipative timescales: a fast one, given by $\Gamma_S^{-1}$, and a slow one, given by the inverse of an effective rate $\Gamma_\mathrm{eff}^{-1}$. 
This hierarchy of timescales allows us to apply the hierarchical adiabatic elimination (HAE) method introduced in the previous Chapter---see \refsec{sec:3}---,  based on the successive application of a series of adiabatic elimination procedures. The application of our method allows us to obtain the expression of the relaxation rate towards the antisymmetric state, given by---see \refap{appendix:D} for details---
\begin{equation}
\colorboxed{Maroon}{
	\Gamma_{\text{eff}}=\frac{4\Gamma_S \delta^2}{\Gamma_S^2+24\Omega^2},
}
	\label{eq:Gamma_eff}
\end{equation}
which was obtained under the assumptions $\Omega \gg \delta \gg J$ and $\Gamma_A \approx 0$. The corresponding population of the state $|A\rangle$ that develops under these conditions is given by---see \refap{appendix:D}---
\begin{equation}
\colorboxed{Maroon}{
	\rho _{A,\mathrm{ss}}=\frac{2\Omega^2}{\delta^2+2\Omega^2}.
}
\end{equation}

For this mechanism to be efficient, we need the timescale of formation of the entanglement to be much faster than the decay of the state $|A\rangle$; in other words, we require $\Gamma_\mathrm{eff}\gg \Gamma_A $, which in the limit $\Omega \gg \Gamma_S$ can be rewritten as 
\begin{equation}
\colorboxed{Maroon}{
	\Gamma_S \gg 6\Gamma_A \left(\frac{\Omega}{\delta}\right)^2.
}
	\label{eq:conditionMechanismII}
\end{equation}
The essential role of the cavity in this mechanism is precisely to enable this condition; since given that we already required that $\Omega/\delta\gg 1$, it is unlikely that the condition is met without a significant Purcell enhancement of $\Gamma_S$ given by $\Gamma_P$ in \eqref{eq:Gamma_S}.

\subsection{Summary of conditions}
These analytical results allow us to unambiguously establish the regimes of parameters in which Mechanism II is efficient. To summarize, we require 
\begin{enumerate}[label=\textcolor{Maroon}{(\roman*)}]
\item The cavity linewidth must be the largest energy scale in the system, disabling the possibility of resolving any energy transition occurring within the emitters. This translates into
\begin{equation} 
\kappa \gg R\approx \delta,  \quad \text{and} \quad  \kappa \gg \Omega,
\end{equation}
for the cavity to provide a collective decay.
\item We must \textit{virtualize} the antisymmetric state $|A\rangle$  such that it is  almost decoupled from the rest of the system. This requires
\begin{equation}
\Omega \gg \delta
\end{equation}
This means that the strongest condition for the cavity decay rate is simply $\kappa \gg \Omega$.
\item The timescale of formation of  the entanglement must be faster than the decay of the state $|A\rangle$, imposing the condition
\begin{equation}
	\Gamma_\mathrm{eff}\gg \Gamma_A.
\end{equation}
Inspecting \eqref{eq:Gamma_eff}, we see that this will impose both a lower and an upper limit to $\Gamma_S$ and, consequently, to the cavity Purcell rate $\Gamma_P$, since for values of the cooperativity $C>1$, we have $\Gamma_S\approx 2\Gamma_P$. Given a fixed $\Gamma_S$, the condition $\Gamma_\mathrm{eff}\gg \Gamma_A$ also poses lower and upper limits to $\delta$ and $\Omega$, respectively. In particular, it is important to notice that this condition makes it necessary to have a finite detuning between the QEs, since $\delta \approx 0$ would imply $\Gamma_\mathrm{eff}\approx 0$, making the timescale of formation of entanglement divergently long.
\end{enumerate}
All these three main conditions are depicted as lines in the phase diagram in the right panel of \reffig{fig:Fig9_EntanglementRegimes_Coherent}~\textcolor{Maroon}{(b)}. These lines effectively outline the frontier of the region where Mechanism II is observed, supporting our analytical results.

\subsection{Optical signatures of entanglement}

In contrast to the optical signatures of entanglement reported for the Mechanism I in \refsec{sec:observation}, in the regime of Mechanism II, the same set of steady-state measurements shown in \reffig{fig:Fig10_Observability} reveals no features clearly correlated with the formation of entanglement. The absence of signatures is shown in \reffig{fig:FigAppendixE_Observability_CollectivePurcell}, where we depict---from top to bottom---the stationary concurrence, the mean intensity, and the second order correlations. 
We observe that, as $\Omega$ is increased, the concurrence increases from zero, reaching a maximum value and eventually vanishing. Contrary to the case of Mechanism I, these features are not correlated with any of the observables shown here, and therefore we conclude that, by themselves, they do not serve as any type of indicator of the stabilization of entanglement.
	\addtocounter{figure}{-1}
 \begin{SCfigure}[0.67][h!]
	\includegraphics[width=0.6\textwidth]{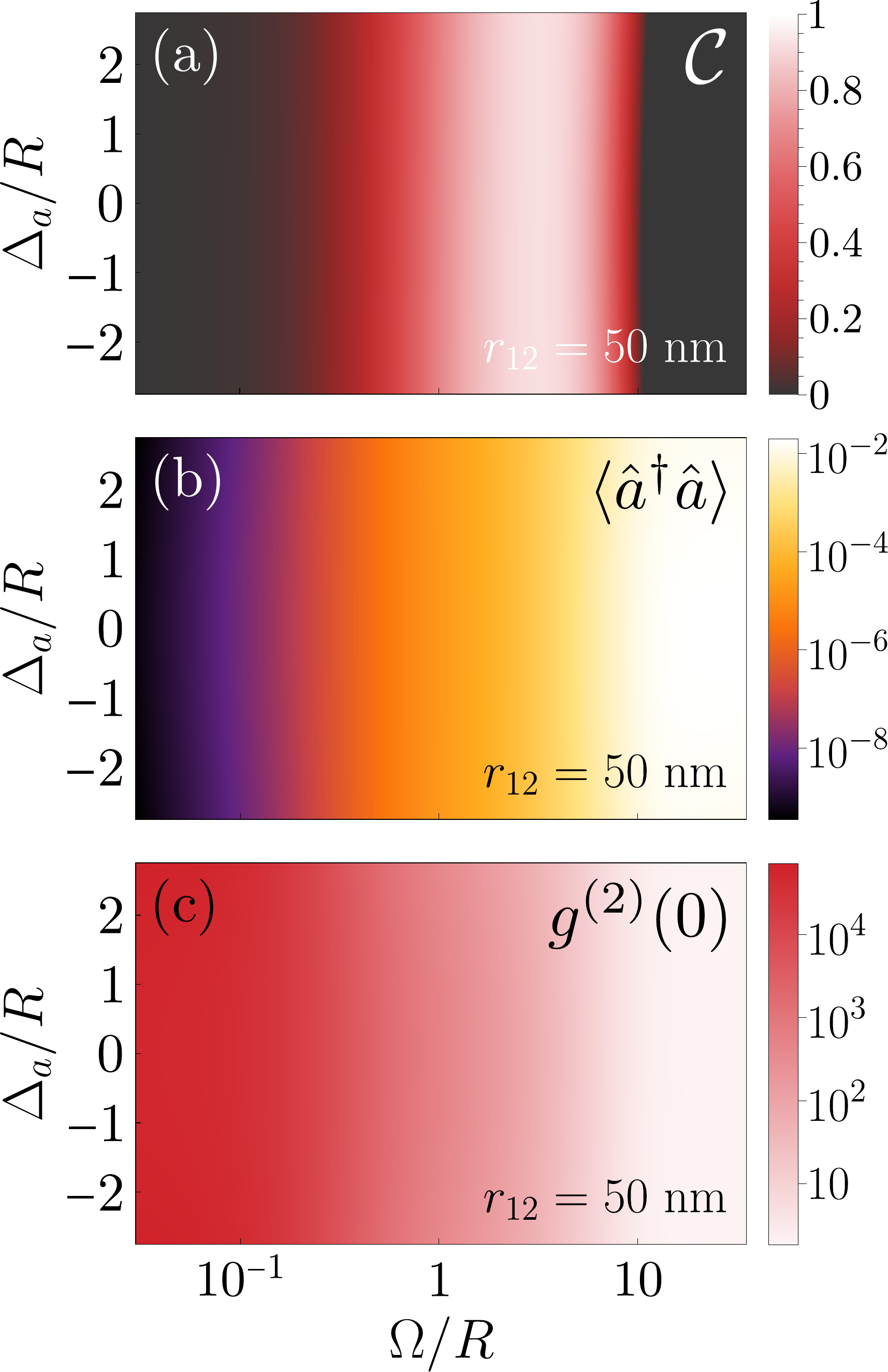}
	\captionsetup{justification=justified}
	\caption[Entanglement detection and control for Mechanism II.]{
	\label{fig:FigAppendixE_Observability_CollectivePurcell}
	\textbf{Entanglement detection and control for Mechanism II.}
	Stationary concurrence and optical properties of the emitted light by the cavity.
	(a), (b), (c) Steady-state concurrence, transmission intensity and photon correlations versus the Rabi frequency of the laser $\Omega$ and the laser-cavity detuning $\Delta_a$. Parameters:  $r=50$ nm, $k=2\pi/780\ \text{nm}^{-1}$, $J=10.65\gamma$, $\gamma_{12}=0.967\gamma$,  $\Delta=0$, $\delta=10^2 J$, $R=10^3\gamma$, $\kappa=10^4 \gamma$, $g =10^{-1} \kappa$.
	}
\end{SCfigure}

\paragraph{Ring-down measurement. }
An alternative approach to observe evidences of the stabilization of an entangled state is a \textit{ring-down measurement} of the output emission from the cavity. This time-resolved emission dynamics measurement~\cite{TiranovCollectiveSuper2023} would involve reaching the steady state under continuous excitation, then switching off the drive and recording the output emission as the system decays towards the ground state. This ring-down measurement also makes the output emission easier to interpret, since, once the laser drive is switched off,  the energy levels of the system and the resulting evolution are simpler than the dressed-state  dynamics of the continuously-driven system, as depicted in \reffig{fig:Fig_EnergyDiagramPurcell}. 
For the simulation of this measurement, we chose realistic parameters reported in state-of-the-art cavity-QED experiments with individual molecules, $\gamma/2\pi = 40$~MHz~\cite{ToninelliSingleOrganic2021}, $g/\pi = 1.54$~GHz~\cite{PschererSingleMoleculeVacuum2021} and $\Omega/2\pi = 370$~MHz~\cite{GerhardtCoherentState2009}, assuming molecules separated enough to consider $J=\gamma_{12}\approx 0$, but collectively coupled to the same cavity mode. We set the remaining parameters as $\delta/2\pi = 306$~MHz and $\kappa/2\pi = 2.32$~GHz, which are  the values that maximize the entanglement with the rest of the parameters fixed. This value of $\kappa$, which yields a cooperativity $C = 25$, is also realistic, being merely twice the value reported in \colorref{GerhardtCoherentState2009}, as required to fulfil the condition $\kappa\gg \Omega$. The value of the concurrence obtained for this set of parameters is $\mathcal C \approx 0.51$.

 Our simulation of the proposed measurement of the emission intensity dynamics is shown in \reffig{fig:Fig11-Ringdown}~\textcolor{Maroon}{(a)} as the time-dependent cavity population versus the laser-qubit detuning $\Delta$. A main feature that correlates with the formation of entanglement is that, in the steady state ($t=0$), the stabilization of the entangled antisymmetric state $|A\rangle$ at $\Delta=0$ manifests as a dip in the cavity population. 
 More importantly, the population of $|A\rangle$---which is not directly coupled to the cavity---decays at a characteristic rate $\Gamma_A^\mathrm{eff}$,  much slower from the rest of timescales given its subradiant nature. We can obtain an effective description of this rate by performing a Hamiltonian adiabatic elimination where the Hilbert space is constrained within the one-excitation manifold $\{ |A,0\rangle, |S,0\rangle, |gg,1\rangle \}$,  and assuming $\gamma_{12}=0$, so that the non-Hermitian Hamiltonian reads
 \begin{equation}
 	\tilde{ H}= \left( \begin{array}{c|c}
 		E_0 & \hat V  \\ \hline
 		\hat V^\dagger&  \hat h
 	\end{array} \right)=
 	\left( \begin{array}{c|cc}
 		0 & -\delta & 0 \\ \hline
 		-\delta & 0 & \sqrt{2}g \\ 
 		0 & \sqrt{2}g &  - i\kappa/2
 	\end{array} \right).
 	\label{eq:NonHermitianHamiltonian_Cavity}
 \end{equation}
Then, following the Hamiltonian perturbative expansion method outlined in \refsec{Section:QuantumTraject}, we derive the effective cavity-induced dissipative rate for the antisymmetric state,
 \begin{equation}
 \colorboxed{Maroon}{
 	\Gamma_A^\mathrm{eff}\approx -2\text{Im}[\hat V (E_0-\hat h)^{-1}\hat V^\dagger]=\gamma + \frac{4\delta^2}{\Gamma_S},
 }
 \end{equation}
 where the second term, which can be comparable or larger than $\gamma$, describes the effective decay of $|A\rangle$ through its coupling to the symmetric state $|S\rangle$ with coupling rate $\delta$. 
 In the regime of high cooperativity $C\gg 1$, the decay of the symmetric state is dominated by the cavity Purcell rate $\Gamma_S \approx 2\Gamma_P$, and therefore $\Gamma_A^\mathrm{eff} \ll \Gamma_S$. This separation of timescales provides unambiguous information about the respective occupation of $|A\rangle$ and $|S\rangle$ via this time-resolved measurement. At $\Delta=0$, the high occupation of the $|A\rangle$ state is slowly released through the cavity in a timescale $\sim 1/\Gamma_A^\mathrm{eff}$ yielding a characteristic bump in the emission intensity at long times, as seen in the solid curve in the top panel of \reffig{fig:Fig11-Ringdown}~\textcolor{Maroon}{(a)}. In contrast, away from the resonance $\Delta=0$, all the emission is released in the much shorter timescale $\sim 1/\Gamma_S$, as we show via the blue dashed curve in the top panel of \reffig{fig:Fig11-Ringdown}~\textcolor{Maroon}{(a)}.

{ 
	\sidecaptionvpos{figure}{t}
  \begin{SCfigure}[0.55][t!]
 	\includegraphics[width=0.68\textwidth]{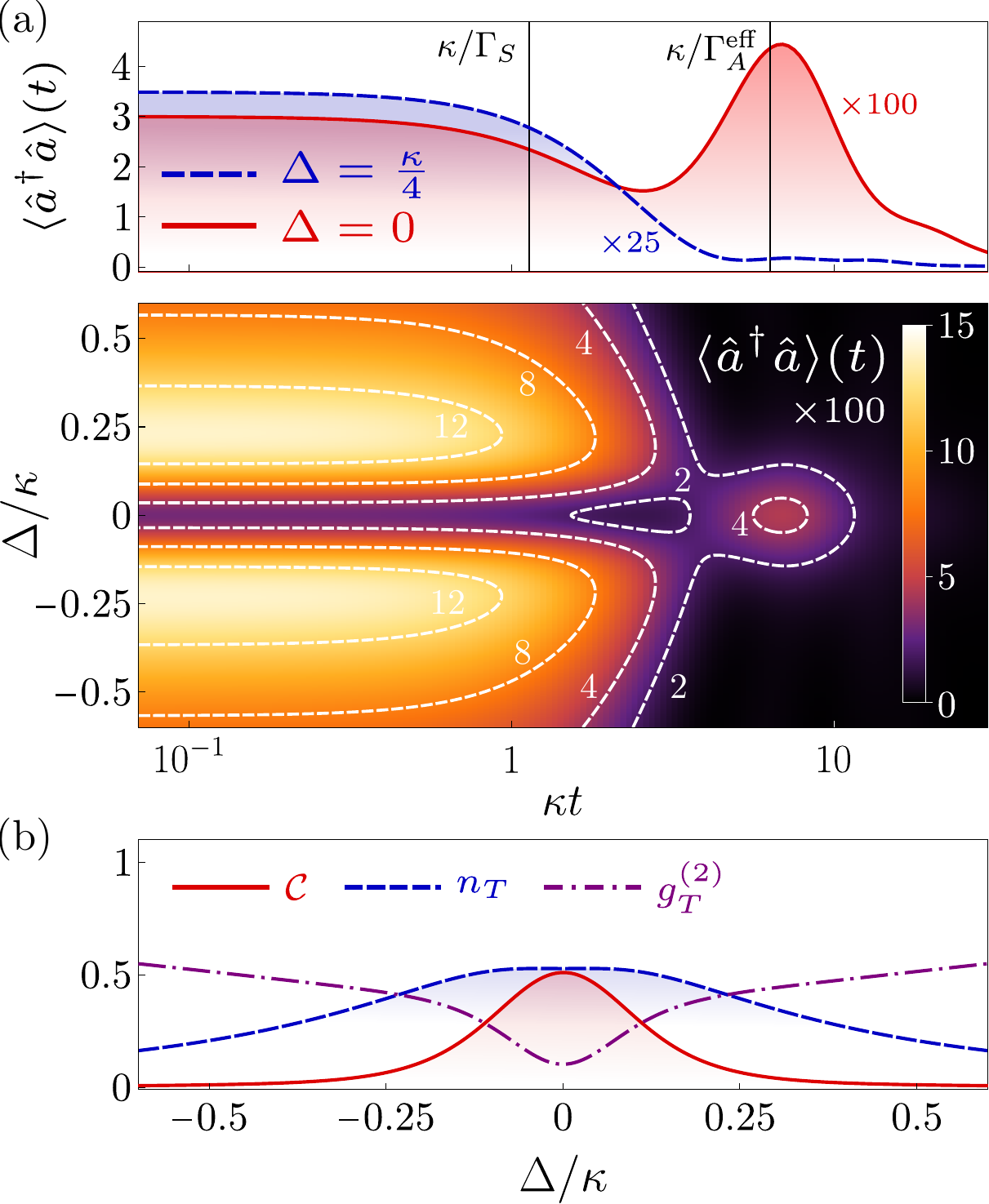}
 	\captionsetup{justification=justified}
 	\caption[Ring-down measurement for the detection of entanglement via Mechanism II.]{\textbf{Ring-down measurement for the detection of entanglement via Mechanism II.}  
 		(a) Time evolution of the emission intensity $\langle \hat a^\dagger \hat a\rangle(t)$ versus the laser-qubit detuning $\Delta$ once the laser drive is switched off. Top panel corresponds to a cut of the emission intensity at $\Delta=0$ (red dashed line) and $\Delta=\kappa/4$ (blue solid line).
 		(b)
 		Steady-state concurrence ($\mathcal{C}$), integrated photon emission ($n_T$) and integrated second-order correlation function ($g_T^{(2)}$) in terms of the laser-qubit detuning $\Delta$.
 		Parameters: $J=\gamma_{12}\approx 0$, $\delta/2\pi=306$ MHz, $\Omega/2\pi=370$ MHz, $\gamma/2\pi=40$ MHz,  $\kappa/2\pi=2.32$ GHz, $g/\pi=1.54$ GHz. }
 	\label{fig:Fig11-Ringdown}
 \end{SCfigure}}

 When the decay of the antisymmetric state dominantly occurs through the cavity channel, the total number of photons emitted from the cavity gives valuable quantitative information about the population of $|A\rangle$, which in turn, is strongly correlated with the value of the concurrence $\mathcal C$. This is shown in \reffig{fig:Fig11-Ringdown}~\textcolor{Maroon}{(b)}, where we plot both the steady-state concurrence and the total number of photons emitted during the ring-down process, given by the integral 
 \begin{equation}
n_T\equiv \kappa \int_0^\infty dt  \langle \hat a^\dagger \hat a\rangle (t) =-\sum_{\mu=0}^\infty \frac{\kappa}{\Lambda_\mu} \text{Tr}[\hat \Lambda_\mu^L \hat \rho(0)] \text{Tr}[\hat a^\dagger \hat a \hat \Lambda_\mu^R].
\label{eq:integrated_number}
 \end{equation}
We observe that, 
\graffito{
	Note that the analytical expressions for $n_T$ and $g_T^{(2)}$ have been derived using the spectral decomposition of the emitters-cavity system---see \refsec{sec:LiouvProp}---, where $\{\Lambda_\mu, \mu =1, \ldots,\infty\}$ are the Liouvillian eigenvalues, each associated with their corresponding left- and right-Liouvillian eigenvectors $\hat \Lambda_\mu^{L/R}$. The initial state is set to the steady state of the system prior to disabling the coherent driving, $\rho(0)= \rho_{\text{ss}}[\Omega\neq 0]$.
}
in the resonant condition of maximum entanglement at $\Delta=0$, both $\mathcal C$ (red solid curve) and the total number of photons emitted (blue dashed curve) have an almost perfect match. Another strong indication of entanglement is the second-order correlation function of the integrated emission$^{\textcolor{Maroon}{*}}$, 
 \begin{multline}
 	g_T^{(2)} = \left(\frac{\kappa}{n_T}\right)^2\int_0^\infty dt dt' \langle : \mathcal{T}[\hat a^\dagger(t)\hat a^\dagger(t')\hat a(t')\hat a(t)]:\rangle \\
 	= \left(\frac{\kappa}{n_T}\right)^2 \sum_{\mu,\nu=0}^\infty \frac{\Lambda_\mu- \Lambda_\nu}{\Lambda_\mu^2 \Lambda_\nu-\Lambda_\mu \Lambda_\nu^2} \text{Tr}[\hat \Lambda_\mu^L \hat \rho(0)] \text{Tr}[\hat \Lambda_\nu^L (\hat a \hat \Lambda_\mu^R \hat a^\dagger)] \text{Tr}[\hat a^\dagger \hat a \hat \Lambda_\nu^R].	
 	\label{eq:integrated_g2}
 \end{multline}
 The second-order correlation function of the integrated emission, $g_T^{(2)}$, shows a stronger degree of antibunching at the entanglement resonance, as shown in \reffig{fig:Fig11-Ringdown}~\textcolor{Maroon}{(b)}. This occurs because the stabilization of the state $|A\rangle$ leads to a zero occupation of the state $|ee\rangle$. In a scenario in which both emitters are populated, this situation is impossible unless the state is non-separable, i.e., entangled.

 The situation in which the decay of $|A\rangle$ takes place mainly through the cavity occurs when $4\delta^2/\Gamma_S\gg \gamma$. Since we require in general $C\gg 1$, meaning that $\Gamma_S \approx 2\Gamma_P$, this condition can be expressed as a requirement for the detuning between emitters given by
 \begin{eqnarray}
\colorboxed{Maroon}{
 	2(\delta/\gamma)^2 \gg C \gg 1.
}
 \end{eqnarray}

\subsection{Effect of additional decoherence channels}

Similarly to the analysis in \refsec{sec:decoherence} for Mechanism I, we now examine the effects of additional decoherence on Mechanism II by considering extra spontaneous decay, local and collective pure dephasing [see \eqref{eq:Additional_Decoh}]. 
In \reffig{fig:Fig11_EntanglementProtection_MechII}, we show the concurrence versus the cooperativity and the additional decoherence rates. These results confirm that the general conditions stated in \cref{eq:decoh_cond1,eq:decoh_cond2}, 
\begin{equation}
		\Gamma_P \gg \gamma_{\text{d}}\quad  \text{and}\quad 	\Gamma_{\text{eff}}^{-1} \ll \gamma_{\text{d}}^{-1} \quad \text{with}\quad \gamma_{\text{d}}=\{ \Gamma,\gamma_\phi, \Gamma_\phi \},
\end{equation}
clearly delimit the regime in which entanglement is generated, showing that $\Gamma_{\text{eff}}$ [\eqref{eq:Gamma_eff}] can be large enough to make the concurrence robust even for large values of extra decoherence rates, e.g., $\Gamma \sim 10^2\gamma$ for $C\sim 10^4$.
 \begin{SCfigure}[0.75][h!]
	\includegraphics[width=0.55\textwidth]{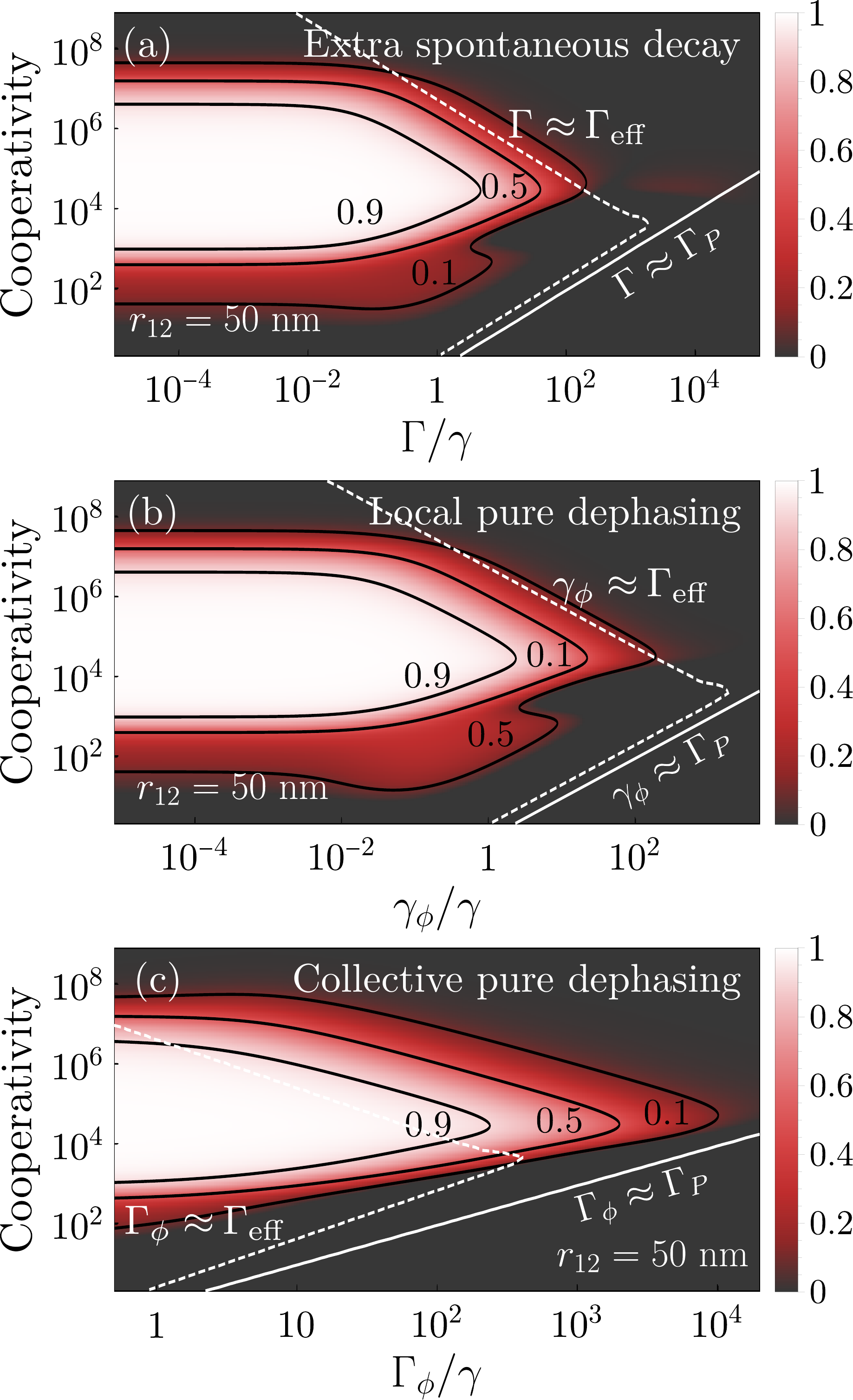}
	\captionsetup{justification=justified}
	\caption[Robustness test of mechanism II]{
		\label{fig:Fig11_EntanglementProtection_MechII}
			\textbf{Robustness test of mechanism II. }
The considered additional channels are: (a) extra spontaneous decay; (b) local pure dephasing; and (c) collective pure dephasing. Each panel depicts the concurrence as a function of the cavity cooperativity $C$ and the corresponding extra decoherence rate.
	Parameters: 
	$J=10.65\gamma,\ \gamma_{12}=0.967\gamma,\ \delta=10^{3}\gamma,\ \ g=10^{-1}\kappa,\ \Delta_a\approx R,\ \Delta=0,\ \Omega=10^4 \gamma$.
	}
\end{SCfigure}

\subsection{Application in a waveguide QED setup }
We have demonstrated that the effective model of collective decay, as described by \eqref{eq:CollectivePurcell}, successfully captures the essence of Mechanism II. 
This model not only provides an accurate representation of QEs coupled to a common cavity, but it is also applicable to other structures, such as waveguides, which also lead to collective decay and Purcell enhancement phenomena~\cite{Gonzalez-TudelaEntanglementTwo2011,TiranovCollectiveSuper2023,SheremetWaveguideQuantum2023}. 
Indeed, waveguide QED is the context in which this mechanism was originally reported~\cite{PichlerQuantumOptics2015}. Therefore, we can conclude that the scope of our results regarding Mechanism II extend beyond cavity-based configurations.

\begin{SCfigure}[0.85][b!]
	\includegraphics[width=0.65\textwidth]{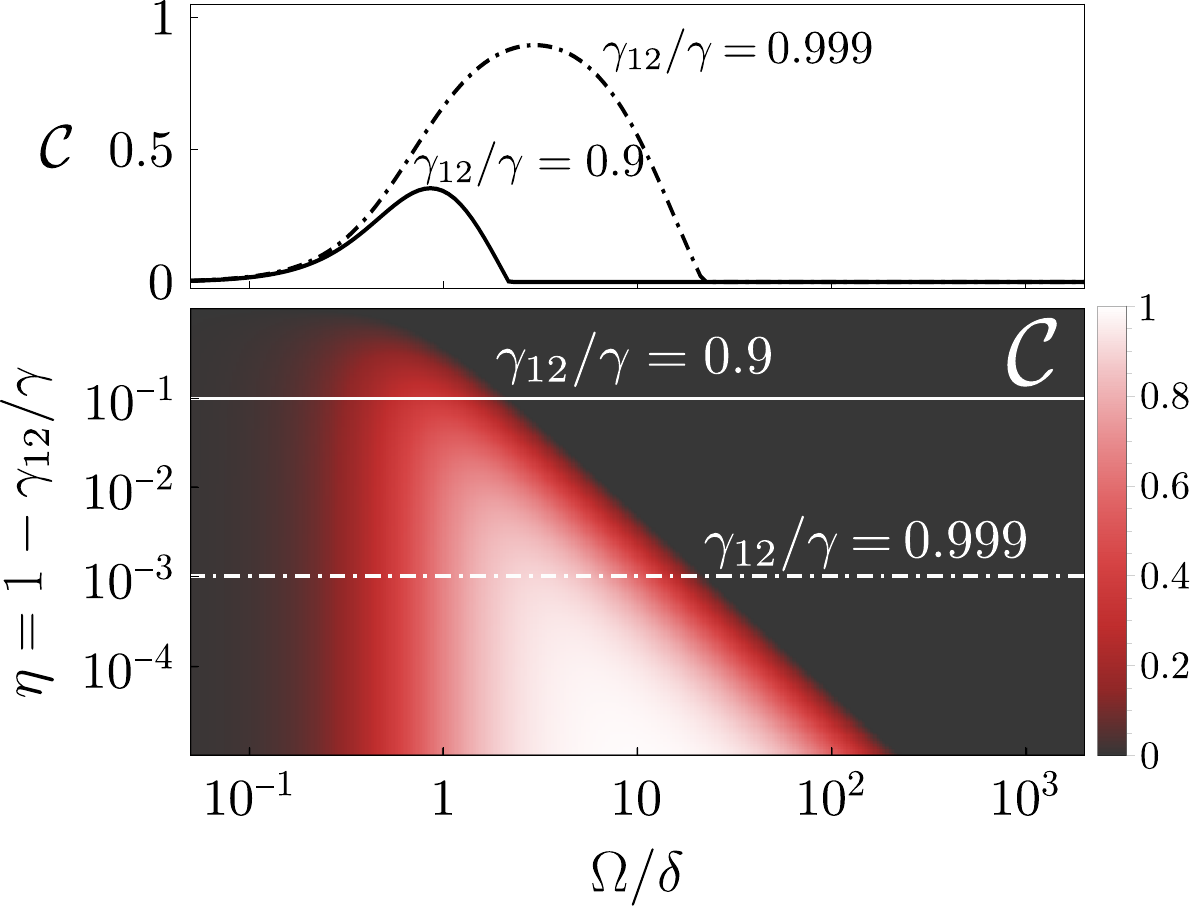}
	\captionsetup{justification=justified}
	\caption[Generation of entanglement via Mechanism II in a waveguide QED setup.]{ 
		\textbf{Generation of entanglement via Mechanism II in a waveguide QED setup.}
		Stationary concurrence for a general system of two non-identical emitters with a collective decay parametrized by $\gamma_{12}$.  The dipole-dipole coupling is disabled in this configuration and thus any dependence on the emitter-emitter distance $r_{12}$.
		Top panel corresponds to a cut of the concurrence when $\gamma_{12}=0.9\gamma$ (solid line) and $\gamma_{12}=0.999\gamma$ (dot-dashed line) versus the driving strength.
		Parameters: $J = 0$, $\Delta=0$, $\delta = 10^2\gamma$.}
	\label{fig:Fig_Waveguide}
\end{SCfigure}

To explore further the possibilities of this mechanism in general platforms, we consider a simpler parametrization of the model in which $\Gamma_A = \gamma-\gamma_{12}$ and $\Gamma_S = \gamma+\gamma_{12}$, so that the ratio between $\Gamma_S$ and $\Gamma_A$ is purely controlled by $\gamma_{12}$, which can encode the dissipative coupling enabled by a structure, such as a waveguide. 
For instance, \colorref{Gonzalez-TudelaEntanglementTwo2011} establishes the relation 
\begin{equation}
	\gamma_{12}/\gamma = \tilde\beta \exp[-d/(2L)],
\end{equation}
where $\tilde\beta$ is the standard factor that measures the fraction of the emitted radiation captured by the waveguide, $d$ is the distance between emitters, and $L$ is the propagation length of the waveguide mode. 
 We notice that values of $\tilde\beta > 0.9$ are within reach in state-of-the-art platforms~\cite{Gonzalez-TudelaEntanglementTwo2011,ArcariNearUnityCoupling2014}.
In \reffig{fig:Fig_Waveguide} we compute the concurrence using this parametrization, versus  $\Omega$ and $\gamma_{12}$, for detuning between emitters set at $\delta = 10^2\gamma$, and ignoring any coherent coupling between qubits by setting $J=0$. 
These results shows that a value $\gamma_{12} = 0.9\gamma$---which would be achieved provided $\tilde\beta=0.9$ and $d\ll L$---can stabilize states with a significant stationary concurrence, reaching maximum values $\mathcal C_\mathrm{max}\approx 0.35$. Notice that this is comparable to the results of stationary concurrence reported in \colorref{Gonzalez-TudelaEntanglementTwo2011}, with the added advantage that the approach considered here enables a detuning between the emitters hundreds of times larger than their linewidths. This  holds relevant implications for stabilizing entangled states among spatially separated emitters featuring different emission energies, without the need of tuning them in resonance by external means such as induced Stark shifts, which poses significant challenges~\cite{KoongCoherenceCooperative2022}.

\section[Mechanism III. Two-photon resonance fluorescence]{Mechanism III. Two-photon resonance \protect\newline fluorescence}
\label{sec:MechanismIII}

\subsection{Identification of the mechanisms}

In this section, we present an alternative mechanism---referred to as Mechanism III---for stabilizing entanglement, which targets a different type of entangled state than the ones discussed so far. Specifically, this process generates states exhibiting a coherent superposition between the ground state $|gg\rangle$ and the doubly excited state $|ee\rangle$. 
We note that this process may be present in two variants, which we label as 
\begin{enumerate}[label=\textcolor{Maroon}{(\roman*)}]
	\item \textcolor{Maroon}{Mechanism III$_\mathrm{sp}$, } when  the main decay channel is spontaneous emission.
	\item \textcolor{Maroon}{Mechanism III$_\mathrm{cav}$, } when the main decay channel is  cavity-enhanced decay.
\end{enumerate}
Examples of entanglement achieved via Mechanism III can be observed in \reffig{fig:Fig4_ConcurrenceQubitCavityDetuning}~\textcolor{Maroon}{(a, c)}, where a finite degree of entanglement is observed at large qubit-qubit detuning $\delta \approx 10^2 J$. Moreover, this mechanism is also depicted in the diagrams of the right column of \reffig{fig:Fig9_EntanglementRegimes_Coherent}~\textcolor{Maroon}{(a)}. Mechanism III results in lower levels of entanglement and reduced purity compared to the practically pure states achieved through Mechanisms I and II.  
Regarding previous reports of this mechanism in the literature, we note that the observation of stationary entanglement between non-identical emitters enabled by a plasmonic structure reported in \colorref{HaakhSqueezedLight2015} is of a similar nature as Mechanism III. However, the specific nanophotonic environment considered in that work lead to a parametrization of the decay rates  in the effective master equation of the emitters slightly different from the one obtained in this work, preventing us from drawing a full parallelism between both situations. 

\subsection{Origin of the mechanisms}

Regarding the underlying physics, Mechanism III is a two-photon analogue of the formation of stationary coherences of the form $\langle \hat\sigma\rangle\neq 0$ in resonance fluorescence, i.e., in a single two-level system excited by a coherent drive~\cite{SanchezMunozSymmetriesConservation2019}. In this analogue case of a single two-level system, one would find that
\begin{equation}
\langle \hat\sigma\rangle =\frac{-2i \Omega\gamma}{\gamma^2+8\Omega^2} ,
\end{equation}
which is zero for both $\Omega \ll\gamma$ and $\Omega \gg \gamma$, but non negligible when $\Omega\sim\gamma$, reaching its maximum at $\Omega_\mathrm{max} = \gamma/(2\sqrt 2)$. In Mechanism III, a similar interplay between the coherent two-photon drive of the transition $|gg\rangle \leftrightarrow |ee\rangle$, characterized by the Rabi frequency $\Omega_\mathrm{2p}$, and losses, with a characteristic rate $\gamma$, 
can enable the formation of stationary coherences between $|gg\rangle$ and $|ee\rangle$ when $\Omega_\mathrm{2p}\approx \gamma$. The resulting state takes the form:
\begin{equation}
\colorboxed{Maroon}{
	\hat\rho_{\text{III}} \approx  (1-\epsilon)\hat\rho_\mathrm{mix} + \epsilon  |\psi_{\text{Bell}}\rangle \langle \psi_{\text{Bell}}|,
}
\end{equation}
where $\hat\rho_\mathrm{mix}$ is a mixed density matrix, and 
\begin{equation}
	|\psi_{\text{Bell}}\rangle \equiv \frac{1}{\sqrt{2}}(|gg\rangle+ i |ee\rangle).
	\label{eq:Bell_state}
\end{equation}

The first variant of this effect, Mechanism III$_\mathrm{sp}$, is solely enabled by the two-photon drive and spontaneous emission. On the other hand, the second variant, Mechanism III$_\mathrm{cav}$, requires stimulated, collective emission provided by the cavity.

\subsection{Mechanism III$_\text{sp}$  }

 This process is  inherent of the qubit-laser system and does not involve the cavity. It is, therefore, independent of the laser-cavity detuning $\Delta_a$. This mechanism is highlighted in yellow in the diagrams at the right columns of \reffig{fig:Fig9_EntanglementRegimes_Coherent}, and can also be observed at large detuning in \reffig{fig:Fig4_ConcurrenceQubitCavityDetuning}~\textcolor{Maroon}{(b)}.  
These results can be explained with a model that does not include the cavity; this simplification allows us to 
effectively describe the dynamics by a reduced system of just three energy levels, 
which yields analytical expressions for the density matrix elements and concurrence, as was shown in \refch{ch:TwoPhotonResonance}. These results are presented in detail in  \refap{appendix:C}, and they align well with our numerical calculations. Our analytical treatment allows us to obtain the exact expression of the optimum detuning that maximizes the concurrence
\begin{equation}
\colorboxed{Maroon}{
	\delta_{\text{max}}=\Omega\sqrt{2(1+\sqrt{5})J/\gamma}.
}
\end{equation}
This expression closely resembles the expression $\delta\approx \Omega\sqrt{2J/\gamma}$ that is directly derived from the condition $\Omega_\mathrm{2p}\approx \gamma$.  While we emphasize that this mechanism is not influenced by the cavity and thus remains mostly independent of the cavity detuning $\Delta_a$, it is important to note that the cavity can indeed interfere with it and degrade the mechanism in the case in which its frequency matches the antisymmetric resonance. This effect is illustrated in the blue curves representing the resonances at  $\Delta_a\approx \pm R$ in \reffig{fig:Fig4_ConcurrenceQubitCavityDetuning}~\textcolor{Maroon}{(a, c)}, which do not present entanglement at high detuning, $\delta \gtrsim  J$.

\subsection{Mechanism III$_\text{cav}$}   
Contrary to Mechanism III$_\mathrm{sp}$, this variant of Mechanism III, which is highlighted in blue in the diagrams at the right panels of \reffig{fig:Fig9_EntanglementRegimes_Coherent}~\textcolor{Maroon}{(a)}, is enabled by the cavity. The role of the cavity here is to enhance the spontaneous decay rates, thereby altering the conditions under which the interplay of two-photon pumping and losses can stabilize coherences between $|gg\rangle$ and $|ee\rangle$.

Providing a systematic and analytical description of the parameter regimes where Mechanism III$_\mathrm{cav}$ is activated is a complex task beyond the scope of this text, as this feature can appear in non-perturbative limits $\Omega\sim R$ where analytical expressions of the eigenstates are unavailable. Nevertheless, we can provide general guidelines based on numerical evidence. We find that a cavity-induced collective decay---like the one responsible for Mechanism II---is necessary, i.e., 
$
	\kappa \gg (\Omega, \delta, J).
$
This observation is supported by the yellow dot-dashed contours in \reffig{fig:Fig9_EntanglementRegimes_Coherent}, which show that this feature is approximately described by the master equation with collective decay [\eqref{eq:CollectivePurcell}]. 	
Also, when $\Omega \gtrsim J$, we find that this mechanism emerges when the cavity-enabled Purcell rate is larger than the drive amplitude, $\Gamma_P > \Omega$. However, it is important to note that once $R$ becomes perturbative with respect to $\Omega$, i.e., $\Omega \gg R$,  Mechanism III$_\mathrm{cav}$ is no longer enabled, with Mechanism II being activated instead.
Some regions in parameter space may fulfil simultaneously the conditions for Mechanism III$_\mathrm{sp}$ and III$_\mathrm{cav}$. In this case, the result is a competition between processes in which the overall formation of entanglement is inhibited.

Notably, Mechanism III$_\mathrm{cav}$, which occurs at high values of the cooperativity, is very robust to increases in the local decay rate of the emitters, as can be seen in the right panel in \reffig{fig:Fig11_EntanglementProtection}~\textcolor{Maroon}{(a)}. This is expected since the dissipative dynamics is completely dominated by the losses through the cavity channel, which is four orders of magnitude larger than the spontaneous decay rate of the emitters in the regions in which this mechanism is relevant. We notice that the extra local decay can also activate Mechanism III$_\mathrm{sp}$. This is observed as a feature that is independent of the cooperativity and that emerges when $\Gamma$ is large enough. This activation occurs since the extra decay can help fulfilling the condition $\Omega_\mathrm{2p} \approx \gamma$ necessary for Mechanism III to take place. 
\section[Mechanism IV. Incoherently driven collective Purcell \protect\newline Enhancement]{Mechanism IV. Incoherently driven  \protect\newline Collective Purcell Enhancement} 
\sectionmark{Mechanism IV. Incoherently driven Collective Purcell Enhancement}
\label{sec:MechanismIV}
\subsection{Identification of the mechanism}
In this section, we report the last \textit{stationary} mechanism of entanglement generation, which, apart from the previously discussed Mechanism I [see \refsec{sec:FreqResolvedMech}], is the only additional regime of high concurrence that emerges under incoherent excitation. This mechanism is highlighted in red in the bottom row of  \reffig{fig:Fig9_EntanglementRegimes_Incoherent}~\textcolor{Maroon}{(a)}.
Similarly to Mechanism II [see \refsec{sec:MechanismII}] and Mechanism III$_{\text{cav}}$ [see \refsec{sec:MechanismIII}], mechanism IV is activated when $\kappa \gg R$, enhancing the spontaneous decay rate through the symmetric channel while maintaining the subradiant nature of the antisymmetric state. However, in contrast to  Mechanism III$_{\text{cav}}$, where the emitters  are coherently driven at the two-photon resonance, the incoherent excitation here leads the QEs to stabilize in the long-time limit into a combination of the ground $|gg\rangle$ and the antisymmetric state $|A\rangle$,
\begin{equation}
\colorboxed{Maroon}{
	\hat \rho_{\text{IV}} \approx \frac{1}{2} (|gg\rangle \langle gg|+|A\rangle \langle A|).
}
\end{equation}
Consequently, due to the significant presence of the antisymmetric state, the concurrence reaches a stationary value of  $\mathcal{C} \approx 0.5$.

\subsection{Origin of the mechanism}

This mechanism originates from the contrast between the subradiant nature of the antisymmetric state and the Purcell-enhanced dissipative nature of the symmetric channel. As discussed in Mechanism I [see \refsec{sec:FreqResolvedMech}], incoherent excitation makes the antisymmetric resonance inefficient due to the depletion of the antisymmetric state by incoherent repumping. However, when the cavity cannot resolve any qubit-qubit transition ($\kappa\gg R$), it enhances the symmetric dissipative channel---instead of just enhancing the transition from $|ee\rangle$ towards $|S\rangle$ experienced in Mechanism I via \eqref{eq:SingleLindbladEff} [see \reffig{fig:Fig6_DiagramPurcellMechanisms}]---,
\begin{equation}
	|ee\rangle \rightarrow |S\rangle \rightarrow |gg\rangle, \quad \text{with rate}\quad \Gamma_S=\gamma+\gamma_{12}+2\Gamma_P,
\end{equation}
while the antisymmetric dissipative channel is suppressed,
\begin{equation}
	|ee\rangle \rightarrow |A\rangle \rightarrow |gg\rangle, \quad \text{with rate}\quad \Gamma_A= \gamma - \gamma_{12}.
\end{equation}
Then, in the limit of $\beta \ll 1$---almost completely subradiant state---, and for a pumping rate $P$ that is faster than the natural emitter lifetime but slower than the cavity-induced decay rate ($\Gamma_P \gg P \gg \gamma$),
the system predominantly occupies the states $|gg\rangle$ and $|A\rangle$, as the doubly-excited state $|ee\rangle$ and the symmetric state $|S\rangle$ rapidly decay due to the collective Purcell enhancement, and that the antisymmetric state essentially cannot decay.

We  analytically confirm this idea by considering the Collective Purcell master equation from \eqref{eq:CollectivePurcell}, and derive the stationary expression for the antisymmetric state (assuming $\beta \approx 0$ and $\gamma_{12}\approx 1$)
\begin{equation}
\colorboxed{Maroon}{
	\rho_{A,\mathrm{ss}}^{\mathrm{IV}}\approx \frac{2(\Gamma_P+\gamma)^2}{P^2+3P(\Gamma_P+\gamma)+4(\Gamma_P+\gamma)^2},
}
\end{equation}
where we assumed $\beta \approx 0$ and $\gamma_{12}\approx \gamma$. We trivially observe that in the limit $\Gamma_P \gg P$, this expression yields a population of $0.5$. We also note that this mechanism is disabled  when $\delta\approx J$, as shown in \reffig{fig:Fig9_EntanglementRegimes_Incoherent}~\textcolor{Maroon}{(a)}. At this detuning,  the subradiant nature of the antisymmetric state is lost, and the interaction with the symmetric state via the qubit-qubit detuning $\delta$ prevents population build-up in the antisymmetric state, resulting in zero concurrence when $\delta\gg J$.
\section[Mechanism V. Metastable two-photon entanglement]{Mechanism V. Metastable two-photon \protect \newline entanglement} 
\label{sec:MechanismV}
We conclude our picture of the entanglement generation processes in this system by describing a \textit{metastable} mechanism that arises when the cavity frequency is resonant with the direct two-photon dressed-state transitions, i.e., when
\begin{subequations}
	\begin{align}
		&\Delta_a=\omega_{23}=2\Omega_\mathrm{2\mathrm p}:\quad \ \ \  |A_2\rangle \rightarrow |S_2\rangle, \\
		&\Delta_a=\omega_{32}=-2\Omega_\mathrm{2\mathrm p}:\quad |S_2\rangle \rightarrow |A_2\rangle
	\end{align}
\end{subequations}
 In these resonant conditions, the cavity enhances said transitions and drives the system into a metastable entangled state where the final state of the transition, either $|A_2\rangle$ or $|S_2\rangle$, is highly populated. We remind the reader that these states correspond to coherent superpositions of the ground state and the doubly-excited state, $|S_2/A_2\rangle=1/\sqrt{2}(|gg\rangle \pm | ee \rangle)$.
For instance, when $\Delta_a = \omega_{23}$, a metastable state of the following form is generated
\begin{equation}
\colorboxed{Maroon}{
	\hat  \rho_{\text{V}} \approx (1-\varepsilon)|S\rangle\langle S|  +\varepsilon |S_2\rangle\langle S_2|.
}
\end{equation}
Alternatively, when 
$\Delta_a = \omega_{32}$, the metastable state takes the form
\begin{equation}
	\colorboxed{Maroon}{
	\hat  \rho_{\text{V}} \approx (1-\varepsilon)|S\rangle\langle S|  +\varepsilon |A_2\rangle\langle A_2|.
	}
\end{equation}
We stress that $\epsilon$ can reach sizeable magnitudes, as evidenced by the high values of the concurrence shown in \reffig{fig:Fig8_TwoPhotonEntanglement}, which for that choice of parameters corresponds to $\epsilon \approx 0.66$. We observe that the lifetime of this metastable state can reach very long values of $\Delta t\sim 10^6/\gamma$.

We must note the extreme value of the inter-emitter distance chosen in \reffig{fig:Fig8_TwoPhotonEntanglement}, $r_{12}=0.5\ \text{nm}$. As previously discussed in Mechanism I [see \ref{sec:FreqResolvedMech}], inter-emitter distances $r_{12}\ll \lambda_0$ would require a more detailed description of the emitter-emitter interaction beyond the point-dipole approximation, e.g., using an extended dipole moment~\cite{CzikklelyExtendedDipole1970}. Nevertheless, this high value is chosen to unambiguously show the effect.
\begin{SCfigure}[0.5][h!]
	\includegraphics[width=0.8\textwidth]{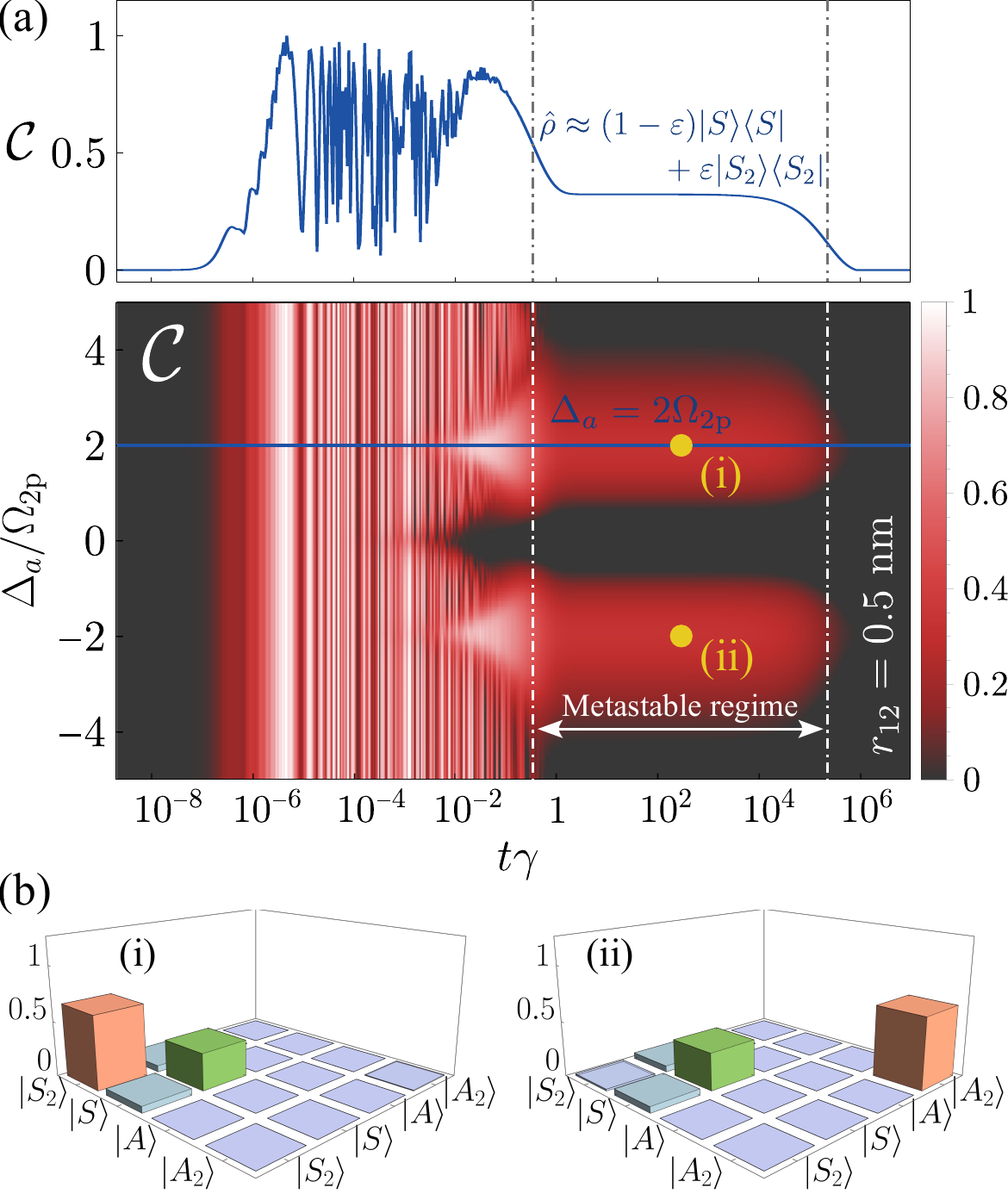}
	\captionsetup{justification=justified}	
	\caption[Generation of metastable entanglement via Mechanism V.]{
		\textbf{Generation of metastable entanglement via Mechanism V.}
		Time evolution of steady-state concurrence versus laser-cavity detuning $\Delta_a$. The system reaches a metastable entangled state when the cavity is in resonance with two-photon dressed energy transitions, i.e., $\Delta_a=\pm\omega_{23}=\pm 2\Omega_\mathrm{2p}$. Top panel corresponds to a cut of the concurrence when the cavity is in resonance with a dressed energy transition (blue curve). (b) Absolute value of the metastable density matrix corresponding to points (i) and (ii) in panel (a).
		Parameters: $r_{12}=0.5\ \text{nm}$, $k=2\pi/780\ \text{nm}^{-1}$, $J=1.15\times 10^7 \gamma$, $\gamma_{12}=0.999\gamma$, $\delta=10^{-4}J$, $\Delta=0$, $\Omega=10^6 \gamma$, $\Omega_{\mathrm{2p}}=1.74\times 10^5 \gamma$, $\kappa=10^5 \gamma$, $g=10^{-1}\kappa$. }
	\label{fig:Fig8_TwoPhotonEntanglement}
\end{SCfigure}

\newpage
\section{Conclusions}

In this Chapter, we have shown that a system of $N$ interacting quantum emitters embedded within a lossy cavity---operating in the bad-cavity limit---can generate high stationary entanglement among the emitters via driven-dissipative processes. 
Specifically, under coherent driving at the two-photon resonance, or incoherent excitation, and properly tuning the single-mode cavity, the system exhibits up to five distinct mechanisms.

The most significant contribution of this work is the identification and characterization of the first mechanism, referred to as the \textit{frequency-resolved Purcell enhancement}. This mechanism emerges when the emitters form a dimer structure, with the cavity selectively enhancing specific transitions within the emitters.  Crucially, this effect occurs when the cavity is in resonance with the symmetric or antisymmetric transitions that emerge from the interacting emitters, thereby approximately stabilizing the emitters into the corresponding entangled eigenstates, $|+\rangle$ and $|-\rangle$, respectively.
We have shown that this mechanism can operate under either coherent two-photon driving ($N=2$) or single-photon incoherent excitation ($N\geq 2$), extending its applicability to the case of $N$ all-to-all interacting emitters. 

A systematic exploration of the parameter phase space revealed that this effect belongs to a larger landscape of five mechanisms of entanglement generation: four leading to stationary entanglement and one yielding metastable entanglement.
In particular, we revisited Mechanism II, originally reported in~\cite{RamosQuantumSpin2014,PichlerQuantumOptics2015}, and provided deeper insights into how the antisymmetric state is stabilized. Using the hierarchical adiabatic elimination (HAE) method introduced in the previous Chapter, we derived analytical expressions for the steady-state populations and entanglement formation timescales.
For each of the five mechanisms, we provided a comprehensive description, including analytical insights---when possible---that define the regimes of parameters where these are activated and their associated timescales. 
This includes exploring their feasibility within state-of-the-art technological platforms, such as color centres~\cite{EvansPhotonmediatedInteractions2018,LukinTwoEmitterMultimode2023}, quantum dots~\cite{SomaschiNearoptimalSinglephoton2016,TommBrightFast2021} or molecules~\cite{GerhardtCoherentState2009,WangCoherentCoupling2017,WangTurningMolecule2019,PschererSingleMoleculeVacuum2021}. 
We also described how the formation of entanglement correlates with specific measurable properties of the light radiated by the cavity.

We have summarized the essential details of all mechanisms in \reffig{fig:Fig9_EntanglementRegimes_Coherent} and \reffig{fig:Fig9_EntanglementRegimes_Incoherent}, and  \textcolor{Maroon}{Table}~\ref{tab:EntanglementTab},  serving as a concise reference for the reader.
%


While this Chapter we has focused on entanglement generation---quantified via concurrence or fidelity to the 
$|W_N\rangle$ state---and its detection through emitted light properties, an intriguing open question remains: \textit{what is the structure of the density matrix of the emitted light, and how could it be captured?} 
In the next Chapter, we introduce a method to explore the fundamental properties of quantum states via the cascaded formalism  and explore two particular scenarios: the generation of entanglement between two photonics modes and quantum metrology using spectral measurements.

\cleardoublepage
\chapter[Frequency correlations for entanglement and \protect\newline quantum metrology]{Frequency correlations for entanglement and quantum metrology}
\chaptermark{Frequency correlations for entanglement and quantum metrology}
\label{chapter:Outlook}

To conclude this Thesis, we present in this Chapter our latest results on the generation of entanglement between photonic modes, and a brief perspective on our ongoing work on quantum metrology through spectral measurements.
This Chapter serves as an outlook, focusing on the fundamental properties of the quantum state---via its density matrix---of the output field emitted by a quantum emitter.
Specifically, using the cascaded formalism~\cite{GardinerQuantumNoise2004}---already introduced in \refsec{Section:Cascaded}---, we explore how \textit{capturing} these fields enables the identification of entangled states between two temporally overlapping, but spectrally orthogonal, photonic modes. Furthermore, we show that frequency correlations between photonic modes yield metrological enhancements compared to single-photon measurements. 
To this end, we describe the quantum emitter as a coherently driven two-level system, which has been a central model throughout this Thesis.  
When the emitter is driven close to resonance,  it exhibits resonance fluorescence~\cite{KimbleTheoryResonance1976,FlaggResonantlyDriven2009,AstafievResonanceFluorescence2010,LodahlInterfacingSingle2015}, a cornerstone of quantum optics and a source of perfect antibunched photons. 
Despite its formal simplicity, recent studies of frequency-filtered modes of the emission spectrum have revealed a rich landscape of multi-photon processes, evidencing the generation of non-classical correlations~\cite{UlhaqCascadedSinglephoton2012,PeirisTwocolorPhoton2015,PeirisFransonInterference2017,LopezCarrenoPhotonCorrelations2017,StrandbergNumericalStudy2019, ZubizarretaCasalenguaConventionalUnconventional2020,LopezCarrenoEntanglementResonance2024}.
In this Chapter, we extend these results, harnessing the non-classical states of light emitted by the emitter to achieve entanglement generation and explore applications in quantum metrology.
%
%

The Chapter is organized into three main sections: 
\textcolor{Maroon}{(i)} A brief introduction to the quantum source, where we discuss the Heitler and Mollow regimes, and outline the landscape of multi-photon processes in the frequency domain. 
%
%
\textcolor{Maroon}{(ii)}  Summary of the theoretical and experimental results on the generation of entanglement between photonic modes, with an emphasis on the theoretical aspects. 
And  \textcolor{Maroon}{(iii)} an outlook of the
recent theoretical results on quantum metrology using spectral measurements to estimate system parameters via the classical Fisher information, upgrading this quantity to its frequency-resolved version.

The results presented in the second section have been published in npj Quantum Information~\cite{YangEntanglementPhotonic2025}, while those in the third section represent ongoing work and might offer potential directions for future research$^{\textcolor{Maroon}{*}}$. 
\graffito{
	$^{*}$  This work is now published as an arXiv preprint~\cite{Vivas-VianaQuantumMetrology2025}.
}
\vspace{-2.5mm}
\section{Model: a coherently driven TLS}
\label{Chapt6_Section_Model}
\vspace{-2.5mm}

All the works discussed in this Chapter---entanglement generation and quantum metrology---are based on the same physical model: a cascaded setup where the source is a coherently driven two-level system (TLS) that feeds the target(s), modeled as fiducial single-mode cavities. To provide context, this section briefly introduces the model and outlines its main features.

\subsection{Resonance fluorescence}
\label{sec:MollowTriplet}

The TLS spans a basis $\{|g\rangle, |e\rangle \}$, denoting the ground and excited states, respectively, such that the lowering operator gets defined as $\hat \sigma\equiv |g\rangle \langle e|$. 
The emitter is driven by a coherent field characterized by a Rabi frequency $\tilde  \Omega$ and a laser frequency $\omega_L$. In the rotating frame of the laser, and under the rotating wave approximation, the total Hamiltonian reads
\begin{equation}
	\hat H_\sigma= \Delta_\sigma \hat \sigma^\dagger \hat \sigma + (\tilde \Omega \hat \sigma  +\tilde \Omega^* \hat \sigma^\dagger),
	\label{eq:TLS_Hamiltonian}
\end{equation}
where $\Delta_\sigma \equiv \omega_\sigma- \omega_L$ is the qubit-laser detuning, with $\omega_\sigma$ being its natural frequency, and $^*$ denoting the complex conjugate.
The quantum emitter undergoes incoherent processes, essentially de-excitation through spontaneous emission at a rate $\gamma$, resulting from its interaction with the vacuum electromagnetic field or a structured environment such as a coplanar waveguide~\cite{BreuerTheoryOpen2007, CarmichaelOpenSystems1993, GarciaRipollQuantumInformation2022}. The dissipative dynamics of the TLS, described by its reduced density matrix $\hat{\rho}$, is governed by the following master equation:
\begin{equation}
	\frac{d \hat \rho }{dt}= -i [\hat H_\sigma, \hat \rho] + \frac{\gamma}{2} \mathcal{D}[\hat \sigma] \hat \rho,
	\label{eq:MasterEqTLS}
\end{equation}
where $\mathcal{D}[\hat \sigma]$ is the Lindblad superoperator, defined as in \eqref{eq:LindbladTerms}.

Considering the master equation in \eqref{eq:MasterEqTLS} under the resonance condition ($\Delta_\sigma = 0$) and setting the Rabi frequency purely real ($\tilde \Omega\in \mathbb{R}$), the fluorescence spectrum can be straightforwardly computed by applying the theoretical results from \refsec{sec:LiouvProp}.  Specifically, this involves the computation of the non-zero$^{\textcolor{Maroon}{*}}$
\graffito{
$^{*}$Note that since the steady-state is unique, there is only one Liouvillian eigenvalue fulfilling the condition
$
\Lambda_1=0,
$
resulting in $Z_1=|\langle \hat \sigma \rangle|^2$.
}
 Liouvillian eigenvalues, $	\Lambda_\mu$,
\begin{equation}
\Lambda_2=- \frac{\gamma}{2}, \quad \Lambda_{3,4}= -\frac{3\gamma}{4}\mp  \gamma_M,
\label{eq:LiouvEigenvals}
\end{equation}
where $\gamma_M\equiv i \sqrt{4 \Omega^2-\gamma^2/16}$, and the coefficients $Z_\mu$ [see \eqref{eq_spectral_coeff}],
\begin{equation}
	Z_2= \frac{2\Omega^2}{\gamma^2+8\Omega^2}, \quad Z_{3,4}= \frac{\pm \gamma^2 \Omega^2(\gamma \mp 4\gamma_M)+8\Omega^4( 4\gamma_M \mp 5\gamma)}{4\gamma_M (\gamma^2+8\Omega^2)^2}.
\end{equation}
Then, the fluorescence spectrum $S_\sigma(\omega)$ of the coherently driven
\graffito{
	${}^{**}$In case a Lorentzian filter is included in the description, we must replace the real part of the Liouvillian eigenvalue, $\lambda_\mu\equiv \text{Re}[\Lambda_\mu]$, by
	$$
	\lambda_\mu \rightarrow \lambda_\mu + \Gamma/2,
	$$
	while the $\delta$-peak transforms into a Lorentzian-shape curve
	$$
	\delta(\omega) \rightarrow \frac{\Gamma/2}{(\omega-\omega_L)^2+(\Gamma/2)^2}.
	$$
	For a more detailed information on the filtering process, we refer the reader to \refsec{Section:SpectralProperties}.
}
 TLS reads$^{\textcolor{Maroon}{**}}$
\begin{equation}
	\colorboxed{Maroon}{
	\begin{aligned}[b]
	&S_\sigma(\omega)= \frac{1}{\pi} 
\frac{\gamma^2}{\gamma^2+8\Omega^2} \delta(\omega-\omega_L)
+\frac{1}{2\pi} 
\frac{\gamma/2}{(\omega-\omega_L)^2+(\gamma/2)^2}
\\
&+\frac{1}{4\pi}\frac{\alpha_+}{[\omega-(\omega_L-\gamma_M)]^2+(3\gamma/4)^2} 
+\frac{1}{4\pi}\frac{\alpha_-}{[\omega-(\omega_L+\gamma_M)]^2+(3\gamma/4)^2} ,
	\end{aligned}
	}
	\label{eq:FluorescenceSpectrum}
\end{equation}
where 
\begin{equation}
	\alpha_\pm\equiv \frac{3\gamma}{4} \frac{8\Omega^2-\gamma^2}{8\Omega^2+\gamma^2}  + \frac{\gamma}{4 \gamma_M}\frac{(40 \Omega^2-\gamma^2)}{ (8\Omega^2+ \gamma^2)} [\omega-(\omega_L\mp \gamma_M)].
\end{equation}

This model exhibits two fundamental regimes depending on the excitation intensity with respect the emission rate of the TLS: \textcolor{Maroon}{(i)} the Heitler regime~\cite{HeitlerQuantumTheory1984} when $\Omega<\gamma/8$, and \textcolor{Maroon}{(ii)} the Mollow regime~\cite{MollowPowerSpectrum1969} when $\Omega>\gamma/8$. 
\paragraph{Heitler regime.} 
	When the TLS is weakly driven ($\Omega < \gamma/8$)---commonly refereed to as \textit{Heitler regime}~\cite{HeitlerQuantumTheory1984}---, the emission spectrum is dominated by the Rayleigh-scattered light of the laser,
	consisting of elastically scattered photons. This contribution appears as a $\delta$-peak centered at the laser frequency.
	In addition to this elastic scattering, the spectrum also exhibits a Lorentzian peak of width $\gamma$, emerging from the fluorescence of the emitter. Together, these components form the characteristic spectral structure of the Heitler regime, as illustrated in \reffig{FigChapt6_HeitlerRegime}.
			\begin{SCfigure}[1][h!]
		\includegraphics[width=0.75\columnwidth]{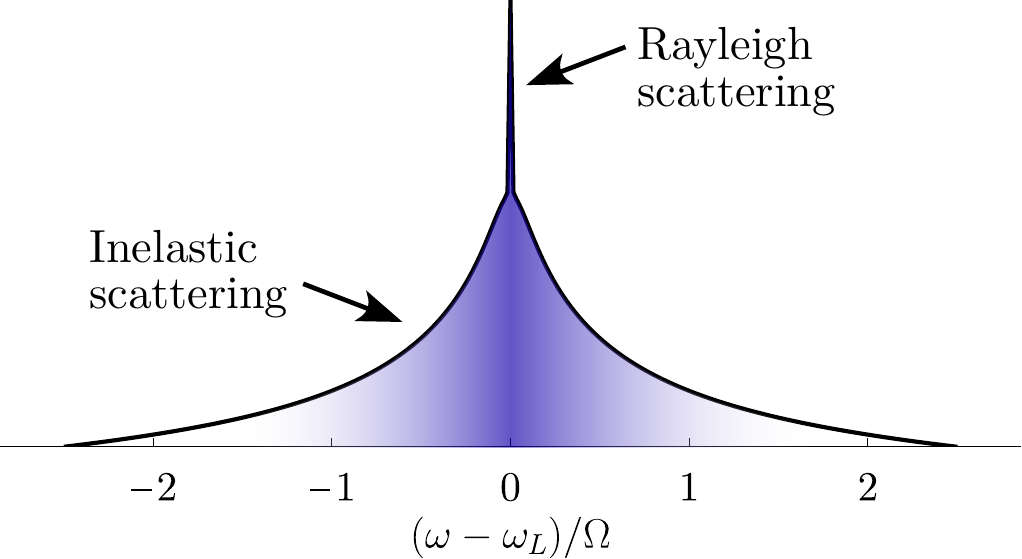}
		\captionsetup{justification=justified}
		\caption[Emission spectrum in the Heitler regime.]{ \label{FigChapt6_HeitlerRegime}	\textbf{Emission spectrum in the Heitler regime. } 
			Illustration of the emission spectrum of a coherently driven (at resonance, $\Delta_\sigma=0$) TLS in the Heitler regime ($\Omega \ll \gamma$). The fluorescence spectrum exhibits a delta peak around the laser frequency, corresponding to the Rayleigh scattering, over a weaker Lorentzian profile given by the inelastic scattering. The spectrum is depicted in logarithmic scale to unambiguously show both contributions in the fluorescence spectrum. 
		}
	\end{SCfigure}

	In the weak driving limit ($\Omega \ll \gamma/8$) and under the resonance condition ($\Delta_\sigma=0$), the spectrum in \eqref{eq:FluorescenceSpectrum} simplifies to
	\begin{equation}
		S_\sigma(\omega)\approx \frac{1}{\pi} \delta(\omega-\omega_L)+ \frac{1}{\pi} \frac{\Omega^2}{(\gamma/2)^2}\frac{\gamma/2}{[(\omega-\omega_L)^2+(\gamma/2)^2]^2}.
	\end{equation}
	Although the resonance fluorescence fraction---the inelastic scattering---appears negligible compared to the scattered laser light, it is the interference between these two processes that fundamentally defines the optical properties of the two-level system~\cite{ZubizarretaCasalenguaTuningPhoton2020}.

\paragraph{Mollow regime.}
	In the opposite scenario, when the emitter is strongly driven ($\Omega > \gamma/8$)---commonly refereed to as the \textit{Mollow regime}~\cite{MollowPowerSpectrum1969}---, 
	the emission spectrum exhibits a characteristic triplet structure known as the \textit{Mollow triplet}, as a consequence of the light-matter hybridization: the Liouvillian eigenvalues $\Lambda_{3,4}$ [see \eqref{eq:LiouvEigenvals}] develop a non-zero imaginary contribution, $\pm \gamma_M$.
	The spectrum consists of a central elastic term (the Rayleigh peak) and three inelastic components. The inelastic terms correspond to Lorentzian-shaped peaks that form the triplet structure, dominating over the elastic contribution when the driving laser is sufficiently strong, as illustrated in \reffig{FigChapt6_MollowRegime}~\textcolor{Maroon}{(a)}. 

	In the strong driving limit ($\Omega\gg \gamma/8$) and under the resonance condition ($\Delta_\sigma=0$), the spectrum in \eqref{eq:FluorescenceSpectrum} simplifies to
	%
	%
	%
	\begin{multline}
		S_\sigma(\omega)\approx \frac{1}{2\pi } \frac{\gamma/2}{(\gamma/2)^2+(\omega-\omega_L)^2} 
		\\
		+ \frac{1}{4\pi }\frac{3\gamma/4}{ [\omega-(\omega_L-2\Omega) ]^2+(3\gamma/4)^2} 
				+\frac{1}{4\pi }\frac{3\gamma/4}{ [\omega-(\omega_L+2\Omega) ]^2+(3\gamma/4)^2} .
	\end{multline}

	Phenomenologically, the physics of the Mollow regime can be better understood through the \textit{dressed-atom approach}~\cite{Cohen-TannoudjiAtomPhotonInteractions1998, DalibardCorrelationSignals1983}.
	In this framework, the energy levels of the TLS become dressed by the laser field, forming an infinite ladder of excitation manifolds comprised of hybrid light-matter states, known as \textit{dressed states}~\cite{JaynesComparisonQuantum1963}.
	 Each manifold consists of a pair of eigenstates, $|\pm\rangle_n$, formed by a superposition of the bare states of the TLS-laser system $\{|g,n+1\rangle, |e,n\rangle \}$, with $n$ denoting the number of excitations. 
	In the strong-driving regime, where $n\gg 1$, the energy splitting between eigenstates from the same manifold, $|+\rangle_n$ and $|-\rangle_n$, is essentially determined by the energy splitting $\omega_S=2|E_\pm|$, where $E_\pm$ represents their respective eigenenergies
	\begin{equation}
		E_\pm=\pm \sqrt{\Omega^2+(\Delta_\sigma/2)^2}.
		\label{eq:MollowEnergy}
	\end{equation}
	In this limit, we simplify the description by dropping the explicit dependence on $n$ (treating it conceptually) and visualizing the energy spectrum as a ladder of doublets, $|\pm\rangle$, each split by an energy $\omega_S$, thereby defining different excitation manifolds. Additionally, the nonlinear behaviour of the Jaynes-Cummings model, in which the energy splitting of the $n$-th rungs is proportional to $\sqrt{n}$, becomes negligible. As a result, the energy separation between adjacent manifolds is constant, determined solely by the laser frequency $\omega_L$.  
	
\begin{SCfigure}[1][t!]
		\includegraphics[width=1.\columnwidth]{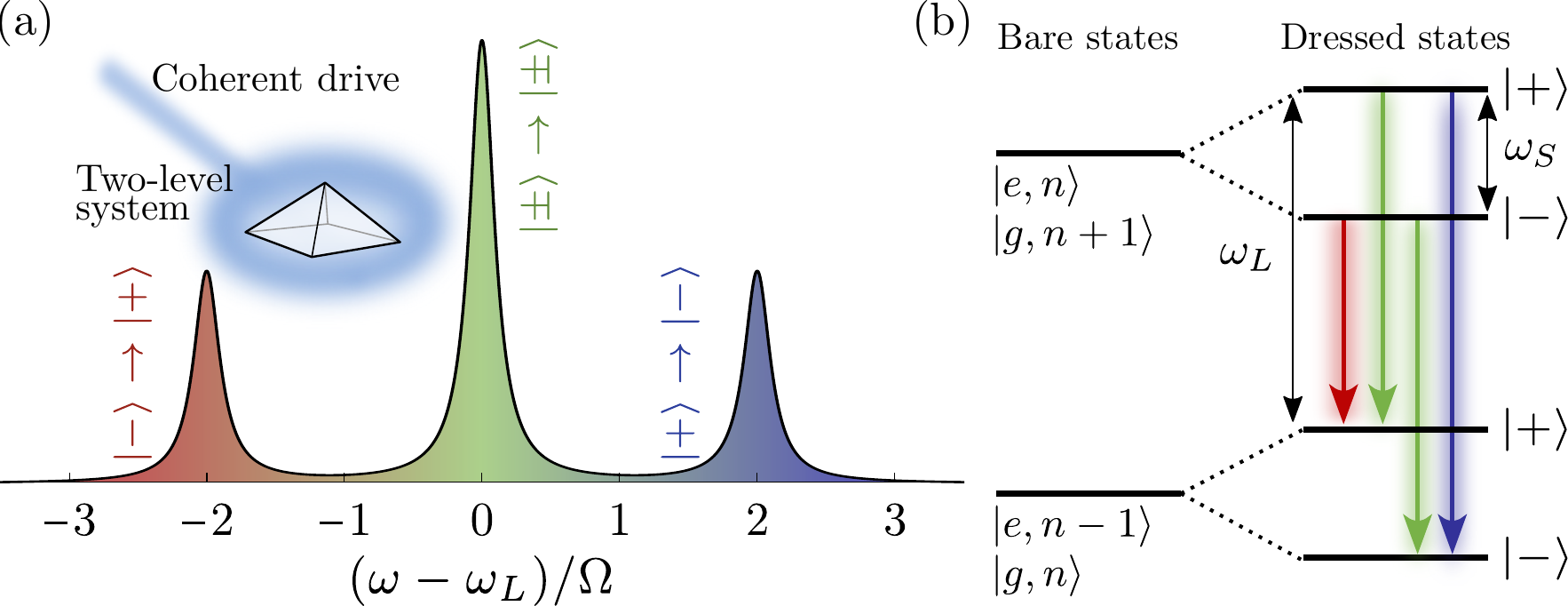}
		\captionsetup{justification=justified}
\end{SCfigure}
	\graffito{\vspace{-16.0cm}
		\captionof{figure}[Emission spectrum in the Mollow regime.]{	\label{FigChapt6_MollowRegime}
			\textbf{Emission spectrum in the Mollow regime. } 
			(a)  Illustration of the emission spectrum of a resonantly driven TLS in the Mollow regime ($\Omega > \gamma/8$).
			(b)	Single photon transitions between adjacent excitation manifolds: $|-\rangle \rightarrow |+\rangle$ (in red), $|+\rangle \rightarrow |-\rangle$ (in blue), and $|\pm\rangle \rightarrow |\pm\rangle$ (in green).  
		}
	}

	Considering the Hamiltonian in \eqref{eq:TLS_Hamiltonian}, the dressed states can be readily expressed as$^{\textcolor{Maroon}{*}}$
	\graffito{
	$^{*}$The symbols $c$ and $s$ derive their names from their association with the cosine and sine of a rotation angle,
	$$
	\theta=\text{arctan} (\xi).
	$$
This relationship arises from the fact that any real symmetric matrix  $S$ can be diagonalized through a rotation~\cite{TungGroupTheory1985}. For a $2\times 2$ matrix, this rotation is represented as
	$$
	R(\theta)= \begin{pmatrix}
		\cos \theta & -\sin \theta \\
		\sin \theta & \cos \theta
	\end{pmatrix},
	$$
such that the diagonalized matrix $D$ is given by
	$$
	D=R^T(\theta)S R(\theta).
	$$
	}
	\begin{subequations}
	\begin{align}
			|+\rangle &=  c |g\rangle + s |e\rangle,  \\ 
		|-\rangle &= s |g\rangle - c |e\rangle,
	\end{align}
	\end{subequations}
	where $c$ and $s$ determine the proportional weights of the TLS bare states $\{|g\rangle, |e\rangle\}$ in each dressed state, with the weights depending on both the Rabi frequency of the laser $\Omega$ and the qubit-laser detuning $\Delta_\sigma$:
\begin{subequations}
	\begin{align}
	c&=\frac{1}{\sqrt{1+\xi^{-2}}}, \\
	s&=\frac{1}{\sqrt{1+\xi^{2}}}, \\
	\xi&= \frac{\Omega}{\Delta_\sigma/2+\omega_S/2}.
	\end{align}
\end{subequations}

	From this perspective, the Mollow structure can be understood as transitions that can take place between adjacent excitation manifolds. The central peak corresponds to a doubly-degenerate transition,
	\begin{equation}
		|\pm \rangle \rightarrow |\pm\rangle \quad \text{with} \quad \omega=\omega_L,
	\end{equation}
	leaving the state of the TLS unchanged, while the two side peaks correspond to transitions between different eigenstates,
	\begin{equation}
		|\pm \rangle \rightarrow |\mp \rangle \quad \text{with}\quad \omega=\omega_L \pm \omega_S,
	\end{equation}
	changing the state of the TLS. These transitions are depicted in \reffig{FigChapt6_MollowRegime}~\textcolor{Maroon}{(b)}: $|\pm \rangle \rightarrow |\pm \rangle$ in green, $	|+ \rangle \rightarrow |- \rangle $ in blue, and $	|- \rangle \rightarrow |+ \rangle $ in red. 
	From this dressed atom approach, we can better understand why the central peak exhibits twice the weight of the two sidebands, as shown in~\reffig{FigChapt6_MollowRegime}~\textcolor{Maroon}{(a)}: two of the four possible one-photon transitions are degenerate, $	|\pm \rangle \rightarrow |\pm\rangle $.

\subsection{Multi-photon processes and cascaded configuration}
\label{sec:cascaded}
The Mollow triplet provides an excellent platform to explore the correlations of photons emitted from the three distinct peaks, as well as the cross-correlations between photons originating from different peaks. Although the emission occurs from a single mode, the unique spectral structure naturally raises questions about the correlations both within and between the three peaks. 

Studies of frequency-filtered modes of the emission spectrum have unveiled a rich landscape of multi-photon processes, evidencing the generation of non-classical correlations~\cite{DelValleTheoryFrequencyFiltered2012,UlhaqCascadedSinglephoton2012,SanchezMunozViolationClassical2014,PeirisTwocolorPhoton2015,PeirisFransonInterference2017,LopezCarrenoPhotonCorrelations2017,ZubizarretaCasalenguaConventionalUnconventional2020,LopezCarrenoEntanglementResonance2024}. 
In the time domain, it has been predicted that under certain conditions, selected temporal
modes from the resonance-fluorescence emission exhibit
a negative Wigner function---a hallmark of nonclassicality---which has been theoretically proposed~\cite{StrandbergNumericalStudy2019,QuijandriaSteadyStateGeneration2018} and experimentally verified~\cite{LuPropagatingWignerNegative2021}. 
For instance, to reveal potential nonclassical correlations [see \reffig{FigChapt6_FreqResolved}~\textcolor{Maroon}{(a)}], we can inspect the cross-correlations between two different frequencies, $\omega_1$ and $\omega_2$, via the second order correlation function~\cite{DelValleTheoryFrequencyFiltered2012,Gonzalez-TudelaTwophotonSpectra2013} $g^{(2)}_\Gamma(\omega_1,\omega_2)$---already defined in \refsec{Section:SpectralProperties} [\eqref{eq:FreqResolvedFun}]---, or via violations of the Cauchy-Schwarz inequality (CSI)~\cite{LoudonQuantumTheory2000,SanchezMunozViolationClassical2014,PeirisTwocolorPhoton2015}, reformulated as 
\begin{equation}
		R_\Gamma(\omega_1,\omega_2)=\frac{g_\Gamma^{(2)}(\omega_1,\omega_2)}{[g_\Gamma^{(2)}(\omega_1,\omega_1)g_\Gamma^{(2)}(\omega_2,\omega_2)]}\leq 1.
\end{equation}

Here, another class of transitions takes place in the same ladder presented above in \reffig{FigChapt6_MollowRegime}~\textcolor{Maroon}{(b)}, the so-called \textit{leapfrog processes}~\cite{Gonzalez-TudelaTwophotonSpectra2013}. It consists of transitions from a real state to another but jumping over an intermediate virtual state, in which two-photon emission occurs$^{\textcolor{Maroon}{*}}$.
\graffito{
	${}^{*}$This kind of transition can be generalized to $N$-photon leapfrog processes. In the general case, the transition from one state to another occurs by jumping over $N$ intermediate virtual states, and thus involving $N$-photon emission~\cite{LopezCarrenoPhotonCorrelations2017}. 
} 
The two-photon leapfrog processes in the Mollow triplet are determined by the conditions~\cite{Gonzalez-TudelaTwophotonSpectra2013,LopezCarrenoPhotonCorrelations2017}:
\begin{subequations}
	\begin{align}
		&|- \rangle  \rightrightarrows |+\rangle   \quad \text{when} \quad     \Delta_1+ \Delta_2 = -\Delta_+,\\
		&|\pm \rangle  \rightrightarrows |\pm\rangle   \quad \text{when} \quad  \Delta_1 + \Delta_2=0, \\
		&|+ \rangle  \rightrightarrows |-\rangle   \quad \text{when} \quad  \Delta_1+ \Delta_2=\Delta_+,
	\end{align}
\end{subequations}
where $\Delta_i\equiv \omega_i-\omega_L$ for $i=1,2$, with $\omega_i$ denoting the frequency of the $i$th emitted photon. These two-photon transitions are the antidiagonal lines drawn in \reffig{FigChapt6_FreqResolved}~\textcolor{Maroon}{(a)}, corresponding to regions of superbunching [red regions in the upper-left triangle] and strong CSI violations [green regions in the lower-right triangle], showing their nonclassical character. 
For a detailed analysis on photon correlations from the Mollow triplet, we refer the reader to \colorref{LopezCarrenoPhotonCorrelations2017}.
 \begin{SCfigure}[1][h!]
	\includegraphics[width=1.\columnwidth]{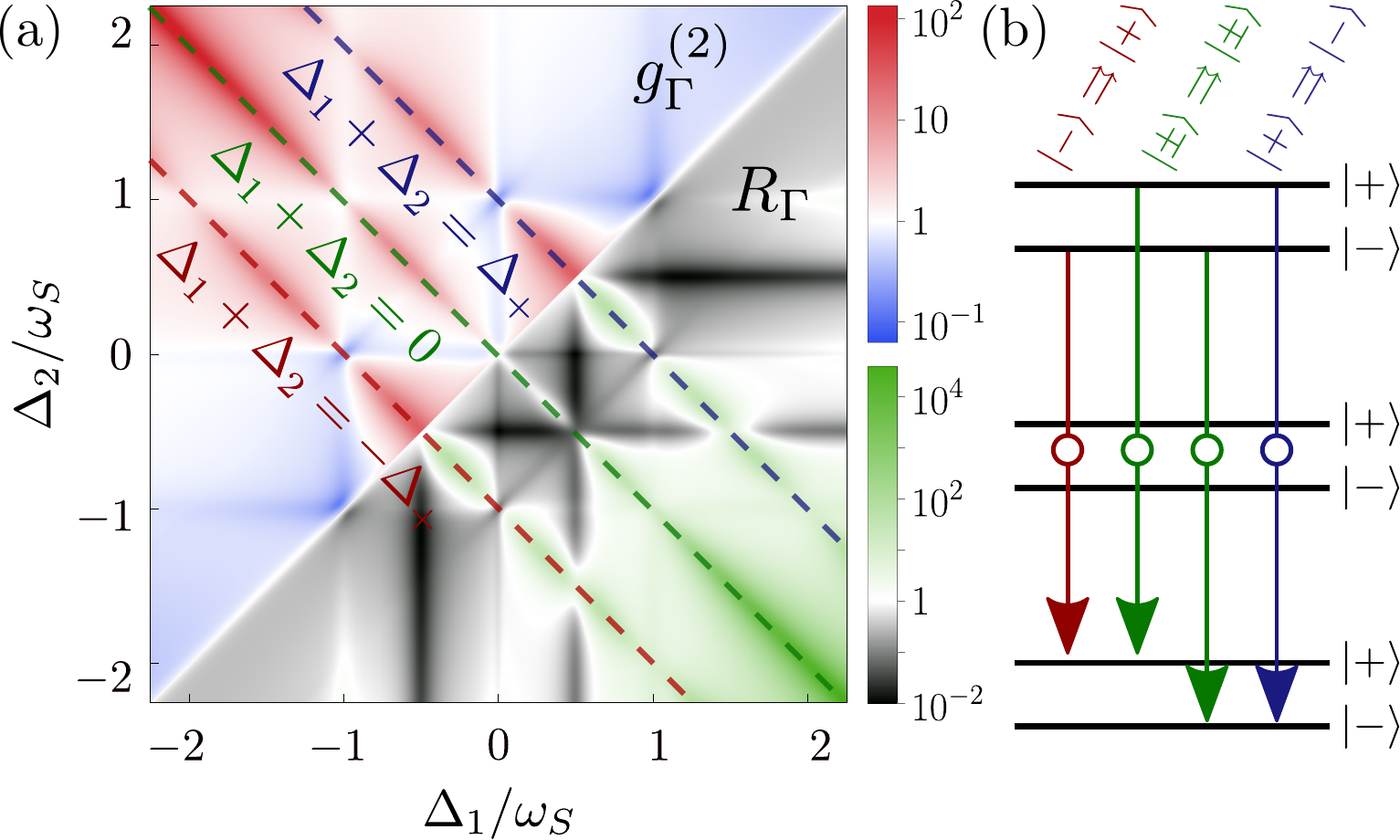}
\end{SCfigure}
\graffito{\vspace{-8.0cm}
	\captionof{figure}[Frequency-resolved measurements.]{
		\label{FigChapt6_FreqResolved}
		\textbf{Frequency-resolved measurements. }
		(a)	 $g^{(2)}_\Gamma$ (upper-left triangle) and $R_\Gamma$ (lower-right triangle) in the frequency  domain. 
		The signatures of nonclassicality (bunching and CSI violation) are observed along the antidiagonals (red, green and blue dashed lines), corresponding to emission involving virtual states. 
		Both figures are symmetric with respect the diagonal.
		(b) Schematic representation of leapfrog processes in the Mollow ladder: two-photon transitions between real states that occur by bypassing intermediate manifolds.
	}
}

We will recourse to the cascaded-sensor method, already introduced in \refsec{sec:cascaded_sensor}, to extract the photon information from any frequency window of a quantum source using all the signal theoretically available, while physically accounting the effect of detection.
In the case of a single sensor, we remind the reader that the cascaded-sensor master equation enables the extraction of information within a single frequency window. In other words, it allows the exploration of phenomena involving only a single frequency---one-photon processes---, such as the fluorescence spectrum~\cite{DelValleTheoryFrequencyFiltered2012,CarrenoExcitationQuantum2016,CarrenoExcitationQuantum2016a}. This master equation is given by
\begin{multline}
	\frac{d \hat \rho }{dt}= -i [\hat H_\sigma+ \Delta_\xi \hat \xi^\dagger \hat \xi, \hat \rho] + \frac{\gamma}{2} \mathcal{D}[\hat \sigma] \hat \rho 
	\\
+  \frac{\Gamma_\xi}{2} \mathcal{D}[\hat \xi]\hat \rho	-
	 \sqrt{\gamma \Gamma_\xi} \left\{ [\hat \xi^\dagger, \hat \sigma  \hat \rho ] + [\hat \rho \hat \sigma^\dagger, \hat \xi] \right\},
	 \label{eq:cascadedmaster1}
\end{multline}
where the sensor is described by a single bosonic mode with annihilation operator $\hat \xi$, sensor-laser detuning $\Delta_\xi \equiv \omega_\xi - \omega_L$ (thus, $\omega_\xi$ is the sensor frequency), and linewidth $\Gamma_\xi$. We also note the reader that the cascaded coupling cannot be arbitrary since it exclusively depends on the dissipative rates of both the source and the target~\cite{GardinerQuantumNoise2004,CarrenoExcitationQuantum2016,CarrenoExcitationQuantum2016a}. The coupling can only be smaller or at most equal to $\sqrt{\gamma \Gamma_\xi}$ in order to maintain the \textit{physicality} of the cascaded process and not lead to unphysical states for the target. 
For instance, \reffig{ExperimentalMollowTriplet} shows the mean photon number of a fiducial single-mode cavity---that follows a similar cascaded master equation to \eqref{eq:cascadedmaster1}---featuring the same structure of the Mollow triplet spectrum as a consequence of the theoretical link between spectral measurements and the sensor method~\cite{DelValleTheoryFrequencyFiltered2012,CarrenoExcitationQuantum2016,CarrenoExcitationQuantum2016a}. 
{
	\sidecaptionvpos{figure}{t}
 \begin{SCfigure}[1][h!]
	\includegraphics[width=0.78\columnwidth]{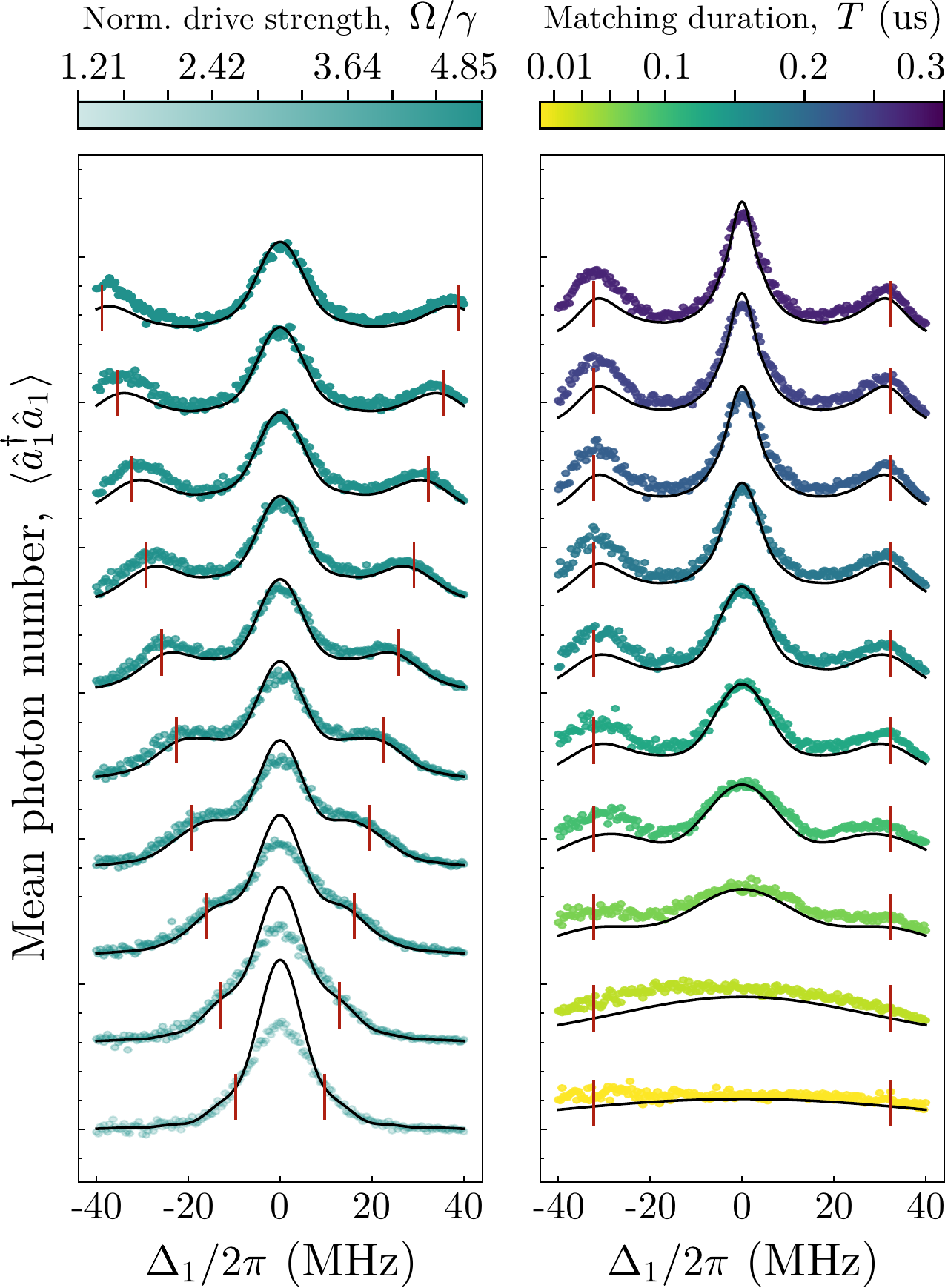}
	\captionsetup{justification=justified}
	\caption[Experimental realization of the Mollow triplet.]{ \label{ExperimentalMollowTriplet} \textbf{Experimental realization of the Mollow triplet.}
		Mean photon number from a single-mode cavity which collects the output field of a coherently driven superconducting qubit at a specific driving intensity $\Omega$ and measurement duration $T$. 
		In the left panel, the measurement time is fixed at $T=100$ ns while the driving intensity $\Omega$ is varied. In the right panel, the driving intensity is set to $\Omega=4.04\gamma$ (Mollow regime), and the measurement time is swept. In both panel, the $x$-axis represents the frequency of a single mode cavity acting as a sensor~\cite{DelValleTheoryFrequencyFiltered2012}.
		The red vertical lines mark the positions of $E_\pm$ in \eqref{eq:MollowEnergy}, which
		match the location of the side peaks of the corresponding curves.
		This figure is adapted from our experimental collaboration in \colorref{YangEntanglementPhotonic2025}. For a more detailed analysis of the experiment, we refer the reader to \cite{YangEntanglementPhotonic2025,YangMicrowavePhoton2024}.
	}
\end{SCfigure}}

While the one sensor case is straightforward, the case of two or more cascaded sensors requires additional attention since the physical implementation actually plays a crucial role. In this Chapter, we will explore two distinct setups based on different experimental realizations. In \refsec{Chapt6_SecEntanglement}, we consider a situation in which the photonic modes are digitally filtered by post-processing the source signal, while in  in \refsec{Chapt6_SecMetrology} we consider a description in which, prior to the capture of each mode, the output signal is physically split (e.g. by a beam splitter). As we will see, the essential difference between these two approaches relies on a factor $1/\sqrt{2}$ in the cascaded coupling, corresponding to vacuum contributions after splitting the signal in the beam splitter.
%
%
%
%
%
%
%
%

\sectionmark{Entanglement of photonic modes from a continuously ...}
\section[Entanglement of photonic modes from a continuously driven two-level system]{Entanglement of photonic modes from a \protect\newline continuously driven two-level system}
\sectionmark{Entanglement of photonic modes from a continuously ...}

\label{Chapt6_SecEntanglement}

In quantum optics, entangled photons have been traditionally
\graffito{${}^{*}$Nonlinear process by which a high-energy photon, called \textit{pump photon}, splits into a pair of lower-energy photons, called signal and idler, that are quantum mechanically entangled.}
generated through spontaneous parametric down-conversion (SPDC)$^{\textcolor{Maroon}{*}}$, often combined with beam splitters and photodetectors for heralding~\cite{KwiatNewHighIntensity1995,RamelowHighlyEfficient2013}. However, the probabilistic nature of this approach can pose challenges for practical applications.
Recent advances have enabled the deterministic generation of entangled photonic states, producing photonic qubits on demand~\cite{EsmannSolidStateSinglePhoton2024}. 
These photonic qubits are typically encoded in either the polarization~\cite{MullerOndemandGeneration2014,IstratiSequentialGeneration2020,ThomasEfficientGeneration2022} or time-bin~\cite{BesseRealizingDeterministic2020,KnautEntanglementNanophotonic2024,FerreiraDeterministicGeneration2024} degree of freedom.
The time-frequency degree of freedom can be a valuable tool for high-dimensional quantum information processing.
Generally, optical time- or frequency-bin entangled photon pairs are produced via parametric downconversion or spontaneous four-wave mixing~\cite{OlislagerFrequencybinEntangled2010,ZhangHighperformanceQuantum2021,AndreaSabattoliSiliconSource2022,ClementiProgrammableFrequencybin2023}, where the temporal mode structure is determined by the spectral profile of the pump~\cite{AnsariTailoringNonlinear2018}.
For instance, in the microwave frequency range, frequency-entangled states have been produced via Josephson junction circuit elements~\cite{EichlerObservationTwoMode2011,SchneiderObservationBroadband2020,GasparinettiCorrelationsEntanglement2017, PeugeotGeneratingTwo2021, EspositoObservationTwoMode2022,PerelshteinBroadbandContinuousVariable2022,JolinMultipartiteEntanglement2023}.
However, besides time-bin qubit encoding has been explored in microwave quantum communication~\cite{KurpiersQuantumCommunication2019,IlvesOndemandGeneration2020}, the temporal degree of freedom has been overlooked.

In this section, we present the results of a joint experimental-theoretical collaboration with the experimental group led by Prof. Simone Gasparinetti~\cite{YangEntanglementPhotonic2025}.  We theoretically propose and experimentally verify a method for generating time-frequency entangled photonic states from the steady-state emission of a quantum emitter coupled to a waveguide.
While similar ideas have been explored previously in \colorref{LopezCarrenoEntanglementResonance2024}, which theoretically verified the generation of entangled photons by off-resonantly driving a two-level system and analysing the emission from the Mollow sidebands [see \reffig{FigChapt6_MollowRegime}], our approach introduces distinct and complementary contributions.
In contrast to the analysis in \colorref{LopezCarrenoEntanglementResonance2024}, we focus on exploring the generation of entanglement between two temporally overlapping but spectrally orthogonal modes from the continuous emission of the quantum emitter when it is resonantly driven. This is achieved through resonant coherent driving combined with digitally implemented temporal mode filters.
This method is theoretically platform-independent and can be applied to quantum systems with diverse level structures and pumping schemes, offering a new avenue for generating entanglement from continuously driven quantum systems.

In the following, we provide a concise overview of the experimental setup and summarize both the theoretical predictions and experimental results. This discussion is adapted from \colorref{YangEntanglementPhotonic2025}.  For a more detailed description of the experimental configuration and techniques, we refer readers to \colorref{YangMicrowavePhoton2024}.

\subsection{Experimental setup}

The experimental realization of the proposed theoretical model---a coherently driven quantum emitter---is implemented in superconducting circuits. The quantum emitter, considered as a TLS,  is achieved by using an X-mon-type superconducting circuit capacitively coupled to a coplanar waveguide, such that it enables the interaction with the propagating microwave modes [see \reffig{Device}].
The waveguide is connected with a reflection setup, where the coherent and continuous drive is sent to the qubit through the waveguide, while from the output the propagating modes can be measured, and characterize the entanglement between the two photonic modes. 
{
	\sidecaptionvpos{figure}{t}
	 \begin{SCfigure}[1][h!]
	\includegraphics[width=0.9\columnwidth]{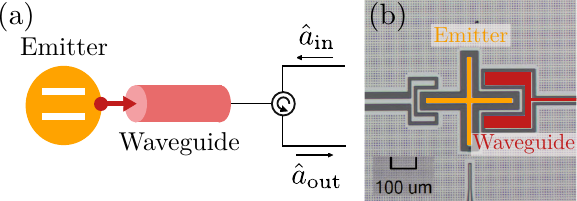}
	\captionsetup{justification=justified}
	\caption[Experimental device.]{ \label{Device}	
		\textbf{Experimental device.}
		Superconducting device used for the generation of entanglement from the steady-state emission of a coherently driven qubit. (a) Schematic representation of the device and the reflection measurement setup. (b) Illustration of a transmon qubit (orange) that is capacitively coupled to a waveguide (red).
	}
\end{SCfigure}}
\newpage 

The qubit is coupled to the waveguide with a decay rate of $\Gamma= 8$ MHz, making the relaxation time of the emitter $T_1=1/\Gamma_1$ of the qubit to be around $20$ ns. The dissipative dynamics of the qubit is described by the same master equation as in \eqref{eq:MasterEqTLS}, with a complex Rabi frequency given by
$
	\tilde \Omega= i\Omega/2,
$
and the qubit being resonantly driven ($\Delta_\sigma=0$), such that the Hamiltonian from \eqref{eq:TLS_Hamiltonian} reduces to
\begin{equation}
	\hat H_\sigma=i\frac{\Omega}{2}(\hat \sigma- \hat \sigma^\dagger).
\end{equation}
The driving term features now an additional phase due to experimental requirements. The pulse sequence employed in the experiment is illustrated in \reffig{Measurement}. The qubit is continuously driven at its resonant frequency, $\omega_L = \omega_{\text{ge}}$, with $\omega_{\text{ge}}/2\pi=4.94$ GHz being the transition frequency of the transmon qubit. After initiating the qubit drive for $200$ ns---a duration significantly longer than the relaxation time of the qubit  ($T_1$)---, it is ensured that the qubit reaches its steady-state. At this point, two temporal mode filters are applied, $f_1(t)$ and $f_2(t)$, on the reflected time trace, in post-processing.

 \begin{SCfigure}[1][b!]
	\includegraphics[width=0.9\columnwidth]{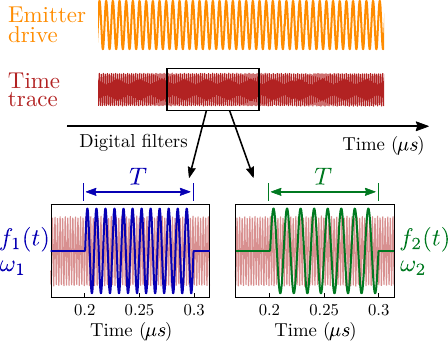}
	\captionsetup{justification=justified}
	\caption[Measurement scheme.]{ \label{Measurement}	\textbf{Measurement scheme.} Pulse sequence for entanglement generation. An on-resonance drive is applied to the waveguide qubit to bring it into the steady state. Two temporal filters, $f_1(t)$ and $f_2(t)$, are then applied simultaneously to match with the reflected time trace in post-processing, as shown in the insets. The same temporal filters are used to match both cases.
	}
\end{SCfigure}

\subsection{Capturing the temporal light mode}

Temporal mode matching is used to isolate the emitted modes from the continuum of modes in the waveguide. Specifically,  digital temporal filters, $f_k(t)$, are applied in order to extract single propagating photon modes, $\hat a_k$, such that
\begin{equation}
	\hat a_k= \int_{-\infty}^\infty dt f_k(t) \hat a_{\text{out}}(t),
\end{equation}
where $ \hat a_{\text{out}}(t)$ is the time-dependent output field measured from the output line connected to the waveguide. This output field is easily obtained from the input-output relation~\cite{LoudonQuantumTheory2000}, already introduced in \eqref{eq:InputOutputRelation}, such that
\begin{equation}
	 \hat a_{\text{out}}(t)= \sqrt{\Gamma} \hat \sigma (t) - \hat a_{\text{in}}(t),
\end{equation}
where $\hat \sigma (t) $ is the qubit lowering operator and $\hat a_{\text{in}}(t)$ is the input field. We note that the extracted modes fulfil bosonic commutation relations, $[\hat a_k, \hat a_k^\dagger]=1$, ensured by the normalization condition for the filter function 
$
	\int_{-\infty}^\infty dt |f(t)|^2=1.
$

Generally, it is possible to consider correlations between any pair of the propagating modes, including non-orthogonal ones. However, the orthogonality of the two temporal modes ensures that they are independent. For this reason, in the following, we will study correlations for overlapping modes, but restricted to orthogonal modes when discussing entanglement. Two temporal filters are orthogonal if they satisfy the condition
\begin{equation}
	\int f_1^*(t)f_2(t) dt=0,
\label{eq:orthogonality}
\end{equation}
where $f_1(t)$ and $f_2(t)$ are temporal filters with carrier frequencies $\omega_1$ and $\omega_2$ that fulfil
\begin{equation}\label{eq:temporal_filter}
	f_k(t)=v(t)\cdot e^{i(\omega_k t+\phi)}, \quad k=1,2. 
\end{equation}
Here, $f_k(t)$ are complex temporal filters, with $v(t)$ representing the real wavepacket profile for both filters, and $\phi$ a constant phase.  The frequency of the $k$th temporal filter, $\omega_k$, is set to the carrier frequency of the emitted radiation$^{\textcolor{Maroon}{*}}$.
\graffito{
${}^{*}$Note that this is equivalent to introduce the frequency dependency in the Hamiltonian via the usual $\Delta_k \hat a_k^\dagger \hat a_k$ terms, with cavity-laser detuning $\Delta_k\equiv \omega_k -\omega_L$.
} 
The wavepacket profile $v(t)$ should match the temporal shape of the photon field to achieve optimal matching efficiency~\cite{KiilerichInputOutputTheory2019,KiilerichQuantumInteractions2020}. However, this might be technologically challenging if the output field exhibits a multi-mode structure:  the number of filters required would equal the number of eigenmodes, with each filter tailored to match the specific temporal profile of a corresponding eigenmode~\cite{KiilerichInputOutputTheory2019}.
Here, $v(t)$ is set to be a boxcar function of duration $T$,
\begin{equation}
	v(t)= \frac{1}{\sqrt{T}}[\Theta(t-t_0) - \Theta (t-t_0-T) ]
	\label{eq:BoxcarFun}
\end{equation}
with $\Theta(t)$ being the Heaviside step function and $t_0$ the starting time of the temporal filters.
For our case with two filters of the form \eqref{eq:temporal_filter}, the orthogonality condition reduces to 
\begin{equation}
	\frac{\omega_2-\omega_1}{2\pi}= \frac{m}{T},
	\label{eq:orthogonalityCond1}
\end{equation}
where $m$ is an arbitrary integer. In this work, we use $T = 100$ ns, so the condition for orthogonality between the two modes is $(\omega_2-\omega_1)/2\pi = m \cdot 10$ MHz. Additionally, we also note that modes spectrally separated by a large frequency detuning can also be considered orthogonal to a good extent.

\subsection{Joint quantum state of temporal modes }

The density matrix describing the joint quantum states
of both modes can be computed by using the recent \textit{input-output theory for quantum pulses}~\cite{KiilerichInputOutputTheory2019,KiilerichQuantumInteractions2020}. This approach treats the system and the emission as a cascaded quantum system, in which the desired temporal mode is captured via a fiducial single-mode cavity coupled non-reciprocally to the system with a time-dependent coupling. Similarly, we can also tailor an incident pulse that will interact with the system by tuning the time-dependent coupling of another fiducial single-mode cavity. See \reffig{InputOutputMolmer} for a schematic representation of the \textit{input-output theory for quantum pulses}.
\begin{SCfigure}[1][h!]
	\includegraphics[width=1.0\columnwidth]{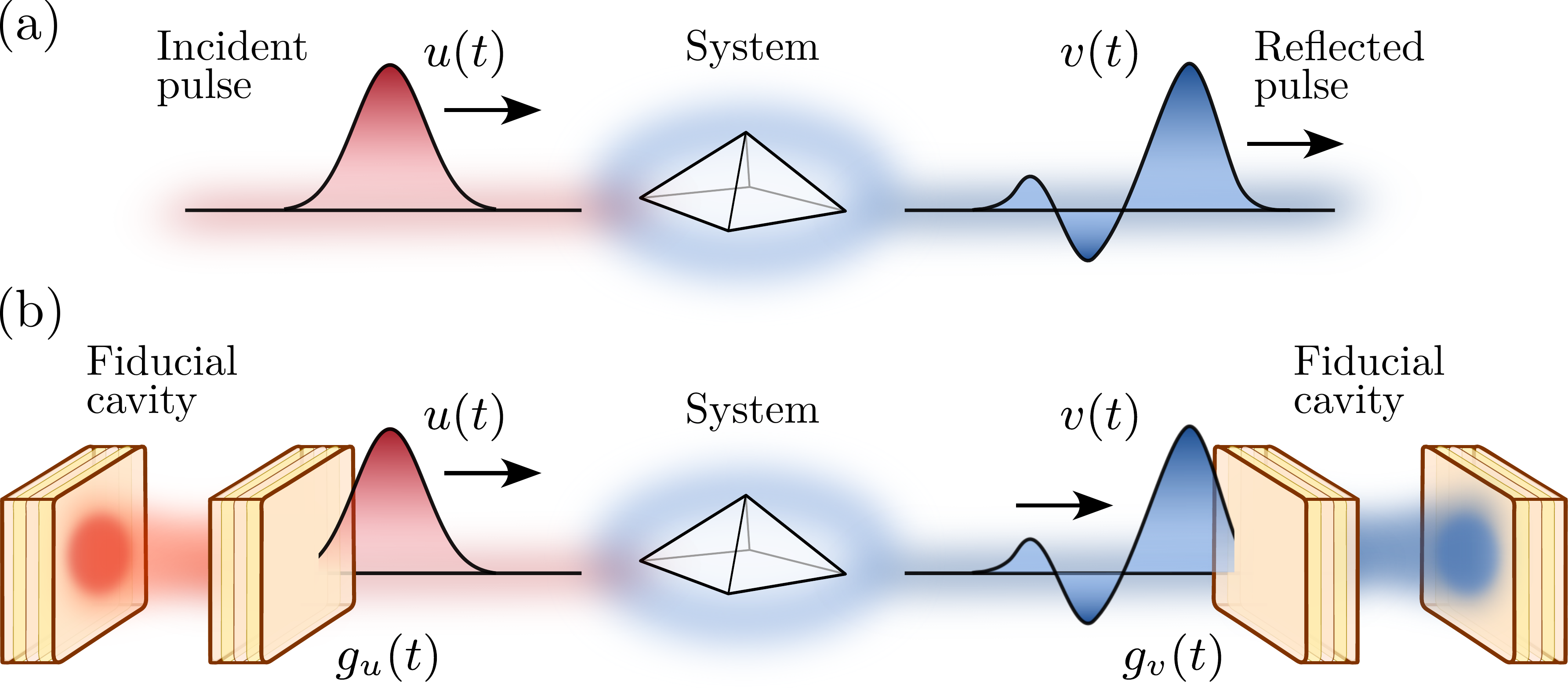}
	\captionsetup{justification=justified}
\end{SCfigure}
\graffito{\vspace{-5.50cm}
	\captionof{figure}[Input-output theory for quantum pulses.]{
		\label{InputOutputMolmer}
		\textbf{Input-output theory for quantum pulses.}
		Schematic representation of the input-output theory for quantum pulses~\cite{KiilerichInputOutputTheory2019,KiilerichQuantumInteractions2020}.
		(a) This theory provides a framework for designing and controlling the generation and absorption of incident [$u(t)$] and reflected [$v(t)$] quantum pulses with arbitrary profile envelopes. (b)This is achieved by employing fiducial cavities with time-dependent couplings, $g_u(t)$ and $g_v(t)$, respectively .
	}
}

Here, we extend this approach to capture simultaneously several temporal modes by introducing two virtual cavities, resulting in the following cascaded master equation in the rotating frame of the
drive: 
\begin{equation}
	\label{eq:molmermastereq}
	\colorboxed{Maroon}{
	\begin{aligned}[b]
		\frac{d \hat \rho }{dt}= -i [\hat H_\sigma, \hat \rho] + &\frac{\Gamma}{2} \mathcal{D}[\hat \sigma] \hat \rho +	\sum_{k=1}^2 \frac{|g_k(t)|^2}{2} \mathcal{D}[\hat a_i]\hat \rho
		\\
		&-
		\sum_{k=1}^2 \sqrt{\Gamma } \left\{ g_k^*(t)[\hat a_i^\dagger, \hat \sigma  \hat \rho ] + g_k(t) [\hat \rho \hat \sigma^\dagger, \hat a_i] \right\}.
	\end{aligned}
	}
\end{equation}
Note that in this cascaded master equation, the dissipative rates of the fiducial cavities are replaced by time-dependent couplings, $\Gamma_k \rightarrow |g_k(t)|^2$. These time-dependent quantities encode the information of the temporal filters via~\cite{KiilerichInputOutputTheory2019,KiilerichQuantumInteractions2020}: 
\begin{equation}
	g_k(t)= - \frac{f_k(t)}{\sqrt{\int_0^t dt' |f_k(t')|^2}} , \quad k=1,2.
\end{equation}
Then, considering the definition of the temporal filter in \eqref{eq:BoxcarFun}, the time-dependent couplings (in the rotating frame of the drive)  are given by
\begin{equation}
\colorboxed{Maroon}{
		g_k(t)=-\frac{e^{i(\Delta_k t + \phi )}}{\sqrt{t-t_0}}, \quad k=1,2, 
}
\end{equation}
where $\Delta_k\equiv\omega_k-\omega_\sigma$ is the $k$th sensor-laser detuning. Note that in \reffig{ExperimentalMollowTriplet}, we showed the application of this formalism for the observation of the Mollow triplet. 
For this example, only one of the fiducial cavities, such as  the one labelled as $1$, is relevant, while the other can be traced-out since only photon processes---those involving a single frequency---are considered. 
The filled circles are the measured data, while the black solid lines are the simulation results, showing both a good agreement.

We note that related previous works~\cite{LopezCarrenoFrequencyresolvedMonte2018,LopezCarrenoEntanglementResonance2024} used a
version of the master equation in \eqref{eq:molmermastereq} that includes a factor $1/\sqrt{2}$ in the last term---the cascaded coupling term---. Such a factor would stem
from a description in which, prior to the capture of each mode, the output is physically split (e.g. by a beam splitter). In that description---which will be studied in the next section devoted to quantum metrology---, the state obtained for each mode depends strongly on the total number of modes included, since the physical splitting has the effect of introducing important vacuum contributions. This would not accurately describe the digital filtering of the modes performed in this experiment, where the particular choice of $f_1$ should not affect the results obtained when filtering $f_2$. 
%
%
Despite the master equation in \eqref{eq:molmermastereq} does not follow the cascaded multi-mode setup proposed in \colorref{KiilerichInputOutputTheory2019,KiilerichQuantumInteractions2020}, it ensures independence of the filtered modes, and therefore serves as a good description of the
type of mode-matching performed in this experiment


\subsection{Generation of entanglement: Theory}
\label{sec:Gen_ent_theory}

To verify the generation of entanglement between two selected and orthogonal propagating modes, we match the measured time trace simultaneously using two temporal filters, $f_1(t)$ and $f_2(t)$, whose temporal profiles are perfectly overlapping $v_1(t)=v_2(t)$ while their frequencies, $\omega_1$ and $\omega_2$, are independently varied. 

In order to quantify the degree of entanglement between the two filtered temporal modes, we make use of the logarithmic negativity~\cite{VidalComputableMeasure2002,PlenioLogarithmicNegativity2005,PlenioIntroductionEntanglement2007,HorodeckiQuantumEntanglement2009}---see \refsec{sec:Conc_and_Negat}---. We recall that given a general bipartite state, composed by systems $A$ and $B$, the logarithmic negativity is defined as
\begin{equation}
	E_\mathcal{N}\equiv \log_2(||\rho^{T_A}||_1),
\end{equation}
where $\text{T}_A$ denotes the partial transpose operation over system $A$, and $||\cdot||_1$ is the trace norm operation. 
%
%
%
%
Additionally,  we also compute the photon-photon correlations of the filtered emission via the zero-delay two-photon cross-correlation$^{\textcolor{Maroon}{*}}$~\cite{DelValleTheoryFrequencyFiltered2012,Gonzalez-TudelaTwophotonSpectra2013}
\graffitoshifted[-2cm]{
	${}^{*}$Note that we have slightly modified the notation for the zero-delay two-photon cross-correlation function introduced in \eqref{eq:FreqResolvedFun}, 
	 $$
	 g_\Gamma^{(2)}(\omega_1,\omega_2 ) \rightarrow g_{1,2}^{(2)}.
	 $$
	This change is made purely for convenience in this section.
}
\begin{equation}
	g_{1,2}^{(2)}=\frac{\langle  \hat a_1^\dagger  \hat a_2^\dagger \hat a_1 \hat a_2\rangle}{\langle \hat a_1^\dagger \hat a_1\rangle \langle \hat a_2^\dagger \hat a_2\rangle}.
\end{equation}

By means of these magnitudes, $E_{\mathcal{N}}$ and $g_{1,2}^{(2)}$, we study the effects of varying the driving strength of the microwave pulse ($\Omega$), the frequency of the filters ($\Delta_k$), and the duration of the measurement ($T$). Additionally, we also explore the possibility of performing non-overlapping temporal measurements.

\paragraph{Entanglement in the frequency domain. }
By studying the range of frequencies in the filtering process, we unveil a rich structure in the generation of entanglement and two-photon cross-correlations, as depicted in \reffig{FigChapt6_EntanglementMap}~\textcolor{Maroon}{(a, b)}, respectively. In the frequency map, we observe two main regions of non-zero logarithmic negativity:
%
%
\begin{enumerate}[label=\textcolor{Maroon}{(\roman*)}]
	\item The diagonal, corresponding to the condition $\Delta_2= \Delta_1$.
	\item The antidiagonal, corresponding to the $\Delta_2= - \Delta_1$ .
\end{enumerate}
However, as we will see below, only the latter is relevant to this work.

\begin{SCfigure}[1][h!]
	\includegraphics[width=1.\columnwidth]{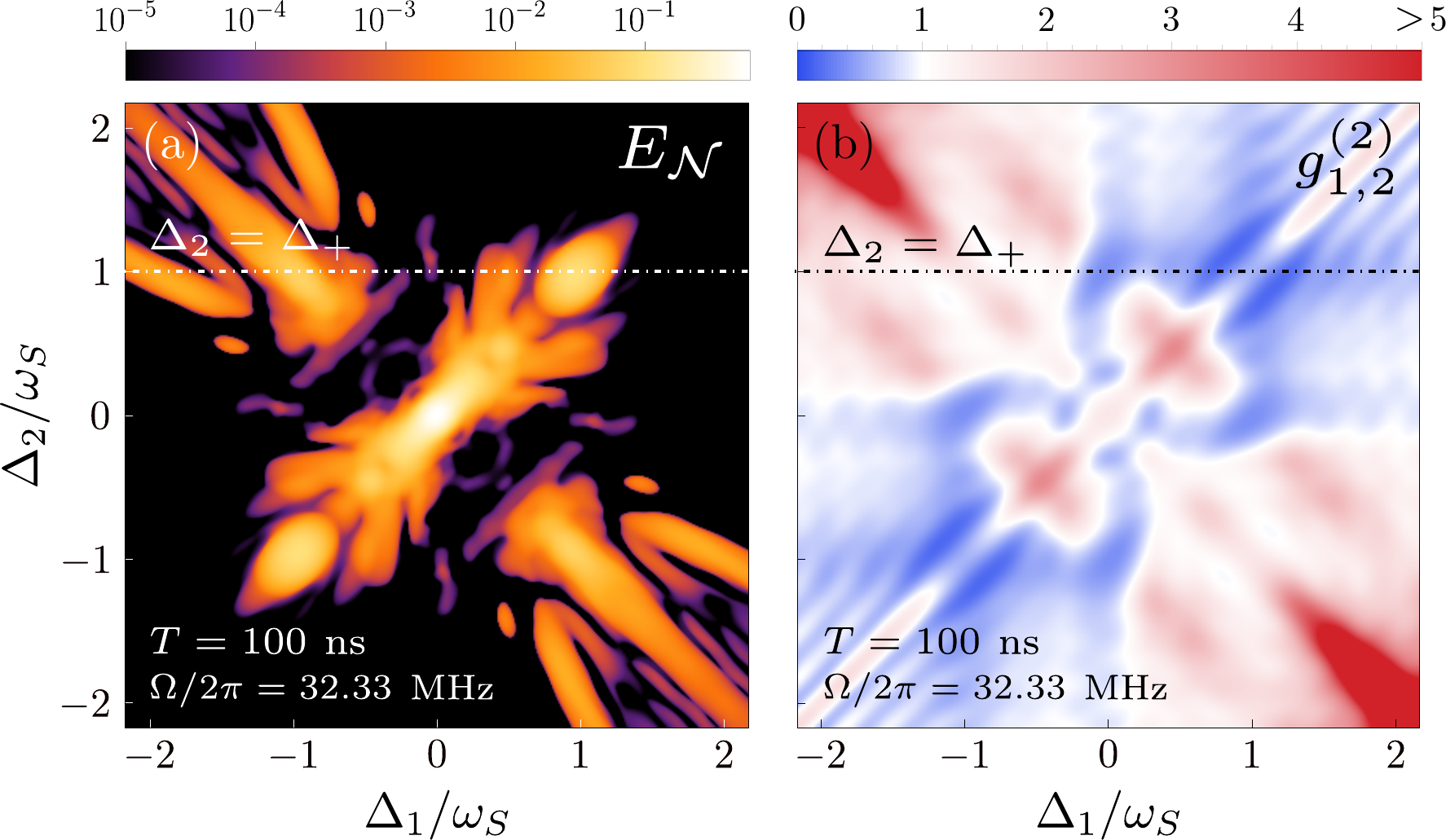}
	\captionsetup{justification=justified}
\end{SCfigure}
\graffito{\vspace{-7.6cm}
	\captionof{figure}[ Frequency-resolved logarithmic negativity  $E_{\mathcal{N}}$ and two-photon correlation function $g^{(2)}_{1,2}$.]{	\label{FigChapt6_EntanglementMap}
		\textbf{Frequency-resolved logarithmic negativity  $E_{\mathcal{N}}$ and two-photon correlation function $g^{(2)}_{1,2}$. } 
		(a) $E_{\mathcal{N}}$ and (b) $g^{(2)}_{1,2}$ in the frequency domain.
		The white (black) dot-dashed lines, respectively, in these plots denote the resonant frequency $\Delta_2 = \Delta_+=\Omega$ (Mollow sideband) which is analysed in \reffig{Entanglement_Map_1D_v1}.
		Parameters: $\Gamma/2\pi=8$ MHz, $\Omega/2\pi=32.33$ MHz, $\Delta=0$, $t_0=200$ ns, $T=100$ ns.
	}
}

The diagonal section entails a breakdown of the orthogonality assumption between modes showed in \eqref{eq:orthogonalityCond1}, which we consider a basic requirement for discussing entanglement between two different \textit{systems}.
More generally, this assumption will be compromised in sections where
\begin{equation}
|\Delta_1-\Delta_2|<\Delta \omega \quad \text{or} \quad \Omega < \Gamma,
\end{equation}
where $\Delta \omega$ is the frequency resolution inherited by the filter. In such regions, the reported values of entanglement cannot be understood as \textit{actual measurement} of entanglement, since spectral resolution is lost and the collected modes become indistinguishable in frequency.  
Nevertheless, this information reveals the amount of entanglement achievable if the whole output field, $\hat a_\text{out}$, is physically split into two orthogonal modes (e.g., by a beam splitter) prior to filtering, in which case entanglement between any two pair of temporal modes extracted from each of the split outputs would be well-defined. As noted earlier when we introduced the cascaded master equation in \eqref{eq:molmermastereq}, this alternative setup requires a slightly different master equation accounting for the vacuum contributions introduced by splitting the signal. Therefore, regimes in which $\Omega >\Gamma$ and $|\Delta_1-\Delta_2|>\Delta \omega$ are the most interesting since they provide a well-resolved emission in different spectral lines, resulting in well-defined entanglement. 

\begin{SCfigure}[1][h!]
	\includegraphics[width=0.7\columnwidth]{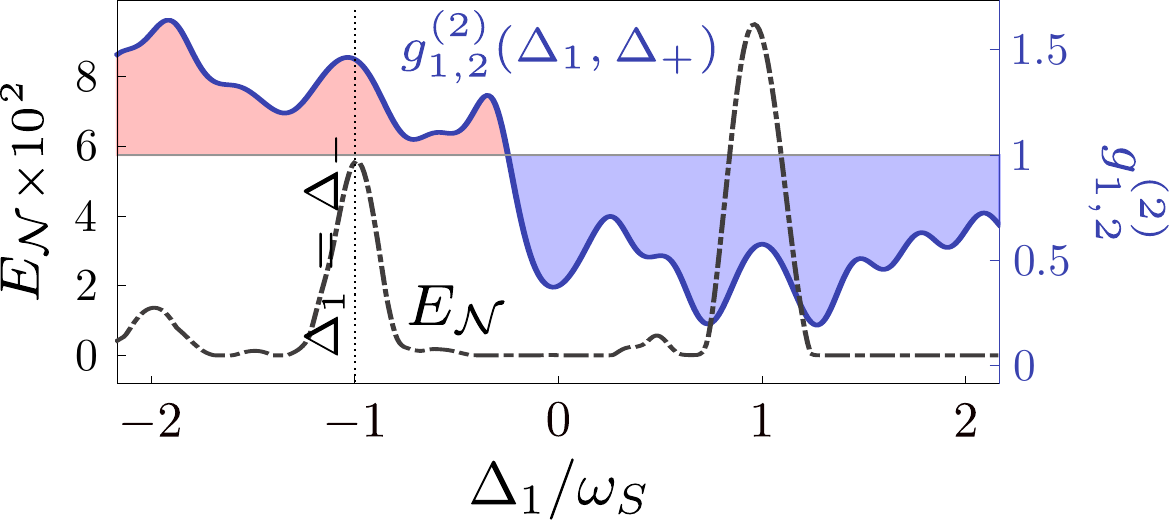}
	\captionsetup{justification=justified}
	\caption[ Correspondence between logarithmic negativity  $E_{\mathcal{N}}$ and two-photon correlation function $g^{(2)}_{1,2}$.]{ \label{Entanglement_Map_1D_v1}
		\textbf{ Correspondence between logarithmic negativity  $E_{\mathcal{N}}$ and two-photon correlation function $g^{(2)}_{1,2}$. } 
		$E_\mathcal{N}$ (black dotted-dashed line) and $g_{1,2}^{(2)} (\Delta_1, \Delta_+)$ (blue solid line) in terms of the filter detuning $\Delta_1$ [$\Delta_2=\Delta_+$,
		dotted-dashed line in \reffig{FigChapt6_EntanglementMap}]. The vertical dotted line denotes the resonant frequency, $\Delta_1=\Delta_-$.
	}
\end{SCfigure}
%
Following the discussion presented in \refsec{sec:cascaded}, the frequency-resolved two-photon cross-correlation map depicted in \reffig{FigChapt6_EntanglementMap}~\textcolor{Maroon}{(b)} exhibits the typical frequency pattern of a coherently driven TLS, where the antidiagonal features a complex structure of correlations due to the multi-photon processes occurring at these frequencies. In fact, in  \reffig{Entanglement_Map_1D_v1}, we observe a clear correspondence between entanglement and bunching around the frequency resonance$^{\textcolor{Maroon}{*}}$
\graffito{$^{*}$We remind that the energy splitting is defined as $\omega_S\equiv 2 |E_\pm|$, where $E_\pm$ are the dressed-state eigenenergies of the coherently driven TLS [see \eqref{eq:MollowEnergy}]. In this case, having set $\tilde{\Omega}=i\Omega/2$ in \eqref{eq:TLS_Hamiltonian}, the energy splitting reads $\omega_S=\Omega$. }
, specifically at $(\Delta_1,\Delta_2)=(\Delta_-,\Delta_+)$, where $\Delta_\pm \equiv \pm \omega_S=\pm \Omega$.

In the following, we focus on the frequency points
\begin{equation}
	(\Delta_1, \Delta_2)= (\Delta_-,\Delta_+), \quad \text{or} \quad 	(\Delta_1, \Delta_2)= (\Delta_+,\Delta_-),
\end{equation}
where the two modes belong to opposite side peaks of the Mollow triplet [see \reffig{FigChapt6_MollowRegime}~\textcolor{Maroon}{(a)}] and satisfy the frequency orthogonality condition in \eqref{eq:orthogonalityCond1}.

\begin{SCfigure}[1][h!]
	\includegraphics[width=0.72\columnwidth]{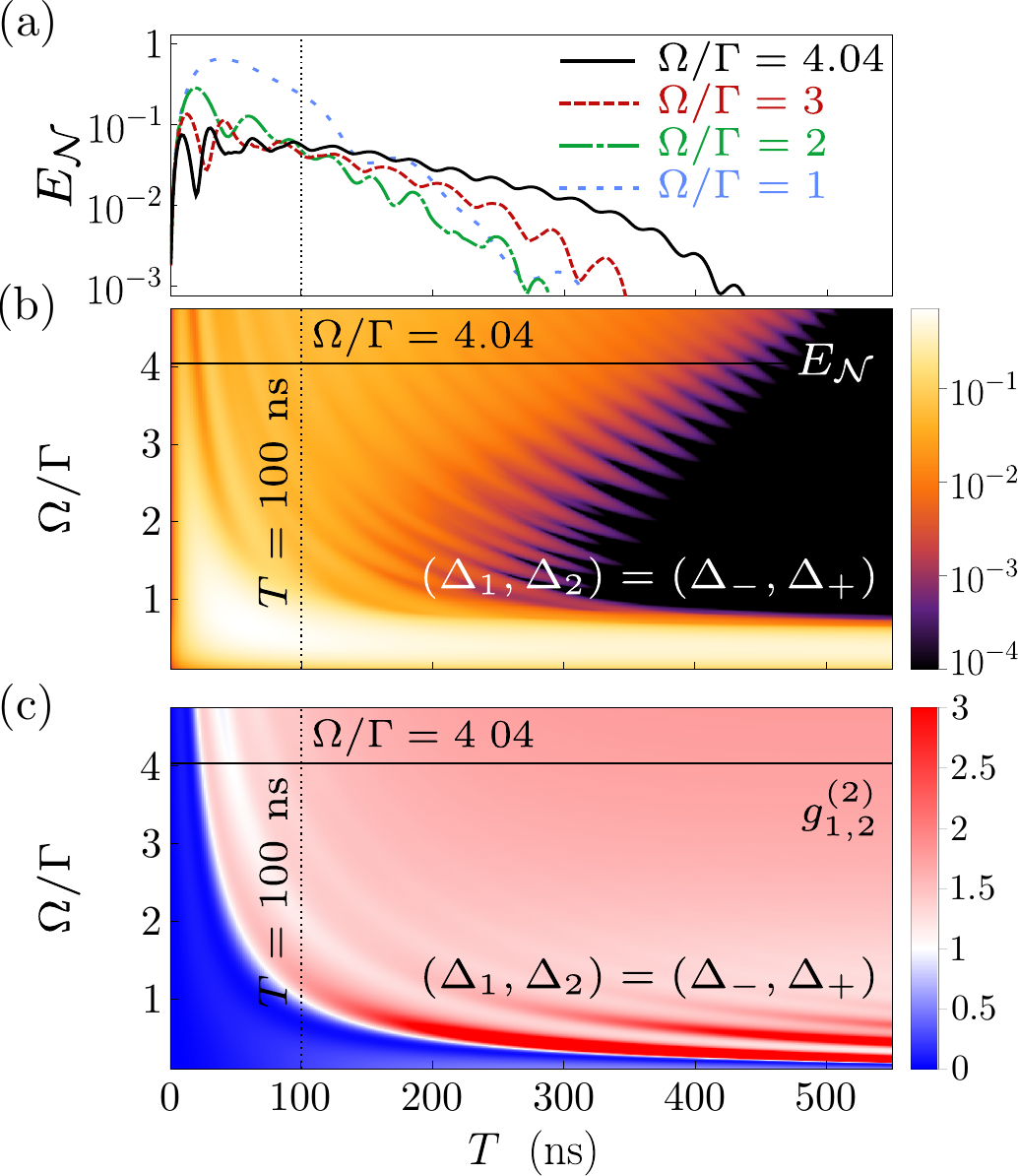}
	\captionsetup{justification=justified}
	\caption[Temporal behaviour of the logarithmic negativity and two-photon correlation function in terms of the driving strength.]{
		\label{FigChapt6_EntanglementTemporal_Omega}
		\textbf{Temporal behaviour of the logarithmic negativity and two-photon correlation function in terms of the driving strength.}
		(a) $E_\mathcal{N}$ at opposite sidebands [$(\Delta_1,\Delta_2)=(\Delta_-,\Delta_+)$, equivalently in (b, c)] in terms of $T$ for several pulse intensities.
		(b, c) $E_\mathcal{N}$ and $g_{1,2}^{(2)}$, respectively, in terms  of $\Omega$ and  $T$.
		The vertical dot-line and horizontal solid line correspond to the experimental values used in the experiment: $(\Omega,T)=(32.32\ \text{MHz},100 \ \text{ns})$.\textbf{} 
	}
\end{SCfigure}

\paragraph{Effects of the filtering time. }
In addition to the frequency dependency on the generation of entanglement,
 there is another crucial aspect: the filtering time $T$ in the measurement process [see \reffig{Measurement}].

The dependence on $T$ in all the results in \reffig{FigChapt6_EntanglementTemporal_Omega} and \reffig{FigChapt6_EntanglementTemporal_Detuning} exhibits similar behaviour. The boxcar
 filter---introduced in \eqref{eq:BoxcarFun}---imposes an approximate bandwidth$^{\textcolor{Maroon}{*}}$ of
 \graffito{
 	${}^{*}$The Fourier transform of the boxcar filter can be easily derived using the properties of the Heaviside step function:
 	$$
 	v(\omega)= \sqrt{\frac{T}{2\pi }} \text{sinc}\left(\frac{\omega T}{2}\right).
 	$$
 	The sinc function, defined as $\text{sinc}(x)\equiv \sin(x)/x$, has its first zeros at $x=\pm \pi$. Consequently, the bandwidth of the central peak is approximately $\Delta \omega \sim 2\pi /T$.
 }
\begin{equation}
	\Delta\omega\sim  \frac{2\pi }{T},
	\label{eq:bandwidth}
\end{equation}
corresponding to the width of the central peak of the sinc function, such that in the limit of infinite frequency precision, $T\rightarrow \infty$, the filtered modes lose their time definition, causing entanglement to vanish.
Conversely, very short measurement times, $T\rightarrow 0$, correspond to broadband detection (colour-blind) [see right-panel in \reffig{ExperimentalMollowTriplet}], resulting in a non-zero logarithmic negativity that does not reflect an actual measure of entanglement, as we previously discussed. Additionally, in this limit, we recover the expected antibunching of a two-level system~\cite{Gonzalez-TudelaTwophotonSpectra2013}.
We also note that higher values of $\Omega$ extend the lifetime of the entanglement [see  \reffig{FigChapt6_EntanglementTemporal_Omega}~\textcolor{Maroon}{(a, b)}]. 

In the limit of long $T$ (i.e., very narrow frequency filtering), we observe that the photon statistics is not completely uncorrelated, $\lim_{ T \rightarrow \infty} g^{(2)}
_{1,2} (\Delta_-, \Delta_\pm) \neq 1$, but bunched $(\lesssim 2)$ [\reffig{FigChapt6_EntanglementTemporal_Omega}~\textcolor{Maroon}{(c)} and \reffig{FigChapt6_EntanglementTemporal_Detuning}~\textcolor{Maroon}{(c)}]. This feature was already reported in \textcolor{Maroon}{Refs.}~\cite{AspectTimeCorrelations1980, DalibardCorrelationSignals1983, BelTheoryWavelengthResolved2009,Gonzalez-TudelaTwophotonSpectra2013}, where it is shown that correlations between opposite sidebands feature bunching when a perfect laser is used to drive the TLS. To recover the expected limit of uncorrelated photons when 
$\Delta_1 = \pm \Delta_2$, a more realistic model
of a laser must be considered, such as an one-atom laser~\cite{MuOneatomLasers1992}.

\begin{SCfigure}[1][h!]
	\includegraphics[width=0.72\columnwidth]{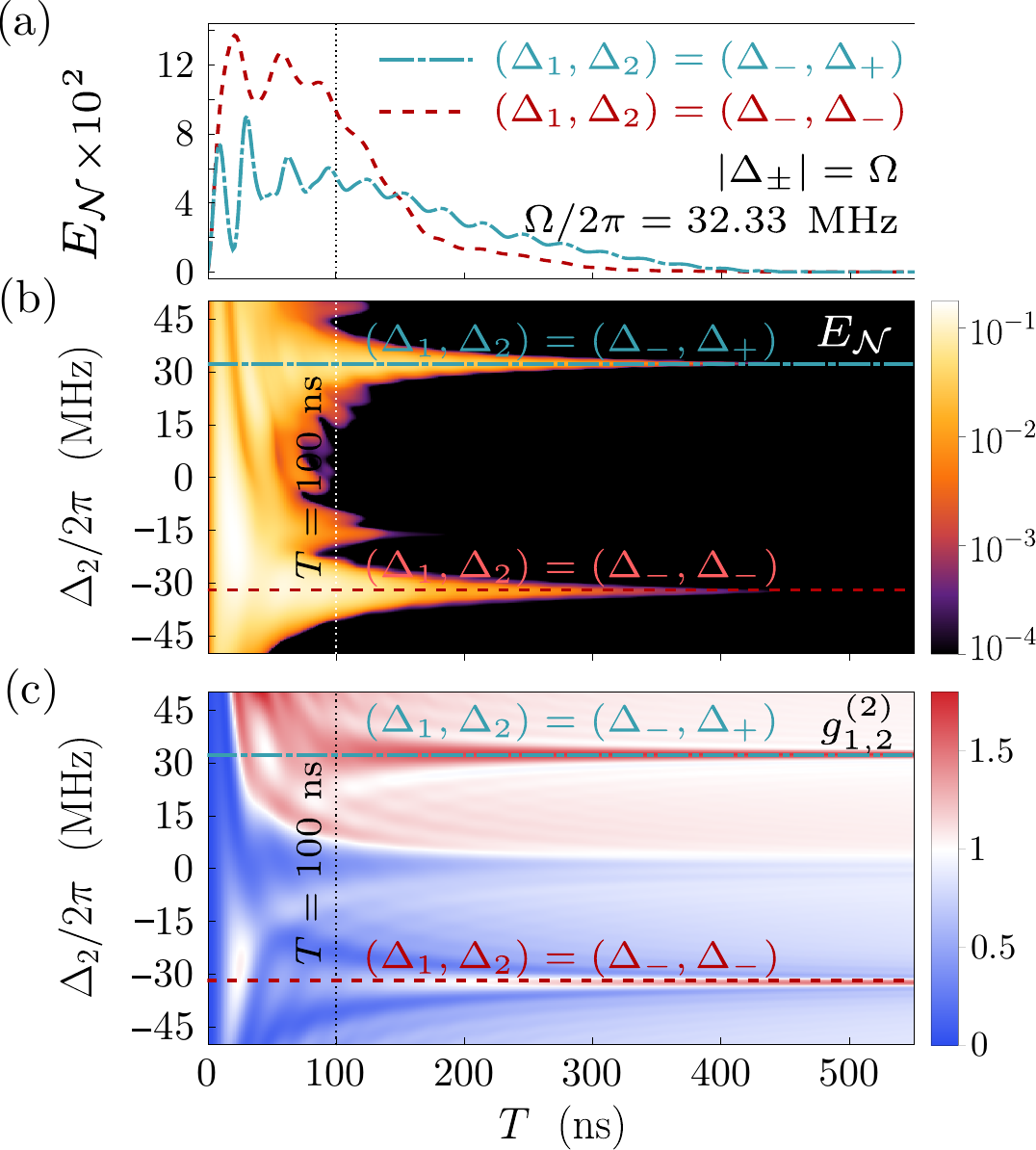}
	\captionsetup{justification=justified}
	\caption[Temporal behaviour of the logarithmic negativity and two-photon correlation function in terms of the sensor detuning.]{
		\label{FigChapt6_EntanglementTemporal_Detuning}
		\textbf{Temporal behaviour of the logarithmic negativity and two-photon correlation function in terms of the sensor detuning.}
		(a) $E_\mathcal{N}$ in terms of $T$ for the two sideband resonances, $(\Delta_1,\Delta_2)=(\Delta_-,\Delta_\pm)$, in blue dot-dashed and red dashed lines, respectively.
		(b, c) $E_\mathcal{N}$ and $g_{1,2}^{(2)}(\Delta_-,\Delta)$, respectively, in terms of $\Delta_2$ and $T$.
		The vertical dot-line and horizontal solid line correspond to the experimental values used in the experiment: $(\Omega,T)=(32.32\ \text{MHz},100 \ \text{ns})$.
		The vertical dot-line and horizontal solid line correspond to the experimental values used in the experiment: $(\Omega,T)=(32.32\ \text{MHz},100 \ \text{ns})$.\textbf{} 
	}
\end{SCfigure}

In \reffig{FigChapt6_EntanglementTemporal_Detuning}~\textcolor{Maroon}{(a, b)}, we see that when one of the filters is fixed on a Mollow sideband, e.g., $\Delta_1 = \Delta_-$, the collected emission by the other filter features two main resonances that contribute to the generation of entanglement. These resonances correspond to: \textcolor{Maroon}{(i)} opposite sidebands, $(\Delta_1, \Delta_2)=(\Delta_-,\Delta_+)$ (in blue), and \textcolor{Maroon}{(ii)} identical sidebands,
$(\Delta_1, \Delta_2)=(\Delta_-,\Delta_-)$ (in red). However, as previously discussed, the nature of these entangled states differs since only the first case involves entanglement between two distinct, well-resolved light modes

\paragraph{Non-overlapping temporal measurements.} 
Throughout this discussion, we have assumed that the temporal profiles of the two modes, $v_1(t)$ and $v_2(t)$, were perfectly overlapping, with their frequencies independently tunable.
Here, we explore how relaxing the condition of temporal overlap affects the entanglement structure. Specifically, we consider a scenario where the boxcar modes in the time domain are temporally delayed relative to each other. To do this, we perform simulations where the starting time of the filter for the second temporal mode (TM2) is adjusted to 
$
	t_0  + T_\text{Delay},
$
with $T_\textrm{Delay}$ varied within the time range $[-0.125, 0.125] \, \mu\text{s}$.
The effect of this temporal delay on the entanglement is shown in \reffig{FigTimeDelay}, where we compute the logarithmic negativity as a function of $T_\text{Delay}$ and the measurement time $T$, while fixing the filters at opposite sidebands $(\Delta_1, \Delta_2) = (\Delta_-, \Delta_+)$.

\begin{SCfigure}[1][h!]
	\includegraphics[width=0.9\columnwidth]{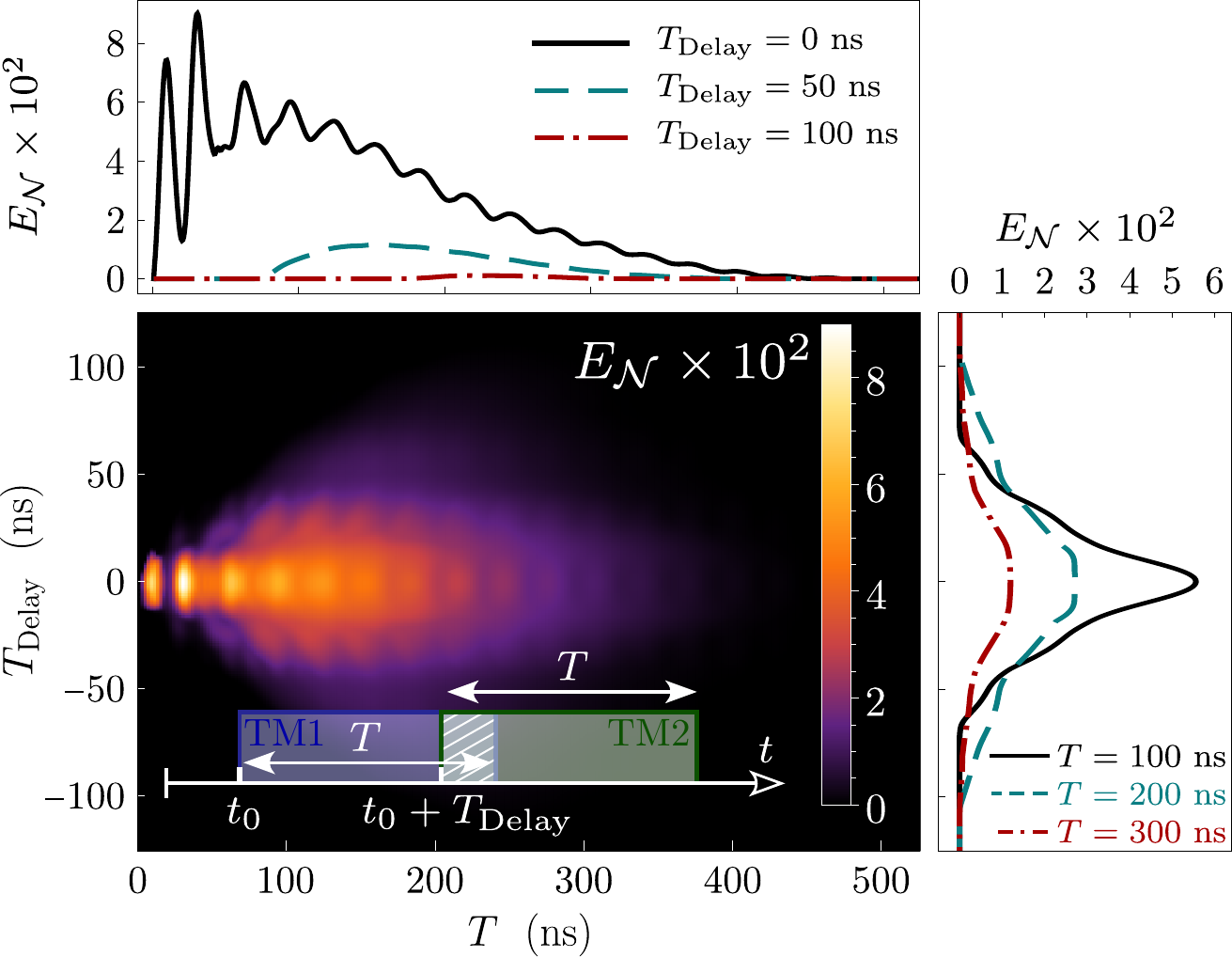}
	\captionsetup{justification=justified}
	\caption[Non-overlapping temporal measurements.]{
			\label{FigTimeDelay}
			\textbf{Non-overlapping temporal measurements.}
 Logarithmic negativity $E_\mathcal{N}$ versus the time delay $T_{\text{Delay}}$ between two temporal boxcar filters and their durations $T$.  The inset illustrates the time sequence of the applied temporal filters: the first filter, TM1, starts at a fixed time $t_0=200$ ns, while the starting time of the second filter, TM2, is varied. Top panel shows the dependence of $E_\mathcal{N}$ on $T$ for selected values of $T_\textrm{Delay}=0, \ 50, \ 100$ ns. (c) Dependence of $E_\mathcal{N}$ on $T_\textrm{Delay}$ for selected values of $T=100, \ 200, \ 300$ ns. }
\end{SCfigure}

As can be observed in the central panel in \reffig{FigTimeDelay}, the logarithmic negativity displays a symmetric decaying pattern centered at zero delay ($T_\text{Delay} = 0$), corresponding to the case of perfect temporal overlap. This behavior can be intuitively understood using the physical insight illustrated in the inset: as one boxcar mode is delayed backward, or forward, in time relative to the other, the temporal overlap decreases, resulting in weaker correlations between the selected modes. This is illustrated in the upper panel of \reffig{FigTimeDelay}.

Furthermore, we observe that when the orthogonality condition is satisfied (e.g., for $T \gtrsim 100$ ns, as shown in \reffig{FigTimeDelay}), and the photonic modes are fully resolvable, the survival of entanglement with respect to the time delay $T_{\text{Delay}}$ becomes more robust for larger measurement times. However, as the measurement time $T$ approaches infinity, the entanglement eventually vanishes.  This phenomenon can be explained by the time-frequency relation imposed by the boxcar filter bandwidth in \eqref{eq:bandwidth}: as the measurement time $T$ increases, one of the boxcar filters narrows while the other remains broader as a consequence of the time delay.  Consequently, the broader filter can continue to overlap with the delayed mode as long as $T_\text{Delay} < T$.

\subsection{Entanglement in the frequency domain: Experiment}


\paragraph{Measurements of moments and correlations. }To experimentally validate the generation of entanglement between these two photonic modes, perfectly overlapping temporal profiles are applied while varying their frequencies independently within the range $[\omega_{\text{ge}} - 40\ \text{MHz}, \omega_{\text{ge}} + 40\ \text{MHz}]$.
For each pair of filter frequencies, first- and second-order moments of the two propagating modes are obtained, as shown in \reffig{Fig2Dmoments}. The left columns show  the simulated results derived from the theoretical model in \eqref{eq:molmermastereq}, while the right columns display the experimentally measured results. 
%
%
\begin{SCfigure}[1][h!]
	\includegraphics[width=1\columnwidth]{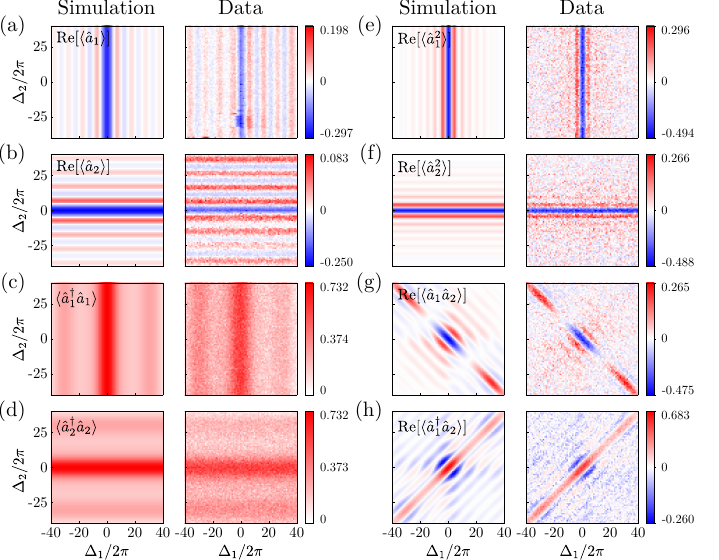}
	\captionsetup{justification=justified}
	\caption[Moments and correlations of temporally matched modes.]{
		\label{Fig2Dmoments}
		\textbf{Moments and correlations of temporally matched modes.}
		Moments up to second order for the temporal modes defined in \eqref{eq:temporal_filter} as a function of frequency detunings $\Delta_k$  ($k = 1, 2$). The left columns shows the simulations and the right columns the data measured
		from the experiments.
	}
\end{SCfigure}

In the frequency regions around $(\Delta_1,\Delta_2)=(\Delta_-, \Delta_+)$ [with $|\Delta_\pm|\approx 32.33$ MHz], that is, near the upper-left anti-diagonal corners of the frequency maps, the first order moments  $\langle \hat a_1 \rangle$ and $\langle \hat a_2 \rangle$ are close to zero [see \reffig{Fig2Dmoments}~\textcolor{Maroon}{(a, b)}], while the cross-second-order moment $\left \langle \hat a_1 \hat a_2 \right \rangle$ shows a peak [see \reffig{Fig2Dmoments}~\textcolor{Maroon}{(g)}]. These measurements signal the generation of a \textit{two-mode squeezing type of entanglement} between the two photonic modes, that is, an entangled state of the form:
\begin{equation}
	|\psi \rangle  \sim c_{00} |00\rangle + c_{11} |11\rangle, \quad \text{(Two-mode squeezed entangled state)}
	\label{eq:TwoModeEntangled}
\end{equation}
where $|ij\rangle \equiv |i\rangle \otimes |j\rangle$ represents the number of photons in each mode, and $c_{ij}$ the corresponding coefficients such that $\sum_{ij} |c_{ij}|^2=1$.
On the other hand, in the region near $(\Delta_1,\Delta_2)=(\Delta_+, \Delta_+)$ or $(\Delta_-, \Delta_-)$, along the diagonal of the frequency moment maps, the cross-second-order moment $\langle \hat a_1 ^\dagger \hat a_2 \rangle$ shows a peak [see \reffig{Fig2Dmoments}~\textcolor{Maroon}{(h)}]. In this case, the measurements of these moments indicate a \textit{two-mode beam-splitter (BS) type of entanglement}, 
\begin{equation}
 	|\psi \rangle  \sim c_{01} |01\rangle + c_{10} |10\rangle, \quad \text{(Two-mode BS entangled state)}.
\end{equation}
Note, however, that in this scenario, even if the cross-second order moment signals the formation of entanglement,
it does not satisfy the frequency orthogonality condition in \eqref{eq:orthogonalityCond1}, and thus it cannot be interpreted as an \textit{actual} measurement of entanglement.  
Then,  the frequency point $(\Delta_1,\Delta_2)=(\Delta_-,\Delta_+)$ is the most promising region for entanglement generation, satisfying the necessary conditions for establishing a well-defined measure of entanglement---as we discussed before in \refsec{sec:Gen_ent_theory}---.


\paragraph{Tomography. }
In order to demonstrate that the two modes are entangled around the frequency point of interest, $(\Delta_1,\Delta_2)=(\Delta_-, \Delta_+)$,  joint quantum state tomography is performed with $n =
2 \times 10^7$ repetitions across the frequency ranges $\Delta_1 \in [-40, -10]$ MHz and $\Delta_2\in [10, 40]$ MHz. Particularly, 27 moments---excluding conjugation redundancy---are computed:  $\langle (\hat{a}_1^{\dagger})^{m_1} \hat{a}_1^{n_1} (\hat{a}_2^{\dagger})^{m_2} \hat{a}_2^{n_2} \rangle$ for $m_1, n_1, m_2, n_2 \in \left \{  0, 1, 2\right \}$ and constraining up to $m_1+n_1+m_2+n_2 \leq 4.$
\begin{SCfigure}[1][b!]
	\includegraphics[width=1.\columnwidth]{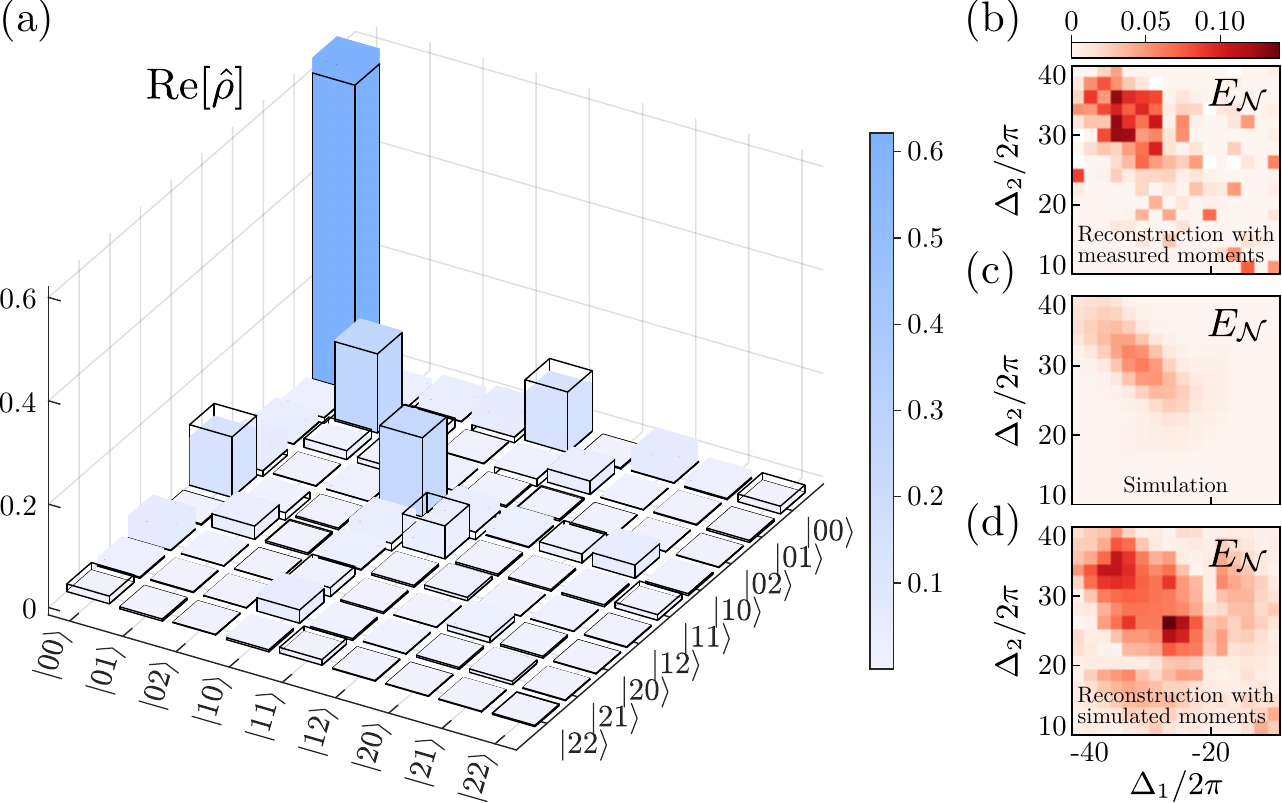}
	\captionsetup{justification=justified}
	\caption[Tomography of the entangled photonic modes.]{	\label{FigChapt6_Tomography}
		\textbf{Tomography of the entangled photonic modes.} 
		(a) Reconstructed density matrices at the point of maximum $E_\mathcal{N}$.
		(b) $E_\mathcal{N}$ reconstructed from $27$ measured
		moments over the frequency ranges $\Delta_1 \in [-40,-10]$ MHz and
		$\Delta_2 \in [10,40]$MHz. 
		(c) Simulated $E_\mathcal{N}$ over the same frequency range. 
		(d) $E_\mathcal{N}$ reconstructed from $27$ simulated moments over the same frequency ranges.
	}
\end{SCfigure}

\newpage

In the reconstructed density matrix [see \reffig{FigChapt6_Tomography}~\textcolor{Maroon}{(a)}], multiple photon states are involved. While the component $\langle 00| \hat \rho |11 \rangle$ is the largest, additional off-diagonal elements are non-zero, along with extra diagonal elements, $\langle 01| \hat \rho |01 \rangle$ and $\langle 10| \hat \rho |10 \rangle$. These contributions disrupt the formation of a purely entangled state of the \textit{two-mode squeezing type} described in \eqref{eq:TwoModeEntangled}. The fidelity~\cite{JozsaFidelityMixed1994}---the state overlap---between the reconstructed and the simulated density matrices is $96.6\%$
%
%
%
%
%

From the reconstructed density matrix, the logarithmic negativity $E_\mathcal{N}$ is straightforwardly obtained [see \reffig{FigChapt6_Tomography}~\textcolor{Maroon}{(b)}]. 
The maximum value of $E_\mathcal{N}$ from the reconstruction near the frequency point $(\Delta_-, \Delta_+)$ takes the value $E_\mathcal{N}=0.128$, while the simulation yields a maximum value of $0.062$ at the same point [see \reffig{Fig2Dmoments}~\textcolor{Maroon}{(c)}].  
The computation of $E_\mathcal{N}$ is very sensitive to small values of coherences in the density matrix, specially when its value is relatively low, as in our case. Therefore, extracting the theoretical value of $E_\mathcal{N}$ from a reconstructed density matrix is particularly challenging.
We hypothesize that the discrepancy between the simulated and measured $E_\mathcal{N}$ arises from using only $27$ moments (up to fourth order) instead of the full $325$ moments required for a Fock-space cutoff at $N=5$, which includes all cases fulfilling $m_1, n_1, m_2, n_2 \in \left \{  0, 1, 2, 3, 4\right \}$. 
Notably, when reconstructing the state using the same $27$ moments as in the experiment but with simulated values, the resulting $E_\mathcal{N}$ is consistently higher and closely matches the experimental value [see \reffig{FigChapt6_Tomography}~\textcolor{Maroon}{(d)}].
In contrast,  if all 325 combinations of simulated moments are used [not shown here], the  $E_\mathcal{N}$ distribution obtained directly from the simulation is reproduced~\cite{YangEntanglementPhotonic2025}. Based on these observations, we conclude that the maximum $E_\mathcal{N}$ observed in the experiment closely match the simulated value of $0.062$, and the higher experimental value likely reflects limitations imposed by the reduced set of moments used for the reconstruction process and the noise of the measurements.


\vspace{-3.5mm}
\subsection{Conclusion}
\vspace{-3.5mm}

In this work, we have introduced an approach for generating entanglement in the time-frequency domain between propagating bosonic modes. 
Specifically, we theoretically proposed and experimentally confirmed the generation of entanglement from steady-state resonance fluorescence in a continuously and resonantly driven quantum emitter. This was achieved by temporally matching two photonic modes corresponding to the opposite sidebands of the Mollow triplet.
%
%
Our study provides distinct and complementary contributions to the ones previously presented in \colorref{LopezCarrenoEntanglementResonance2024}.

Compared to previous works~\cite{GasparinettiCorrelationsEntanglement2017,WangDeterministicEntanglement2011,LangCorrelationsIndistinguishability2013,NarlaRobustConcurrent2016,KannanGeneratingSpatially2020,BesseRealizingDeterministic2020}, our approach has two main advantages: \textcolor{Maroon}{(i)} it uses a simple system; and \textcolor{Maroon}{(ii)} the entangled modes are generated at the steady-state, eliminating timing constraints for applying temporal filters.
Additionally, our method naturally incorporates multi-photon states, in which the number of photons is determined by the mode shape and output intensity of the quantum emitter. As such, there is potential to explore and optimize the mode shape to obtain states with higher entanglement or other desired properties, e.g., using Hermite-Gauss modes instead of boxcar modes~\cite{BrechtPhotonTemporal2015,RaymerTemporalModes2020}. 

\section[Quantum metrology through spectral measurements in \protect \newline quantum optics]{Quantum metrology through spectral \protect \newline measurements in quantum optics}
\sectionmark{Quantum metrology through spectral measurements in ...}
\label{Chapt6_SecMetrology}

As we have seen throughout this Thesis, even the simplest quantum optical systems can emit light with highly complex spectral properties. For instance, in \refch{ch:TwoPhotonResonance}, we discussed how the emergence of dressed states in coherently driven quantum emitters lead to rich fluorescence spectra with multiple peaks, reflecting the complex structure of eigenstates. These spectra exhibit equally complex dependencies on the parameters that govern the dynamics of the system, providing a pathway to enhance the inference of unknown parameters by frequency-filtering the emitted signal.

Building on the idea that hybridized light-matter systems exhibit complex correlations in frequency space~\cite{Gonzalez-TudelaTwophotonSpectra2013,SanchezMunozViolationClassical2014} [see \refsec{sec:cascaded}], we discuss the role of quantum correlations and quantify their impact on the precision by which unknown atomic parameters can be estimated. Specifically, we explore the potential of frequency-resolved correlation measurements for the task of parameter estimation in driven-dissipative quantum optical systems, focusing on a single coherently driven TLS [see \refsec{sec:MollowTriplet}].

In this section, we provide a brief overview of ongoing work, introducing the concept of \textit{frequency-resolved quantum metrology}. The results presented here are preliminary$^{\textcolor{Maroon}{*}}$. 
\graffito{
	$^{*}$  This work is now published as an arXiv preprint~\cite{Vivas-VianaQuantumMetrology2025}.
}

\subsection{Model}

In this section we still make use of the sensor-cascaded formalism to analyse the properties of the emitted light by the quantum emitter. 
While the one-sensor cascaded master equation remains identical to that presented in \eqref{eq:cascadedmaster1}, the two-sensor case introduces a slight variation compared to the model discussed in the previous section [see \eqref{eq:molmermastereq}]. 
In the earlier setup, the photonic mode-capturing process was implemented in a post-processing stage. Here, instead, we consider a physical implementation where, before capturing each mode,  the output signal is physically split (e.g., by a beam splitter),
\begin{equation}
\colorboxed{Maroon}{
\begin{aligned}[b]
	\frac{d \hat \rho }{dt}= -i [&\hat H_\sigma+ \Delta_1 \hat a_1^\dagger \hat a_1+ \Delta_2 \hat a_2^\dagger \hat a_2, \hat \rho] + \frac{\gamma}{2} \mathcal{D}[\hat \sigma] \hat \rho 
	\\
	&+ \sum_{k=1}^2 \frac{\Gamma_k}{2} \mathcal{D}[\hat a_k]\hat \rho	-
	\sqrt{\gamma \Gamma_k/2} \left\{ [\hat a_k^\dagger, \hat \sigma  \hat \rho ] + [\hat \rho \hat \sigma^\dagger, \hat a_k] \right\}.
\end{aligned}
}
	\label{Eq_Chapt6_CascadedMetrology}
\end{equation}
The Hamiltonian $\hat{H}_\sigma$ is the one presented in \eqref{eq:TLS_Hamiltonian}, with $\tilde{\Omega}=\Omega$ purely real ($\Omega \in \mathbb{R}$). Note that, besides the time-dependent coupling coming from the implementation of the \textit{input-output theory for quantum pulses}~\cite{KiilerichInputOutputTheory2019,KiilerichQuantumInteractions2020}, the essential difference between \eqref{eq:molmermastereq} and \eqref{Eq_Chapt6_CascadedMetrology} relies on the factor $1/\sqrt{2}$ in the cascaded coupling term, corresponding to vacuum contributions after splitting the signal in the beam splitter.

\subsection{Quantum vs classical fluctuations.}

In \refsec{sec:4-parameter-estimation}, we already introduced a metrological application based on photon counting measurement via a fiducial sensor~\cite{DelValleTheoryFrequencyFiltered2012}, such that, in practice, we were considering fluorescence spectrum measurements as our POVM~\cite{ParisQUANTUMESTIMATION2009,Demkowicz-DobrzanskiQuantumLimits2015,PolinoPhotonicQuantum2020,BarbieriOpticalQuantum2022} [see \reffig{MetrologicalProtocolOneSensor} for an illustration of the protocol]. Specifically, in that section, we followed the approach presented in \colorref{DelaubertQuantumLimits2008}, where the detector is assumed to follow a conditioned Poissonian probability distribution, 
\begin{equation}
		p(n|\theta)= \frac{\bar n(\theta)^n}{n!} e^{-\bar n(\theta)},
		\label{eq:PoissDistrib}
\end{equation}
with $\bar n (\theta)= \text{Tr}[\hat n \hat \rho(\theta)]= \sum_{n=0}^\infty n p (n|\theta)$ denoting the mean photon distribution conditioned on the parameter $\theta$. 
Intuitively, we can understand this approach if we consider a situation in which we cannot access the full information of the density matrix due to technological limitations in the detector, such as time resolution, giving access to only the classical properties of the field.

\begin{SCfigure}[1][h!]
	\includegraphics[width=1.\columnwidth]{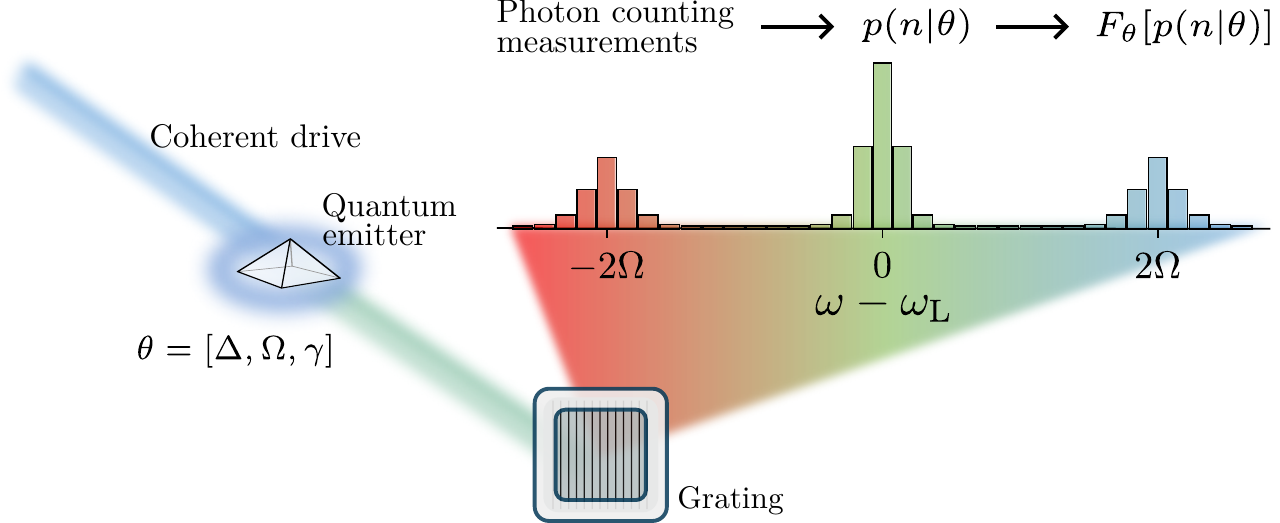}
	\captionsetup{justification=justified}
\end{SCfigure}
\graffito{\vspace{-5.50cm}
	\captionof{figure}[Metrological protocol for one sensor.]{		\textbf{Metrological protocol for one sensor.}
		\label{MetrologicalProtocolOneSensor}
		Illustration of the metrological protocol: a coherently driven quantum emitter (the source) whose output it's analyzed in the frequency domain with a grating, e.g., a Fabry-Perot filter (the target).
		The grating allows the extraction of spectral information, e.g., the fluorescence spectrum, and application of quantum parameter estimation techniques via photon-counting measurements.
		
	}
}

By introducing this conditional probability distribution, \eqref{eq:PoissDistrib}, in the general definition of the classical Fisher information, already introduced in \eqref{eq:FisherSpec},
\begin{equation}
	F_P= \sum_{n=0}^\infty \frac{1}{p(n|\theta)} [\partial_\theta p(n|\theta)]^2 = 
	\sum_{n=0}^\infty p(n|\theta) \left(\frac{n}{\bar n}-1\right)^2 [\partial_\theta \bar n(\theta)]^2,	
\end{equation}
we obtain the formal solution for what we term \textit{Poissonian Fisher information} (PFI):
\begin{equation}
	\colorboxed{Maroon}{
	F_P\equiv \frac{[\partial_\theta \bar n(\theta)]^2}{\bar n}.
	}	
	\label{eq:PoissonFisher}
\end{equation}

In the results presented in \refsec{sec:4-parameter-estimation}, we neglected the intrinsic nature of the quantum state, i.e., its quantum fluctuations. Hence, we ask ourselves if there is an actual enhancement in the acquisition of information when the full state description state is considered in contrast to considering a Poissonian description. To give an answer to this, we consider a simpler model to the one used in \refch{ch:TwoPhotonResonance}: \textit{a single two-level system that is coherently driven}, already presented in \refsec{sec:MollowTriplet}, whose output field is analyzed via the cascaded formalism [see \eqref{eq:cascadedmaster1} and \eqref{Eq_Chapt6_CascadedMetrology}].  Particularly, we are interested in the ratio between the standard classical Fisher information, $F_\theta$, and the Poissonian Fisher information, $F_\theta^P$, and explore if there are advantages in the frequency domain. In other words, do the quantum fluctuations enhance the acquisition of information? 
\begin{equation}
		F_\theta \overset{?}{\gtrless}	F_\theta^P.
\end{equation}

\begin{SCfigure}[1][h!]
	\includegraphics[width=0.67\columnwidth]{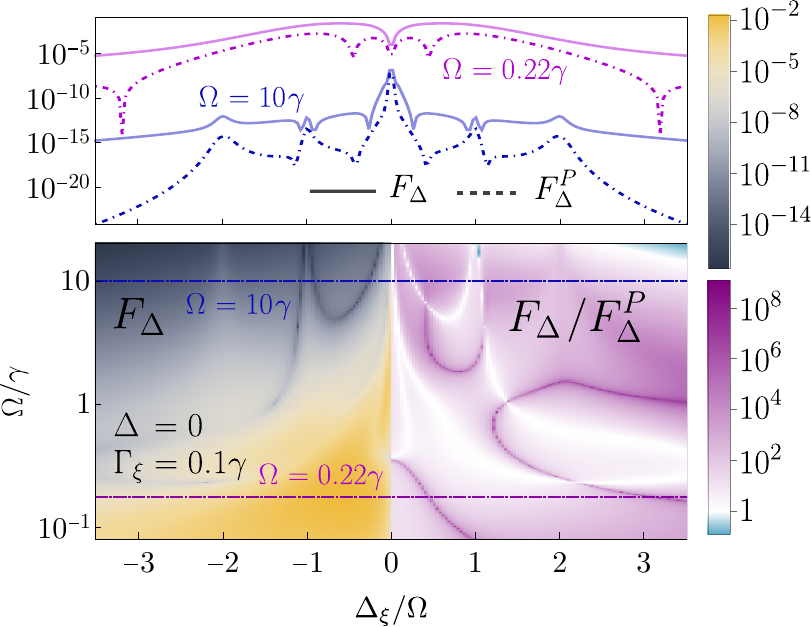}
	\captionsetup{justification=justified}
	\caption[Relevance of quantum fluctuations over classical fluctuations in quantum parameter estimation.]{
		\label{OneSensorResults}
		\textbf{Relevance of quantum fluctuations over classical fluctuations in quantum parameter estimation.}
		FI, $F_\Delta$, (left-hand side) and ratio between the FI and PFI, $F_\Delta/F_\Delta^P$, for estimating the the qubit-laser detuning $\Delta$ in terms of the Rabi frequency $\Omega$ and the sensor frequency $\Delta_\xi$. 
		Both magnitudes are vertically symmetric with respect the zero detuned condition $\Delta_\xi =0$. 
		The upper panel corresponds to cuts at $\Omega=0.22 \gamma$ (Heitler regime) and $\Omega=10 \gamma$ (Mollow regime), depicting the FI (solid line) and the PFI (dot-dashed line).
		Parameters: $\Delta=0$, and $\Gamma_{\xi}=0.1 \gamma$
	}
\end{SCfigure}

From \reffig{OneSensorResults}, we observe that, in general, using the full information contained in the output density matrix offers a significant metrological advantage compared to relying solely on Poissonian fluctuations. This is particularly evident in the purple regions of the lower panel in \reffig{OneSensorResults}, where the characteristic ratios reach differences of up to $8$ orders of magnitude. Although the ratio  $F_\Delta/F_\Delta^P\gg 1$ holds across most of the $\{\Omega, \Delta_\xi\}$-space, some values of this ratio may be misleading. For instance, let us consider the cut at 
$\Omega=10\gamma$ shown in the upper panel of \reffig{OneSensorResults} (blue curve). This case demonstrates a notable metrological advantage in using the full quantum state, yielding differences of approximately $\sim 3-5$ orders of magnitude. However, both the Fisher information (FI) and Poissonian Fisher information (PFI) are relatively small compared to the maximum values of the Fisher information at the optimal parameter regimes. In such cases, the apparent advantage may be exaggerated or impractical since the ratio involves dividing a small quantity by an even smaller one—close to numerical zero—resulting in an artificially large value. Additionally, we also highlight that when the PFI takes large values, the FI generally takes even larger values when quantum fluctuations are included, as can be observed in the cut at $\Omega=0.22\gamma$ in the upper panel of \reffig{OneSensorResults} (purple curve).

Finally, we note that while the Heitler regime ($\Omega<\gamma$) provides higher absolute values for both the Fisher and Poissonian Fisher information, the Mollow regime ($\Omega>\gamma$) exhibits a frequency-resolved structure. This occurs because the detector linewidth in this regime can resolve the energy transitions associated with the dressed states, enabling potential frequency-selective metrological applications that might be enhanced via tailored environment, as we used in \refch{chapter:Entanglement} for generating high entanglement between emitters.

\paragraph{Poissonian Fisher information bound.} Despite the intuitive idea that quantum fluctuations provide higher sensitivity in metrological protocols~\cite{PolinoPhotonicQuantum2020}, it is not mathematically guaranteed that the classical Fisher information---considering the full state---is metrologically better than just considering the Poissonian fluctuations [see the small blue regions in the lower panel in \reffig{OneSensorResults}, denoting $F_\Delta/F_\Delta^P<1$].   

Generally, we can prove this by applying the
\graffito{
	$^{*}$Let $a_1,a_2, \ldots, a_n$ be real numbers and $b_1,b_2, \ldots, b_n$ positive real numbers. Then, the Sedrakyan's inequality reads
	$$
	\sum_{k=1}^n \frac{a_k^2}{b_k} \geq \frac{(\sum_{k=1}^n a_k)^2}{\sum_{k=1}^n b_k}.
	$$
	We note that this inequality is just an alternative version of the Cauchy-Schwarz inequality (CSI),
	\begin{equation*}
		\begin{split}
			\left(\sum_{k=1}^n x_k^2 \right) &\left(\sum_{k=1}^n y_k^2 \right)\\
			&\geq \left(\sum_{k=1}^n x_k y_k \right)^2,
		\end{split}
	\end{equation*}
	which can be derived by setting $x_k=a_k/\sqrt{b_k}$ and $x_k=\sqrt{b_k}$, that is, $a_k\in \mathbb{R}$ and $b_k\in \mathbb{R}^+-\{0\}$. Under these definitions, the CSI simplifies to the Sedrakyan's inequality.
}
 Sedrakyan's inequality$^{\textcolor{Maroon}{*}}$~\cite{SedrakyanAlgebraicInequalities2018}
in the definition of the Poissonian Fisher information in \eqref{eq:PoissonFisher},
\begin{equation}
	F_\theta^P =\frac{[\sum_n n  \partial_\theta  p(n|\theta)]^2}{\sum_n n p(n|\theta)} \leq \sum_n n \frac{[\partial_\theta p(n|\theta)]^2}{p(n|\theta)}=\sum_n n F^{(n)},	
	\label{eq:cond1}
\end{equation}
where we can define $F^{(n)}\equiv [\partial_\theta p(n|\theta)]^2/p(n|\theta)$ as the $n$th order term in the expansion of the Fisher information, $F_\theta= \sum_n [\partial_\theta p(n|\theta)]^2/p(n|\theta)= \sum_n F^{(n)}$. From this definition, it is trivial the consequence $n F^{(n)} \geq F^{(n)}$ since $n\in \mathbb{N}^+$, and thus get the inequality
\begin{equation}
	F_ \theta= F^{(0)}+ \sum_{n=1} F^{(n)} \leq F^{(0)} + \sum_{n=1} n F^{(n)}. 
	\label{eq:cond2}
\end{equation} 
Considering these two conditions, \cref{eq:cond1,eq:cond2}, we get the relation
\begin{equation}
	\colorboxed{Maroon}{
	|F_\theta - F_\theta^P| \leq F^{(0)}.
	}
\end{equation}
Hence, the intuitive notion that $F_\theta$ must always be greater than $F_\theta^P$ fails, since it may occur that the variance of a given operator, along with a general quantum probability distribution, is greater than the one associated to a Poissonian probability distribution, and thus it may occur $F_ \theta<F_\theta^P$.

\subsection{Frequency-resolved Fisher information} 

Following the fundamental idea from previous works on frequency-resolved measurements to explore the rich landscape of multi-photon processes~\cite{DelValleTheoryFrequencyFiltered2012,UlhaqCascadedSinglephoton2012,SanchezMunozViolationClassical2014,PeirisTwocolorPhoton2015,PeirisFransonInterference2017,LopezCarrenoPhotonCorrelations2017,ZubizarretaCasalenguaConventionalUnconventional2020,LopezCarrenoEntanglementResonance2024, YangEntanglementPhotonic2025} [see \refsec{Chapt6_SecEntanglement} for an application on the generation of entanglement], we extend this framework into the domain of quantum metrology. Particularly, we explore the potential quantum metrological advantages of considering cross-correlations in quantum parameter estimation, introducing the notion of \textit{frequency-resolved fisher information}. 
\begin{SCfigure}[1][h!]
	\includegraphics[width=1.\columnwidth]{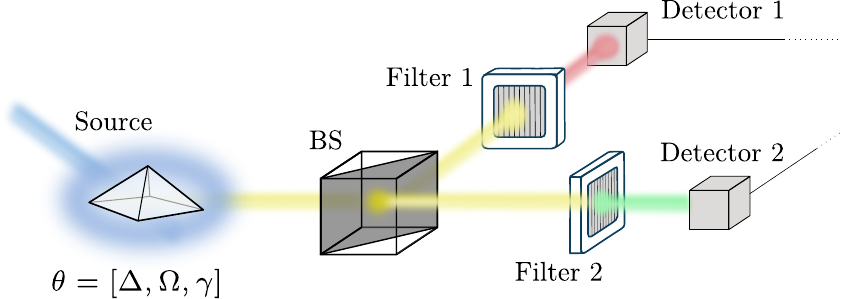}
	\captionsetup{justification=justified}
\end{SCfigure}
\graffito{\vspace{-8.0cm}
	\captionof{figure}[Metrological protocol for two sensors.]{	\label{MetrologicalProtocolTwoSensors}
		\textbf{Metrological protocol for two sensors.} 
	Illustration of the metrological protocol: a coherently driven quantum emitter (the source) whose output field is physically split into two separate modes using an optical beam splitter. Each mode is then redirected to a different sensor (the targets). Before reaching each sensor, a grating is placed to extract spectral information at different frequencies, enabling the analysis of cross-correlations.
	
	}
}

To this end, we consider the cascaded master equation presented in \eqref{Eq_Chapt6_CascadedMetrology} [see \reffig{MetrologicalProtocolTwoSensors} for an illustration of the protocol] and compute the classical Fisher information for the joint quantum state of both sensors, 
\begin{equation}
	F_\theta [p(n_1,n_2)]= \sum_{n_1,n_2} \frac{1}{p(n_1,n_2|\theta) } [\partial_\theta p(n_1,n_2|\theta)]^2,
	\label{eq:jointFisher}
\end{equation}
where $p(n_1,n_2)=\text{diag}[\hat \rho_{1,2}]$ denotes the joint probability distribution of the density matrix $\hat \rho_{1,2}$ describing the two sensors. The corresponding marginal probability distributions are given by $p(n_1)=\sum_{n_2} p(n_1,n_2)$ and $p(n_2)=\sum_{n_1} p(n_1,n_2)$.  
In the simplest scenario of uncorrelated variables, where joint probability distribution factorizes as $p(n_1,n_2)=p(n_1)p(n_2)$, the Fisher information simplifies to the sum of its marginals$^{\textcolor{Maroon}{*}}$
\graffito{
${}^{*}$The classical Fisher information is additive for independent conditional probability distributions~\cite{JaynesProbabilityTheory2003}.
}
\begin{equation}
	F_\theta[p_1(n)p_2(n)]= F_\theta[p_1(n)] + F_\theta[p_2(n)],
	\label{eq:uncorrFisher}
\end{equation}
This result naturally raises the question: \textit{Does the joint density matrix provide more information than the marginal distributions?} Or equivalently, \textit{Do the cross-correlations enhance the acquisition of information?} This question can be mathematically formulated as:
\begin{equation}
F_\theta [p(n_1,n_2)]  \overset{?}{\gtrless} F_\theta[p_1(n)p_2(n)]
\end{equation}
\vspace{-3mm}
\begin{SCfigure}[1][h!]
	\includegraphics[width=0.68\columnwidth]{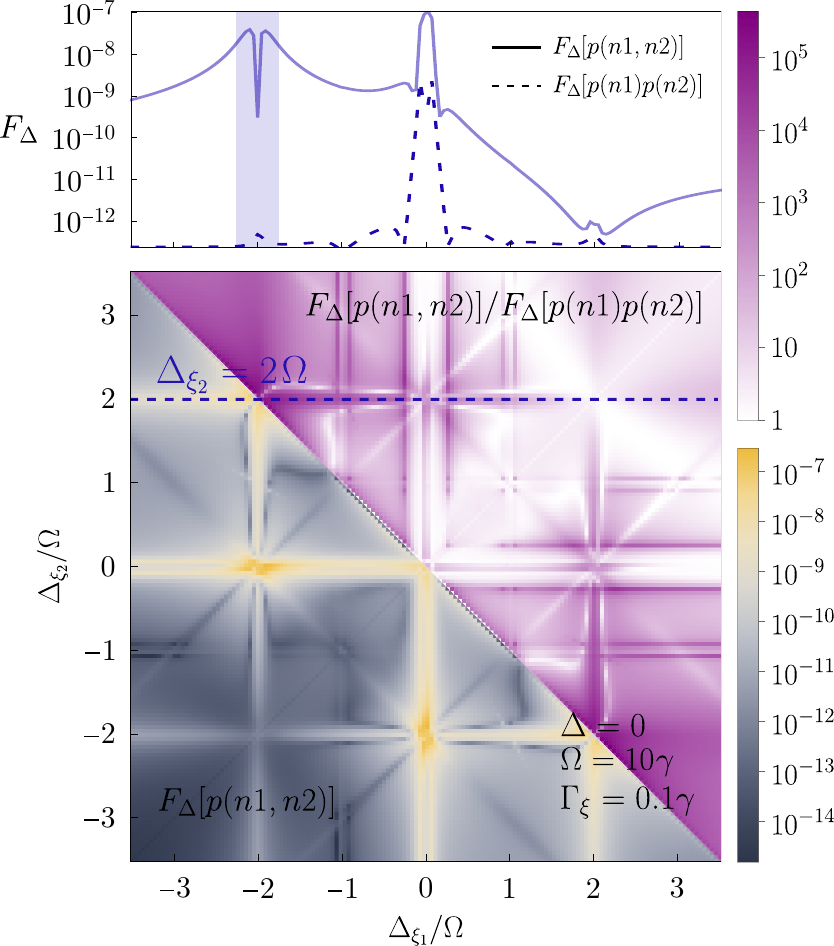}
	\captionsetup{justification=justified}
	\caption[Relevance of cross-correlations in quantum parameter estimation]{
		\label{TwoSensorResult}
		\textbf{Relevance of cross-correlations in quantum parameter estimation.}
		Fisher information for the joint quantum state, $F_\Delta [p(n_1,n_2)]$ (left-lower triangle) and its ratio relative to the case of uncorrelated variables, $F_\Delta [p(n_1,n_2)]/F_\Delta [p(n_1)p(n_2)]$ (right-upper triangle), for estimating the qubit-laser detuning $\Delta$.
		The upper panel shows a cut at $\Delta_{\xi_2}=2\Omega$, comparing $F_\Delta [p(n_1,n_2)]$ (blue solid curve) and $F_\Delta [p(n_1)p(n_2)]$ (blue dashed curve). This highlights a significant metrological enhancement—exceeding 5 orders of magnitude—achieved by preserving information about cross-correlations.
		Parameters: $\Delta=0$, $\Omega=10\gamma$, and $\Gamma_\xi=0.1\gamma$.
	}
\end{SCfigure}

To address this question, we compute the classical Fisher information for the joint probability distribution $p(n_1,n_2)$ for estimating the qubit-laser detuning $\Delta$ in the frequency domain $\{\Delta_1, \Delta_2\}$, a measure we refer to as the \textit{frequency-resolved Fisher information}. 
Specifically, we compute both the joint and uncorrelated Fisher information, as defined in \eqref{eq:jointFisher} and \eqref{eq:uncorrFisher}, respectively. 
In the upper panel of \reffig{TwoSensorResult}, both magnitudes, $F_\theta[p(n_1,n_2|\theta)]$ (solid line) and $F_\theta[p(n_1|\theta)p(n_2|\theta)]$ (dashed line),  are plotted as functions of the sensor-laser detuning, $\Delta_1$, while the other is fixed at one of the Mollow sidebands, $\Delta_2=\Delta_+\approx2\Omega$ [see \refsec{Chapt6_Section_Model}].

From this figure, we can clearly observe the relevance of considering cross-correlations: there is an enhancement of up to $4-5$ orders of magnitude, as highlighted in the marked region. 
This metrological advantage, quantified as  $F_\theta[p(n_1,n_2|\theta)]/F_\theta[p(n_1|\theta)p(n_2|\theta)] \gg 1$, is not limited to a specific frequency window but extends across the entire frequency domain, as illustrated in the lower panel of \reffig{TwoSensorResult}. In this figure, the lower-left triangle represents the joint Fisher information, while the upper-right triangle shows the ratio between joint and uncorrelated Fisher information.
Notably, regions of high joint Fisher information---indicated in yellow---correspond to areas of significant metrological advantage, highlighted in purple.

As in the case of a single sensor, there are regions where the ratio becomes particularly large, $F_\theta[p(n_1,n_2|\theta)]/F_\theta[p(n_1|\theta)p(n_2|\theta)]\sim 10^4$ (e.g., the upper-left region). However, these extreme values may reflect an impractical enhancement, arising from the ratio of a small magnitude to an even smaller one.

We also observe that regions of high Fisher information and significant metrological advantage are located near the antidiagonals [see discussion in \ref{Chapt6_Section_Model} and \reffig{FigChapt6_FreqResolved}]. This is reasonable, as these regions correspond to where cross-correlations signal the emergence of non-classical states of light that violate the Cauchy-Schwarz or Bell inequalities~\cite{SanchezMunozViolationClassical2014}, or where they generate entanglement~\cite{LopezCarrenoEntanglementResonance2024} [see also the previous section, \ref{Chapt6_SecEntanglement}].

\paragraph{Cross-correlations Fisher information bound. } Intuitively, one might quickly conclude that the joint probability distribution is always metrologically advantageous with respect the uncorrelated case, as it \textit{captures} more complete information. While this holds true in \reffig{TwoSensorResult}, it is not a general rule.
Here below, following the same spirit of the mathematical derivation for the single-sensor case, we provide a theoretical bound by which the joint Fisher information is not always greater than the uncorrelated case.

To prove this, we consider the case of a general joint density matrix $\hat \rho_{1,2}$ describing the quantum state of both bosonic sensors. Although in practice the Hilbert space of these sensors is typically restricted to a few excitations, we will consider here the general case up to $(N_1,N_2)$-excitations. The computation of the Fisher information in \eqref{eq:jointFisher} requires the use of the joint probability distribution, $p(n_1,n_2|\theta)$,
\begin{equation}
	p(n_1,n_2|\theta)= 
	\begin{pmatrix}
			p(0,0|\theta) & p(0,1|\theta) & p(0,2|\theta) & \ldots & p(0,N_2|\theta) \\
			p(1,0|\theta) & p(1,1|\theta) & p(1,2|\theta) & \ldots & p(1,N_2|\theta) \\
			p(2,0|\theta) & p(2,1|\theta) & p(2,2|\theta) & \ldots & p(2,N_2|\theta) \\
			\vdots & \vdots & \vdots & \ddots & \vdots \\
			p(N_1,0|\theta) & p(N_1,1|\theta) & p(N_1,2|\theta) & \ldots & p(N_1,N_2|\theta)
	\end{pmatrix}
\end{equation}
such that the marginal probability distributions are given by
\begin{equation}
	p(n_1|\theta)= \sum_{n_2} p(n_1,n_2|\theta) =
		\begin{pmatrix}
p(0,0|\theta)+p(0,1|\theta)+ \ldots + p(0,N_2|\theta) \\
p(1,0|\theta)+p(1,1|\theta)+ \ldots + p(1,N_2|\theta)\\
\vdots \\
p(N_1,0|\theta)+p(N_1,1|\theta)+ \ldots + p(N_1,N_2|\theta)
	\end{pmatrix},
\end{equation}
and equivalently for $p(n_2|\theta)$.
%
%
		%
		%
		%
		%
		%
		%
%
%
%
Consequently, by inserting the definition of $p(n_1,n_2|\theta)$ into \eqref{eq:jointFisher} we get
\begin{multline}
	F_\theta[p(n_1,n_2|\theta)] =
	\frac{ [\partial_\theta p(0,0|\theta)]^2}{p(0,0|\theta)} +	\frac{[\partial_\theta p(0,1|\theta)]^2}{p(0,1|\theta)}  \\
	+\frac{ [\partial_\theta p(1,0|\theta)]^2}{p(1,0|\theta)}  +
	\frac{[\partial_\theta p(1,1|\theta)]^2}{p(1,1|\theta)}  + \ldots +	\frac{[\partial_\theta p(N_1,N_2|\theta)]^2}{p(N_1,N_2|\theta)} ,
	\label{eq:JointFisher_Ext}
\end{multline}
while inserting the expressions for $p(n_1|\theta)$ and $p(n_2|\theta)$ into the marginal Fisher information terms in \eqref{eq:uncorrFisher}, we get
\begin{multline}
	F_\theta[p(n_1|\theta)] =
	\frac{[\partial_\theta p(0,0|\theta)+\partial_\theta p(0,1|\theta)+ \ldots + \partial_\theta p(0,N_2|\theta)]^2}{p(0,0|\theta)+p(0,1|\theta)+ \ldots + p(0,N_2|\theta)} \\
	+\ldots + 	\frac{[\partial_\theta p(N_1,0|\theta)+\partial_\theta p(N_1,1|\theta)+ \ldots + \partial_\theta p(N_1,N_2|\theta)]^2}{p(N_1,0|\theta)+p(N_1,1|\theta)+ \ldots + p(N_1,N_2|\theta)},
	\label{eq:MarginalFisher1}
\end{multline}
and analogously to $F_\theta[p(n_2|\theta)]$.
%
%
%
%
Now, if we rearrange appropriately the terms in \eqref{eq:JointFisher_Ext},
\begin{multline}
	F_\theta[p(n_1,n_2|\theta)] =
	\frac{1}{2}\left(
	\frac{ [\partial_\theta p(0,0|\theta)]^2}{p(0,0|\theta)}+ \ldots +\frac{ [\partial_\theta p(N_1,N_2|\theta)]^2}{p(N_1,N_2|\theta)}
	\right)
	\\
	+
	\frac{1}{2}\left(
	\frac{ [\partial_\theta p(0,0|\theta)]^2}{p(0,0|\theta)}+ \ldots +\frac{ [\partial_\theta p(N_1,N_2|\theta)]^2}{p(N_1,N_2|\theta)}
	\right),
\end{multline}
and apply the Sedrakyan's inequality, we obtain the following condition
\begin{multline}
	F_\theta[p(n_1,n_2|\theta)] \geq  
	\frac{1}{2} \left(
	\frac{[\partial_\theta p(0,0|\theta)+ \ldots + \partial_\theta p(0,N_2|\theta)]^2}{p(0,0|\theta)+ \ldots + p(0,N_2|\theta)}
	\right. \\
	\left. +\ldots + 	\frac{[\partial_\theta p(N_1,0|\theta)+ \ldots + \partial_\theta p(N_1,N_2|\theta)]^2}{p(N_1,0|\theta)+ \ldots + p(N_1,N_2|\theta)}
	\right)
	\\+
	\frac{1}{2} \left(
	\frac{[\partial_\theta p(0,0|\theta)+ \ldots + \partial_\theta p(N_1,0|\theta)]^2}{p(0,0|\theta)+ \ldots + p(N_1,0|\theta)}
	\right. \\
	\left. +\ldots + 	\frac{[\partial_\theta p(0,N_2|\theta)+ \ldots + \partial_\theta p(N_1,N_2|\theta)]^2}{p(N_1,N_2|\theta)+ \ldots + p(N_1,N_2|\theta)}
	\right).
\end{multline}
Finally, by recalling the definition of the marginal Fisher information in \eqref{eq:MarginalFisher1}, we obtain the bound
\begin{equation}
	F_\theta[p(n_1,n_2|\theta)] \geq \frac{1}{2} \left(F_\theta[p(n_1|\theta)] +F_\theta[p(n_2|\theta)]\right),
\end{equation}
which can be rewritten as
\begin{equation}
\colorboxed{Maroon}{
F_\theta[p(n_1,n_2|\theta)] \geq  \frac{1}{2}F_\theta[p(n_1|\theta)p(n_2|\theta)].
}
\end{equation}
In other words, we cannot assert that knowledge of the full joint probability distribution always guarantees a metrological advantage over performing uncorrelated measurements.  While this result might seem counter-intuitive, it can be understood by considering the limiting case of two perfectly correlated (or anticorrelated) random variables: once one variable is known the other is straightforwardly inferred. Therefore, in such cases, $p(n_1,n_2|\theta)$ does not provide more information than the product $p(n_1|\theta)p(n_2|\theta)$, which describes uncorrelated events.

\subsection{Outlook}

In this short section, we have briefly presented our latest results on quantum metrology through spectral measurements, highlighting two key aspects: \textcolor{Maroon}{(i)} the relevance of quantum over classical fluctuations, and \textcolor{Maroon}{(ii)} the role of cross-correlations in enhancing the acquisition of information for quantum parameter estimation. Both perspective yield a better metrological performance in concrete regions of the parameter space, paving the way for novel quantum metrological protocols based on spectral properties.

As noted in the introduction, these results are part of ongoing work, and a deeper analysis is required to better understand the underlying mechanisms. Several important questions remain to be addressed: Why is the Heitler regime more sensitive than the Mollow regime, as shown in \reffig{OneSensorResults}? Could a tailored environment enhance these results, in a similar manner to the results on entanglement generation presented in \refch{ch:TwoPhotonResonance}? Additionally, what is the impact of spectral resolution---i.e., the sensor linewidth---on the effectiveness of this metrological protocol? Answering these question is currently work in progress.
\cleardoublepage

\cleardoublepage


\chapter{Conclusions}

\section{English}
In this Thesis,  we have studied the dynamics of collective quantum optical systems with the aim of exploring both fundamental aspects and technological applications of the cooperative phenomena emerging from light-matter dressing and its interaction with structure photonic environments.
Specifically, we have considered systems consisting of ensembles of interacting quantum emitters that are coherently driven by a laser field and coupled to a single photonic mode.

Most theoretical studies have typically focused on systems of quantum emitters with identical natural frequencies---overlooking practical limitations in state-of-the-art solid-state platforms, such as inhomogeneous broadening---, or assuming weak emitter-emitter coupling regimes and low driving strengths---constraining the theoretical results to a very specific parametric regime---.
Here, instead, we considered the simplest minimal---while general---model that exhibits cooperative effects: 
\textit{two quantum emitters, described as two-level systems, with non-identical frequencies interacting via either free-space interactions or through a photonic structure that mediates their interaction.}
Additionally, the emitters could be subject to either coherent or incoherent excitation through the injection of a laser field, which, in fact, played a central role for the emergence of cooperative effects as it induces  light-matter dressing among the interacting emitters. 
These elements constitute some of the fundamental building blocks in quantum optics. 
Despite the apparent simplicity of the model, it revealed a rich landscape of cooperative phenomena to be explored, specifically when the two interacting emitters are coherently driven at the two-photon resonance, enabling a coherent, nonlinear excitation of the transition from the ground state to a doubly-excited state via a two-photon process.
This effect constituted the backbone of the Thesis,  offering intriguing insights into both fundamental aspects and technological applications, such as an enhanced sensitivity for estimating inter-emitter distances through two-photon effects, a reconceptualization of virtual states in the presence of dissipation, and mechanisms for generating long-lived entanglement between quantum emitters.
Additionally, the results of this Thesis addressed some of the fundamental challenges in generating cooperative effects in solid-state platforms. These include the fabrication of strongly interacting quantum emitters, inhomogeneous broadening, and intrinsic decoherence processes inherent in any realistic quantum system. 

During the first part of the Thesis, we examined in detail this minimal model,
where two coherently driven quantum emitters interact through the vacuum electromagnetic field. 
%
For the first time, we derived general analytical solutions for the steady-state density matrix elements of non-identical interacting emitters that are valid for an ample regime of parameters, subject only to the condition that the energy splitting between one-excitation eigenstates must be the largest energy scale in the system. These analytical expressions reveal how optical observables, such as the emission intensity, the photon statistics, or the fluorescence spectrum, are shaped by the intricate interplay of emitter-emitter coupling, emitter-emitter energy detuning and driving strength.
Of particular interest is the appearance of two-photon effects when driving the emitters at the two-photon resonance---commonly illustrated by a central peak in the emission intensity---, signalling the emergence of cooperative effects within the emitters. 
These two-photon effects exhibit high sensitivity to the parameters governing the emitter-emitter interactions, such as the inter-emitter distance and energy detuning.
This strong dependency allowed us to identify the onset of two-photon effects as the most sensitive regime for estimating parameters such as the inter-emitter distance,  potentially enabling sub-wavelength imaging techniques that surpass the diffraction limit by exploiting cooperative effects.

The exploration of this minimal model under the influence of two-photon driving raised the question of whether we could harness the two-photon driving to stabilize the system into a stationary entangled state by incorporating a photonic structure.
Our initial approach to this problem led to an unexpected discovery: a mechanism by which off-resonant virtual states may exhibit an unconventional behaviour, where they become populated in the long-time limit even without being connected to any dissipative channel. 
This implies that the regime in which the virtual state remains unpopulated is merely metastable,  contradicting the conventional intuition regarding virtual states.
For a better understanding of this phenomenon, we developed a novel \textit{hierarchical adiabatic elimination} (HAE) technique that allowed us to determine analytical insights into the system dynamics, including the time-dependent density matrix elements and the associated relaxation timescale.
The implications of this \textit{de-virtualization} process extend beyond theoretical interest. Virtual states and metastable dynamics can be exploited for practical purposes, such as generating long-lived entanglement between quantum emitters in more sophisticated light-matter models. 
Through the application of the HAE method, we were also able to determine the duration of this entanglement, providing a key tool for designing and optimizing the generation of long-lived entanglement. This approach, in fact, was crucial for developing strategies of entanglement generation in the next chapter.

Following the question of generating entanglement between non-identical interacting emitters, we demonstrated that coupling the two interacting emitters to a lossy cavity unveils a rich landscape of mechanisms of entanglement generation, with up to five distinct driven-dissipative mechanisms capable of producing both stable and metastable entanglement.
The lossy cavity, acting as a tailored environment, induces cooperative effects within the emitter-emitter system, such that when properly tuned, it stabilizes the quantum emitters into nearly perfect entangled states.
These cavity-induced phenomena are remarkably robust against decoherence, operating across a wide range of parameters and driving strengths, demonstrating their validity in current state-of-the-art solid-state platforms.
Interestingly, we revealed that entanglement formation correlates with measurable properties of the light emitted by the cavity, such as subradiance in the photon emission or  antibunching in the photon statistics, offering clear experimental signatures. 
Among the five mechanisms, we highlight two in particular. The first is the \textit{frequency-resolved Purcell enhancement}, where the cavity selectively enhances specific transitions within the emitters when they form a dimer, stabilizing them into either superradiant or subradiant states, which are well-known signatures of collective behaviour between interacting emitters. Notably, this mechanism can be scaled up to multi-emitter systems ($N>2$), effectively addressing the scalability problem when generating long-lived entanglement. 
The second is the \textit{Collective Purcell enhancement}, which stabilizes the system into the subradiant state even when the emitters are not directly interacting through coherent coupling.
Although this effect was already reported in the literature, we provided deeper insights into how the antisymmetric state is stabilized via the unconventional mechanism of virtual state population that we unveiled in the previous chapter.
In particular, our HAE method allowed us to derive exact expressions for the density matrix elements and the stabilization timescale, offering a more general and accurate description than previous reports of this effect.

Throughout this Thesis, we have focused on the properties of the quantum emitters and optical observables, seeking cooperative signatures such as  two-photon effects, super- and subradiance. However, open questions remain regarding the structure of the density matrix of the emitted light and its potential applications for entanglement generation and quantum metrology.
As a first step in this direction, we briefly addressed these aspects by studying the light emitted from a coherently driven two-level system, a fundamental model in quantum optics that serves as an ideal test-bed for exploring light-matter interactions. 
Our theoretical analysis demonstrated that spectrally distinct but temporally overlapping photonic modes can become entangled---a result that was confirmed through an experimental collaboration in which we participated, and that is reported as part of this Thesis---.
Additionally,  we also explored the potential of the emitted light for quantum metrology, emphasizing the crucial role of quantum and frequency correlations between photonic modes in enhancing the precision of spectral measurements.
By exploiting these correlations, we showed that information acquisition can be significantly improved, pointing toward practical applications in quantum sensing and parameter estimation.
Although these results are part of ongoing work, they already pave the way for novel quantum metrological protocols based on the spectral properties of the emitted light by a quantum system in dissipative scenarios, which may include collective quantum systems as the ones we studied throughout this Thesis.  

In conclusion, this Thesis is a humble contribution to the understanding of collective light-matter interactions in quantum optics while offering practical possibilities for using quantum matter and light in both fundamental research and technological applications. 
From precise inter-emitter distance estimation enabled by two-photon processes to driven-dissipative entanglement generation mediated by tailored photonic environments or quantum metrological protocols based on spectral measurements, the results presented here lead to new approaches for understanding and controlling collective light-matter interactions in both ideal and state-of-the-art technological configurations. 
Notably, we emphasize that even the simplest realizations of light-matter systems reveal an exceptional rich world of phenomena with potential application in quantum technologies.
This aspect naturally raises a fundamental question: \textit{what new theoretical models and technological advances could emerge in more complex and scalable light-matter systems?} By addressing this question, future research could uncover novel physical mechanisms and further expand the potential of cooperative quantum systems.

This Thesis opens several promising directions for future research. One ongoing line  involves analysing  the fundamental properties of the quantum state---via its density matrix---of the output field emitted by a quantum system, such as coherently driven qubit, as we briefly introduced in the last Chapter.
Additionally, throughout this Thesis, we have assumed that the nature of the environments is Markovian, where memory effects are neglected. However, in many realistic systems, particularly those involving strong coupling or structured photonic environments, non-Markovian effects can play a crucial role in the system dynamics. Exploring how memory effects influence the results presented here, such as the \textit{de-virtualization} of virtual-states, or the driven-dissipative generation of entanglement, could offer deeper insights into the behaviour of open quantum systems in more realistic scenarios.
Connected to this idea, an immediate next step would be to consider structured baths,  such as photonic crystals or topological waveguides, which may exhibit exotic properties like chiral emission.
On the matter side, our description was constrained to quantum emitters within the dipole approximation, interacting with light through a single coupling point. By relaxing this approximation and considering more generalized models, such as giant atoms---which couple to light at multiple discrete points---, new phenomena could arise, e.g., the complete avoidance of decoherence in nested configurations. These systems offer an exciting platform for exploring novel collective behaviours, potentially leading to breakthroughs in quantum control and noise-resilient quantum devices.

\cleardoublepage

\section{Español}
En esta Tesis hemos estudiado la dinámica de sistemas colectivos de óptica cuántica con el objetivo de explorar tanto aspectos fundamentales como aplicaciones tecnológicas de los fenómenos cooperativos emergentes del “vestimiento” luz-materia y su interacción con entornos fotónicos estructurados. 
Específicamente, hemos considerado sistemas compuestos por conjuntos de emisores cuánticos interactuantes que son excitados de forma coherente por un láser y acoplados a un único modo fotónico. 

La mayoría de los estudios teóricos se han centrado típicamente en sistemas de emisores cuánticos con frecuencias naturales idénticas---pasando por alto limitaciones prácticas en plataformas actuales de estado sólido, como el ensanchamiento inhomogéneo---o asumiendo regímenes de acoplamiento débil entre emisores y bajas intensidades de excitación---limitando los resultados teóricos a un régimen paramétrico muy específico---. 
Aquí, en cambio, consideramos el modelo más sencillo y general que exhibe efectos cooperativos: 
\textit{dos emisores cuánticos, descritos como sistemas de dos niveles, con frecuencias no idénticas que interactúan ya sea mediante interacciones a través del vacío o de una estructura fotónica que media su interacción.} 
Adicionalmente, los emisores podían estar sometidos a excitación coherente o incoherente mediante la inyección de un láser, el cual, de hecho, desempeñó un papel central para la aparición de efectos cooperativos al inducir un “vestimiento” luz-materia entre los emisores interactuantes. 
Estos elementos constituyen algunos de los elementos fundamentales en la óptica cuántica. 
A pesar de la aparente simplicidad del modelo, éste reveló un amplio abanico de fenómenos cooperativos por explorar, específicamente cuando los dos emisores interactuantes son excitados de forma coherente en la resonancia de dos fotones, lo que habilita una excitación coherente y no lineal de la transición desde el estado fundamental hacia un estado doblemente excitado mediante un proceso de dos fotones. 
Este efecto constituyó la columna vertebral de la Tesis, ofreciendo perspectivas intrigantes tanto en aspectos fundamentales como en aplicaciones tecnológicas, tales como una mayor sensibilidad para estimar distancias entre emisores a través de efectos de dos fotones, una re-conceptualización de los estados virtuales en presencia de disipación y mecanismos para generar entrelazamiento de larga duración entre emisores cuánticos. 
Adicionalmente, los resultados de esta Tesis abordaron algunos de los desafíos fundamentales en la generación de efectos cooperativos en plataformas de estado sólido. Éstos incluyen la fabricación de emisores cuánticos fuertemente interactuantes, el ensanchamiento inhomogéneo y los procesos intrínsecos de decoherencia inherentes a cualquier sistema cuántico realista.

Durante la primera parte de la Tesis, examinamos en detalle este modelo mínimo, donde dos emisores cuánticos excitados coherentemente interactúan a través del campo electromagnético del vacío. 
Por primera vez, derivamos soluciones analíticas generales para los elementos de la matriz densidad de emisores interactuantes no idénticos en el estado estacionario, las cuales son válidas para un amplio régimen de parámetros, sujetas únicamente a la condición de que la separación de energía entre autoestados de una excitación debe ser la mayor escala de energía del sistema. Estas expresions revelan cómo los observables ópticos, tales como la intensidad de emisión, la estadística de fotones o el espectro de fluorescencia, se ven afectados por la compleja interrelación del acoplamiento entre emisores, la diferencia de energía entre ellos y la intensidad del bombeo. 
De particular interés es la aparición de efectos de dos fotones al excitar los emisores en la resonancia de dos fotones, ilustrado comúnmente por un pico central en la intensidad de emisión, lo cual señala la emergencia de efectos cooperativos entre los emisores. 
 Estos efectos de dos fotones exhiben una alta sensibilidad a los parámetros que gobiernan las interacciones entre emisores, tales como la distancia interemisores y la diferencia de energía. 
Esta fuerte dependencia nos permitió identificar el inicio de los efectos de dos fotones como el régimen de mayor sensibilidad para la estimación de parámetros como la distancia entre emisores, lo que potencialmente habilita técnicas de obtención de imágenes a sublongitud de onda que superan el límite de difracción mediante el aprovechamiento de efectos cooperativos.

La exploración de este modelo mínimo bajo la influencia de excitación de dos fotones planteó la pregunta de si podríamos aprovechar dicha excitación para estabilizar el sistema en un estado estacionario entrelazado mediante la incorporación de una estructura fotónica. 
Nuestro enfoque inicial a este problema condujo a un descubrimiento inesperado: un mecanismo mediante el cual estados virtuales fuera de resonancia pueden exhibir un comportamiento poco convencional, en el que se vuelven poblados en el límite estacionario, incluso sin estar conectados a ningún canal disipativo. 
Esto implica que el régimen en el que el estado virtual permanece sin poblar es meramente metastable, lo que contradice la intuición convencional respecto a los estados virtuales.
Para una mejor comprensión de este fenómeno, desarrollamos una novedosa técnica de \textit{eliminación adiabática jerárquica} (HAE, por sus siglas en inglés) que nos permitió obtener resultados analíticos sobre la dinámica del sistema, incluyendo los elementos dependientes del tiempo de la matriz densidad y la escala de tiempo asociada a la relajación. 
Las implicaciones de este proceso de \textit{de-virtualización} se extienden más allá del interés teórico. Los estados virtuales y las dinámicas metastables pueden ser aprovechados para fines prácticos, tales como la generación de entrelazamiento de larga duración entre emisores cuánticos en modelos de luz-materia más sofisticados. 
Mediante la aplicación del método HAE, también pudimos determinar la duración de este entrelazamiento, proporcionando una herramienta clave para diseñar y optimizar la generación de entrelazamiento de larga duración. Este enfoque, de hecho, fue crucial para desarrollar estrategias de generación de entrelazamiento en el capítulo siguiente.

Siguiendo la cuestión de generar entrelazamiento entre emisores interactuantes no idénticos, demostramos que acoplar los dos emisores interactuantes a una cavidad disipativa desvela un amplío panorama de mecanismos de generación de entrelazamiento, con hasta cinco mecanismos distintos activados por disipación capaces de producir tanto entrelazamiento estacionario como metastable. 
 La cavidad disipativa, actuando como un entorno diseñado, induce efectos cooperativos dentro del sistema de emisores, de tal forma que, cuando se ajusta adecuadamente, estabiliza a los emisores cuánticos en estados casi perfectamente entrelazados. 
 Estos fenómenos inducidos por la cavidad son notablemente robustos contra la decoherencia, operando en un amplio rango de parámetros e intensidades de excitación, lo que demuestra su validez en las plataformas actuales de estado sólido. 
  También demostramos que la formación de entrelazamiento se correlaciona con propiedades medibles de la luz emitida por la cavidad, tales como la subradiancia en la emisión de fotones o la \textit{anti-correlación} en la estadística de fotones, ofreciendo patrones experimentales claross. 
  Entre los cinco mecanismos, destacamos dos en particular. El primero es el \textit{efecto Purcell con resolución en frecuencia}, donde la cavidad mejora selectivamente transiciones específicas dentro de los emisores cuando estos forman un dímero, estabilizándolos en estados superradiantes o subradiantes, los cuales son marcas bien conocidas del comportamiento colectivo entre emisores interactuantes. Cabe destacar que este mecanismo se puede escalar a sistemas con múltiples emisores ($N>2$), abordando el problema de escalabilidad en la generación de entrelazamiento de larga duración. 
    El segundo es el \textit{efecto Purcell colectivo}, que estabiliza el sistema en el estado subradiante incluso cuando los emisores no están interactuando directamente mediante acoplamiento coherente. 
     Aunque este efecto ya había sido reportado en la literatura, proporcionamos conocimientos más profundos sobre cómo el estado antisimétrico se estabiliza mediante el mecanismo poco convencional de población de estados virtuales que desvelamos en el capítulo anterior. 
      En particular, nuestro método HAE nos permitió derivar expresiones exactas para los elementos de la matriz  densidad y la escala de tiempo de estabilización, ofreciendo una descripción más general y precisa que los estudios previos sobre este efecto.


A lo largo de esta Tesis nos hemos enfocado en las propiedades de los emisores cuánticos y en los observables ópticos, buscando rasgos de cooperativad tales como los efectos de dos fotones, la super- y subradiancia. Sin embargo, quedan preguntas abiertas respecto a la estructura de la matriz densidad de la luz emitida y sus potenciales aplicaciones para la generación de entrelazamiento y la metrología cuántica. 
Como primer paso en esta dirección, abordamos brevemente estos aspectos mediante el estudio de la luz emitida por un sistema de dos niveles excitado coherentemente, un modelo fundamental en la óptica cuántica que sirve como plataforma ideal para explorar las interacciones luz-materia. 
Nuestro análisis teórico demostró que modos fotónicos espectralmente distintos, pero temporalmente superpuestos, pueden entrelazarse---un resultado que fue confirmado mediante una colaboración experimental en la que participamos y que se reporta como parte de esta Tesis---. 
Adicionalmente, exploramos también el potencial de la luz emitida para la metrología cuántica, enfatizando el papel crucial de las correlaciones cuánticas y de frecuencia entre los modos fotónicos en la mejora de la precisión de las mediciones espectrales. 
 Al explotar estas correlaciones, mostramos que la adquisición de información puede mejorar significativamente, apuntando hacia aplicaciones prácticas en el campo de la detección cuántica y la estimación de parámetros. 
   Aunque estos resultados forman parte de trabajos en curso, ya allanan el camino para nuevos protocolos metrológicos cuánticos basados en las propiedades espectrales de la luz emitida por un sistema cuántico en escenarios disipativos, que podrían incluir sistemas cuánticos colectivos como los que estudiamos a lo largo de esta Tesis.

En conclusión, esta Tesis es una humilde contribución a la comprensión de las interacciones colectivas luz-materia en la óptica cuántica, a la vez que ofrece posibilidades prácticas para el uso de la materia y la luz cuántica tanto en investigación fundamental como en aplicaciones tecnológicas. 
Desde la estimación precisa de la distancia entre emisores habilitada por procesos de dos fotones hasta la generación de entrelazamiento activada por disipación mediado por entornos fotónicos diseñados o protocolos metrológicos cuánticos basados en mediciones espectrales, los resultados presentados aquí conducen a nuevos enfoques para comprender y controlar las interacciones colectivas luz-materia en configuraciones tanto ideales como en plataformas tecnológicas actuales. 
Cabe destacar que incluso las realizaciones más simples de sistemas luz-materia revelan un mundo excepcionalmente rico en fenómenos con potencial aplicación en tecnologías cuánticas. 
Este aspecto plantea de forma natural una pregunta fundamental: \textit{¿qué nuevos modelos teóricos y avances tecnológicos podrían surgir en sistemas luz-materia más complejos y escalables?} Al abordar esta pregunta, la investigación futura podría descubrir nuevos mecanismos físicos y ampliar aún más el potencial de los sistemas cuánticos cooperativos. 

Esta Tesis abre varias direcciones prometedoras para investigaciones futuras. Una línea en curso involucra analizar las propiedades fundamentales del estado cuántico---mediante su matriz densidad---de la luz emitida por un sistema cuántico, como un qubit excitado coherentemente, tal como introdujimos brevemente en el último capítulo. 
Adicionalmente, a lo largo de esta Tesis hemos asumido que la naturaleza de los ambientes es Markoviana, donde se descartan los efectos de memoria. Sin embargo, en muchos sistemas realistas, particularmente aquellos que involucran acoplamientos fuertes o ambientes fotónicos estructurados, los efectos no Markovianos pueden desempeñar un papel crucial en la dinámica del sistema. Explorar cómo los efectos de memoria influyen en los resultados presentados aquí, tales como la \textit{de-virtualización} de los estados virtuales o la generación de entrelazamiento activada por disipación, podría ofrecer perspectivas más profundas sobre el comportamiento de los sistemas cuánticos abiertos en escenarios más realistas. 
  Relacionado con esta idea, un siguiente paso inmediato sería considerar baños estructurados, tales como cristales fotónicos o guías de onda topológicas, que pueden exhibir propiedades exóticas como la emisión quiral. 
    En el ámbito de la materia, nuestra descripción se limitó a emisores cuánticos dentro de la aproximación dipolar, interactuando con la luz a través de un único punto de acoplamiento. Al relajar esta aproximación y considerar modelos más generalizados, tales como átomos gigantes---los cuales se acoplan a la luz en múltiples puntos discretos---, podrían surgir nuevos fenómenos, por ejemplo, la eliminación completa de la decoherencia en configuraciones anidadas. Estos sistemas ofrecen una plataforma emocionante para explorar nuevos comportamientos colectivos, lo que potencialmente conduciría a avances en el control cuántico y dispositivos cuánticos resistentes al ruido.

\appendix
\cleardoublepage
\chapter{Appendix: Chapter $3$}\label{ch:Appendix_TwoPhotonResonance}

\section{Derivation of the master equation for two emitters interacting through vacuum}
\sectionmark{Derivation of the master equation for two emitters ...}

In this appendix we derive the master equation used in \eqref{eq:LMEemitters}, which the describe the effective dynamics of two independent quantum emitters are weakly coupled to the same environment, an electromagnetic field in vacuum.
To address this, we follow the same procedure used for the derivation of the master equation of a single quantum emitter interacting with a vacuum electromagnetic environment in \refsec{Section:SpontaneousEmission}.

The total Hamiltonian of the system is given by 
\begin{equation}
	\hat H=\hat H_S+ \hat H_E +\hat H_{\text{int}},
\end{equation}
where $\hat H_S$ is the free Hamiltonian of the quantum emitters (system), $\hat H_E$ is the free Hamiltonian of the electromagnetic field (environment), and $\hat H_{\text{int}}$ is the light-matter interaction term derived in \eqref{eq:JaynesCummingsHamiltonian}:
\begin{align}
	\hat H_S&=\sum_\alpha\omega_{\alpha} \hat \sigma^\dagger_\alpha \hat \sigma_\alpha, \\
	\hat H_E&=\sum_{\mathbf{k},s}\omega_k \hat a_{\mathbf{k},s}^\dagger \hat a_{\mathbf{k},s} , \\
	\hat H_{\text{int}}&= -\hat{\bm \mu}  \cdot  \mathbf{\hat E}\approx- \sum_{\mathbf{k},s} \sum_\alpha g^{(\alpha)}_{\mathbf{k},s} (\hat \sigma_\alpha^\dagger \hat a_{\mathbf{k},s}+ \hat \sigma_\alpha \hat a^\dagger_{\mathbf{k},s}),
\end{align}
where each emitter spans a basis $\{ |g_\alpha\rangle, |e_\alpha\rangle \}$, such that the total dipole operator reads
\begin{equation}
	\hat{\bm \mu}  \equiv \sum_\alpha \bm \mu_\alpha (\hat \sigma_\alpha + \hat \sigma_\alpha^\dagger),\quad  \text{with}\quad \bm \mu_\alpha\equiv \langle g_\alpha | \hat{\bm \mu} |e_\alpha\rangle \in \mathbb{R},
\end{equation}
and $\hat \sigma_\alpha \equiv |g_\alpha \rangle \langle e_\alpha|$. We note that each emitter preserves the properties inherited from the $\mathfrak{su}(2)$ algebra, such that
\begin{equation}
	[\hat \sigma_\alpha^\dagger,\hat \sigma_\beta]=2 \hat \sigma_{z,\alpha} \delta_{\alpha \beta}, \quad 	
	[\hat \sigma_{z,\alpha},\hat \sigma^{(\dagger)}_\beta]=\pm \hat \sigma_{\alpha}^{(\dagger)}\delta_{\alpha \beta}, \quad 
	\{\hat \sigma_\alpha^\dagger,\hat \sigma_\beta \}= \delta_{\alpha \beta},
\end{equation}
where $\hat \sigma_{z,\alpha}\equiv 2 \hat \sigma_\alpha^\dagger \hat \sigma_\alpha - \mathbb{\hat I}_\alpha$ is the $z$-Pauli matrix for the $\alpha$th emitter. The light-matter coupling, $g^{(\alpha)}_{\mathbf{k},s}$, is equivalently defined as in \eqref{eq:lightmattercoupling},
\begin{equation}
	g^{(\alpha)}_{\mathbf{k},s}\equiv \sqrt{\frac{\omega_k}{2\varepsilon_0 V}}e^{i \mathbf{k}\cdot \mathbf{r}_\alpha} \mathbf{e}_{\mathbf{k},s} \cdot \bm \mu_\alpha, \quad \text{with} \quad g^{(\alpha)}_{\mathbf{k},s}\in \mathbb{R}
\end{equation}
where $\mathbf{r}_\alpha$ denotes the position, and $\bm \mu_\alpha$ the dipole moment element of the $\alpha$th emitter. As we noted in \refsec{Section:RabiJaynesModels}, we assume that the light-matter couplings are purely real for all emitters, but we allow the possibility of being different among them $g^{(\alpha)}_{\mathbf{k},s} \neq g^{(\beta)}_{\mathbf{k}',s'}$ for $\alpha\neq \beta$.
Finally, we identify the system and environment operators following the general description of the interaction Hamiltonian (before the RWA is applied) in \eqref{eq:spectralIntHamiltonian}:
\begin{equation}
	\hat H_{\text{int}}=\sum_\alpha \sum_{m, \tilde \omega_\alpha} e^{-i \tilde \omega_\alpha t} \hat S_m( \tilde  \omega_\alpha) \otimes \hat E_m=\sum_{\alpha, \tilde  \omega_\alpha}  e^{-i \tilde  \omega_\alpha t} \mathbf{\hat S}_\alpha( \tilde  \omega_\alpha) \otimes \mathbf{\hat E}(t), 
	\label{eq:identificationNemitter}
\end{equation}
where $ \mathbf{\hat E}(t)$ is the electric field operator defined in \eqref{eq:ElectricFieldOp}, and $\tilde \omega_\alpha=\pm \omega_\alpha$, such that the system operators satisfy
\begin{equation}
	\mathbf{\hat S }_\alpha( \omega_\alpha)= \bm \mu_\alpha \hat \sigma_\alpha, \quad 	\text{and}\quad \mathbf{\hat S }_\alpha(- \omega_\alpha)= \bm \mu_\alpha^* \hat \sigma_\alpha^\dagger.
	\label{eq:identificationNemitters2}
\end{equation}

We can use the techniques proved in\refsec{Section:OPS} in order to obtain an effective description of the quantum emitter in terms of the following Bloch-Redfield master equation: 
\begin{multline}
	\frac{d \hat \rho^{\text{I}}_S (t)}{dt}= \sum_{\alpha,\beta} \sum_{\tilde \omega_\alpha,\tilde \omega_\beta}\sum_{i,j} 
	\left[ e^{i( \tilde \omega_\alpha-\tilde \omega_\beta)t} \Gamma_{ij}( \tilde \omega_\alpha) [\hat S_{j,\beta}(\tilde\omega_\beta) \hat \rho^{\text{I}}_S (t), \hat S^\dagger_{i,\alpha}(\tilde \omega_\alpha)  ] \right.\\
	\left.	+ e^{i( \tilde \omega_\beta-\tilde \omega_\alpha)t} \Gamma_{ji}^*( \tilde \omega_\beta) [\hat S_{j,\beta}(\tilde\omega_\beta) \hat \rho^{\text{I}}_S (t), \hat S^\dagger_{i,\alpha}(\tilde \omega_\alpha)  ]  \right],
	\label{eq:MasterEqExample2}
\end{multline}
where $S_{i,\alpha}(\omega_\alpha) \equiv  \mu_{i,\alpha} \hat \sigma_{\alpha}$, with $d_{i,\alpha}$ the $i$th element of $\bm \mu_\alpha$, and $\Gamma_{ij}( \tilde \omega_\alpha)$ defined in a similar way to \eqref{eq:spectralcorrtensor}. Next, we perform the limit to the continuum and a rotating wave approximation that neglects all terms oscillating at frequencies $2 \omega_\alpha$ and $\omega_\alpha+ \omega_\beta$. Hence, by expanding the frequency sum and after some algebra, the master equation simplifies to~\cite{FicekQuantumInterference2005}:
\begin{multline}
	\frac{d \hat \rho^{\text{I}}_S (t)}{dt}= \sum_{\alpha,\beta} \left\{
	Y_{\alpha\beta} [\hat S_\beta^\dagger \hat \rho^{\text{I}}_S (t), \hat S_\alpha] + 	Y_{\beta \alpha}^* [\hat S_\beta^\dagger ,\hat \rho^{\text{I}}_S (t) \hat S_\alpha]
	\right. \\
	\left.
	X_{\alpha \beta} [\hat S_\beta \hat \rho^{\text{I}}_S (t), \hat S^\dagger_\alpha] + X_{\beta \alpha}^* [\hat S_\beta ,\hat \rho^{\text{I}}_S (t) \hat S^\dagger_\alpha]
	\right\},
\end{multline}
where we have defined $X_{\alpha \beta}$ and $Y_{\alpha \beta}$ as
\begin{align}
	X_{\alpha \beta}&= \frac{1}{(2\pi c)^3}\int_0^\infty d\omega_{\mathbf{k},s} \omega_{\mathbf{k},s}^2 e^{i(\omega_\alpha-\omega_\beta)t} \int_{\Omega_k} d\Omega_k \mathcal{K}_{\alpha \beta}^{(-)} (\mathbf{k},s) , 
	\label{eq:Xalphabeta}\\
	Y_{\alpha \beta}&= \frac{1}{(2\pi c)^3}\int_0^\infty d\omega_{\mathbf{k},s} \omega_{\mathbf{k},s}^2 e^{-i(\omega_\alpha-\omega_\beta)t} \int_{\Omega_k} d\Omega_k \mathcal{K}_{\alpha \beta}^{(+)} (\mathbf{k},s),
	\label{eq:Yalphabeta}
\end{align}
with $\mathcal{K}_{\alpha \beta}^{(\pm)} (\mathbf{k},s)$ denoting the sum over $(\mathbf{k},s)$ and the spectral correlation terms:
\begin{equation}
	\begin{aligned}[b]
		\mathcal{K}_{\alpha \beta}^{(\pm)}(\mathbf{k},s)=
		&(\bar n(\omega_{\mathbf{k},s}) +1)\sum_s [\bm \mu_\alpha \cdot \mathbf{e}_{\mathbf{k},s}(\mathbf{r}_\alpha)] [\bm \mu^*_\beta \cdot \mathbf{e}^*_{\mathbf{k},s}(\mathbf{r}_\beta)]\\
		&\quad\quad \times
		\left(
		\pi \delta(\omega_{\mathbf{k},s}\pm \omega_\beta)- \text{P.V.}\frac{1}{\omega_{\mathbf{k},s}\pm \omega_\beta}
		\right) \\
		&+ \bar n \sum_s [\bm \mu^*_\alpha \cdot \mathbf{e}_{\mathbf{k},s}(\mathbf{r}_\alpha)] [\bm \mu_\beta \cdot \mathbf{e}^*_{\mathbf{k},s}(\mathbf{r}_\beta)]\\
		&\quad\quad\times \left(
		\pi \delta(\omega_{\mathbf{k},s}\mp \omega_\beta)+ \text{P.V.}\frac{1}{\omega_{\mathbf{k},s}\mp \omega_\beta}
		\right).
	\end{aligned}
\end{equation}

To perform the sum over polarizations, $s$, we consider that the direction of the $z$ axis coincides with the direction of the interatomic vector $\mathbf{r}_{ij}\equiv\mathbf{r}_i-\mathbf{r}_j$, and assume that the dipole moments of the atoms are parallel and polarized in the $xz$ planes (for instance, there is an external electric field that align the dipoles and confine the polarization in a plane) as
\begin{align}
	\bm \mu_\alpha &= |\bm \mu_\alpha| (\sin \zeta,0,\cos \zeta),\\
	\bm \mu_\beta &= |\bm \mu_\beta| (\sin \zeta,0,\cos \zeta), 
\end{align}
where $\zeta$ is the angle between dipole moments and $\mathbf{r}_{\alpha \beta }$. Using the spherical representation of the propagation vector $\mathbf{k}$, and the unit polarization vectors, $\mathbf{e}_{\mathbf{k},1}$ and $\mathbf{e}_{\mathbf{k},2}$, 
\begin{align}
	\mathbf{k}&= |\mathbf{k}| (\sin \theta_k \cos \phi_k, \sin \theta_k \sin \phi_k, \cos \theta_k), \\
	\mathbf{e}_{\mathbf{k},1}&=(-\cos \theta_k \cos \phi_k,-\cos \theta_k \sin \phi_k,\sin \theta_k), \\
	\mathbf{e}_{\mathbf{k},2}&=(\sin \phi_k,-\cos \phi_k,0),
\end{align}
we find that the sum over polarizations results in
\begin{multline}
	\sum_s [\bm \mu_\alpha \cdot \mathbf{e}_{\mathbf{k},s}(\mathbf{r}_\alpha)] [\bm \mu^*_\beta \cdot \mathbf{e}^*_{\mathbf{k},s}(\mathbf{r}_\beta)] = \sum_s [\bm \mu^*_\alpha \cdot \mathbf{e}_{\mathbf{k},s}(\mathbf{r}_\alpha)] [\bm \mu_\beta \cdot \mathbf{e}^*_{\mathbf{k},s}(\mathbf{r}_\beta)] \\
	= \frac{\omega_{\mathbf{k},s} |\bm \mu_\alpha| |\bm \mu_\beta|}{2\varepsilon_0 V} 
	\left\{
	\sin^2\zeta (\cos^2 \theta_k \cos^2\phi_k + \sin^2 \phi_k) 
	\right. \\
	\left.
	+\cos^2 \zeta \sin^2 \theta_k- \frac{1}{4}\sin 2\zeta\sin 2\theta_k \cos \phi_k
	\right\} e^{i k r_{\alpha \beta} \cos \theta_k},
\end{multline}
where $k=|\mathbf{k}|$ and $r_{\alpha \beta}=|\mathbf{r}_{\alpha \beta }|$ is the distance between atoms. Therefore, the integral over the solid angle $d\Omega$ yields
\begin{multline}
	\int d\Omega	\sum_s [\bm \mu_\alpha \cdot \mathbf{e}_{\mathbf{k},s}(\mathbf{r}_\alpha)] [\bm \mu^*_\beta \cdot \mathbf{e}^*_{\mathbf{k},s}(\mathbf{r}_\beta)] \\
	= 	\int d\Omega	\sum_s \sum_s [\bm \mu^*_\alpha \cdot \mathbf{e}_{\mathbf{k},s}(\mathbf{r}_\alpha)] [\bm \mu_\beta \cdot \mathbf{e}^*_{\mathbf{k},s}(\mathbf{r}_\beta)] \\
	= \frac{2\pi \omega_{\mathbf{k},s} |\bm \mu_\alpha||\bm \mu_\beta|}{\varepsilon_0 V} \left\{
	\left[1-( \bar{\bm \mu} \cdot  \mathbf{\bar r}_{\alpha\beta})^2 \right]  \frac{\sin (k r_{\alpha \beta})}{kr_{\alpha \beta}} + \right. \\
	\left.
	\left[1- 3( \bar{\bm \mu}\cdot  \mathbf{\bar r}_{\alpha \beta})^2 \right] \left[\frac{\cos (k r_{\alpha \beta})}{(kr_{\alpha \beta})^2}-\frac{\sin (k r_{\alpha \beta})}{(kr_{\alpha \beta})^3}\right]
	\right\},
\end{multline}
where $\bar{\bm \mu}=\bar{\bm \mu}_\alpha=\bar{\bm \mu}_\beta$ and $\mathbf{\bar r}_{\alpha \beta}$ are unit vectors along dipole moments and the vector $\mathbf{r}_{\alpha \beta}$, respectively. 
Hence, the integration over $\omega_{\mathbf{k},s}$ in~\cref{eq:Xalphabeta,eq:Yalphabeta} results in
\begin{align}
	X_{\alpha \beta}&= \left[(1+\bar n) \left(\frac{1}{2} \gamma_{\alpha \beta}- i \Lambda_{\alpha \beta}^{(-)}\right) +i \bar n \Lambda_{\alpha \beta}^{(+)}\right] e^{i(\omega_\alpha-\omega_\beta)t} , 
	\label{eq:XalphabetaSolved}\\
	Y_{\alpha \beta}&= \left[\bar n \left(\frac{1}{2} \gamma_{\alpha \beta}+ i \Lambda_{\alpha \beta}^{(-)}\right) -i (1+\bar n )\Lambda_{\alpha \beta}^{(+)}\right] e^{-i(\omega_\alpha-\omega_\beta)t},
	\label{eq:YalphabetaSolved}
\end{align}
where we have approximated $\bar n(\omega_{\mathbf{k},s})\approx \bar n(\omega_0)\equiv \bar n$, with $\omega_0\equiv(\omega_\alpha+\omega_\beta)/2$. 
From~\cref{eq:Xalphabeta,eq:Yalphabeta}, we identify the spontaneous decay rate of the $\alpha$th emitter, $\gamma_{\alpha \alpha}$, and the collective spontaneous emission rates, $\gamma_{\alpha \beta}$:
\begin{align}
	\gamma_{\alpha \alpha} &= \gamma_\alpha= \frac{\omega_\alpha^3 |\bm \mu_\alpha|^2}{3\pi \varepsilon_0 c^3}, \\ 	\gamma_{\alpha \beta}&=\gamma_{\beta \alpha}=\sqrt{\gamma_\alpha \gamma_\beta }F(k_0 r_{\alpha \beta }),
\end{align}
with
\begin{multline}
	F(k_0 r_{\alpha \beta})\equiv \frac{3}{2}\left\{
	\left[1-(\bar{\bm \mu}\cdot  \mathbf{\bar r}_{\alpha \beta})^2 \right]  \frac{\sin (k_0 r_{\alpha \beta})}{k_0r_{\alpha \beta}} + \right. \\
	\left.
	\left[1- 3(\bar{\bm \mu}\cdot  \mathbf{\bar r}_{\alpha \beta})^2 \right] \left[\frac{\cos (k_0 r_{\alpha \beta})}{(k_0r_{\alpha \beta})^2}-\frac{\sin (k r_{\alpha \beta})}{(k_0r_{\alpha \beta})^3}\right]
	\right\},
	\label{eq:Fk0}
\end{multline}
where $k_0 \equiv \omega_0/c$, having assumed that $\omega_0\gg |\omega_\alpha-\omega_\beta|$. The parameters $\Lambda_{\alpha \beta}^{(\pm)}$ will contribute to atomic shifts and emitter-emitter vacuum-induced interactions, given by the Cauchy principal value integral:
\begin{equation}
	\Lambda_{\alpha \beta}^{(\pm)}= \frac{\sqrt{\gamma_\alpha \gamma_\beta}}{2\pi \omega_0^3}\text{P.V.}\int_0^\infty d\omega_{\mathbf{k},s} \frac{\omega_{\mathbf{k},s}^3 F(\omega_{\mathbf{k},s}r_{\alpha \beta}/c) }{\omega_{\mathbf{k},s}\pm \omega_\beta}
\end{equation}
where $F(\omega_{\mathbf{k},s}r_{\alpha \beta}/c)$ is given by \eqref{eq:Fk0} with $k_0$ replaced by $\omega_{\mathbf{k},s}r_{\alpha \beta}/c$. From this, we obtain the Lamb shifts of the atomic levels from the diagonal terms, and a vacuum-induced coherent (dipole-dipole) interaction between the atoms from the off-diagonal terms:
\begin{align}
	\Lambda_{\alpha \alpha}&\equiv \Lambda_\alpha= (1+2\bar n) (\Lambda_{\alpha \alpha}^{(+)}-\Lambda_{\alpha \alpha}^{(-)}),\quad &&(\alpha=\beta) \\
	\Lambda_{\alpha \beta }&=-(\Lambda_{\alpha \beta }^{(+)}+\Lambda_{\alpha \beta }^{(-)}), \quad &&(\alpha\neq \beta).
\end{align}
By using contour integration, we find the explicit form of the vacuum-induced dipole-dipole interaction $\Lambda_{\alpha \beta }$~\cite{LehmbergRadiationAtom1970}:
\begin{multline}
	\Lambda_{\alpha \beta }= \frac{3}{4}\sqrt{\gamma_\alpha \gamma_\beta}\left\{-
	\left[1-( \bar{\bm \mu}\cdot  \mathbf{\bar r}_{\alpha \beta })^2 \right]  \frac{\cos (k_0 r_{\alpha \beta })}{k_0r_{\alpha \beta }}  \right. \\
	\left.
	+\left[1- 3(\bar{\bm \mu}\cdot  \mathbf{\bar r}_{\alpha \beta })^2 \right] \left[\frac{\sin (k_0 r_{\alpha \beta })}{(k_0r_{\alpha \beta })^2}+\frac{\sin (k r_{\alpha \beta })}{(k_0r_{\alpha \beta })^3}\right]
	\right\}.
\end{multline}

Finally, the master equation of a two non-identical QEs interacting with a vacuum electromagnetic fields is given by:
\begin{equation}
	\colorboxed{Maroon}{
		\begin{aligned}[b]
			\frac{d \hat \rho_S (t)}{dt}&= -i[\hat H, \hat \rho_S(t)]\\
			&+ \sum_{\alpha, \beta }^N \frac{\gamma_{\alpha \beta }(1+\bar n)}{2}  \mathcal{D}[\hat \sigma_\alpha,\hat \sigma_\beta]\hat \rho_S(t) +  \frac{ \gamma_{\alpha \beta } \bar n}{2} \mathcal{D}[\hat \sigma^\dagger_\alpha,\hat \sigma^\dagger_\beta]\hat \rho_S(t),
		\end{aligned}
	}
\end{equation}
where we have extended the definition of dissipator given in \eqref{eq:LindbladTerms} to:
\begin{equation}
	\mathcal{D}[\hat O_i,\hat O_j]\hat \rho \equiv 2 \hat O_i \hat \rho \hat O_j^\dagger - \{\hat O_j^\dagger \hat O_i,\hat \rho\},
\end{equation}
and the Hamiltonian $\hat H$ is now given by
\begin{equation}
	\colorboxed{Maroon}{
		\begin{aligned}[b]
			\hat H=\hat H_S+&\hat H_{LS}+\hat H_{\Lambda}
			\\
			&=\sum_\alpha^N (\omega_\alpha+\Lambda_\alpha) \hat \sigma_\alpha^\dagger \hat \sigma_\alpha +\sum_{\alpha\neq \beta}^N \left(\Lambda_{\alpha \beta } \hat \sigma_\alpha^\dagger \hat \sigma_\beta + \text{H.c.}\right)
		\end{aligned}
	}
\end{equation}
\cleardoublepage
\chapter{Appendix: Chapter $5$}
\label{Apendix:Entanglement}

\section{Incoherent pumping}
\label{Appendix:IncoherentPump}

Incoherent pumping is understood as a process that injects excitation (e.g., photons, or phonons) due to some external interaction that we do not have information of, or that it effectively reduces to a flow of random incoming excitations. 
To understand the microscropic origin of this mechanism of excitation we should consider a specific model.
For instance, it is known that thermal/caothic light induces incoherent pumping with a rate that depends on the temperature of the environment~\cite{CarmichaelStatisticalMethods1999}. 
Or in semiconductor quantum dots (QDs) embedded in a cavity, which are typically excited far above resonance and electron-hole pairs relax incoherently to excite the QD in a continuous flow of excitations~\cite{KavokinMicrocavities2017,LaussyStrongCoupling2008}.
Nevertheless, independently of the platform, this excitation mechanism, introduced by a Lindblad-like term in the master equation, is just an effective description of a more intricate physical interaction between the system and some other source. Consequently, there is not any frequency-dependent character in the pumping term since it is treated as a stochastic mechanism that just introduces excitations into the system and in an irreversible manner.

We can understand this better by going to a particular example (a quantum optics toy model) where the incoherent pumping emerges as an effective description of another process.  
Let us consider a three level system, denoted as $\{ |1\rangle,|2\rangle,|3\rangle  \}$ (ordered in increasing energy), in which the transition from $|1\rangle$ to $|3\rangle$ is coherently driven (with driving strength $\Omega$ and frequency $\omega_L$), $|3\rangle$ decays towards $|2\rangle$ with a rate $\Gamma$, and $|2\rangle$ decays towards $|1\rangle$ with a rate $\gamma$.
This toy model is reproduced by a master equation of the form:
\begin{equation}
	\frac{d \hat \rho}{dt}=-[\hat H, \hat \rho]+ \frac{\Gamma}{2}\mathcal{D}[|2\rangle \langle 3 |] \hat \rho + \frac{\gamma}{2}\mathcal{D}[|1\rangle \langle 2|] \hat \rho,
\end{equation}
with a Hamiltonian $\hat H= \sum_i \omega_i |i\rangle \langle i |+ \Omega (|1\rangle \langle 3 | e^{i\omega_L t} +\text{H.c.})$. Moving to the rotating frame of the laser, we get the following differential equations provided by the previous master equation,
\begin{subequations}
	\begin{align}
		\dot \rho_{2,2}&=-\gamma \rho_{2,2}+\Gamma \rho_{3,3}, \\
		\dot \rho_{3,3}&=-\gamma \rho_{3,3}-i\Omega(\rho_{1,3}-\rho_{3,1}), \\
		\dot \rho_{1,3}&=-\frac{1}{2}(\Gamma-2i\Delta) \rho_{1,3}+i\Omega (1-\rho_{2,2}-2\rho_{3,3}),
	\end{align}
\end{subequations}
where $\Delta \equiv \omega_3-\omega_1$. In the case of weak driving, $\Gamma \gg \Omega$ and large detuning, $\Gamma \gg \Delta$, we can treat $|3\rangle$ as a virtual state that mediates interaction between $|1\rangle$ and $|2\rangle$. Therefore, $|3\rangle$ is adiabatically eliminated from the dynamics by setting $\dot \rho_{3,3}, \dot \rho_{1,3}=0$, and thus obtaining
\begin{equation}
	\rho^{\text{ss}}_{3,3}\approx \frac{2i\Omega}{\Gamma-2i\Delta}(1-\rho_{2,2}-2\rho^{\text{ss}}_{3,3}), \quad \quad  \rho^{\text{ss}}_{1,3}\approx  \frac{4\Omega^2}{\Gamma^2 +8\Omega^2+4\Delta^2}(1-\rho_{2,2}).
\end{equation}
In the limit $\Gamma \gg \Omega, \Delta$, and reintroducing these expressions into $\dot \rho_{2,2}$, we obtain an effective two-level system where $|2\rangle$ is incoherently pumped from $|1\rangle$ with a rate 
\begin{equation}
\colorboxed{Maroon}{
	P\equiv \frac{4\Omega^2}{\Gamma}, \quad \text{(Incoherent pumping rate)}
}
\end{equation}
such that
\begin{equation}
	\dot \rho_{2,2}\approx P-(P+\gamma)\rho_{2,2}.
\end{equation}

\begin{SCfigure}[0.55][h!]
	\includegraphics[width=0.75\columnwidth]{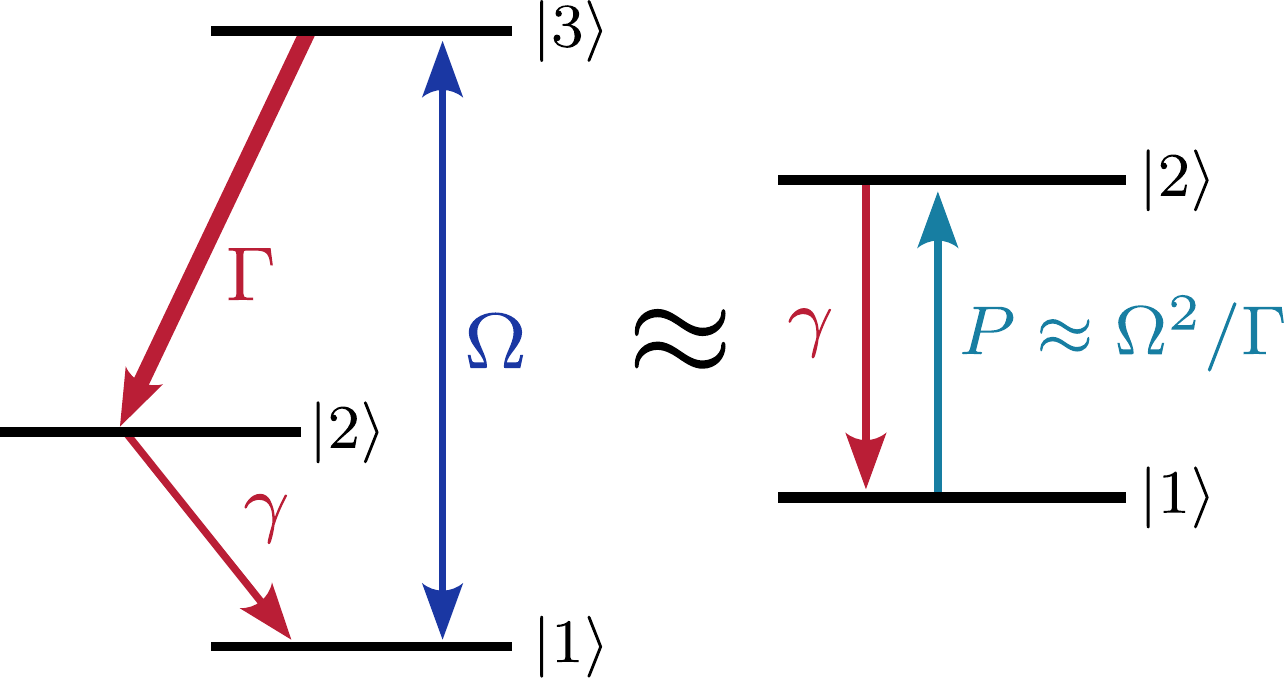}
	\captionsetup{justification=justified}
	\caption[Mechanism of incoherent excitation.]{\label{fig:IncohPump}	\textbf{Mechanism of incoherent excitation. } Effective description of an incoherent excitation channel from a more complex system involving the coherent excitation of a higher energy level that is very lossy.}
\end{SCfigure}

\section{Validity of the rotating-wave approximation}
\label{sec:SM_Validity_RWA}
In \eqref{eq:full_master_eq}, the three terms that compose the Hamiltonian, $\hat H_q$, $\hat H_a$, and $\hat H_d$, are written under a rotating-wave approximation (RWA), wherein the contribution to the dynamics of fast counter-rotating terms is neglected~\cite{Cohen-TannoudjiAtomPhotonInteractions1998}. 
Here, we prove numerically the validity of such approximation by including the neglected counter-rotating terms into the Hamiltonian and comparing the results with those obtained from the Hamiltonian used in the main text. Specifically, the qubit-qubit Hamiltonian ($\hat H_q$), the cavity Hamiltonian ($\hat H_a$), and the drive Hamiltonian ($\hat H_d$) are now characterized, in the laboratory frame, by   
\begin{subequations}
	\begin{align}
		\hat H_q&= (\omega_0-\delta)\hat \sigma_1^\dagger \hat \sigma_1+(\omega_0+\delta)\hat \sigma_2^\dagger \hat \sigma_2 + J (\hat \sigma_1^\dagger+\hat \sigma_1)(\hat \sigma_2^\dagger+\hat \sigma_2),  \\
		\hat H_a&= \omega_a \hat a^\dagger \hat a + g(\hat a^\dagger +\hat a)[ (\hat \sigma_1^\dagger+\hat \sigma_1)+(\hat \sigma_2^\dagger+\hat \sigma_2) ] \\
		\hat H_d&=2\Omega \cos (\omega_L t) [ (\hat \sigma_1^\dagger+\hat \sigma_1)+(\hat \sigma_2^\dagger+\hat \sigma_2) ]. 
		\label{eq:Ht_d}
	\end{align}
\end{subequations}
The presence of counter-rotating terms in $\hat H_q$ and $\hat H_a$ prevents the elimination of the time-dependency introduced by the driving term in \eqref{eq:Ht_d} by moving to a rotating frame, and consequently, the Liouvillian is now explicitly time-dependent. In this scenario, we can write the Liouvillian superoperator as
\begin{equation}
	\mathcal{L}(t)= \mathcal{L}_0+\mathcal{L}_{+}e^{i\omega_L t}+\mathcal{L}_{-}e^{-i\omega_L t},
\end{equation}
where $\mathcal{L}_{\pm}$ corresponds to time-independent Liouvillian terms which are multiplied by the exponential $e^{\pm i\omega_L t}$. As a result, a fully stationary solution cannot be obtained from the time-dependent master equation, 
\begin{equation}
	 \frac{d \hat \rho(t)}{dt}=\mathcal{L}(t) \hat \rho(t).
\end{equation}
To address this problem, we can make use of Floquet theory and the method of continued fractions~\cite{MajumdarProbingSingle2011,PapageorgeBichromaticDriving2012,MaragkouBichromaticDressing2013}. This method allows us to derive a time-averaged steady-state solution by identifying the nullspace of the matrix
\begin{equation}
 \mathcal{M}=\mathcal{L}_0+\mathcal{L}_{+}\mathcal{S}_{-1}+\mathcal{L}_{-}\mathcal{S}_1,
\end{equation}
where the matrices $\mathcal{S}_n$ are obtained by solving the recursive equation
\begin{equation}
	\mathcal{S}_{\pm n}=-[(\mathcal{L}_0-i n \omega_L \mathbb{I})+\mathcal{L}_\mp S_{\pm(n+1)}]^{-1} \mathcal{L}_\pm.
\end{equation}
A converged solution is achieved for a sufficiently large $n_{\text{max}}$, for which one truncates setting $\mathcal{S}_{\pm n_{\text{max}}}=0$. In our case, we set $n_{\text{max}}=3$, which provided a converged solution. The computation of the time-averaged steady-state density matrix results in a linear matrix problem, $\mathcal{M}\vec \rho_{ss}=0.$ 
\begin{SCfigure}[1][b!]
	\includegraphics[width=0.75\textwidth]{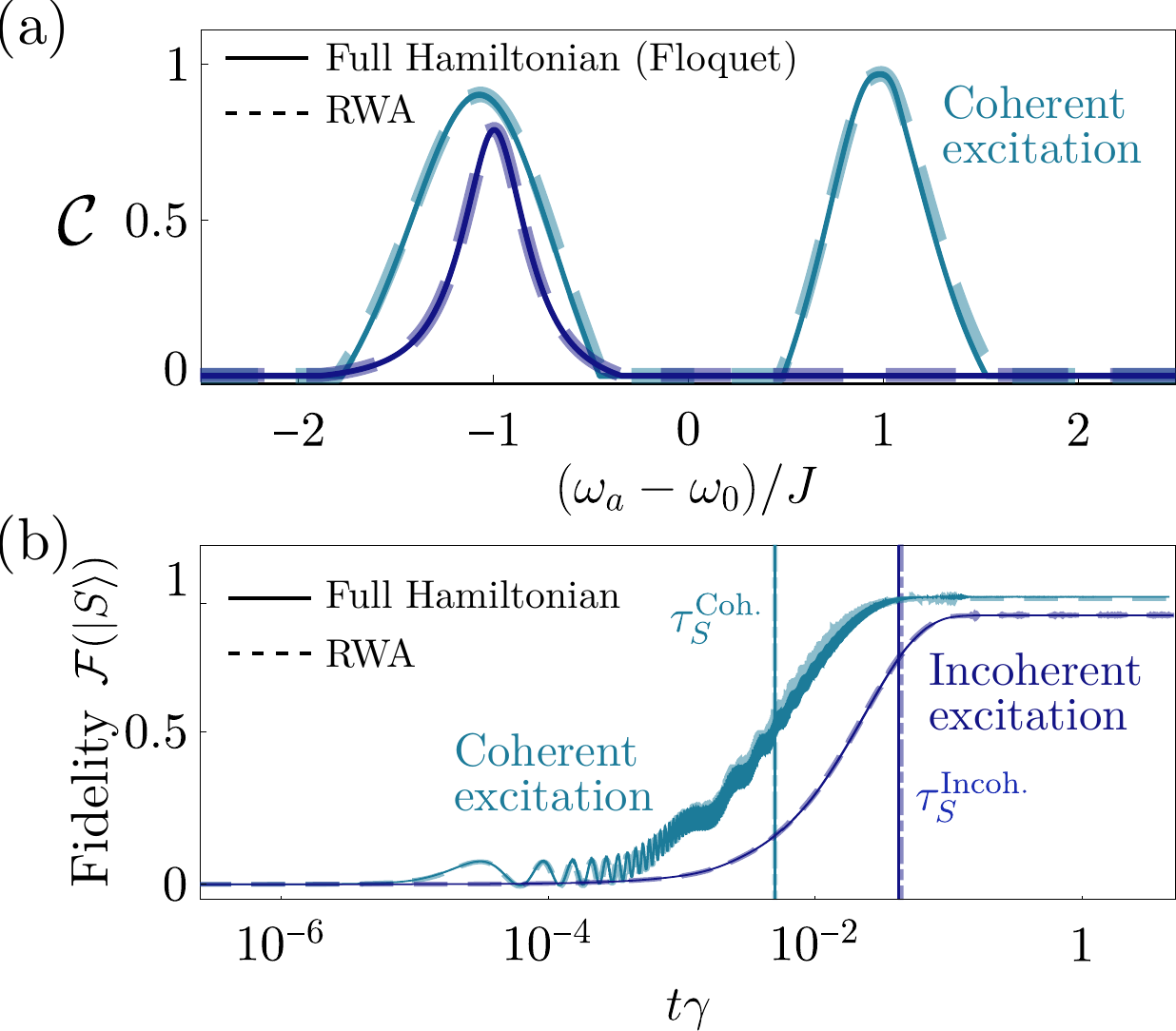}
	\captionsetup{justification=justified}
	\caption[Validity of the rotating-wave approximations (RWA)]{
		\textbf{Validity of the rotating-wave approximations (RWA).} (a) Stationary concurrence versus the cavity frequency $\omega_a$. (b) Time evolution of the fidelity $\mathcal{F}(|S\rangle)$ when $\omega_a=\omega_0-J$. In both plots, solid lines correspond to numerical simulations using the full Hamiltonian, while dashed lines correspond to numerical simulation using the RWA Hamiltonian outlined in the main text.   
		Parameters: 
		$\omega_0=10^7\gamma,\ \omega_L=\omega_0,\ J=9.18\times 10^4\gamma,\ \gamma_{12}=0.999\gamma,\ \delta=10^{-2}J,\ \ g=10^{-1}\kappa,\ \omega_a=\omega_0 -J,\ P=40\gamma,\ \Omega=10^4 \gamma$.}
	\label{fig:Validity_RWA}
\end{SCfigure}

In \reffig{fig:Validity_RWA}~\textcolor{Maroon}{(a)}, we simulate the concurrence $\mathcal C$ versus cavity frequency $\omega_a$ using the full-Hamiltonian and the RWA-Hamiltonian. The Floquet method was used for the case of coherent drive (the incoherent case is simulated straightforwardly). We observe that both models exhibit perfect agreement, thus proving the validity of the RWA in steady-state simulations.   
Additionally, in \reffig{fig:Validity_RWA}~\textcolor{Maroon}{(b)}, we analyze the time evolution of the fidelity $\mathcal{F}(|S\rangle)$ by simulating the dynamics under a time-dependent Hamiltonian with QuTiP \cite{ShammahOpenQuantum2018,JohanssonQuTiPOpensource2012,JohanssonQuTiPPython2013}. We observe that, beyond small discrepancies in the transient evolution towards the stationary solution, the RWA accurately describes the actual dynamics. 

These numerical results are consistent with the high ratio between interaction rates and bare frequencies that justify our use of the RWA. Since we are considering solid-state emitters with optical transition frequencies, $\omega_0$, in the range of hundreds of THz (with linewidths of a few MHz), a resonant coherent driving, $\omega_L=\omega_0$, and a single mode cavity that remains in vacuum (bad cavity limit), the different RWAs applied in the Hamiltonians ($\hat H_q$, $\hat H_a$, and $\hat H_d$) are perfectly valid within the parameters used across this work.
For instance, the set of parameters widely used in \refch{chapter:Entanglement}:
\begin{equation}
	(J,\Omega,g)=(10^5,10^4,10^3)\gamma \approx (1,10^{-1},10^{-2}) \  \text{THz},
\end{equation} 
where $\gamma\approx 30$ MHz is the typical linewidth for molecules with optical transitions~\cite{ToninelliSingleOrganic2021}, and assuming the resonant driving condition $\omega_L=\omega_0\approx 100$ THz, we can readily verify that
\begin{enumerate}[label=\textcolor{Maroon}{(\roman*)}]
	\item atom-atom RWA: $\omega_0/J\approx 10^2 \gg 1$ and $\omega_0 \gg \delta$.
	\item light-atom RWA: $\omega_0/g\approx 10^4 \gg 1$ and $\langle \hat a^\dagger \hat a \rangle \approx 0 $.
	\item driving RWA: $\omega_0/\Omega \approx 10^3 \gg 1$ and $\omega_L=\omega_0$.
\end{enumerate}
Hence, the validity of the RWA in our work is completely guaranteed.

\section{Different spontaneous decay rates}
\label{Appendix:diff_gamma}
If we let the spontaneous decay rates to be different, $\gamma_1\neq \gamma_2$, the dissipative part of the qubits acquires extra terms in the collective basis basis $\{|gg\rangle , |+\rangle, |-\rangle , |ee\rangle \}$, leading to
\begin{multline}
	\sum_{i,j=1}^2 \frac{\gamma_{ij}}{2}\mathcal{D}[\hat \sigma_i,\hat \sigma_j]\hat \rho  \longrightarrow     \frac{\tilde \gamma_+}{2}\mathcal{D}[\hat S_+]\hat \rho +\frac{\tilde \gamma_-}{2}\mathcal{D}[\hat S_-]\hat \rho \\
	+\frac{\tilde \gamma_{\text{col}}}{2}(\mathcal{D}[\hat S_+,\hat S_-] +\mathcal{D}[\hat S_-,\hat S_+])\hat \rho,
\end{multline}
where $\hat S_\pm$ are symmetric/antisymmetric collective ladder operators,  $\tilde \gamma_\pm$ are the generalized version of the super- and subradiant decay rates presented in the text, $ \gamma_\pm=\gamma\pm \gamma_{12}\cos \beta$, given by
\begin{equation}
	\tilde \gamma_\pm = (\gamma_1 + \gamma_2) \pm \gamma_{12}\cos \beta \pm (\gamma_2-\gamma_1)\sin \beta,
\end{equation}
while $\gamma_{\text{col}}$ is a collective damping rate between the collective states, $|\pm\rangle$, expressed as:
\begin{equation}
	\gamma_{\text{col}}=(\gamma_1-\gamma_2)\cos \beta + \gamma_{12} \sin \beta.
\end{equation}
Nevertheless, we note that most of the effects discussed in this work are enabled by the cavity, with a characteristic rate that is obtained after performing the adiabatic elimination of the cavity mode. As we discuss in the text, the super- and subradiant decay rates, $\tilde \gamma_\pm$, get modified by an additional term proportional to $\Gamma_P$. We focus on scenarios in which this is much larger that the natural and collective linewidths at high cooperativity ($C>1$), 
\begin{equation}
	\Gamma_P\gg  \gamma_i, \tilde \gamma_\pm, |\gamma_1-\gamma_2|.
\end{equation}
Therefore, we generally state that, in the present scenario where both emitters are coupled to a lossy cavity, different spontaneous decay rates would not drastically affect  the presented results. Hence, we can assume the approximation of similar decay rates without loss of generality, even if we do not consider identical emitters, if the following conditions are satisfied: 
\begin{equation}
	|\gamma_1-\gamma_2|\ll (\gamma_1+\gamma_2), \gamma_{12}.
\end{equation}
This additional condition ensures that $|-\rangle$ retains its population in the long-time limit, preventing any transfer to the symmetric state $|+\rangle$. We consider this condition is fulfilled even considering the different decay rates emerging from the energy detuning, since we impose that $\omega_0\gg \delta$.

\sectionmark{Adiabatic elimination of a two-qubit system coupled ...}
\section{Adiabatic elimination of a two-qubit system coupled to a lossy bosonic mode}
\sectionmark{Adiabatic elimination of a two-qubit system coupled ...}
\label{appendix:AdiabaticElimination}

Here, we provide a step-by-step demonstration of the adiabatic elimination for the cavity degrees of freedom to obtain the Bloch-Redfield Master Equation for the QEs presented in \eqref{eq:Nakajima}. In addition, we illustrate how this equation reduces to the Collective Purcell master equation in \eqref{eq:CollectivePurcell} when the cavity linewidth surpasses the maximum value of the dressed energy transitions.
In order to proceed, we make use of the projective adiabatic elimination method \cite{GardinerQuantumNoise2004,BreuerTheoryOpen2007,RivasOpenQuantum2012}.

Now, we apply this method in our case, consisting of a system of two interacting nonidentical QEs that are both driven by a coherent field and coupled to a single mode cavity. Here, our starting point is the master equation in \eqref{eq:full_master_eq}, which we can expand as~\cite{Navarrete-BenllochIntroductionQuantum2022}:
\begin{equation}
	\frac{d \hat  \rho(t)}{dt}= \mathcal{\hat L}_S[\hat  \rho]+ \mathcal{\hat L}_E[\hat  \rho] +\mathcal{\hat L}_{\text{int}}[\hat  \rho],
\end{equation}
where
\begin{subequations}
	\begin{align}
		\mathcal{\hat L}_S[\hat  \rho]&= -i[\hat H_\mathrm q+\hat H_\mathrm d,\hat  \rho(t)]+\sum_{i,j=1}^2 \frac{\gamma_{ij}}{2}\mathcal{D}[\hat \sigma_i,\hat \sigma_j]\hat  \rho(t)+\frac{P_i}{2} \mathcal{D}[\hat \sigma_i^\dagger]\hat \rho(t), \\
		\mathcal{\hat L}_E[\hat  \rho]&=-[\Delta_a \hat a^\dagger \hat a,\hat \rho(t)]+ \frac{\kappa}{2} \mathcal{D}[\hat a]\hat \rho(t), \\
		\mathcal{\hat L}_{\text{int}}[\hat \rho]&= -i[g[\hat a^\dagger\otimes(\sigma_1+\sigma_2)+\hat a\otimes(\hat \sigma_1^\dagger+\hat \sigma_2^\dagger)],\hat \rho].
	\end{align}
\end{subequations}
Because the system is assumed to be dressed, it is desirable to change to the eigenstate basis of the qubit-laser system [see \refch{ch:TwoPhotonResonance}]:
\begin{multline}
	\hat H_{\text{int}}=g[\hat a^\dagger\otimes(\hat \sigma_1+\hat \sigma_2)+\text{H.c.}]\\
	=\sum_{i,j} g_{ij} \hat a^\dagger \otimes \hat \sigma_{ij} +\text{H.c.}=\hat a^\dagger \otimes \hat \xi+\text{H.c.},
\end{multline}
where we have defined $\sigma_{ij}\equiv |j\rangle \langle i|$, $g_{ij}\equiv \langle j | \hat \sigma_1+\hat \sigma_2 |i\rangle$, and $\hat \xi\equiv \sum_{i,j} g_{ij} \hat \sigma_{ij}$ ($|i\rangle$, $i=1,\ldots,4$, denote the eigenstates of the qubit-laser system). In comparison with the general formula for the interaction Hamiltonian, $\hat H_{\text{int}}=\sum_m g_m \hat S_m\otimes \hat E_m$, we identify: $g_1=g_2=1$, $\hat E_1=\hat a^\dagger=\hat E_2^\dagger$, and $\hat S_1=\hat \xi=\hat S_2^\dagger$ (note that $\hat E_1$ is already the Hermitian conjugate of the bosonic operator). 
When $1/\kappa$ is much shorter than any other timescale in the problem, we can assume that no reversible dynamics occur, and then the cavity remains in the vacuum state $\hat  \rho_E\approx |0\rangle \langle 0|$. Therefore, in this scenario we can adiabatically eliminate the cavity and get an effective master equation for the two QEs of the same structure as the one in \eqref{eq:EffectiveMasterEq}.

Firstly, we need to compute the two-time correlation functions $C_{nm}$ and $K_{mn}$. We do this by neglecting any effect of the system into the environment. To make it clearer, we include a subscript $E$ (denoting \textit{environment}) under the expected value. Then, the bosonic operator forms a closed set,
\begin{equation}
	\frac{d\langle \hat a (\tau)\rangle_E}{d\tau}=\trace{\hat a \mathcal{L}_E[\hat  \rho(t)]}{E}=-(\kappa/2+i\Delta_a)\langle \hat a (\tau)\rangle_E,
\end{equation}
that can be easily integrated in time
\begin{equation}
	\langle \hat a (\tau)\rangle_E=\langle \hat a (0)\rangle_Ee^{-(\kappa/2+i\Delta_a)\tau}.
\end{equation}
Invoking the Quantum Regression Theorem~\cite{CarmichaelStatisticalMethods1999}, we can compute all the possible correlators $C_{nm}$ and $K_{mn}$,
\begin{subequations}
	\begin{align}
		\langle \hat A(t) \hat B (t+\tau )\rangle &=\trace{\hat B e^{\mathcal{L}\tau }[\hat \rho(t)\hat A]}{}, \\
		\langle \hat A(t+\tau) \hat B (t)\rangle &=\trace{\hat A e^{\mathcal{L}\tau }[\hat B \hat \rho(t)]}{}.
	\end{align}
\end{subequations}
Taking into account that $\hat E_1=\hat a^\dagger=\hat E_2^\dagger$, we obtain 
\begin{subequations}
	\begin{align}
		C_{11}(\tau)&=\lim_{t\rightarrow \infty}\langle \hat a^\dagger(t) \hat a^\dagger(t+\tau)\rangle_E=0, \\
		C_{22}(\tau)&=\lim_{t\rightarrow \infty}\langle \hat a(t) \hat a(t+\tau)\rangle_E=0,\\
		C_{12}(\tau)&=\lim_{t\rightarrow \infty}\langle \hat a^\dagger(t) \hat a(t+\tau)\rangle_E=0, \\
		C_{21}(\tau)&=\lim_{t\rightarrow \infty}\langle \hat a(t) \hat a^\dagger(t+\tau)\rangle_E=e^{-(\kappa/2-i\Delta_a)\tau},
	\end{align}
\end{subequations}
where we used the canonical commutation relation for bosonic operators $[\hat a,\hat a^\dagger]=1$.  Analogously for the $K_{mn}(\tau)$ we get
\begin{subequations}
	\begin{align}
		K_{11}(\tau)&=\lim_{t\rightarrow \infty}\langle \hat a^\dagger(t+\tau) \hat a^\dagger(t)\rangle_E=0, \\
		K_{22}(\tau)&=\lim_{t\rightarrow \infty}\langle \hat a(t+\tau) \hat a(t)\rangle_E=0, \\
		K_{12}(\tau)&=\lim_{t\rightarrow \infty}\langle \hat a^\dagger(t+\tau) \hat a(t)\rangle_E=0,\\
		K_{21}(\tau)&=\lim_{t\rightarrow \infty}\langle \hat a(t+\tau) \hat a^\dagger(t)\rangle_E=e^{-(\kappa/2+i\Delta_a)\tau}.
	\end{align}
\end{subequations}
In consequence, we obtain the following general expressions for the two-time correlators
\begin{subequations}
	\begin{align}
		C_{nm}(\tau)=e^{-(\kappa/2-i\Delta_a)\tau}\delta_{n,2}\delta_{m,1}, \\
		K_{mn}(\tau)=e^{-(\kappa/2+i\Delta_a)\tau}\delta_{m,2}\delta_{n,1}.
	\end{align}
\end{subequations}
We note that the indices between $C_{nm}$ and $K_{mn}$ are interchanged, which will be relevant later.

We recall that we are assuming that the correlation functions are either zero or decay at a much faster rate than any other dissipative process affecting the system. This means that $1/\kappa$ is the shortest dissipative timescale, and that in that timescale the only relevant dynamics of the system is given by its coherent evolution.
Then, by defining the action of the qubit-laser Hamiltonian $\hat H_q+\hat H_d$ onto the eigenstate basis as $(\hat H_q+\hat H_d) |i\rangle =\lambda_i |i\rangle $, we get
\begin{align}
	\hat \xi(\tau)&\approx e^{-i(\hat H_q+\hat H_d) \tau }\hat \xi e^{i(\hat H_q+\hat H_d) \tau } \notag\\
	&= \sum_{i,j} g_{ij}e^{-i(\hat H_q+\hat H_d) \tau } |j\rangle \langle i| e^{i(\hat H_q+\hat H_d) \tau }\notag\\
	&=\sum_{i,j} g_{ij}e^{i (\lambda_i-\lambda_j) \tau } |j\rangle \langle i| =\sum_{i,j} g_{ij}e^{i \omega_{ij} \tau } \hat \sigma_{ij},
\end{align}
where we have defined $\omega_{ij}\equiv \lambda_i- \lambda_j$. 

With these expressions, we can formally integrate \eqref{eq:EffectiveMasterEq} and obtain an effective master equation for the dressed-dimer system
\begin{equation}
	\partial_t \hat \rho_S(t)\approx \mathcal{L}_S[\hat \rho_S]	+ \mathcal L_\mathrm{eff}[\hat \rho_S] 
\end{equation}
where
\begin{equation}
	\colorboxed{Maroon}{
		\mathcal L_\mathrm{eff}[\hat \rho_S] =  \sum_{i,j,m,n} \frac{g_{ij} g_{mn}^*}{\kappa/2+i(\Delta_a-\omega_{ij})} [\hat \sigma_{ij} \hat \rho_S(t),\hat \sigma_{mn}^\dagger] + \text{H.c.}  
	}
	\label{eq:EffectiveMasterEqsol}
\end{equation}
We note that \eqref{eq:EffectiveMasterEqsol} has the form of a Redfield equation instead of a Lindblad form. Indeed, labeling the index pairs with greek letters $\alpha\equiv(i,j)$, $\nu\equiv (m,n)$, we see that \eqref{eq:EffectiveMasterEqsol} has the form:
\begin{multline}
	\mathcal L_\mathrm{eff}[\hat \rho_S] =  \sum_{\alpha,\nu} \frac{a_{\alpha \nu}}{2}\left(\hat\sigma_\alpha \hat \rho \hat\sigma_\nu^\dagger - \hat\sigma_\nu^\dagger\hat\sigma_\alpha \rho \right) 
	+ \frac{a_{\nu \alpha}^*}{2}\left( \hat\sigma_\alpha \hat \rho \hat\sigma_\nu^\dagger - \hat \rho \hat\sigma_\nu^\dagger   \hat\sigma_\alpha \right),
	\label{eq:Redfield_greek}
\end{multline}
where
\begin{equation}
	a_{\alpha\nu}\equiv \frac{2 g_\alpha g_\nu^*}{\kappa/2 + i(\Delta_a - \omega_\alpha)}.
	\label{eq:a_alpha_beta}
\end{equation}
This Redfield equation only acquires a Lindblad form provided that
\begin{equation}
	a_{\alpha\nu} = a^*_{\nu\alpha},
	\label{eq:condition_lindlad}
\end{equation}
in which case we obtain
\begin{equation}
	\mathcal L_\mathrm{eff}[\hat \rho_S] =  \sum_{\alpha,\nu} a_{\alpha\nu}\left(\hat\sigma_\alpha \hat  \rho \hat\sigma_\nu^\dagger  - \frac{1}{2}\left\{\hat\sigma_\nu^\dagger\hat\sigma_\alpha, \hat  \rho \right\}\right),
\end{equation}
which can be written in a Lindblad form by diagonalizing the matrix $a_{\alpha \nu}$~\cite{BreuerTheoryOpen2007}. However, the condition in \eqref{eq:condition_lindlad} will only be approximately true for certain choices of the system parameters.

For example, let us consider that the cavity leakage rate is much greater than any dressed energy transition, i.e., $\kappa\gg \omega_{ij}$. In this scenario the cavity is not able anymore to resolve the internal energy levels of the laser-dressed dimer, and thus we can approximate the denominators as $\kappa/2 \pm i(\Delta_a-\omega_{ij})\approx \kappa/2$. 
Defining the \textit{Purcell rate} as $\Gamma_P\equiv 4g^2/\kappa$, we recover the collective Purcell master equation described in \eqref{eq:CollectivePurcell}, which in this case  has a Lindblad structure,
\begin{equation}
	\colorboxed{Maroon}{
		\mathcal L_\mathrm{eff}[\hat  \rho_S] \approx \frac{\Gamma_P}{2} \mathcal{D}[(\hat \sigma_1+\hat\sigma_2)]\hat  \rho_S(t).
	}
\end{equation}

\section{Stationary entanglement at $\Delta=\pm R$}
\label{appendix:B}
In this section, we investigate the potential for entanglement generation using an alternative approach to the excitation at the two-photon resonance ($\Delta=0$) studied thorough this work: the direct excitation of the super/subradiant-like eigenstates of a dimer of interacting quantum emitters ($J \gg \gamma, \delta$) at the one-photon resonances $\Delta = \mp R$. 
We demonstrate that, in the best possible scenario, this approach  achieves a maximum steady-state concurrence of $\mathcal{C} = 0.5$. This is because the resonant excitation isolates the ground state and one entangled eigenstate, effectively behaving as a coherently driven two-level system that cannot achieve population inversion.
This upper limit is significantly lower than the maximum possible value of $\mathcal{C} = 1$, which, as we have demonstrated in this work, is attainable through excitation at the two-photon resonance. This difference justifies the importance of the regime of two-photon driving that has been the focus of this work.
In \refch{ch:TwoPhotonResonance}, we demonstrated that two strongly coupled emitters with a sufficiently weak drive ($J\gg \delta,\gamma, \Omega$) can be described as effective two-level systems composed by $\{ |gg\rangle, |\pm\rangle \}$ when fixing the drive in the corresponding one-photon resonances ($\Delta=\mp R$). These transitions are driven with a Rabi frequency $\Omega_{\pm}\equiv \Omega \sqrt{1\pm\cos \beta}$, respectively. In this two-level regime, the maximum steady-state population of the states $|\pm\rangle$ is the same as for a coherently driven two-level system, i.e., $1/2$.
If we now consider coupling to a cavity in the bad-cavity regime providing collective dissipation ($\kappa \gg J$), a similar analysis as in \refap{appendix:AdiabaticElimination} shows that the cavity modifies the super/subrradiant decay rates with additional Purcell rates:
\begin{equation}
	\Gamma_{\pm}=\gamma \pm \gamma_{12}\cos \beta + (1\pm\cos\beta)\Gamma_P.
\end{equation}
Plugging these rates into the analytical expressions for the stationary populations obtained from  the two-level description [see \refch{ch:TwoPhotonResonance}, \eqref{eq:rho_SymAntisym_1PModel}], we get 
\begin{equation}
	\colorboxed{Maroon}{
		\rho^{\text{eff}}_{\pm,\pm}\approx \frac{4\Omega^2 (1\pm\cos \beta )}{\Gamma_{\pm}^2+4(\Delta \pm R)^2+8\Omega^2(1\pm\cos \beta)}.
	}
	\label{eq:AnalyticSymAntisym}
\end{equation}
\begin{SCfigure}[1][t!]
	\includegraphics[width=0.9\textwidth]{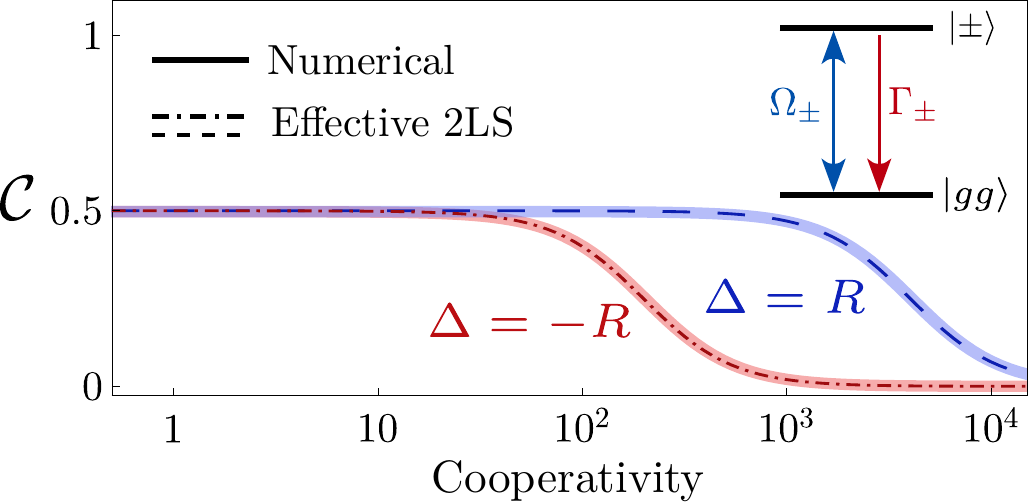}\\
	\captionsetup{justification=justified}
	\caption[Stationary entanglement at $\Delta=\pm R$ versus cooperativity.]{\textbf{Stationary entanglement at $\Delta=\pm R$ versus cooperativity.}
		Solid lines correspond to exact calculations from the full model, and red dot-dashed and blue dashed lines are analytical prediction from an effective two-level system composed by $\{|gg\rangle, |\pm\rangle\}$, respectively (depicted in the figure). 
		Parameters:
		$r=2.5$ nm,
		$k=2\pi /780\ \text{nm}^{-1}$,
		$J=9.18\times 10^4\gamma$,
		$\gamma_{12}=0.999$, $\delta=10^{-2} J$, $R=9.18\times 10^4\gamma$, $\Delta=\pm R$, $\Delta_a=-R$,
		$\Omega=10^2 \gamma$, $\kappa=10^6 \gamma$ .}
	\label{fig:Fig17_Appendix_ConcurrenceDeltaR}
\end{SCfigure}
The effect of the cavity in this regime is thus increasing $\Gamma_{\pm}$, which in turn decreases the population of the entangled states according to \eqref{eq:AnalyticSymAntisym}. The reduction will be significant when $\Gamma_{\pm}>\Omega$.
This dependence of the entanglement on the cavity coupling is demonstrated in \reffig{fig:Fig17_Appendix_ConcurrenceDeltaR}, where we compute the concurrence in terms of the cooperativity. Solid lines correspond to results obtained with the full model given by \eqref{eq:full_master_eq}, and dot-dashed ($\Delta=-R$) and dashed ($\Delta=R$) to analytical prediction of \eqref{eq:AnalyticSymAntisym} from the effective two-level system models described above. 
We note that in the latter case, the concurrence is simply related to the steady-state populations as:
\begin{equation}
	\colorboxed{Maroon}{
		\mathcal{C}\approx \text{Max}\left[0,\rho_{\pm,\pm}^{\text{eff}}\right].
	}
\end{equation}
As shown in \reffig{fig:Fig17_Appendix_ConcurrenceDeltaR}, this expression perfectly matches the full numerical simulations.

Finally, we remark that, in the regime of weakly-coupled emitters ($J\ll \delta$), the eigenstates are no longer entangled states but bare qubit states,  $|-\rangle_{\beta\approx \pi/2} \approx |eg\rangle$ and $|+\rangle_{\beta\approx \pi/2} \approx |ge\rangle$. Therefore, excitation at the one-photon resonances will not generate any type of entanglement via one-photon excitation. The main mechanism of entanglement generation in that regime is Mechanism II, provided the corresponding conditions (c.f. \textcolor{Maroon}{Table}~\ref{tab:EntanglementTab}) are fulfilled.

\sectionmark{Dependence of the entanglement generation on the ...}
\section{Dependence of the entanglement generation on the coupling rate between emitters}
\sectionmark{Dependence of the entanglement generation on the ...}
\label{sec:SM_Validity}

The finite value of $J$ can be a limiting factor for the generation of entanglement, as we just saw in \eqref{eq:rhoS_max_J}. Assuming optimal driving intensities, $P_{\text{opt}}$ and $\Omega_{\text{opt}}$, we extend here the numerical analysis presented in the main text to different values of the coupling rate between emitters, $J$.

In the simplest scenario, $J$ is set by dipole-dipole interactions, and therefore different value of $J$ correspond to different inter-emitter distances. With this situation in mind, we also vary $\gamma_{12}$ accordingly~\cite{ReitzCooperativeQuantum2022,FicekQuantumInterference2005}. Nevertheless, we note that $\gamma_{12}$ is not a critical parameter (since we work in a regime in which it is overcome by the Purcell-enhanced decay rate $\Gamma_P\gg \gamma$) and that, in general, $J$ could be provided by mechanisms other than dipole-dipole interaction, e.g., mediated by photonic structures~\cite{GoldsteinDipoledipoleInteraction1997,EvansPhotonmediatedInteractions2018,ChangColloquiumQuantum2018}.

In \reffig{fig:FigSM1_Validity}, we show the stationary concurrence, $\text{max}_{P/\Omega}(\mathcal{C})$ and the fidelity to $|S\rangle$, $\text{max}_{P/\Omega}[\mathcal{F}(|S\rangle)]$, both maximized over the driving strength, plotted against $J$ and the cavity cooperativity $C$. 
In this plot, we see wide regions of values of $J$ and $C$ where high values of stationary entanglement are obtained, consistent with the regions where the symmetric state $|S\rangle$ is highly populated. We observe that these regions are well bounded by the condition $\kappa \ll J$ that allows for the frequency-resolved Purcell effect to take place.
In both schemes, we observe that interaction rates of at least $J\sim 10^3\gamma$ are necessary to achieve sizable values of the concurrence, with values as high as $\mathcal C\sim 0.7$ within reach for cooperativities in the range $C\sim 10-100$. 
These results underscore the generality of the reported mechanism and its potential application for obtaining maximum entanglement in different solid-state cavity QED platforms.

\begin{SCfigure}[1][t!]
	\includegraphics[width=1.0\textwidth]{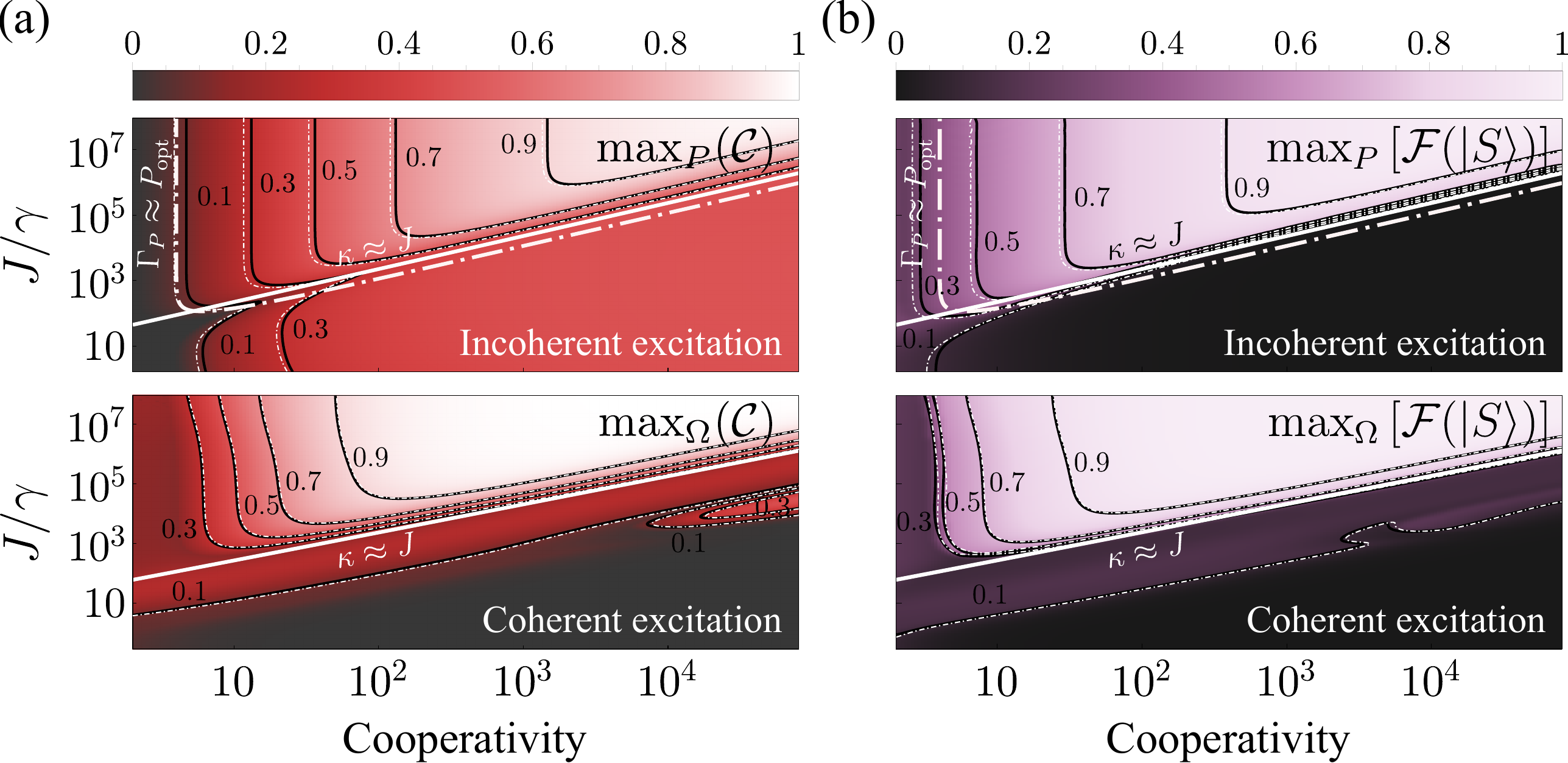}
	\captionsetup{justification=justified}
\end{SCfigure}
\graffito{\vspace{-13.0cm}
	\captionof{figure}[Maximum achievable entanglement in the parameters space]{	\label{fig:FigSM1_Validity}
		\textbf{Maximum achievable entanglement in the parameters space.}
		(a) Maximum achievable stationary concurrence and (b) fidelity with $|S\rangle$ versus cooperativity $C$ and dipole-dipole coupling $J$. 
		In each panel the driving intensity optimizes the generation of entanglement, that is, $P=P_{\text{opt}}$ and $\Omega=\Omega_{\text{opt}}$, respectively.
		The black straight lines correspond to numerical computations, and grey-dashed lines are numerical predictions from the Bloch-Redfield master equation.
		Parameters: $\delta=10^{-3}J$, $\Omega=\Omega_{\text{opt}}$, $P=P_{\text{opt}}$.
	}
}	

\section[Mechanism I: Derivation of a master  equation in Lindblad form]{Mechanism I: Derivation of a master \protect\newline equation in Lindblad form}
\label{appendix:MechanismI}

The frequency-resolved Purcell regime, where mechanism I is activated, is set by the condition $R\gg \kappa$, which means the cavity is able to resolve the different sets of transitions taking place among the excitonic levels of the dimer, split by $2R$ [see \reffig{fig:Fig1_FailureTwoPhotonDecay}]. 
We now show that, in this case, \eqref{eq:EffectiveMasterEqsol} can be cast in a Lindblad form, providing valuable insights into the mechanism underlying the stabilization of entanglement.

\subsection{Antisymmetric transition $\Delta_a =R$ (Coherent excitation)}
\label{appendix:DeltapR_Coh}
By setting $\Delta_a = R$, the sum over $i$ and $j$ in \eqref{eq:EffectiveMasterEqsol} will be dominated by terms in which $\omega_{ij} \approx R$. Following the notation that we used to define the eigenstates in \refch{ch:TwoPhotonResonance}, and the corresponding list of eigenvalues in \eqref{eq_eigenenergies_twophoton}, we observe that the set of pairs $(i,j)$ dominating the sum are 
\begin{equation}
	\mu_A \equiv \{(1,2), (1,3), (2,4),(3,4) \},
	\label{eq:mu_A}
\end{equation}
corresponding to the transitions $|+\rangle \rightarrow |A_2\rangle$, $|+\rangle \rightarrow |S_2\rangle$, $|A_2\rangle \rightarrow |-\rangle$ and $|S_2\rangle \rightarrow |-\rangle$, respectively. The rest of terms in the sum will be proportional to $\propto g^2/R$ and, based on the assumption $R\gg \kappa \gg g$, they can be neglected. Similarly, all the terms in the sum over $(m,n)$ that do not belong to the set $\mu_A$ will give rise to terms that, in the Heisenberg picture, will rotate with a frequency proportional to $R$. Based on the same assumptions, we can perform a rotating wave approximation and neglect all the terms in the sum over $(m,n)$ that do not belong to $\mu_A$. 

Since we will restrict the sum in \eqref{eq:Redfield_greek} to $\alpha,\beta\in \mu_A$, the value of $(\Delta_a - \omega_\alpha)$ in the denominator of \eqref{eq:a_alpha_beta} will, at most, proportional to $\Omega_\mathrm{2p}$. If $\kappa$ is such that the cavity cannot resolve the different transitions in $\mu_A$, i.e., if $\kappa\gg \Omega_\mathrm{2p}$, we may set 
\begin{equation}
	a_{\alpha \nu} \approx \frac{4 g_\alpha g_\nu^*}{\kappa}.
	\label{eq:a_matrix}
\end{equation}
Under this approximation, $a_{\alpha \nu}$ fulfills the condition in \eqref{eq:condition_lindlad}, meaning we can write $\mathcal L_\mathrm{eff}$ in the first standard form.
The values of the coefficients $g_{ij}$ can be obtained from the expressions of the eigenstates of the laser-dressed dimer given in \refsec{sec:perturbative_regime} [\cref{eq:SymAntiSym_beta0,eq:two_photondressed_eig}].
To first order in $\beta$ (we remind the reader that this discussion applies to the case in which the emitters are strongly interacting, so $\beta \ll 1$)  these are:
\begin{subequations}
	\begin{align}
		g_{1,2} &= g_{1,3} = g,\\
		g_{2,4} &= -g_{3,4} \approx g\frac{\beta}{2},
	\end{align}
\end{subequations}
which allows us to straightforwardly build the coefficient matrix $a_{\alpha \nu}$ using \eqref{eq:a_matrix}. The resulting matrix has a single non-zero eigenvalue with value $\lambda =2\Gamma_{P}+ \mathcal O (\beta^2)$, where $\Gamma_P$ is the standard Purcell rate
\begin{equation}
	\Gamma_P=  \frac{4g^2}{\kappa}.
\end{equation} 
The operator built from the corresponding eigenstate~\cite{BreuerTheoryOpen2007} is given by 
\begin{equation}
	\hat \xi_A = \frac{1}{\sqrt 2}\left[\hat \sigma_{12} + \hat\sigma_{13} +\frac{\beta}{2}\left(\hat\sigma_{24} - \hat\sigma_{34} \right)  \right].
\end{equation}
Using again the expression of the eigenstates in\cref{eq:SymAntiSym_beta0,eq:two_photondressed_eig}, we can see that $\hat\sigma_{12} + \hat\sigma_{13} = \sqrt{2}|gg\rangle\langle +|$, and $\hat\sigma_{24} - \hat\sigma_{34}  = -\sqrt{2} |-\rangle\langle ee|$, so 
\begin{equation}
	\hat\xi_A = |gg\rangle\langle + | - \frac{\beta}{2}|-\rangle \langle ee|.
	\label{eq:jump_xiA}
\end{equation}

As a result, we can write the effective dynamics induced by the cavity in a simple Lindblad form, which consists only of a single jump operator:
\begin{equation}
	\mathcal L_\mathrm{eff}[\hat \rho] = \Gamma_P\mathcal D[\hat\xi_A ]\hat\rho.
\end{equation}

The combination of this jump operator and the two-photon driving eventually stabilizes the subradiant state $|-\rangle \approx |A\rangle$.
This can be intuitively understood in the following way: first, the two-photon excitation populates the state $|ee\rangle$; then, the term $\frac{\beta}{2}|-\rangle \langle ee|$ in the jump operator $\hat\xi_A$ takes the system from $|ee\rangle$ into $|-\rangle$. The latter process occurs with a rate 
\begin{equation}
	\Gamma_{\mathrm I,A} = \beta^2\Gamma_P/2,
\end{equation}
which means that, since the system leaves state $|-\rangle$ with a rate $\gamma_- = \gamma-\gamma_{12}\cos \beta$, this state will be stabilized with high occupation provided that $\Gamma_{\mathrm I,A}  \gg \gamma_-$.

In the regime of parameters considered in this text we will typically find that  $\Omega_{\mathrm{2p}}\gg \Gamma_{\mathrm I,A}$. In this limit, the resulting rate equations can be reduced to that of a two-state system, and the evolution of the population $\rho_{-}\equiv \langle -|\hat\rho|-\rangle$ is well described by the equation
\begin{equation}
	\rho_{-}(t)  = \rho_{A,\mathrm{ss}}\left[1-e^{-\frac{1}{2}(\gamma_- + \Gamma_{\mathrm I,A})t}\right],
\end{equation}
where the steady-state value is given by
\begin{equation}
	\rho_{A,\mathrm{ss}} = \frac{\Gamma_{\mathrm I,A}}{\Gamma_{\mathrm I,A}+\gamma_-}.
\end{equation}
These equations provide the timescale of stabilization $\tau^{\text{Coh}}_{A}=2(\Gamma_{\mathrm{I},A}+\gamma_-)^{-1}$, which in the case $\Gamma_{\mathrm I,A}\gg \gamma_-$ the process is efficient, is given by
\begin{equation}
	\tau^{\text{Coh}}_{A}\approx 2/\Gamma_{\mathrm I,A} = \frac{4}{\beta^2 \Gamma_P}.
\end{equation}

Notice that the presence of the factor $\beta$ in the rate $\Gamma_{\mathrm I,A}$ indicates that a certain nonzero detuning $\delta$ is required to ensure that $\Gamma_{\mathrm I,A}\neq 0$. This finite detuning is necessary to prevent the state $|-\rangle$ from being entirely subradiant and uncoupled from the cavity.

\subsection{Symmetric transition $\Delta_a =-R$ (Coherent excitation)}
\label{appendix:DeltamR_Coh}

When the cavity is placed close to the symmetric resonance, the following set of transitions are enhanced.
\begin{equation}
	\mu_S \equiv \{(2,1), (3,1), (4,2),(4,3) \}.
	\label{eq:mu_S}
\end{equation}
Repeating the procedure described above for this set, we also obtain an effective dynamics described by a single Lindblad term with the following jump operator 
\begin{equation}
	\hat\xi_S = \frac{1}{\sqrt 2}\left[\hat\sigma_{21} - \hat\sigma_{31}  +\frac{\beta}{2}\left(\hat\sigma_{42} + \hat\sigma_{43} \right)   \right].
\end{equation}
Using the expression of the eigenstates, one finds that  $\hat\sigma_{21} - \hat\sigma_{31} = -\sqrt{2}|+\rangle \langle ee|$, and $\hat\sigma_{42} + \hat\sigma_{43} = \sqrt{2}|gg\rangle \langle -|$, which allows us to write the jump operator as
\begin{equation}
	\hat\xi_S = -|+\rangle \langle ee| + \frac{\beta}{2}|gg\rangle\langle -|.
\end{equation}
The effective dynamics is then given by
\begin{equation}
	\mathcal L_\mathrm{eff}[\hat\rho]=\Gamma_P\mathcal D[\hat\xi_S]\hat\rho.
\end{equation}
The process of stabilization of the superradiant state $|+\rangle$ is analogous to the subradiant case, with the primary difference being the decay rate from the doubly excited state $|ee\rangle$, which is notably enhanced in the superradiant case and given by
\begin{equation}
	\Gamma_{\mathrm I,S}=2\Gamma_P.
\end{equation}

Another difference  in this case is that, for the typical parameters considered, the induced decay rate from $|ee\rangle$ may be comparable to the two-photon driving strength  $\Gamma_{\mathrm I,S} \sim \Omega_{\mathrm{2p}}$ or even much larger $\Gamma_{\mathrm I,S} \gg \Omega_{\mathrm{2p}}$. In the second case, we may adiabatically eliminate $|ee\rangle$ from the dynamics and describe the combination of two-photon pumping and cavity-induced decay as an effective incoherent pumping of the state $|S\rangle$ from $|gg\rangle$, with a rate
\begin{equation}
	P_S = \frac{4\Omega_\mathrm{2p}^2}{\Gamma_{\mathrm I,S}} = \frac{2\Omega_\mathrm{2p}^2}{\Gamma_P}.
\end{equation}
Since $\Gamma_{\mathrm I,S}\gg \gamma_-$, we can also neglect the subradiant state $|-\rangle$, and obtain a description of the dynamics corresponding to a incoherently-pumped two-level system---consisting of states  $|gg\rangle$ and $|S\rangle$---with pumping rate $P_S$ and decay rate $\gamma_+=\gamma+\gamma_{12}\cos \beta$. The resulting population of superradiant state $\rho_+ \equiv \langle +|\hat\rho|+\rangle$ as a function of time is then given by
\begin{equation}
	\rho_+(t) = \rho_{S,\mathrm{ss}}\left[1-e^{- (\gamma_+ + P_{S})t} \right],
\end{equation}
where the steady-state value is 
\begin{equation}
	\rho_{S,\mathrm{ss}} = \frac{P_S}{P_S + \gamma_+}.
\end{equation}
This steady state is stabilized in a timescale $\tau_{\mathrm I,S}$ given by 
\begin{equation}
	\tau^{\text{Coh}}_{S} = \frac{2}{\gamma_+ + P_S} \approx \frac{\Gamma_P}{\Omega_\mathrm{2p}^2},
	\label{eq:tauIS-appendix}
\end{equation}
where the last inequality is taken in the limit $P_S \gg \gamma_+$ where the state is stabilized with high probability.

If we cannot use the approximation $\Gamma_{\mathrm I, S} \gg \Omega_\mathrm{2p}$, the dynamics cannot be straightforwardly reduced to that of a two-level system by adiabatically eliminating $|ee\rangle$. However, although more involved, a three-level system description of the problem (in which we only eliminated $|-\rangle$ based on the assumption $\Gamma_{\mathrm I,S}\gg\gamma_-$) can still give useful analytical insights. In particular, we find that a more general expression of the steady state is given by
\begin{equation}
	\rho_{S,\mathrm{ss}} = \frac{1}{1 + \gamma_+\left(P_S^{-1}+\Gamma_P^{-1}\right)}.
\end{equation}
Full diagonalization of the	Liouvillian also gives a full expression of the timescale of the relaxation, given by:
\begin{equation}
	\tau^{\text{Coh}}_{S} =  
	\frac{2}{\Gamma_P - \text{Re} \sqrt{\Gamma_P^2-4\Omega_\mathrm{2p}^2}}
	, 
\end{equation}
which neatly recovers the result in \eqref{eq:tauIS-appendix} in the limit $\Gamma_P \gg \Omega_\mathrm{2p}$.

\subsection{Symmetric transition $\Delta_a= -R$ (Incoherent excitation)}
\label{appendix:DeltamR_Incoh}

In order to keep the discussion as general as possible, we consider in this Section  the possibility of a finite qubit-qubit detuning $\delta$, so that the single-excitation eigenstates are given by 
\begin{equation}
	|\pm \rangle \equiv 1/\sqrt{2}(\sqrt{1\mp \sin \beta} |eg\rangle\pm \sqrt{1\pm \sin \beta}|ge\rangle) ,
\end{equation}
where $\beta\equiv\arctan (\delta/J)$ denotes a mixing angle. The corresponding eigenvalues are  $\omega_\pm =\Delta\pm R$, where $R=\sqrt{J^2 +\delta^2}$ is the Rabi frequency of the emitter-emitter coupling. We assume in the following that: \textcolor{Maroon}{(i)} $\delta$ is finite but much smaller than $J$, so that $R\approx J$, \textcolor{Maroon}{(ii)} the dressing of the qubits in the case of coherent drive is negligible, and \textcolor{Maroon}{(iii)} the rate of incoherent pumping $P$ is low enough to be disregarded in the derivation of \eqref{eq:Nakajima}. In that case, we label the eigenstates of the qubit-qubit system as
\begin{equation}
	|1\rangle = |gg\rangle;\quad
	|2\rangle = |-\rangle;\quad
	|3\rangle = |+\rangle;\quad
	|4\rangle = |ee\rangle.
\end{equation}

By setting the cavity frequency at the symmetric resonance, the terms where $\omega_{ij}\approx \Delta_a \approx -J$ will be dominant in the sum over $(i,j)$ in \eqref{eq:Nakajima}.
These terms are given by the following set of enhanced transitions,
\begin{equation}
	\mu_S \equiv \{ (2,1), (4,3) \},
\end{equation}
corresponding to  $|-\rangle \rightarrow |gg\rangle$, and $|ee\rangle \rightarrow |+\rangle$, respectively. The rest of the terms in the sum can be neglected since they are proportional to $g^2/J\rightarrow 0$, provided the conditions for a frequency-resolved Purcell effect hold,
\begin{equation}
	J\gg \kappa \gg g.
	\label{eq:conditions_freq_Purcell}
\end{equation}
By performing a rotating-wave approximation, terms not belonging to $\mu_S$ in the sum over $(m,n)$
can also be neglected, since these will rotate at frequencies $J\gg g^2/\kappa$ in the Heisenberg picture [the inequality being a consequence of \eqref{eq:conditions_freq_Purcell}].  

Under these conditions, the cavity contribution in \eqref{eq:Nakajima} can be reduced to the sum
\begin{equation}
	\mathcal{L}_{\text{cav}}^{\text{eff}}\approx  \frac{g_{21}}{\kappa/2}[\hat \sigma_{21} \hat \rho,\hat \Lambda]+ \frac{g_{43}}{\kappa/2}[\hat \sigma_{43}\hat \rho,\hat \Lambda]+ \text{H.c.},
\end{equation}
where 
\begin{multline}
	\hat  \Lambda\equiv \sum_{(m,n) \in \mu_S} g_{mn}^* \hat \sigma_{mn}^\dagger=g_{21}^* \hat \sigma_{21}^\dagger+g_{43}^* \hat \sigma_{43}^\dagger \\
	\approx \beta\frac{g}{\sqrt{2}} |-\rangle \langle gg | + \sqrt{2}g |ee\rangle \langle +|,
\end{multline}
where $g_{21}\approx \beta g/\sqrt{2}$ (assuming $\delta \ll J$), $g_{43}\equiv \sqrt{2}g$, $\hat \sigma_{21}\equiv |gg\rangle \langle - |$, and $\hat \sigma_{43}\equiv |+\rangle \langle ee|$. 
Rearranging these terms, we obtain an effective Lindblad-like term of the form,
\begin{equation}
	\mathcal{L}_{\text{cav}}^{\text{eff}} \approx \Gamma_P \mathcal{D}[\hat \xi_{S}]\hat \rho,
	\label{eq:liouvillian_cav}
\end{equation}
with $\Gamma_P \equiv 4g^2/\kappa$ and a single jump operator given by
\begin{equation}
	\hat \xi_S\equiv |+\rangle \langle ee|+ \beta |gg\rangle \langle -|.
	\label{eq:xi_S}
\end{equation}
Notice that, in the limit $\beta\rightarrow 0$ considered in the main text, we obtain $\hat\xi_S \approx |S\rangle \langle ee|$. The effective dynamics $\mathcal L^\text{eff}_\text{cav}$ thus induces an enhanced superradiant transition from the doubly-excited state to the symmetric state, $|ee\rangle \rightarrow |S\rangle$, with a rate $2\Gamma_P$. For non-neglibible $\delta$, the second term in \eqref{eq:xi_S} corresponds to a weakly enhanced  transition from the subradiant state to the ground state, $|-\rangle \rightarrow |gg\rangle$, with a rate $\beta^2 \Gamma_P$.
Given its quadratic dependence with the small parameter $\beta$, this second type of transition will clearly be negligible in the dimer configuration $\delta \ll J$, where $\beta=\tan^{-1}(\delta/J)\ll 1$.

\sectionmark{Analytical solutions of the master equation under ...}
\section{Analytical solutions of the master equation under incoherent excitation}
\sectionmark{Analytical solutions of the master equation under ...}
\label{sec:SM_AnalyticalDensityMatrix}
In the case of an incoherent drive and $N=2$, we can obtain good analytical approximations of the dynamics based on the adiabatic elimination of the cavity discussed above. A first approach is the direct diagonalization of the Bloch-Redfield equation in \eqref{eq:Nakajima}, which allows to obtain analytical solutions of the steady state of the emitters, provided $\kappa \gg g$. 
In the limit $\beta\approx0$ and setting $\Delta_a\approx -J$, the steady-state populations read (assuming $P\gtrsim \gamma$)
\begin{subequations}
	\begin{align}
		\label{eq:rho_S_general}
		\rho_{S,\text{ss}}&\approx \frac{P^2 J^2 (8g^2 + \gamma_S \kappa)}{8 g^2 P J^2(P+\gamma_S) + 4g^4 P \kappa +J^2\kappa P^3}, \\
		\rho_{A,\text{ss}}&\approx \frac{P \left[ 4g^2J^2 \gamma_S+2g^4 \kappa+J^2 \kappa \gamma_S \gamma    \right]}{8 g^2 P J^2(P+\gamma_S) + 4g^4 P \kappa +J^2\kappa P^3}, \\
		\rho_{ee,\text{ss}}&\approx \frac{P^2 J^2 \kappa (P+\gamma)}{8 g^2 P J^2(P+\gamma_S) + 4g^4 P \kappa +J^2\kappa P^3}.
	\end{align}
\end{subequations}
We can rewrite \eqref{eq:rho_S_general}, which corresponds to the relevant fidelity $\mathcal F(|S\rangle)$ used as a figure of merit in the text, in terms of the Purcell rate $\Gamma_P$. In the limit $C\gg 1$, it takes the following transparent form
\begin{equation}
	\label{eq:rho_S_general_simplified}
	\rho_{S,\text{ss}} \approx \left[1 + \frac{P}{2\Gamma_P} + \frac{\gamma_S}{P} + \frac{1}{8}\frac{\Gamma_P}{P}\left(\frac{\kappa}{J} \right)^2 \right]^{-1}.
\end{equation}
This expression clearly shows how, under the conditions established in the main text for the efficient generation of entanglement, $J \gg \kappa \gg \Gamma_P \gg P \gg \gamma$, the population saturates to its maximum possible value $\rho_{S,\text{ss}}\approx 1$. 

The expressions above are general and apply beyond the conditions of the frequency-resolved Purcell effect in \eqref{eq:conditions_freq_Purcell}. However, using the simple form of the cavity-induced Lindblad term in \eqref{eq:liouvillian_cav}, which applies in the frequency-resolved Purcell regime, allows us to obtain simpler expressions for the steady state and the dynamics. In particular, in the dimer configuration $\beta\approx 0$, the system can be reduced to a cascaded three-level system composed of $\{ |gg\rangle, |S\rangle, |ee\rangle\}$, since the subradiant state, $|A\rangle$, can be neglected as it is essentially decoupled from the system dynamics due to its dark nature~\cite{FicekQuantumInterference2005}. Consequently, the system is effectively described by a density matrix $\hat \rho_{\text{eff}}$ whose dynamics is governed by the following master equation
\begin{equation}
	\frac{d \hat \rho_{\text{eff}}}{dt}=  \left(
	\frac{\gamma_S}{2} \mathcal{D}[\hat \sigma_{gg,S}] +
	\Gamma_P \mathcal{D}[\hat \sigma_{S,ee}] +
	\frac{P}{2} \mathcal{D}[\hat \sigma_{gg,S}^\dagger] +
	\frac{P}{2} \mathcal{D}[\hat \sigma_{S,ee}^\dagger]\right) \hat \rho_{\text{eff}}.
\end{equation}
where we have defined $\hat \sigma_{gg,S}\equiv |gg\rangle \langle S|$ and $\hat \sigma_{S,ee}\equiv |S\rangle \langle ee|$.
From this equation, we can straightforwardly compute analytical expressions for the qubit-qubit dynamics, such as the symmetric state population $\rho_{S}(t)\equiv\langle S | \hat \rho_{\text{eff}} (t) |S\rangle$,
\begin{equation}
	\rho_{S}(t)\approx \rho_{S,\text{ss}} \left( 1- e^{-t/\tau^{\text{Coh}}_{S} }\right),
\end{equation}
where $\rho_{S,\text{ss}}$ is its stationary population,
\begin{equation}
	\rho_{S,\text{ss}}\approx \left[1+ \frac{P}{2 \Gamma_{P}}+\frac{\gamma_S}{P}\right]^{-1},
	\label{eq:rhoS_purcell_ap}
\end{equation}
and $\tau_S$ denotes the stabilization timescale,
\begin{multline}
	1/\tau_S^{\text{Incoh.}} \approx  P+\Gamma_P+\frac{\gamma_S}{2}-\sqrt{P\gamma_S+\frac{1}{4}(\gamma_S-2\Gamma_P)^2} \\
	\approx P+\Gamma_{P}-\sqrt{\Gamma_{P}^2+P\gamma_S}.
\end{multline}
The last approximation corresponds to the expression presented in the text under the assumption $\Gamma_P\gg \gamma_S$. We note  that \eqref{eq:rho_S_general_simplified} corresponds to the limit $J \gg \kappa$ of \eqref{eq:rhoS_purcell_ap}.

These results allow us to obtain an analytical expression for the optimal incoherent pumping rate $P_{\text{opt}}$, that maximizes the population of the symmetric state $|S\rangle$. 
To do so, we find the $P$ that maximizes the expression for $\rho_{S,\text{ss}}$ in \eqref{eq:rho_S_general}, which has a wider range of validity than  \eqref{eq:rhoS_purcell_ap}. The optimal pumping rate thus reads, assuming $\gamma_S = 2\gamma$ and the cooperativity $C\equiv \Gamma_P/\gamma$,
\begin{equation}
	P_{\text{opt}}=\frac{\Gamma_P}{2}\sqrt{\left(\frac{\kappa}{J}\right)^2+\frac{16}{C}}.
\end{equation}
Under the conditions that we established for the efficient generation of entanglement, $J\gg \kappa$ and $C\gg 1$ (equivalent to $\Gamma_P \gg \gamma$), this equation  shows that the optimum pumping consistently fulfills the condition $P\ll \Gamma_P$. By substituting $P_\text{opt}$ back in \eqref{eq:rho_S_general_simplified}, we can obtain $\rho_{S,\text{ss}}^\text{max}$, the maximum achievable population of $|S\rangle$:
\begin{equation}
	\rho_{S,\text{ss}}^\text{max} = \left[1+\frac{1}{2}\sqrt{\left(\frac{\kappa}{J}\right)^2+\frac{16}{C}} \right]^{-1}.
\end{equation}

There are two different limits to this maximum value that can be considered when approaching the optimum conditions of entanglement, namely (i) $C\gg \frac{J}{\kappa} \gg 1 $ or (ii) $\frac{J}{\kappa}\gg C \gg 1$. In the first case, we find that the maximum population for the entangled state is given by
\begin{equation}
	\label{eq:rhoS_max_J}
	\rho_{S,\text{ss}}^\text{max}(C\gg J/\kappa) = \left(1+\frac{\kappa}{2J} \right)^{-1},
\end{equation}
while in the second case, 
\begin{equation}
	\label{eq:rhoS_max_C}
	\rho_{S,\text{ss}}^\text{max}(J/\kappa \gg C) = \left(1+2/\sqrt{C} \right)^{-1}.
\end{equation}
\eqref{eq:rhoS_max_J} and  \eqref{eq:rhoS_max_C} serve as guidelines that inform one about the maximum fidelity to $|S\rangle$ attainable (and thus the maximum values of stationary entanglement) when the limiting factor are given by the inter-emitter coupling rate, $J$, or the cavity cooperativity $C$, respectively.

\section{Removing the coherent contribution}
\label{Appendix:CoherentDisp}

Let us consider a reduced model to exemplify this point: a TLS coupled to a coherently driven cavity. The resulting Hamiltonian in the rotating frame of the laser reads
\begin{equation}
	\hat H= \Delta \hat \sigma^\dagger \hat \sigma +\Delta_a \hat a^\dagger \hat a + g(\hat a^\dagger \hat \sigma +\hat a \sigma^\dagger ) + \Omega_a (\hat a+\hat a^\dagger),
\end{equation}
with $\Omega_a$ the driving strength of the laser. The system is described by master equation of the form,
\begin{equation}
	\frac{d\hat \rho}{dt}=-i[\hat H,\hat \rho]+\frac{\gamma}{2}\mathcal{D}[\hat \sigma]\hat \rho +\frac{\kappa}{2}\mathcal{D}[\hat a]\hat \rho.
	\label{eq:Master_Eq_Test} 
\end{equation}
Now, we can coherently displace the cavity in phase space a quantity $\alpha$, applying a displacement transformation~\cite{WallsQuantumOptics2008}, $\hat D(\alpha)$, such that
\begin{equation}
	\hat D^\dagger(\alpha) \hat a\hat D(\alpha)= \hat a+\alpha,
\end{equation}
where now we can identify $\hat a\sim \delta \hat a$ as the quantum fluctuations and $\alpha$ as the classical, coherent contribution of the field. Performing this transformation to the master equation in \eqref{eq:Master_Eq_Test}, we can \textit{eliminate} the coherent driving term of the cavity by fixing the coherent displacement $\alpha$ to
\begin{equation}
	\alpha=- \frac{\Omega_a}{\Delta_a+i\kappa/2}.
\end{equation}
Consequently, we obtain an analogous master equation to \eqref{eq:Master_Eq_Test}, with $\hat a$ replaced by the quantum fluctuations $\delta \hat a$, and a Hamiltonian that contains a coherent driving term for the TLS by identifying $\Omega\equiv g \alpha= -g \frac{\Omega_a}{\Delta_a+i\kappa/2}$:
\begin{equation}
	\hat H= \Delta \hat \sigma^\dagger \hat \sigma +\Delta_a \delta \hat a^\dagger \delta \hat a + g(\delta \hat a^\dagger \hat \sigma +\delta \hat a \sigma^\dagger ) + \Omega (\hat \sigma+\hat \sigma^\dagger).
\end{equation}

\section{Entanglement in large emitter ensembles - Post-selection measurements}
\sectionmark{Entanglement in large emitter ensembles ...}
\label{sec:SM_AddingNEmitters}
\begin{SCfigure}[0.75][b!]
	\includegraphics[width=0.55\textwidth]{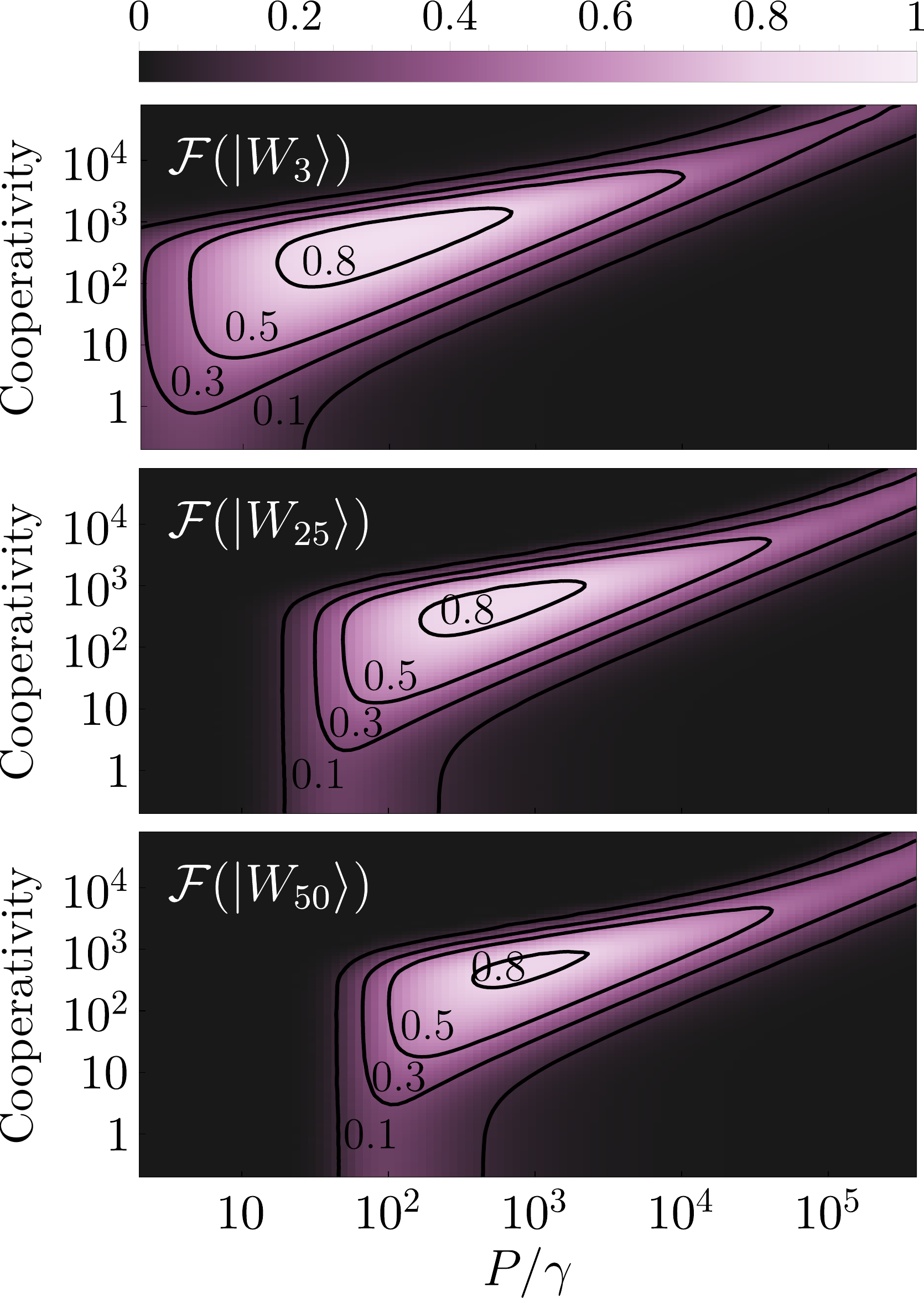}
	\captionsetup{justification=justified}
	\caption[Dependence of the fidelity $\mathcal{F}(|W_N\rangle)$ on the number of emitters $N$]{
		\label{fig:FigSM3_Nqubits}
		\textbf{Dependence of the fidelity $\mathcal{F}(|W_N\rangle)$ on the number of emitters $N$.}
		Fidelity to the $W$ state for $N=3, 25, 50$ emitters versus cavity cooperativity and incoherent pumping rate.  This illustrates how the fidelity evolves as the system size increases, providing insights into the scalability and robustness of the  $|W_N\rangle$ state in the parameter space.
		Parameters: $J=10^5 \gamma,\ \gamma_{\text{col}}=0.999\gamma,\ g=10^{-1}\kappa,\ \tilde \Delta_a=J(2-N)$.
	}
\end{SCfigure}
In this Section, we provide further numerical evidence of the prospects of the proposed mechanism to generate entangled $W$ states in large emitter ensembles $N\gg 1$. To do this, we make use of the master equation  generalized to $N$ emitters,
\begin{equation}
	\frac{d \hat \rho}{dt}=-i[\hat H, \hat \rho]+\frac{\kappa}{2}\mathcal{D}[\hat a]\hat \rho +\sum_{i,j}^N \frac{\gamma_{ij}}{2}\mathcal{D}[\hat \sigma_i, \hat \sigma_j]\hat \rho + \sum_i^N \frac{P_i}{2} \mathcal{D}[\hat \sigma_i^\dagger]\hat \rho,
\end{equation}
where the Hamiltonian is now given by
\begin{equation}
	\hat H= J \hat S^+ \hat S^- + \Delta_a \hat a^\dagger \hat a +g(\hat a^\dagger \hat S^- +\hat  a \hat  S^+),
\end{equation}
having defined $S^-\equiv \sum_i^N \hat \sigma_i$. As it is considered in the case of $N=2$ emitters, $\gamma_{ii}=\gamma$, $P_i=P$, and $\gamma_{ij}=\gamma_{ji}$ (with $i\neq j$) for all $(i,j)=1,\ldots, N$, then we denote the collective dissipative coupling simply as $\gamma_{ij} \rightarrow \gamma_{\text{col}}$.

In \reffig{fig:FigSM3_Nqubits}, the fidelity $\mathcal{F}(|W_N\rangle)$ for $N=3,25, 50$ emitters is plotted versus cooperativity and the incoherent pumping rate $P$ when the cavity frequency is placed in resonance at $\tilde \Delta_a=J(N-2)$. These calculations make use of the PIQS library in QuTiP~\cite{ShammahOpenQuantum2018,JohanssonQuTiPOpensource2012,JohanssonQuTiPPython2013}).
Our numerical simulations confirm the stabilization of $W$ states with fidelities exceeding $0.8$ for numbers of emitters as high as $N=50$, which is the case shown in the main text.
The addition of extra emitters increases the minimum pumping rate required to generate entanglement. For instance, in the case of $N=3$, we observe that $P>\gamma$, while for $N=50$, a much higher pumping intensity is needed, $P\gtrsim 10^2\gamma$. This observation is consistent with the intuition that introducing more emitters increases the pumping intensity required to bring the system into a fully excited state $|ee\ldots e\rangle$, from where the relevant cavity-enhanced decay takes place.
\begin{SCfigure}[1][b!]
	\includegraphics[width=0.92\textwidth]{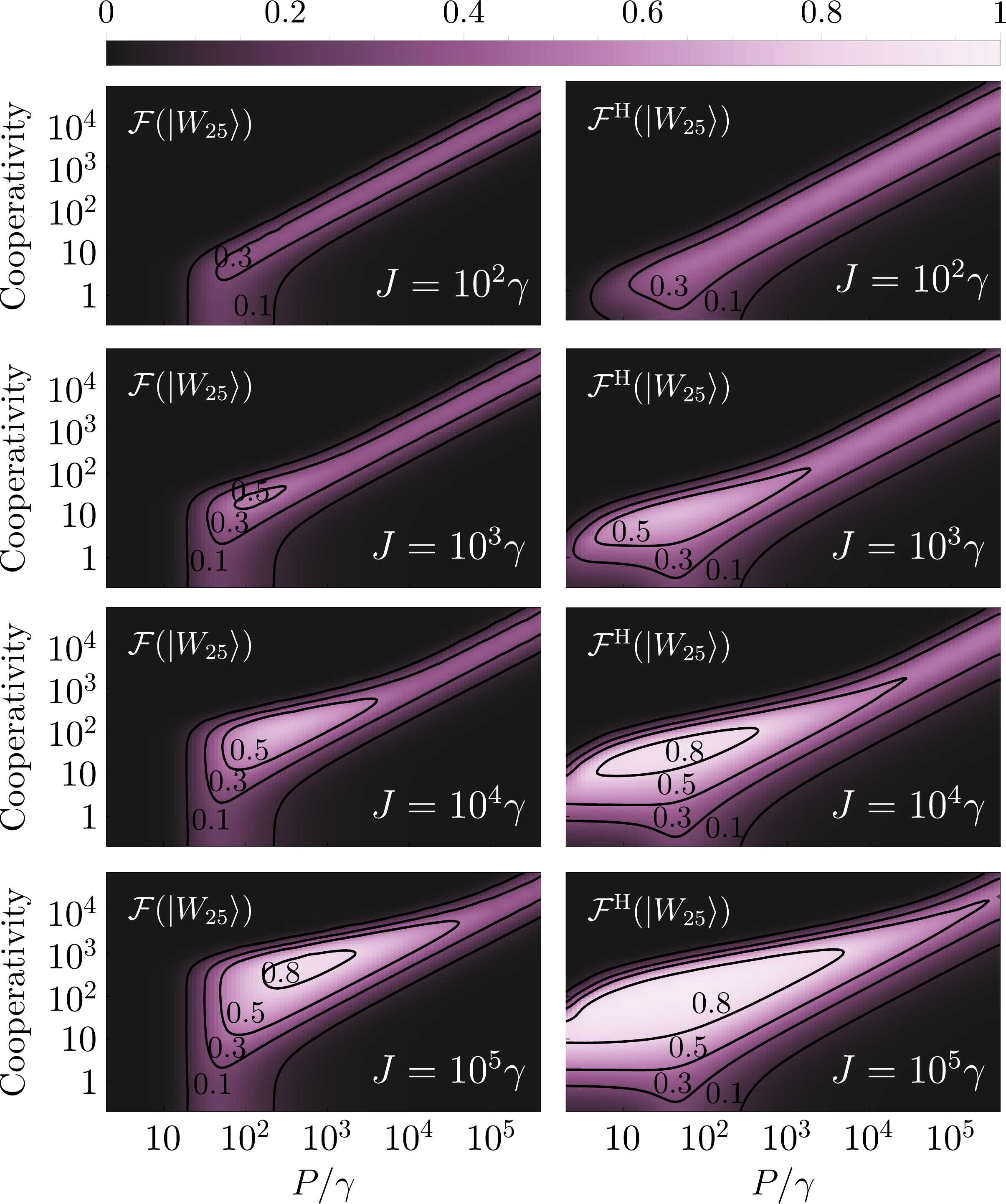}
	\captionsetup{justification=justified}
	\caption[Dependence of $\mathcal{F}(|W_N\rangle)$ on the dipole-dipole coupling $J$ and the effect of post-selection]{
		\label{fig:FigSM_3_ComparisonNormalHeralded}
		\textbf{Dependence of $\mathcal{F}(|W_N\rangle)$ on the dipole-dipole coupling $J$ and the effect of post-selection. }
		Comparison of fidelities between non- (upper row, $\mathcal{F}(|W_{25}\rangle)$) and post-selection (lower row, $\mathcal{F}^\text{H}(|W_{25}\rangle)$) measurement schemes for detecting $W$ states.
		In both rows, the fidelity to the $W$ state for $N=25$ emitters is depicted versus cavity cooperativity and incoherent pumping rate for different coherent coupling rates are considered, from left to right, $J/\gamma=10^2, 10^3,10^4,10^5$.
		Parameters: $\gamma_{\text{col}}=0.999\gamma,\ g=10^{-1}\kappa,\ \tilde \Delta_a=J(2-N)$.
	}
\end{SCfigure}
Here we explore further the prospects of enhancing entanglement via post-selection---where the entangled state is heralded  by the detection of a photon emitted by the cavity---and how this allows to relax the condition of high coherent coupling between emitters.
For this, we perform calculations with a fixed number of emitters, $N=25$, and vary the coherent coupling rate from a relatively small value of $J=10^2\gamma$ to the case presented in the text of $J=10^5\gamma$. We then compare the fidelity to $|W\rangle$ obtained in the steady-state and heralded by the detection of a photon from the cavity.
\textcolor{Maroon}{Figure}~\ref{fig:FigSM_3_ComparisonNormalHeralded} illustrates the fidelity to the $W$ state versus cooperativity and incoherent pumping rate. The left and right columns represent the non-post-selected ($\mathcal{F}(|W_{25}\rangle)$) and post-selected ($\mathcal{F}^{H}(|W_{25}\rangle)$) detection schemes, respectively.
These results highlight the significant advantage of using post-selection measurement techniques to achieve higher fidelities to the $W$ state. For instance, in the case of $J=10^4$, the non-post-selected fidelity reaches a maximum achievable value of approximately $0.5$, while in the post-selected measurement, this quantity increases to around $0.8$.
%
%

%
\section[Mechanism II: Hierarchical Adiabatic Elimination]{Mechanism II: Hierarchical Adiabatic \protect\newline Elimination}
\label{appendix:D}

Here, we show step-by-step the application of the hierarchical adiabatic elimination (HAE) technique introduced in \refch{chapter:Unconventional} in order to obtain effective, analytical time-dependent formulae for the density matrix elements of the emitter-emitter system when Mechanism II is activated, c.f. \textcolor{Maroon}{Table}~\ref{tab:EntanglementTab}. 
To proceed, we consider the following assumptions: \textcolor{Maroon}{(i)} the dipole-dipole interaction rate is almost negligible in comparison with the other system terms $J\ll \delta, \Omega$, allowing us to set $J\approx0$; \textcolor{Maroon}{(ii)} since $\Gamma_S\gg \Gamma_A$, the antisymmetric state is approximately decoupled from any dissipative channel, i.e., $\Gamma_A\approx 0$, and it is only coupled to the dynamics via the qubit-qubit detuning term $\delta$. We recall that $\Gamma_S=2\Gamma_P + \gamma+\gamma_{12}$, where $\Gamma_P$ si the Purcell rate.

Under these assumptions, the effective system is described by the following master equation
\begin{equation}
	\partial_t \hat \rho \approx -i[\hat H_{\text{eff}}, \hat \rho]	+ \frac{\Gamma_S}{2}\mathcal{D}[\tilde \sigma_{1}]\hat  \rho+ \frac{\Gamma_S}{2}\mathcal{D}[\tilde \sigma_{2} ] \hat \rho,
\end{equation}
where we defined the ladder operators $\tilde \sigma_{1} \equiv |gg\rangle \langle S|$ and $\tilde \sigma_{2} \equiv |S\rangle \langle ee|$, and the Hamiltonian
\begin{equation}
	\hat H_{\text{eff}}=\sqrt{2}\Omega ( |gg\rangle \langle S| + |S\rangle \langle ee| +  \text{H.c.}  )-\delta(|S\rangle\langle A|+\text{H.c.}).
\end{equation}

In the language of the hierarchical adiabatic elimination (HAE) method [see \refsec{sec:3}], we identify the ``real'' states $\{|gg\rangle, |S\rangle, |ee\rangle\}$, and the ``virtual'' state that mediates the interaction $|A\rangle = |V\rangle$. 

\subsection{First adiabatic elimination}
The first step of the hierarchical adiabatic elimination consists in adiabatically eliminating the ``virtual'' state, assuming the dynamics is completely contained within the ``real'' Hilbert space, $\mathcal{H}_R=\{|gg\rangle, |S\rangle, |ee\rangle\}$. In other words, in a timescale $\sim 1/\Gamma_S$ the antisymmetric state plays the role a virtual state: highly detuned from the subspace where the dynamics take place and weakly coupled to it. Therefore, the virtual state remains unpopulated and its only impact is to provide effective energy shifts to the states in the real subspace. Here, this definition translates into  $\delta \gg J$. Then, we can set $\dot \rho_{i,A}=0$ with $i=gg,S,ee$ and replace the virtual coherence terms by their steady-state values.
This results in the following set of differential equations
\begin{subequations}
	\begin{align}
		&\text{ \fontsize{7.0}{10}\selectfont   $  
			\dot \rho_{gg,S}  \approx -\frac{\Gamma_S (\delta^2+2\Omega^2)}{4\Omega^2}\rho_{gg,S}(t)+\frac{2\delta^2 }{\Gamma_S}(\rho_{S,ee}^*(t) -\rho_{gg,S}(t))  \notag 
			$}\\
		&\text{ \fontsize{7.0}{10}\selectfont   $  \qquad   +\frac{i}{\sqrt{2}\Omega}[2\Omega^2 +\delta^2(\rho_{S,S}(t)-\rho_{A,A}(t))+2\Omega^2(\rho_{gg,ee}(t)-2\rho_{S,S}(t)-\rho_{A,A}(t)-\rho_{ee,ee}(t))],
			$} \\
		&\text{ \fontsize{7.0}{10}\selectfont   $  \dot \rho_{gg,ee} \approx -\frac{1}{2}\Gamma_S \rho_{gg,ee}(t) + i\sqrt{2}\Omega (\rho_{gg,S}(t)-\rho_{S,ee}(t))
			$} \\
		&\text{ \fontsize{7.0}{10}\selectfont   $  	\dot \rho_{S,S}  \approx -\Gamma_S (\rho_{S,S}(t)-\rho_{ee,ee}(t))-\frac{\sqrt{2}}{\Omega} [(\delta^2-2\Omega^2)\text{Im}[\rho_{gg,S}(t)]+2\Omega^2\text{Im}[\rho_{S,ee}(t)]] 
			$} \\
		&\text{ \fontsize{7.0}{10}\selectfont   $  
			\dot \rho_{S,ee}  \approx \frac{1}{\Gamma_S}[2\delta^2 \rho_{gg,S}^*(t)-(\Gamma_S^2+2\delta^2)\rho_{S,ee}(t)-i\sqrt{2}\Gamma_S \Omega(\rho_{gg,ee}(t)-\rho_{S,S}(t)+\rho_{ee,ee}(t))]
			$} \\
		&\text{ \fontsize{7.0}{10}\selectfont   $  
			\dot \rho_{ee,ee}  \approx -\Gamma_S \rho_{ee,ee}(t) +2\sqrt{2}\Omega \text{Im}[\rho_{S,ee}(t)],
			$}
	\end{align}
\end{subequations}
where $\text{Im}[(*)]$ denotes imaginary part.
The differential equation that governs the dynamics of the antisymmetric state after the first adiabatic elimination takes the form
\begin{equation}
	\dot \rho_{A,A}\approx \frac{\sqrt{2} \delta^2}{\Omega} \text{Im}[\rho_{gg,S}].
	\label{eq:DiffEqAntisymHAE1}
\end{equation}

In the limit $\Omega\gg \delta$, these equations can be simplified to
\begin{subequations}
	\begin{align}
		&\text{ \fontsize{8.35}{10}\selectfont   $  	\dot \rho_{gg,S}  \approx -\frac{\Gamma_S}{2}\rho_{gg,S}(t)+i\sqrt{2}\Omega[1+\rho_{gg,ee}(t)-2\rho_{S,S}(t)-\rho_{A,A}(t)-\rho_{ee,ee}(t)], 
			$}\\
		&\text{ \fontsize{8.35}{10}\selectfont   $  
			\dot \rho_{gg,ee}  \approx -\frac{1}{2}\Gamma_S \rho_{gg,ee}(t) + i\sqrt{2}\Omega (\rho_{gg,S}(t)-\rho_{S,ee}(t))
			$} \\
		&\text{ \fontsize{8.35}{10}\selectfont   $  
			\dot \rho_{S,S}  \approx -\Gamma_S (\rho_{S,S}(t)-\rho_{ee,ee}(t))+\sqrt{2}\Omega \text{Im}
			[\rho_{gg,S}(t)-\rho_{S,ee}(t)]  $}\\
		&\text{ \fontsize{8.35}{10}\selectfont   $  
			\dot \rho_{S,ee}  \approx -\Gamma_S \rho_{S,ee}(t)-i\sqrt{2}\Omega (\rho_{gg,ee}(t)-\rho_{S,S}(t)+\rho_{ee,ee}(t)) $} \\
		&\text{ \fontsize{8.35}{10}\selectfont   $  
			\dot \rho_{ee,ee}  \approx -\Gamma_S \rho_{ee,ee}(t) +2\sqrt{2}\Omega \text{Im}[\rho_{S,ee}(t)],$}
	\end{align}
\end{subequations}
which describe a cascaded three level system driven with an Rabi frequency $\sqrt{2}\Omega$, and a decay rate $\Gamma_S$. The additional term $\rho_{A,A}$ evolves in a much slower timescale; so here it is considered as a time-independent parameter. For instance, we may set $\rho_{A,A}=0$ without loss of generality.

\subsection{Second adiabatic elimination}
In a timescale longer than $\sim 1/\Gamma_S$, the ``real'' variables $\rho_{i,j}$ with $i,j=gg,S,ee$ are considered as fast variables, since they relax to a steady-state in a very short time in comparison with the ``virtual state'' $\rho_{A,A}$. Therefore, we can perform a second adiabatic elimination.
Assuming that the antisymmetric state is barely unchanged in a timescale $\sim 1/\Gamma_S$, we can compute a steady-state solution for the real variables that will depend on the fixed value of $\rho_{A,A}$. 
In other words, we will obtain a steady-state solution that will follow any slow change of the antisymmetric state.
The resulting $\rho_{A,A}$-dependent steady-state for the real variables is obtained by solving a linear system of the form $\hat M.\vec{\rho}+\vec{b}=0$ for the vector $\vec{\rho}=\left( \rho_{gg,S}^{SS}, \rho_{gg,ee}^{SS}, \rho_{S,gg}^{SS}, \rho_{S,S}^{SS}, \rho_{S,ee}^{SS}, \rho_{ee,gg}^{SS}, \rho_{ee,S}^{SS}  ,\rho_{ee,ee}^{SS}\right)^T$, where $\hat M$ and $\vec{b}$ are given by
\begin{equation}
	M=	\text{ \fontsize{4.0}{10}\selectfont   $  
		\left(
		\begin{array}{cccccccc}
			-\frac{\Gamma_S  \left(\delta ^2+2 \Omega ^2\right)}{4 \Omega ^2}-\frac{2
				\delta ^2}{\Gamma_S } & i \sqrt{2} \Omega  & 0 & \frac{i \left(\delta ^2-4
				\Omega ^2\right)}{\sqrt{2} \Omega } & 0 & 0 & \frac{2 \delta ^2}{\Gamma_S } &
			-i \sqrt{2} \Omega  \\
			i \sqrt{2} \Omega  & -\frac{\Gamma_S }{2} & 0 & 0 & -i \sqrt{2} \Omega  & 0 & 0
			& 0 \\
			0 & 0 & -\frac{\Gamma_S  \left(\delta ^2+2 \Omega ^2\right)}{4 \Omega
				^2}-\frac{2 \delta ^2}{\Gamma_S } & -\frac{i \left(\delta ^2-4 \Omega
				^2\right)}{\sqrt{2} \Omega } & \frac{2 \delta ^2}{\Gamma_S } & -i \sqrt{2}
			\Omega  & 0 & i \sqrt{2} \Omega  \\
			\frac{i \left(\delta ^2-2 \Omega ^2\right)}{\sqrt{2} \Omega } & 0 & -\frac{i
				\left(\delta ^2-2 \Omega ^2\right)}{\sqrt{2} \Omega } & -\Gamma_S  & i
			\sqrt{2} \Omega  & 0 & -i \sqrt{2} \Omega  & \Gamma_S  \\
			0 & -i \sqrt{2} \Omega  & \frac{2 \delta ^2}{\Gamma_S } & i \sqrt{2} \Omega  &
			-\frac{2 \delta ^2}{\Gamma_S }-\Gamma_S  & 0 & 0 & -i \sqrt{2} \Omega  \\
			0 & 0 & -i \sqrt{2} \Omega  & 0 & 0 & -\frac{\Gamma_S }{2} & i \sqrt{2} \Omega 
			& 0 \\
			\frac{2 \delta ^2}{\Gamma_S } & 0 & 0 & -i \sqrt{2} \Omega  & 0 & i \sqrt{2}
			\Omega  & -\frac{2 \delta ^2}{\Gamma_S }-\Gamma  & i \sqrt{2} \Omega  \\
			0 & 0 & 0 & 0 & -i \sqrt{2} \Omega  & 0 & i \sqrt{2} \Omega  & -\Gamma_S  \\
		\end{array}
		\right),
		$}
\end{equation}
\begin{multline}
	\vec{b}=\begin{pmatrix}
		-i\frac{\delta^2+2\Omega^2}{\sqrt{2}\Omega} &
		0 &
		i\frac{\delta^2+2\Omega^2}{\sqrt{2}\Omega} &
		0&
		0&
		0&
		0&
		0
	\end{pmatrix}^T \rho_{A,A}(t)\\
	+
	\begin{pmatrix}
		i\sqrt{2}\Omega &
		0&
		-i\sqrt{2}\Omega &
		0&
		0&
		0&
		0&
		0
	\end{pmatrix}^T.
\end{multline}
Then, by solving this linear system, we obtain the analytical quasistationary $\rho_{A,A}$-dependent solution for the real variables,
\begin{subequations}
	\begin{align}
		&\text{ \fontsize{7.0}{10}\selectfont   $  
			\rho_{gg,S}^{SS}[\rho_{A,A}(t)]=\frac{4 i \sqrt{2} \Gamma_S  \Omega ^3 \left(\Gamma_S^2+2 \delta ^2+8 \Omega
				^2\right)}{\chi}-\frac{2 i
				\sqrt{2} \Gamma_S  \Omega   \left(\delta ^2+2 \Omega ^2\right)
				\left(\Gamma_S^2+2 \delta ^2+8 \Omega ^2\right)}{\chi}\rho _{A,A}(t)
			$}
		\\
		&\text{ \fontsize{7.0}{10}\selectfont   $  
			\rho_{gg,ee}^{SS}[\rho_{A,A}(t)]=-\frac{16 \Omega ^4 \left(\Gamma_S^2+6 \delta ^2\right)}{\chi}+\frac{8 \Omega ^2  \left(\Gamma_S^2+6 \delta ^2\right) \left(\delta
				^2+2 \Omega ^2\right)}{\chi}\rho _{A,A}(t),
			$}
		\\
		&\text{ \fontsize{7.0}{10}\selectfont   $  
			\rho_{S,S}^{SS}[\rho_{A,A}(t)]=-\frac{8 \Omega ^2 \left(\delta ^2-2 \Omega ^2\right)\left(\Gamma_S^2+2 \delta ^2+8 \Omega ^2\right)}{\chi}+\frac{4  \left(\delta ^4-4 \Omega ^4\right) \left(\Gamma_S^2+2
				\delta ^2+8 \Omega ^2\right)}{\chi} \rho _{A,A}(t),
			$}
		\\
		&\text{ \fontsize{7.0}{10}\selectfont   $  
			\rho_{S,ee}^{SS}[\rho_{A,A}(t)]=	-\frac{16 i
				\sqrt{2} \Gamma_S  \Omega ^3 \left(\delta ^2-2 \Omega ^2\right)}{\chi}+\frac{8 i \sqrt{2} \Gamma_S \Omega  \left(\delta ^4-4 \Omega
				^4\right)}{\chi}\rho _{A,A}(t) ,
			$}
		\\
		&\text{ \fontsize{7.0}{10}\selectfont   $  
			\rho_{ee,ee}^{SS}[\rho_{A,A}(t)]=	-\frac{64 \Omega ^4 \left(\delta
				^2-2 \Omega ^2\right)}{\chi}+\frac{32 \Omega ^2  \left(\delta ^4-4 \Omega ^4\right)}{\chi}\rho _{A,A}(t) ,
			$}
	\end{align}
\end{subequations}
with 
\begin{multline}
	\chi\equiv \Gamma_S^4 \left(\delta ^2+2 \Omega ^2\right)+\Gamma_S^2 \left(6
	\delta ^4-4 \delta ^2 \Omega ^2+64 \Omega ^4\right) \\
	+8 \left(\delta ^6-4
	\delta ^4 \Omega ^2+4 \delta ^2 \Omega ^4+48 \Omega ^6\right).
\end{multline}

Now, we can substitute the pseudostationary value of $\rho_{gg,S}$ back into \eqref{eq:DiffEqAntisymHAE1}, and obtain the effective differential equation for the antisymmetric state
\begin{multline}
	\dot \rho_{A,A}(t)=
	\frac{8 \Gamma_S  \delta ^2 \Omega ^2 \left(\Gamma_S^2+2 \delta ^2+8 \Omega
		^2\right)}{\chi} \\
	-\frac{4
		\Gamma_S  \delta ^2  \left(\delta ^2+2 \Omega ^2\right)
		\left(\Gamma_S^2+2 \delta ^2+8 \Omega ^2\right)}{\chi}\rho _{A,A}(t).
\end{multline}
Solving this equation, we obtain the analytical expression for the evolution of $\rho_{A,A}(t)$
\begin{equation}
	\rho_{A,A}(t)=\rho _{A,A}^{SS}(1-e^{-\Gamma_{\text{eff}}t}),
\end{equation}
where $\rho _{A,A}^{SS}$ is the steady-state value 
\begin{equation}
	\rho _{A,A}^{SS}=\frac{2\Omega^2}{\delta^2+2\Omega^2},
\end{equation}
and $\Gamma_{\text{eff}}$ stands for the effective relaxation rate,
\begin{equation}
	\Gamma_{\text{eff}}=\frac{4 \Gamma_S  \delta ^2 \left(\delta ^2+2 \Omega ^2\right) \left(\Gamma_S^2+2
		\delta ^2+8 \Omega ^2\right)}{\chi},
	\label{eq:RelaxRate}
\end{equation}
which serves as an analytical estimation of the Liouvillian gap. We can simplify this expression further under the assumption $\Omega\gg \delta$, which is a natural condition for this mechanism to take place.The relaxation rate then reduces to
\begin{equation}
	\Gamma_{\text{eff}}=\frac{4\Gamma_S \delta^2}{\Gamma_S^2+24\Omega^2},
	\label{eq:EffectiveRelaxRate}
\end{equation}
which is the expression shown in the main text.

\begin{SCfigure}[0.75][b!]
	\includegraphics[width=0.65\textwidth]{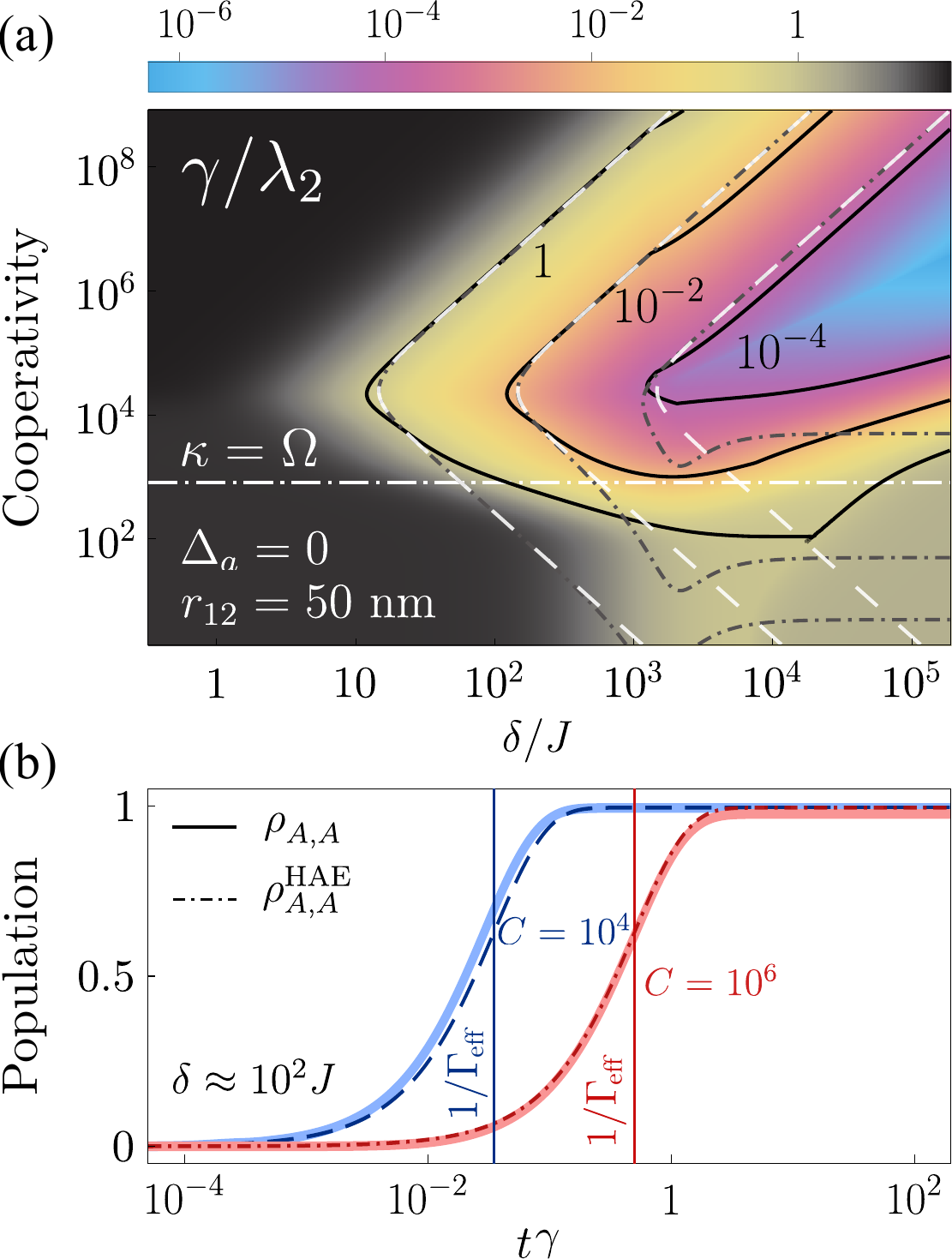}
	\captionsetup{justification=justified}
	\caption[Validity of the  effective relaxation rate $\Gamma_{\text{eff}}$ for Mechanism II in the parameter space]{
		\textbf{Validity of the  effective relaxation rate $\Gamma_{\text{eff}}$ for Mechanism II in the parameter space.}
		(a) Liouvillian gap in terms of the cooperativity $C$ and qubit-qubit detuning $\delta$. Black solid lines are exact computations, black dot-dashed lines are analytical prediction from the relaxation rate, \eqref{eq:RelaxRate}, and white dashed lines are analytical predictions from the approximated relaxation rate, \eqref{eq:EffectiveRelaxRate}. (b) Dynamics of the antisymmetric state for two different values of cooperativity: $C=10^4$ (blue) and $C=10^6$ (red). Solid lines correspond to exact computations, and blue dashed and red dot-dashed lines are analytical predictions from the hierarchical adiabatic elimination considering the full relaxation rate when $C=10^4, 10^6$, respectively. 
		Parameters: 
		(a-b) 
		$r_{12}=50\ \text{nm}$,
		$k=2\pi /780\ \text{nm}^{-1}$,
		$J=10.65\gamma$,
		$\gamma_{12}=0.967$,
		$\Delta=0$,
		$\Omega=10^4 \gamma$,
		$g=10^{-1}\kappa$;
		(b)
		blue curve $C=10^4$, $\delta=10^3\gamma$;  red $C=10^6$, $\delta=10^3\gamma$.}
	\label{fig:FigAppendixD_HAE}
\end{SCfigure}
\subsection{Validity of the relaxation rate $\Gamma_{\text{eff}}$}
Finally, we analyse the validity of the analytical relaxation rates in \eqref{eq:RelaxRate} and \eqref{eq:EffectiveRelaxRate} obtained under the application of the HAE method to estimate the Liouvillian gap.

In \reffig{fig:FigAppendixD_HAE}\textcolor{Maroon}{(a)} we show the exact computation of the Liouvillian gap in terms of the cooperativity and the qubit-qubit detuning $\delta$. Through the different contour lines included, we compare the result obtained with the full model \eqref{eq:full_master_eq} (black solid lines) with the analytical estimations of the Liouvillian gap given by the full expression in \eqref{eq:RelaxRate} (black dot-dashed lines), and the simplified expression in \eqref{eq:EffectiveRelaxRate} (white dashed lines).
Both analytical expression feature a good agreement with the numerical computation provided that $\kappa \geq \Omega$, which is precisely the condition required for Mechanism II to take place. Therefore, the deviations below this limit are expected.

In addition, we may observe the agreement between the exact (solid lines) and the analytical (dot-dashed and dashed lines) computations in \reffig{fig:FigAppendixD_HAE}~\textcolor{Maroon}{(b)}, where we depict the time evolution of the antisymmetric state for two values of the cooperativity: \textcolor{Maroon}{(i)} $C=10^4$ (blue lines), and \textcolor{Maroon}{(ii)} $C=10^6$ (red lines). We note that both the analytical expressions for the density matrix element for $\rho_{A,A}(t)$ and the relaxation rate $\Gamma_{\text{eff}}$ provide a good approximation to exact computation.  
We note that $\gamma/\lambda_2\approx 1$ approximately coincides with the onset of entanglement generation.


\section{Mechanism III$_\text{sp}$: Analytical expression for the density matrix}
\label{appendix:C}
In this appendix, we derive analytical expressions for the density matrix elements  of the dimer system when Mechanism III$_\mathrm{sp}$ is activated.
As we observed in \reffig{fig:Fig4_ConcurrenceQubitCavityDetuning}, a sizeable degree of entanglement emerges at the onset of saturation  $\Omega_{\text{2p}}\sim\gamma$. This stationary concurrence is independent of the cavity frequency, suggesting that the cavity plays no role in its formation.

Based on this hypothesis, we describe the system by a reduced three level system composed of the ground and the doubly-excited states, and an intermediate single-photon state, as it is described in \refch{ch:TwoPhotonResonance}. Under the assumptions: \textcolor{Maroon}{(i)} $\Delta \approx 0$, \textcolor{Maroon}{(ii)} $R\gg \Delta$, and \textcolor{Maroon}{(iii)} $\Omega \ll R$, the state of the system $\xi^{\text{2p}}$ in the reduced space $\{ |gg\rangle, |ee\rangle, |1\rangle \}$ is described by the master equation 
\begin{equation}
	\partial_t \hat  \xi^{\text{2p}}=-i[\hat H_{\text{2p}},\hat  \xi^{\text{2p}}]
	+\frac{2\gamma}{2}\mathcal{D}[\sigma_\alpha]\hat \xi^{\text{2p}}+\frac{\gamma}{2}\mathcal{D}[\sigma_\beta]\hat \xi^{\text{2p}},
	\label{eq:Model2p}
\end{equation}
where $\sigma_\alpha\equiv |1\rangle \langle ee|$ and $\sigma_\beta \equiv |gg\rangle \langle 1|$, and $\hat H_{\text{2p}}$ is the effective two-photon Hamiltonian 
\begin{multline}
	\hat H_{\text{2p}}=(2\Delta -\Omega_{\text{2p}})|ee\rangle \langle ee| -\Omega_{\text{2p}}|gg\rangle \langle gg| \\
	-\Omega_{\text{2p}}(|ee\rangle \langle gg|+ |gg\rangle \langle ee ),
\end{multline}
where $\Omega_{\text{2p}}\equiv 2\Omega^2/R \cos \beta$ can be understood as a two-photon Rabi frequency. Now, establishing the relations: $\rho_{ee,ee}^{(2)}\equiv \xi^\text{2p}_{ee,ee}$, $\rho_{gg,gg}^{(2)}\equiv \xi^\text{2p}_{gg,gg}$, $\rho_{gg,ee}^{(2)}\equiv \xi^\text{2p}_{gg,ee}$, and $\rho_{S,S}^{(2)}=\rho_{A,A}^{(2)}\equiv \xi^\text{2p}_{1,1}/2$, the steady-state solution of \eqref{eq:Model2p} yields,
\begin{subequations}
	\begin{align}
		\rho_{gg,gg}^{(2)}&=\frac{R^2(\gamma^2+4\Delta^2)+4\Omega^4 \cos^2 \beta}{R^2(\gamma^2+4\Delta^2)+16\Omega^4 \cos^2 \beta},  \\
		\rho_{ee,ee}^{(2)}&=\frac{4\Omega^4 \cos^2 \beta}{R^2(\gamma^2+4\Delta^2)+16\Omega^4 \cos^2 \beta}, \\
		\rho_{gg,ee}^{(2)}&=\frac{2R(-i\gamma+2\Delta)\Omega^2 \cos \beta }{R^2(\gamma^2+4\Delta^2)+16\Omega^4 \cos^2 \beta}, \\
		\rho_{S,S}^{(2)}&=\rho_{A,A}^{(2)}=\frac{4\Omega^4 \cos^2 \beta}{R^2(\gamma^2+4\Delta^2)+16\Omega^4 \cos^2 \beta} \, .
	\end{align}
\end{subequations}
We note that $\rho_{i,j}^{(2)}$ denotes density matrix elements resulting from two-photon processes. 

In order to compute an analytical solution for the concurrence, we need to write this reduced density matrix in the full Hilbert space of the emitters. We do so by considering the following ansatz density matrix
\begin{equation}
	\hat 	\rho \approx 
	\begin{pmatrix}
		\rho_{gg,gg}^{(2)} & 0 & 0 & \rho_{gg,ee}^{(2)} \\
		0								& 	\rho_{S,S}^{(2)} & 0 &0 \\
		0&0&	\rho_{A,A}^{(2)}&0 \\
		\rho_{ee,gg}^{(2)} &0 &0& \rho_{ee,ee}^{(2)}
	\end{pmatrix}.
\end{equation}
In order to compute the concurrence, we calculate the eigenvalues $ \lambda_i$ in \eqref{eq:eigenvalues-concurrence}. In decreasing order, these are given by
\begin{subequations}
	\begin{align}
		\lambda_1&=\sqrt{	\rho_{gg,gg}^{(2)} 	\rho_{ee,ee}^{(2)}}+\text{Abs}[	\rho_{ee,gg}^{(2)}], \\
		\lambda_2&=\left(\rho_{S,S}^{(2)} \right)^2, \\
		\lambda_3&= \left(\rho_{A,A}^{(2)} \right)^2, \\
		\lambda_4&=\sqrt{	\rho_{gg,gg}^{(2)} 	\rho_{ee,ee}^{(2)}}-\text{Abs}[	\rho_{ee,gg}^{(2)}], 
	\end{align}
\end{subequations}
where $\text{Abs}[(*)]$ denotes the absolute value operation.
Since the region of applicability is located at $\delta > J$, we can approximate $\beta\approx \pi/2$ and obtain a simple expression for the concurrence, given by \eqref{eq:Concurrence}. The result is
\begin{equation}
	\mathcal{C}\approx \text{Max} \left[ 0,\frac{4J\Omega^2 (\gamma \delta^2 -2J\Omega^2)}{\gamma^2\delta^4+16J^2 \Omega^4} \right].
\end{equation}
This expression matches well our numerical results presented in the main text.  In addition, from this expression we can obtain the analytical value of the qubit-qubit detuning that maximizes the concurrence, 
\begin{equation}
	\delta_{\text{max}}=\Omega\sqrt{2(1+\sqrt{5})J/\gamma}.
\end{equation}

Finally, we remark that, in the regime of weakly-coupled emitters ($J\ll \delta$), the eigenstates are no longer entangled states but bare qubit states,  $|-\rangle_{\beta\approx \pi/2} \approx |eg\rangle$ and $|+\rangle_{\beta\approx \pi/2} \approx |ge\rangle$. Therefore, excitation at the one-photon resonances will not generate any type of entanglement via one-photon excitation. The main mechanism of entanglement generation in that regime is Mechanism II, provided the corresponding conditions (c.f. \textcolor{Maroon}{Table}~\ref{tab:EntanglementTab}) are fulfilled.


\cleardoublepage
\pdfbookmark[1]{Publications}{Publications}
\chapter*{Publications}
\addcontentsline{toc}{chapter}{Publications}  
\newcolumntype{L}{>{\raggedleft}p{0.1\textwidth}}
\newcolumntype{R}{p{0.8\textwidth}}
\newcommand\VRule{\color{lightgray}\vrule width 0.5pt}

\newcolumntype{L}{>{\raggedleft}p{0.0\textwidth}}
\newcolumntype{R}{p{0.9\textwidth}}
\renewcommand\VRule{\color{lightgray}\vrule width 0.pt}

Works published during the realization of this Thesis (2020-2025). \\

\textsc{Peer-reviewed journals}\\[10pt]
\begin{tabular}{L!{\VRule}R}
	
	1.&\emph{ Two-photon resonance fluorescence of two interacting non-identical quantum emitters.} 
	
	\underline{A. Vivas-Viaña}, and C. Sánchez Muñoz.\\
	
	&\href{https://journals.aps.org/prresearch/abstract/10.1103/PhysRevResearch.3.033136}{\textbf{Phys. Rev. Research 3, 033136 (2021)}}  \\ 
	
	2.&\emph{Unconventional mechanism of virtual-state population through dissipation.}
	
	\underline{A. Vivas-Viaña}, A. González-Tudela,  and C. Sánchez Muñoz.\\
	
	&\href{https://journals.aps.org/pra/abstract/10.1103/PhysRevA.106.012217}{\textbf{Phys. Rev. A 106, 012217 (2022)}}\\[0pt]

	3.&\emph{Resolving nonclassical magnon composition of a magnetic ground state via a qubit.}
	
	ALE. Römling, \underline{A. Vivas-Viaña}, C. Sánchez Muñoz, and A. Kamra.\\
	
	&\href{https://journals.aps.org/prl/abstract/10.1103/PhysRevLett.131.143602}{\textbf{Phys. Rev. Lett. 131, 143602 (2023)}}\\ [5pt]
	
	4.&\emph{Dissipative stabilization of maximal entanglement between nonidentical emitters via two-photon excitation.} 
	
	\underline{A. Vivas-Viaña}, D. Martín-Cano,  and C. Sánchez Muñoz.\\
	
	&\href{https://doi.org/10.1103/PhysRevResearch.6.043051}{\textbf{Phys. Rev. Research 6, 043051 (2024)}}\\[5pt]
	
	5.&\emph{Frequency-Resolved Purcell Effect for the Dissipative Generation of Steady-State Entanglement.}
	
	\underline{A. Vivas-Viaña}, D. Martín-Cano,  and C. Sánchez Muñoz.\\
	
	&\href{https://doi.org/10.1103/PhysRevLett.133.173601}{\textbf{Phys. Rev. Lett. 133, 173601 (2024)}}\\[5pt]
	
	6.&\emph{Entanglement of photonic modes from a continuously driven two-level system. }

	J. Yang, I. Strandberg, \underline{A. Vivas-Viaña}, A. Gaikwad, C. Castillo-Moreno, A. Frisk Kockum, M. Asad Ullah, C. Sánchez Muñoz, A. Martin Eriksson, S. Gasparinetti.

	\href{https://doi.org/10.1038/s41534-025-00995-1}{\textbf{npj Quantum Inf 11, 69 (2025)} }\\[5pt]

	  \end{tabular}\\[5pt]

\textsc{Preprints}\\[10pt]
\begin{tabular}{L!{\VRule}R}
	7.&\emph{Quantum metrology through spectral measurements in quantum optics.}
	
		\underline{A. Vivas-Viaña}, and C. Sánchez Muñoz.\\
		
		&\href{https://arxiv.org/abs/2509.04300}{\textbf{arXiv:2509.04300} }
\end{tabular}

\cleardoublepage

\renewcommand{\bibname}{Bibliography}
\def\bibpreamble{}

\cleardoublepage
\defbibheading{bibintoc}[\bibname]{%
  \phantomsection
  \manualmark
  \markboth{\spacedlowsmallcaps{#1}}{\spacedlowsmallcaps{#1}}%
  \addtocontents{toc}{\protect\vspace{\beforebibskip}}%
  \addcontentsline{toc}{chapter}{#1}%
  \chapter*{#1}%
}

\printbibliography[heading=bibintoc]



\end{document}